\documentclass{ws-ijmpe}
\usepackage[super,compress]{cite}
\usepackage[usenames,dvipsnames]{color}
\usepackage{nicefrac}
\usepackage{float,ulem}
\usepackage{epsfig,epstopdf,amsmath,amssymb}

\usepackage{xcolor}
\usepackage{subfig}
\newcommand{\be}{\begin{equation}}
	\newcommand{\ee}{\end{equation}}
\newcommand{\ba}{\begin{eqnarray}}
	\newcommand{\ea}{\end{eqnarray}}
\newcommand{\nn}{\nonumber\\}

\newcommand{\bas}{\begin{eqnarray*}}
	\newcommand{\eas}{\end{eqnarray*}}

\usepackage{scalerel}
\newcommand{\paral}{\stretchrel*{\parallel}{\perp}}

\usepackage{soul}
\usepackage{hyperref}
\usepackage{wrapfig}

\newcommand{\snn}{$\sqrt{s_{NN}}~$}
\newcommand{\pt}{$p_T~$}

\usepackage{slashed}
\usepackage{comment}
\usepackage{float}
\usepackage{graphicx}
\begin{document}
\setcounter{tocdepth}{1} 

\markboth{S. K. Das et al.}{Dynamics of Hot QCD Matter - Current status and developments}

\catchline{}{}{}{}{}

 \title{Dynamics of Hot QCD Matter -- Current status and developments}

\author{
Santosh K. Das$^1$\footnote{santosh@iitgoa.ac.in}~,
	Prabhakar Palni$^2$\footnote{prabhakar.palni@unigoa.ac.in}~,
	Jhuma Sannigrahi$^1$,
	Jan-e Alam$^3$,
	Cho Win Aung$^4$, 
	Yoshini Bailung$^5$,
	Debjani Banerjee$^6$,
    Gergely G\'abor Barnaf\"oldi$^7$, 
	Subash Chandra Behera$^8$, 
	Partha Pratim Bhaduri$^3$,
	Samapan Bhadury$^9$, 
	Rajesh Biswas$^9$, 
	Pritam Chakraborty$^{10}$,
	Vinod Chandra$^{11}$,
	Prottoy Das$^6$,
	Sadhana Dash$^{10}$,
	Saumen Datta$^{12}$,
	Sudipan De$^{13}$,
	Vaishnavi Desai$^2$,
	Suman Deb$^5$, 
	Debarshi Dey$^{14}$,
	Jayanta Dey$^5$, 
	Sabyasachi Ghosh$^4$,
	Najmul Haque$^9$,
	Mujeeb Hasan$^{14}$,
	Amaresh Jaiswal$^9$, 
	Sunil Jaiswal$^{12}$,
	Chitrasen Jena$^{15}$,
	Gowthama K K$^{11}$,
	Salman Ahamad Khan$^{14}$,
	Lokesh Kumar$^{16}$,
	Sumit Kumar Kundu$^5$, 
	Manu Kurian$^{11}$,
	Neelkamal Mallick$^5$, 
	Aditya Nath Mishra$^{7}$, 
	Sukanya Mitra$^{12}$,
	Lakshmi J. Naik$^{17}$,
	Sonali Padhan$^{10}$
	Ankit Kumar Panda$^9$,
	Pushpa Panday$^{14}$,
	Suvarna Patil$^1$,
	Binoy Krishna Patra$^{14}$,
	Pooja$^1$,
	Raghunath Pradhan$^8$, 
	Girija Sankar Pradhan$^5$,
	Jai Prakash$^1$,
	Suraj Prasad$^5$,
	Prabhat R. Pujahari$^8$, 
	Shubhalaxmi Rath$^{10}$,
	Sudhir Pandurang Rode$^{18}$,
	Ankhi Roy$^5$,
	Victor Roy$^9$,
	Marco Ruggieri$^{19}$,
	Rohan V S$^{20}$,
	Raghunath Sahoo$^5$, 
	Nihar Ranjan Sahoo$^{21}$,
	Dushmanta Sahu$^5$, 
	Nachiketa Sarkar$^{9}$,
	Sreemoyee Sarkar$^{22}$,
	Sarthak Satapathy$^{13}$,
	Captain R. Singh$^5$, 
	V. Sreekanth$^{17}$, K. Sreelakshmi$^{17}$,
	Sumit$^{14}$,
	Dhananjaya Thakur$^{23}$,
	Sushanta Tripathy$^{24}$,
	Thandar Zaw Win$^4$, 
	(authors)\footnote{
		The contributors on this author list have contributed only to those sections of the report, which they cosign with their name. Only those have collaborated together, whose names appear together in the header of a given section.}}

  \address{
 	$^{1}$ School of Physical Sciences, Indian Institute of Technology Goa, Ponda-403401, Goa, India\\
$^{2}$ School of Physical and Applied Sciences, Goa University, Goa, India \\
$^{3}$ Variable Energy Cyclotron Centre, HBNI, 1/AF Bidhan Nagar, Kolkata 700064, India\\
$^{4}$ Indian Institute of Technology Bhilai, GEC Campus, Sejbahar, Raipur 492015, Chhattisgarh, India\\ 
$^{5}$ Department of Physics, Indian Institute of Technology Indore, Simrol, Indore 453552, India\\ 
$^{6}$ Department of Physics, Bose Institute, Kolkata - 700091, WB, India\\ 
$^{7}$ Wigner Research Center for Physics, 29-33 Konkoly-Thege Miklós Str., H-1121 Budapest, Hungary\\ 
$^{8}$ Indian Institute Of Technology, Madras, Chennai, Tamilnadu 600036, India\\ 
$^{9}$ School of Physical Sciences, National Institute of Science Education and Research, An OCC of Homi Bhabha National Institute, Jatni-752050, India\\
$^{10}$ Department of Physics, Indian Institute of Technology Bombay, Mumbai, Maharashtra, Pin-400076, India\\       
$^{11}$ Indian Institute of Technology Gandhinagar, Gandhinagar-382355, Gujarat, India\\
$^{12}$ Tata Institute of Fundamental Research, Homi Bhabha Road, Mumbai 400005, India\\
$^{13}$ Department of Physics, Dinabandhu Mahavidyalaya, Bongaon, West Bengal State University, P.O. - Bongaon, North 24 Parganas, West Bengal, PIN - 743235, India\\
$^{14}$ Department of Physics, Indian Institute of Technology Roorkee, Roorkee, Uttarakhand-247667, India\\ 
$^{15}$ Indian Institute of Science Education and Research, Tirupati, Rami Reddy Nagar, Mangalam, Tirupati, Andhra Pradesh, PIN - 517507, India\\
$^{16}$ Department of Physics, Panjab University, Chandigarh 160014, India\\
$^{17}$ Department of Sciences, Amrita School of Physical Sciences, Coimbatore Amrita Vishwa Vidyapeetham, India\\
$^{18}$ Veksler and Baldin Laboratory of High Energy Physics, Joint Institute for Nuclear Research, Dubna 141980, Moscow region, Russian Federation\\
$^{19}$ School of Nuclear Science and Technology, Lanzhou University 222 South Tianshui Road, Lanzhou 730000, China\\
$^{20}$ Amrita School of Physical Sciences, Amrita Vishwa Vidyapeetham, Amritapuri Kollam, Kerala 690525, India\\
$^{21}$ Institute of Frontier and Interdisciplinary Science, Shandong University, Qingdao, Shandong, 266237, China and 
Key Laboratory of Particle Physics and Particle Irradiation, Shandong University, Qingdao, Shandong, 266237, China\\
$^{22}$ Mukesh Patel School of Technology Management and Engineering, NMIMS University, Mumbai-56, India\\
$^{23}$ Institute of Modern Physics, Chinese Academy of Sciences, Lanzhou, Gansu, 730000, China \\
$^{24}$ INFN - sezione di Bologna, via Irnerio 46, 40126 Bologna BO, Italy
 }%
\maketitle

\begin{abstract}
The discovery and characterization of hot and dense QCD matter, known as Quark-Gluon Plasma (QGP), remains the most international collaborative effort and synergy between theorists and experimentalists in modern nuclear physics to date. The experimentalists around the world not only collect an unprecedented amount of data in heavy-ion collisions, at Relativistic Heavy Ion Collider (RHIC), at Brookhaven National Laboratory (BNL) in New York, USA, and the Large Hadron Collider (LHC), at CERN in Geneva, Switzerland but also analyze these data to unravel the mystery of this new phase of matter that filled a few microseconds old universe, just after the Big Bang. In the meantime, advancements in theoretical works and computing capability extend our wisdom about the hot-dense QCD matter and its dynamics through mathematical equations. The exchange of ideas between experimentalists and theoreticians is crucial for the progress of our knowledge. 
The motivation of this first conference named ``HOT QCD Matter 2022" is to bring the community together to have a discourse on this topic. In this article, there are 36 sections discussing various topics in the field of relativistic heavy-ion collisions and related phenomena that cover a snapshot of the current experimental observations and theoretical progress. This article begins with the theoretical overview of relativistic spin-hydrodynamics in the presence of the external magnetic field, followed by the Lattice QCD results on heavy quarks in QGP, and finally, it ends with an overview of experiment results.

\end{abstract}


\ccode{PACS numbers:12.38.-t, 12.38.Aw}

\tableofcontents 

\section{Relativistic Dissipative Spin-hydrodynamics and spin-magnetohydrodynamics}
	\author{Samapan Bhadury and Amaresh Jaiswal}	

\bigskip

\begin{abstract}
We write semi-classical kinetic equations for a relativistic fluid of spin-\nicefrac{1}{2} particles within the relaxation time approximation, with and without magnetic field. Building on these, we then go on to formulate the theory of relativistic hydrodynamics in both cases. Consequently, we obtain the theories of relativistic dissipative spin-hydrodynamics and spin-magnetohydrodynamics. While in the former case, we find, for the first time, the dissipation mechanism of spin degrees of freedom, in the latter case, we note effects analogous to Einstein-de Haas and Barnett effects at the dissipation level.
\end{abstract}

\subsection{Introduction}

In the ultra-relativistic non-central heavy-ion collisions at the collider facilities such as Relativistic Heavy-Ion Collider (RHIC) and Large Hadron Collider (LHC), a large angular momentum \cite{Becattini:2007sr} as well as a large magnetic field \cite{Tuchin:2013ie} are generated at the early stages of the evolution, which can couple with the intrinsic spin of the constituent particle via the processes similar to Einstein-de Haas and Barnett effects. It was predicted that such couplings lead to the spin polarization of the medium \cite{Liang:2004ph, Liang:2004xn}, which can be observed in particles emitted during freeze-out. This was latter confirmed experimentally \cite{STAR:2017ckg, STAR:2018gyt, STAR:2019erd, ALICE:2019aid, STAR:2020xbm, STAR:2021beb} giving a significant boost to the study of spin-polarization.

The hydrodynamic models assuming equilibrated spin degrees of freedom successfully explained these observations of global spin-polarization but failed to show similar success in the case of longitudinal spin-polarization \cite{Li:2017dan, ALICE:2019aid}. This indicated different possible origins of spin-polarization, which then led to the belief that the spin degrees of freedom in the transverse plane may not achieve equilibration at the time of freeze-out \cite{Bhadury:2020cop}, and consequently, we may require dissipative theories of hydrodynamics.

Hence, we formulated a theory of relativistic dissipative spin-hydrodynamics first without the influence of an external field and then in the presence of a magnetic field from the relativistic kinetic theory that is consistent with macroscopic conservation laws. In the latter case, we found it may be possible to observe effects similar to Einstein-de Haas and Barnett effects if the fluid is magnetizable in addition to being spin-polarizable.



\subsection{Relativistic Spin-hydrodynamics\label{RSH}}

The theory of hydrodynamics is built on the basis of conservation laws of the fluid under consideration. Such conservation laws lead to the equations of motion that govern the evolution dynamics of that specific fluid. Noting that the origin of spin-polarization is traced back to the rotation of the fluid, the relevant conservation laws for a spin-polarizable fluid are - (i) Particle/Baryon/Charge four-current, (ii) Stress-energy tensor, (iii) Angular momentum tensor. In this section, we will first provide the equations of motion of relativistic spin-polarized fluid of a single species, without a magnetic field and then with a magnetic field.

\subsection{Without magnetic field\label{RSH_B=0}}

The particle four-current of a dissipative relativistic fluid is given by,
\begin{align}
    N^\mu = n_0\, u^\mu + n^\mu, \label{RSH1:N^m_decomp}
\end{align}
where $n_0$ is the equilibrium particle number density, $n^\mu$ is particle diffusion current and $u^\mu$ is the fluid four-velocity. In writing Eq.~\eqref{RSH1:N^m_decomp} we have assumed the out-of-equilibrium number density is, $n = n_0$. This is one of the Landau matching conditions. Then the conservation law is,
\begin{align}
    \partial_\mu N^\mu = 0. \label{RSH1:del.N}
\end{align}
The next conservation law is for the stress-energy tensor of the fluid, which is given by,
\begin{align}
    T_{\rm f}^{\mu\nu} &= \epsilon_0 u^\mu u^\nu - \left( P + \Pi \right) \Delta^{\mu\nu} + \pi^{\mu\nu}, \label{RSH1:T^mn_decomp}
\end{align}
where $\epsilon_0$ is the equilibrium energy density, $P$ is the equilibrium pressure, $\Pi$ is the bulk viscous pressure, and $\pi^{\mu\nu}$ is the shear viscous pressure. As before, we have assumed in writing Eq.~\eqref{RSH1:T^mn_decomp} that the out-of equilibrium energy density is $\epsilon = \epsilon_0$, i.e., the other matching condition. Additionally, we also assumed the Landau frame definition for the fluid four-velocity, which satisfied the relation, $T^{\mu\nu} u_{\nu} = \epsilon u^{\mu}$. The associated conservation law is given by,
\begin{align}
    \partial_\mu T_{\rm f}^{\mu\nu} = 0. \label{RSH1:del.T}
\end{align}
The final relevant conservation law is the one for angular momentum tensor. This becomes important for rotating fluids and is given by,
\begin{align}
    \partial_\lambda J_{\rm f}^{\lambda,\mu\nu} = 0, \label{RSH1:del.J}
\end{align}
where $J_{\rm f}^{\lambda,\mu\nu}$ is the total angular momentum tensor of the fluid that consists of two parts, an orbital part ($L_{\rm f}^{\lambda,\mu\nu}$) and a spin part ($S^{\lambda,\mu\nu}$) i.e. $J_{\rm f}^{\lambda,\mu\nu} = L_{\rm f}^{\lambda,\mu\nu} + S^{\lambda,\mu\nu}$. As the orbital part is defined as the moment of the stress-energy tensor (i.e. $L_{\rm f}^{\lambda,\mu\nu} = x^\mu T_{\rm f}^{\lambda\nu} - x^\nu T_{\rm f}^{\lambda\mu}$), it does not carry any new information that was not already present in the stress-energy tensor. Therefore, the good conservation law to describe system is the one for spin tensor and for a symmetric tress-energy tensor, we have,
\begin{align}
    \partial_\lambda S^{\lambda,\mu\nu} = 0. \label{RSH1:del.S}
\end{align}
Thus, Eqs.~\eqref{RSH1:del.N}, \eqref{RSH1:del.T} and \eqref{RSH1:del.S} are the conservation laws satisfied by a relativistic spin-polarizable fluid in absence of any external field.

\subsection{ With magnetic field\label{RSH_B_neq_0}}

The next step is to consider the effect of the magnetic field. The conservation law for the particle four-current remains the same, and since we consider only a single species, the charge four-current of the fluid is simply $J^\alpha_{\rm f} = \mathfrak{q} N^\mu$ ($\mathfrak{q}$ being the particle's electric charge). Since it is the charge currents that generate the magnetic field, it is important to note that other charge currents may exist ( say, $J^\mu_{\rm ext}$). In fact, in the case of heavy-ion collisions, the charged current that produces the majority of the magnetic field is that of the spectator particles. The field strength tensor due to the total current, $J^\mu$ ($= J^{\mu}_{\rm f} + J^{\mu}_{\rm ext}$) is $F^{\mu\nu}$. While the conservation law for the particle four-current still holds true, the laws for energy and momentum (both linear and angular) for the fluid are no longer satisfied due to interaction with the produced field. Consequently, the fluid stress-energy tensor ($T^{\mu\nu}_{\rm f}$) and angular momentum tensor ($J^{\lambda,\mu\nu}_{\rm f}$) do not remain conserved. One should note that the conservation laws for the whole system are still satisfied \cite{Bhadury:2022qxd}. Hence for a spin-polarizable magnetizable fluid, we have,
\begin{align}
    \partial_\nu T^{\mu\nu}_{\rm f} &= F^{\mu}_{~\, \alpha} J^{\alpha}_{\rm f} + \frac{1}{2} \left( \partial^\mu F^{\alpha\beta} \right) M_{\alpha\beta}, \label{RSH2:del.T}\\
    \partial_\lambda J^{\lambda,\mu\nu}_{\rm f} &= -\tau^{\mu\nu}, \label{RSH2:del.J}
\end{align}
where $M^{\alpha\beta}$ is the magnetization tensor and $\tau^{\mu\nu}$ is the torque generated by the force term appearing on the R.H.S. of Eq.~\eqref{RSH2:del.T}. Owing to the definition of the orbital angular momentum, we can show the spin tensor is still conserved, and we get back Eq.~\eqref{RSH1:del.S}. Hence, the only relevant modification in conservation laws for the formulation of spin-magnetohydrodynamics as compared to field-free case is in Eq.~\eqref{RSH2:del.T}. Therefore, the equations of motion of relativistic spin-magnetohydrodynamics are give by Eqs.~\eqref{RSH1:del.N}, \eqref{RSH1:del.S} and \eqref{RSH2:del.T}. Next, we describe the kinetic theory formulation of spin hydrodynamics, with and without magnetic field.

\subsection{Relativistic Kinetic Theory\label{RKT}}

Spin, being a purely quantum mechanical effect, must be introduced in the kinetic theory via quantum field theory. Hence, our starting point is the Wigner function for spin-$\nicefrac{1}{2}$ particles and its Clifford-algebra decomposition \cite{Elze:1986hq, Elze:1986qd, Vasak:1987um, Weickgenannt:2019dks, Weickgenannt:2020aaf, Sheng:2021kfc},
\begin{align}
    \mathcal{W}(x,k) &= \frac{1}{4} \left[ \mathcal{F}(x,k) \!+\! i \gamma_5 \mathcal{P}(x,k) \!+\! \gamma^\mu {\cal V}_{\mu}(x,k) \!+\! \gamma_5 \gamma^\mu {\cal A}_{\mu}(x,k) \!+\! \Sigma^{\mu\nu} {\cal S}_{\mu \nu}(x,k) \right]\!, \label{RSH:wig_decomp}
\end{align}
where $x$ is the space-time coordinate, $k^\mu = (k^0, \textbf{k})$ denotes the particle momentum and, $\gamma^\mu$ are the Dirac gamma matrices with $\gamma^5 = i \gamma^0 \gamma^1 \gamma^2 \gamma^3$. It is possible to show that not all the coefficients of the right-hand side of Eq.~\eqref{RSH:wig_decomp} are totally independent. For the formulation of spin-hydrodynamics, it suffices to work with only two of the components, which in our case are, the scalar component ($\mathcal{F}$) and the axial component ($\mathcal{A}$). Their kinetic equations can be obtained from the Dirac equation to be given by,
\begin{align}
    k^\mu \partial_\mu {\cal F}(x,k) = C_{\cal F},
    \qquad
    k^\mu \partial_\mu \, {\cal A}^\nu(x,k) = C^\nu_{\cal A}, 
    \qquad
    k_\nu \,{\cal A}^\nu(x,k) = k_\nu C^\nu_{\cal A} = 0, \label{RSH:kineqAC1}
\end{align}
where $\mathcal{C}_{\mathcal{F}}$ and $\mathcal{C}_{\mathcal{A}}^\mu$ are the collision kernels. In the absence of an external field, following the approach as described in Ref.~\cite{Bhadury:2020puc}, we can obtain a Boltzmann-like equation of scalar phase-space distribution functions, $f^\pm(x,p,s)$ with an extended phase-space for particles and anti-particles under the relaxation time approximation as,
\begin{align}
    p^\mu \partial_\mu f^\pm(x,\textbf{p},s) = \left(u \cdot p\right) \frac{f^\pm_{\rm eq}(x,\textbf{p},s)-f^\pm(x,\textbf{p},s)}{\tau_{\rm eq}}, \label{RSH:Beq1}
\end{align}
where $\tau_{\rm eq}$ is the relaxation time. We can use this distribution function, $f(x,\textbf{p},s)$ to express the scalar and axial components of the Wigner function. The zeroth and first moment of Eq.~\eqref{RSH:Beq1} will result in the conservation laws of Eqs.~\eqref{RSH1:del.N} and \eqref{RSH1:del.T} and the spin moment will lead to Eq.~\eqref{RSH1:del.S} provided we use appropriate frame and matching conditions. One can easily check this by noting the definitions of the conserved currents from kinetic theory i.e.,
\begin{align}
    N^{\mu} &= \int dP dS p^\mu \left( f^+ - f^- \right), \label{RSH:N^m_defKT} \\
    T^{\mu\nu}_{\rm f} &= \int dP dS p^\mu p^\nu \left( f^+ + f^- \right), \label{RSH:T^mn_defKT} \\
    S^{\lambda,\mu\nu} &= \int dP dS s^{\mu\nu} \left( f^+ + f^- \right), \label{RSH:S^lmn_defKT}
\end{align}

where $dP \equiv  d^3p/[p^0 (2 \pi )^3]$ and $dS \equiv m/(\pi \mathfrak{s}) \,  d^4s \, \delta(s \cdot s + \mathfrak{s}^2) \, \delta(p \cdot s)$ are the momentum and spin integral measures respectively with $p^0$ being the zeroth component of the momentum four-vector and $\mathfrak{s}$ being the eigenvalue of the Casimir operator. In the presence of a magnetic field, however, we obtain a modified Boltzmann equation under RTA as,
\begin{align}
    \left( p^\alpha \dfrac{\partial }{\partial x^\alpha} + m\,\mathcal{F}^\alpha \dfrac{\partial }{\partial p^\alpha} 
    \right) f^\pm &= - \left(u\cdot p\right) \frac{\left(f^\pm - f^\pm_{\rm eq}\right)}{\tau_{\rm eq}}\,,\label{RSH:Beq2}
\end{align}
where the force term is given by,
\begin{align}
    \mathcal{F}^\alpha &= \frac{d p^\alpha}{d \tau} = \frac{\mathfrak{q}}{m}\,  F^{\alpha\beta} p_\beta + \frac{1}{2} \left( \partial^\alpha F^{\beta\gamma} \right) m_{\beta\gamma}, \label{RSH:Beq_Force}
\end{align}
with $m^{\alpha\beta}$ being the magnetic dipole moment that is related to the internal angular momentum, $s^{\alpha\beta}$ as, $m^{\alpha\beta} = \chi s^{\alpha\beta}$. The quantity, $\chi$ is analogous to the gyromagnetic ratio. We can define the magnetization tensor, $M^{\alpha\beta}$ using this $m^{\alpha\beta}$ as,
\begin{align}
    M^{\alpha\beta} = m \int dP dS m^{\alpha\beta} \left( f^+ + f^- \right), \label{RSH:M_def}
\end{align}
where $m$ is the mass of the particle. It should be noted that in Eq.~\eqref{RSH:Beq2} we have left out one term that we call the `pure torque' term (associated to $d\mathcal{S}^{\alpha\beta}/d\tau$). A derivation of spin-magnetohydrodynamics with this term will be provided in some other publications. It is straightforward to recover Eq.~\eqref{RSH:Beq1} from Eq.~\eqref{RSH:Beq2} by setting $F^{\alpha\beta} \to 0$. We can arrive at the macroscopic conservation laws for spin-magnetohydrodynamics by starting from Eq.~\eqref{RSH:Beq2} and taking the zeroth, first and spin moments i.e. Eqs.~\eqref{RSH1:del.N}, \eqref{RSH2:del.T} and \eqref{RSH1:del.S}.

\subsection{Transport Coefficients\label{TC}}

We can now solve the Eqs.~\eqref{RSH:Beq1} and \eqref{RSH:Beq2} to obtain the non-equilibrium correction to the phase-space distribution functions and use them to get the dissipative currents. For this purpose, we consider a Chapman-Enskog like iterative expansion, where the out-of-equilibrium phase-space distribution function can be expressed as a sum of the equilibrium part ($f^\pm_{\rm eq}$) and the non-equilibrium correction part ($\delta f^\pm_{\rm eq}$) i.e.,
\begin{align}
    f^\pm (x,\textbf{p},s) = f_{\rm eq}^\pm (x,\textbf{p},s) + \delta f^\pm (x,\textbf{p},s). \label{RSH:CE_f}
\end{align}
We can use Eq.~\eqref{RSH:CE_f} in Eqs.~\eqref{RSH:Beq1} and \eqref{RSH:Beq2} to obtain the expressions of $\delta f^\pm (x,\textbf{p},s)$ in both cases of with and without magnetic field.

In the field-free case, we find the dissipative corrections to the currents, $N^\mu$ and $T^{\mu\nu}$ remain unaffected by the spin-polarization. However, we obtained the dissipation mechanism of the spin degrees of freedom for the first time to be given by \cite{Bhadury:2020puc},
\begin{align}
    \delta S^{\lambda,\mu\nu} &= \tau_{\rm eq} \Big[ B^{\lambda,\mu\nu}_{\Pi}\, \theta + B^{\kappa\lambda,\mu\nu}_{n}\, (\nabla_\kappa \xi) + B^{\alpha\kappa\lambda,\mu\nu}_{\pi}\, \sigma_{\alpha\kappa} + B^{\kappa\lambda\alpha\beta,\mu\nu}_{\Sigma}\, (\nabla_\kappa \omega_{\beta\alpha}) \Big]. \label{RSH:deltaS}
\end{align}
where, $\theta = \partial\cdot u$ is the scalar expansion, $\sigma^{\mu\nu} = \Delta^{\mu\nu}_{\alpha\beta} \partial^\alpha u^\beta$ is the shear stress tensor and, $\nabla^\alpha \xi$ and $\nabla^\alpha \omega^{\mu\nu}$ are the particle and spin diffusion respectively. Here we have defined the spacelike derivative as, $\nabla^\mu \equiv \Delta^\mu_\alpha \partial^\alpha$, with $\Delta^{\mu\nu} = g^{\mu\nu} - u^\mu u^\nu$ being the projection operator orthogonal to the fluid four-velocity. We have defined another projection operator is defined as, $\Delta^{\mu\nu}_{\alpha\beta} = \left( \Delta^\mu_\alpha \Delta^\nu_\beta + \Delta^\mu_\beta \Delta^\nu_\alpha \right)/2 -  \Delta^{\mu\nu} \Delta_{\alpha\beta}/3$ that is symmetric and traceless in the indices $\mu,\nu$ and $\alpha,\beta$.

In the presence of magnetic field, the dissipative currents for the spin-polarizable and magnetizable relativistic fluid are given by \cite{Bhadury:2022qxd},
\begin{align}
    n^{\mu} &= \tau_{\mathrm{eq}} \Big[ \beta_{n\Pi}^{\langle\mu\rangle}\, \theta 
    + \beta_{na}^{\langle\mu\rangle\alpha} \dot{u}_{\alpha}
    + \beta_{nn}^{\langle\mu\rangle\alpha} \left( \nabla_\alpha\xi \right) 
    + \beta_{nF}^{\langle\mu\rangle\alpha\beta} \left( \nabla_{\alpha} B_{\beta} \right) \nonumber\\
    &\qquad\quad+ \beta_{n\pi}^{\langle\mu\rangle\alpha\beta} \sigma_{\alpha\beta}
    + \beta_{n\Omega}^{\langle\mu\rangle\alpha\beta} \Omega_{\alpha\beta} 
    + \beta_{n\Sigma}^{\langle\mu\rangle\alpha\beta\gamma} \left( \nabla_\alpha \omega_{\beta\gamma} \right) \Big], \label{RSH2:n}\\
    \Pi &= \tau_{\mathrm{eq}} \Big[ \beta_{\Pi\Pi}\, \theta 
    + \beta_{\Pi a}^{\alpha} \dot{u}_{\alpha} 
    + \beta_{\Pi n}^{\alpha} \left( \nabla_\alpha\xi \right) 
    + \beta_{\Pi F}^{\alpha\beta} \left( \nabla_{\alpha} B_{\beta} \right) 
    + \beta_{\Pi\pi}^{\alpha\beta} \sigma_{\alpha\beta} 
    + \beta_{\Pi\Omega}^{\alpha\beta} \Omega_{\alpha\beta} \nonumber\\
    &\qquad\quad+ \beta_{\Pi\Sigma}^{\alpha\beta\gamma} \left( \nabla_\alpha \omega_{\beta\gamma} \right) \Big], \label{RSH2:Pi}\\
    \pi^{\mu\nu} &= \tau_{\mathrm{eq}} \Big[ \beta_{\pi\Pi}^{\langle\mu\nu\rangle} \theta 
    \!+ \beta_{\pi a}^{\langle\mu\nu\rangle\alpha} \dot{u}_{\alpha} 
    \!+ \beta_{\pi n}^{\langle\mu\nu\rangle\alpha}\! \left(\! \nabla_\alpha\xi \right)
    \!+\! \beta_{\pi F}^{\langle\mu\nu\rangle\alpha\beta}\! \left(\! \nabla_{\alpha} B_{\beta}\! \right) 
    \!+\! \beta_{\pi\pi}^{\langle\mu\nu\rangle\alpha\beta} \sigma_{\alpha\beta} \nonumber\\
    &\qquad\quad+ \beta_{\pi\Omega}^{\langle\mu\nu\rangle\alpha\beta} \Omega_{\alpha\beta}
    \!+\! \beta_{\pi\Sigma}^{\langle\mu\nu\rangle\alpha\beta\gamma}\! \left(\! \nabla_\alpha \omega_{\beta\gamma} \right) \Big]\!, \label{RSH2:pi}\\
    \delta S^{\lambda,\mu\nu} &= \tau_{\mathrm{eq}}\, \Big[ B_{\Pi}^{\lambda,[\mu\nu]}\, \theta 
    + B_{a}^{\lambda,[\mu\nu]\alpha} \dot{u}_{\alpha} 
    + B_{n}^{\lambda,[\mu\nu]\alpha} \!\left( \nabla_\alpha \xi \right) 
    + B_{F}^{\lambda,[\mu\nu]\alpha\beta}\! \left( \nabla_{\alpha} B_{\beta} \right) \nonumber\\
    &\qquad\quad+ B_{\pi}^{\lambda,[\mu\nu]\alpha\beta} \sigma_{\alpha\beta} 
    + B_{\Omega}^{\lambda,[\mu\nu]\alpha\beta} \Omega_{\alpha\beta} + B_{\Sigma}^{\lambda,[\mu\nu]\alpha\beta\gamma}\! \left( \nabla_\alpha \omega_{\beta\gamma} \right) \Big], \label{RSH2:delS}
\end{align}
where, $\Dot{u}^\mu \equiv \left(u\cdot\partial\right) u^\mu$ is the acceleration four-vector, $\Omega_{\alpha\beta} = \left(\nabla_\alpha u_\beta - \nabla_\beta u_\alpha\right)/2$ is the vorticity tensor and, $B^\alpha$ is the magnetic four-vector that is related to the field strength tensor as, $F^{\mu\nu} = \epsilon^{\mu\nu\alpha\beta} u_\alpha B_\beta$. In Eqs.~\eqref{RSH:deltaS}-\eqref{RSH2:delS} the transport coefficients are denoted by $\beta_{i}$'s and $B_i$'s.

\subsection{Summary and Conclusion\label{S&C}}

In this work, we demonstrated how to formulate the relativistic dissipative hydrodynamics for spin-polarizable particles with and without magnetic field starting from the kinetic theory. First, we obtained the dissipation mechanism of the spin degrees of freedom of a relativistic fluid in absence of any magnetic field for the first time and found it to depend on multiple hydrodynamic gradients. Next, we extended the framework to study the effect of magnetic field on a spin-polarizable relativistic fluid at the dissipation level. We found that it is necessary to consider the system to be magnetizable if we want to observe the coupling between spin and magnetic field. In this case, we found that all the dissipative currents depend on multiple hydrodynamic gradients, showing resemblance to the Einstein-de Haas and Barnett effects for the first time.

\newcommand\bef{\begin{figure}}
\newcommand\eef[1]{\label{#1}\end{figure}}
\newcommand\beq{\begin{equation}}
\newcommand\eeq[1]{\label{#1}\end{equation}}
\newcommand\bea{\begin{eqnarray}}
\newcommand\eea{\end{eqnarray}}
\newcommand\bet{\begin{table}}
\newcommand\eet[1]{\label{tbl:#1}\end{table}}

\newcommand\fgn[1]{Figure \ref{#1}}
\newcommand\eqn[1]{eq.\ (\ref{#1})}
\newcommand\scn[1]{Section \ref{sec.#1}}
\newcommand\apx[1]{Appendix \ref{sec.#1}}
\newcommand\tbn[1]{Table \ref{tbl:#1}}

\newcommand\incfig[2]{\includegraphics[scale=#1]{#2}}
\newcommand\aul[3]{#1^{+#2}_{-#3}}
\newcommand\auls[4]{#1^{+#2}_{-#3}{(\pm#4)}}

\newcommand\jhep{{\sl J.\ H.\ E.\ P.\/}\ }
\newcommand\np{{\sl Nucl.\ Phys.\/}\ }
\newcommand\npps{{\sl Nucl.\ Phys.\ Proc.\ Suppl.\/}\ }
\newcommand\pos{{\sl PoS\/}\ }
\newcommand\pr{{\sl Phys.\ Rev.\/}\ }
\newcommand\prlt{{\sl Phys.\ Rev.\ Lett.\/}\ }
\newcommand\plt{{\sl Phys.\ Lett.\/}\ }
\newcommand\prt{{\sl Phys.\ Rept.\/}\ }

\newcommand\cdof{\chi^2/\mathrm{DOF}}
\newcommand\ie{{\sl i.e.\/}}
\newcommand\etc{{\sl etc\/.}}
\newcommand\tr{\mathrm{Tr}\;}
\newcommand{\GeV}{\mathrm{GeV}}
\newcommand\formul{formul{\ae}}
\newcommand\etal{{\sl et al.\/}}
\newcommand{\txt}{\textstyle}

\newcommand{\etad}{\eta}
\newcommand{\nt}{N_\tau}
\newcommand{\ns}{N_\sigma}
\newcommand{\ttc}{T/T_c}
\newcommand{\tc}{T_c}
\newcommand{\ti}{t_{\rm imp}}
\newcommand{\tmin}{t_{\rm imp}^{\rm min}}
\newcommand{\kt}{\kappa/T^3}
\newcommand{\ke}{\kappa_{\scriptscriptstyle E}}
\newcommand{\kb}{\kappa_{\scriptscriptstyle B}}
\newcommand{\ket}{\kappa_{\scriptscriptstyle E}/T^3}
\newcommand{\kbt}{\kappa_{\scriptscriptstyle B}/T^3}
\newcommand{\rom}{\rho_T(\omega)}
\newcommand{\rir}{\rho_{\scriptscriptstyle IR}(\om)}
\newcommand{\ruv}{\rho_{\scriptscriptstyle UV}(\om)}
\newcommand{\vx}{\vec{x}}
\newcommand{\vy}{\vec{y}}
\newcommand{\vxo}{\vec{x_O}}
\newcommand{\vsq}{\langle v^2 \rangle}
\newcommand{\md}{m_{\scriptscriptstyle D}}
\newcommand{\mq}{m_{\scriptscriptstyle Q}}
\newcommand{\pq}{p_{\scriptscriptstyle Q}}
\newcommand{\mqi}{\frac{\textstyle 1}{\textstyle \mq}}
\newcommand{\mufit}{\mu_{\rm fit}}
\newcommand{\cf}{C_f}
\newcommand{\ds}{2 \pi \, T \, D_s}
\newcommand{\zee}{Z_{\scriptscriptstyle EE}}
\newcommand{\raa}{R_{\scriptscriptstyle AA}}
\newcommand{\taut}{\tau_{\rm int}}
\newcommand{\cosec}{{\rm cosec}}
\newcommand{\gbsq}{g_{\scriptscriptstyle B}^2}
\newcommand{\lms}{\Lambda_{\overline{\scriptscriptstyle \rm MS}}}

\newcommand{\gcont}{G^{\scriptscriptstyle \rm Cont}_{\scriptscriptstyle \rm norm}}
\newcommand{\glat}{G^{\scriptscriptstyle \rm Lat}_{\scriptscriptstyle \rm norm}}
\newcommand{\gecont}{G_{\scriptscriptstyle EE}^{\scriptscriptstyle \rm cont}}
\newcommand{\gelat}{G_{\scriptscriptstyle EE}^{\scriptscriptstyle \rm bare}}
\newcommand{\gern}{G_{\scriptscriptstyle EE}^{\scriptscriptstyle \rm renorm}}
\newcommand{\gfr}{G_{norm}}
\newcommand{\get}{G_{\scriptscriptstyle EE}}
\newcommand{\gbt}{G_{\scriptscriptstyle BB}(\tau)}
\newcommand{\ggfr}{\frac{\textstyle \get}{\textstyle \gfr} (\tau)}
\newcommand{\zelp}{Z_{\scriptscriptstyle EE}^{1lp}}
\newcommand{\puv}{\rho_{\scriptscriptstyle UV}(\omega)}
\newcommand{\pir}{\rho_{\scriptscriptstyle IR}(\omega)}
\newcommand\omi[1]{\mathcal{O}(\mq^{-#1})}
\newcommand{\mkin}{m_{\rm kin}}
 
\section{Heavy Quarks in QGP: Results From Lattice QCD}
\author{Saumen Datta}	

\bigskip

\begin{abstract}
  The energy loss of low and intermediate momentum charm and bottom
  quarks in quark-gluon plasma can be understood in a Langevin
  framework. Lattice QCD can play a crucial role in this
  understanding, by calculating the relevant transport coefficients.
  Here I summarize the current status of the results obtained from
  lattice QCD.
\end{abstract}

\vskip1cm

Heavy quarks are among the most interesting probes of the medium
created in ultrarelativistic heavy ion collisions (URHIC).  If $\mq
\gg T$, the heavy quarks are expected to be created only by hard gluons
within $\tau \sim \tfrac{1}{2 \mq}$ of the collision. \footnote{This
  condition may not hold for the charm in LHC.} Therefore a heavy
quark probe is expected to carry information about the whole history
of the fireball. Besides, the heavy mass $\mq \gg \lms$ gives an
additional small parameter which allows use of various theoretical
tools.

The energy loss of heavy quark in the plasma has been of great
interest in recent times. The energy loss mechanisms of heavy quarks
are different from that of light quarks: gluon radiation is suppressed
\cite{dk} in a cone of angle $\tfrac{\mq}{E}$. For quarks of moderate
energy, collision with thermal particles is the
leading energy loss mechanism \cite{mt}. Since even for a
near-thermalized heavy
quark, its momentum $\pq \sim \sqrt{\mq T} \gg T$, individual
collisions with light thermal particles with momentum $\sim T$ do not
change the momentum of a heavy quark significantly. Therefore, a
Langevin description of the heavy quark energy loss for
low-intermediate energy heavy quarks was proposed: 
\cite{svetitsky,mt,mustafa}
\beq
\frac{dp_i}{dt} \; = \; - \etad \, p_i \; +  \; \xi_i(t), \qquad
\langle \xi_i(t) \, \xi_j(t^\prime) \rangle \; = \; \kappa \;
\delta_{ij} \; \delta(t-t^\prime)
\eeq{langevin}
where the second term denotes a white noise. Since $\etad$ is related
to $\kappa$ by Einsten's relation, the motion can be described by 
only one parameter $\kappa$. This formalism has been very successful in
explaining the $\raa$ and $v_2$ of $D$ mesons
\cite{annrev}. It is, however, difficult to calculate $\kappa$ from
thermal QCD: the leading order (LO) result \cite{svetitsky,mt} is far
too small for phenomenology. Perturbation theory for this object
is also quite ill-behaved: for temperatrues of interest, the next to
leading order (NLO) result is an order of magnitude larger than LO \cite{cm}.
A nonperturbative estimation of the diffusion coefficient is therefore
called for.

In phenomenological studies, one often tweaks the LO calculation to
get a value suitable for explaining the experimental data: for
example, by altering the gluon propagator from its perturbative form,
by changing the running of the coupling, or by calculating the
scattering cross-sections through a tuned potential \cite{cao,das}. A direct
extraction
of $\kappa$ from data using a Bayesian analysis has also been
attempted \cite{duke}. But of course, for a satisfactory understanding
of the formalism, one would like to calculate $\kappa$ from QCD. This
is where lattice calculations have made a significant progress. I will
describe the status of such calculations in this report. Except for a
comment near the end, all the results I will describe are for a gluonic
plasma only, i.e., the plasma has no thermal quarks. The results are
still very useful, as they give us an idea about the size of the
nonperturbative effects. 

Standard relations connect the drag
coefficient $\eta$ in \eqn{langevin} and the spatial diffusion
coefficient $D_s$ to $\kappa$:
\beq
\eta \approx \frac{\kappa}{2 \, M \, T}, \qquad D_s \approx
\frac{2 T^2}{\kappa} \, .
\eeq{fluc}
A fluctuation-dissipation theorem relates $D_s$
to the vector current correlator $\langle J_i(\vec{p}=0, t+\tau) \;
J_i(\vec{p}=0, t) \rangle$ \cite{kapusta}. This correlator can be
easily calculated on the lattice. From this, an extraction of $D_s$
has been attempted \cite{ding}. This extraction though is quite
difficult: $D_s$ is related to the width of the transport peak of the
spectral function of the correlator. The spectral function is
connected to the correlator by an integral transform similar to a
Laplace transform. It is very difficult to invert the transform accurately
enough to extract
the width of its transport peak. The transport peak is in the infrared
and it further gets affected by the infrared cutoffs put in the
calculation, like size of the system.

\begin{figure}[h]
  \begin{center}
    \centerline{\includegraphics[width=6.5cm,height=5cm]{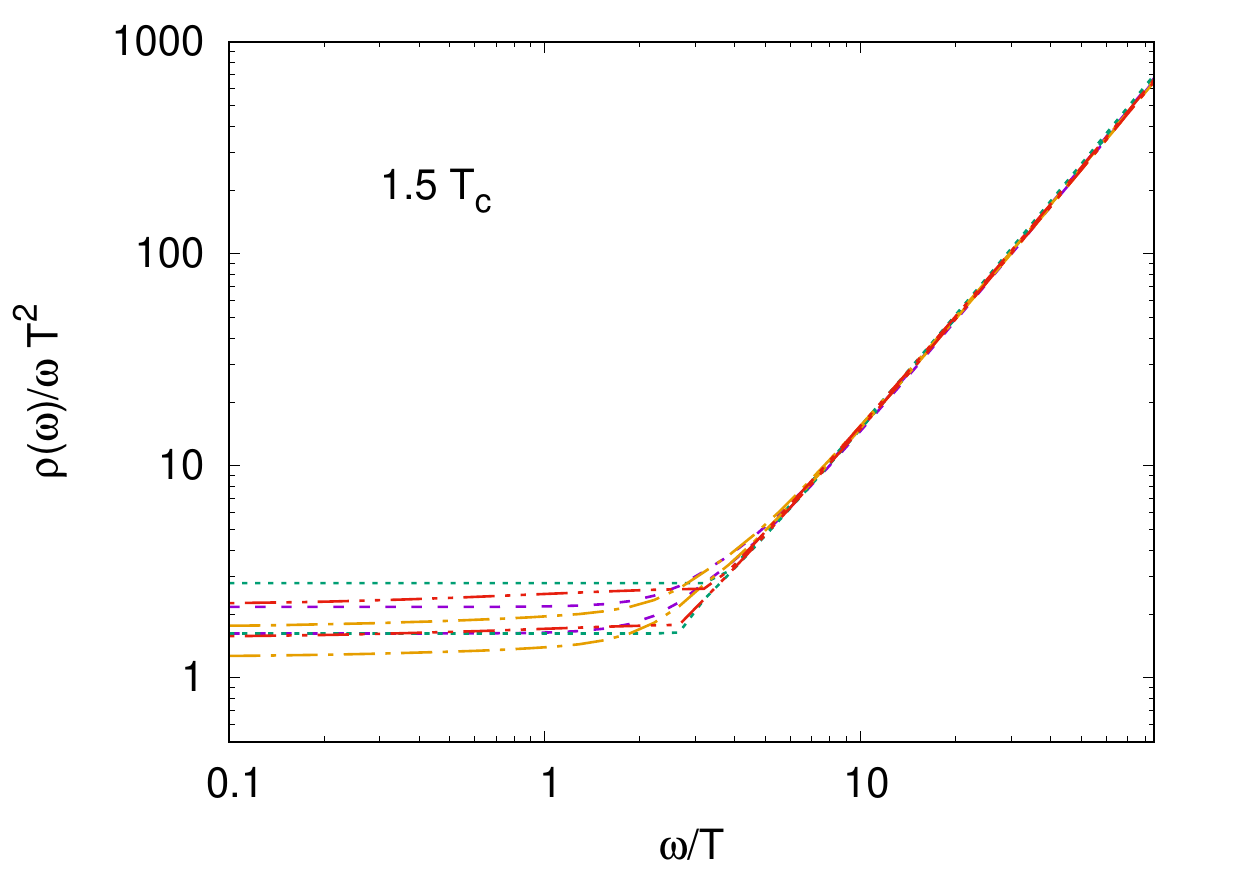}
    \includegraphics[width=6.5cm,height=5cm]{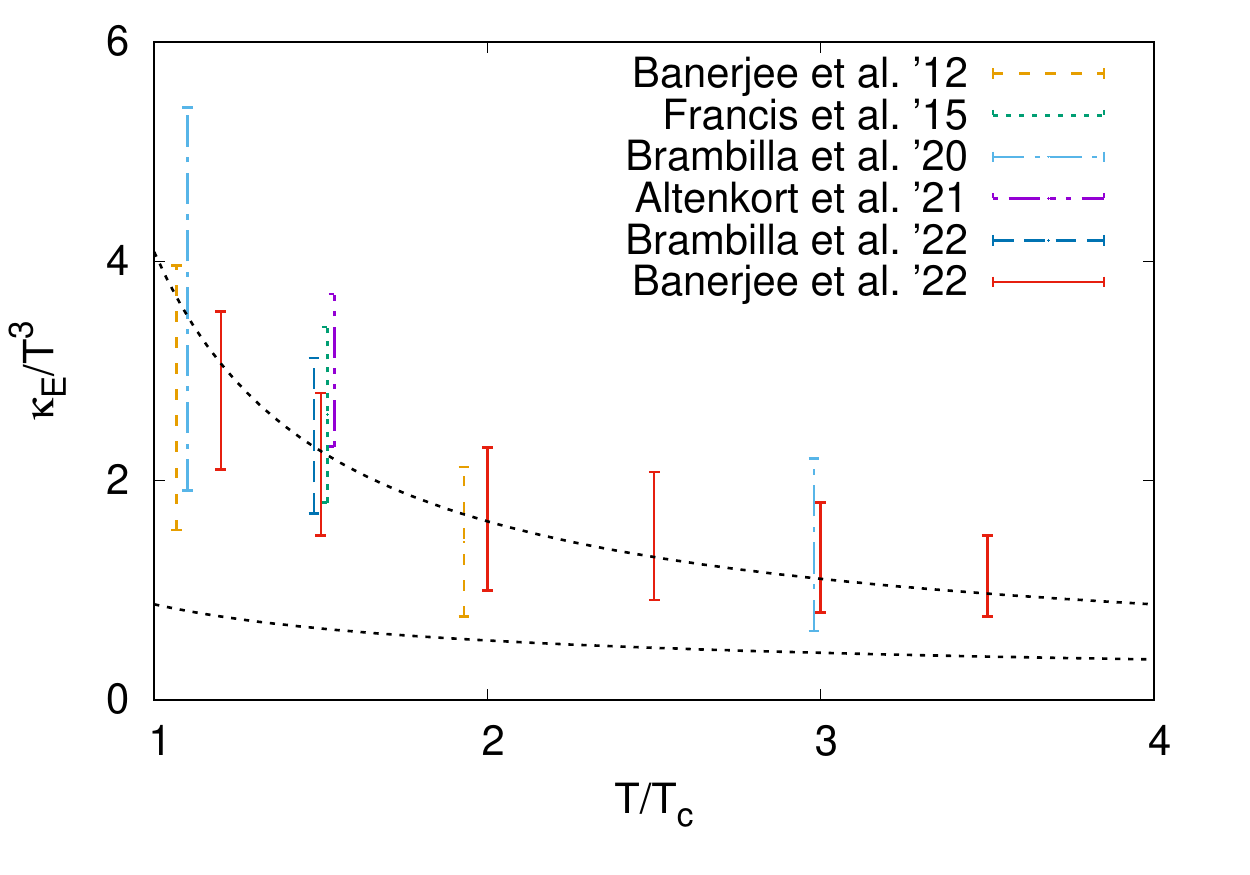}}
    \caption{(Left) Structure of the spectral function at 1.5
      $\tc$\cite{hight}. Different line styles show the results of using
      different
      models of $\rom$. $\rom$ has no transport peak: it goes
      $\sim \ke \, \omega$ at low $\omega$. (Right) A collation of the
      available lattice results for $\ke$ in a gluon plasma. The results shown
      are from Banerjee et al. \cite{prd11,hight},
      Francis et al. \cite{bielefeld}, Brambilla et al. \cite{tum1, tum2} and
      Altenkort et al. \cite{gflow}.
    }
    \label{fg.ke}
  \end{center}
\end{figure}

A far more promising approach has been to calculate $\kappa$ directly
from its defining relation: the force-force correlator \cite{ct,clm}.
In the leading order in $\mqi$, the force term $\xi$ in
\eqn{langevin} is $g E_i$, the color electric field. So one can
extract $\kappa$ from the spectral function of the color electric
field correlator: $\langle (g E_i)( \vec{p}=0, t+\tau) \; (g E_i)(
\vec{p}=0, t) \rangle$. The advantage is that there is no transport
peak in this spectral function $\rom$ (see Fig. \ref{fg.ke}). In the infrared 
$\rom \, \sim \, \omega$, and the coefficient of this linear
term, $\ke$, is the estimate for $\kappa$ for static quarks
\cite{clm}. The $\mqi$ correction to this term has also been
estimated: under certain assumptions, \cite{blaine}
\beq
\kappa \ = \ \ke \; + \; \frac{2}{3} \, \vsq \, \kb \; + \; ...
\qquad \vsq \ = \ \frac{3 T}{\mkin} \; \left(1 \, - \,
\frac{5 T}{2 \mkin} \right) 
\eeq{expansion}
where $\kb$ is obtained from the color magnetic field correlator in an
analogous way to $\ke$. Note that treating the expansion as a series in
$\vsq$, the average velocity squared of the heavy quark, leads to better
stability \cite{blaine}.

There have been several \cite{prd11,bielefeld,tum1,gflow,tum2,hight}
lattice calculations for $\ke$ in a gluonic plasma, estimated from the
$EE$ correlator. All these works use a model of $\rom$ to extract
$\ke$ from the $EE$ correlator. We know the form of $\rom$ at very
high and low energies: it approaches the perturbative spectral
function at very high $\omega$ and a dispersive form $\sim \ke \omega$
in the infrared. For intermediate values of $\omega$, different
calculations use different models. The calculations also differ in how
they renormalize the lattice correlators. The left panel of Fig. \ref{fg.ke}
shows the results for $\rom$ obtained at 1.5 $\tc$ in a recent
calculation \cite{hight} when using different models (stylistically I have
followed here a similar plot of an earlier calculation\cite{bielefeld}).
The right panel of Fig. \ref{fg.ke} shows a compilation of the results
obtained in these calculations, in the 1-4 $\tc$ range. To reduce clutter,
where the same group has published two calculations, I have only kept the
new one. There seems to be a good agreement between the different
calculations, within the large error bars (which are dominated by
systematics: the effect of using different models of $\rom$).

Figure \ref{fg.ke} also shows the NLO result for $\kappa$ \cite{ct}. For
the scale of the coupling, I have taken $\mu_{\rm opt} \sim 6.7 \, T$
\cite{burnier}, which is the scale obtained using the {\em principle of
  minimum sensitivity}, and used the range $[0.5,2] \, \mu_{\rm opt}$ to get the
band shown in the figure. The lattice data seems to agree quite well with
the perturbative result. Note, however, that perturbation theory is
inherently unstable here. The LO result (with the same scale) is very
different: it even turns negative at moderately high temperatures.

What are the missing ingredients for comparing these results with
phenomenology? Within the gluonic plasma approximation, one needs an
estimate of the size of the $\mqi$ correction. And the major next step
is the inclusion of the dynamical quarks. 

\begin{figure}[ht]
  \begin{center}
    \centerline{\includegraphics[width = 6.5cm,height=5cm]{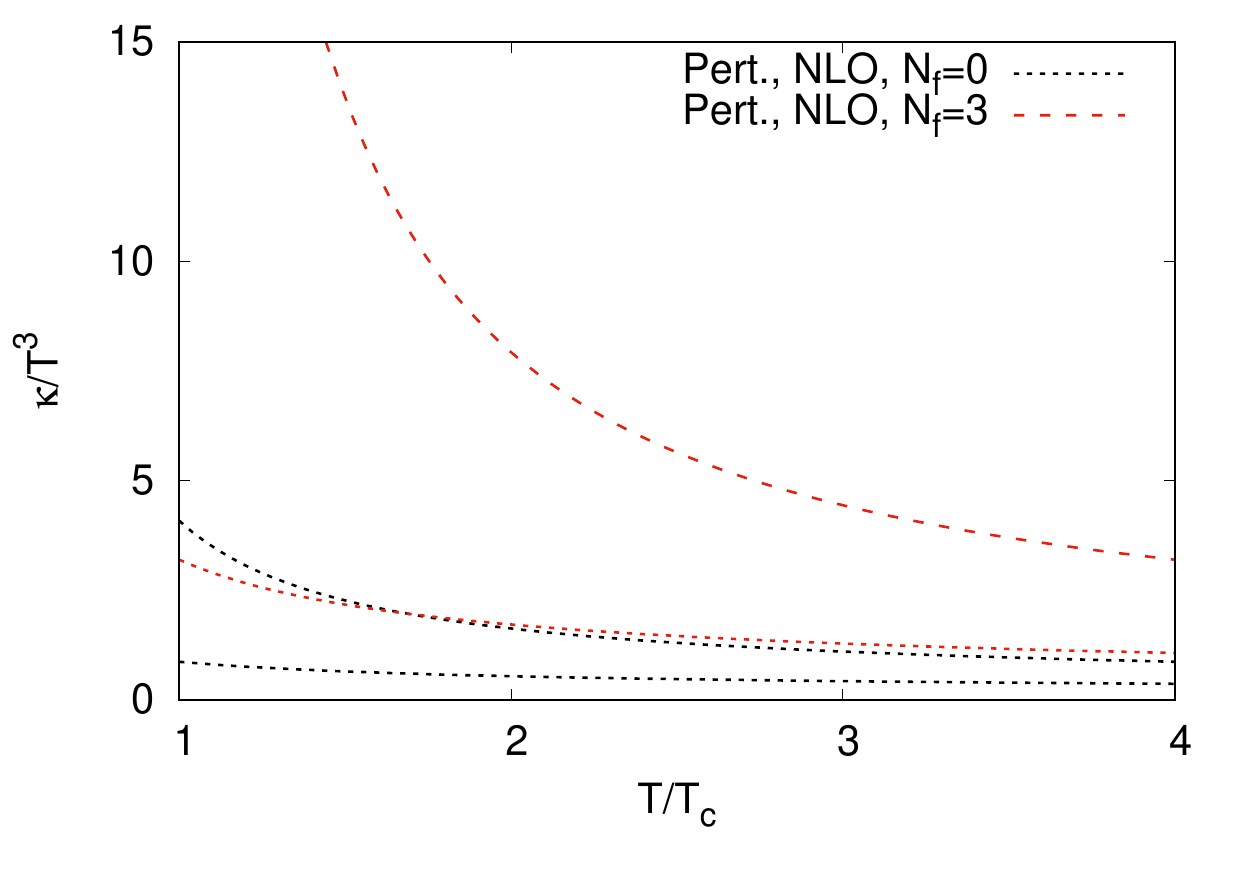}\includegraphics[width = 6.5cm,height=5cm]{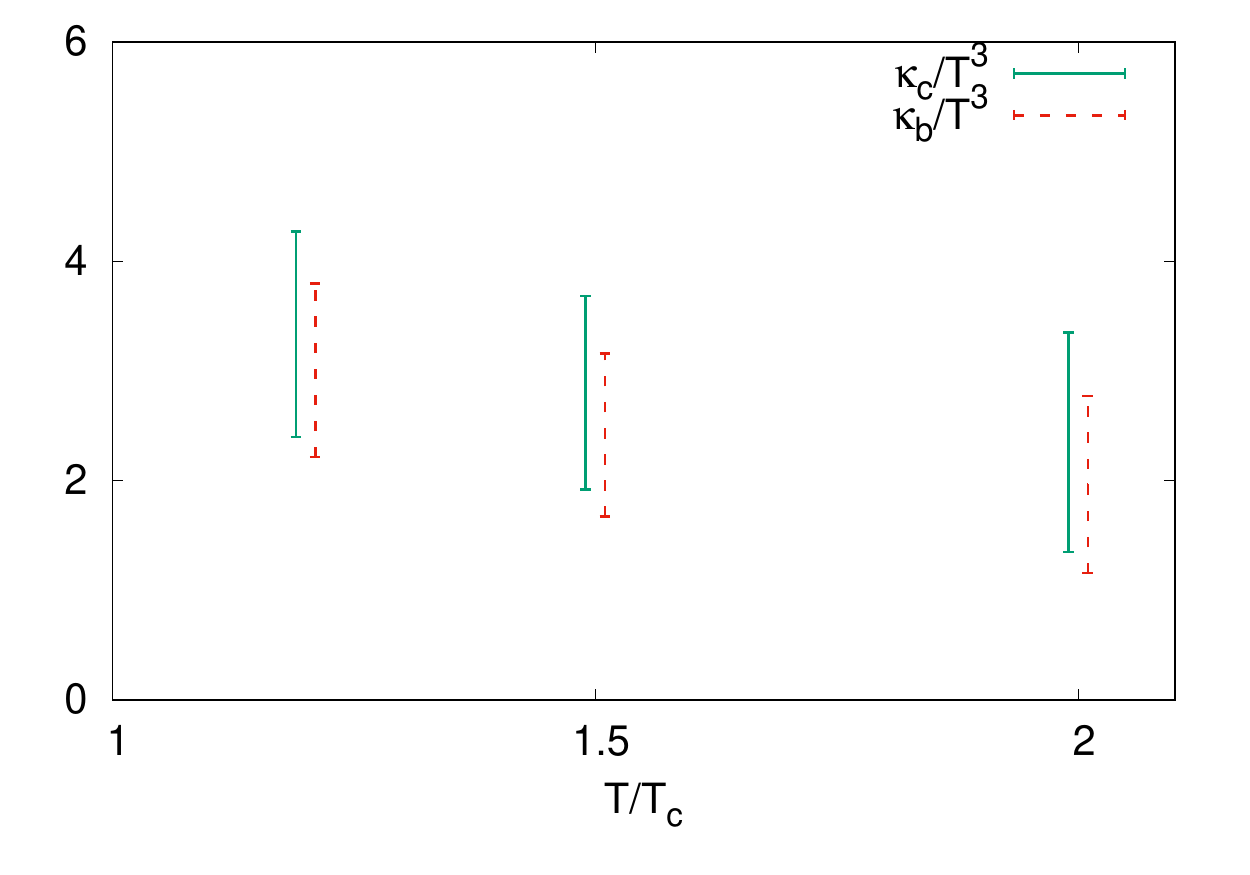}}
    \caption{(Left) A comparison of the NLO results \cite{cm} for $\ket$ for the gluon plasma and for QGP with 3 light flavors. (Right) $\kappa_b$ and $\kappa_c$ for the gluon plasma \cite{bb} .}
    \label{fg.oth}
  \end{center}
\end{figure}

The $\mqi$ correction coefficient in \eqn{expansion} has been
calculated recently: $\kb$ has been computed in the temperature range
1.2-2 $\tc$\cite{bb}. A second calculation at 1.5 $\tc$
\cite{tum2} agrees very well within errorbars. Using
\eqn{expansion} and an estimate of $\vsq$ from ratio of
susceptibilities, $\kappa_b$ and $\kappa_c$ have been estimated
\cite{bb}. These results are shown in the right panel of
Fig. \ref{fg.oth}, where I have updated the figure (as well as Fig.
\ref{fg.pheno} below) with a recent
estimate \cite{hight} of $\ke$.

Perturbation theory would suggest a large effect of the dynamical quarks:
the left panel of Fig. \ref{fg.oth} compares the NLO results of the theory
with three light quark flavors with that of the gluon plasma.
There are no published lattice results with dynamical quarks, but 
preliminary results from a dynamical study was presented in
Quark Matter \cite{altenkort}. The unquenching effect, from the first
results, is expected to be large.

In what follows, we will stick to the gluon plasma results.
Other quantities of interest can be calculated from the
results for $\kappa$ using \eqn{fluc}. Fig. \ref{fg.pheno} shows
the results \cite{bb} for $\eta$ and $\ds$.

\begin{figure}[ht]
  \begin{center}
    \centerline{\includegraphics[width = 6.5cm,height=5cm]{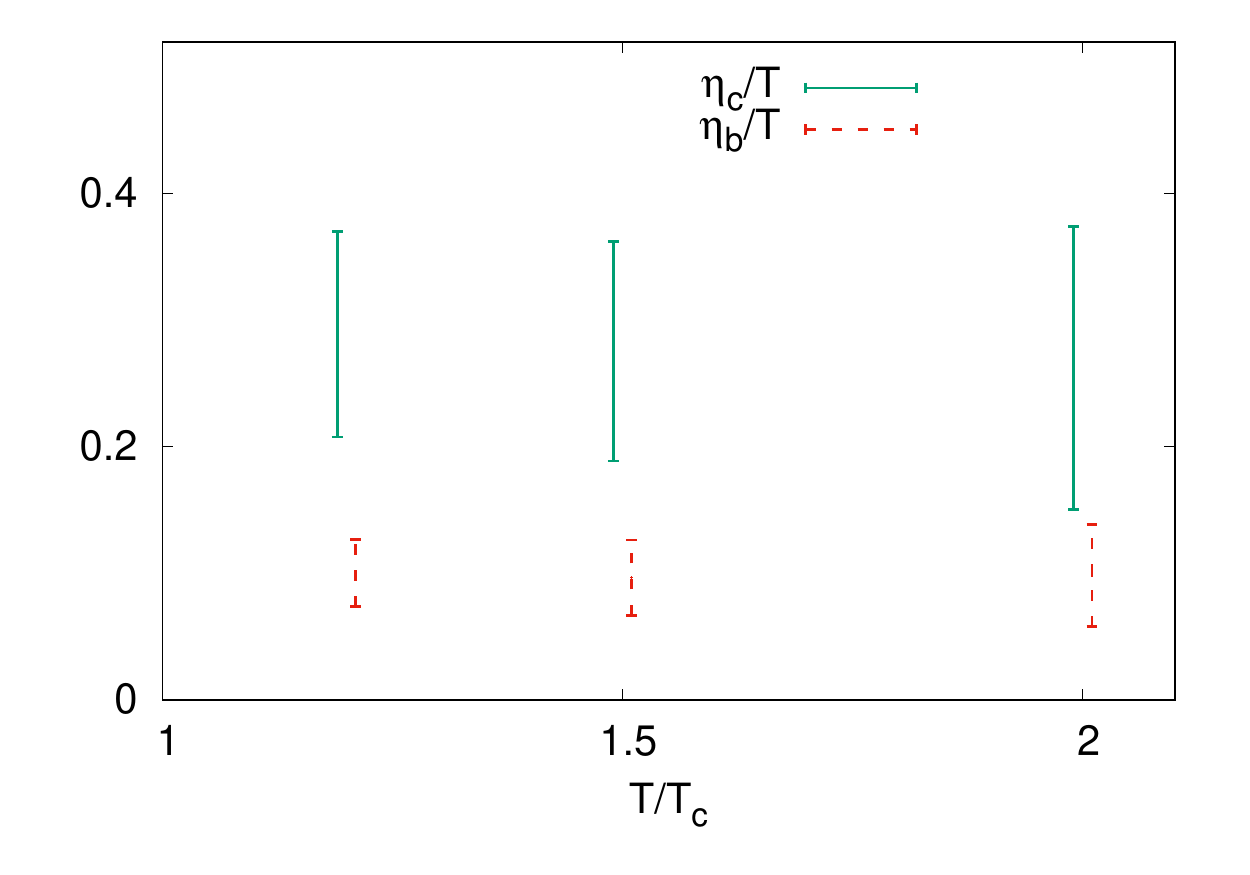}\includegraphics[width = 6.5cm,height=5cm]{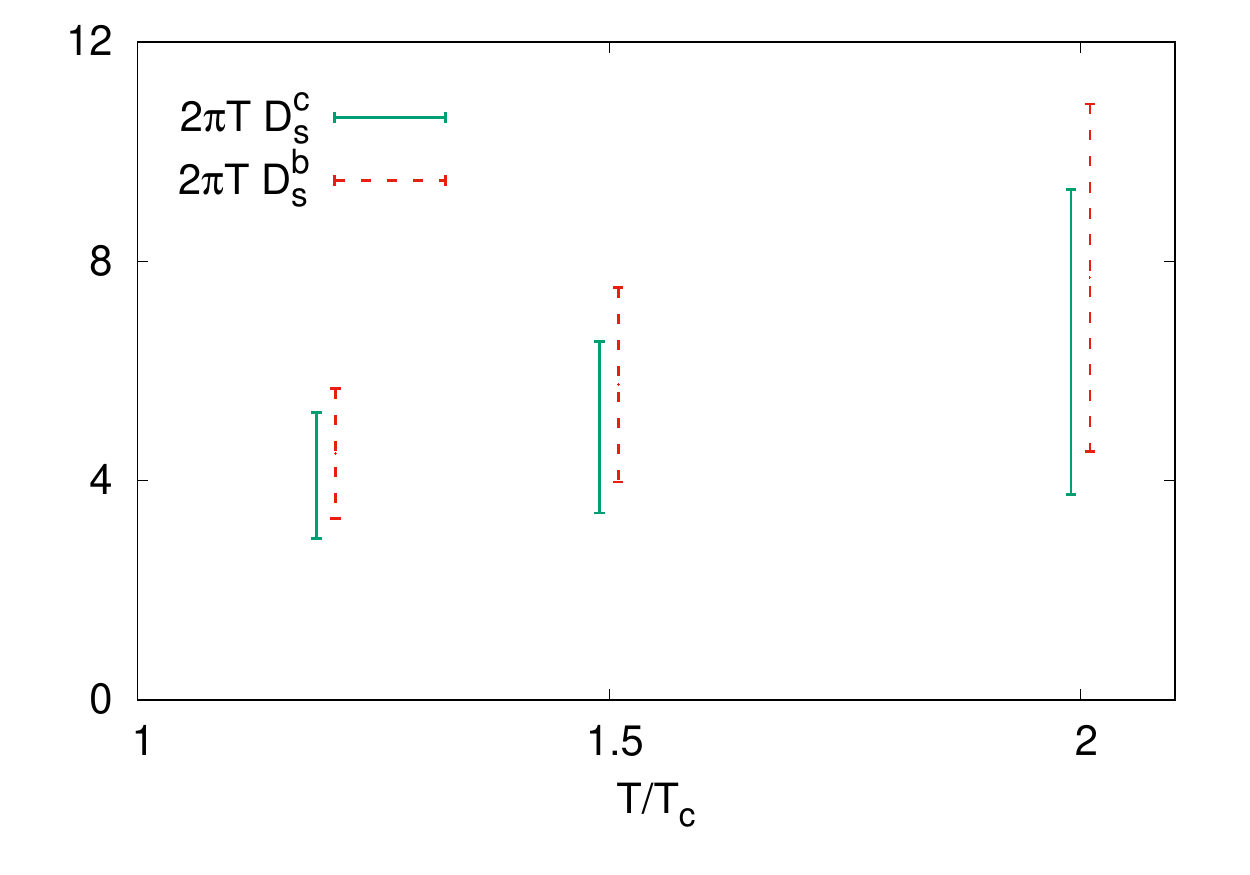}}
    \caption{(Left)Estimate \cite{bb} for $\eta_{c,b}$ and (right)
      $\ds$ for charm and bottom in a gluon plasma.}
    \label{fg.pheno}
  \end{center}
\end{figure}

The relaxation time is given by the inverse of $\eta$. From the figure
one sees $\tau_b \sim 10 fm > \tau_c \sim 3 fm$. So $b$ quark may
only be partially thermalized. The
spatial diffusion coefficient $\ds$ does not play any direct role in
the Langevin calculation. However, the results of the diffusion
calculations are often expressed in terms of $\ds$ rather than
$\kappa$ \cite{cao}. An analysis of experimental data prefers $\ds
\sim 1.5 - 4.5$ in the LHC, which is consistent with Fig. \ref{fg.pheno}.

To summarize, lattice QCD calculations have provided important
ingredients for an understanding of the energy loss pattern of heavy
quarks in terms of a Langevin description. The static result has been
available for a while; the first correction terms $\mathcal{O}(v^2 \sim
\tfrac{T}{M})$ have also been calculated recently. These results are
for quenched lattices; the first unquenched studies are ongoing.

\section{Relativistic Hydrodynamics with momentum-dependent relaxation time approximation}
\author{Sukanya Mitra}	

\bigskip
 
\begin{abstract}
A relativistic hydrodynamic theory up to second order in gradient expansion, has been derived using momentum dependent relaxation time approximation (MDRTA) in the relativistic transport equation. A correspondence 
has been established between the out of equilibrium system dissipation and the thermodynamic field redefinition of the macroscopic variables through MDRTA. It is found that the result from the numerical solution of 
the Boltzmann equation lies somewhere in between the two popular extreme limits: linear and quadratic ansatz, indicating a fractional power of momentum dependence in relaxation time to be appropriate. Finally, the 
causality and stability of the first-order relativistic theory with hydrodynamic field redefinition via MDRTA have been analyzed. 

\end{abstract}
\subsection{Introduction}

Relativistic dissipative hydrodynamic theory has been proved to be reasonably successful in describing the out of equilibrium dynamics of a system in the long wavelength limit. The evolution of the relevant 
macroscopic quantities such as temperature and charge chemical potential is given by a set of coupled differential equations where the dissipative effects are included by the transport coefficients such as 
viscosity and conductivities. However, the macroscopic thermodynamic quantities such as energy density and particle number 
density in these equations are usually set to their equilibrium values even in the dissipative medium by imposing certain matching or fitting conditions. In this work, these dissipative corrections have 
been included in the out of equilibrium thermodynamic fields from gradient expansion technique of solving the relativistic transport equation using momentum dependent relaxation time approximation (MDRTA)
\cite{Teaney:2013gca,Kurkela:2017xis,Rocha:2021zcw,Mitra:2020gdk,Mitra:2021owk,Mitra3}.

The manuscript is organised as follows. Section II contains the formal framework for hydrodynamic field redefinition 
obtained from relativistic transport equation using MDRTA upto second order of gradient correction. Section III provides a quantitative estimation of the effect of momentum dependent relaxation time on the pressure 
anisotropy of the system. In section IV the causality and stability of a first order theory with fields redefined under MDRTA have been analyzed. Finally, in section V the work has been summarized with prior 
conclusions and useful remarks.

\subsection{Field redefinition with MDRTA}
\begin{figure}
    \centering
    \subfloat[Scaled $\zeta$ and $\lambda$ as a function of $z$.]{{\includegraphics[width=6cm]{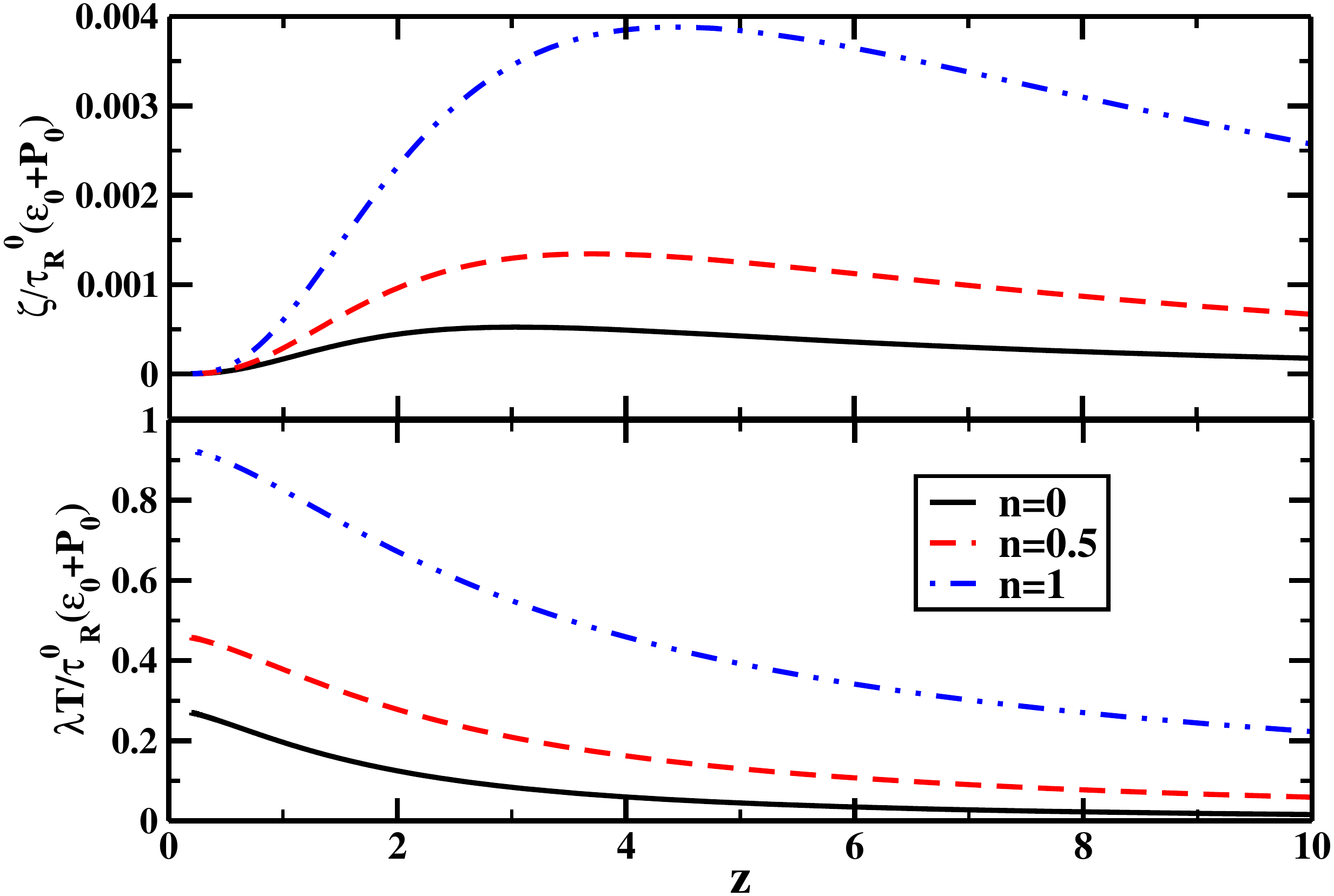} }}
    \quad
    \subfloat[Correction coefficients as function of $n$.]{{\includegraphics[width=5.5cm]{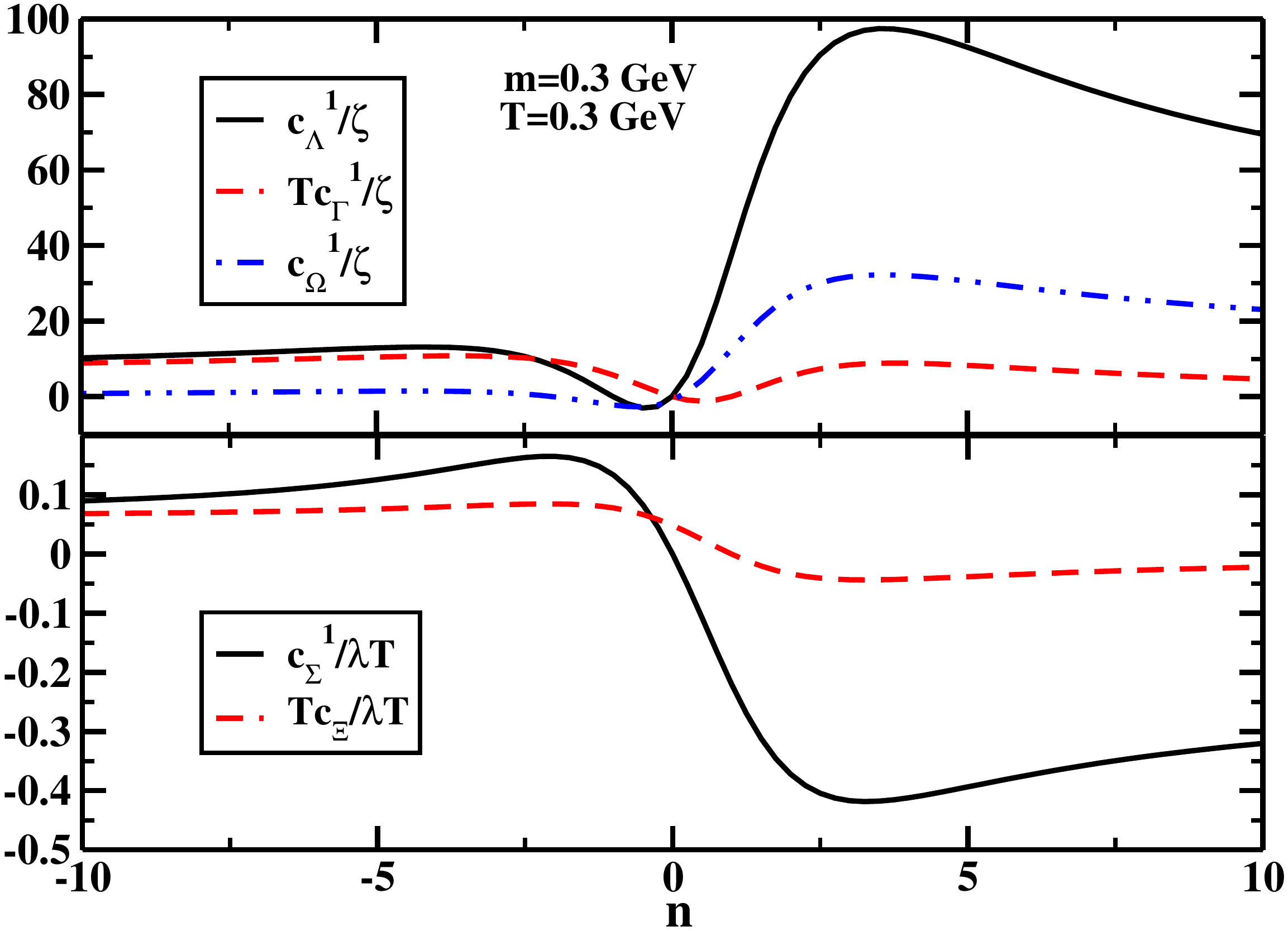} }}
    \caption{MDRTA modified field corrections and transport coefficients}
    \label{figMDRTA}
\end{figure}

\begin{figure}
    \centering
    \subfloat[$\tau_{\pi}/\tau_R^0$ and $\tau_{\Pi}/\tau_R^0$ as a function of $z$.]{{\includegraphics[width=6cm]{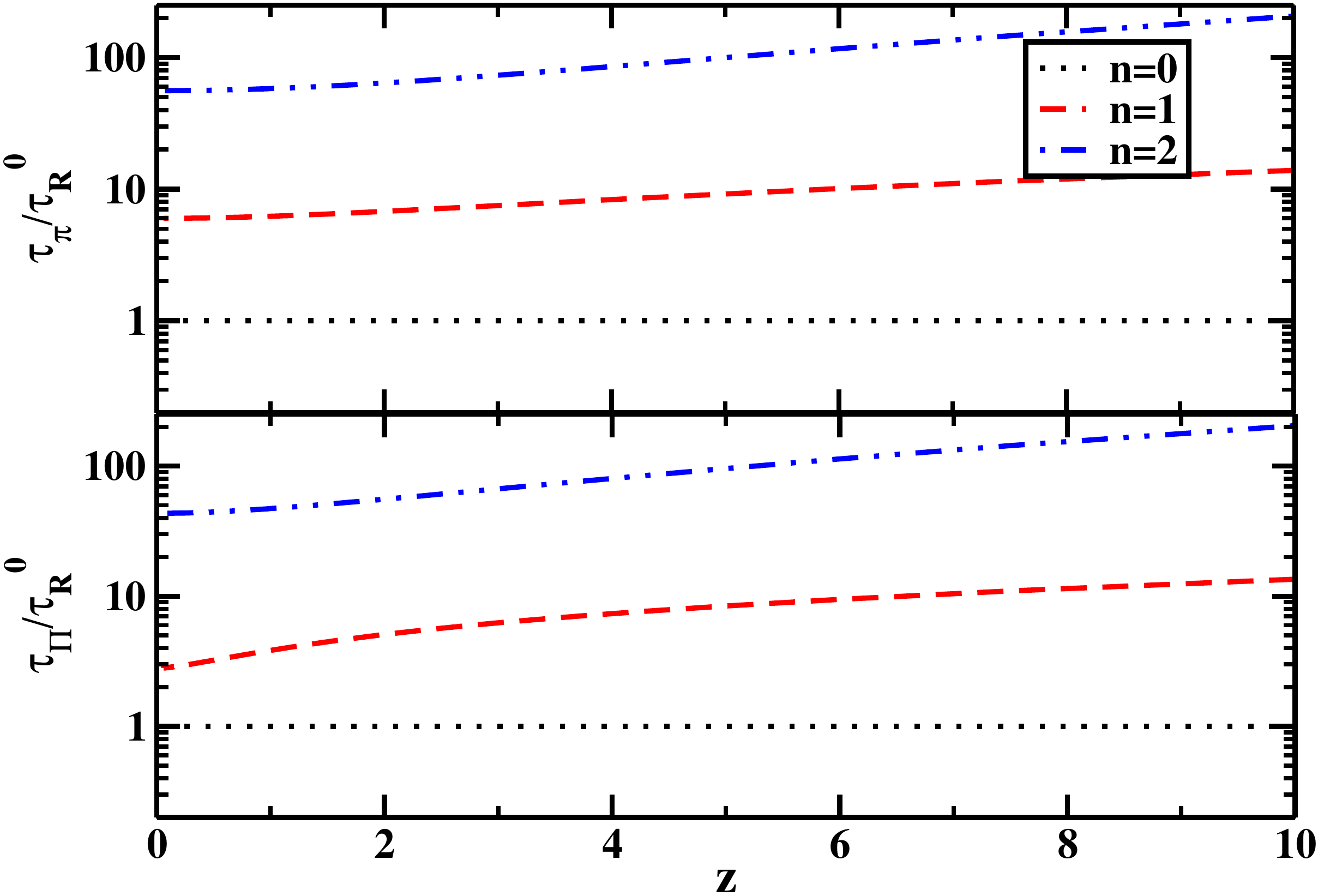} }}
    \quad
    \subfloat[Time evolution of pressure anisotropy.]{{\includegraphics[width=6cm]{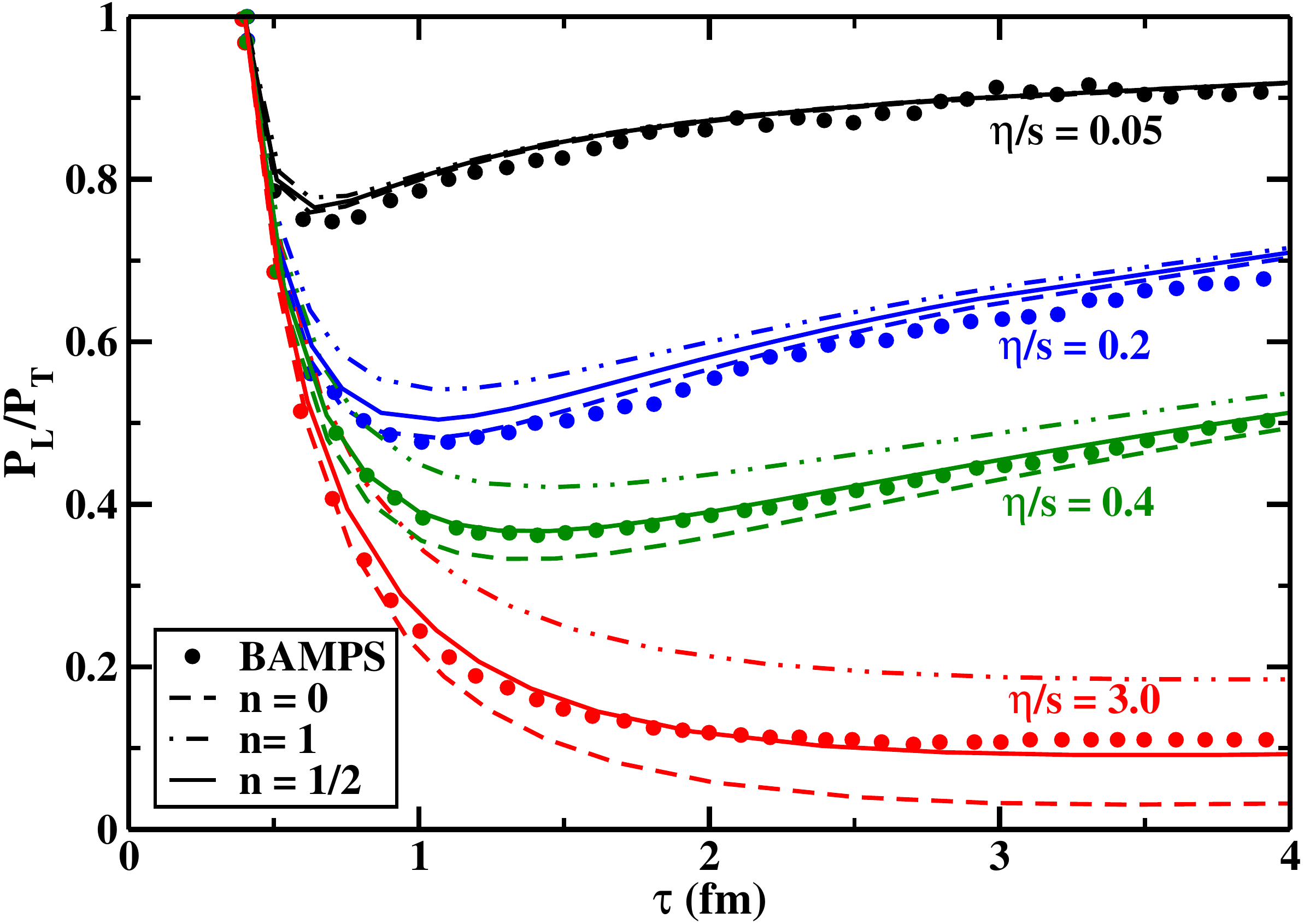} }}
    \caption{Effect of MDRTA on macroscopic quantities.}
    \label{figMDRTA1}
\end{figure}

In relaxation time approximation, the relativistic transport equation for the single particle distribution function $f(x,p)$ with particle four-momenta $p^{\mu}$ and space-time variable $x^{\mu}$ is given by, 
\be
{p}^{\mu}\partial_{\mu}f=-\frac{(p\cdot u)}{\tau_R}f^{(0)}(1\pm f^{(0)})\phi ~,~~~~\tau_R(x,p)=\tau_R^0(x) \left(\frac{p\cdot u}{T}\right)^n~,
\label{RTE1}
\ee
with $f^{(0)}$ as the equilibrium distribution, $\phi$ as the out of equilibrium distribution deviation, $\tau_R^0$ as the momentum independent part and $n$ as a number specifying the power of the scaled energy. 
In order to solve Eq.(\ref{RTE1}), here the well known iterative technique of gradient expansion, the Chapman-Enskog (CE) method has been adopted \cite{Degroot}. With this, the first order corrections in energy 
density, particle number density, pressure, particle flux and energy flow are respectively given by,
\begin{align}
&\delta \epsilon^{(1)}=c_{\Lambda}^1(\partial\cdot u)~,~~\delta {\rho}^{(1)}=c_{\Gamma}^1(\partial\cdot u)~,~~\delta P^{(1)}=c_{\Omega}^1(\partial\cdot u)~,\nn
&W^{(1)\alpha}=-c_{\Sigma}\hat{h}(\nabla^{\alpha}T/T-Du^{\alpha})~,~~V^{(1)\alpha}=c_{\Xi}(\nabla^{\alpha}\tilde{\mu}).
\label{field1}
\end{align}
Including field corrections, the expressions for particle four-flow and energy-momentum tensor read,
\ba
N^{\mu}=&&({\rho}_0+\delta {\rho})u^{\mu}+V^{\mu}~,
\label{numberflow}\\
T^{\mu\nu}=&&(\epsilon_0+\delta\epsilon)u^{\mu}u^{\nu}-(P_0+\delta P)\Delta^{\mu\nu}+(W^{\mu}u^{\nu}+W^{\nu}u^{\mu})+\pi^{\mu\nu}~.
\label{enmomflow}
\ea
Here, $\pi^{\mu\nu}=\Delta^{\mu\nu}_{\alpha\beta}\delta T^{\alpha\beta}=2\eta\sigma^{\mu\nu}$ is the shear stress tensor. The expressions of first order field correction coefficients and the physical transport 
coefficients bulk viscosity ($\zeta$), thermal conductivity ($\lambda$) and shear viscosity ($\eta$) in MDRTA are given in \cite{Mitra:2021owk}. It has been checked that $\zeta/\tau_R^0,\lambda/\tau_R^0,
\eta/\tau_R^0>0$ for all values of $n$ for various combinations of $z(=m/T), T$ and ${\mu}$. They have been plotted in Fig.(\ref{figMDRTA}) (left panel) as a function of the scaled mass $z$ for several $n$ values. 
Next, we observe that for any $n$ value, $c^1_{\Omega}-c_{\Lambda}^1\big(\frac{\partial P_0}{\partial\epsilon_0}\big)_{{\rho}_0}-c^1_{\Gamma}\big(\frac{\partial P_0}{\partial {\rho}_0}\big)_{\epsilon_0}=-\zeta,~
c^1_{\Sigma}-\frac{(\epsilon_0+P_0)}{{\rho}_0}c^1_{\Xi}=-\frac{\lambda T}{\hat{h}}$, such that, 
$\delta P^{(1)}-(\frac{\partial P_0}{\partial\epsilon_0})_{{\rho}_0}\delta\epsilon^{(1)}-(\frac{\partial P_0}{\partial {\rho}_0})_{\epsilon_0}\delta {\rho}^{(1)}=\Pi^{(1)},~
W^{(1)\mu}-\hat{h}T V^{(1)\mu}=q^{(1)\mu}$, with $\Pi^{(1)}$ and $q^{(1)\alpha}$ as first order bulk viscous and diffusion flow. This shows that the individual correction coefficients combine to give the physical 
transport coefficients as predicted by 
\cite{Kovtun:2019hdm}. The coefficients of first order dissipative correction (scaled by physical transport coefficients) have been plotted for $m=0.3$ GeV and $T=0.3$ GeV as a function of $n$ in right panel of 
Fig.(\ref{figMDRTA}). For momentum independent case $n=0$, $c_{\Lambda}^1=c_{\Gamma}^1=c_{\Sigma}^1=0$. $c_{\Xi}^1$ vanishes for $n=1$ with $c_{\Lambda}^1=3c_{\Omega}^1$. Fig.(\ref{figMDRTA}) shows that the 
individual field corrections take how much fractional part of the dissipative flux, is decided by the value of $n$.
The second order hydrodynamic evolution equations for bulk and shear viscous pressure with MDRTA is respectively given by.
\ba
&\Pi=-\zeta\partial\cdot u -\tau_{\Pi}D\Pi+c_{\Pi}^{\sigma}\pi^{\mu\nu}\sigma_{\mu\nu}+c_{\Pi}^{\theta}\Pi(\partial\cdot u)~,
\label{bulkhydro2}\\
&\pi^{\mu\nu}=2\eta\sigma^{\mu\nu}-\tau_{\pi}D\pi^{\langle\mu\nu\rangle}+c_{\pi}^{\omega}\pi_{\rho}^{\langle\mu}\omega^{\nu\rangle\rho}+c_{\pi}^{\sigma}\pi_{\rho}^{\langle\mu}\sigma^{\nu\rangle\rho}
+c_{\pi}^{\theta}\pi^{\mu\nu}(\partial\cdot u)+c_{\pi}^{\zeta}\Pi\sigma^{\mu\nu}.
\label{shearhydro2}
\ea
The explicit expression of the associated coefficients can be found in \cite{Mitra:2021owk}. It can be observed that $\tau_{\pi}=\tau_{\Pi}=\tau_R^0$ holds only for $n=0$. For all other $n$, the three time 
scales are evidently separate (Fig (\ref{figMDRTA1}) left panel).

\subsection{Phenomenological application}

To have a quantitative idea about how MDRTA affects the physical observables, the second order hydro equations have been solved for a conformal, boost invariant 
Bjorken system with ultrarelativistic equation of state $(\epsilon=3P)$ as the following \cite{Mitra:2020gdk},
\be
\frac{d\epsilon}{d\tau}=-\frac{\epsilon +P}{\tau}+\frac{\pi}{\tau}~,~~~~~~\frac{d\pi}{d\tau}=-\frac{\pi}{\tau_{\pi}}+\beta_{\pi}\frac{4}{3\tau}-\frac{(4+\lambda)}{3}\frac{\pi}{\tau}~,
\label{eqhydro}
\ee
with initial time, temperature and viscous pressure at $\tau_i=0.4$ fm, $T_i=0.5$ GeV and $\pi_i=0$. Fig.(\ref{figMDRTA1}) (right panel) shows the proper time $(\tau)$ evolution of pressure anisotropy 
$P_L/P_T=(P-\pi)/(P+\pi/2)$ for the three values of $n$ and four sets of $\eta/s$ ratio. The $n=0$ case as shown in \cite{Jaiswal:2013npa} under predicts the BAMPS data \cite{El:2009vj} which becomes prominent 
for large values of viscosity. $n=1$ case clearly over predicts the data a good deal showing even larger deviation from BAMPS for high $\eta/s$. However, the $n=1/2$ situation 
remarkably agrees with BAMPS results even with large viscosity like $\eta/s=3.0$ throughout the evolution range. In \cite{Dusling:2009df} the $n=1/2$ momentum dependence has been related to the dynamics of a 
two-flavoured quark-gluon gas where the BAMPS data has been extracted for the same by a parton cascade model \cite{Xu:2004mz}. This reasonable agreement of numerical data with fractional power of momentum dependence 
is very illuminating in the context of Ref \cite{Dusling:2009df} which argued that most of the interaction theories relevant for QGP lie between the two extreme limits of linear ($n=0$) and quadratic ($n=1$) ansatz. 

\subsection{Stability and causality of a first order theory}
Recently, studies have been carried out that propose a causal and stable first order hydrodynamic theory~\cite{Bemfica:2017wps,Bemfica:2019knx,Kovtun:2019hdm,Hoult:2020eho,Bemfica:2020zjp}. Motivated from these 
studies, in this work, the stability and causality of the first order theory derived in previous section has been tested. At local rest frame for small wave number ($k$) we have the frequency modes,
$\omega^{\paral}_{1,2}=\frac{i}{2}\left[\frac{4\eta/3+\zeta}{(\epsilon_0+P_0)}\right]k^2\pm k c_s~,~
\omega^{\paral}_3=i \left[\frac{\lambda T}{(\epsilon_0+P_0)}\right] k^2
\label{mode12}$,
which are always stable because of the positive imaginary parts of all the modes. At large $k$, the modes and the associated group velocity $v_g$ become,
\be
\omega^{\paral}_{1,2}=\frac{i}{2}\bigg\{\frac{B}{A}-\frac{E}{D}\bigg\}\pm k\sqrt{\frac{D}{A}}~,~~~
\omega^{\paral}_3=i\frac{E}{D}~,
\label{mode22}~~~
v_g=\lim_{k\rightarrow\infty}\bigg\vert\frac{\partial \textrm{~Re}({\omega})}{\partial k}\bigg\vert=\sqrt\frac{D}{A}~.
\ee
(detailed expressions are given in \cite{Mitra3}). $v_g$ turns out to be subluminal with small mass and non-zero values of the exponent $n$ of MDRTA (Eq.\eqref{RTE1}), giving rise to a causal propagating mode 
previously absent in Navier-Stokes (NS) theory. However, with a boosted background with arbitrary velocity $\textbf{v}$, at shear channel with small $k$ we have,
$\omega^{\perp}_1=\textbf{v} k + {\cal{O}}(k^2)~,~~~~~~\omega^{\perp}_2=-\frac{i}{\gamma \Gamma  \textbf{v}^2}+\frac{(2-\textbf{v}^2)}{\textbf{v}}k+{\cal{O}}(k^2)$,
with $\Gamma=\eta/(\epsilon_0+P_0)$ and $\gamma = 1/\sqrt{1-\textbf{v}^2}$. At large $k$  the shear modes becomes, $\omega^{\perp}_{1,2}=\frac{1}{\textbf{v}}k$. For a background velocity $0<\textbf{v}<1$, these modes
are both acausal and unstable. In small $k$ limit, the sound modes become,
\begin{align}
&\omega^{\paral}_1=\textbf{v} k+{\cal{O}}(k^2)~,~~\omega^{\paral}_{2,3}=\frac{1}{2}\left[M\pm \sqrt{M^2-4N}\right]k+{\cal{O}}(k^2)~,\nn
&\omega^{\paral}_{4,5}=\frac{i}{2}\left[Q\pm \sqrt{Q^2+4R}\right]+{\cal{O}}(k)~.
\label{sound3}
\end{align}
The additional modes $\omega^{||}_{4,5}$ are unstable for any combination of the field correction coefficients. In the limit of large wave numbers, the roots of $v_g^{\paral}$ are obtained as,
\be
v_{g,1}^{\paral}=\textbf{v}~,~~~v_{g,2,3}^{\paral}=\frac{\left[\textbf{v}(A-D)\pm \sqrt{AD-2AD\textbf{v}^2+AD\textbf{v}^4}\right]}{A-D\textbf{v}^2}~,~~~v_{g,4,5}^{\paral}=\pm \frac{1}{\textbf{v}}~.
\ee
where the two new roots $v_{g,4,5}^{\paral}$ are always acausal for $0<\textbf{v}<1$. So although at local rest frame the asymptotic causality condition and stability criteria are maintained, the new modes of 
shear and sound channels due to the boosted background are conclusively showing that the theory is acausal and unstable.

\subsection{Summary and discussions}
In this work, momentum dependent relaxation time approximation has been used to redefine the thermodynamic fields in order to include the out of equilibrium dissipative effects up to second order in gradient 
correction. The key finding is that, these corrections are not independent but constrained to give the dissipative flux of same tensorial rank where the associated coefficients are sensitive to the interaction. 
The derived equations have been applied for hydrodynamic simulation for a conformal system which demonstrates that the pressure anisotropy for the fractional power of MDRTA shows an impressive agreement with the 
numerical solution of Boltzmann equation. Finally, motivated from a series of recent studies, the causality and stability of the first order theory have been analyzed. In local rest frame, the equations of motion 
give a causal propagating mode which was previously absent in the usual NS theory but with a boosted background new modes appear which are both acausal and unstable. Hence, in order to establish a first order 
relativistic, stable-causal theory, an alternate microscopic approach of field redefinition is required which has been recently explored in \cite{Biswas:2022cla}.


\section{Recent results in small systems from CMS}
\author{Prabhat R. Pujahari (for the CMS collaboration)}	

\bigskip

\begin{abstract}
The observation of a wide variety of physical phenomena in the context of the formation of a 
strongly interacting QCD matter in heavy-ion nuclear collisions at the LHC has drawn significant 
attention to the high energy heavy-ion physics community. The appearance of a varieties of similar 
phenomena as in heavy-ion in the high multiplicity proton-proton and proton-nucleus collisions at 
the LHC energies has triggered further investigation to understand the dynamics of particle production 
mechanism in a highly dense and small QCD medium. The CMS 
collaboration uses many different probes in these studies ranging from the particle production cross 
section to multi-particle correlations. In this proceeding, I report a few selected recent CMS results 
from the small systems with the main focus on the measurement of collective phenomena in high 
multiplicity pp and pPb collisions.

\end{abstract}
\subsection{Introduction}
In the context of high energy heavy-ion physics, the collisions between protons or a proton with a 
nucleus is commonly referred to as small
system and they can provide baseline measurements for heavy-ion collisions. 
Traditionally, it is thought that such small systems do not show characteristics of QGP formation a priori.
However, in the recent few years, this simplistic view of a small system has been
challenged at the LHC -- thanks to the new frontier in energies and state-of-the-art
instrumentations.  The individual events in a high multiplicity pp and pPb
collisions can have very high charged particle multiplicity and energy density which is
comparable to that of $AA$ collisions~\cite{dEtapaper}. 

With the advent of the LHC, high multiplicity pp and pPb collisions show unexpected phenomena which
have never been observed before in such small systems. 
The observation of a long range rapidity ridge in the measurement
of two-particle angular correlation in heavy-ion collisions is no surprise to us and this can be well explained by
hydrodynamical collective flow of a strongly interacting and expanding medium~\cite{PLB2013}. However, the appearance of similar structures
in a high multiplicity pp and pPb collisions has drawn a lot of attention and prompted studies to understand the cause of such behaviour
in small systems. In particular, the discovery of the ridge by CMS collaboration~\cite{CMS} in high multiplicity pp collisions is
one of such intriguing results observed in small systems~\cite{PLB2013pp}.
Long-range, near-side angular correlations in particle production emerged in pp and subsequently in pPb collisions
paved the way for a systematic investigation of the existence of the collective phenomena.
Much information can also be gained by focusing on collective properties of each event, such as multi-particle correlations, or event-by-event fluctuations of such quantities.
We observe signatures traditionally attributed to a collective behaviour not only in PbPb collisions but also in small systems. Since then, a wealth of new,
unexpected phenomena has been observed with striking similarities to heavy-ion observations.

\subsection{Transverse energy density}
The total transverse energy, $E_{T}$, is a measure of the energy
liberated by the ``stopping'' of the colliding nucleons in a heavy-ion or
proton-nucleus collision.
From Figure~\ref{fig1} it can be seen that
$dE_{T}/d\eta$ $\mid_{\eta=0}$ $\approx 22$ GeV.
This is 1/40  of the value observed
for the 2.5$\%$ most central PbPb collisions. However, since the cross sectional
area of pPb collisions is much smaller than that of central PbPb collisions,
this result implies that the maximum energy density in pPb collisions is comparable
to that achieved in PbPb collisions~\cite{dEtapaper}. 
Several modern generators are compared to these results but none is
able to capture all aspects of the $\eta$ and centrality dependence of the data~\cite{dEtapaper}.
\begin{figure}[!ht]
\begin{center}$
\begin{array}{cc}
\includegraphics[width=2.3in,height=14em]{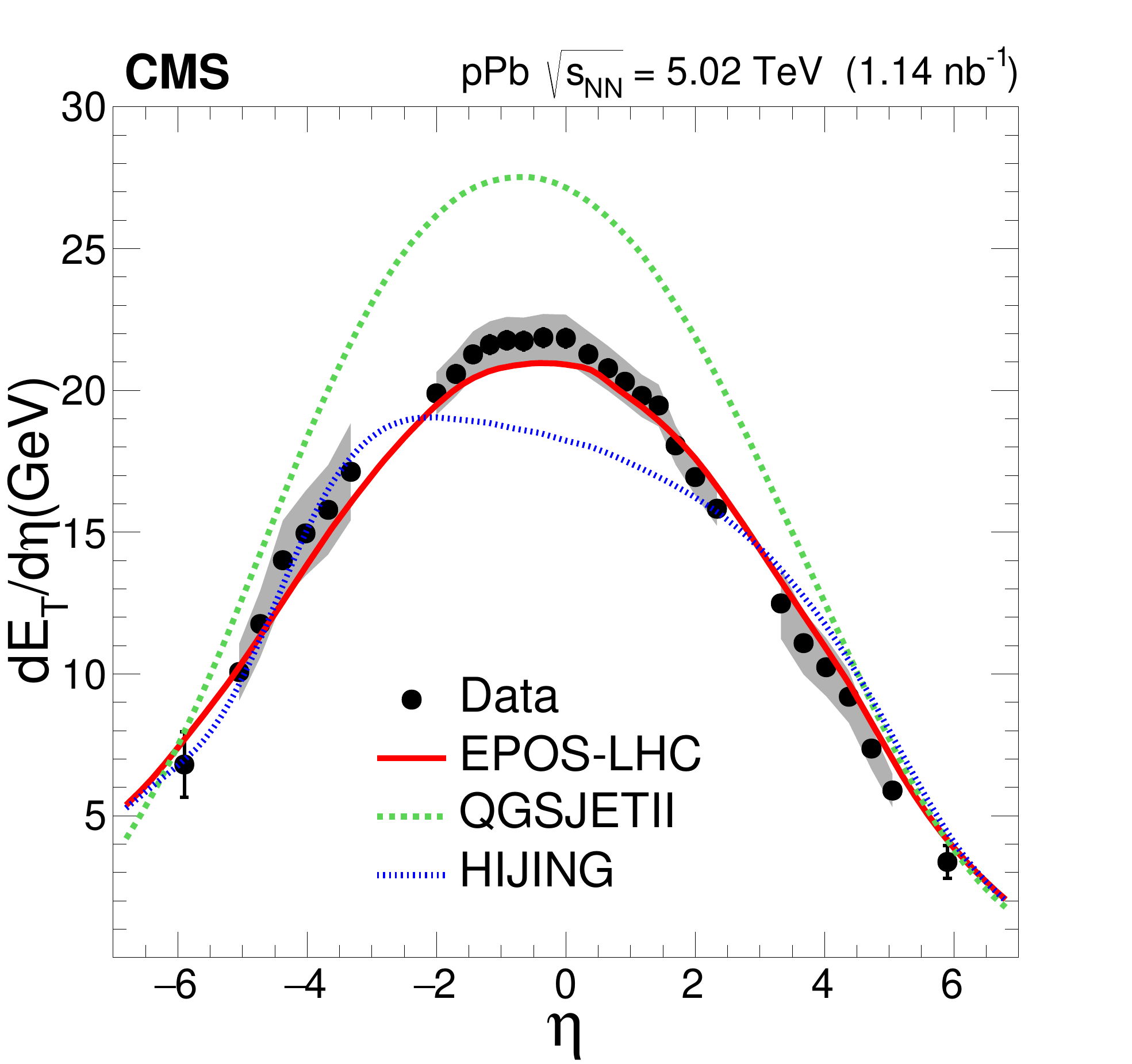} &
\includegraphics[width=2.3in,height=14em]{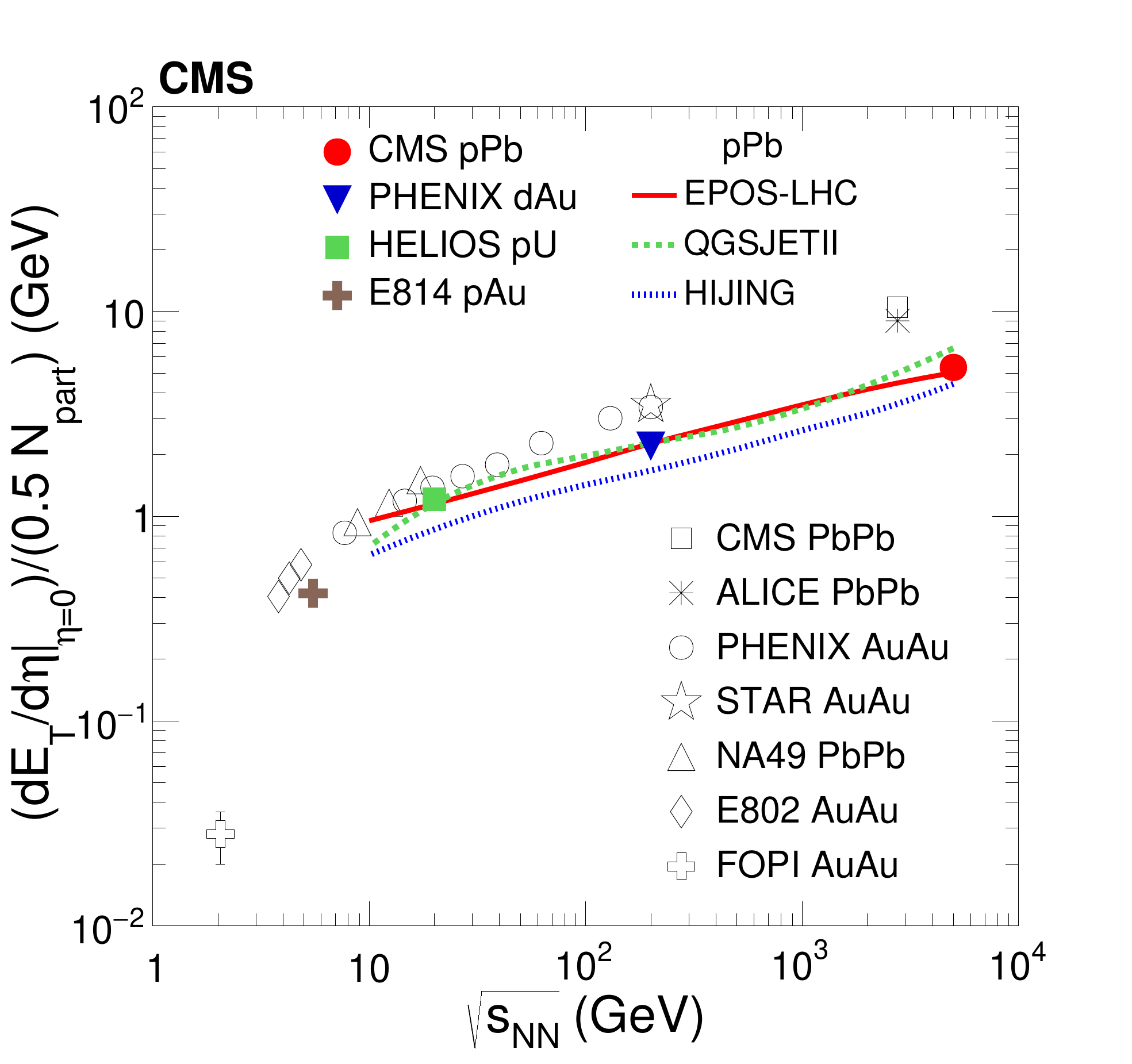}
\end{array}$
\end{center}
\caption{(left) Transverse energy density versus $\eta$ from minimum bias pPb collisions at $\sqrt{s_{NN}}=$5.02 TeV. (right) Transverse energy density per participating nucleon-nucleon pair evaluated at various $\sqrt{s_{NN}}$ for minimum bias pAu, pU, dAu, and pPb collisions~\cite{dEtapaper}.}
\label{fig1}
\end{figure}
\subsection{Collectivity in small systems at the LHC}
The $p_{\rm T}$ distributions of identified hadrons are one of the important tools to probe the collective behaviour of particle production.
The $p_{\rm T}$ distributions in pp and pPb collisions show a clear evolution, becoming harder as the multiplicity increases~\cite{spectra}.
As it is shown in Figure~\ref{fig2}, models including hydrodynamics describes the data better for the $p_{\rm T}$ spectra.
Data-to-model agreement is good at higher charged particle multiplicity, $N_{ch}$. In addition, the evolution of the $p_{\rm T}$ spectra with multiplicity can be compared more directly by
measuring the average transverse kinetic energy, $\langle KE_{T} \rangle$~\cite{spectra}.
If collective radial flow develops, this would result in a characteristic dependence of the shape of the transverse momentum distribution on the particle mass.

\begin{figure}[!ht]
\begin{center}$
\begin{array}{cc}
\includegraphics[width=2.5in,height=14em]{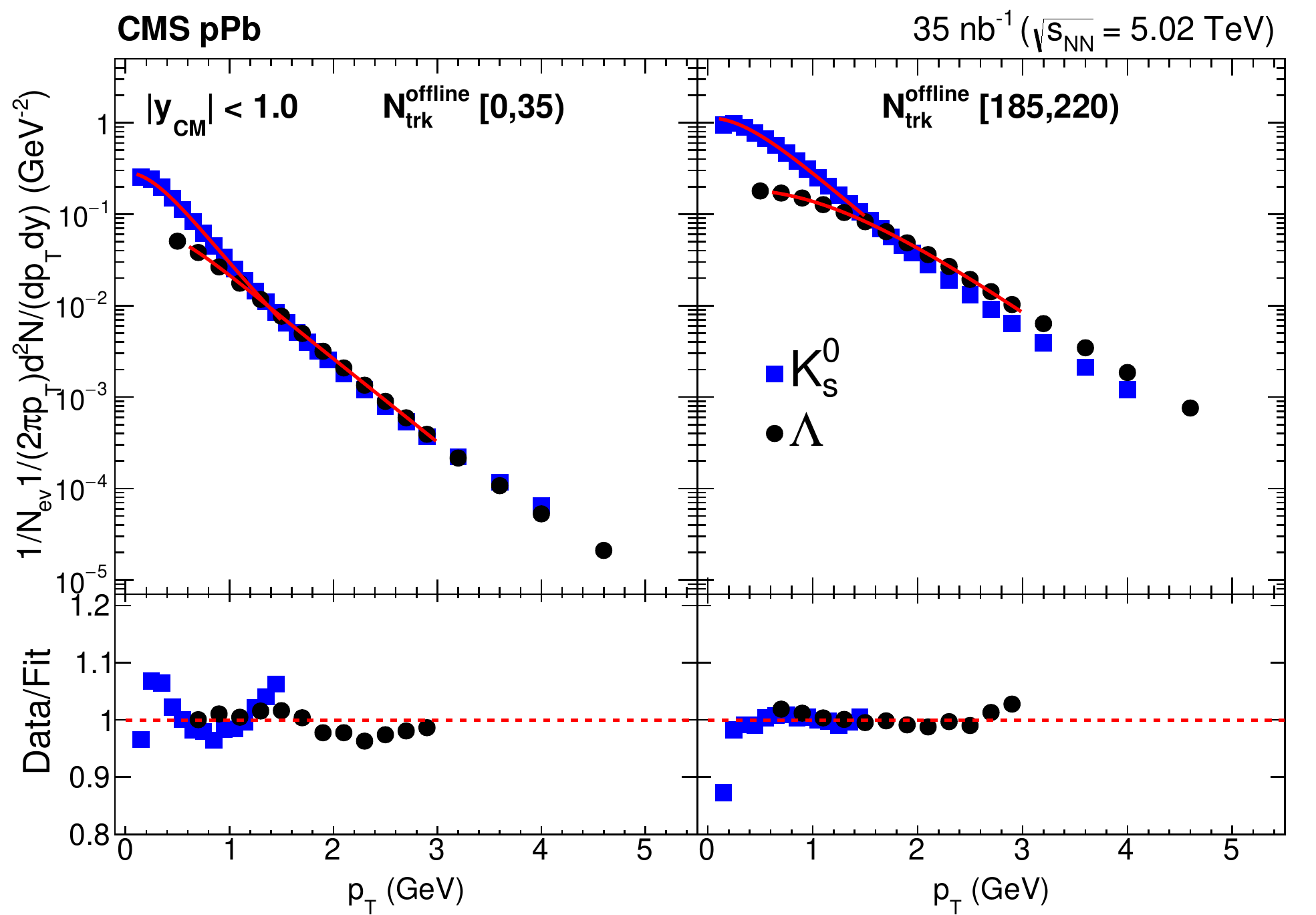} &
\includegraphics[width=2.4in,height=14em]{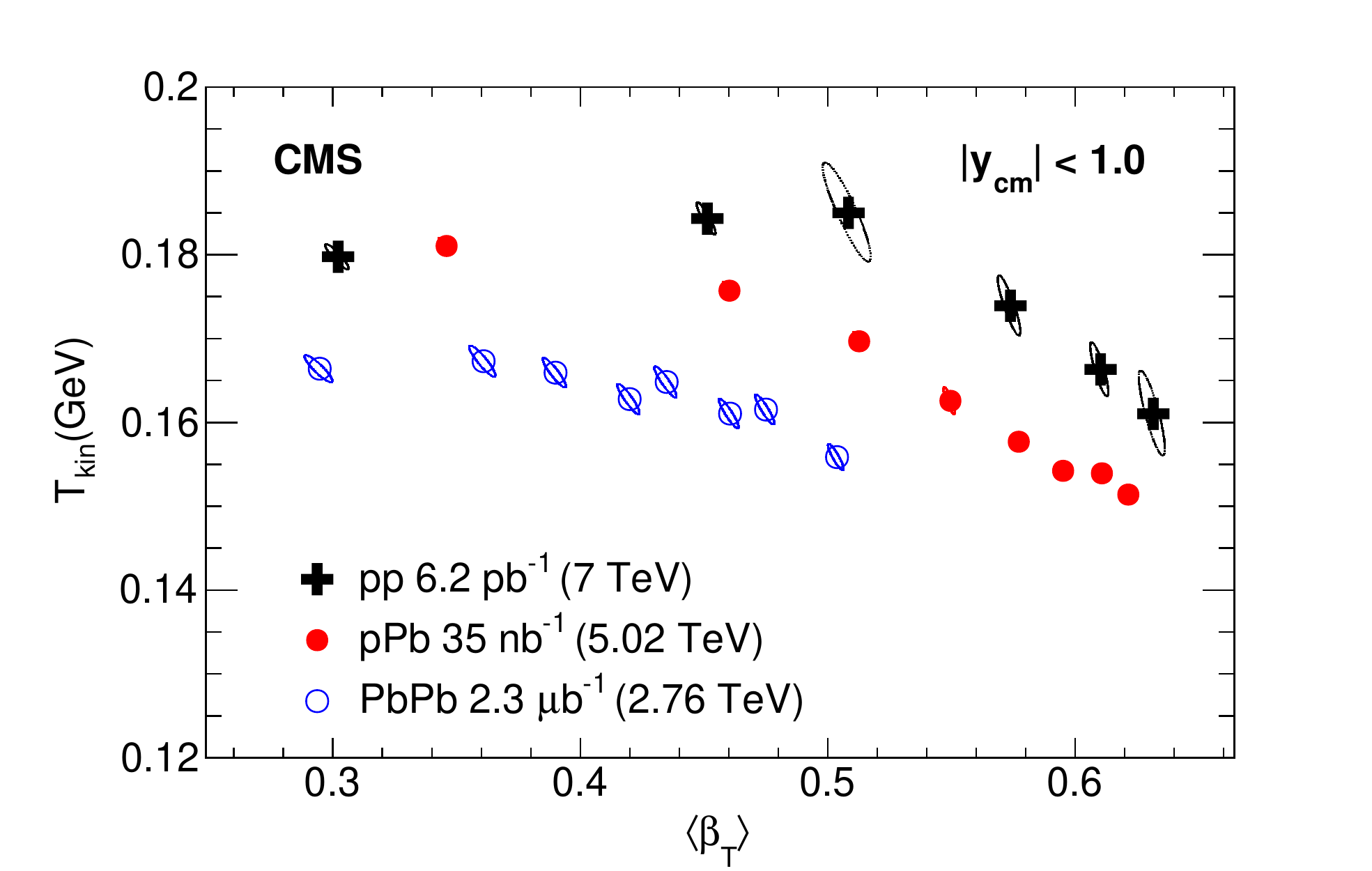}
\end{array}$
\end{center}
\caption{(left) Simultaneous blast-wave fits of the $p_{\rm T}$ spectra of ${K_{s}}^{0}$ and $\Lambda $ in low- and high-multiplicity pPb events.
(right) The extracted kinetic freeze-out temperature, $T_{kin}$, versus the average radial-flow velocity, $\langle \beta_{T} \rangle $, from a simultaneous 
blast-wave fit to the ${K_{S}}^{0}$ and $\Lambda $ $p_{\rm T}$ spectra at $|y_{cm}| < $ 1 for different multiplicity intervals in pp , pPb , and PbPb collisions.}
\label{fig2}
\end{figure}
The $\langle KE_{T} \rangle $ for ${K_{s}}^{0}$, $\Lambda $ and $\Xi$ particles as a function of multiplicity are shown in Figure~\ref{fig3}.
For all particle species,  $\langle KE_{T} \rangle $ increases with increasing multiplicity. A theoritical Blast-wave model~\cite{Blast} fits have also been performed to the $p_{\rm T}$ spectra of strange particles 
in several multiplicity bins as shown in Figure~\ref{fig2}. The interpretation of the parameters of these fits, such as kinetic freeze-out temperature, $T_{kin}$ and 
transverse radial flow velocity, $\beta_{T}$, are model dependent.
In the context of the Blast-Wave model, when comparing the parameters of different systems at similar $dN_{ch}/d\eta $, 
it was found that $\beta_{T} $ is larger for small systems i.e., $\beta_{T} {\rm (pp)} >$ $\beta_{T} {\rm (pPb)} >$ $\beta_{T}{\rm(PbPb)} $.
This could be an indication of a larger radial flow in small systems as a consequence of stronger pressure gradients due to a more explosive system.
However, a similar decreasing trend is observed for $T_{kin}$ and $\beta_{T}$  as a function of multiplicity in all three collision systems. 
\begin{figure}[ht!]
\includegraphics[height=15em]{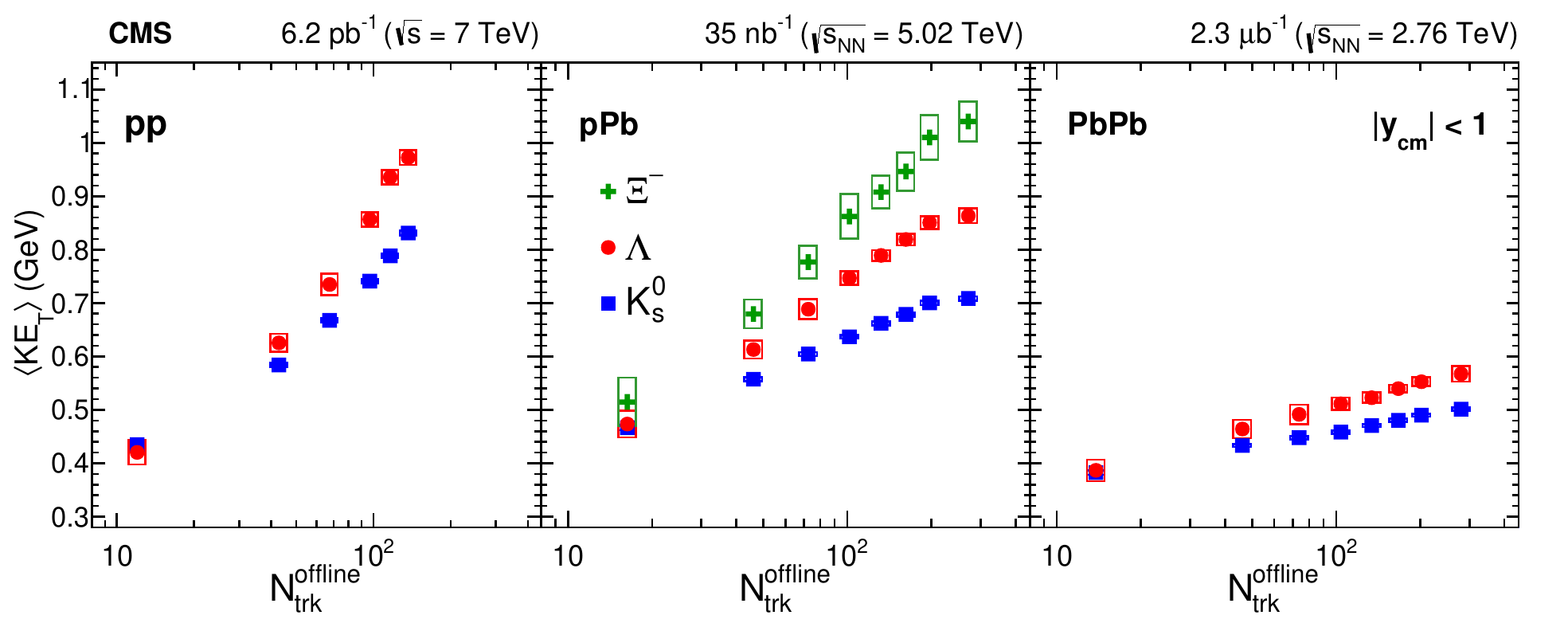}
\caption{The average transverse kinetic energy for ${K_{s}}^{0}$, $\Lambda $ and $\Xi$ particles as a function of multiplicity in pp, pPb, and PbPb collisions~\cite{spectra}. }
\label{fig3}
\end{figure}
One of the key questions about the nature of the ridge and its collectivity is whether the two-particle azimuthal correlation structures observed at large relative pseudorapidity in 
pp and pPb collisions result from correlations exclusively between particle pairs, or if it is a multi-particle genuine collective effect, needs to be further understood.
A strong hint for multi-particle correlations in high multiplicity pp and pPb collisions was reported by the CMS collaboration~\cite{multipart,prl115}. 
Figure~\ref{fig4} shows the second-order azimuthal anisotropy Fourier harmonics ($v_{2}$) measured in pp, pPb and PbPb collisions over a wide pseudorapidity 
range based on correlations calculated up to eight particles. The $v_{2}$ values stay high and show similar trends in all three systems. 
The $v_{2}$ computed from two-particle correlations is found to be larger than that obtained with four-, six- and eight-particle correlations, as well as the Lee-Yang zeroes method.
However, the $v_{2}$ obtained from multi-particle correlations, 
all yield to similar $v_{2}$ values i.e., $v_{2}\{4\} \approx v_{2}\{6\} \approx v_{2}\{8\} \approx v_{2}\{\rm LYZ\}$~\cite{multipart}.
These observations support the interpretation 
of a collective origin for the observed long-range correlations in high-multiplicity pp and pPb collisions.
\begin{figure}[ht!]
\includegraphics[height=15em, width=38em]{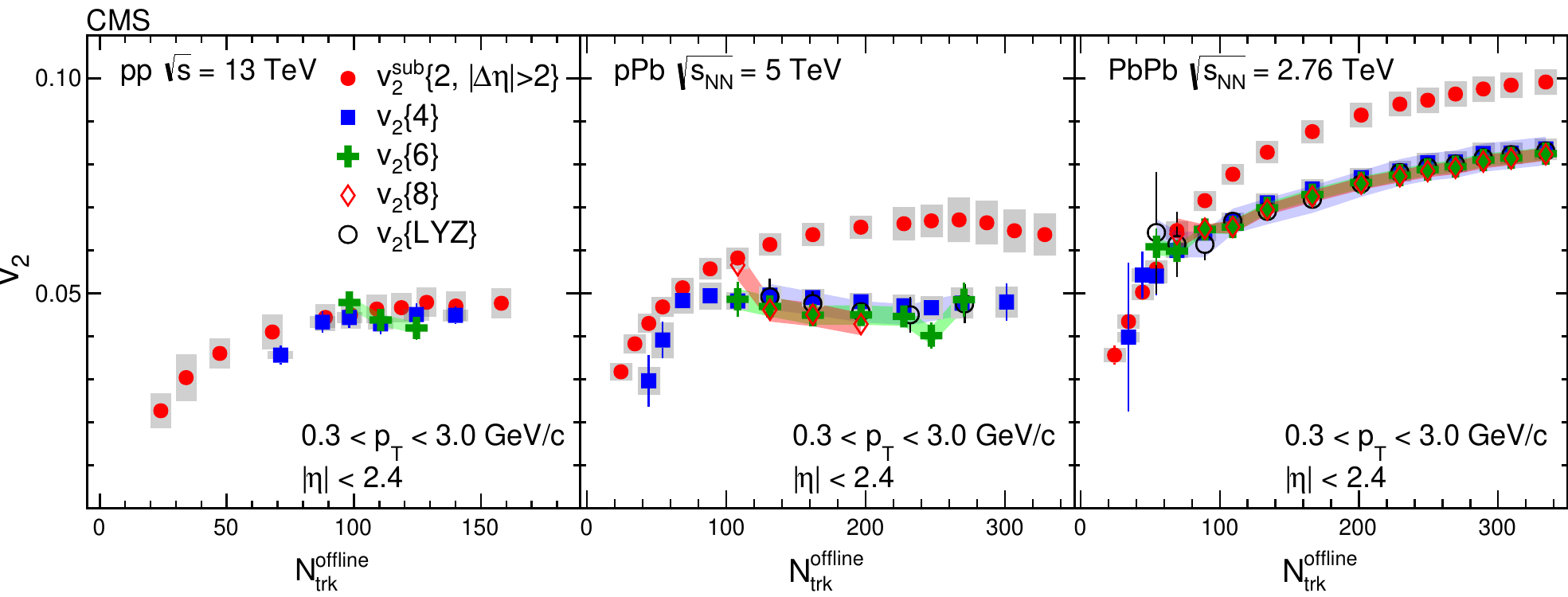}
\caption{ Second-order azimuthal anisotropy Fourier harmonics, $v_{2}$ measured by CMS in pp, pPb and PbPb collisions based on multi-particle correlations~\cite{multipart}.}
\label{fig4}
\end{figure}
Another useful observable in the study of collectivity is the event-by-event correlation between Fourier harmonics of different order flow coefficients. 
The CMS Collaboration has measured these normalized symmetric cumulants, $\rm SC(m, n)$, where $m$ and $n$ are different order flow coefficients, 
in pp, pPb and PbPb collisions, as a function of track multiplicity~\cite{prl120}.
Similar observations are made in all three systems. In the case of $\rm SC(2, 3)$, which gauges the correlation between $v_{2}$ and $v_{3}$, an anti-correlation 
is found at high track multiplicity, as shown in Figure~\ref{fig5}. On the contrary, $\rm SC(2, 4) >$ 0: the $v_{2}$ and $v_{4}$ values are positively correlated event-by-event. 
Similar trends are observed in pPb and PbPb collisions, and high multiplicity pp collisions, regarding the trend of these observables as a function of track multiplicity.
A long-range near-side two-particle correlation involving an identified particle is also observed~\cite{multipart,prl121}. 
Results for both pPb and pp collisions are shown in Figure~\ref{fig6}. 
Moving to high-multiplicity events for both systems, a particle species dependence of $v_{2}$ is observed. 
The mass ordering of $v_{2}$ was first seen in $AA$ collisions
at RHIC and LHC energies~\cite{STAR,PHENIX}, which can be understood as the effect of radial flow pushing heavier particles towards higher $p_{T}$. 
This behavior is found to be qualitatively
consistent with both hydrodynamic models~\cite{Hydro} and an alternative initial state interpretation~\cite{CGC}.
\begin{figure}[ht!]
\includegraphics[height=16em, width=38em]{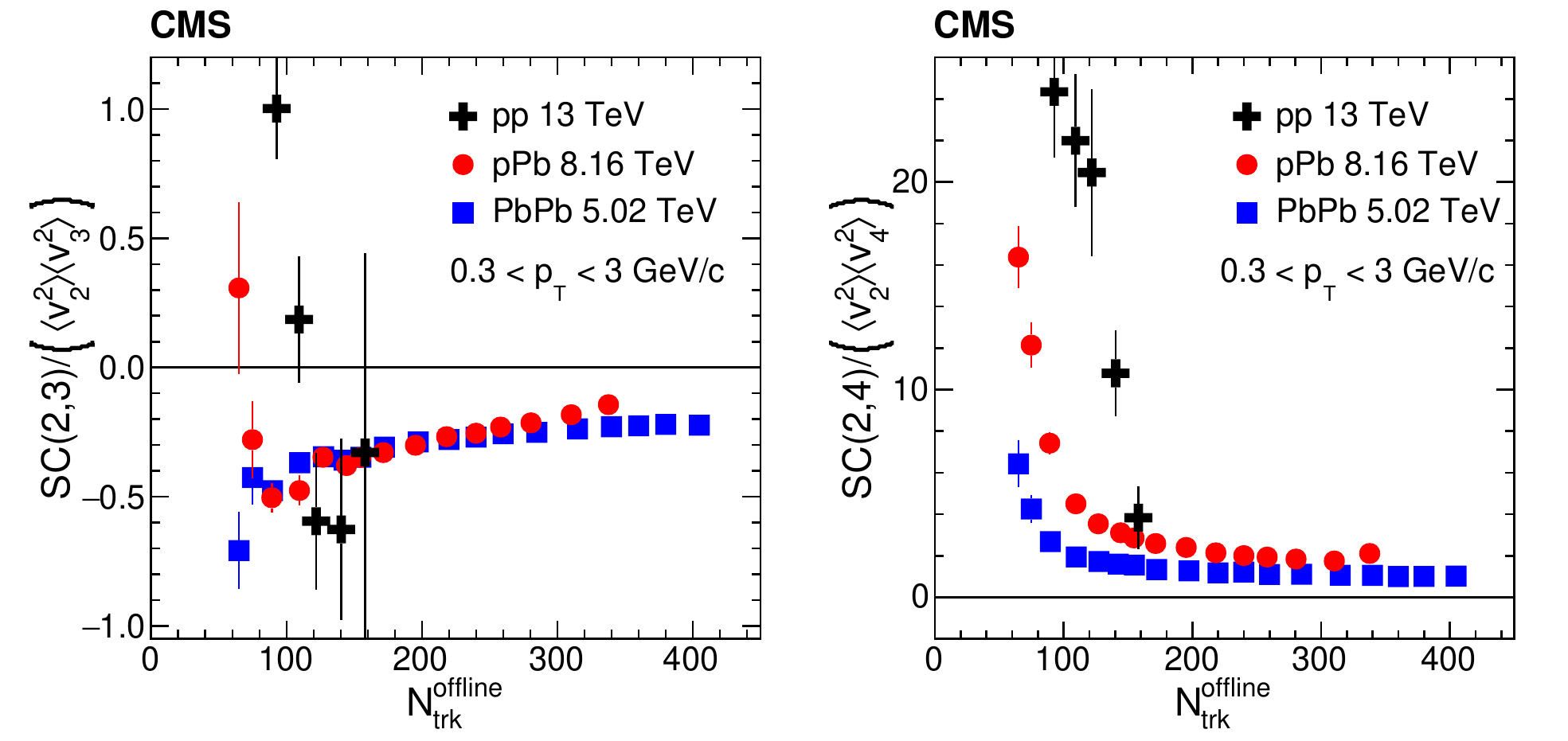}
\caption{ The normalized symmetric cumulant for the second and third coefficients (left) and the second and fourth coefficients (right) 
are shown for pp (black cross), pPb (red circle), and PbPb (blue square). Tracks with transverse momentum between 0.3 and 3.0 GeV are used~\cite{prl120}.} 
\label{fig5}
\end{figure} 
A measurement of the elliptic flow of prompt J/$\Psi$ meson in high-multiplicity
pPb collisions is reported by the CMS experiment~\cite{jpsi}. 
The prompt J/$\Psi$ results are compared with the $v_{2}$ values for open charm mesons ($D^{0}$) and strange
hadrons. As shown in Figure~\ref{fig6}, positive $v_{2}$ values are observed for
the prompt J/$\Psi$ meson, as extracted from long-range two-particle correlations with
charged hadrons, for $2 < p_{T} < 8 $ GeV. 
\begin{figure}[!ht]
\begin{center}$
\begin{array}{cc}
\includegraphics[width=2.4in]{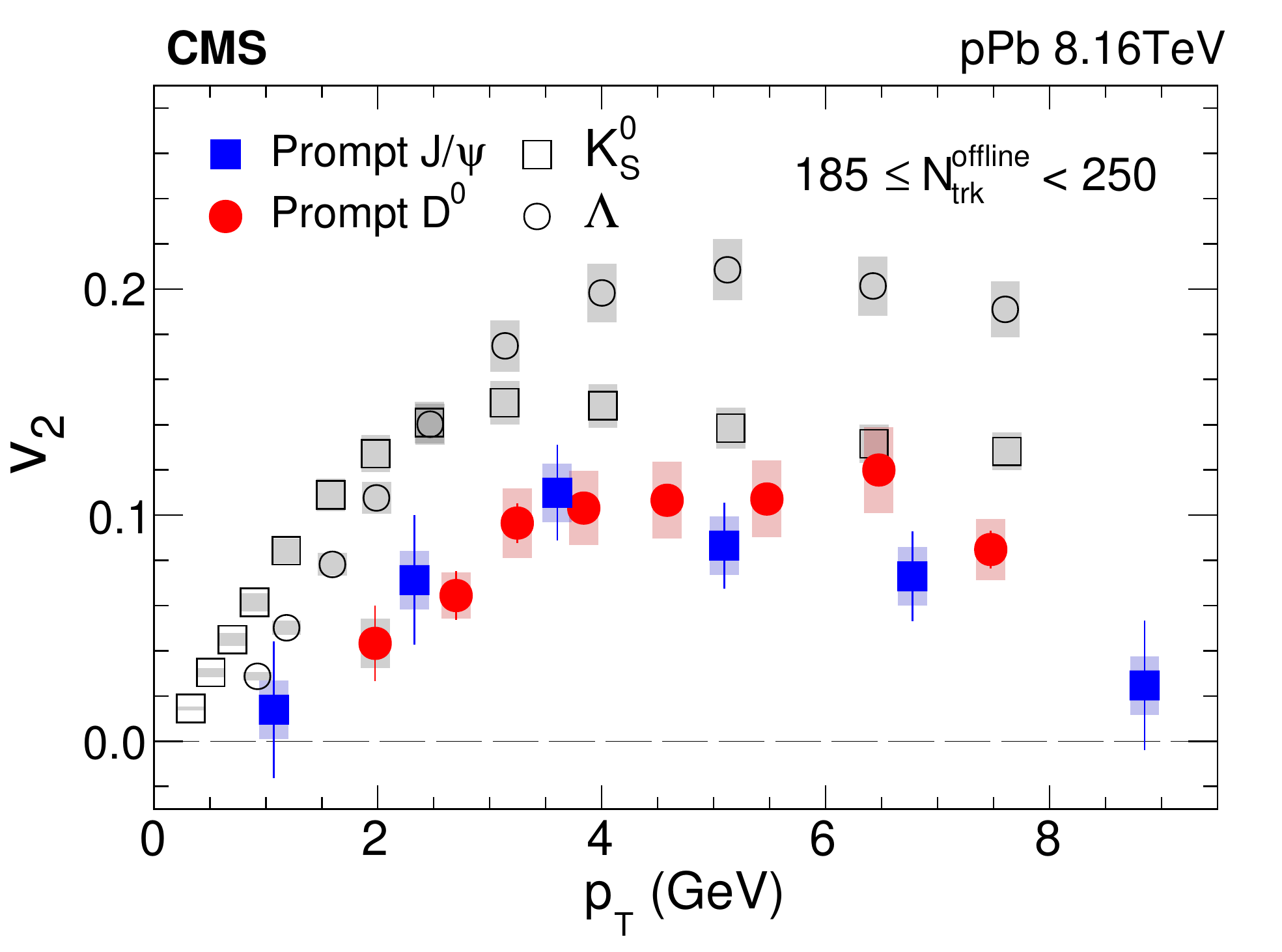} &
\includegraphics[width=2.0in]{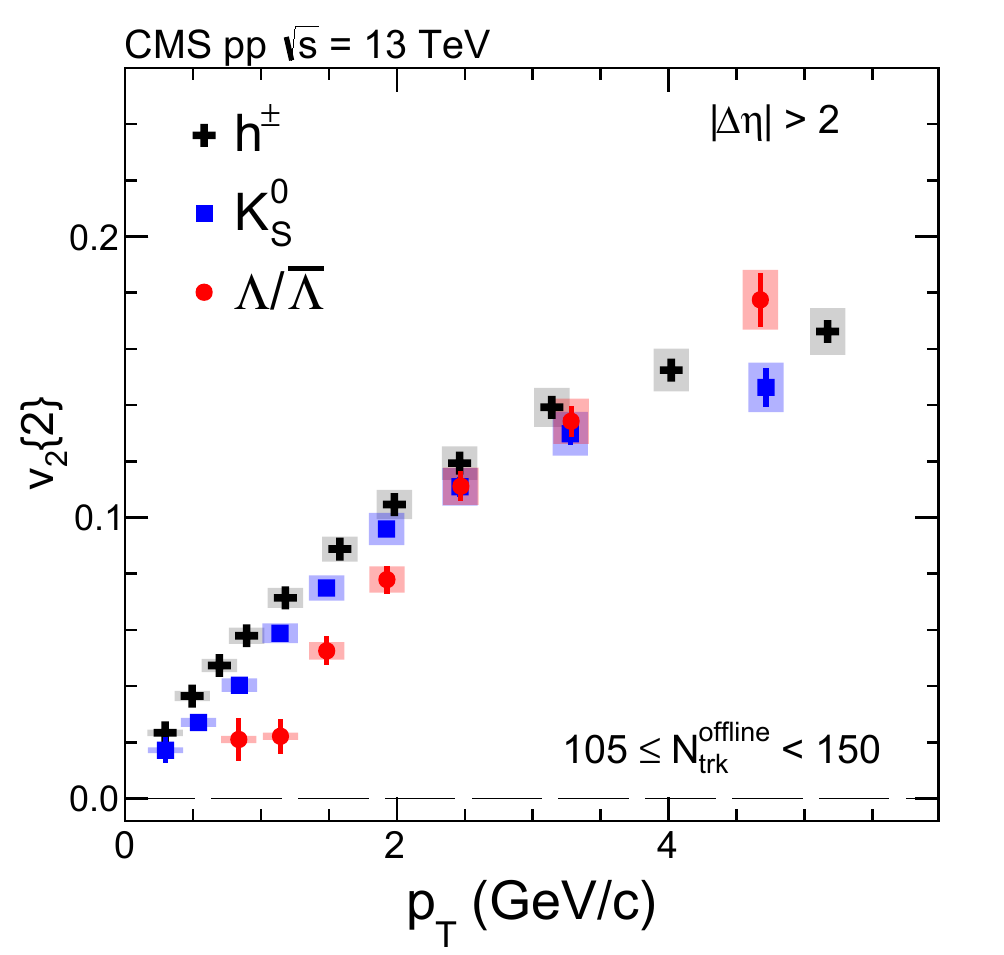}
\end{array}$
\end{center}
\caption{The $v_{2}$ results for ${K_{s}}^{0}$,
and $\Lambda $, prompt $D^{0}$ and prompt J/$\Psi$ in high-multiplicity pPb (left) events. (right) 
The $v_{2}$ results for inclusive charged particles, ${K_{s}}^{0}$,
and $\Lambda $ as a function of $p_{T}$ in pp collisions at $\sqrt{s}=$ 13 TeV.}
\label{fig6}
\end{figure}
The prompt J/$\Psi$ meson results, together
with results for light-flavor and open heavy-flavor hadrons, provide novel insights
into the dynamics of the heavy quarks produced in small systems that lead to high final-state
multiplicities.

\subsection{Conclusions}
Several effects, such as mass-dependent hardening of $p_{\rm T}$ distributions, near-side long-range correlations, multi-particle azimuthal correlations, etc, which in nuclear collisions are 
typically attributed to the formation of a strongly-interacting collectively-expanding quark-gluon medium, have been observed in high-multiplicity pp and pPb collisions at the LHC.
The study of small collision systems at high multiplicity is undoubtedly of considerable interest.
While a lot  of  progress  has  been  made  towards  understanding  the  long-range correlation  phenomena  in  small  colliding  systems,  there  are  still  many  
open questions to be addressed by the experiemental and theoritical communities.

\section{Heavy Quark Diffusion in QCD Matter: Glasma vs Plasma}
\author{Pooja, Marco Ruggieri, and Santosh Kumar Das}	

\bigskip

\begin{abstract}
	Heavy quarks (HQs) are considered potential probes of the quantum chromodynamics (QCD) matter produced in high-energy nuclear collisions. In the pre-equilibrium stage of relativistic heavy-ion collisions, strong quasi-classical gluon fields called Glasma emerge at about $\tau_0=0.08$ fm/c that evolves according to the classical Yang-Mills (CYM) equations. The diffusion of HQs, namely, charm and bottom quarks in the evolving Glasma fields is compared with that of the Markovian-Brownian motion in a thermalized medium. The diffusion of HQs in the evolving Glasma (EvGlasma) is investigated within the framework of Wong equations while we use famous Langevin equations for the Brownian motion with diffusion coefficients evaluated within the pQCD framework. We observe that for a smaller value of saturation scale, $Q_s$, the average transverse momentum broadening is approximately the same for the two cases, but for larger $Q_s$, Langevin dynamics underestimates the $\sigma_p$. This difference is related to the fact that HQs in the Glasma fields experience diffusion in strong, coherent gluon fields that lead to a faster momentum broadening due to memory, or equivalently to a strong correlation in the transverse plane. 
\end{abstract}

\subsection{Introduction}

The initial condition produced in the relativistic high-energy collisions and its evolution to quark-gluon plasma (QGP) is of prime importance to study the QCD matter in the extremum conditions. According to the color-glass condensate (CGC) effective theory, the collision of two nuclei at ultra-relativistic velocities results in strong longitudinal color electric and magnetic fields called Glasma\cite{glsm} which evolves according to classical Yang-Mills\cite{hq3} (CYM) equations. The typical lifetime of this pre-equilibrium Glasma phase is 0.2 -1 fm/c. Heavy quarks\cite{rap2018,Scardina:2017ipo,Das:2010tj,hq3, hq4}, namely, charm and beauty quarks, work as an excellent probe to study the early stages of high-energy collisions.

This research aims to do a systematic comparison of the diffusion of the HQs in the EvGlasma and a hot thermalized medium by fixing the saturation scale, $Q_s$, and the QCD coupling\cite{wong4_g}, $g$ in our calculations. For this, we compute the transverse momentum broadening defined as
\begin{equation}
\sigma_p = \frac{1}{2} \langle (p_x(t)-p_{0x})^2 + (p_y(t)-p_{0y})^2 \rangle.
\label{sigmap_p_definition}
\end{equation} 

\subsection{Formalism}

The diffusion of HQs in the EvGlasma is investigated by the means of Wong equations\cite{hq3,wong1, wong2,wong3_Lngvn,wong4_g,Sun:2019fud}:
\begin{eqnarray}
	\frac{dx^i}{dt}  &=&  \frac{p^i}{E},
	\label{Wong_X}\\
\frac{dp^i}{dt}  &=&  g Q_a F^{i\nu}_a  p_\nu  ,
\label{Wong_P}\\
\frac{dQ_a}{dt} &=& \frac{g}{E} f_{abc} A^{\nu}_b p_\nu Q_c.
\label{Wong_Qa}
\end{eqnarray}

The motion of HQs in the gluonic Plasma is studied using Langevin equations\cite{wong3_Lngvn}. We assume that this is a Markovian process with no drag included.

\subsection{Results}

\subsection{Momentum broadening in the static box}

\begin{figure}
\centering
\begin{minipage}{0.5\textwidth}
  \centering
  \includegraphics[scale=0.25]{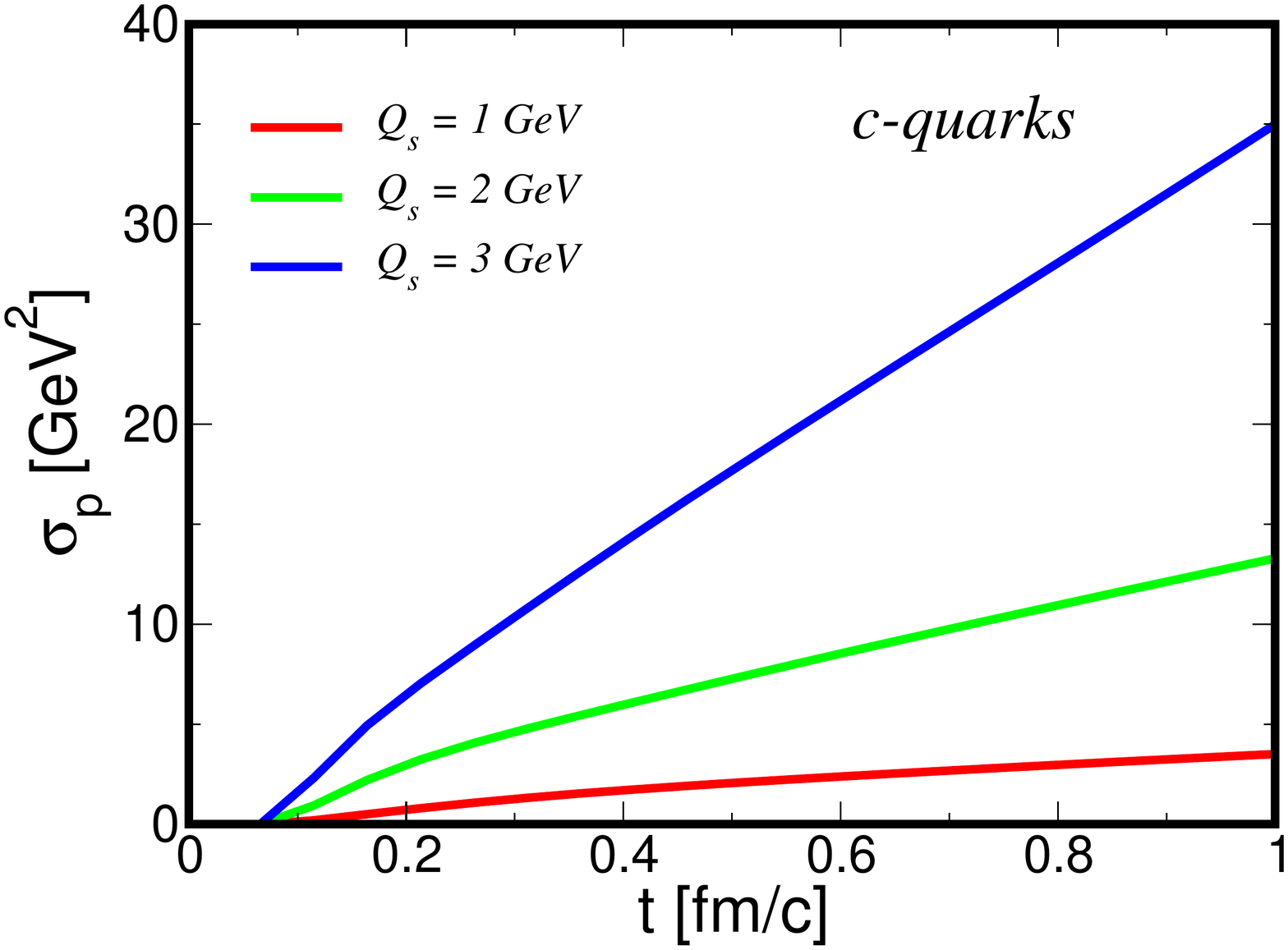}
\end{minipage}%
\begin{minipage}{0.5\textwidth}
  \centering
  \includegraphics[scale=0.25]{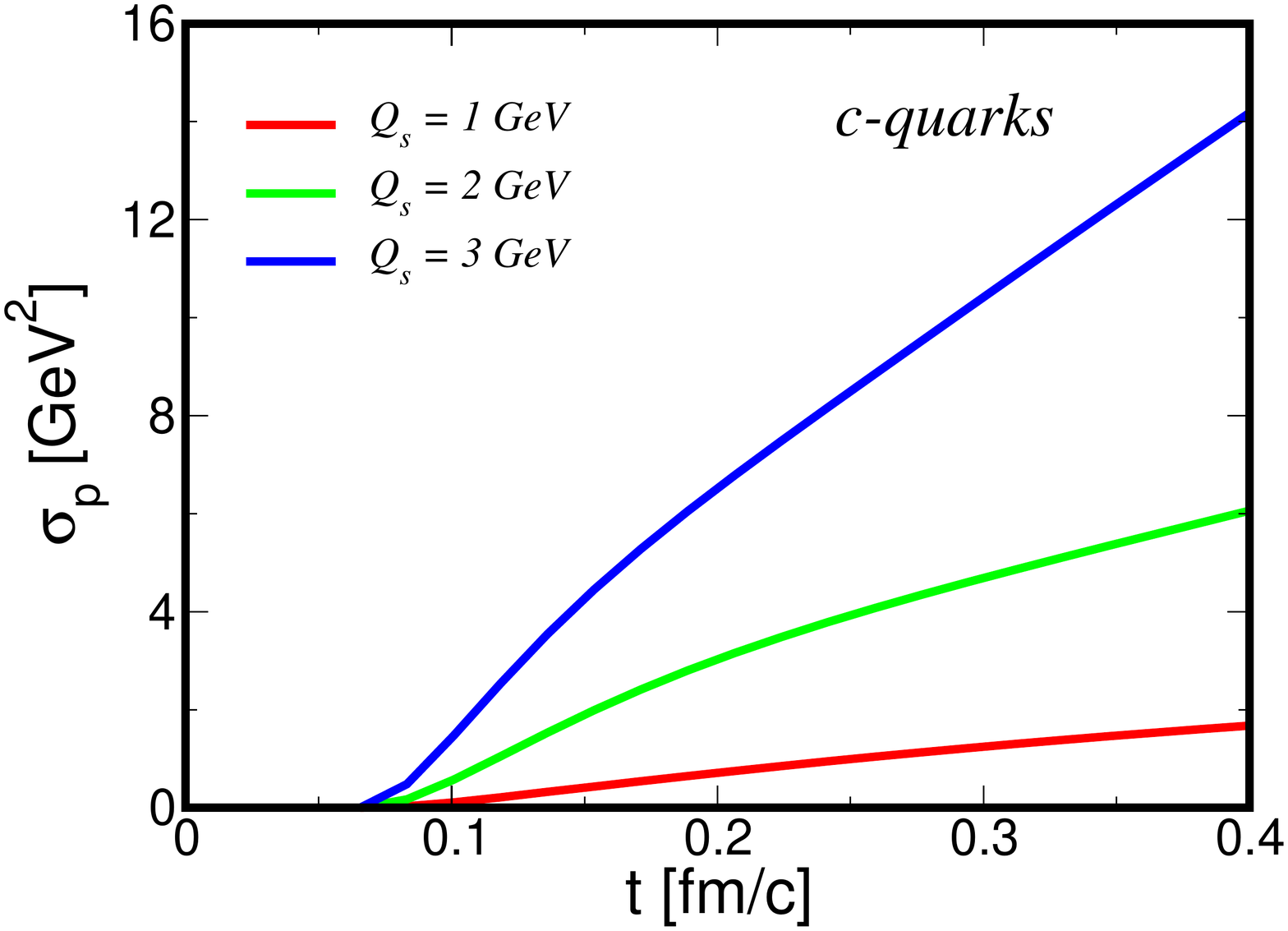}
\end{minipage}
\caption{$\sigma_p$ versus proper time for charm quarks, for the initial $p_T = 0.5$ GeV. The calculations correspond to evolving Glasma fields in a static box. }
\label{fig:Fig_1}
\end{figure}

In Fig.~\ref{fig:Fig_1}, we plot $\sigma_p$ for charm quarks versus proper time, for several values of $Q_s$, up to $\tau = 0.4$ fm/c and $\tau = 1.0$ fm/c. 
During the very early time, $\sigma_p$ doesn't increase linearly due to the correlations of the Lorentz force acting on the charm quarks at different times, namely to the memory of the gluon fields. The memory time, $\tau_{mem}$, has been calculated in Ref.~\citen{wong3_Lngvn}: $\tau_{mem} \approx \frac{1}{Q_s}$.
After the initial transient, $\sigma_p$ rises linearly, similar to the standard Brownian motion without a drag.

\subsection {Comparison with the Langevin dynamics}

\begin{figure}
\centering
\begin{minipage}{0.5\textwidth}
  \centering
  \includegraphics[scale=0.25]{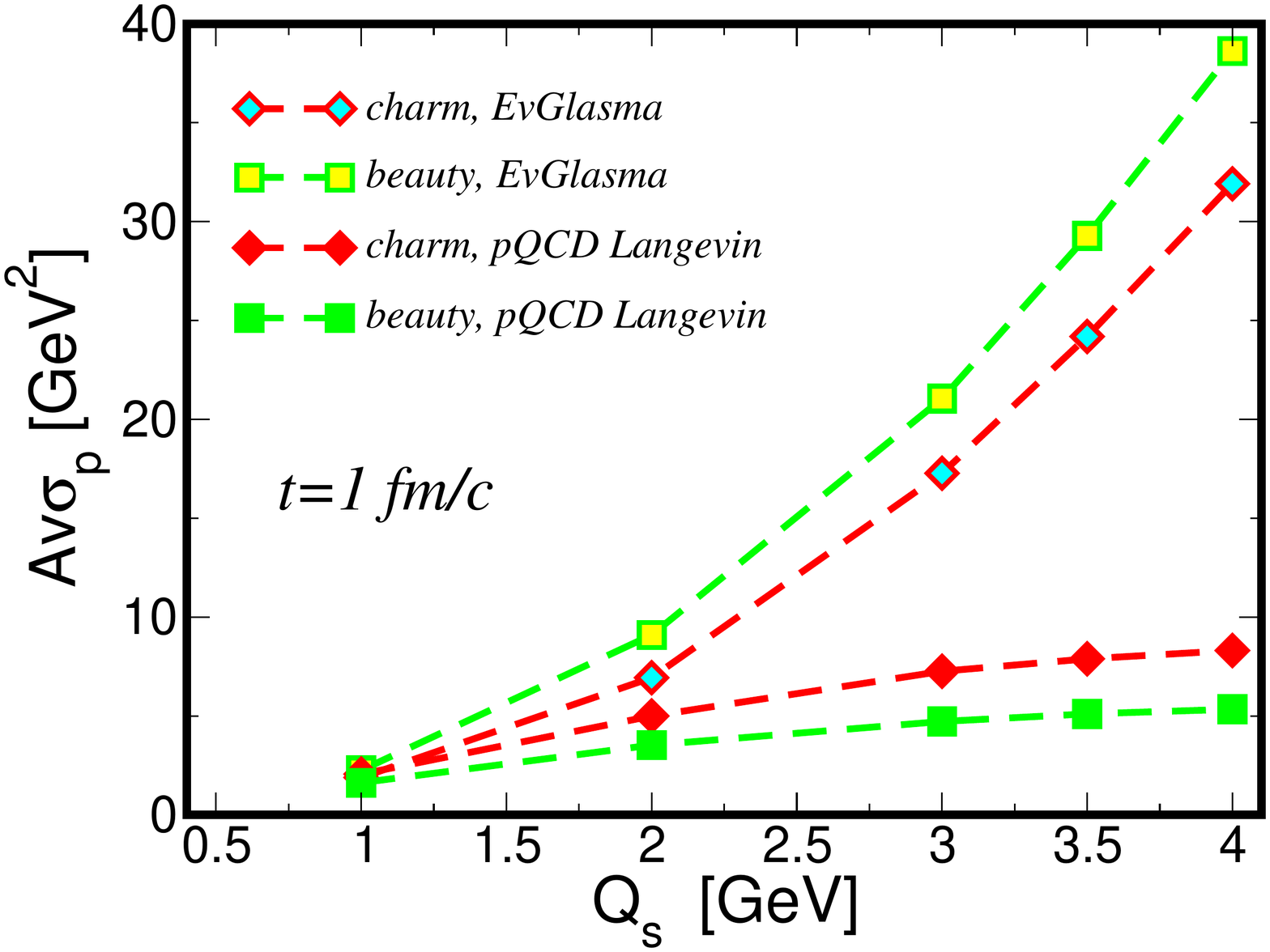}
\end{minipage}%
\begin{minipage}{0.5\textwidth}
  \centering
  \includegraphics[scale=0.25]{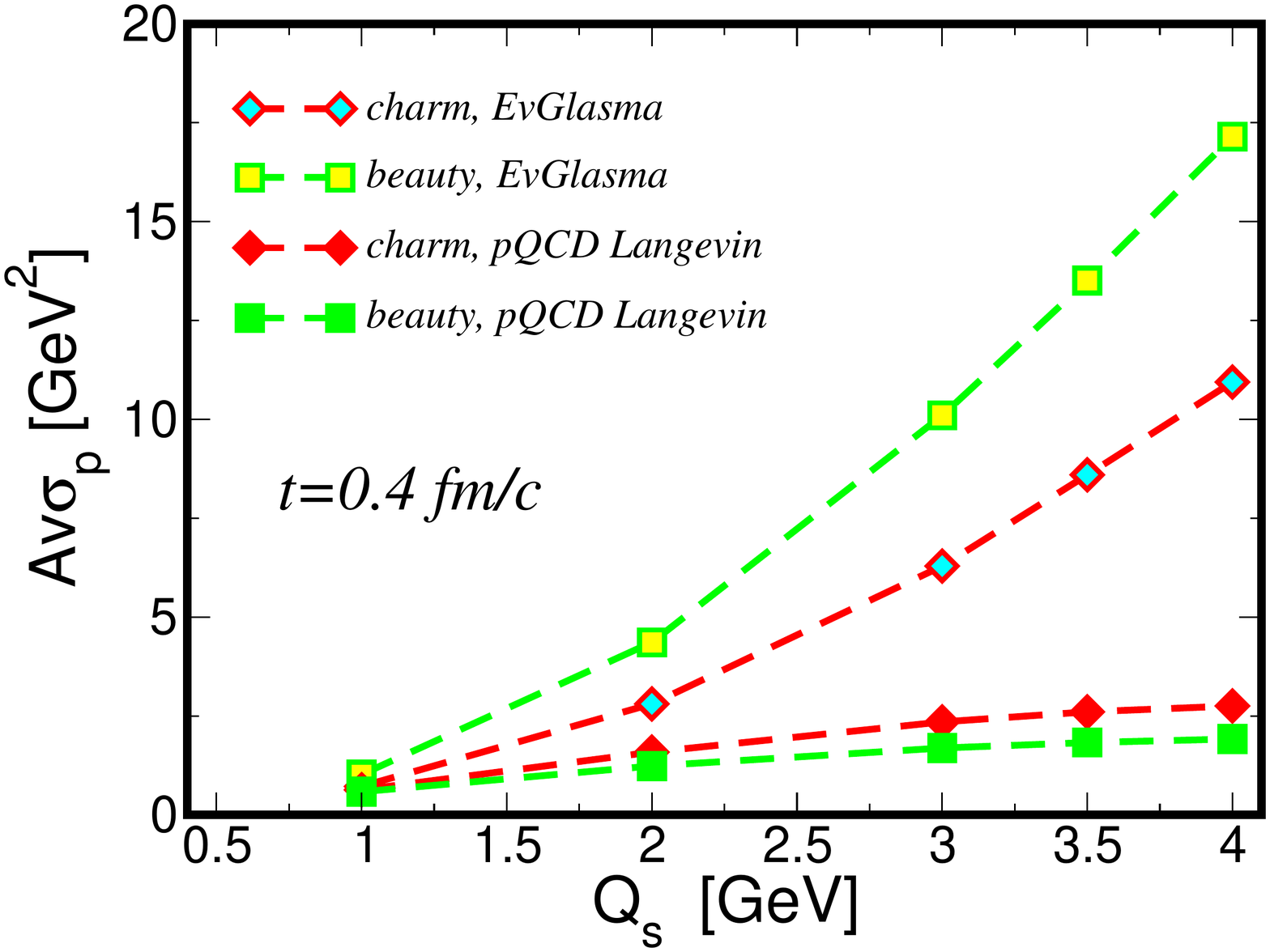}
\end{minipage}
\caption{Time-averaged $\sigma_p$ versus $Q_s$ for charm and beauty quarks, for EvGlasma and pQCD Langevin dynamics. Calculations correspond to the static box geometry. }
\label{fig:Fig_8}
\end{figure}

In Fig.~\ref{fig:Fig_8}, we plot the time-averaged transverse momentum broadening, Av$\sigma_p$ versus $Q_s$ for charm and beauty quarks. We find that for smaller $Q_s$, Av$\sigma_p$ is comparable for EvGlasma and collisional Langevin dynamics. It is because, for smaller $Q_s$, EvGlasma behaves like a system of dilute gluons resulting in momentum diffusion which is similar to the collisional dynamics. On the other hand, for larger $Q_s$, the HQs in the EvGlasma feel strong coherent gluonic fields, while the dynamics remain the same as collisional for pQCD Langevin. Hence, the difference between the two systems is quite substantial.

\begin{figure}[h]
	\begin{center}
		\includegraphics[width = 7cm]{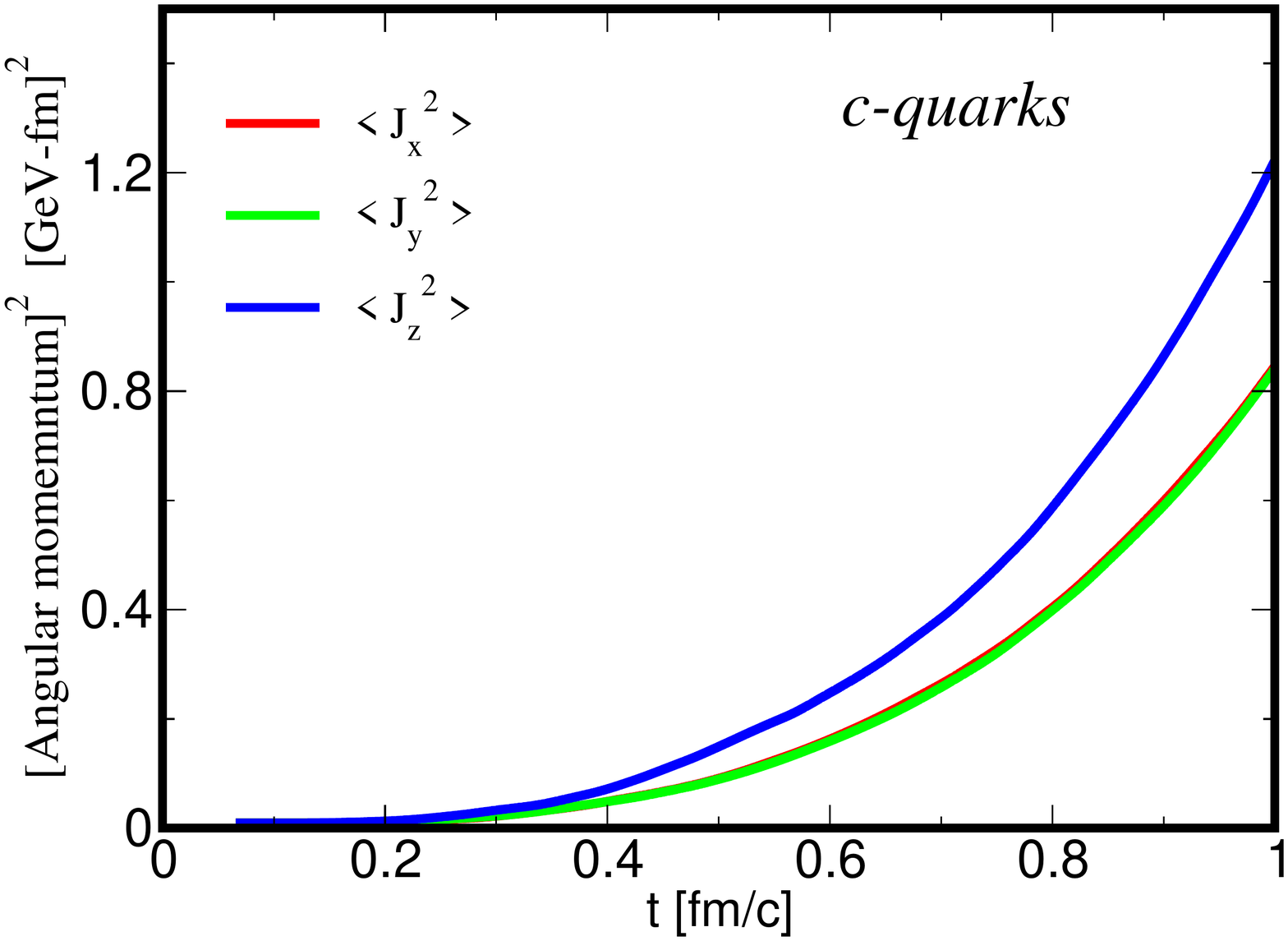}
		\caption{Average of square of angular momentum components versus proper time for charm quarks, for $Q_s = 2$ GeV .}
		\label{fig:Fig_J}
	\end{center}
\end{figure}

\subsection {Spin polarisation of heavy quarks in the evolving Glasma}

In Fig.~\ref{fig:Fig_J}, we show a preliminary result for the evolution of averaged $J_x^2$, $J_y^2$ and $J_z^2$ of HQs in the evolving Glasma in a static box geometry. Starting with a non-polarised system of HQs, the Glasma dynamics result in the polarization of HQs, i.e., HQs have more spin in the longitudinal $z$-direction as compared to the spin in the transverse $x$ and $y$ directions.

\subsection{Conclusions and Outlook}
Glasma, the pre-equilibrium stage of high-energy nuclear collisions, affects the HQs dynamics significantly. We observe that the diffusion of HQs in the early stage of high energy collisions is affected by the strong coherent gluon fields and memory effects become substantial.
The time-averaged momentum broadening of HQs in the EvGlasma is in agreement with the standard pQCD-Langevin for small values of $Q_s$, while differs significantly for larger $Q_s$.

Spin polarization of HQs is another interesting aspect to be explored. The first results in this direction tell that HQs spin is polarized in the longitudinal direction.

\section{The Impact of Memory on Heavy Quarks Dynamics in Hot QCD Medium}
\author{Jai Prakash, Marco Ruggieri, Pooja, Suvarna Patil, Santosh Kumar Das}	

\bigskip

\begin{abstract}
We study the effect of memory on the heavy quarks (HQs) dynamics in the  Quark-Gluon Plasma (QGP) within the scope of integro-differential Langevin where the memory enters through the thermal noise, $\eta$, and the dissipative force. We assume that the time correlations of the $\eta$ decay exponentially over a time scale called the memory time, $\tau$. We have observed the significant impact of memory on transverse momentum broadening, $\sigma_{p}$, and the nuclear modification factor, $R_{AA}$ of the HQs dynamics in QGP. We notice that the HQs dynamics are very sensitive to memory. 
\end{abstract}

\subsection{Introduction}

In an ultra-relativistic heavy-ion collision, the existence of hot and dense nuclear matter, QGP, has been realized at Relativistic Heavy-Ion Collider (RHIC) and the Large Hadron Collider (LHC). The HQs are one of the novel probes~\cite{rap2018,cao, Plumari:2017ntm} to study the evolution of QGP. The estimated thermalization time of the HQs is greater than the QGP lifetime, which makes HQs a witness to the entire evolution of QGP. The dynamics of the HQs in QGP  are usually studied within the framework of the standard Langevin equation, where the time correlation of noise is the delta function. We have studied the dynamics of HQs in the bath of time-correlated thermal noise within the framework of the Langevin equation, where the time correlation of thermal noise is an exponentially decaying function over a particular time span, $\tau$. The drag coefficient is related to the thermal noise through the fluctuation-dissipation theorem. We have observed that the presence of memory slows down the momentum evolution of the HQs in the thermal bath resulting into slowing down of the formation of $R_{AA}$ and evolution of transverse momentum broadening.

\subsection{Formalism}
We can study the momentum evolution of the HQs in the QGP within the ambit of Langevin equation as follow, 
 \begin{align}
  \label{1} \frac{dp_i}{dt}=-\int_0^t \gamma(t-s)p(s)ds + \eta(t),
      \end{align}
where $p$ is the momentum of the particles, the integral term in ``Eq.~(\ref{1})'' is a dissipative force and $\eta(t)$ is stochastic term that governs the noise,  the correlation of thermal noise does not vanish at different time, which is written as follow,
\begin{align}
\langle \eta(t) \eta(t')\rangle=2\mathcal{D}f(|t-t'|).
\end{align}
The drag coefficient, $\gamma(t,t')$ is related to the thermal noise through the fluctuation-dissipation theorem in the relativistic limit as follow, 
\begin{align}
    \label{26}&\gamma(t,t')=\frac{1}{\textit{E}\textit{T}}\langle \eta(t)\eta(t^{\prime})\rangle.
\end{align}
In this model, we have assumed the correlator to be a decaying exponential function of time,
\begin{equation}
    f(|t-t^{\prime}|)=\frac{1}{2\tau}{e^{-|t-t^{\prime}|/{\tau}}},
\end{equation}
where $\tau$ is memory time. We have fixed $t'$, and analyzed the momentum evolution of HQs in QGP for $t \ge t'$.

\subsection{Ancillary process}
We introduce an ancillary process to generate the time-correlated thermal noise in the hot QCD medium for the Langevin equation as follow\cite{1_jai},

 \begin{align}
 \label{h}&\frac{dh}{dt}=-\alpha h +\alpha\xi,
  \end{align}
where $h$ stands for the ancillary process, \ $\xi$ is the uncorrelated noise and dimensionless parameter having the properties, 
 $\langle \xi\rangle=0, \ 
 \langle \xi(t)\xi(t^{\prime})\rangle=\frac{1}{\alpha}\delta(t-t^{\prime})$,
 where $\alpha$ is the inverse of memory time,  $\alpha \equiv \frac{1}{\tau}$, which balances the dimension of time.
 The other study on memory has been made in Ref. \citen{6_jai}.
The approximate solution of  “Eq.~(\ref{h})”, can be written as,
 \begin{align}
    \label{35}&\langle h(t)h(t^{\prime})\rangle\approx\frac{e^{-\alpha|t-t^{\prime}|} }{2}.
 \end{align}
 With the properties $\langle h(t)\rangle=0$,  $h(t)$ is
 a memory process, which we use in the Langevin equation to study the momentum evolution of the HQs.

\subsection{Results}
\subsection{Transverse momentum broadening} 
 The transverse momentum broadening,
 $\sigma_{p}$, is  calculated in the presence of memory in the system, which is written as follows,
 \begin{equation}			
         \sigma_p=\langle (p_T-\langle p_T\rangle)^2\rangle. 
         \end{equation}
         
 \begin{figure}[htb]
\centering
{\includegraphics[scale=0.25]{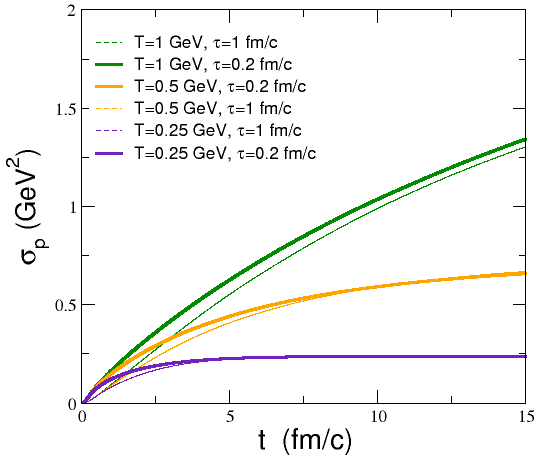}}
\caption{$\sigma_p$ versus versus
time (t), for three values of the temperature (T= 0.25 GeV, 0.5 GeV, 1 GeV) and
for $\tau$ = 1 $fm$ and 0.2 $fm$.}
\label{3_sigma}
 \end{figure}

We have plotted the $\sigma_{p}$ for three different temperatures at $\tau$ = 1 $fm$ and 0.2 $fm$, at a constant diffusion coefficient, $\mathcal{D}$=0.2 $ GeV^2/fm$ for illustrative purposes only, as depicted in fig. (\ref{3_sigma})\cite{1_jai}. The evolution of $\sigma_{p}$ slows down in the presence of memory.

\subsection{Nuclear modification factor}

\begin{figure}[htb]
\centering
\begin{minipage}{.5\textwidth}
  \centering
  \includegraphics[scale=0.25]{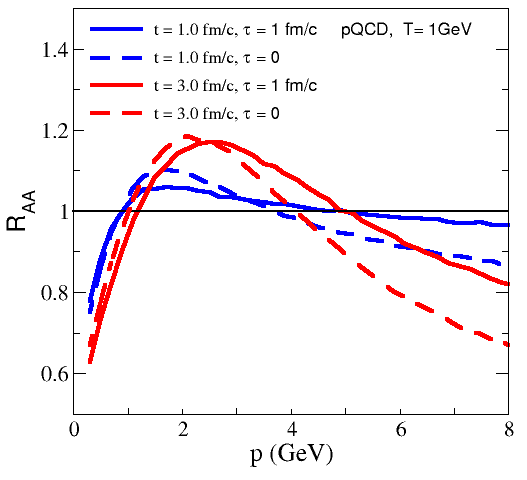}
\end{minipage}%
\begin{minipage}{.5\textwidth}
  \centering
  \includegraphics[scale=0.245]{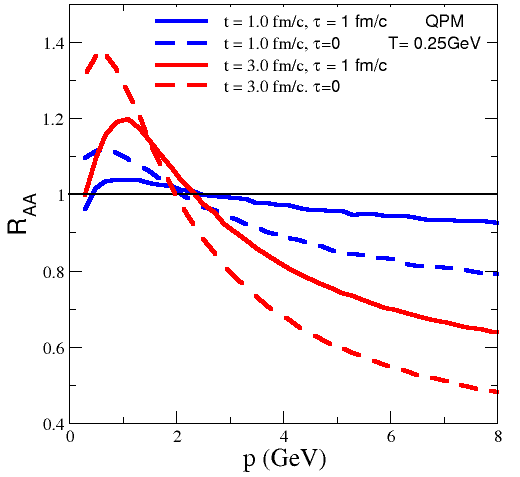}
\end{minipage}
\caption{$R_{AA}$ versus transverse momentum (p) at temperature 1 GeV for pQCD (left panel) and for temperature 0.25 GeV for QPM  (right panel),
time (t) = 3 $fm$ and 1 $fm $.
}
\label{fig:test1}
\end{figure}

  We have studied the impact of memory on the nuclear modification factor, $R_{AA}$, of HQs within perturbative QCD (pQCD) at temperature 1 GeV and quasiparticle model (QPM) at temperature 0.25 GeV.
The impact has been calculated for the evolution time, $t$ = 1 $fm$ and 3 $fm$ with two values of $\tau $ as depicted in the fig.(\ref{fig:test1})  \cite{1_jai}. The effect of memory on $R_{AA}$ is quite significant. The formation of $R_{AA}$ delays when  $\tau$ increases in the system, which means the HQ energy loss will be less than that without memory in the system.

\subsection{Summary and Outlook}
We have studied the effect of time-correlated thermal noise on the momentum evolution of HQs in thermalized QGP. The dissipative force and thermal noise play a role in implementing the memory in the dynamics of HQs in QGP within the integro-differential Langevin equation. In the system, as $\tau\rightarrow0$, the memory disappears and tends to idealize the system \cite{Scardina:2017ipo,Das:2016llg}. We have observed the significant impact of memory on the momentum evolution of HQs in the hot QCD matter and calculated the $R_{AA}$ and $\sigma_{p}$ of the HQs, namely, charm quark under the framework of the stochastic Langevin equation. In the presence of memory, the formation of $R_{AA}$ and the evolution of $\sigma_{p}$ are slowed down, delaying the energy loss and thus increasing the thermalization time of HQs in the medium.

\section{Modification of intra-jet properties in high multiplicity pp collisions at $\sqrt{s}$ = 13 TeV with ALICE}
\author{Debjani Banerjee for the ALICE Collaboration}	

\bigskip

\begin{abstract}
	We present measurements of mean charged-particle multiplicity and jet fragmentation function for leading jets in minimum bias and high multiplicity proton-proton (pp) collisions at $\sqrt{s}$ = 13\,TeV with ALICE. Jets are reconstructed at midrapidity from charged particles using the sequential recombination anti-$k_{\rm T}$ jet finding algorithm for $R$ = 0.4. The results are compared to predictions from PYTHIA8 Monash2013 and EPOS LHC.
\end{abstract}

\subsection{Introduction}
 The partons produced with large transverse momentum in high energy nuclear or hadronic collisions fragment into a collimated spray of final state particles, known as jets. Jets are the key ingredient to test the perturbative quantum chromodynamics (pQCD) predictions. In addition, jet measurement in small collision system such as high-multiplicity pp is important in order to look for the onset of QGP-like effects as a function of particle multiplicity. In this work, we present the measurements of intra-jet properties, the mean charged particle multiplicity and fragmentation functions and their multiplicity dependence for leading jets in pp collisions at $\sqrt{s}$ = 13\,TeV with ALICE. 
\subsection{Analysis details and Jet observables}
The data presented here were recorded by the ALICE detector in 2016, 2017 and 2018 by colliding protons at center-of-mass energy ($\sqrt{s}$) = 13 TeV. Events are rejected if the vertex z-position, $|z_{\rm vtx}| > $ 10\,cm from the nominal interaction point (IP). Minimum Bias (MB) events are selected using ALICE MB trigger condition which requires the coincidence in the V0A and V0C forward scintillator arrays~\cite{V0} whereas high multiplicity (HM) events are selected using HM trigger condition which requires the sum of V0A and V0C amplitudes to be more than 5 times the mean MB signal. The used data samples consist of $\sim$ 1802\,M for MB and $\sim$ 183\,M for HM event classes. 
Charged particles detected by both the Time Projection Chamber (TPC)~\cite{TPC} and the Inner Tracking System (ITS)~\cite{ITS} with $p_{\rm T} >$ 0.15 GeV/$c$ in the pseudorapidity $|\eta| <$ 0.9 and azimuthal angle 0 $ < \varphi <$ 2$\pi$ are considered for this analysis. Jets are constructed from these selected charged particles with FastJet 3.2.1~\cite{FastJet} using the anti-$k_{\rm T}$ algorithm with $p_{\rm T}$ recombination scheme for jet resolution parameter $R$ = 0.4. Mean charged-particle multiplicity,  $\left<N_{\rm ch}\right>$ and jet fragmentation function, $z^{\rm ch} = {p_{\rm T}^{\rm particle}}/{p_{\rm T}^{\rm jet,ch}}$ are measured for leading jets in the range of jet $p_{\rm T}$ from 5--110 GeV/$c$.

\subsection{Correction procedure and estimation of systematic uncertainty}
Instrumental effects such as tracking inefficiency, particle-material interactions and track $p_{\rm T}$ resolution are corrected by performing a 2D unfolding in $p_{\rm T}^{\rm jet,ch}$ and $N_{\rm ch}$ or $z^{\rm ch}$  using the iterative Bayesian unfolding~\cite{Bayesian} algorithm implemented in the RooUnfold package~\cite{Roounfold}. To account for the instrumental effects, PYTHIA8 (version 8.210) Monash2013 and GEANT detector simulation are used to construct a 4D response matrix (R) that describes the response of detector and background in $p_{\rm T}^{\rm jet,ch}$ and $N_{\rm ch}$ or $z^{\rm ch}$ contained within R ($p_{\rm T, det}^{\rm jet,ch}$, $N_{\rm ch, det}/ z^{\rm ch}_{\rm det}$, $p_{\rm T, truth}^{\rm jet,ch}$, $N_{\rm ch, truth}/ z^{\rm ch}_{\rm truth}$), where $p_{\rm T, det}^{\rm jet,ch}$ is detector level jet $p_{\rm T}$ and $p_{\rm T, truth}^{\rm jet,ch}$ is truth level jet $p_{\rm T}$ and analogously for $N_{\rm ch}$ and $z^{\rm ch}$. Underlying events (UE) coming from sources other than jets are estimated using well established perpendicular cone method used by ALICE~\cite{jetpp7TeV}. UE subtraction is performed on a statistical basis after unfolding both the raw distributions and UE contributions separately.

The sources of systematic uncertainties include tracking efficiency, MC dependence, choice of regularization parameter, number of iterations in Bayesian unfolding and change in prior distribution. The total systematic uncertainty for $\left<N_{\rm ch}\right>$ is found to be 2--8\% (3--6\%) for MB (HM) events whereas for $z^{\rm ch}$ ($p_{\rm T}^{\rm jet,ch}$ = 10--20 GeV/$c$) it is found to be 5--12\% (7--15\%) for MB (HM) events. However the uncertainty on $z^{\rm ch}$ distributions in minimum bias also depends on the jet $p_{\rm T}$ range, it varies from 5--20\% for jet $p_{\rm T}^{\rm jet,ch}$ = 20--30 GeV/$c$ whereas it is 12--24\% for higher jet $p_{\rm T}$ (= 40--60 GeV/$c$).

\subsection{Results and discussion}
\begin{figure}[h!]
	\centering
	\begin{minipage}[b]{0.30\linewidth}
		\includegraphics[scale=0.21]{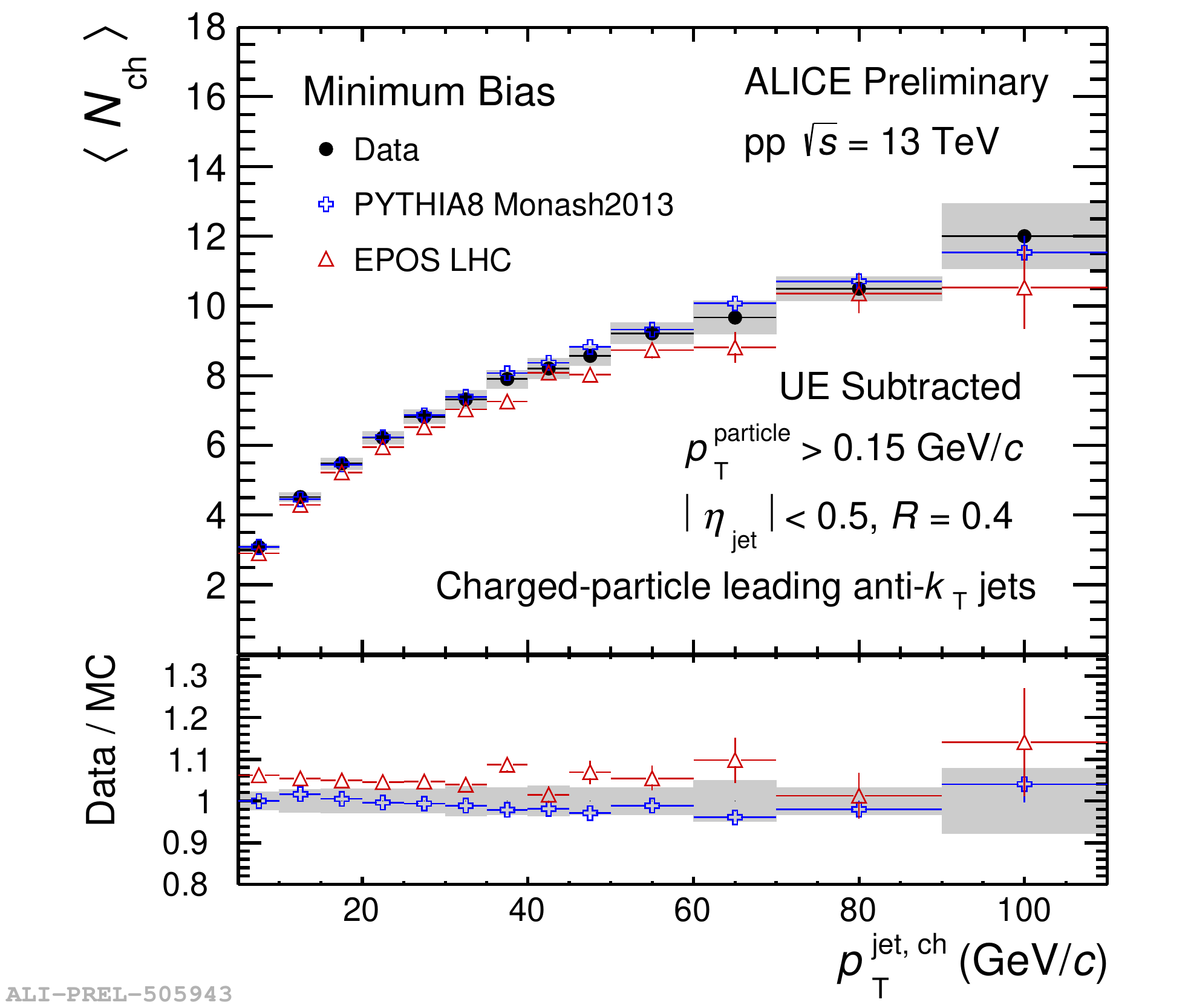}
	\end{minipage}
	\quad
	\begin{minipage}[b]{0.30\linewidth}
		\includegraphics[scale=0.22]{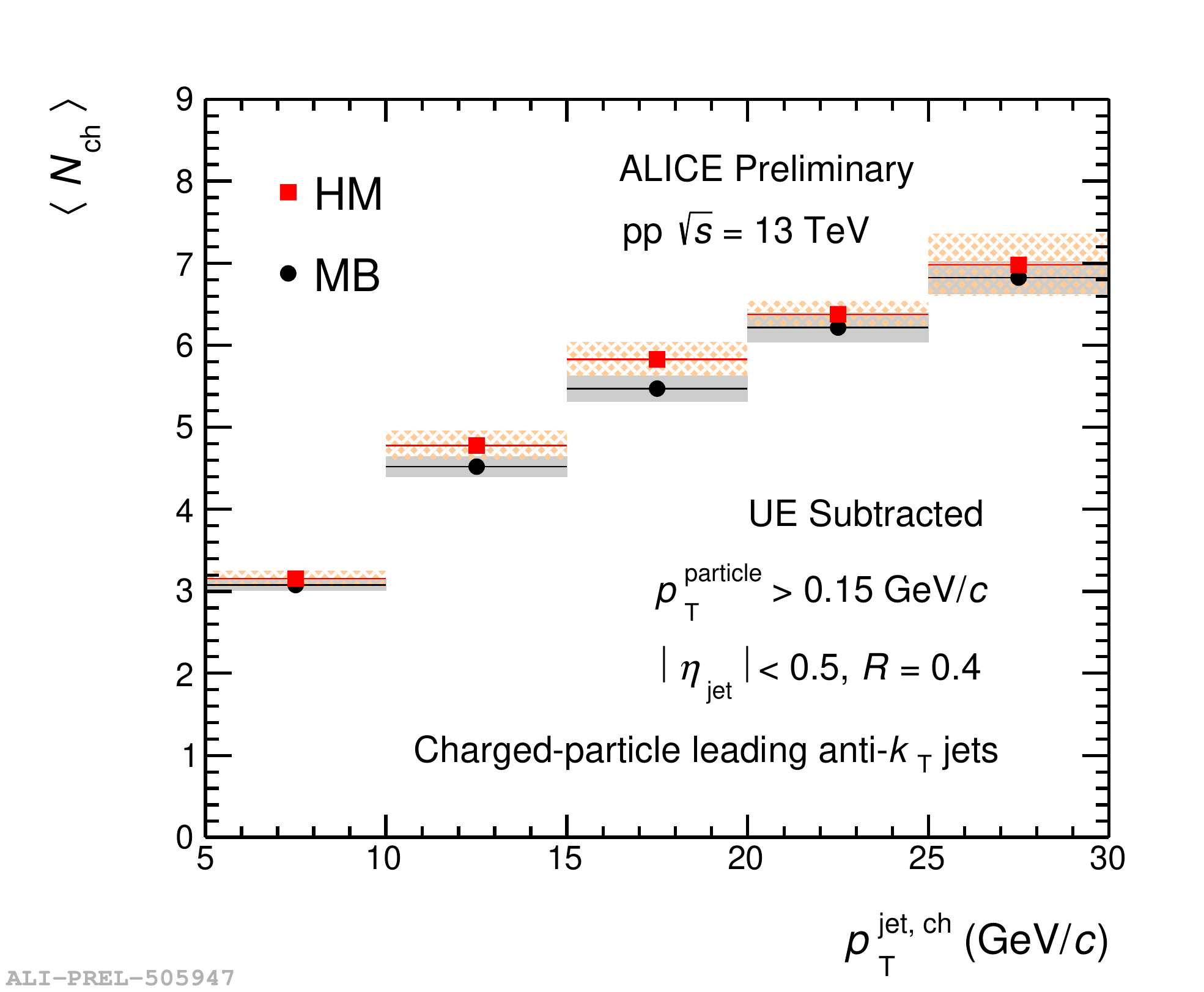}
	\end{minipage}
\quad
\begin{minipage}[b]{0.30\linewidth}
	\includegraphics[scale=0.22]{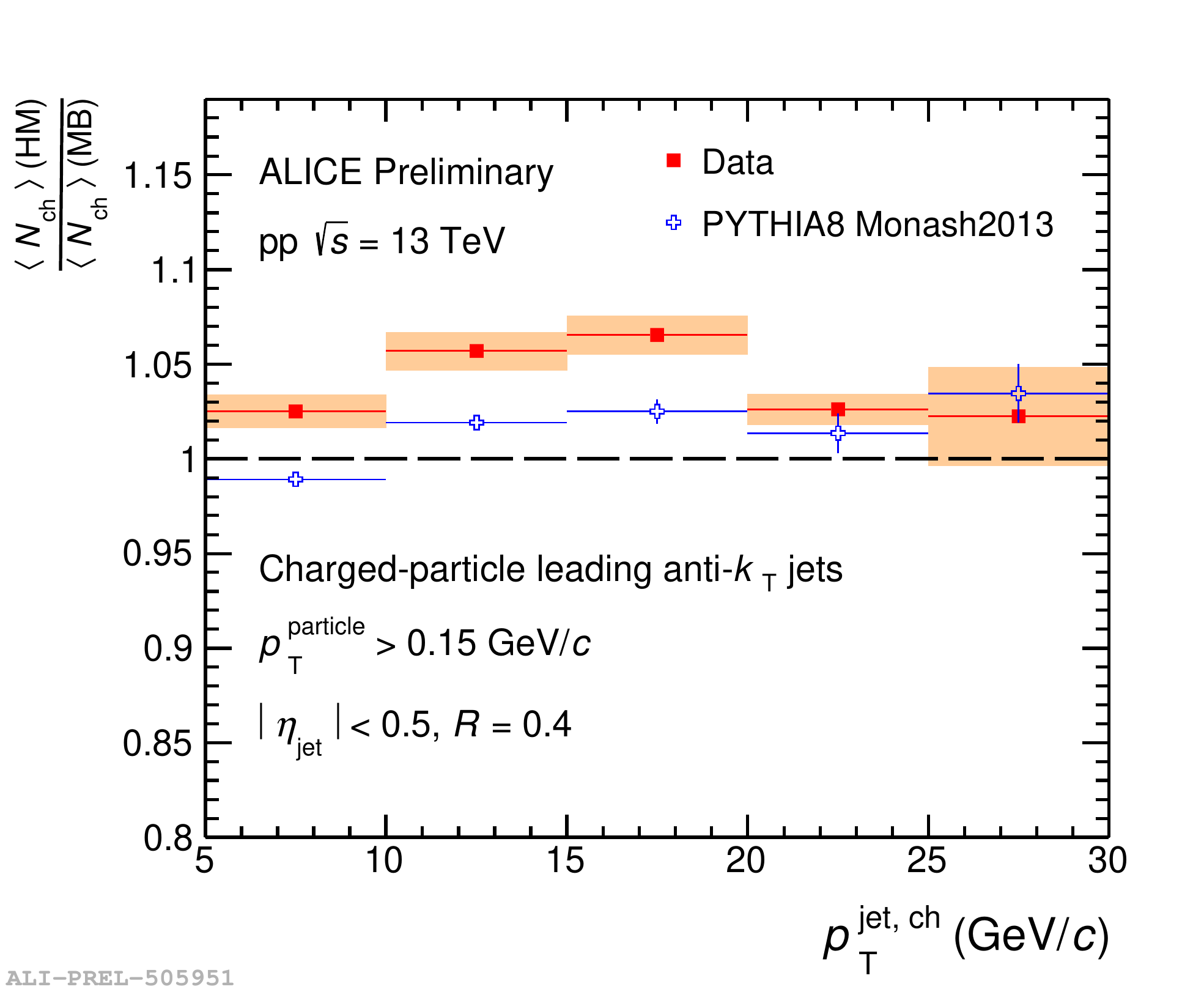}
\end{minipage}
\caption{Fully corrected $\left<N_{\rm ch}\right>$ as a function of $p_{\rm T}^{\rm jet,ch}$. Left: Blue and red markers show PYTHIA8 Monash2013 and EPOS LHC predictions respectively. Bottom panel shows the ratio between data and MC. Middle: Red and Black markers show HM and MB results respectively. Shaded region represents systematic uncertainties. Right: Ratio of $\left<N_{\rm ch}\right>$ in HM to the same in MB as a function of $p_{\rm T}^{\rm jet,ch}$.}
\label{NchHMMB}
\end{figure}
\begin{figure}[h!]
	\centering
	\begin{minipage}[b]{0.30\linewidth}
		\includegraphics[scale=0.22]{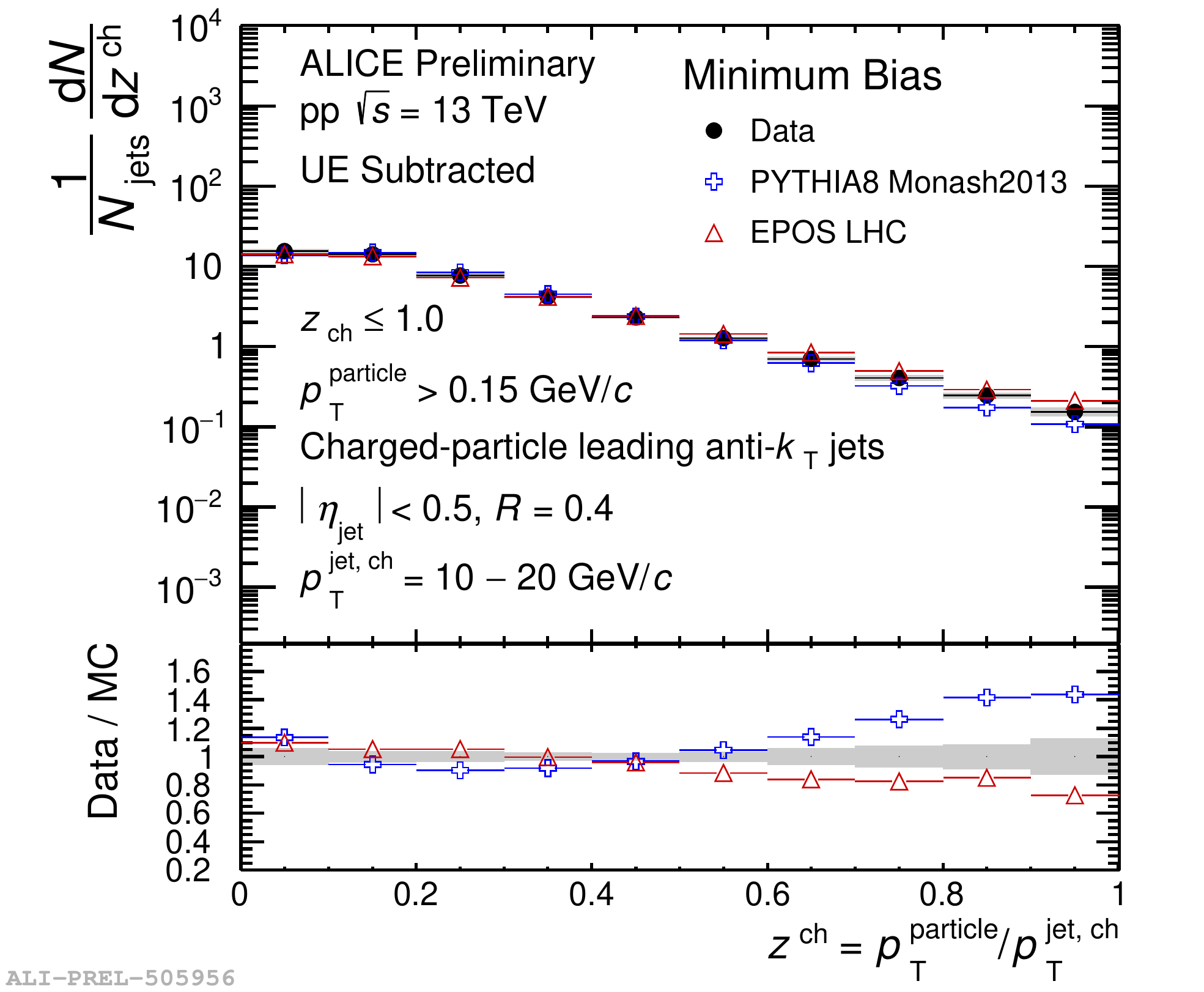}
	\end{minipage}
	\quad
	\begin{minipage}[b]{0.30\linewidth}
		\includegraphics[scale=0.22]{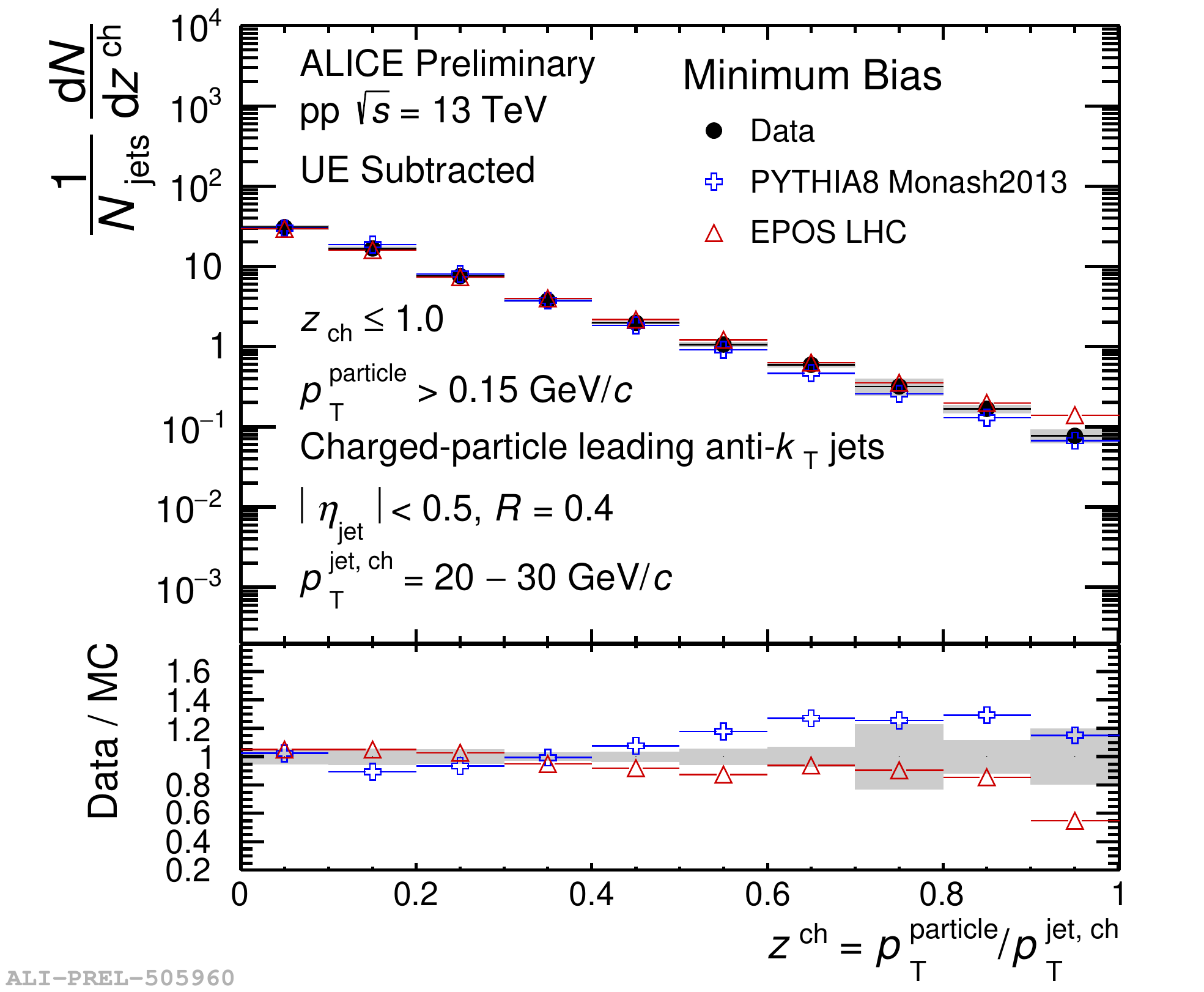}
	\end{minipage}
	\quad
	\begin{minipage}[b]{0.30\linewidth}
		\includegraphics[scale=0.22]{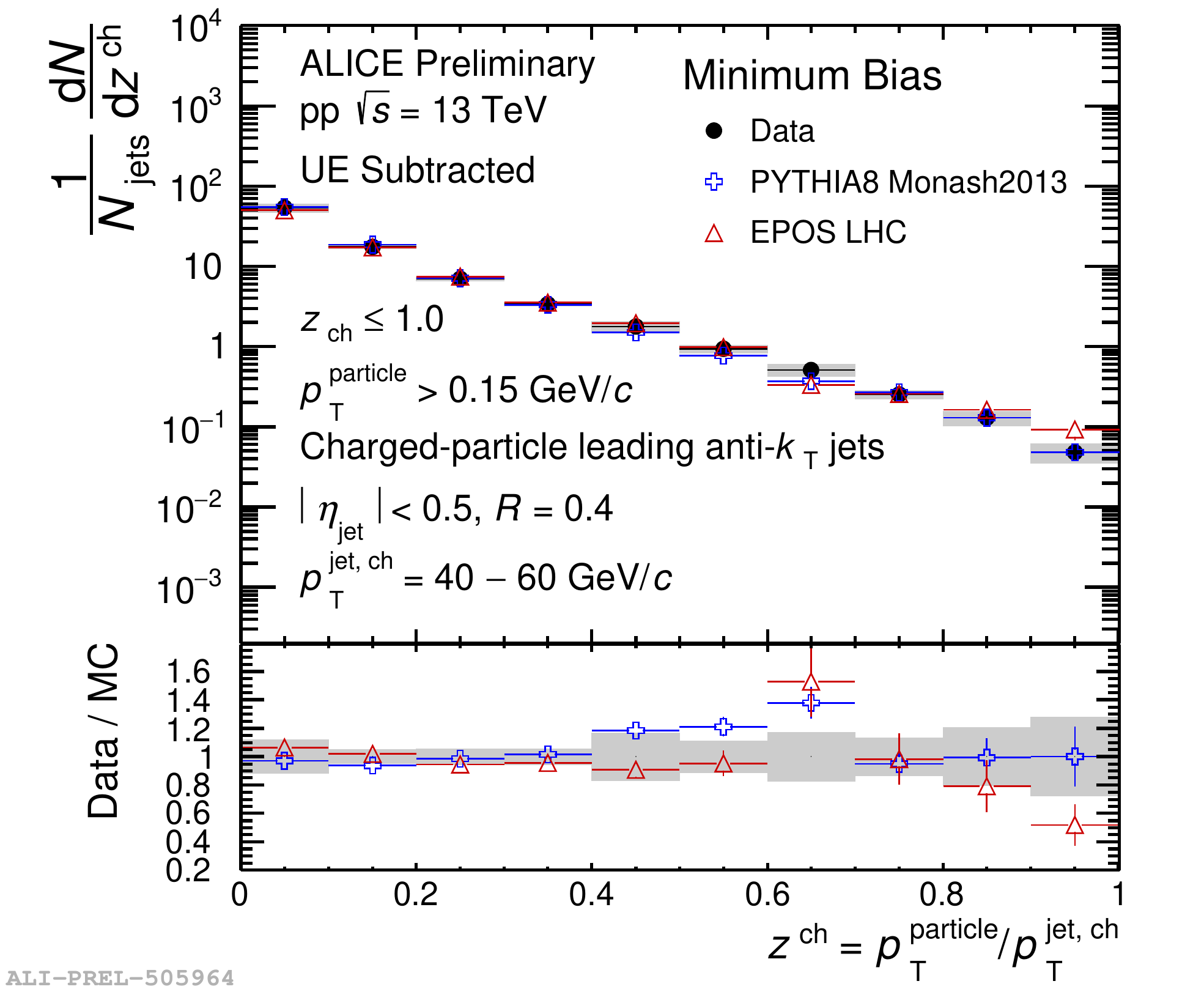}
	\end{minipage}
		\caption{Fully corrected $\rm z^{\rm ch}$ distributions for leading $p_{\rm T}^{\rm jet,ch}$ = 10 - 20 GeV/$c$ (left), 20 - 30 GeV/$c$ (middle) and 40 - 60 GeV/$c$ (right). Blue and red markers show PYTHIA8 Monash2013 and EPOS LHC predictions respectively. Bottom panels show the ratio between data and MC. Shaded region represents systematic uncertainties.}
	\label{zMB}
\end{figure}
Figure~\ref{NchHMMB} (left) shows $\left<N_{\rm ch}\right>$ as a function of leading $p_{\rm T}^{\rm jet,ch}$. The black markers represent the data whereas blue and red markers show PYTHIA8 Monash2013 and EPOS LHC (version 3400) predictions respectively. PYTHIA is a parton based MC generator where the hadronization is treated using the Lund string fragmentation model for collider physics with an emphasis on pp interactions. EPOS is based on perturbative QCD, Gribov-Regge multiple scattering, and string fragmentation for pp and AA collisions. In this figure the ratios between data and MC predictions are shown in the bottom panel. In Fig.~\ref{NchHMMB} (middle), red and black markers show $\left<N_{\rm ch}\right>$ as a function of leading $p_{\rm T}^{\rm jet,ch}$ for HM and MB respectively. The ratio of $\left<N_{\rm ch}\right>$ (HM)/$\left<N_{\rm ch}\right>$ (MB) is shown as a function of leading $p_{\rm T}^{\rm jet,ch}$ in Fig.~\ref{NchHMMB} (right). It is observed that $\left<N_{\rm ch}\right>$ increases with leading $p_{\rm T}^{\rm jet,ch}$ for HM and MB events. EPOS LHC slightly underestimates the data whereas PYTHIA8 Monash2013 describes the data within systematic uncertainty. Figure~\ref{NchHMMB} (right) shows that $\left<N_{\rm ch}\right>$ is slightly larger for HM events and qualitatively reproduced by PYTHIA8 Monash2013 for $p_{\rm T}^{\rm jet,ch} <$ 20 GeV/$c$.

\begin{figure}[h!]
	\centering
	\begin{minipage}[b]{0.30\linewidth}
		\includegraphics[scale=0.22]{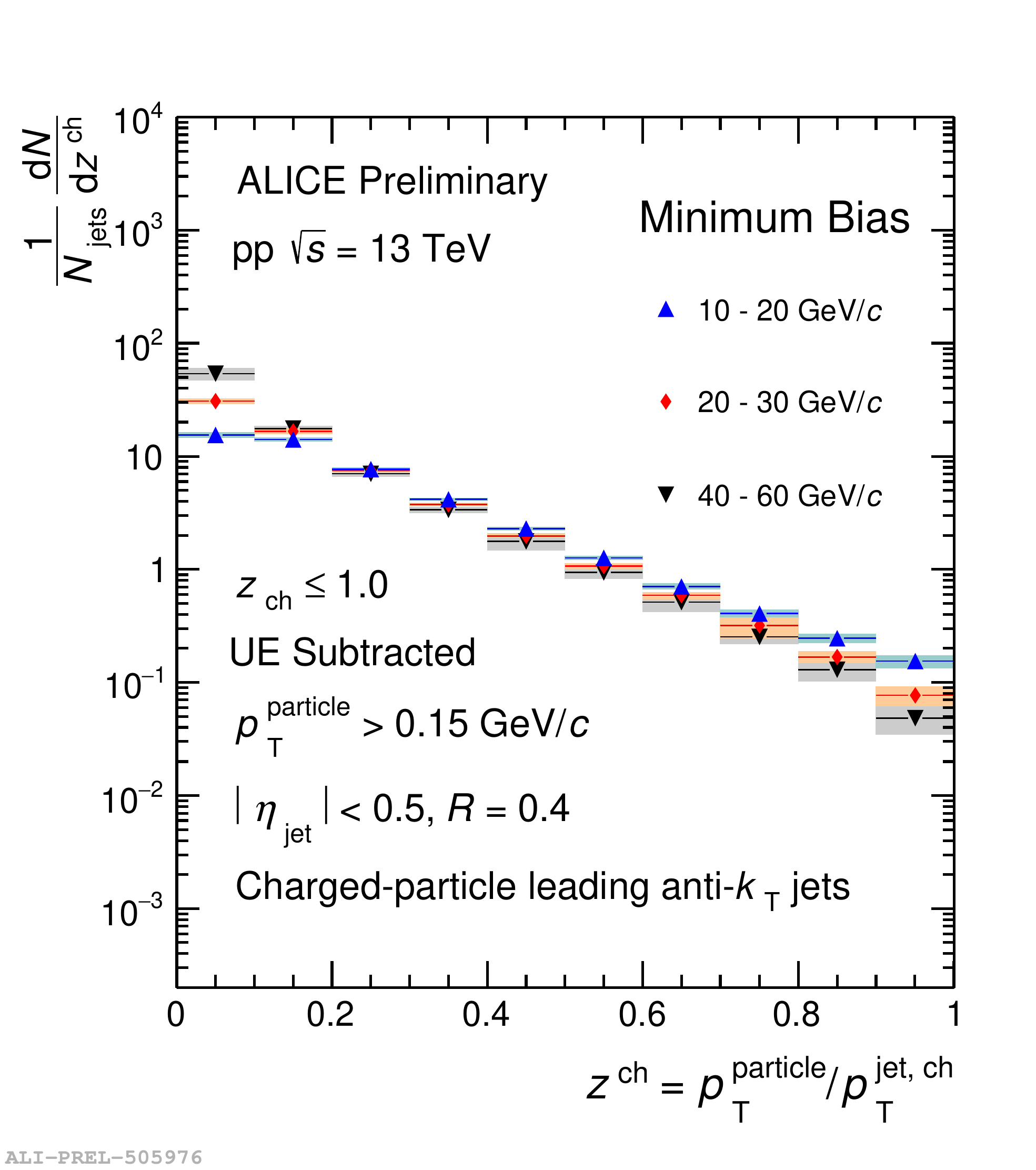}
	\end{minipage}
	\quad
	\begin{minipage}[b]{0.30\linewidth}
		\includegraphics[scale=0.22]{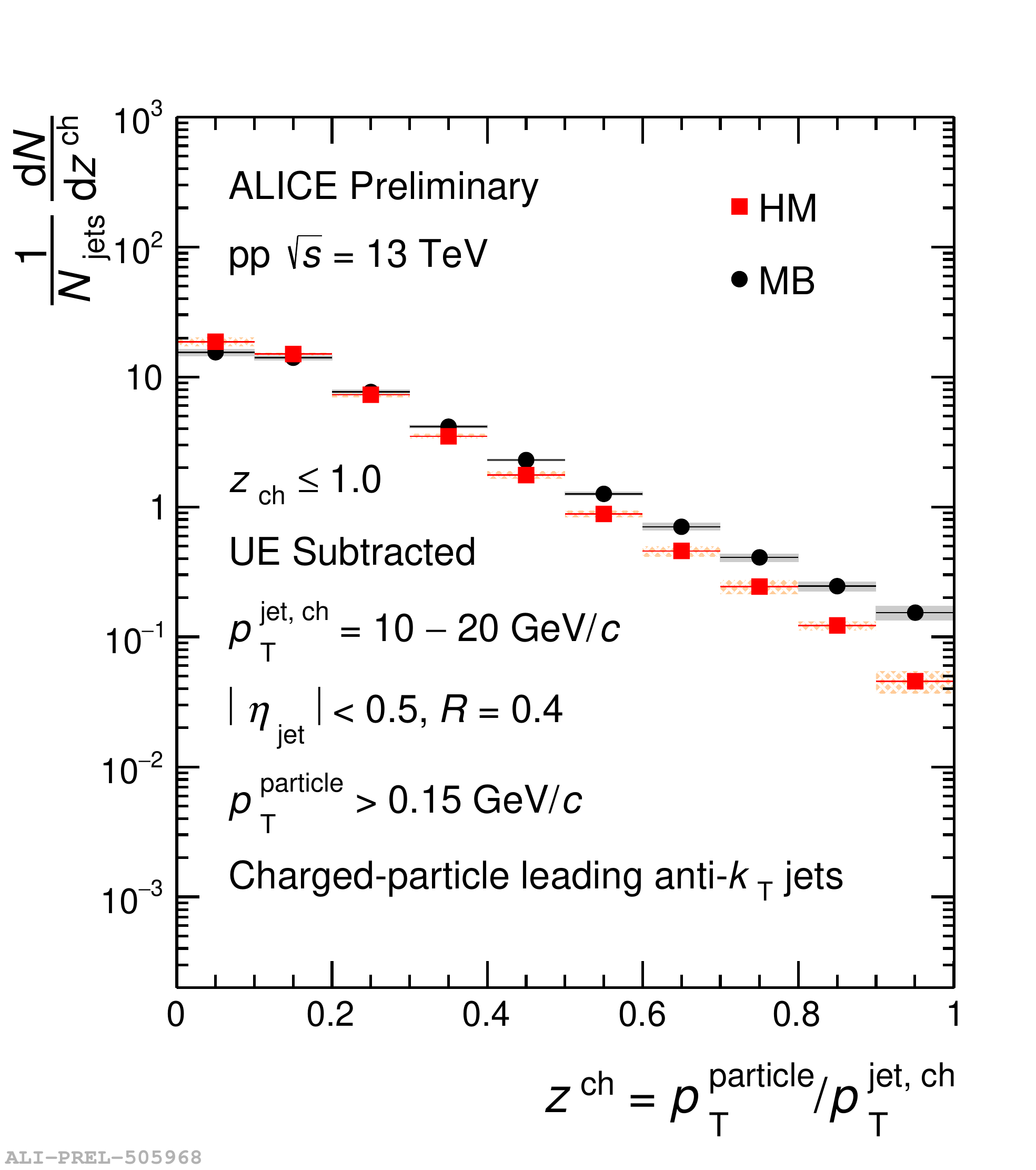}
	\end{minipage}
	\quad
	\begin{minipage}[b]{0.30\linewidth}
		\includegraphics[scale=0.22]{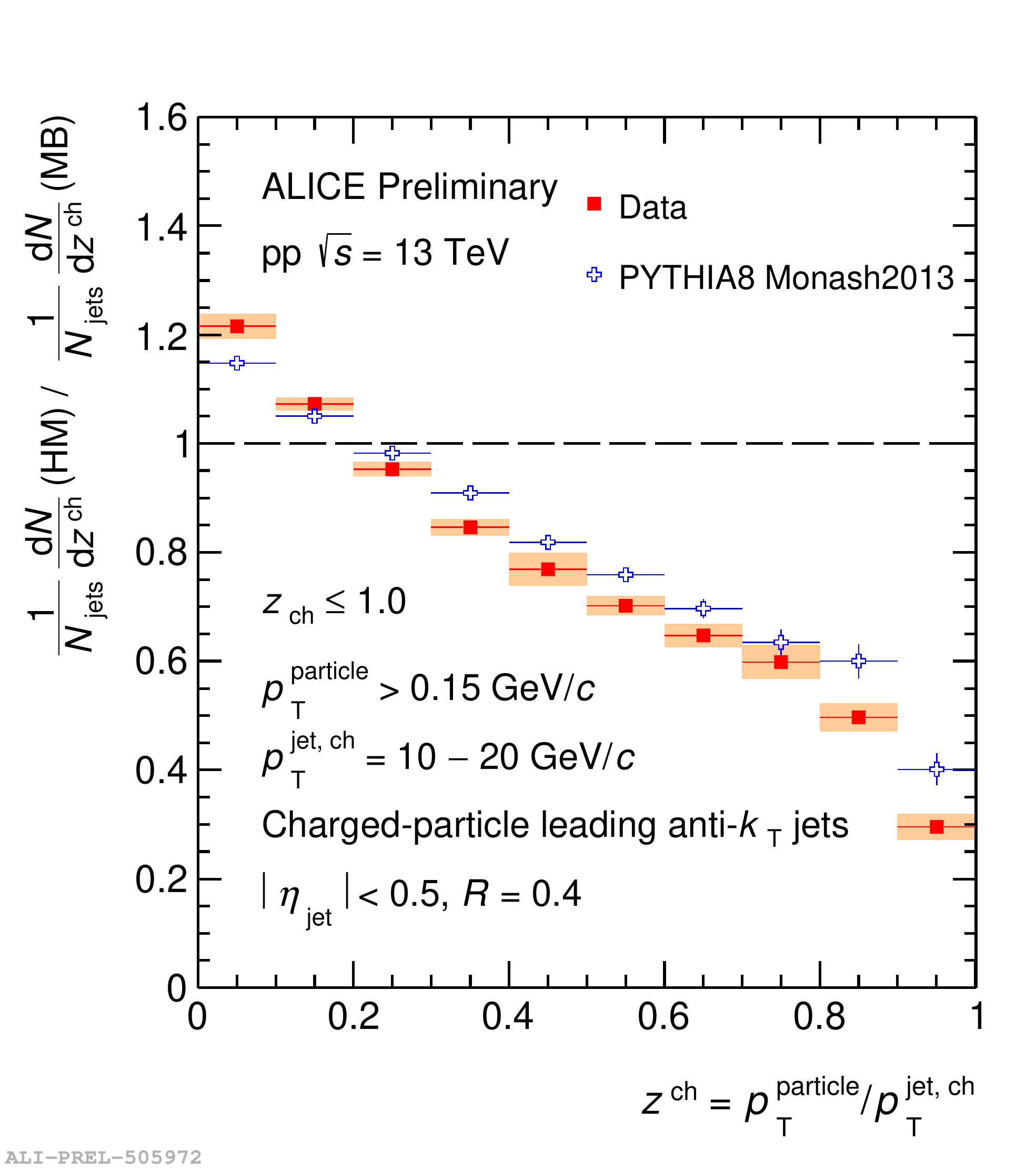}
	\end{minipage}
	\caption{ Left: $z^{\rm ch}$ distributions in MB events for leading $p_{\rm T}^{\rm jet,ch}$ = 10 - 20 GeV/$c$ (blue), 20 - 30 GeV/$c$ (red) and 40 - 60 GeV/$c$ (black). Middle: $z^{\rm ch}$ distributions for MB and HM. Right: Ratio of $z^{\rm ch}$ in HM to the same in MB. Model comparison is superimposed in this figure.}
	\label{zallMBHM}
\end{figure}

 Figure~\ref{zMB} shows $z^{\rm ch}$ distributions for leading $p_{\rm T}^{\rm jet,ch}$ = 10 - 20 GeV/$c$ (left), 20 - 30 GeV/$c$ (middle) and 40 - 60 GeV/$c$ (right). Red and blue markers show EPOS LHC  and PYTHIA8 Monash2013 predictions respectively. The ratios between data and MC predictions are presented in bottom panels. It is observed that for low $z^{\rm ch}$ ($<$ 0.5), both models predict the data well within systematic uncertainties whereas for high $z^{\rm ch}$ ($>$ 0.5) and lower jet $p_{\rm T}$ range, EPOS LHC explains the data better than PYTHIA8 Monash2013. Moreover for high $z^{\rm ch}$ ($>$ 0.5) and higher jet $p_{\rm T}$ range, both models explain the data within systematic uncertainties.

Figure~\ref{zallMBHM} (left) shows $z^{\rm ch}$ distributions in three jet $p_{\rm T}$ domains. Scaling of charged-particle jet fragmentation with jet $p_{\rm T}$ is observed except at highest and lowest $z^{\rm ch}$. In Figure~\ref{zallMBHM} (middle) red and black markers show $z^{\rm ch}$ distributions for HM and MB events respectively. Figure~\ref{zallMBHM} (right) shows the ratio of $z^{\rm ch}$ distributions obtained in HM and MB events. It is interesting to notice that jet fragmentation is softer in HM events explaining more interactions between the jet and the partons.
\subsection{Summary}

We have presented the measurement of intra-jet properties and their multiplicity dependence in pp collisions at 13\,TeV in ALICE. Results are compared with PYTHIA8 Monash2013 and EPOS LHC predictions. We have observed significant modification in $z^{\rm ch}$ distributions in HM events compared to MB events. We have also observed that the jet fragmentation is softer in the HM events.

 \section{Heavy Quarkonia in a hot and dense strongly magnetized
 	QCD medium}
 \author{Salman Ahamad Khan, Mujeeb Hasan,  and  Binoy Krishna Patra}	
 
 \bigskip

\begin{abstract}
We have studied the effects of the finite quark chemical potential
$(\mu)$ on the $Q\bar{Q}$ potential and  have further studied the 
various properties like binding energy, decay width and 
dissociation temperatures for the $J/\psi$ and $\Upsilon$ states.
We have noticed that the real part of the potential
becomes more attractive while magnitude of the imaginary 
part gets reduced. The dissociation temperature slightly gets enhanced
in a medium having finite amount of $\mu$. 
\end{abstract}
\subsection{Introduction}
Heavy quarkonia (bound state of $c\bar{c}$ and $b\bar{b}$)
 is a very promising signature 
of the hot and dense quark matter produced at the
 heavy ion collision experiments at RHIC and LHC. In non-central 
collisions, a very strong magnetic field (around $m_{\pi}^2-15 m_{\pi}^2$) is 
also generated whose life time is elongated by  
the  electrical conductivity of the medium. In addition to this a large quark
 chemical potential  (100-200 MeV) is also 
present near the deconfinement region
 ~\cite{Cleymans:JPG35'2008,Fukushima:PRL117'2016}. 
The influence of the strong magnetic field on the 
properties of the heavy quarkonia immersed in hot QCD medium have
 been studied
in recent years by two of us in strong
~\cite{Mujeeb:EPJC77'2017,Mujeeb:NPA995'2020} as well as
in weak magnetic field~\cite{Mujeeb:PRD102'2020}. In other
works the effect of the magnetic field has been 
investigated for the case of a harmonic interaction 
and for Cornell potential plus a spin spin interaction term 
in~\cite{Alford:PRD88'2013,Bonati:PRD92'2015} and   
 using the generalized Gauss law in~\cite{Balbeer:PRD97'2018}. 
 In present work, we have
 explored the effects of finite $\mu$ on the 
 $Q\bar{Q}$ bound states immersed in strongly magnetized 
 hot QCD medium. In order to incorporate the non-perturbative
 part we have added a phenomenological term induced by dimension
 two gluon condensate in  gluon
 propagator in addition to usual HTL resummed propagator.
 \subsection{Medium modification to the $Q\bar{Q}$ potential}
The medium modification to the heavy quark potential 
can be obtained from the  inverse Fourier transform of the 
resummed gluon propagator 
in the static limit as~\cite{Dumitru:PRD79'2009}
\begin{eqnarray}
V(r;T,B,\mu)= C_F~g^2\int\frac{d^3p}{(2\pi)^3}(e^{ip.r}-1)~D^{00}(p_0=0,p),
\label{pot_defn}
\end{eqnarray} 
where $C_F$ is the cashimir factor and $D^{00}$ is the
static limit of the  temporal
component of the resummed gluon propagator in the 
strongly magnetized hot QCD medium which will be calculated 
in the next subsection.
\subsection{Covariant structure of gluon self energy and resummed propagator
 in magnetic field}
In order to obtain the resummed gluon propagator,
 we need the gluon self energy in the above mentioned environment. In the 
 presence of the magnetic field the rotational invariance is broken
 and a extended tensor basis is required. 
 The covariant tensor structure of the gluon self energy 
in the presence of the magnetic field is given by~
\cite{karmakar:EPJC79'2019} 
\begin{eqnarray}
\Pi^{\mu\nu}(P)=b(P)B^{\mu\nu}(P)+c(P)R^{\mu\nu}(P)+d(P)M^{\mu\nu}(P)
+a(P)N^{\mu\nu}(P),
\label{self_decomposition}
\end{eqnarray}
here the various projection tensors are constructed as 
\begin{eqnarray}
B^{\mu\nu}(P)&=&\frac{{\bar{u}}^\mu{\bar{u}}^\nu}{{\bar{u}}^2},
\quad 
R^{\mu\nu}(P)=g_{\perp}^{\mu\nu}-\frac{P_{\perp}^{\mu}P_{\perp}^{\nu}}
{P_{\perp}^2},\\
M^{\mu\nu}(P)&=&\frac{{\bar{n}}^\mu{\bar{n}}^\nu}{{\bar{n}}^2},
\quad
N^{\mu\nu}(P)=\frac{{\bar{u}}^\mu{\bar{n}}^\nu+{\bar{u}}^\nu{\bar{n}}^\mu}
{\sqrt{{\bar{u}}^2}\sqrt{{\bar{n}}^2}},
\end{eqnarray}
 $u^\mu=(1,0,0,0)$ is the four velocity of the heat bath and 
 $n_\mu=(0,0,0,1)$ represents the direction of $B$.

 The form factors defined in  
\eqref{self_decomposition} can be evaluated  
by taking  the appropriate contractions as 
\begin{eqnarray}
b(P)&=&B^{\mu\nu}(P)~\Pi_{\mu\nu}(P)\quad 
\label{structure_b}
c(P)=R^{\mu\nu}(P)~\Pi_{\mu\nu}(P),
\label{structure_c}\\
d(P)&=&M^{\mu\nu}(P)~\Pi_{\mu\nu}(P), \quad
\label{structure_d}
a(P)=\frac{1}{2}N^{\mu\nu}(P)~\Pi_{\mu\nu}(P)
\label{structure_a}.
\end{eqnarray}
The resummed gluon propagator in magnetized medium in the Landau gauge
 is given as~\cite{karmakar:EPJC79'2019} 
\begin{eqnarray}
D^{\mu\nu}(P)&=&\frac{(P^2-d)B^{\mu\nu}}{(P^2-b)(P^2-d)-a^2}
+\frac{R^{\mu\nu}}{P^2-c}+\frac{(P^2-b)M^{\mu\nu}}{(P^2-b)(P^2-d)-a^2}\nonumber\\
&+&\frac{aN^{\mu\nu}}{(P^2-b)(P^2-d)-a^2}.
\end{eqnarray}
Since we are interested in the static $Q\bar{Q}$ potential,
  $D^{00}(P)$ in the static limit becomes
\begin{eqnarray}
D^{00}(P)=-\frac{1}{(P^2-b)},
\label{propagator_00}
\end{eqnarray}
which requires  the real and imaginary parts of the form
factor $b$ that can be calculated from
\begin{eqnarray}
b(P)&=&B^{\mu\nu}(P)\Pi_{\mu\nu}(P)
=\frac{u^\mu u^\nu}{{\bar{u}}^2}\Pi_{\mu\nu}(P).
\label{form_b}
\end{eqnarray}
We calculate the real and imaginary parts of the form
factor $b$ which  give the real and imaginary 
parts of the resummed gluon propagator in the 
static limit as 
\begin{eqnarray}
{\rm Re}~D^{00}(p_0=0,p)&=&-\frac{1}
{p^2+m_D^2}-\frac{m_G^2}
{(p^2+m_D^2)^2}
\label{real_propagator},\\
{\rm Im}~D^{00}(p_0=0,p)&=&\sum_f \frac{g^2|q_fB|m_f^2 }
{4\pi}\frac{1}{p_3^2(p^2+m_D^2)^2}+
\frac{\pi T m_g^2}{p(p^2+m_D^2)^2}\nonumber\\
&&+\frac{2\pi T m_g^2m_G^2}
{p(p^2+m_D^2)^3}.
\label{imaginary_propagator}
\end{eqnarray}
where $m_G^2=2\sigma/\alpha$. The last term in the above Eqs.
\eqref{real_propagator} and \eqref{imaginary_propagator} are due to 
the dimension two gluon condensate and corresponds to the 
non-perturbative part of the potential. $m_D^2$ is the Debye 
screening mass which reads as
\begin{eqnarray}
m_{D}^2 (T,\mu;B)& =&\sum_f g^2\frac{|q_fB|}
{4\pi^2 T}\int_0^{\infty}{dk_3}~
\big\{n^{+}(E_1)(1-n^{+}(E_1))\nonumber\\
&&+n^{-}(E_1)(1-n^{-}(E_1))\big\}+\frac{N_C}{3}g^2T^2.
\end{eqnarray}
\subsection{Real and imaginary part of the $Q\bar{Q}$
potential}
The real part of the $Q\bar{Q}$ potential has been calculated 
using the real part of the resummed propagator from
  Eq.~\eqref{real_propagator} in Eq.~\eqref{pot_defn} as
\begin{eqnarray}
{\rm{Re}} ~V(r;T,B,\mu)&=&-\frac{4}{3}\alpha_s\left(\frac{e^{-\hat{r}}}
{r}+m_D (T,\mu,B) \right)+\frac{4}{3}\frac{\sigma}{m_D (T,\mu,B)}
\left(1-e^{-\hat{r}}\right),
\label{real_potential}
\end{eqnarray}
similarly using~\eqref{imaginary_propagator},
 we obtain the imaginary part as
\begin{eqnarray}
{\rm Im}~ V(r,T,B,\mu)&=&\sum_f \alpha_s g^2m_f\frac{|q_fB|}{3\pi^2}
\bigg[\frac{\pi}{2m_D^3}-\frac{\pi e^{-\hat{r}}}{2m_D^3}
-\frac{\pi\hat{r}e^{-\hat{r}}}{2m_D^3}
-\frac{2\hat{r}}{m_D}\int_0^{\infty}
\frac{pdp}{(p^2+m_D^2)^2}\nonumber\\
&&\int_0^{pr}\frac{\sin{t}}{t}dt\bigg]
-\frac{4}{3}\frac{\alpha_s Tm_g^2}{m_D^2}\psi_1(\hat{r})
-\frac{16\sigma T m_g^2}{3m_D^4}
\psi_2(\hat{r})
\end{eqnarray}
where the functions $\psi_1(\hat{r})$ and $\psi_2(\hat{r})$ are 
given by
\begin{eqnarray}
\psi_1(\hat{r})&\approx &-\frac{1}{9}{\hat{r}}^2\left(3\ln \hat{r}-
4+3\gamma_E\right), \quad 
\psi_2(\hat{r})\approx \frac{{\hat{r}}^2}{12}+\frac{{\hat{r}}^4}{900}
\left(15\ln\hat{r}-23+15\gamma_E\right)\nonumber
\end{eqnarray}
\subsection{Results and Discussions}
\begin{center}
\begin{figure}
\begin{tabular}{cc}
\includegraphics[width=4cm]{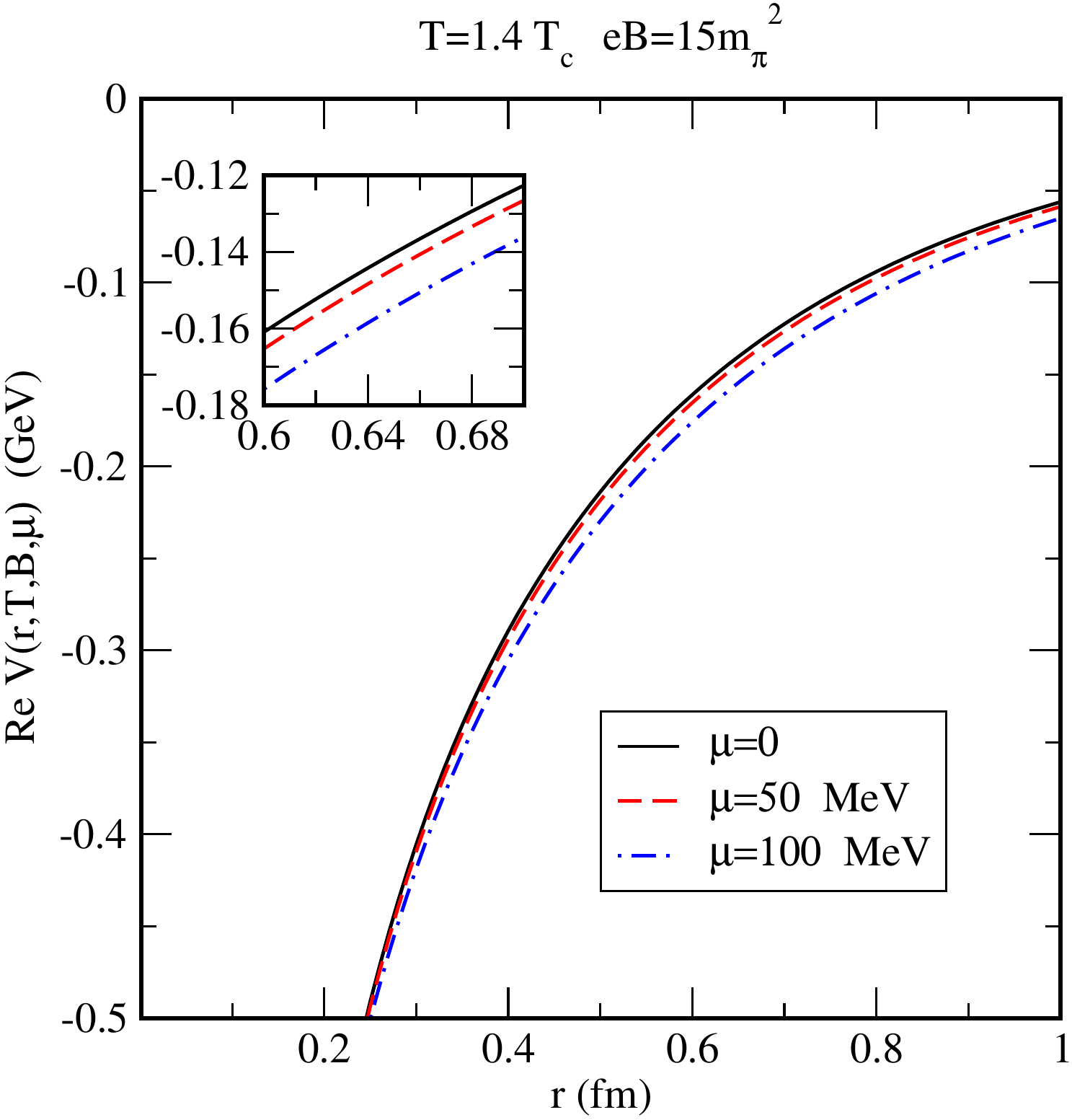}
\hspace{.1cm}
\includegraphics[width=4cm]{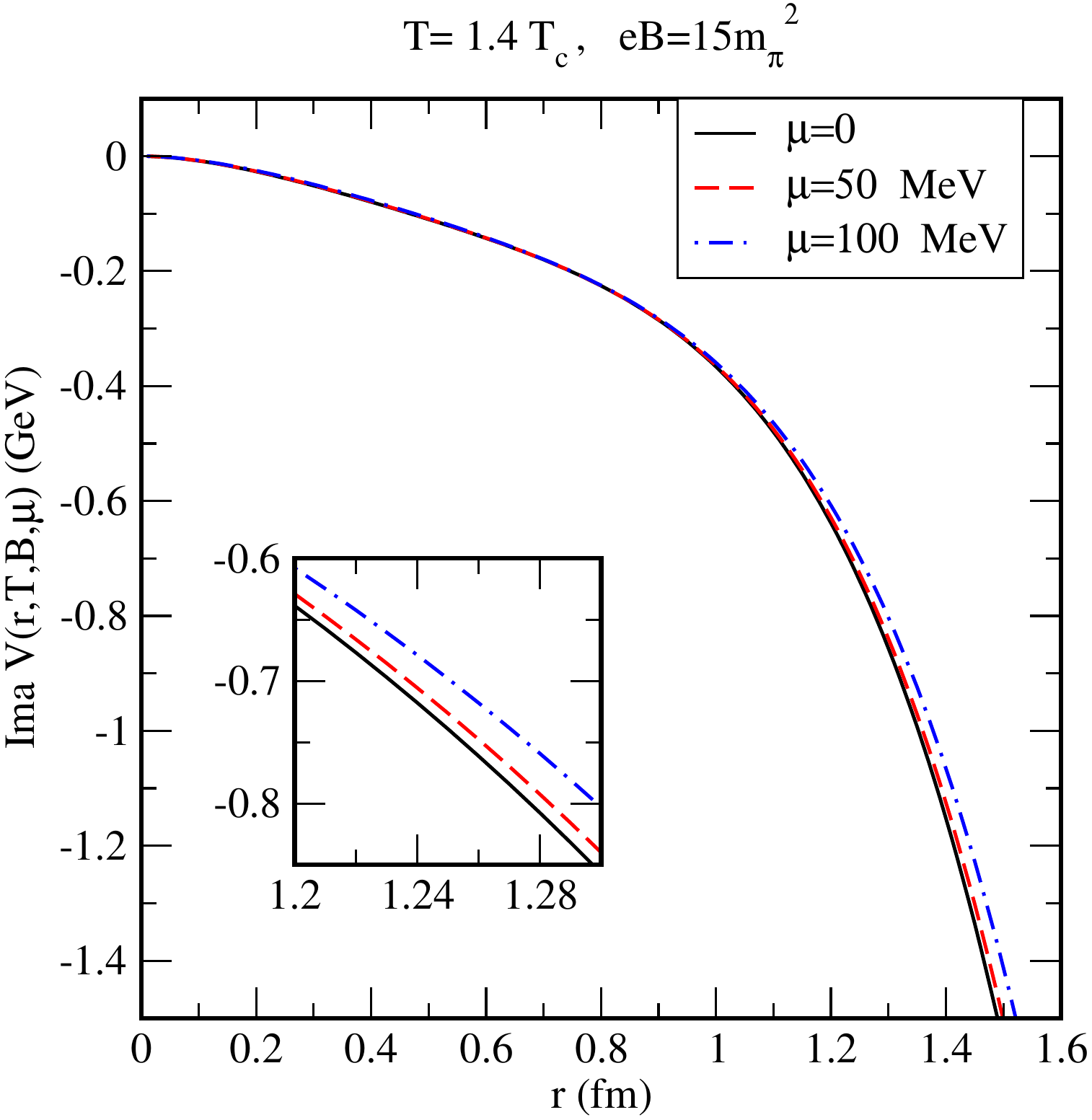}
\hspace{.1cm}
\includegraphics[width=4cm]{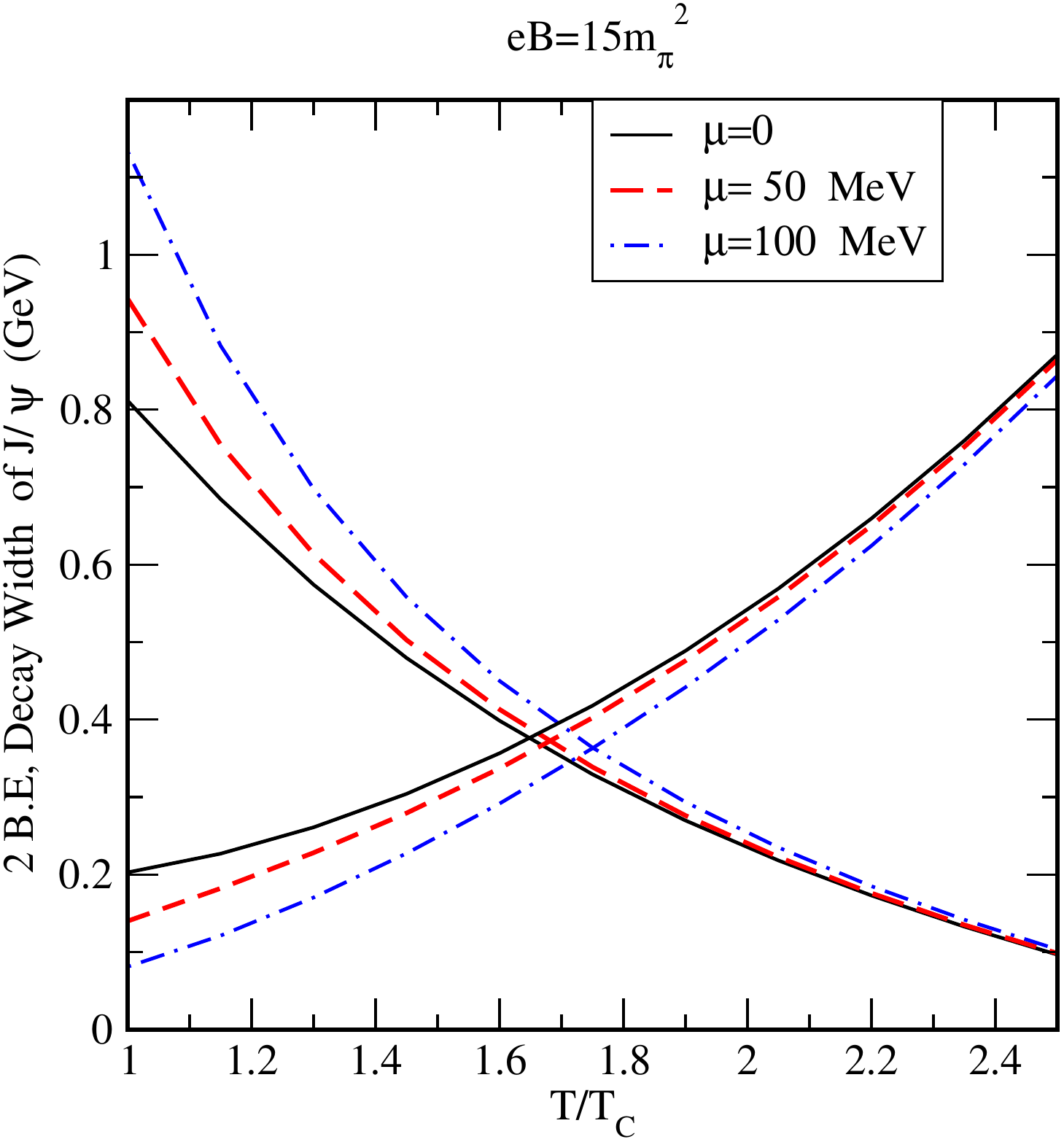}\\
(a) \hspace{4cm}(b)  \hspace{4cm} (c)
\end{tabular}
\caption{{\bf (a)} Variation of real part with inter-quark separation
$r$ {\bf (b)}  Variation of imaginary part with inter-quark separation
$r$ {\bf (c)} Competition between twice the binding energy and decay width of 
$J/\psi$ } 
\label{real_part} 
\end{figure} 
\end{center}

We have observed that the Debye mass gets reduced  
at finite chemical potential but the effect is only visible
at the low $T$ region. The real part of the $Q\bar{Q}$
potential becomes more attractive due to the less screening in the medium
[Fig~\ref{real_part}(a)] while the magnitude of the imaginary 
part gets decreased~[Fig~\ref{real_part}(b)]. 
The binding energy of the $J/\psi$ and $\Upsilon$ 
states is found to be increased at the finite $\mu$ whereas 
the decay width gets reduced. The dissociation
temperature of the $J/\psi$ state has been 
calculated in [Fig~\ref{real_part}(c)] by studying the 
competition between twice the  binding energy and decay 
width and it becomes slightly
 higher in comparison to medium with $\mu=0$.
 The dissociation temperatures for $J/\psi$ are  
found to be $1.64~T_c$ ,$1.68~T_c$,and $1.75~T_c$ 
at the $\mu=0,50$ and $100$ MeV
respectively whereas  $\Upsilon$ is   
dissociated at $1.95~T_c$, $1.97~T_c$ and $2.00~T_c$  
for  $\mu=0,50$ and $100$ MeV
respectively.
\subsection{Summary}
We have examined the effects of finite baryon density 
 on the properties of the 
heavy quarkonia immersed in strongly magnetized hot 
QCD medium. For that purpose we have 
calculated the inverse Fourier transform of the 
(complex) resummed gluon propagator in the static limit.
 We conclude that
quark chemical potential ($\mu$)
 prevents early dissociation
of the quarkonia by reducing the screening 
mass of the medium.

\section{Measurement of charged-particle jet properties in p-Pb collisions at $\sqrt{s_{\rm NN}}=5.02$ TeV with ALICE}
\author{Prottoy Das on behalf of ALICE Collaboration}	

\bigskip

\begin{abstract}
	We present the measurement of charged-particle jet properties in minimum bias p--Pb collisions at $\sqrt{s_{\rm NN}}$~=~5.02 TeV in ALICE. Jets are reconstructed from charged particles at midrapidity using the anti-$k_{\rm T}$ jet finding algorithm with jet resolution parameter $R$~=~0.4. The mean charged particle multiplicity and jet fragmentation function for leading charged-particle jets in the $p_{\rm T}$ interval 10~$<p_{\rm T, jet}^{\rm ch}<$~100 GeV/\textit{c} are measured  and compared with theoretical model predictions.	
\end{abstract}


\subsection{Introduction}
	Jets are collimated sprays of particles produced from the fragmentation and hadronization of hard-scattered partons in high-energy hadronic and nuclear collisions. Jet properties are sensitive to the details of parton showering process and are expected to be modified in the presence of a dense partonic medium. Measurements of intra-jet properties in p--Pb collisions are useful to investigate cold nuclear matter effects~\cite{CNM} and enrich our current understanding of particle production in such collision systems. In this work, we present the measurement of charged-particle jet properties, such as the mean charged particle multiplicity and fragmentation functions, for leading jets in the $p_{\rm T}$ interval 10~$<p_{\rm T, jet}^{\rm ch}<$~100 GeV/\textit{c} at midrapidity in p--Pb collisions at a center of mass energy per nucleon-nucleon pair $\sqrt{s_{\rm NN}}$~=~5.02 TeV with ALICE. Results are compared with theoretical model predictions.
	
\subsection{Analysis details}
The data presented here were collected with the ALICE apparatus in 2016. Detailed information about the ALICE detector can be found in Ref.~\cite{ALICE_det}. Events are selected for this analysis using a minimum bias trigger condition which requires the coincidence of signals in the V0A and V0C forward scintillator arrays~\cite{V0}. Only events with a primary vertex within 10 cm from the nominal interaction point along the beam direction ($|z_{\rm vertex}|$~=~0) are considered which results in total 5.15$\times$10$^8$ events. Charged particles reconstructed with the Inner Tracking System (ITS)~\cite{ITS} and the Time Projection Chamber (TPC)~\cite{TPC} are used for the reconstruction of the primary vertex and jets. These detectors are placed inside a large solenoidal magnet that provides a homogeneous magnetic field $B$~=~0.5~T. 

Charged tracks with $p_{\rm T}>$~0.15 GeV/$c$ within a pseudorapidity range $|\eta|<$~0.9 over the full azimuth are accepted. Charged-particle jets are reconstructed from the selected tracks using anti-$k_{\rm T}$ jet finding algorithm~\cite{antikT} with the $p_{\rm T}$-recombination scheme of FastJet package~\cite{FastJet} with jet resolution parameter, $R$~=~0.4. Only leading jets (jet of highest $p_{\rm T}$ in an event) with 10~$<p_{\rm T, jet}^{\rm ch}<$~100 GeV/\textit{c} are considered for this analysis. The mean charged particle multiplicity in leading jet ($\langle N_{\rm ch} \rangle$) and leading-jet fragmentation function ($z^{\rm ch}=p_{\rm T, track}/p_{\rm T,jet}^{\rm ch}$, where $p_{\rm T, track}$ is the $p_{\rm T}$~of jet constituent inside the leading-jet cone) are studied as a function of jet $p_{\rm T}$.

\begin{figure}
	\vspace*{-0.3cm}
	\begin{center}		
		\includegraphics[width=0.9\linewidth]{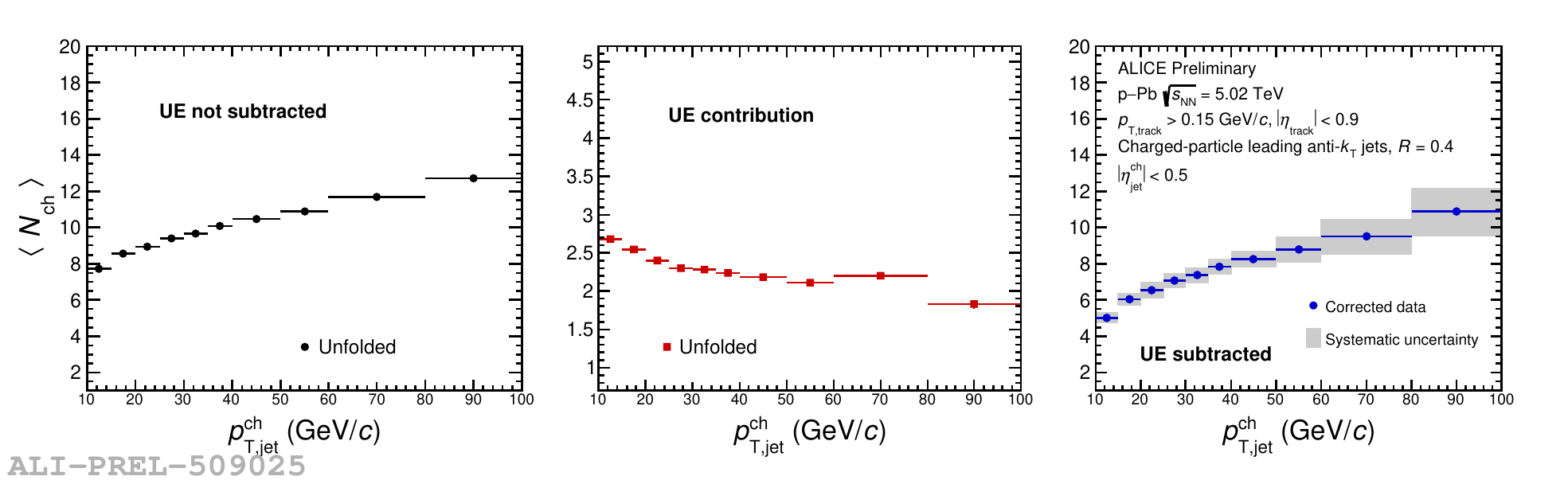}
		\includegraphics[width=0.9\linewidth]{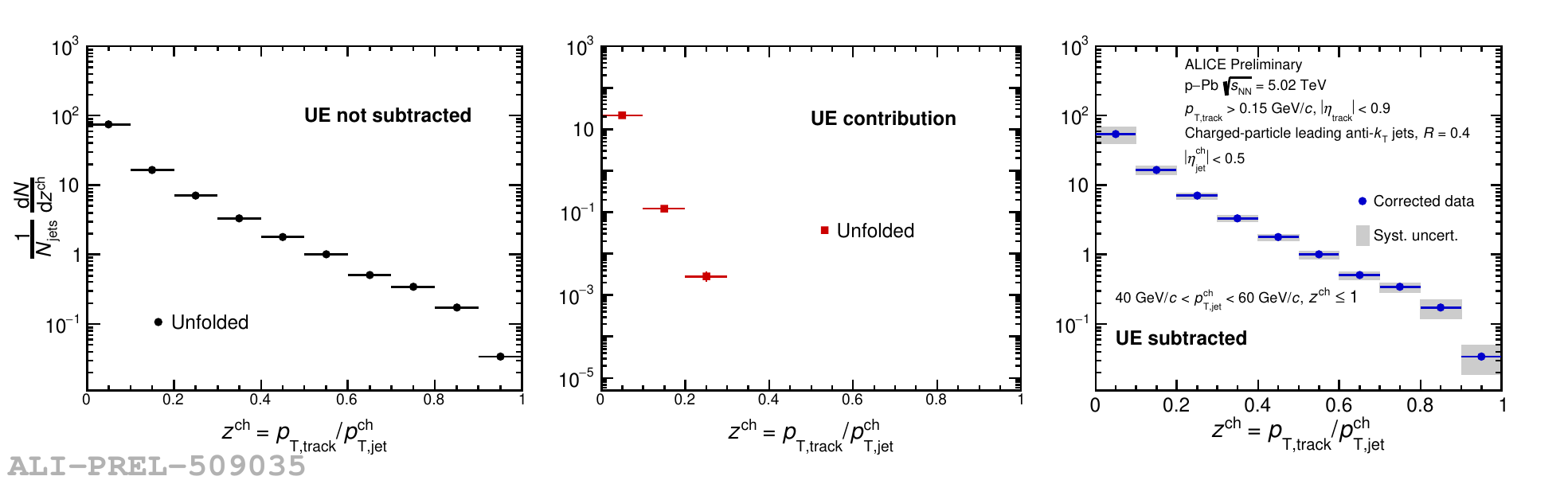}
		\vspace*{-0.3cm}
		\caption{Correction procedure to subtract UE contributions for $\langle N_{\rm ch}\rangle$ and $z^{\rm ch}$ (40~$<p_{\rm T, jet}^{\rm ch}<$~60 GeV/\textit{c}): Unfolded distributions without UE subtraction (left), UE contribution (middle) and after UE subtraction (right).
		}\label{Fig:UECorrMethod}
	\end{center}
\end{figure}

Contribution from the underlying event (UE; coming from sources other than the hard scattered partons) is estimated using the perpendicular-cone method and subtracted on a statistical basis after unfolding as reported in Fig.~\ref{Fig:UECorrMethod}. In the perpendicular-cone method, circular cones of radius $R$~=~0.4 at the same $\eta$ as the leading jet and perpendicular to the leading jet axis are used for the estimation of the contribution from the UE.

A 2-dimensional Bayesian unfolding technique~\cite{Bayesian} (implemented in RooUnfold~\cite{RooUnfold} package) is applied to correct for instrumental effects such as track-reconstruction efficiency and momentum resolution. For each of the jet observables, a 4D response matrix is constructed from a Monte Carlo (MC) simulation performed with the DPMJET~\cite{DPMJET} event generator and the generated particles are transported through the GEANT detector simulation. DPMJET is a multipurpose generator based on the Dual Parton Model (DPM) and is able to simulate a wide variety of collision systems for energies ranging from a few GeV up to the highest cosmic-ray energies. To construct the response matrix, the detector- and particle-level jets are matched geometrically and only the leading detector-level jet and the corresponding matched particle-level jet in an event are considered. 

Systematic uncertainties from various sources such as tracking efficiency, modelling of the jet properties and the detector response in the MC simulation, choice of regularization parameter or number of iterations in Bayesian unfolding, change in prior distribution, and underlying event contribution are estimated and added in quadrature to calculate the total systematic uncertainty. The total systematic uncertainty is found to be 5\%~-~12\% in the $\langle N_{\rm ch} \rangle$ analysis whereas in the $z^{\rm ch}$ analysis, it is estimated to be $\sim$20\% for 20~$<p_{\rm T, jet}^{\rm ch}<$~30 GeV/\textit{c} and 25\%~-~45\% for 40~$<p_{\rm T, jet}^{\rm ch}<$~60 GeV/\textit{c}.


\subsection{Results and discussion}

\begin{figure}[htb]
	\centering
	\begin{minipage}{.5\textwidth}
		\centering
		\includegraphics[scale=0.25]{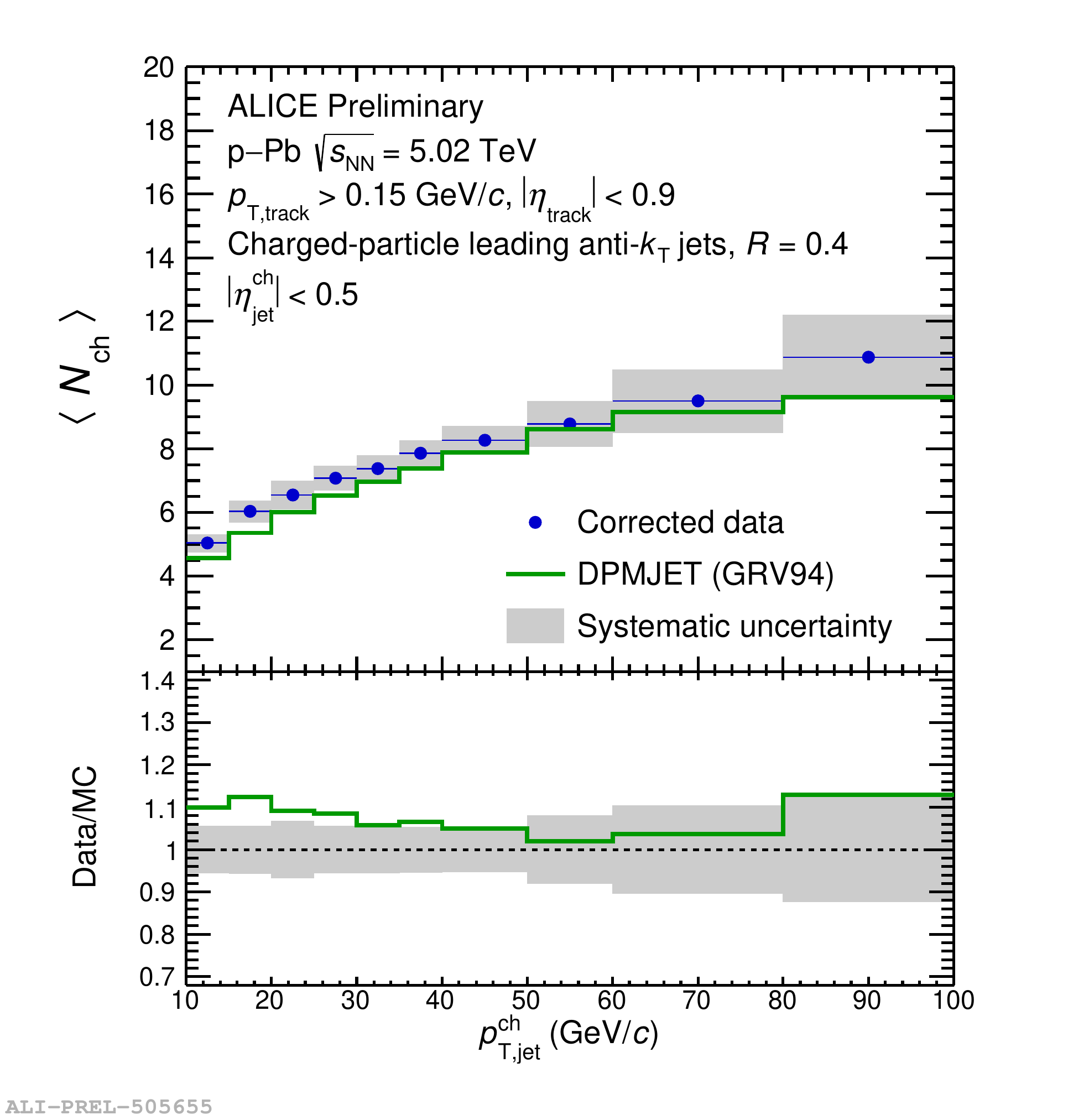}
		\label{Fig:NchMB}
	\end{minipage}%
	\begin{minipage}{.5\textwidth}
		\centering
		\includegraphics[scale=0.245]{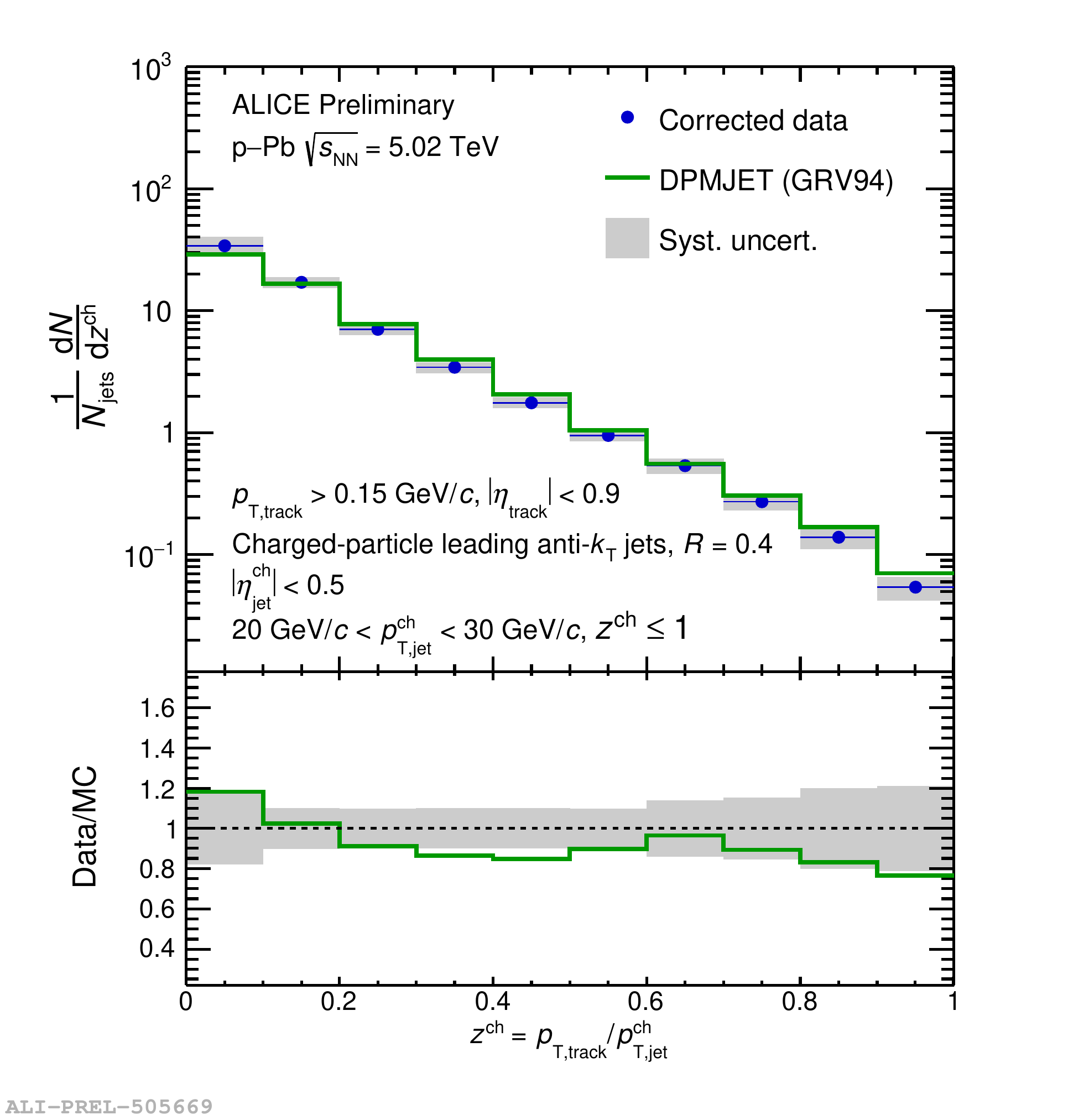}
		\label{Fig:FFMB20to30}
	\end{minipage}
	\centering
	\begin{minipage}{.5\textwidth}
		\centering
		\includegraphics[scale=0.25]{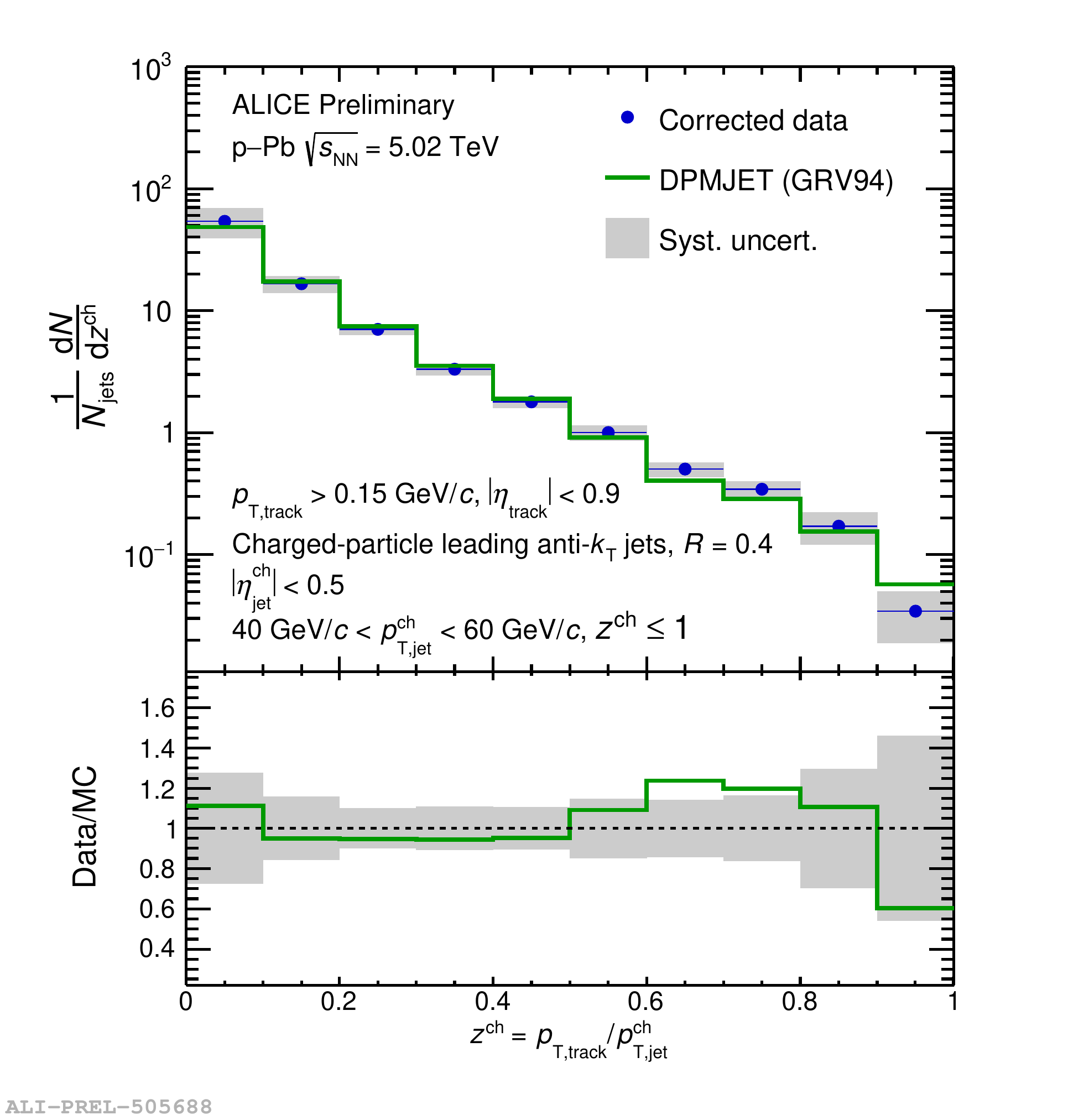}
		\label{Fig:FFMB40to60}
	\end{minipage}%
	\centering
	\begin{minipage}{.5\textwidth}
		\centering
		\includegraphics[scale=0.25]{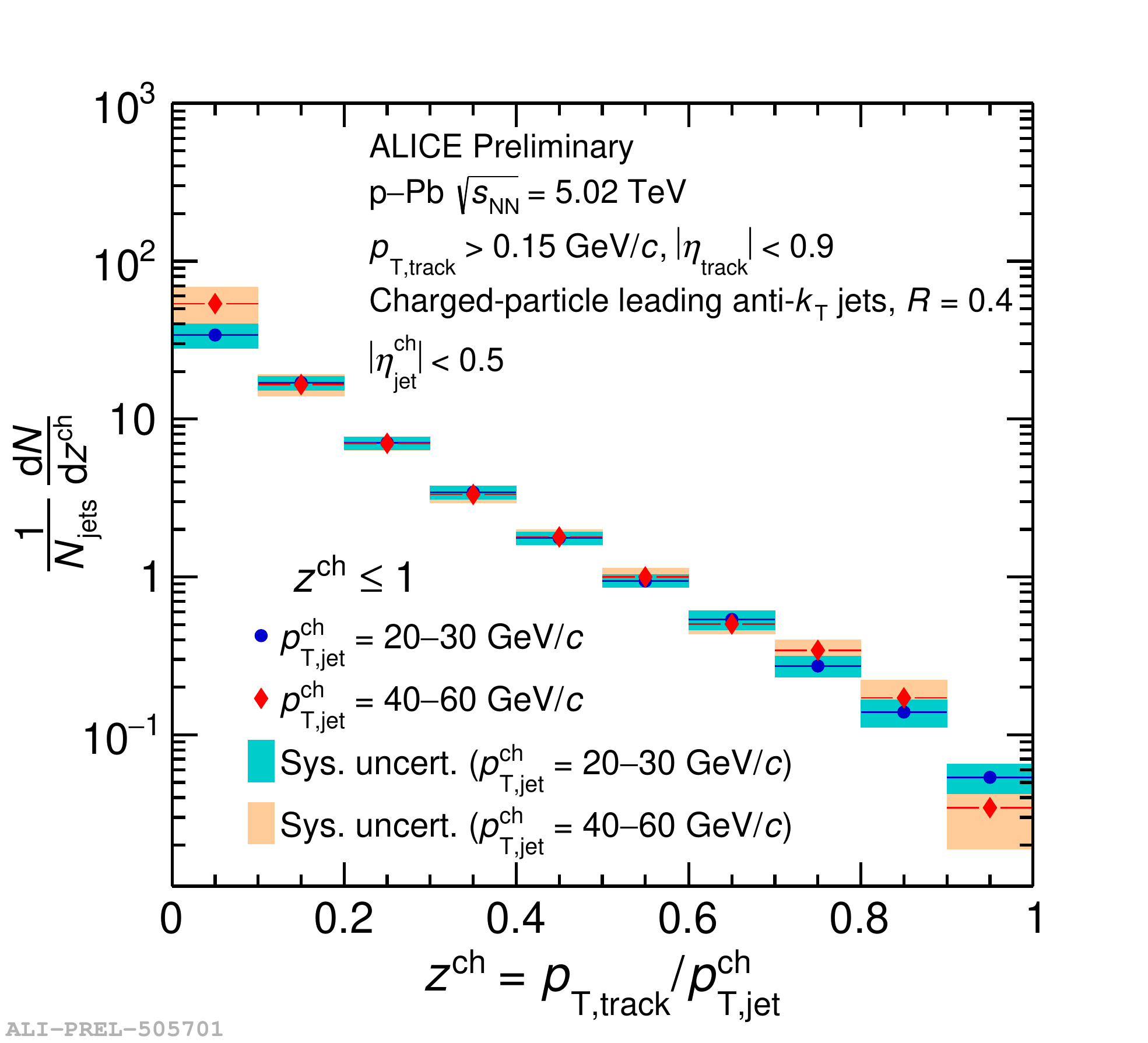}
	\end{minipage}%
	
	\caption{Top panels: (top left) $\langle N_{\rm ch}\rangle$ distribution as a function of leading jet $p_{\rm T, jet}^{\rm ch}$. (top right) and (bottom left) $z^{\rm ch}$ distributions for leading jets with 20~$<p_{\rm T, jet}^{\rm ch}<$~30 GeV/\textit{c} and 40~$<p_{\rm T, jet}^{\rm ch}<$~60 GeV/\textit{c} respectively. Bottom panels: ratio between the data and MC results.  (Bottom right) $z^{\rm ch}$ distributions for leading jets with 20~$<p_{\rm T, jet}^{\rm ch}<$~30 GeV/\textit{c} and 40~$<p_{\rm T, jet}^{\rm ch}<$~60 GeV/\textit{c}.}
	\vspace*{-0.4cm}
	\label{Fig:FFAllMB}
\end{figure}


Figure~\ref{Fig:FFAllMB} (top left) shows the $\langle N_{\rm ch}\rangle$ distribution as a function of leading jet $p_{\rm T, jet}^{\rm ch}$ in the top panel. The blue markers represent the data and the green curve shows the DPMJET (GRV94~\cite{GRV94}) prediction. The ratio between the data and DPMJET is depicted in the bottom panel. It is observed that $\langle N_{\rm ch}\rangle$ increases with $p_{\rm T, jet}^{\rm ch}$ and DPMJET reproduces the data for $p_{\rm T, jet}^{\rm ch}>$~30 GeV/$c$
within uncertainties. Figure~\ref{Fig:FFAllMB} (top right and bottom left) show the $z^{\rm ch}$ distributions for leading jets with 20~$<p_{\rm T, jet}^{\rm ch}<$~30 GeV/\textit{c} and 40~$<p_{\rm T, jet}^{\rm ch}<$~60 GeV/\textit{c} respectively in the to panels. Results are compared with the DPMJET (GRV94) predictions. The ratios between the data and DPMJET predictions are shown in the bottom panels. It is found that DPMJET reproduces the $z^{\rm ch}$ distributions in both $p_{\rm T, jet}^{\rm ch}$ ranges within systematic uncertainties. The comparison of the $z^{\rm ch}$ distributions in the intervals 20~$<p_{\rm T, jet}^{\rm ch}<$~30 GeV/\textit{c} and 40~$<p_{\rm T, jet}^{\rm ch}<$~60 GeV/\textit{c} shown in Fig.~\ref{Fig:FFAllMB} (bottom right) indicates that the $z^{\rm ch}$ distribution follows a scaling behaviour with $p_{\rm T, jet}^{\rm ch}$ within systematic uncertainties.

\subsection{Summary}
We have presented the measurement of charged-particle jet properties in minimum bias p--Pb collisions at $\sqrt{s_{\rm NN}}$~=~5.02 TeV in ALICE. Results are compared with DPMJET (GRV94) predictions, which reproduces both the measured distributions ($\langle N_{\rm ch}\rangle$ and $z^{\rm ch}$) within uncertainties except for $\langle N_{\rm ch}\rangle$ at very low $p_{\rm T, jet}^{\rm ch}$ ($<30$ GeV/$c$). A scaling of jet fragmentation with leading charged-particle jet $p_{\rm T}$ is observed within systematic uncertainties.

%
%
\newcommand {\roots}    {\ensuremath{\sqrt{s}}}
\newcommand {\sqrtsNN}  {\ensuremath{\sqrt{s_{_{\text{NN}}}}}}
\providecommand{\KS}{\ensuremath{\mathrm{K^0_S}}}
\providecommand{\LM}{\ensuremath{\Lambda}}
\providecommand{\ALM}{\ensuremath{\overline{\Lambda}}}
\providecommand{\KSKS}{\ensuremath{\mathrm{K^{0}_{S}K^{0}_{S}}}}
\providecommand{\LLALAL}{\ensuremath{\Lambda\Lambda\oplus\overline{\Lambda}\overline{\Lambda}}}
\providecommand{\LAL}{\ensuremath{\Lambda\overline{\Lambda}}}
\providecommand{\ALMALM}{\ensuremath{\overline{\Lambda}\overline{\Lambda}}}
\providecommand{\LMLM}{\ensuremath{\Lambda\Lambda}}
\providecommand{\LKS}{\ensuremath{\Lambda\mathrm{K^0_S}}}
\providecommand{\ALKS}{\ensuremath{\overline{\Lambda}\mathrm{K^0_S}}}
\providecommand{\LALKS}{\ensuremath{\Lambda\mathrm{K^0_S}\oplus\overline{\Lambda}\mathrm{K^0_S}}}
\providecommand{\VZ}{\ensuremath{\textrm{V}^{0}}}
\providecommand{\roots}{\ensuremath{\sqrt{s}}}
\section{Strange particles femtoscopic correlation in PbPb collisions at \sqrtsNN = 5.02 TeV}
\author{Raghunath Pradhan}	

\bigskip

\begin{abstract}
	Two-particle correlations as a function of relative momentum are measured for \KS, \LM, and $\ALM$ strange hadrons.  The data were obtained for PbPb collisions at \sqrtsNN = 5.02 TeV using the CMS detector at the LHC. Such correlations are sensitive to  quantum statistics and to possible final-state interactions between the particles. The source radii are extracted from $\KSKS$ correlations in different centrality regions and are found to decrease from central to peripheral collisions. The strong-interaction scattering parameters (i.e., scattering length and effective range) are extracted from $\LALKS$ and $\LLALAL$ correlations using the Lednicky-Lyuboshits model and compared with other experimental and theoretical results.
\end{abstract}

\subsection{Introduction}
Identical and nonidentical particle short-range correlations in relative momentum, known as ``femtoscopic" correlations, can be used to study the space-time extent of the particle emitting source created in relativistic heavy ion collisions~\cite{1_Raghunath}. The identical particle correlations are sensitive to quantum statistics (QS) and to possible final-state interactions (FSI), while nonidentical particle correlations are only sensitive to final state interactions. The correlations among the neutral \KS, \LM, and $\ALM$ particles, collectively referred to as $\VZ$ particles, are of specific interest. They can be used to extract the size of the particle-emitting source, and, in a way complementary to dedicated scattering experiments, the strong-interaction parameters, i.e., the scattering length and the effective range. Because of their heavy mass and absence of Coulomb interactions, femtoscopy based on neutral kaon particles (\KS) complements the more commonly studied pion and charged kaon femtoscopy. By studying $\LALKS$ and $\LMLM$ correlations, it is possible to extract the strong interaction scattering parameters for baryon-meson and baryon-baryon systems, which can shed light on the nature of the strong interaction. \par This note presents the \KSKS, $\LALKS$ and $\LLALAL$ femtoscopic correlations in lead-lead (PbPb) collisions at center-of-mass energy per nucleon pair of $\sqrtsNN$ = 5.02 TeV using the data recorded by ``the CMS~\cite{CMS} experiment" at the LHC.  The $\KSKS$ correlations were measured in the extended range of centrality bin ($0\text{--}60\%$), where centrality is defined as the fraction of the total nucleus-nucleus cross section, with $0\%$ denoting the maximum overlap of the colliding nuclei. The $\LALKS$ and $\LLALAL$ correlations were measured in centrality range $0\text{--}80\%$. The size of the particle emitting source is extracted from \KSKS, $\LALKS$, and $\LLALAL$ correlations. The strong interaction scattering parameters are extracted from $\LALKS$ and $\LLALAL$ correlations using Lednicky-Lyuboshits fit~\cite{3_Raghunath} and compared with theoretical calculations and results from other experiments.
\subsection{$\KSKS$ femtoscopic correlation}
\begin{figure}[thbp]
\centering
\includegraphics[width=6.7cm, height = 5.2cm]{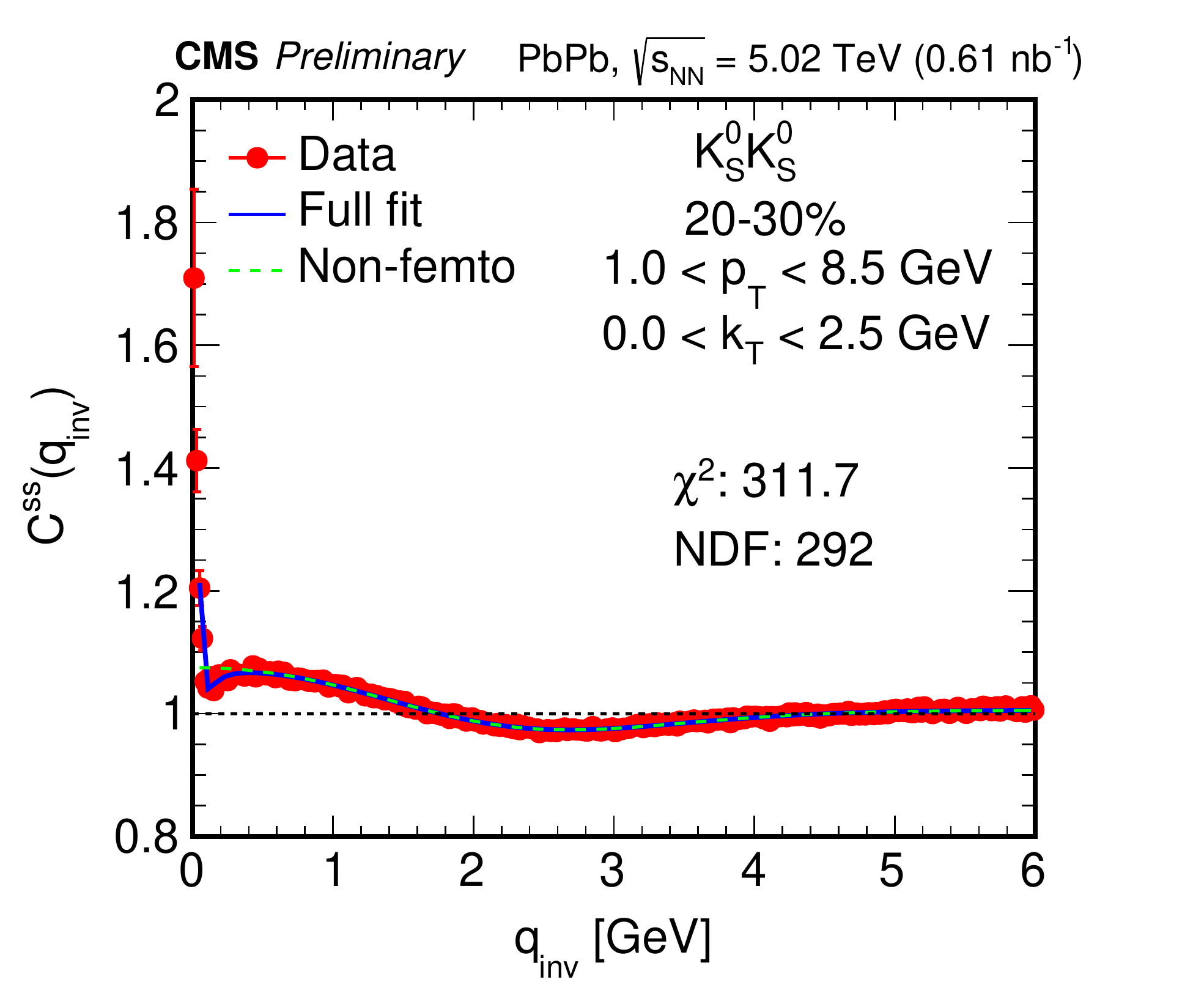}
\includegraphics[width=5.8cm, height = 5cm]{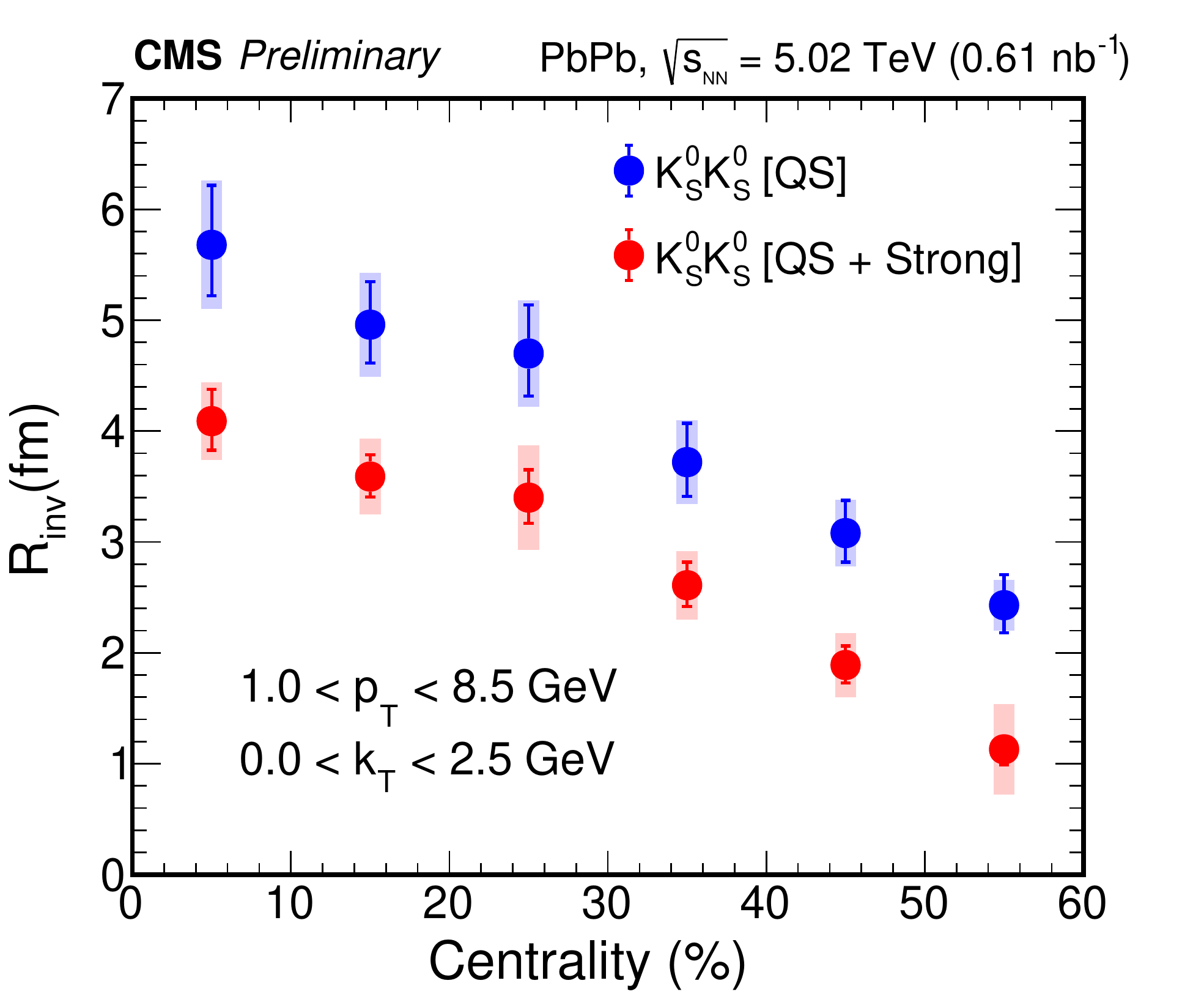}
\caption{Left: An example of correlation measurement and their fit for $\KSKS$ in $20\text{--}30\%$ centrality. In this plots, red filled circles are the experimental results, the blue solid line is the full fit~\cite{3_Raghunath}, and the green dotted line is the nonfemtoscopic background~\cite{3_Raghunath}. Right: $R_{\text{inv}}$ as a function of centrality by considering only the QS (blue circle) and both the QS and strong FSI effects (red circles). For each data point, the line and shaded area indicate the statistical and systematic uncertainty, respectively.}
\label{fig:ks}
\end{figure}

The left plot of figure~\ref{fig:ks} shows the $\KSKS$ correlation measurement as a function of relative momenta of the particles pair $q_{inv}$~\cite{1_Raghunath,3_Raghunath} in $20\text{--}30\%$ centrality with $0 < k_{T} < 2.5$ GeV, where $k_{T} \equiv |\vec{p}_{\mathrm{T},1} + \vec{p}_{\mathrm{T},2}|/2$ is the average transverse momentum of the pair~\cite{3_Raghunath}. The size of the particle-emitting source $R_{\text{inv}}$ (right) extracted from $\KSKS$ correlation using the Lednicky-Lyuboshits fit~\cite{4_Raghunath} together with nonfemtoscopic background~\cite{3_Raghunath} for different centrality ranges and plotted in the right plot of figure~\ref{fig:ks}~\cite{3_Raghunath}. It can be seen that $R_{\text{inv}}$ decreases from central ($0\text{--}10\%$) to peripheral ($50\text{--}60\%$) collisions. The values of $R_{\text{inv}}$ as extracted by considering only the QS effect are larger than those found from considering both QS and FSI effects, which suggests that the FSI effects needs to be consider for the accurate measurement of $R_{\text{inv}}$. 

\begin{figure}[thbp]
\centering
\includegraphics[width=6.2cm, height = 5.2cm]{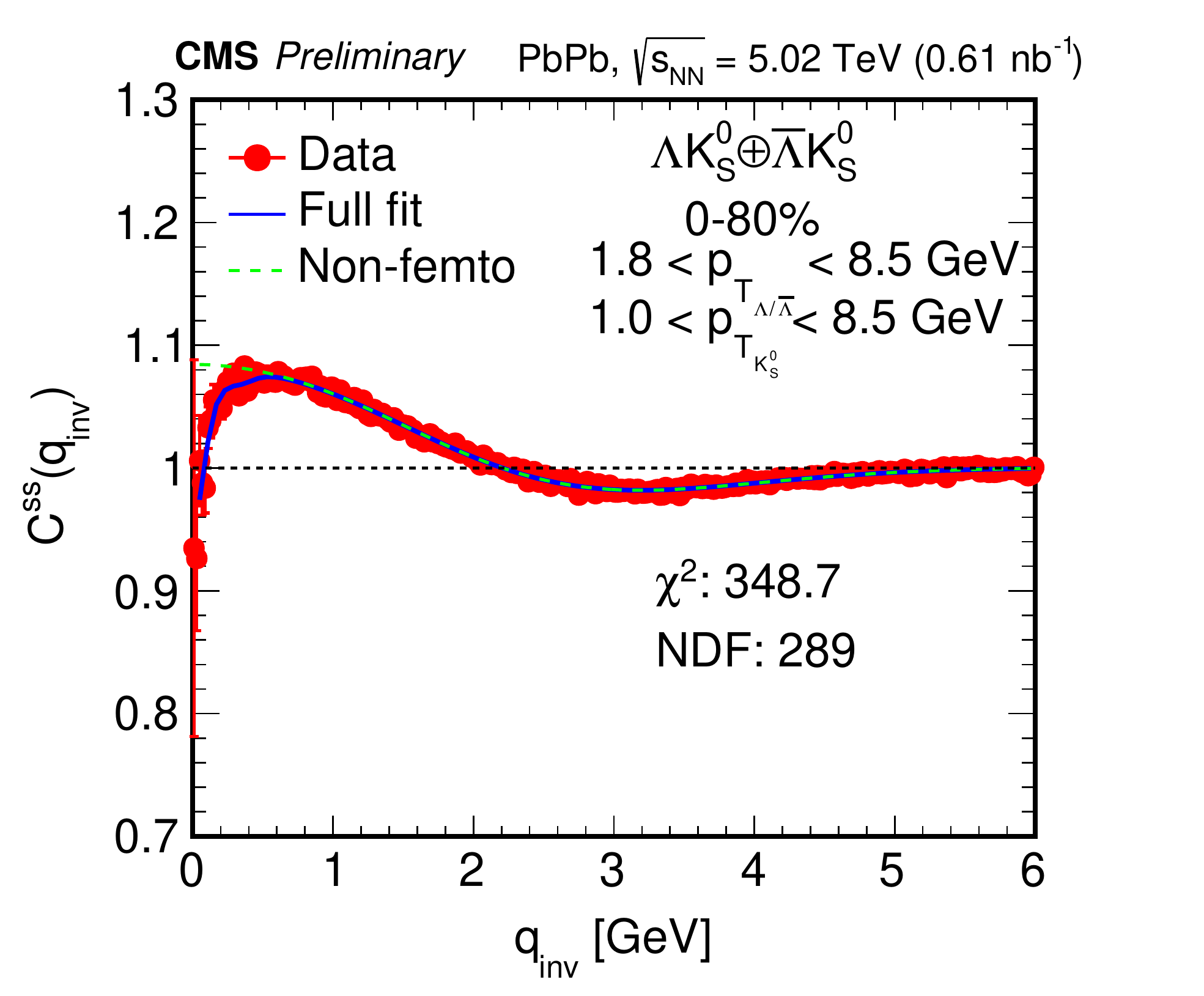}
\includegraphics[width=6.2cm, height = 5.2cm]{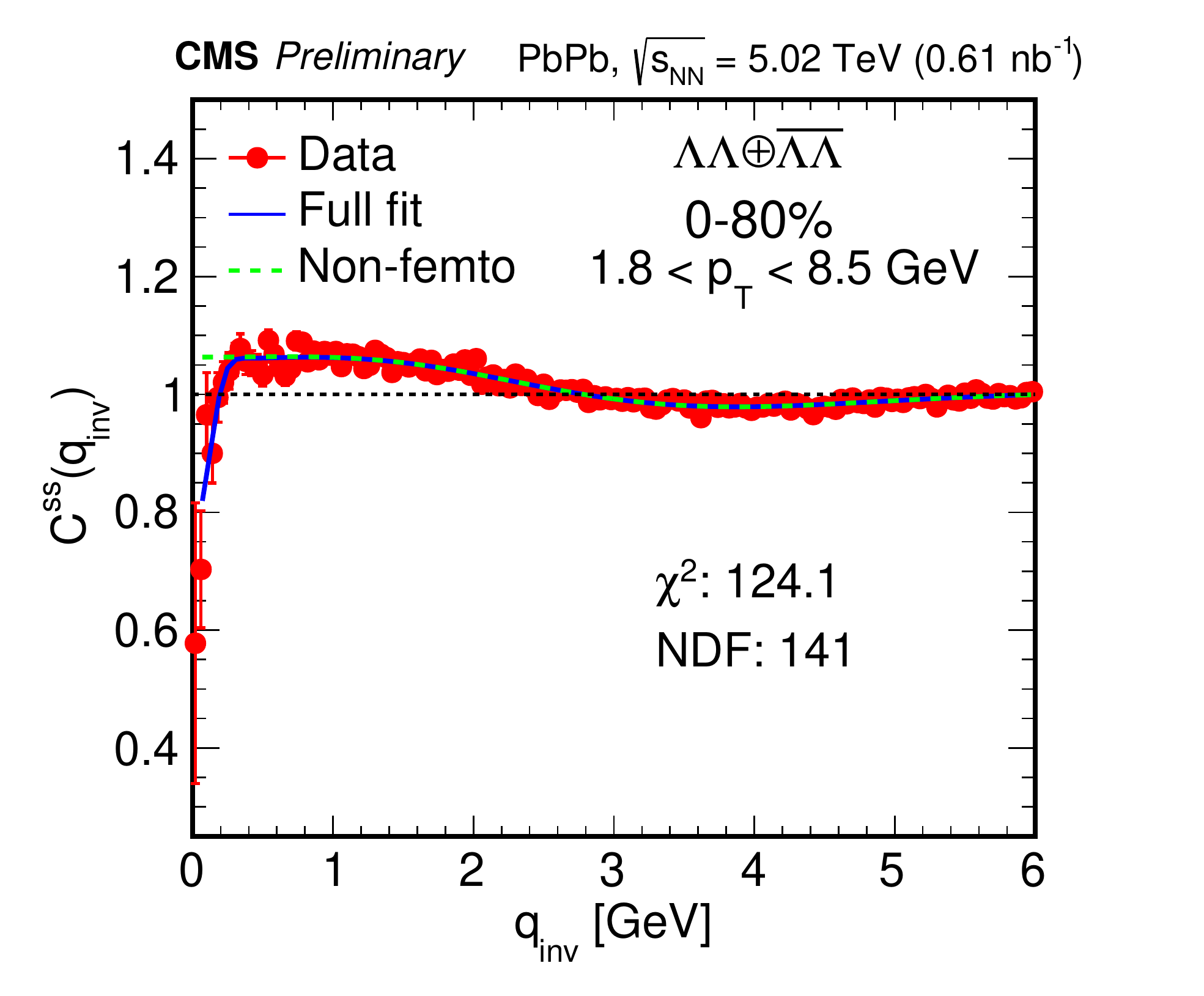}
\caption{The correlation measurements and their fits for $\LALKS$ (left) and $\LLALAL$ (right)in $0\text{--}80\%$ centrality. In these plots, red circles are the experimental results, the blue solid line is the full fit, and the green dotted line is the nonfemtoscopic background fit~\cite{3_Raghunath}. For each data point, the line and shaded area indicate the statistical and systematic uncertainty, respectively.}
\label{fig:lks}
\end{figure}

\begin{figure}[h!]
\centering
\includegraphics[width=6cm, height = 5cm]{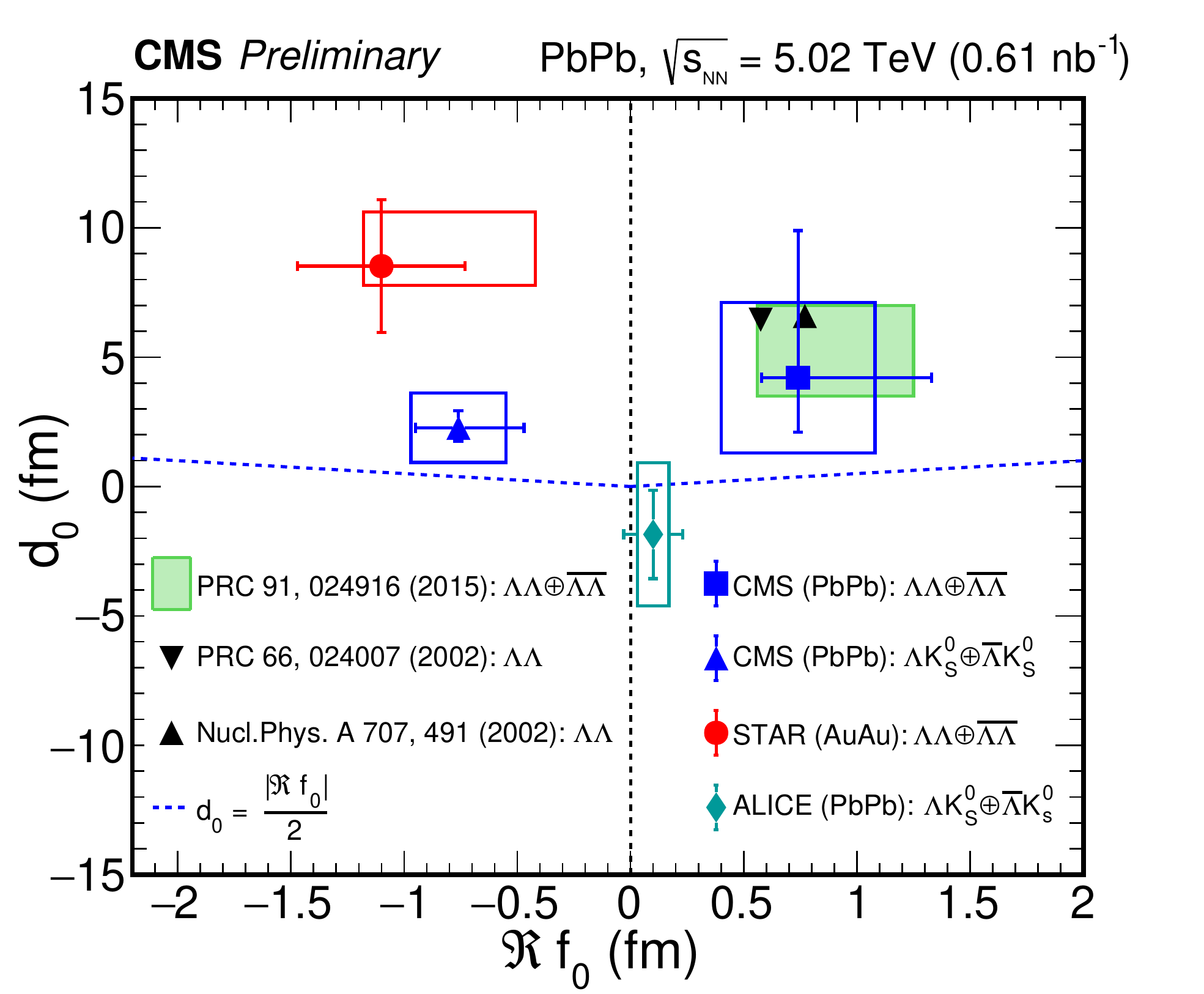}
\includegraphics[width=6cm, height = 5cm]{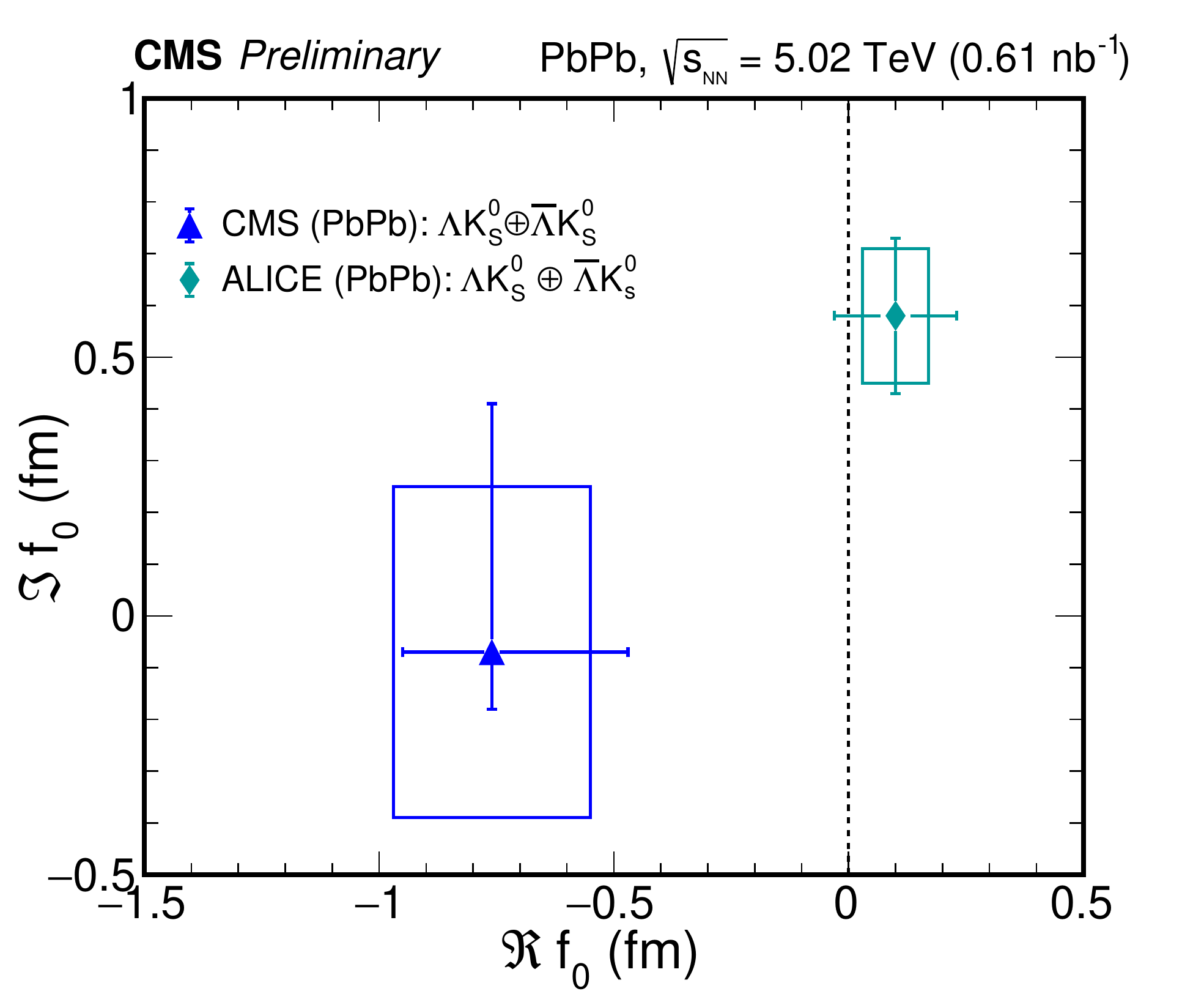}
\caption{Figure shows the values of $d_{0}$ and $\Re f_{0}$ (left) and the values of $\Im f_{0}$ and $\Re f_{0}$ (right). In the left plot, the blue triangle and square marker are for $\LALKS$ and $\LLALAL$ correlations, respectively, and are compared with the $\LLALAL$ results from STAR experiment~\cite{6_Raghunath} and $\LALKS$ result from ALICE experiment~\cite{5_Raghunath}. A reanalysis of STAR data for $\LLALAL$ correlations is shown in the shaded area~\cite{7_Raghunath}. Theory calculations of the $\Lambda\Lambda$ interaction parameters are shown as black triangles~\cite{8_Raghunath,9_Raghunath}. In the right plot, the triangle is for $\LALKS$ correlation, and is compared with ALICE $\LALKS$ results~\cite{5_Raghunath}.For each data point, the two lines and the box indicate the (one-dimensional) statistical and systematic uncertainties, respectively.}
\label{fig:f0d0}
\end{figure}
\subsection{$\LALKS$ and $\LLALAL$ femtoscopic correlations}
Figure~\ref{fig:lks} shows the $\LALKS$ (left) and $\LLALAL$ (right) correlation measurements in $0-80\%$ centrality with no restriction on $k_{T}$~\cite{3_Raghunath}. The $R_{\text{inv}}$ and strong-interaction scattering parameters: real scattering length ($\Re f_{0}$), imaginary scattering length ($\Im f_{0}$), and effective range ($d_{0}$) are extracted from the $\LALKS$ and $\LLALAL$ correlations using Lednicky-Lyuboshits fit~\cite{4_Raghunath} are listed in Table~\ref{tab:scatteringpara} and plotted in Fig.~\ref{fig:f0d0}. Figure~\ref{fig:f0d0} shows $d_{0}$ versus $\Re f_{0}$ (left) and $\Im f_{0}$ versus $\Re f_{0}$ (right). Comparisons are shown to theoretical calculations and results from other experiments~\cite{5_Raghunath,6_Raghunath,7_Raghunath,8_Raghunath,9_Raghunath}. The negative value of $\Re f_{0}$ in $\LALKS$ correlations suggests that the $\mathrm{\Lambda K^{0}_{S}}$ ($\mathrm{\overline{\Lambda} K^{0}_{S}}$) interaction is repulsive while $\Im f_{0}$ is consistent with zero within their uncertainty, preventing us from drawing any conclusion about inelastic processes~\cite{3_Raghunath}. A positive $\Re f_{0}$ value for the $\LLALAL$ correlations indicates that the $\LMLM(\ALMALM)$ interaction is attractive.

\begin{table}[ht]\renewcommand{\arraystretch}{1.2}\addtolength{\tabcolsep}{-1pt}
\caption{Extracted values of the $R_{\text{inv}}$, $\Re f_{0}$, $\Im f_{0}$, and $d_{0}$ from $\LALKS$ and $\LLALAL$ correlations in the $0\text{--}80\%$ centrality.}
\centering
\begin{tabular}{ c  c  c }
\hline
Parameter  & \LALKS & \LLALAL \\
\hline
$R_{inv}$ (fm) & $2.09^{+1.42}_{-0.50}\;(\text{stat}) \pm 0.65\;(\text{syst})$ & $1.33^{+0.37}_{-0.23}\;(\text{stat}) \pm 0.25\;(\text{syst})$ \vspace{0.11cm}\\
$\Re f_{0}$ (fm) &  $-0.76^{+0.29}_{-0.19}\;(\text{stat}) \pm 0.21\;(\text{syst})$ & $0.74^{+0.59}_{-0.16}\;(\text{stat}) \pm 0.34\;(\text{syst})$ \vspace{0.11cm}\\
$\Im f_{0}$ (fm) &  $-0.07^{+0.48}_{-0.11}\;(\text{stat}) \pm 0.31\;(\text{syst})$ & \text{---} \vspace{0.11cm}\\
$d_{0}$ (fm) & $2.27^{+0.66}_{-0.53}\;(\text{stat}) \pm 1.31\;(\text{syst})$ & $4.21^{+5.68}_{-2.11}\;(\text{stat}) \pm 2.91\;(\text{syst})$ \vspace{0.11cm}\\
\hline
\end{tabular}
\label{tab:scatteringpara}
\end{table}

\subsection{Summary}
The $\KSKS$, $\LALKS$, and $\LLALAL$ femtoscopic correlations are presented  in PbPb collisions at a center-of-mass energy per nucleon pair of $\sqrtsNN = 5.02$ TeV, as measured by the CMS experiment at the LHC. This is the first report of $\LLALAL$ correlations in PbPb collisions. The source size $R_{inv}$ is extracted for $\KSKS$ correlations in six centrality bins from $0\text{--}60\%$ centrality and found to decrease toward more peripheral collisions. The values of $R_{\text{inv}}$, based on $\LALKS$ and $\LLALAL$ correlations, are also presented for the $0\text{--}80\%$ centrality range . The Lednicky-Lyuboshits model fit to the data suggests that the $\LLALAL$ interaction is attractive, whereas the $\LALKS$ interaction is repulsive.

%
 
\section{Measurement of exclusive vector meson photoproduction in pPb and PbPb collisions with the CMS experiment}
\author{Subash Chandra Behera}	

\bigskip

\begin{abstract}
	The exclusive photoproduction of vector mesons provides a unique opportunity to constrain the gluon distribution function within protons and nuclei. Measuring vector mesons of various masses over a wide range of rapidity and as a function of transverse momentum provides important information on the evolution of the gluon distribution within nuclei. A variety of measurements, including the exclusive J/$\psi$, $\rho$, and $\Upsilon$ meson production in pPb (at nucleon-nucleon center of mass energies of 5.02 and 8.16 TeV) and PbPb (5.02 TeV) collisions, are presented as a function of squared transverse momentum and the photon-proton center of mass energy. Finally, compilations of these data and previous measurements are compared to various theoretical predictions.
\end{abstract}

\subsection{Introduction}
This article presents the measurement of the exclusive photoproduction of $\Upsilon$ and $J/\psi$ mesons from protons in pPb collisions at a nucleon–nucleon centre-of-mass energy of $\sqrt{s_\text{NN}} = 5.02$ TeV with the CMS detector~\cite{CMS}. Ultraperipheral collisions (UPCs) of protons or ions occur at the impact parameter is larger than the sum of their radii, therefore it is largely suppressed by hadronic interaction~\cite{1_Subash}. In UPCs, one of the incoming hadron emits quasi real photon and they converted into vector meson state by a color singlet gluon exchange process with the other hadrons.  Since incoming hadrons remain intact in the process and vector meson is produced in the event, the process is called “exclusive”. The study of exclusive quarkonia photoproduction can provide a unique probe of the target hadron structure, with the large mass of the $J/\psi$ and $\Upsilon$ mesons providing a hard scale for calculations based on perturbative quantum chromodynamics (pQCD)~\cite{2_Subash}. The study of the photoproduction of $\Upsilon$ and $\rho^{0}$ mesons from proton is sensitive to the generalized parton distributions (GPDs), which can be approximated by the square of the gluon density in the proton. Exclusive vector photoproduction is very interesting because the Fourier transform of the $t$ distribution is related to the two-dimensional distribution of the struck partons in the transverse plane. Here, $t$ indicates the squared four-momentum transfer at the proton vertex. In this proceedings, the $|t| \approx p_\text{T}^{2} $ distributions of $\Upsilon$ and $\rho^{0}$ are presented.\\

In this talk, we discuss the exclusive photoproduction of $\rho^{0}$ that decay to $\pi^{+}\pi^{-}$ and $\Upsilon$ decay to $\mu^{+}\mu^{-}$ channel in ultra-peripheral pPb collisions at $\sqrt{s_\text{NN}} = 5.02$. The cross section is measured as a function of  photon-proton center-of-mass energy, $W_{\gamma p}$ and $t$. This note is organized as follows. In section~\ref{sec:massinv} discusses the invariant mass of the vector meson. Section~\ref{sec:xsec} presents the differential cross section as a function of rapidity, transverse momentum and $W_{\gamma p}$. 

\subsection{Invariant mass}
\label{sec:massinv}
Figure 1 shows the invariant mass distribution for dimuon in the range between 8 and 12 GeV that satisfies the selection criteria described in Ref~\cite{upsilon}. An unbinned likelihood fit to the spectrum is performed using ROOFIT [52] with a linear function to describe the QED $\gamma  + \gamma \to \mu{+} \mu^{-}$ continuum background, where the background slope parameter is fixed to the \textsc{Starlight} $\gamma  + \gamma \to \mu{+} \mu^{-}$ simulation, with three Gaussian functions for the three $\Upsilon$ signal peaks. 

\begin{figure}[thbp]
\centering
\includegraphics[width=6.cm, height = 5.cm]{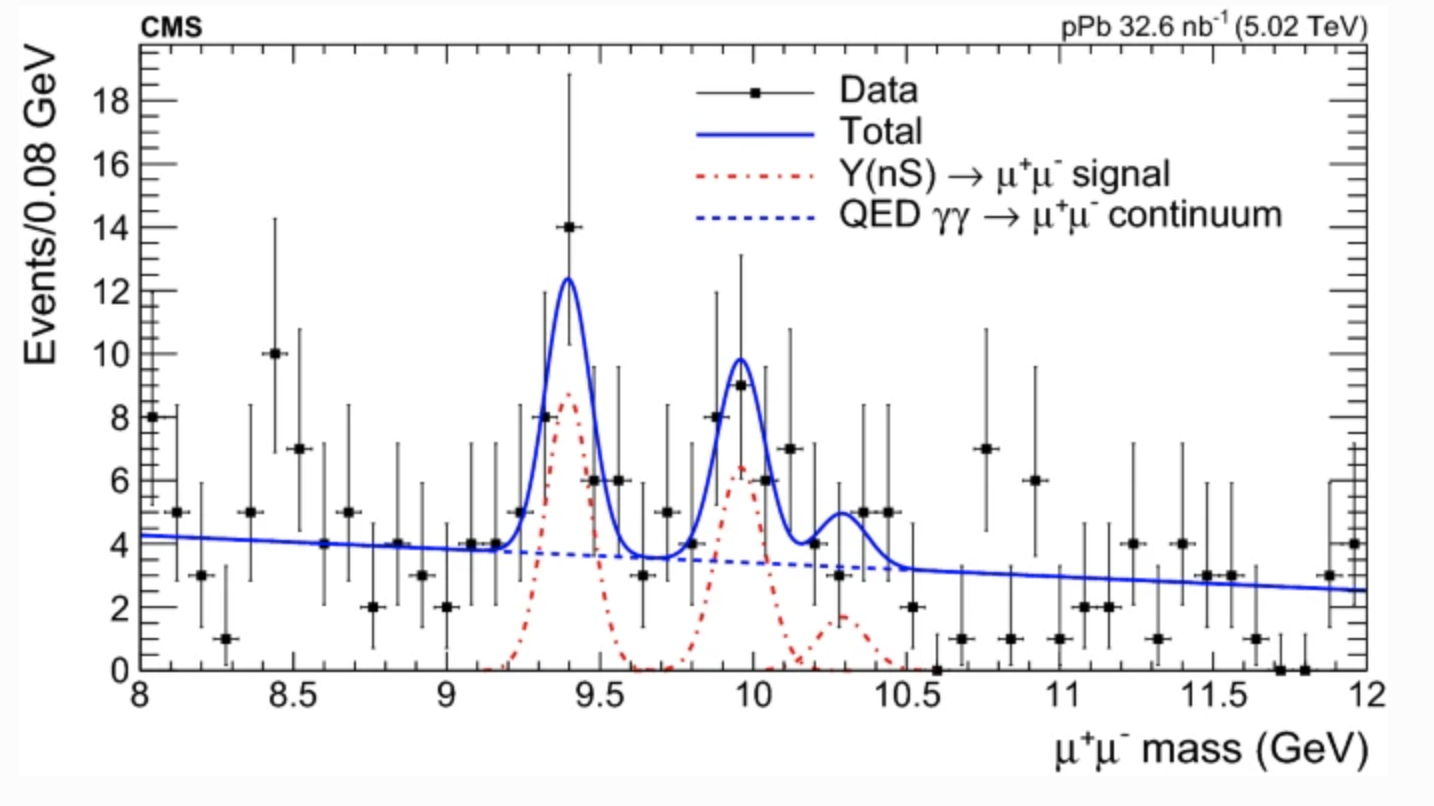}
\includegraphics[width=6.cm, height = 5.cm]{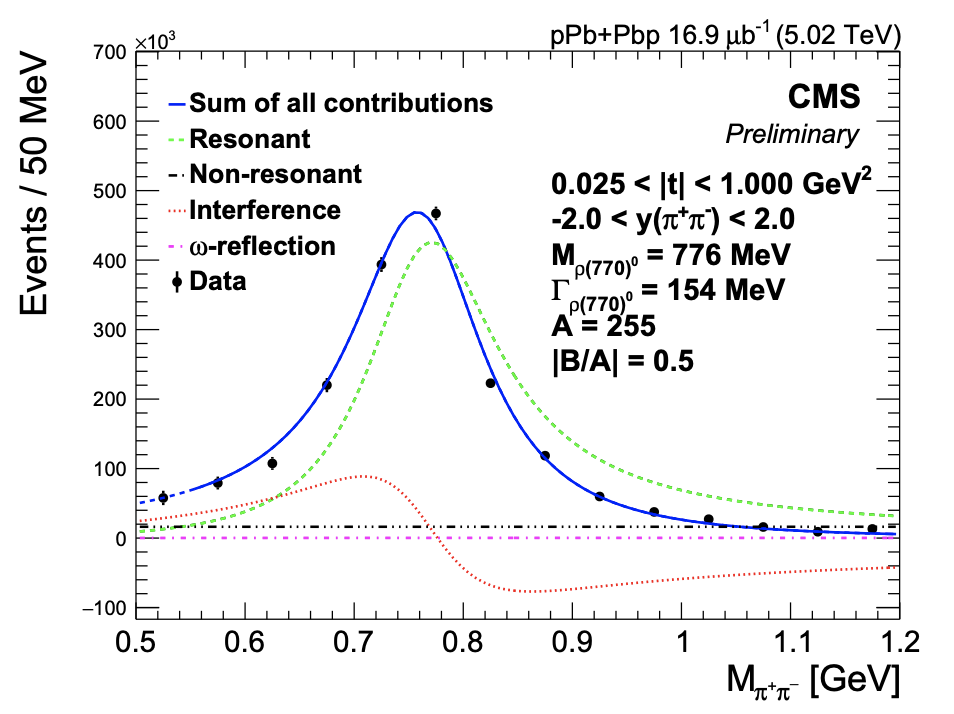}
\caption{Left side figure shows the invariant mass distribution of the exclusive dimuon candidates in the range $8 < m_{\mu^{+}\mu^{-}} < 12$ GeV that satisfied all the selection criteria, fitted to a linear function for the two-photon QED continuum (blue dashed line) plus three Gaussian distributions corresponding to the $\Upsilon(\text{1S}), \Upsilon(\text{2S}), \Upsilon(\text{3S})$ mesons (dashed-dotted-red curves). Right side figure shows unfolded $\pi^{+}\pi^{-}$ invariant mass distribution in the pion pair rapidity interval $|y_{\pi^{+}\pi^{-}}| < 2$  fitted with the Soding model. 
The green dashed line indicates resonant $\rho(770)^{0}$ production, the red dotted line represents the interference term, the magenta dash-dotted line corresponds to the background from $\omega \to \pi^{+}\pi^{-}\pi^{0}$, the black dash-dotted line represents the no resonant contribution, and the blue solid line represents the sum of all these contributions.}
\label{fig:mass}
\end{figure}

Figure ~\ref{fig:mass} (right plot) presents the fit of the unfolded distribution with the modifed Soding model. A least squares fit is performed for the interval $0.6 < M_{\pi^{+}\pi^{-}}< 1.1 $ GeV, with the quantities, $M_{\rho(770)^{0}}$,  $M_{\omega(783)}$,  $\Gamma{\rho(770)^{0}}$, $\Gamma_{\omega(783)}$, $A, B , C$, $\phi_{\omega(783)}$ treated as free parameters.

\begin{figure}[thbp]
\centering
\includegraphics[width=6.cm, height =5.cm]{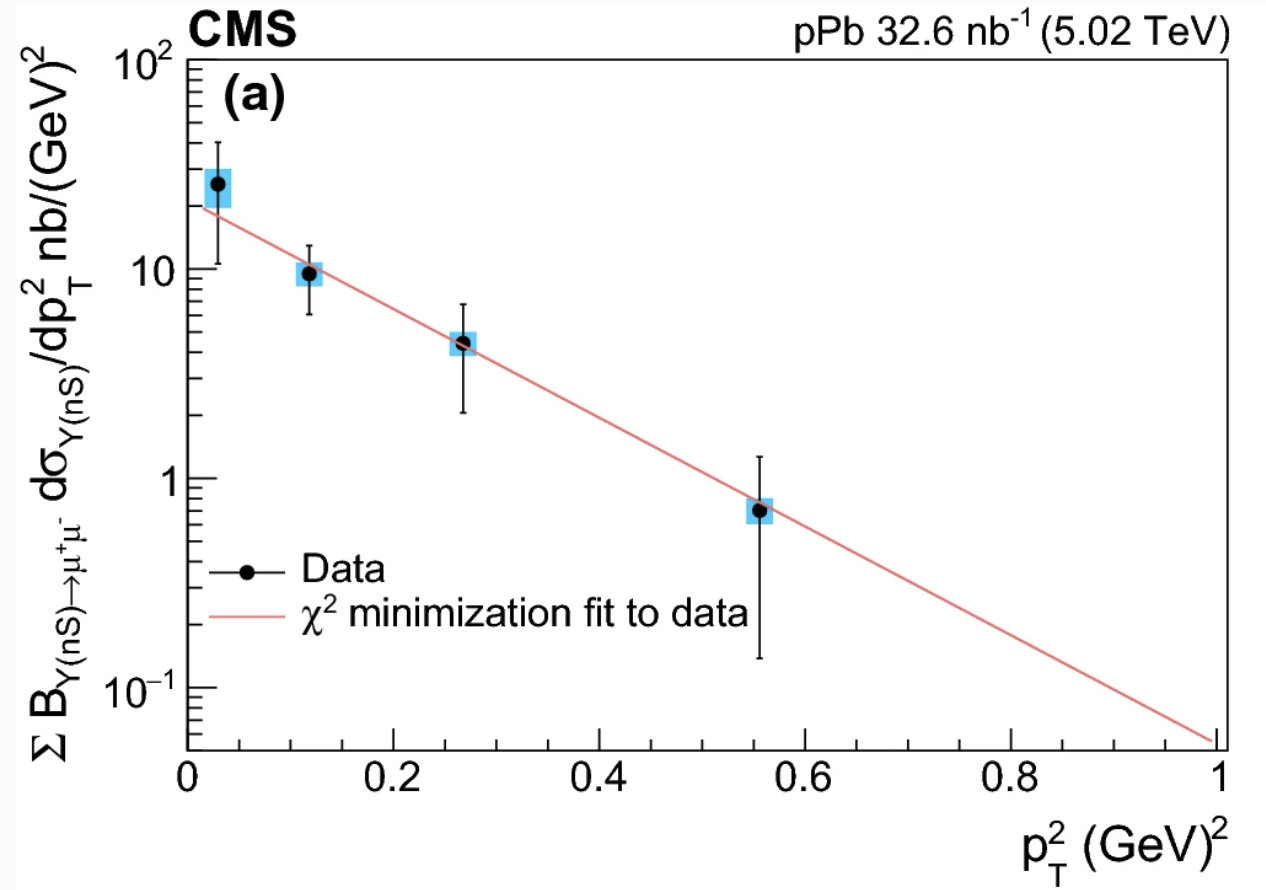}
\includegraphics[width=6.cm, height =5.cm]{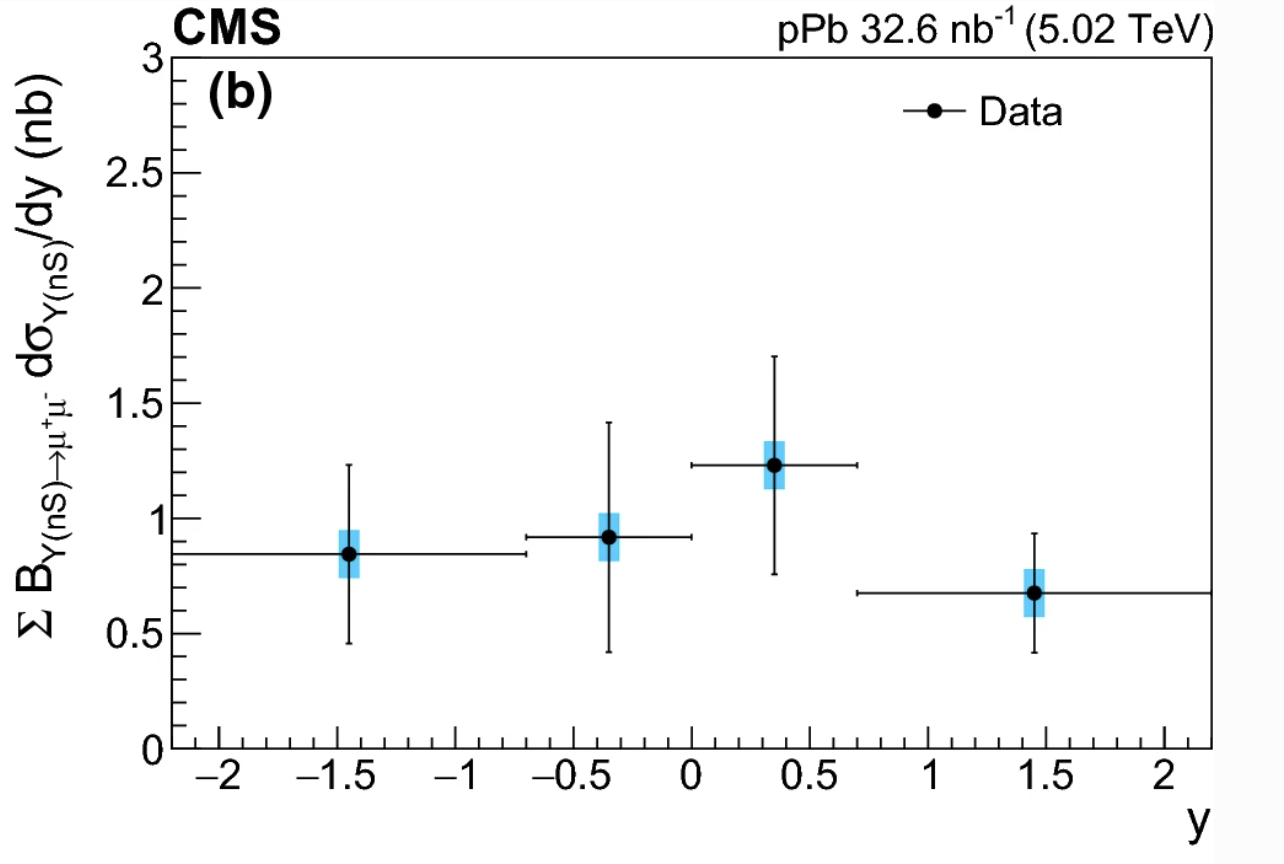}
\caption{Left side shows the differential $\Upsilon(\text{nS}  )\to \mu{+} \mu^{-}$  as a function of  $p_\text{T}^{2}$ and rapidity y  in pPb collisions. The data points are placed along the abscissa following the prescription, and the solid line is an exponential fit. In the right plot, the horizontal bars are  indicating the width of each y bin.}
\label{fig:xsection}
\end{figure}

\subsection{Differential cross section measurement}
\label{sec:xsec}
The differential cross section of the exclusive $\Upsilon(\text{nS})$ is measured as a function of  $p_\text{T}^{2}$ and y over $|y| < 2.2$, are shown in Figure~\ref{fig:xsection}. The $p_\text{T}^{2}$-differential cross section is fitted with an exponential function in the region $0.01 < p_\text{T}^{2} < 1.0$ Ge$\text{V}^{2}$, using a $\chi^{2}$ goodness-of-fit minimization technique. The slope parameter $b = 6.0 \pm 2.1$ (stat) $\pm 0.3$ (syst) Ge$\text{V}^{-2}$ is extracted, and in agreement with  $b = 4.3 ^{2.0}_{-1.3}$ (stat) $^{0.5}_{-0.6}$ (syst) Ge$\text{V}^{-2}$, is measured by the ZEUS in the photon–proton centre-of-mass energy range $60 < W_{\gamma p} < 220$ GeV. The measured results are consistent with the predictions of pQCD-based models.

\begin{figure}[thbp]
\centering
\includegraphics[width=6.cm, height =5.cm]{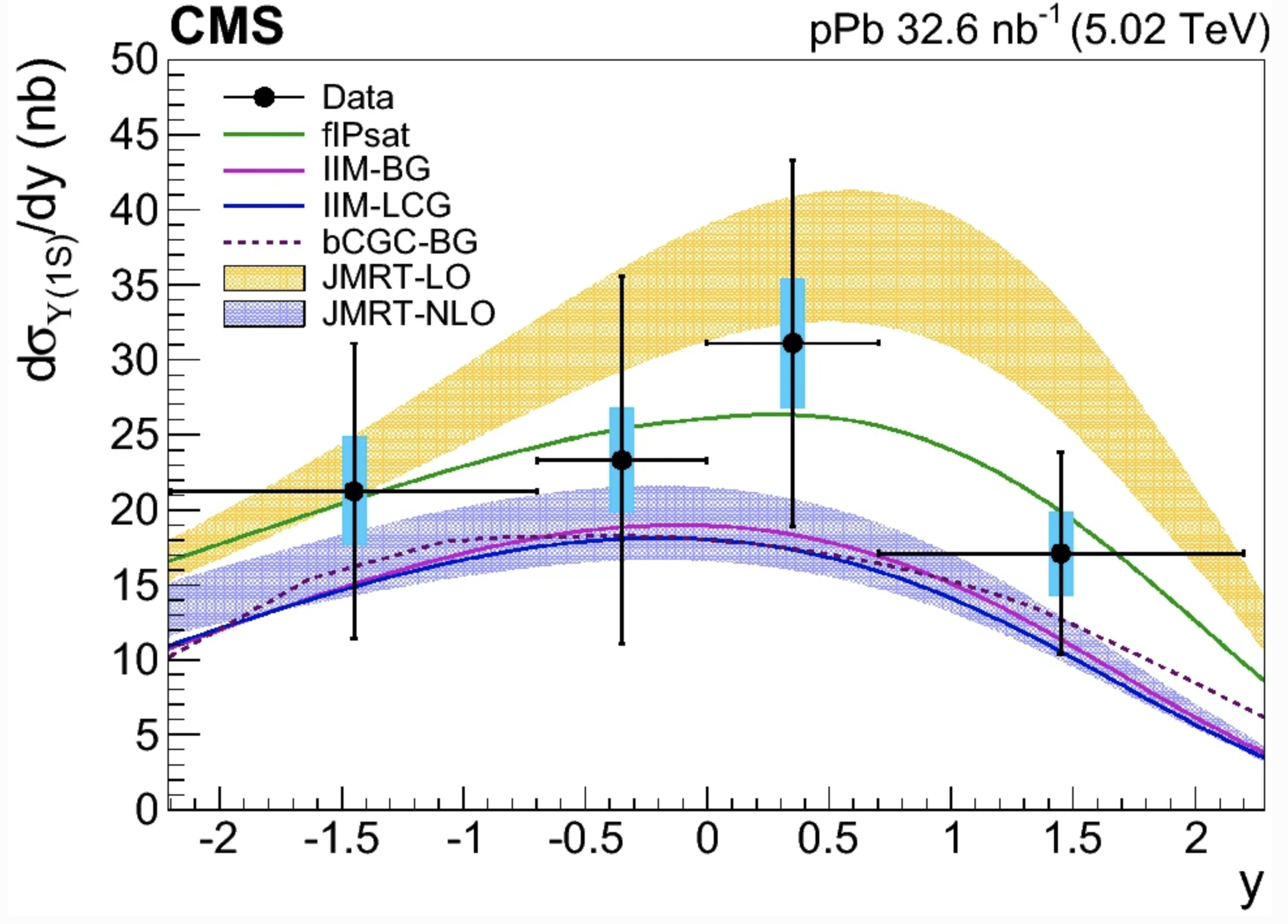}
\includegraphics[width=6.cm, height =5.cm]{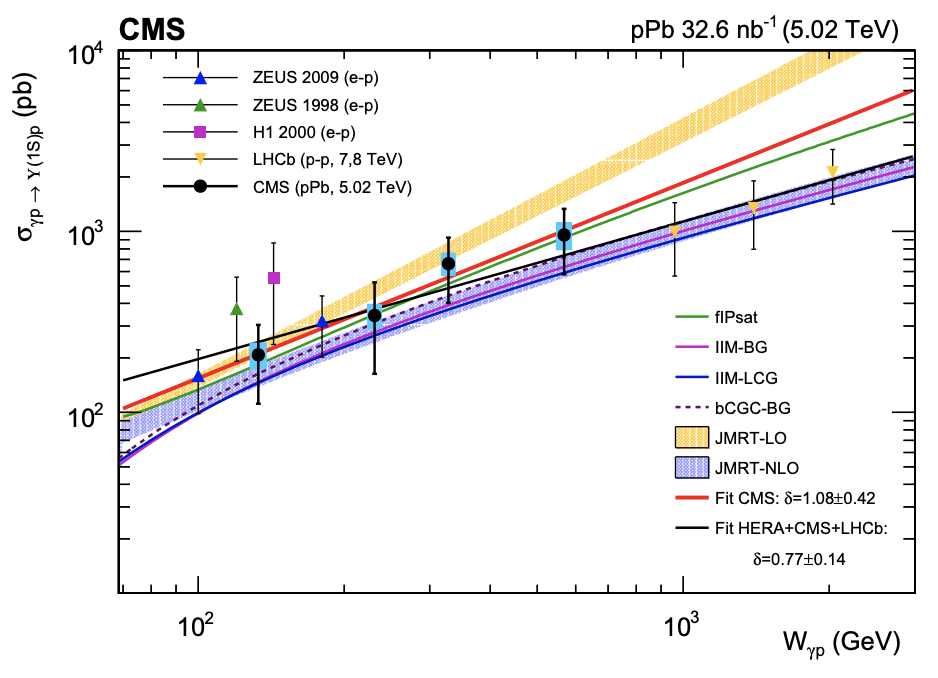}
\caption{Left plot is the differential cross section as a function of rapidity measured in pPb collisions. Right plot is the $W_{\gamma p}$  dependence of cross section for the exclusive photo production.}
\label{fig:modelxsection}
\end{figure}

Figure~\ref{fig:modelxsection} (left plot) shows the rapidity distribution of the $\Upsilon(\text{1S})$. Our results are compared with the various theoretical predictions, and they are consistent with the data within the relatively large experimental uncertainties. The JMRT-LO~\cite{jmrt} results being systematically above the data points as well as all the other calculations.

The data are compared to the various theoretical predictions. A fit with the power-law function in the entire $W_{\gamma p}$ range of data yields $\delta = 1.30$ and $\delta = 0.84$ for the LO and NLO calculations, respectively. We significantly reduced the  uncertainty compared to ZEUS~\cite{zeus} and covered a wide range of $W$.

\subsection{Summary}
\label{sec:summary}
The study of the exclusive photoproduction of $\Upsilon$ and $\rho(770)^{0}$ are measured  in UPC pPb collisions at $\sqrt{s_\text{NN}} = 5.02$ TeV with the CMS. The differential cross section as a function of rapidity, $p_{T}^{2}$ and $W_{\gamma p}$ are calculated and compared with the previous experiment at H1, LHCb and ZEUS. The data, within their currently large statistical uncertainties, are consistent with various pQCD approaches that model the behaviour of the low-x gluon density, and provide new insights on the gluon distribution in the proton in this poorly explored region.
 \section{Investigation of jet quenching effects due to different energy loss mechanisms in heavy-ion collisions using JETSCAPE framework}
\author{Rohan V S}	

\bigskip

\begin{abstract}
Heavy ion collisions produced at relativistic high energies generate a hot, dense medium of strongly interacting nuclear matter known as a quark–gluon plasma (QGP). The jets produced in the QGP medium lead to the expulsion of a large number of particles in a parton shower. The quantity of energy lost by the jet and the shape of the underlying transverse momentum $p_T$ spectrum are the objects of interest to determine the nuclear modification factor ($R_{AA}$) of jets. Jet Energy-loss Tomography with a Statistically and Computationally Advanced Program Envelope (JETSCAPE) is a multi-stage jet evolution framework that provides an integrated depiction of jet quenching which could be used to analyze the multi-stage high-energy jet evolution in QGP medium in great detail.
In this work, Pb-Pb collisions at $\sqrt{s_{NN}} = $ 5.02 TeV  and Au-Au collisions at 200 GeV were selected for various combinations of jet energy loss models including MATTER, LBT, Martini, and AdSCFT and are examined. For centrality classes ranges from 0 to 10$\%$, 30 to 40$\%$ and 60 to 80$\%$ for the different energy loss models were compared in both QGP medium and vacuum to study the nuclear modification factor.
\end{abstract}



\subsection{Introduction}

 Heavy ion collisions at relativistic high energies results in the production of a fireball that creates the QGP\cite{1_Rohan}. When a jet is produced in a heavy ion collision, the partons in the shower must traverse through the droplet of QGP produced in the same collision. The evolution of jets produced in the early stages of collisions provide crucial information about the short-distance-scale interactions of high energy partons within the QGP medium. The jets originating from the collision of ions pass through the QGP medium and lose energy due to jet-medium interactions, a phenomenon termed as jet quenching\cite{2_Rohan}. The energy lost by the partons while propagating through the QGP medium result in the suppression of yield at different transverse momentum values. In this proceeding, we present the nuclear modification factor of inclusive jets.
 
 Simulations of high energy heavy-ion collisions require a multitude of interacting elements, ranging from simulations of the incoming nuclei, to the thermalization of the deposited energy-momentum, viscous fluid dynamical expansion, as well as the production of hard partons, their propagation and interaction with the dense medium,  hadronization and freeze-out, escape and fragmentation into jets\cite{3_Rohan}. To compare with high-statistics experimental data, one requires a modular and extendable event generator, with state-of-the-art components modeling of each aspect of the collision. The JETSCAPE\cite{4_Rohan} used in this study is such a framework for general purpose event generation.
%
%
%
It contains modules that runs simulations of each sector of high energy heavy-ion collisions. This framework includes an initial soft sector event generator for simulations of the incoming nuclei, viscous fluid dynamical model for the medium expansion, hadronization model and freeze-out model. Four different energy loss modules are available; The Modular All Twist Transverse-scattering Elastic-drag and Radiation (MATTER)\cite{5_Rohan}  for modeling high virtuality and high energy parton evolution, Linear Boltzmann Transport (LBT)\cite{6_Rohan} for modeling low virtuality and high energy evolution, Modular Algorithm for Relativistic Treatment for heavy Ion Interactions (MARTINI)\cite{7_Rohan} for modeling low virtuality and high-energy evolution using a gluon radiation process and AdS/CFT  for modeling a low virtuality and low energy parton shower. Initial hard partons generated by PYTHIA  are fed into MATTER and propagated with a virtuality-ordered shower.








\subsection{Jet Observable (Nuclear modification factor)}



The nuclear modification factor ($R_{AA}$) is the ratio of the jet yields in heavy-ion collisions (Pb-Pb or Au-Au) to those in p-p collisions as a function of transverse momentum, normalised by the mean nuclear thickness function.
\begin{equation}
     R_{AA} = \frac{\frac{1}{N_{\mathrm{ev}}} \frac{d^2N_{\mathrm{jet}}}{dp_T dy}}{ \langle T_{AA} \rangle \frac{d^2\sigma_{\mathrm{jet}}}{dp_T dy}}
\end{equation}

where  $N_{\mathrm{ev}}$ is the number of hard scattering events,  $N_{\mathrm{jet}}$ is the number of jets in heavy-ion collisions, ${p_{\mathrm{T}}}$  is the transverse momentum,  $y$ is the rapidity, $\sigma_{jet}$ is the jet cross section in p-p collisions and  $\langle T_{AA} \rangle$ is the mean nuclear thickness function.

\subsection{Results and Summary}
In this work, we have used the JETSCAPE framework to produce Pb-Pb collisions at $\sqrt{s_{NN}} = $ 5.02 TeV and Au-Au collisions at $\sqrt{s_{NN}} = $ 200 GeV. We have compared the results obtained using the JETSCAPE framework to the corresponding results from ATLAS\cite{8_Rohan} and STAR\cite{9_Rohan} experiments. Initial state model TRENTO\cite{10_Rohan} was added, in which, we set the corresponding geometric and kinematic properties of ions i.e. Pb-Pb at 5.02 TeV, cross-section, normalization and centrality etc. A free streaming module Milne was used for bulk evolution. The hard scattering was carried out using the PYTHIA 8\cite{11_Rohan} gun with initial state radiation and multi parton interaction turned on and final state radiation turned off.The hydrodynamic module MUSIC\cite{12_Rohan,13_Rohan,14_Rohan} was used; subsequently, the jets were hadronized using the hadronization module.





\begin{figure}[htbp]
    \centering
    \includegraphics[scale=0.3]{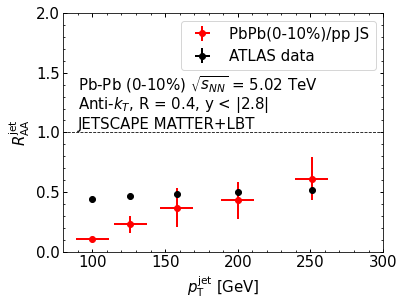}
    \includegraphics[scale=0.3]{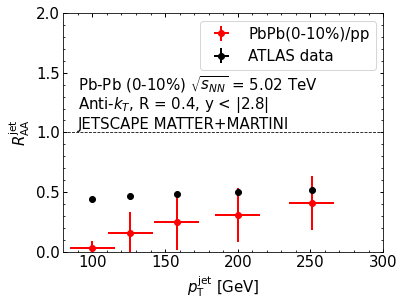}
    \includegraphics[scale=0.3]{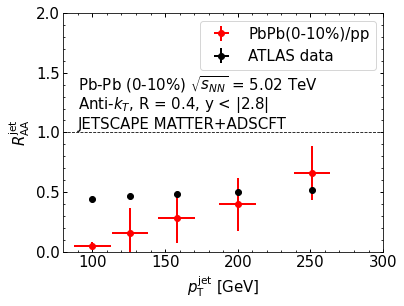}
    \caption{The dependence of different energy loss modules on $R_{AA}$ for Pb-Pb at 5.02 TeV }
    \label{fig1_r}
\end{figure}

The dependence of different energy loss modules on $R_{AA}$ was studied using Pb-Pb collisions at 0-10$\%$ centrality. Fig. \ref{fig1_r} (left plot) shows $R_{AA}$ with MATTER and LBT energy loss modules. Fig. \ref{fig1_r} (middle plot) uses MATTER and MARTINI energy loss modules and Fig. \ref{fig1_r} (right plot)  uses MATTER and ADSCFT for the same centrality interval. The hadrons that are formed were then reconstructed using the anti-kt algorithm\cite{15_Rohan} with radius of cone R = 0.4, minimum $p_T$ of jets  = 100 GeV, $y < |2.8|$ and minimum track $p_T$  = 4 GeV. Figure \ref{fig1_r} shows that for $p_T$ above 150 GeV, the results are consistent with the experimental results for all the three energy loss modules, and for the low $p_T$ region the JETSCAPE results under predict the data (more quenching in this region).

\begin{figure}[htbp]
    \centering
    \includegraphics[scale=0.3]{Rohan/pb-pb-0-10-lbt-raa4.png}
    \includegraphics[scale=0.3]{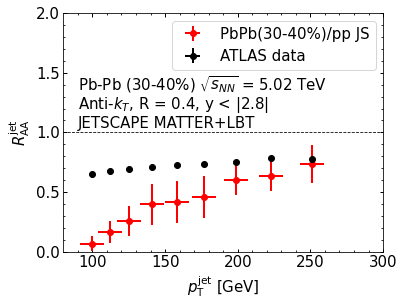}
    \includegraphics[scale=0.3]{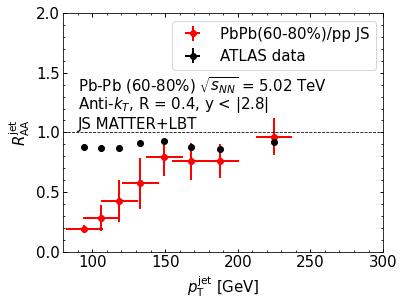}
    \caption{Centrality dependence of  $R_{AA}$ for Pb-Pb collisions at 5.02 TeV}
    \label{fig2_r}
\end{figure}

Fig. \ref{fig2_r} shows the centrality dependence of  $R_{AA}$ for Pb-Pb collisions for 0-10$\%$ 30-40$\%$ and 60-80$\%$ centrality intervals simulated using the MATTER + LBT energy loss modules. The switching virtuality was set to 2 GeV and the strong coupling constant $\alpha_s$ = 3.0 Here we can observe that the energy lost due to jet quenching is the highest for the central collisions 0-10$\%$ (left plot) and less for the  mid-central collisions 30-40$\%$ (middle plot) and it is least for the peripheral collisions 60-80$\%$ (right plot). However, we can also see that in Fig. \ref{fig2_r}, the JETSCAPE results under predict the data in the lower  $p_T$ region.

The Au–Au heavy-ion collisions were simulated for the jets of transverse momentum from 5-30 GeV with TRENTO initial conditions and PYTHIA 8 hard scattering mechanism. The partons evolved from this are sent to MATTER and LBT energy loss modules with the switching virtuality set to 1 GeV and $\alpha_s$ = 3. The MUSIC module was used to describe the hydrodynamic evolution of bulk medium. The hadrons formed were reconstructed with the jet cone radius R = 0.4, minimum $p_T$ of jets = 5 GeV, $y < |1|$ and minimum track $p_T$ = 0.5 GeV.

\begin{figure}[htbp]
\centering
    \includegraphics[scale=0.35]{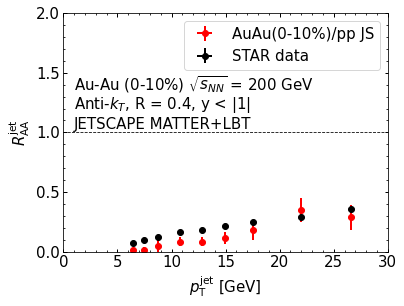}
    \includegraphics[scale=0.35]{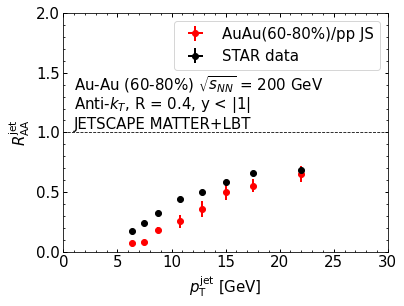}
    \caption{Centrality dependence of $R_{AA}$ for Au-Au collisions at 200 GeV}
    \label{fig3}
\end{figure}
The  centrality dependence of $R_{AA}$ for Au-Au collisions for centrality classes 0-10$\%$ and 60-80$\%$  is shown in Fig. \ref{fig3}. The JETSCAPE results are  consistent with the STAR experimental data and qualitatively explains the data  in both the centrality classes. Fig. \ref{fig3} shows that the jet quenching is more in the most central collisions 0-10$\%$ (left plot) as compared to peripheral collisions 60-80$\%$ (right plot) and the quenching gradually decreases with the transverse momentum of the jets as expected.

 \section{Recent Results from STAR and ALICE Experiments}
\author{Lokesh Kumar}	

\bigskip

\begin{abstract}
	We report recent selected results on particle
        production and fluctuations from the STAR and ALICE
        experiments. The results from the ALICE include baryon-to-meson
        ratios, elliptic
        flow $v_2$, and rapidity asymmetry. 
        The results from the STAR are presented on kinetic freeze-out
        parameters at lower energy, 
        and observables
        related to search of critical point and transition
        temperature. The physics implications of these results are
        discussed. 
\end{abstract}


\subsection{Introduction}

The Solenoidal Tracker At RHIC (STAR)~\cite{STAR:2005gfr} at Relativistic Heavy-Ion
Collider (RHIC), Brookhaven National Laboratory (BNL), USA and A Large Ion Collider
Experiment (ALICE)~\cite{ALICE:2008ngc} at Large Hadron Collider (LHC), CERN, Switzerland, are dedicated experiments for the heavy-ion
collisions. They are built to study the properties
of Quark Gluon  Plasma (QGP), quark-hadron phase transition, and
critical point in the phase diagram of Quantum Chromodynamics
(QCD).
The energy range ($\sqrt{s_{NN}}=$ 2.76 -- 5.44
  TeV) ) at which ALICE experiment works
is towards the region where $\mu_B \sim 0$, while STAR covers the
energy range ($\sqrt{s_{NN}}=$ 3.0 -- 200 GeV) from small $\mu_B \sim 20 $ MeV up to the $\sim$720 MeV that
includes the data collected in the fixed target mode of the STAR 
experiment. The higher energies will lead to
QGP with longer lifetime and hene allow for the detailed study of QGP while the lower energy ranges help to look for the expected first order phase
transition and critical point in the conjectured QCD phase diagram. In
this way, both STAR and ALICE experiments provide important
information regarding the study of QGP and QCD phase diagram. In
addition, the results from small systems (pp or pA) help in understanding the
particle production and the baseline studies.  We
present selected results from the ALICE and STAR experiments.  

\subsection{Results from the ALICE}\label{Sec:alice}

Recently, ALICE has collected data for Xe-Xe collisions at
$\sqrt{s_{NN}}=$ 5.44 TeV. The Xe is a medium sized nucleus and bridges the gap between p-Pb and Pb-Pb systems. Moreover, the
detailed comparison of Xe-Xe and Pb-Pb allows to investigate the
results for the systems at a similar multiplicity but with different
initial eccentricity.
\begin{figure}[h]
                \begin{center}
  \includegraphics[width = 5.0cm]{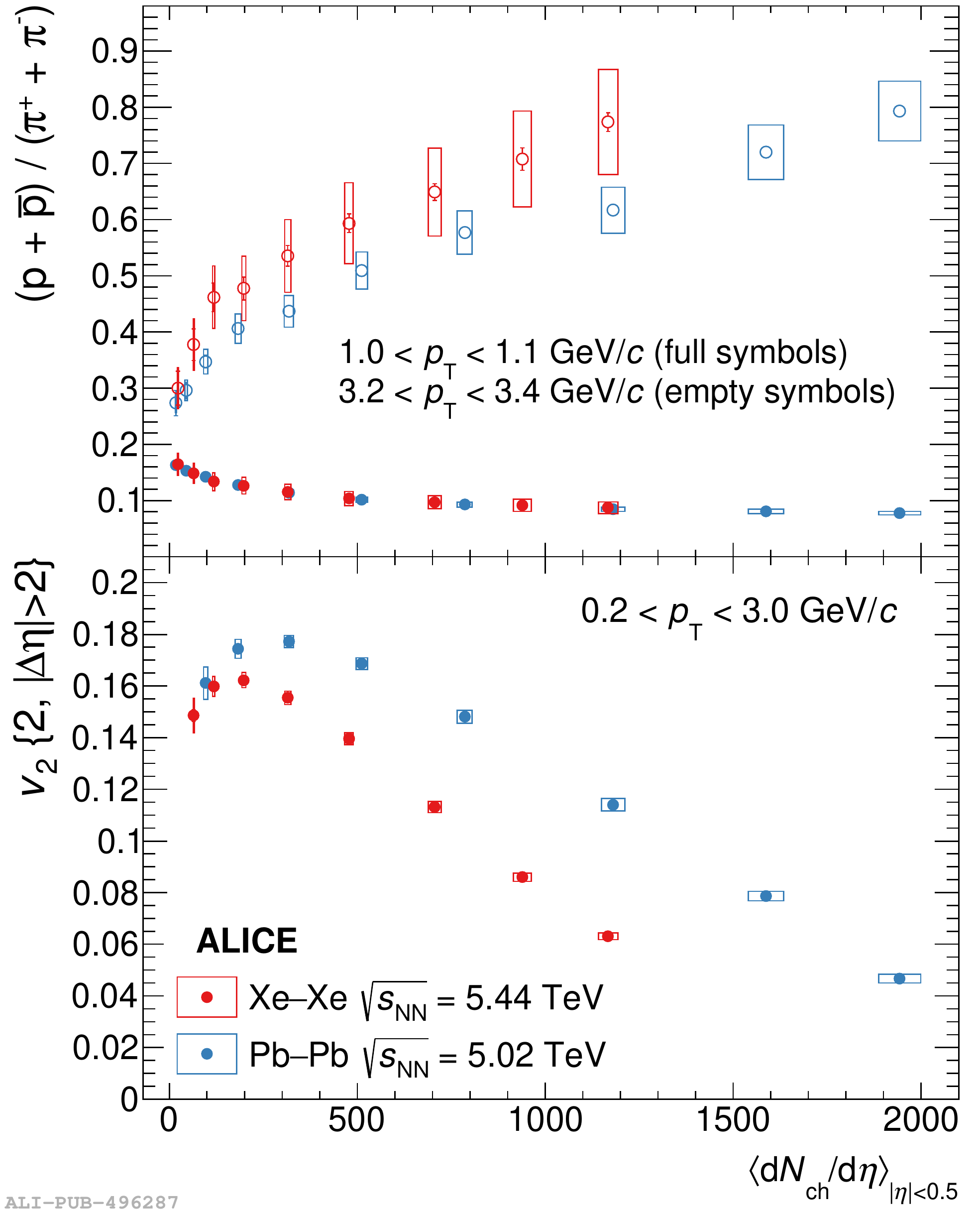}
 \includegraphics[width = 7.5cm]{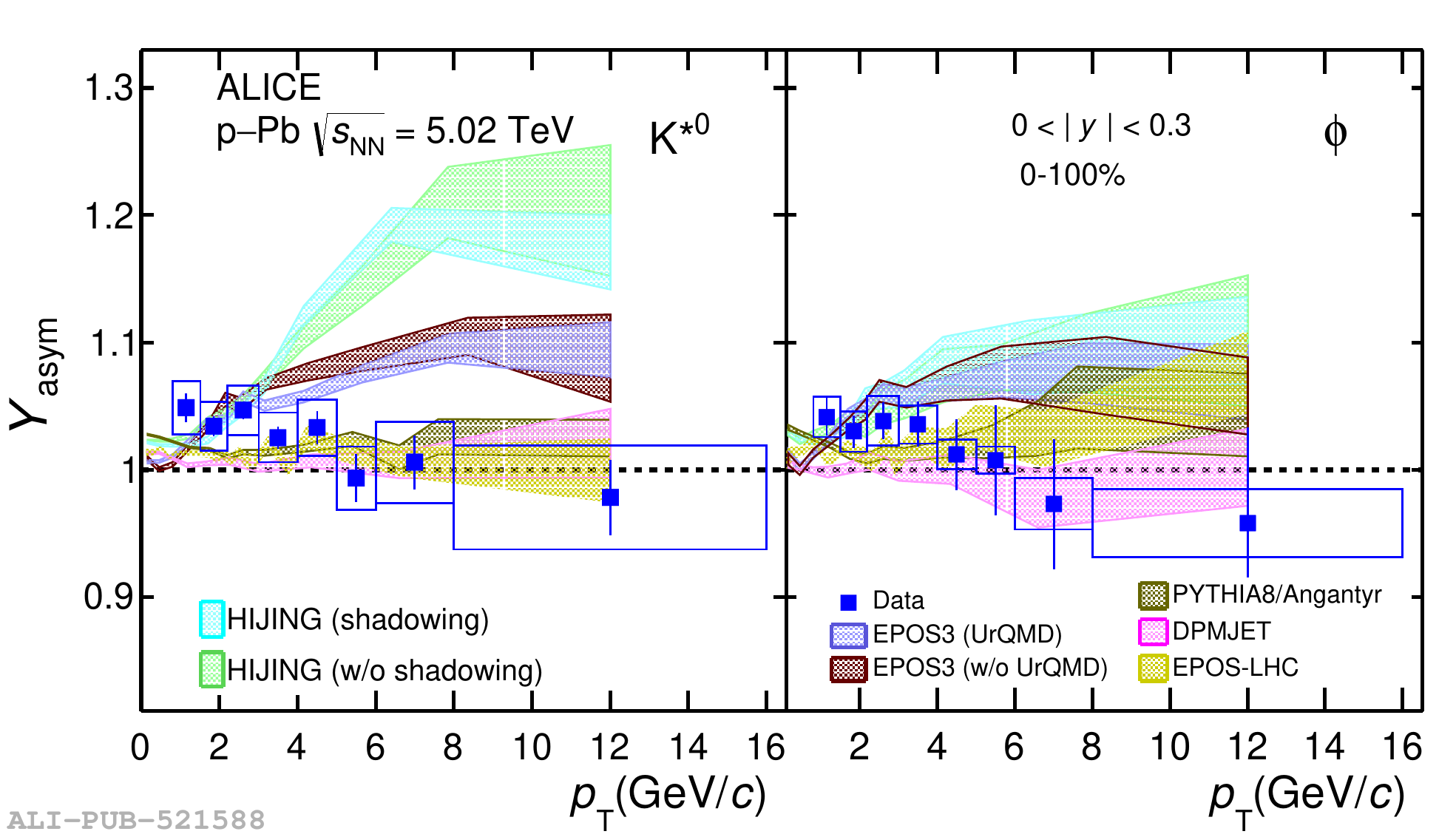}
                \vspace{-0.3cm}
		\caption{Left: Comparison of proton-to-pion ratio (top
                panel) and $v_2$ (bottom panel) as a function of
                  charged particle multiplicity density between Xe–Xe
                  at \snn = 5.44 TeV sand Pb-Pb at \snn = 5.02
                  TeV. Right: The $Y_{\rm{asym}}$  for $K^{*0}$ and
                  $\phi$ meson as a function of \pt in the rapidity
                range $0 < |y| < 0.3$ in p–Pb collisions at \snn $=$
                 5.02 TeV. The results are compared with various
                models. }\label{Fig:xe}
                \end{center}
                 \vspace{-0.7cm}                
  \end{figure}
Figure~\ref{Fig:xe} (left) shows the $p/\pi$ ratio (top panel) and
$v_2$ (bottom panel) comparison between Xe-Xe at  $\sqrt{s_{NN}}=$
5.44 TeV and Pb-Pb 
at $\sqrt{s_{NN}}=$ 5.02 TeV~\cite{ALICE:2021lsv}, as a function of charged
particle multiplicity density $\langle dN_{\rm{ch}}/d\eta \rangle$. The $p/\pi$ ratio
is plotted for the low-transverse momentum \pt  (1.0--1.1 GeV/$c$) and intermediate-\pt
(3.2--3.4 GeV/$c$)  ranges. Both Xe-Xe and Pb-Pb show depletion of $p/\pi$
ratio at low-\pt while enhancement of this ratio at
intermediate-$p_T$. The enhancement of this ratio at intermediate-\pt is
associated with quark-recombination and radial
flow~\cite{ALICE:2019hno,qc}. One observes that at a similar charged
particle multiplicity density, the magnitude of $p/\pi$ ratio is similar in
both Xe-Xe and Pb-Pb systems. This suggests that this ratio depends on the
final state multiplicity. However, the lower panel shows that for a
given $\langle dN_{\rm{ch}}/d\eta \rangle$ (except for very low multiplicity),
the $v_2$ in Pb-Pb is higher than that in Xe-Xe. Since, the $v_2$
originates from the hydrodynamical expansion which in turn depends on
the initial eccentricity, the results here suggests that Pb-Pb and
Xe-Xe have different initial eccentricities. 

The hadron production depends on various effects such as nuclear
modification of parton distribution functions (nuclear shadowing) and
possible parton saturation, multiple scattering, and radial
flow. These effects depend on the rapidity of produced particles. The
p-Pb collisions provide an opportunity to study the rapidity asymmetry ($Y_{\rm{asym}}$)
of hadron production. Figure~\ref{Fig:xe} (right) shows the $Y_{\rm{asym}}$ for $K^{*0}$ and
$\phi$ mesons as a function of \pt for 0--100\% collision centrality
in p–Pb collisions at \snn $=$ 5.02 TeV~\cite{ALICE:2022uac}. The rapidity
asymmetry is defined as: $Y_{\rm{asym}}(p_T) =
\frac{[d^2N/(dp_Tdy)]_{-0.3<y<0}}{[d^2N/(dp_Tdy)]_{0<y<0.3}} $.  
It is observed that the $Y_{\rm{asym}}$ deviates from unity for
both $K^{*0}$ and $\phi$ at low-$p_T$ whereas at high \pt it is
consistent with unity. The deviations from unity at low-\pt suggests
that there is rapidity dependence of nuclear effects in p-Pb collisions. The HIJING
and EPOS3 model calculations suggest $Y_{\rm{asym}}$ at low-$p_T$ but significantly
overestimate the data at high-$p_T$.

\subsection{Results from the STAR}

The blast wave (BW), a hydrodynamic based model, has been successful in explaining the \pt spectra of
various produced particles in heavy-ion collisions. It
provides information on the kinetic freeze-out temperature
$T_{\rm{kin}}$ and average radial
flow velocity $\langle \beta \rangle $ of the system produced in these collisions. 
STAR recently has collected data in a fixed target mode in Au+Au
collisions at \snn = 3.0 GeV. It is interesting to note if the blast
wave model could explain the data at such a low
energy.
\begin{figure}[h]
	\begin{center}
		\includegraphics[width = 5.7cm]{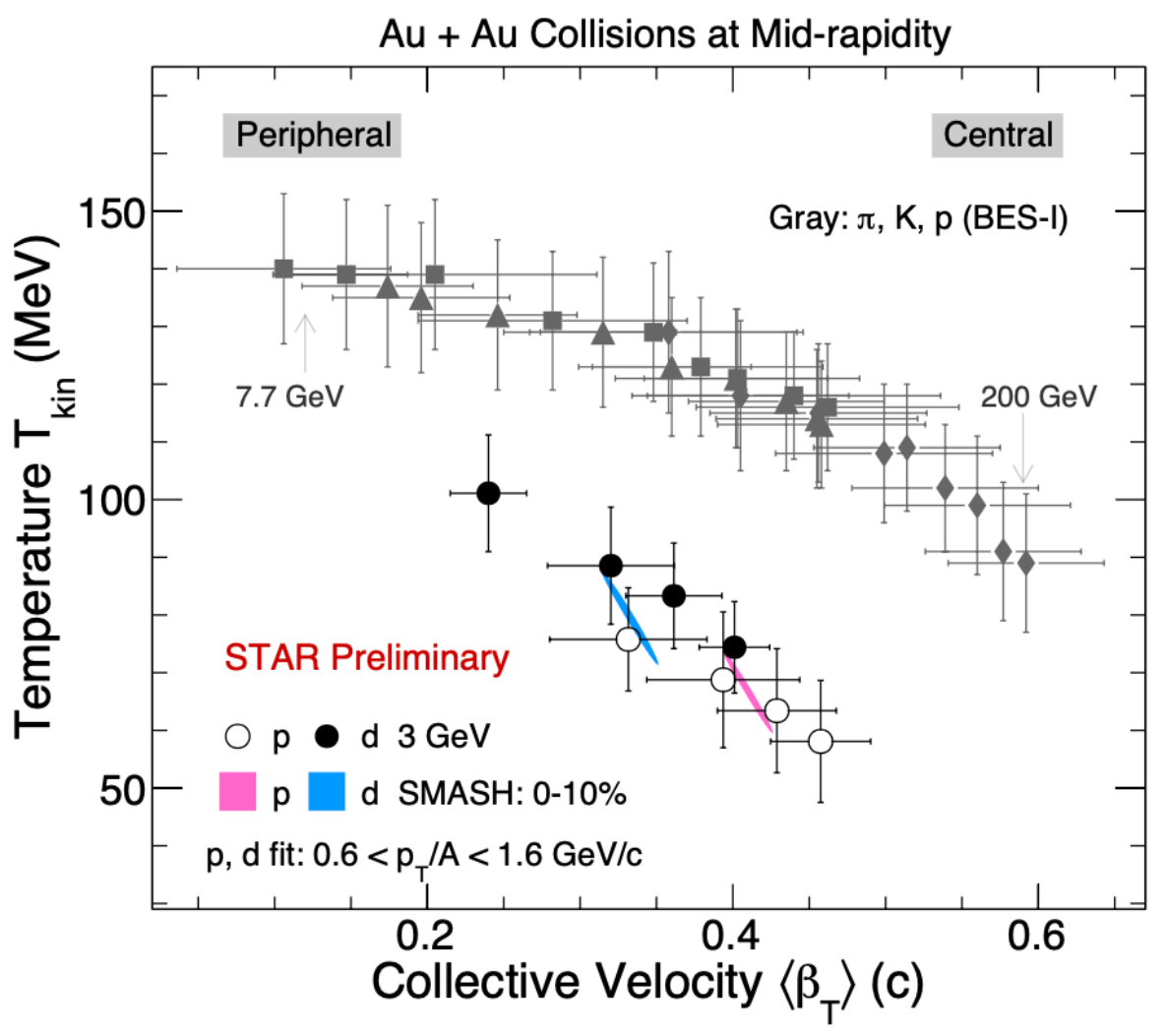}
		\includegraphics[width =6.4cm]{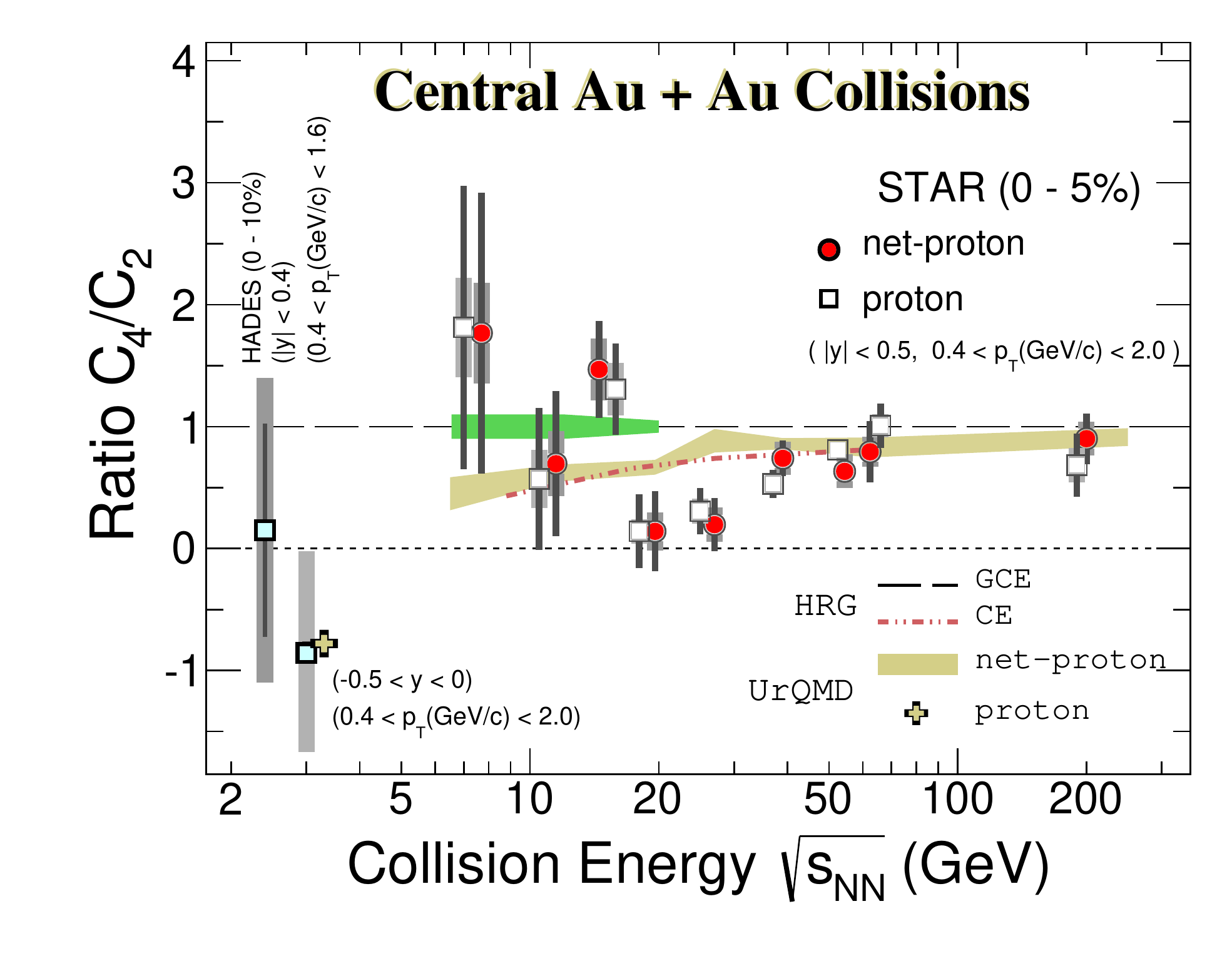}          
                \vspace{-0.3cm}
		\caption{Left: Variation of $T_{\rm{kin}}$ as a function of
                  $\langle \beta \rangle$ for various energies~\cite{hui_qm2022}.
Right: Ratios of cumulants, C4/C2, for protons (squares) and net-protons (red circles)
as a function of collision energy. The vertical black and gray bars
are the statistical and systematic uncertainties,
respectively. Comparison with the HRG model (Canonical Ensemble (CE) and Grand-Canonical Ensemble (GCE)),
and transport model UrQMD are also shown. 
}\label{Fig:3gev_kin}
	\end{center}
                \vspace{-0.7cm}
      \end{figure}
Figure~\ref{Fig:3gev_kin} (left) shows the variation of $T_{\rm{kin}}$ as a
function of $\langle \beta \rangle $ for various energies~\cite{hui_qm2022}. The results
from \snn = 3.0 GeV are obtained by fitting protons and deuterons
using the BW model. The results shown for other STAR energies are obtained
by fitting pion, kaon, and proton and their antiparticles' spectra
simultaneously with BW model. The $T_{\rm{kin}}$ and $\langle \beta
\rangle $ exhibit an anticorrelation behavior.
It is observed that the kinetic
freeze-out parameters at \snn = 3.0  GeV are drastically different
compared to other higher energies. This may suggest a different
equation of state prevailing at 3 GeV compared to that at higher
energies. It is also observed that the $T_{\rm{kin}}$ ($\langle \beta
\rangle $) is higher (lower) for deuterons as compared to protons.   The SMASH model, a hadronic transport model, shows
similar results as data at 3 GeV.

One of the main goals of heavy-ion collisions is to investigate the
phase structure of the QCD matter. At $\mu_B\sim0$, the lattice QCD
calculations suggest a smooth crossover transition from hadronic
medium to the QGP phase and the transition temperature is about 155 MeV. At
finite $\mu_B$,  a critical point followed by a first order phase
transition is expected. At critical point, the correlation length of
the system diverges and can be studied through the higher moments
(skewness $S=\langle(\delta N)^3\rangle/\sigma^3$ and kurtosis
$\kappa=[\langle(\delta N)^4\rangle/\sigma^4]-3$, where $\delta N = N
- \langle N \rangle$) of conserved quantities such as
net-baryon (or net-proton) number which are sensitive to the correlation length. A
non-monotonic behavior of higher moments of conserved quantities as a
function of \snn has been proposed as an experimental signature of
critical point.
\begin{figure}[h]
	\begin{center}
		\includegraphics[width =12cm]{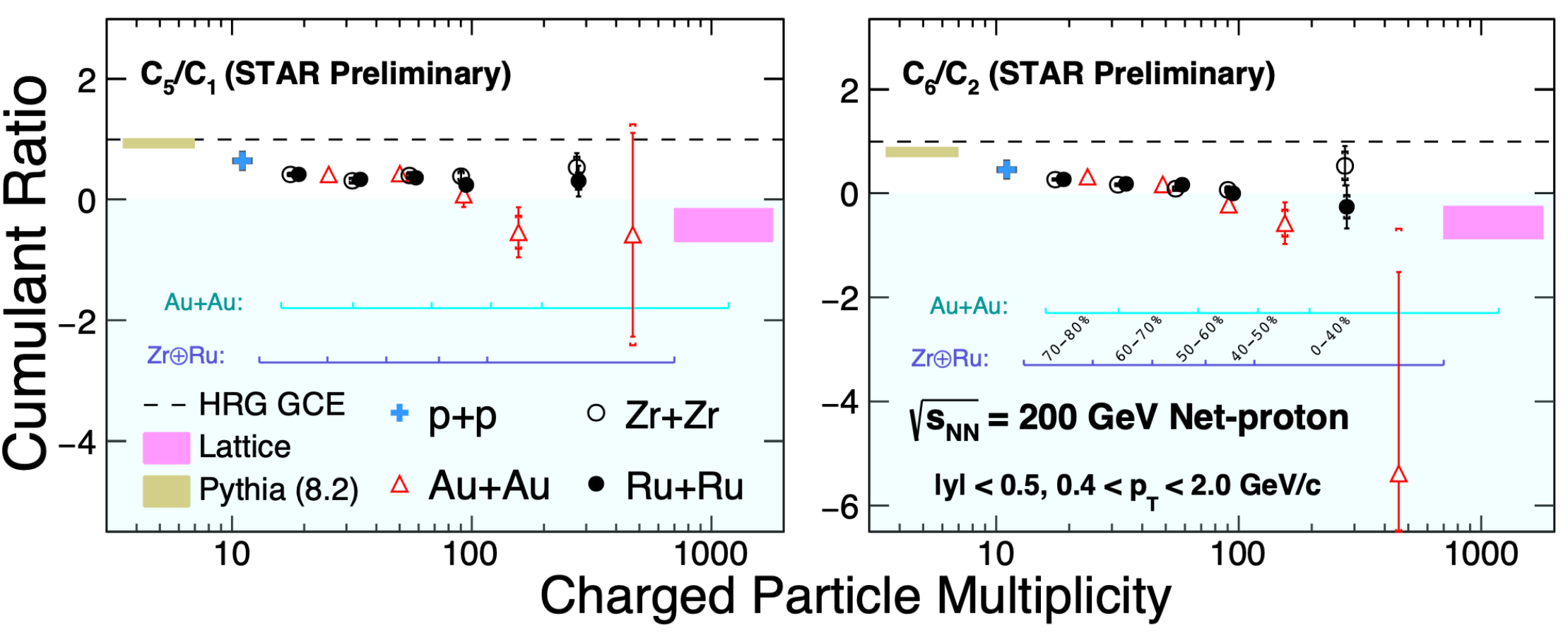}          
                \vspace{-0.3cm}
		\caption{$C_5/C_1$ (left panel) and $C_6/C_2$ (right panel) as a function of charged particle
multiplicity for various systems pp, Au+Au, Ru+Ru, and
Zr+Zr at \snn = 200 GeV. 
		}\label{Fig:chiral}
	\end{center}
                \vspace{-0.7cm}
      \end{figure}
To study the shape of the event-by-event net-proton
distribution, the cumulants ($C_n$) of various orders, $C_1=M,
C_2=\sigma^2, C_3=S\sigma^3$, and $C_4=\kappa \sigma^4$, are
calculated. The ratios of these cumulants such as $C_3/C_2=S\sigma$,
and $C_4/C_2=\kappa \sigma^2$, lead to the cancelation of volume
effects and are also related to the ratio of baryon-number
susceptibilities computed in lattice QCD.
Recent results from the STAR in the energy range \snn=7.7--27 GeV for
0-5\% central collisions, on net-proton fluctuation measurements
$C_4/C_2$ show a 
non-monotonic behavior as a function of \snn with a significance of
3.1$\sigma$ as shown in Fig.~\ref{Fig:3gev_kin} (right)~\cite{STAR:2020tga}. The
calculations from hadron resonance gas models (HRG) and Ultra
Relativistic Molecular Dynamics (UrQMD) show a smooth energy
dependence. These models have the baryon number conservation and do
not include the critical point in their calculations. New results from
STAR for proton cumulant ratios in Au+Au collisions at \snn = 3.0 GeV
are also shown in Fig.~\ref{Fig:3gev_kin} (right)~\cite{STAR:2021fge}. The
$C_4/C_2$  value at 3.0 GeV is around -1 and is reproduced by the
transport model UrQMD. The results at 3.0 GeV are also consistent with
those from HADES experiment. These results can be
explained by fluctuations driven by baryon number conservation at high
baryon density where hadronic interactions are dominant. It is
expected that if the critical point exists, it can be observed in the
region of \snn $>$ 3.0 GeV to 19.6 GeV. The high statistics beam energy
scan-II (BES-II) data will provide a definite answer in this regard.

Figure~\ref{Fig:chiral} shows the cumulant ratios $C_5/C_1$ (left panel) and $C_6/C_2$ (right panel) as a function of charged particle
multiplicity for various systems pp, Au+Au, Ru+Ru, and
Zr+Zr at \snn = 200 GeV~\cite{hosan_qm2022}. Results from HRG GCE model, Pythia (8.2), and lattice QCD
calculation that include smooth crossover phase transition at
$\mu_B=0$, are compared with the experimental data. Both cumulant
ratios decrease as a function of multiplicity and the most central
Au+Au collisions results are consistent with the lattice results for
the thermalized QCD matter and smooth crossover.

\subsection{Summary}
We have reported the latest results from the STAR and ALICE experiments. The
measurement of $p/\pi$ ratio and $v_2$ in Xe-Xe collisions at \snn = 5.44 TeV
compared to that in Pb-Pb collisions at \snn = 5.02 TeV suggests
that $p/\pi$ ratio is sensitive to final state effect while $v_2$ is
sensitive to initial
effect, and Xe-Xe collisions have different initial eccentricity
compared to Pb-Pb collisions. The $K^{*0}$ and  $\phi$ production in
p-Pb collisions show a rapidity asymmetry at
low-$p_T$.
The matter produced in Au+Au collisions
at \snn = 3.0 GeV shows different equation of state compared to that
at higher energies. 
Results on net-proton cumulant ratios
$C_4/C_2$ at \snn = 3.0 GeV suggests that fluctuations at this energy
are consistent with expectation of baryon number conservation at high baryon density
where hadronic interactions are dominant. The higher cumulant ratios
$C_5/C_1$  and $C_6/C_2$ results suggest that most central Au+Au
collisions at \snn = 200 GeV are consistent with the lattice results
for thermalized QCD matter and smooth crossover.


 \section{Impact of Time Varying Electromagnetic Field on Electrical Conductivity of Hot QCD Matter}
\author{K K Gowthama, Manu Kurian, and Vinod Chandra}	

\bigskip
 
\begin{abstract}
	The impact of the time dependence of the electromagnetic fields and collisional aspects of the medium on the induced electric and Hall current densities have been explored using the relativistic Boltzmann equation. The effect of momentum anisotropy on electric charge transport has also been studied. The electric response of the medium is seen to be significantly affected by the inhomogeneity of the fields and the anisotropy of the medium.
\end{abstract}

\subsection{Introduction}
Experiments at Relativistic Heavy Ion Collider (RHIC) and Large Hadron Collider (LHC) have indicated the existence of the strongly interacting matter-Quark Gluon plasma (QGP)~\cite{STAR:2005gfr} and strong magnetic fields in the heavy-ion collision process~\cite{2_Gowthama,3_Gowthama}. Transport coefficients of the hot QCD matter serve as the input parameters for the hydrodynamical description of the evolution of the created medium and act as a key ingredient in exploring the critical properties of the medium~\cite{4_Gowthama}. The response of the QCD medium to the time varying external electromagnetic fields can be quantified in terms of induced electric and Hall current densities and associated conductivities. Here, we have explored (i) a general formalism to study the electric charge transport in the presence of time-varying electromagnetic fields~\cite{5_Gowthama}, (ii) medium response in an anisotropic QCD medium~\cite{5_Gowthama}. To that end, we have estimated the general form of quark degrees of freedom in the presence of time dependent fields within the kinetic theory framework for both isotropic and anisotropic cases.

\subsection{Formalism}
The induced vector current in the QCD medium with a finite quark chemical potential $\mu$ can be defined as,
\begin{align}\label{1}
{\bf j}&=2N_c\sum_f \int \frac{d^3 \textbf{p}}{(2\pi)^3}\,{\bf v}\,\Big(q_q f_q-q_{\bar{q}}f_{\bar{q}}\Big),  
 \end{align}
where $f_k = f^0_k + \delta f_k$ (the subscript k incidates the particle species) is the quark/antiquark distribution function with $f^0_k$ being the fermi Dirac distribution function for quarks and antiquarks, $\textbf{v}$ the velocity vector, and $N_c$ is number of colors. We chose the following ansatz for non-equilibrium part of the distribution function $\delta f_k$ due to the inhomogeneous fields,
\begin{equation}\label{2}
\delta f_k=({\bf{p}}.{\bf \Xi} ) \frac{\partial f^0_k}{\partial \epsilon}, 
\end{equation}
where the vector  $\mathbf{\Xi}$ is related to the strength of electric and magnetic fields and their derivatives and has the form as follows,
\begin{align}\label{3}
\mathbf{\Xi} =& \alpha_1\textbf{E}+ \alpha_2\dot{\textbf{E}}+ \alpha_3(\textbf{E}\times \textbf{B})+ \alpha_4(\dot{\textbf{E}}\times \textbf{B})+ \alpha_5(\textbf{E}\times \dot{\textbf{B}})\nonumber\\&+\alpha_6 ({\pmb \nabla} \times \textbf{E}) +\alpha_7 \textbf{B}+\alpha_8 \dot{\textbf{B}}+\alpha_9 ({\pmb \nabla} \times \textbf{B}).
\end{align} 
With $\alpha_i (i = (1, 2, .., 9))$ are the unknown functions that relate to the respective transport coefficients associated with the electric charge transport. The transport equation that describes the dynamics of the momentum distribution function can be defined as,
\begin{align}\label{4}
\frac{\partial f_k}{\partial t} +{\bf v}.\frac{\partial f_k}{\partial {\bf x}} +q_{f_k} [{\bf E} +{\bf v} \times {\bf B}].\frac{\partial f_k}{\partial {\bf p}} =-\frac{\delta f_k}{\tau_R},
\end{align}
where $\textbf{E}$ and $\textbf{B}$ are the electric and magnetic fields, using the relaxation time approximation (RTA) for the collision kernal with $\tau_R$ being the relaxation time. We have obtained the $\alpha$'s for various cases of electromagnetic fields by solving the Boltzmann equation~\cite{5_Gowthama,6_Gowthama}. 

\subsection{Time varying electromagnetic fields}
In the presence of time-varying electromagnetic fields, the current density take the form as $\textbf{j} = j_e \hat{\textbf{e}} + j_H (\hat{\textbf{e}} \times \hat{\textbf{b}})$ with,
\begin{align}\label{7}
    &j_e = j_e^{(0)}+j_e^{(1)}, &&j_H = j_H^{(0)}+j_H^{(1)}+j_H^{(2)}, 
\end{align}
where $j_e$ corresponds to the electric current in the direction of the electric field $\hat{\textbf{e}}$ and $j_H$ is the electrical current in the direction perpendicular to both electric and magnetic fields, $(\hat{\textbf{e}} \times \hat{\textbf{b}})$ are given in detail in Ref. 5. In the present analysis we have conisdered the form of the electric and magnetic fileds to be exponentially decaying fields with the decay parameters, $\tau_E$ and $\tau_B$ respectively~\cite{7_Gowthama}.

\subsection{Effects of momentum anisotropy}
Momentum anisotropy can be modeled by compressing or expanding the isotropic distribution function $f^0_k$ as,
\begin{align}\label{8}
    f_{{(\text{aniso})_k}}= \sqrt{1+\xi}\,f_k^0\Big(\sqrt{p^2+\xi({\bf p}\cdot{\bf n})^2}\Big),
\end{align}
where $\xi$ is the anisotropic parameter and ${\bf n}$ is the direction of anisotropy. The electric and Hall currents develop additional components due to the anisotropy as,
\begin{align}\label{9}
    (j_{e})_{{\text{aniso}}} =j_e^{(0)} +\delta j_e^{(0)} +j_e^{(1)} +\delta j_e^{(1)},
\end{align}
\begin{align*}
 (j_{H})_{{\text{aniso}}} =j_H^{(0)} +\delta j_H^{(0)} +j_H^{(1)} +\delta j_H^{(1)}+j_H^{(2)} +\delta j_H^{(2)}.
\end{align*}
The forms of the additional components are described in detail in Ref. 5.

\subsection{Results and Observations}
To analyze the impact of time-dependence of the electromagnetic fields on the QCD medium response, we define the following ratios,
\begin{align*}
   &R_{e} =  \frac{j_{e}^{(0)}}{ET} + \frac{j_{e}^{(1)}}{ET},
    &&R_{H} =  \frac{j_{H}^{(0)}}{EBT} + \frac{j_{H}^{(1)}}{EBT} +\frac{j_{H}^{(2)}}{EBT}.
\end{align*}
For the case with anisotropy we have, $ (R_{e})_{aniso} =  \frac{(j_{e})_{aniso}}{ET}$ and $(R_{H})_{aniso} =  \frac{(j_{H})_{aniso}}{EBT}$. In the limiting case of constant electromagnetic fields, the term $j^{(0)}_e /(ET) = \sigma_e/ T$ and the term $j^{(1)}_e /(ET)$ denotes the correction due to the time dependence of the electric field. In the similar way, $j^{(0)}_H$ describes the leading order term and the quantities $j^{(1)}_H$ and $j^{(2)}_H$ denote the corrections to the current density in the direction $(\hat{\textbf{e}} \times \hat{\textbf{b}})$. The strength of the time dependence of the fields is quantified in terms of decay time, $\tau_B, \tau_E$. We have depicted the effect of the inhomogeneity of time of the external electromagnetic fields on the temperature dependence of $R_e$ in Fig. 1. The time dependence of the fields is seen to have a visible impact on the QCD medium response. For the case of a time-varying electric field and a constant magnetic field, the value of $R_e$ is higher in comparison with the case of constant fields due to the additional contribution from $\dot{\textbf{E}}$. The inclusion of time inhomogeneity of the magnetic field further introduces back current in the QGP medium. The same observation holds for $R_H$. In addition, we have observed that $j_e \hat{\textbf{e}}$ and $j_H (\hat{\textbf{e}} \times \hat{\textbf{b}})$ decrease with an increase in momentum anisotropy of the medium. The additional components to current densities may play a vital role in the magnetohydrodynamical framework for the QCD medium in the collision experiments.
 \begin{center}
\begin{figure} 
		\includegraphics[width = 5cm]{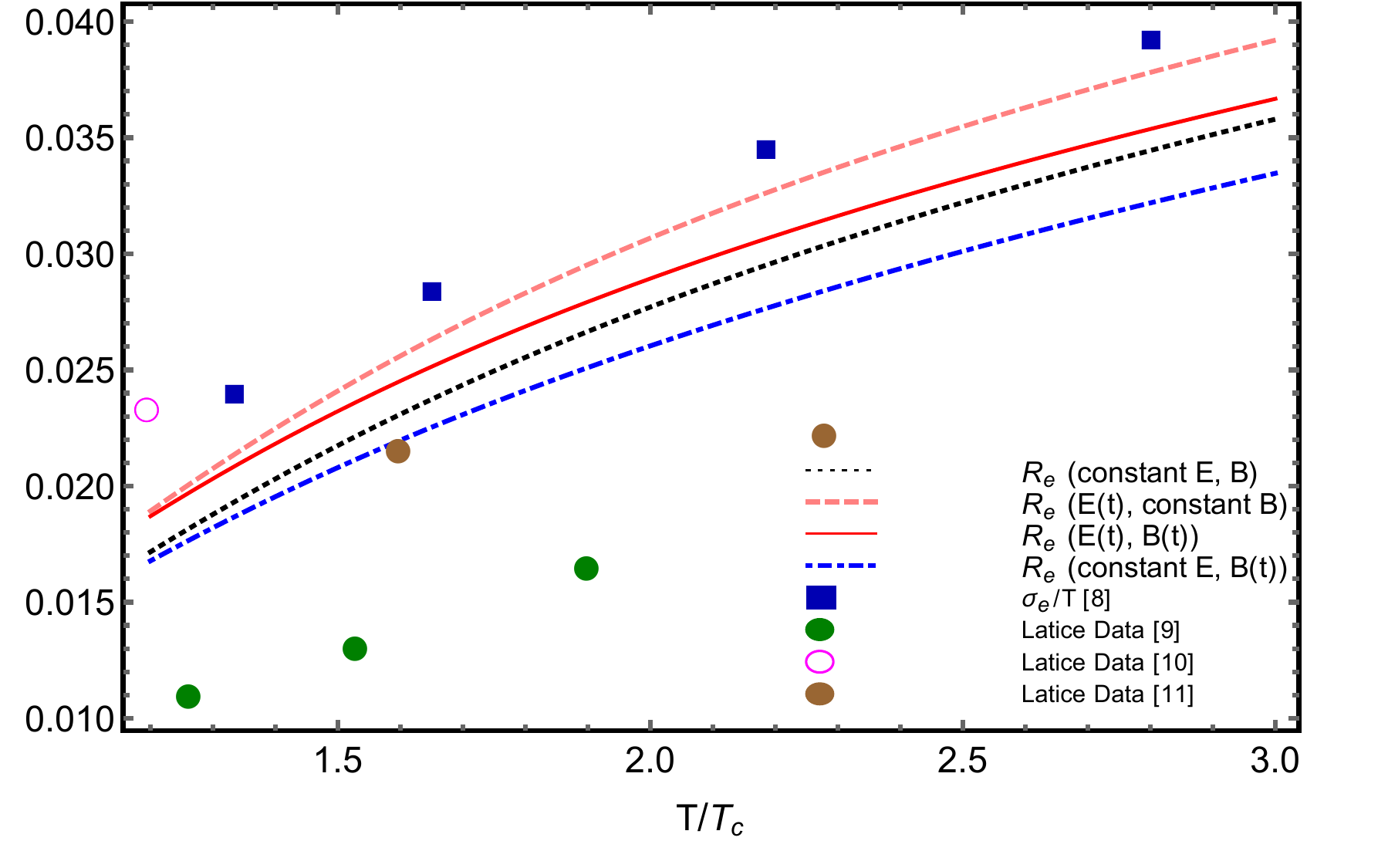}
                \includegraphics[width = 5cm]{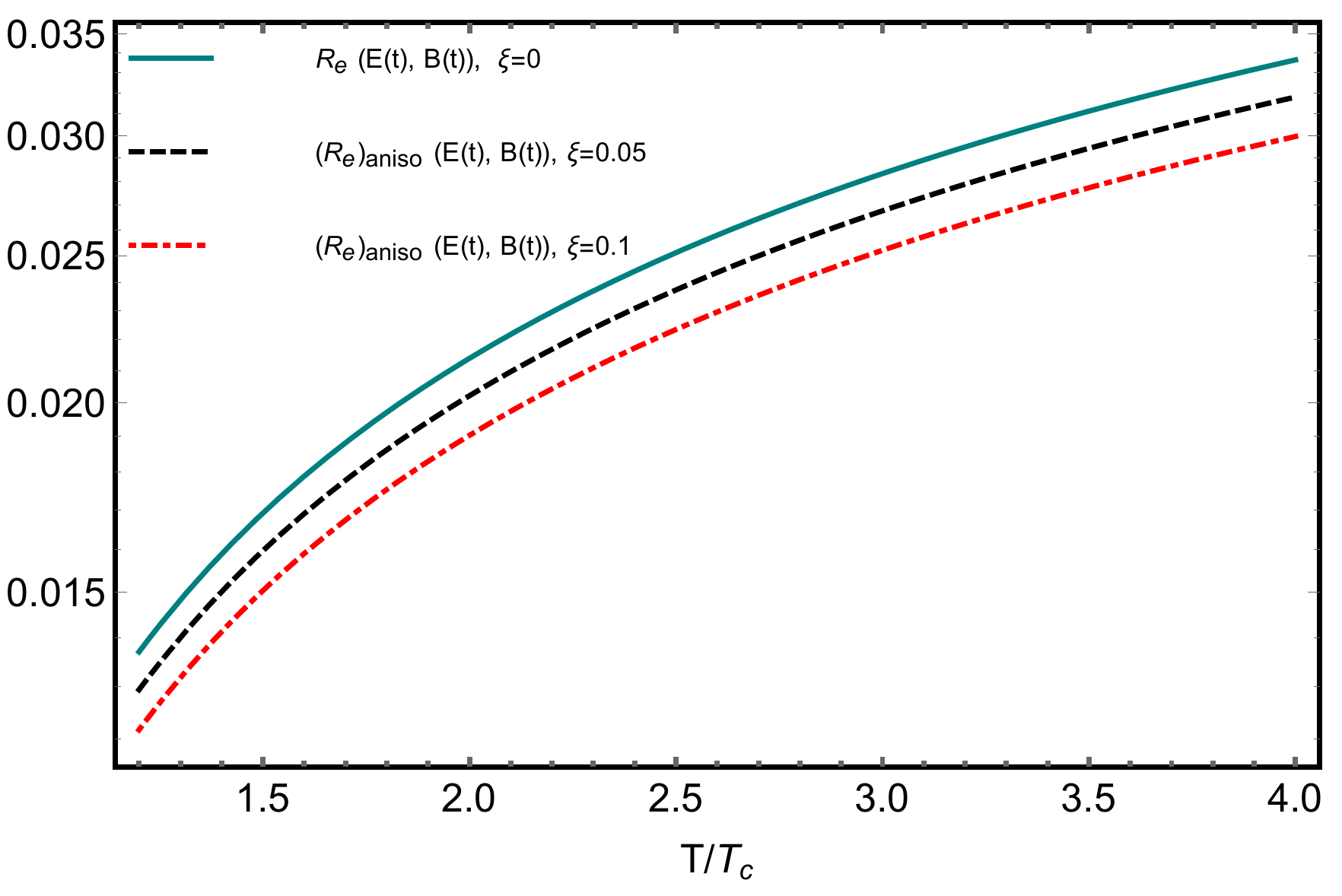}

		\caption{The temperature dependence of $R_e$ (left panel) for various choices of the external electromagnetic fields. For constant magnetic field case $eB = 0.03$ GeV$^2$ and we choose $\tau_B=9$ fm, $\tau_E=7$ fm for the time-varying cases. The effects of aniotropy and equation of state vs temperatre (right panel) for time dependent fields with $\tau_B = 9$ fm.
		}\label{Fig}
\end{figure}
\end{center}

\subsection{Summary}
In summary, we have observed that the time dependent electromagnetic fields and momentum anisotropy of the medium play a significant role in the electric charge transport of the hot QCD medium.

 \section{Study of Electromagnetic Effect by Charged-dependent Directed Flow in Isobar Collisions at  $\sqrt{s_{NN}} = $ 200 GeV using STAR at RHIC }
\author{Dhananjaya Thakur (for the STAR collaboration)}	

\bigskip

\begin{abstract}
In non-central heavy-ion collisions, it is predicted that an initial strong but transient magnetic field ($ \sim 10^{18}$ Gauss) can be generated. The charge-dependent directed flow ($v_{1}$) can serve as the probe to detect this initial magnetic field. In addition,  $v_{1}$ of several identified hadron species with different constituent quarks will help to disentangle the role of produced and transported quarks. In this proceedings, we present the measurements of $v_{1}$ for $\pi^{\pm}$, $K^{\pm}$, $p(\bar{p}$) in Ru+Ru and Zr+Zr collisions at $\sqrt{s_{NN}} = $ 200 GeV as a function of transverse momentum, rapidity, and centrality. The difference of $v_{1}$ slope ($\Delta \rm{d} \it v_{1}$/dy) between protons and anti-protons is observed and is studied as a function of centrality. While the contribution from transported quarks can give positive $\Delta \rm{d} \it v_{1}$/dy, the electromagnetic field is predicted to give negative $\Delta \rm{d} \it v_{1}$/dy. The significant negative $\Delta \rm{d} \it v_{1}$/dy of protons in peripheral collisions is consistent with the prediction from initial strong magnetic field in heavy-ion collisions. 


\end{abstract}


\subsection{Introduction}
A collision of two ultra-relativistic nuclei  forms a strongly interacting matter called the Quark-Gluon Plasma (QGP)~\cite{Shuryak:2014zxa}. Anisotropic flow is quantified by Fourier coefficients of particle’s distribution in azimuthal angle measured with respect to the reaction plane. The first coefficient of Fourier expansion is termed as directed flow ($v_{1}$)~\cite{STAR:2011hyh},
 \begin{equation}
 v_{1} = \langle cos(\phi - \Psi_{R}) \rangle,
\end{equation}
where $\phi$ denotes the azimuthal angle of an outgoing particle and $\Psi_{R}$ is the orientation of the reaction plane defined by the beam axis and the impact parameter vector.  The rapidity-odd component of directed flow
$v_{1}$(y) has been argued to be sensitive to initial strong electromagnetic (EM) fields~\cite{Gursoy:2014aka}.

In the early stages of the collisions, an ultra strong magnetic field is expected to be created (eB $\sim$ $m_{\pi}^{2}$ at top Relativistic Heavy Ion Collider (RHIC) energy)~\cite{Voronyuk:2011jd}~. This magnetic field is generated by the incoming spectator protons in the collision, and may be captured if the medium produced has finite electric conductivity. As the spectator protons recede from the collision zone the produced magnetic field decays with time. This time-varying magnetic field induces an electric field due to the Faraday effect. The Lorentz force results in an electric current perpendicular to expansion velocity of medium and magnetic field, akin to the classical  Hall effect. The interplay of competing Faraday and Hall effects can influence the $v_{1}$. In other words, the EM fields are expected to drive positively-charged and negatively-charged particles in opposite ways, leading to a splitting of $v_{1}$(y)~\cite{Gursoy:2014aka} .

The UrQMD calculations~\cite{Guo:2012qi} at RHIC energies have shown that the transported protons and the spectator nucleons have the same sign of $v_{1}$ and hence they have a positive $v_{1}$ slope ($dv_{1}/dy >$ 0) at the mid-rapidity. On the other hand, the produced protons and anti-protons can have negative $v_{1}$ ($dv_{1}/dy <$ 0) slope due to contribution other than transported quarks, e.g. the tilted source~\cite{Bozek:2010bi} . This results in a positive splitting between protons and anti-protons [$\Delta dv_{1}/dy = dv_{1}/dy (p) - dv_{1}/dy (\bar{p}) > 0$]. Therefore, the transport will affect the splitting between any particle and anti-particle pairs having transported quark content, e.g splitting between $ \pi^{+} (u\bar{d})$ and $ \pi^{-} (d\bar{u})$, and also between $K^{+} (u \bar{s})$ and $K^{-} (\bar{u}s)$. Finally, the interplay between the EM field and transported quark effect determines the sign and magnitude of splitting between particle and anti-particle pairs.

\subsection{Method and Analysis}
This analysis uses Ru+Ru and Zr+Zr collisions at $\sqrt{s_{N N}} =$ 200 GeV, collected by Solenoidal Tracker at RHIC (STAR) during 2018. Details about the event cuts and track selections can be found in Ref~\cite{STAR:2003xyj} . In this analysis, the first order event plane angle is determined using Zero Degree Calorimeter (ZDC)~\cite{Adler:2000bd} .  The description of measuring $v_{1}$ using the ZDC event plane can be found in Ref~\cite{STAR:2003xyj} . The Time Projection Chamber (TPC)~\cite{Anderson:2003ur} was used for charged-particle tracking within pseudorapidity $|\eta| <$ 1, with full $2\pi$ azimuthal coverage. After the vertex selection, we analysed about 1.7 billion Ru+Ru events and 1.8 billion Zr+Zr events. Centrality is defined from the number of charged particles detected by the TPC within $|\eta| <$ 0.5. The directed flow analyses were carried out on tracks that have transverse momenta, $p_{T} > $ 0.2 GeV/$c$ for $\pi^{\pm}$ and $K^{\pm}$, and $p_{T} > $ 0.4 GeV/$c$ for $p$($\bar{p}$). The tracks should pass a requirement to be within 3 cm of distance of closest approach (DCA) to the primary vertex ($V_{z}$), and have at least 15 space points ($N_{hits}$) in the main TPC acceptance. The $\pi^{\pm}$, $K^{\pm}$, $p$ and $\bar{p}$ are identified based on the truncated mean value of the track energy loss ($\langle dE/dx \rangle$) in the TPC and we select $|n\sigma| <$ 2  ($ n\sigma = \frac{1}{\sigma_{R}} ln(\langle dE/dx \rangle / \langle dE/dx \rangle_{[\pi/K/p]}$, $\sigma_{R}$ is the $\langle dE/dx \rangle$ resolution). To ensure the purity of identified particles, we select particles with momentum smaller than 2 GeV/c for protons, and 1.6 GeV/c for pions and kaons. The time-of-flight detector (TOF)~\cite{Llope:2003ti} was used to improve the particle identification and we select particles within mass-square ($m^{2}$) range, -0.01 $< m^{2} <$ 0.1 ((GeV/$c^{2})^2$) for pions, 0.2 $< m^{2} <$ 0.35 ((GeV/$c^{2})^2$) for kaons and 0.8 $< m^{2} <$ 1.0 ((GeV/$c^{2})^2$) for protons.

The systematic uncertainties of the $v_{1}$ measurements are calculated by varying DCA, $V_{z}$, $N_{hits}$, $n\sigma$  etc. within a reasonable maximum range. The absolute difference ($|\Delta_{i}|$) between default cut with the cut variation is taken as systematic uncertainty. In addition, the absolute difference between the $v_{1} (y)$ slopes between forward and backward rapidities is also considered as a source of systematic uncertainty. The final systematic error is the quadrature sum of the systematic errors from all the sources, which are calculated as $|\Delta_{i}|/\sqrt{12}$ assuming uniform probability distribution.

\subsection{Results}
\begin{figure}[th]
\centerline{\includegraphics[width=11.3cm]{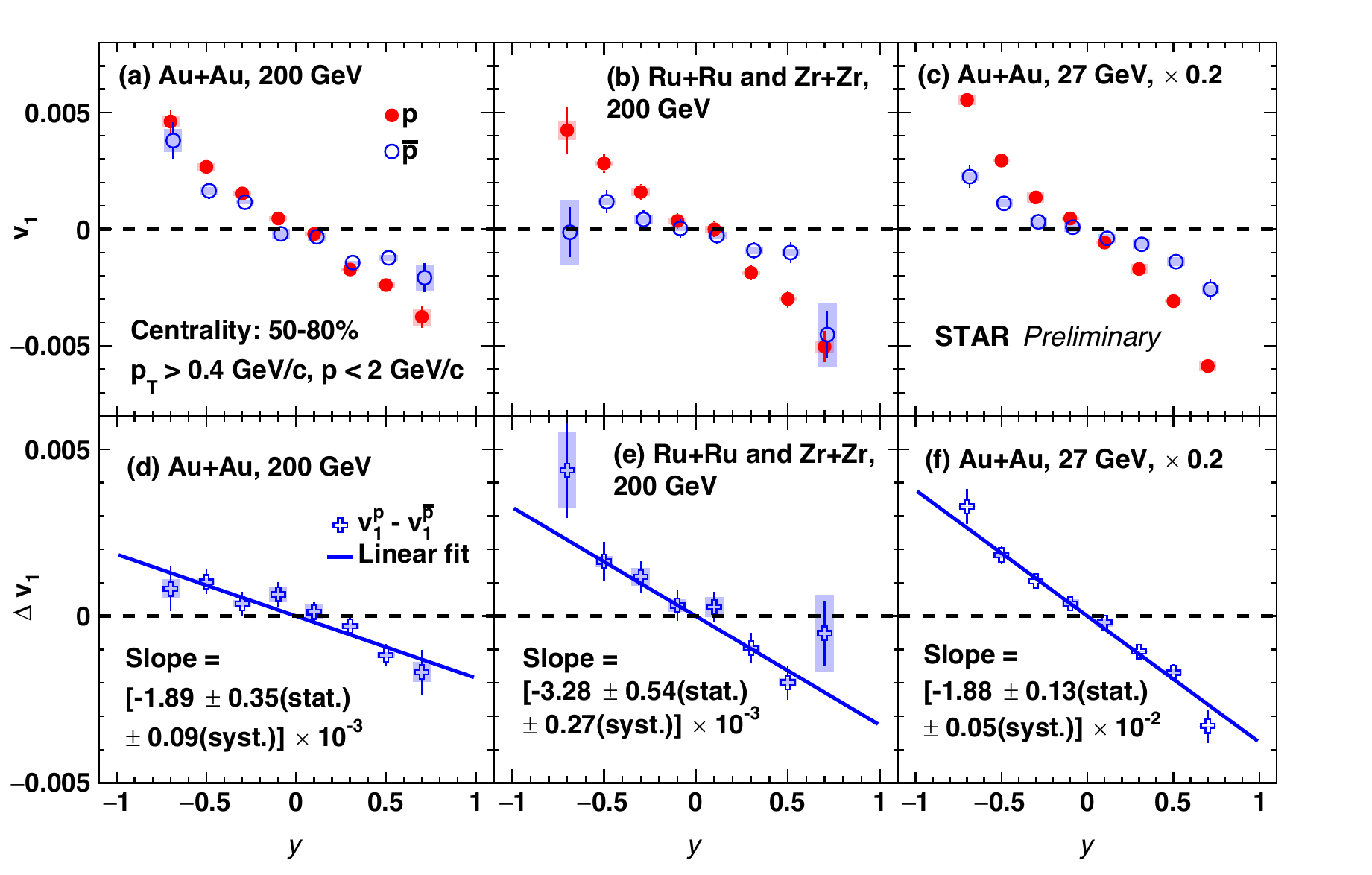}}
\caption{ Directed flow of protons and anti-protons and their difference ($v_{1}^{p} - v_{1}^{\bar{p}}$) as a function of rapidity for Au+Au collisions at $\sqrt{s_{NN}} =$ 27 and 200 GeV, and Ru+Ru and Zr+Zr collisions at $\sqrt{s_{NN}} =$ 200 GeV for 50-80 \% centrality~\cite{Diyu:starv1} . The systematic uncertainties are indicated with shaded bands and the slopes are obtained with linear fits (solid lines). }
\label{isobar_v1}
\end{figure}

\begin{figure}[th]
\centerline{\includegraphics[width=11.3cm]{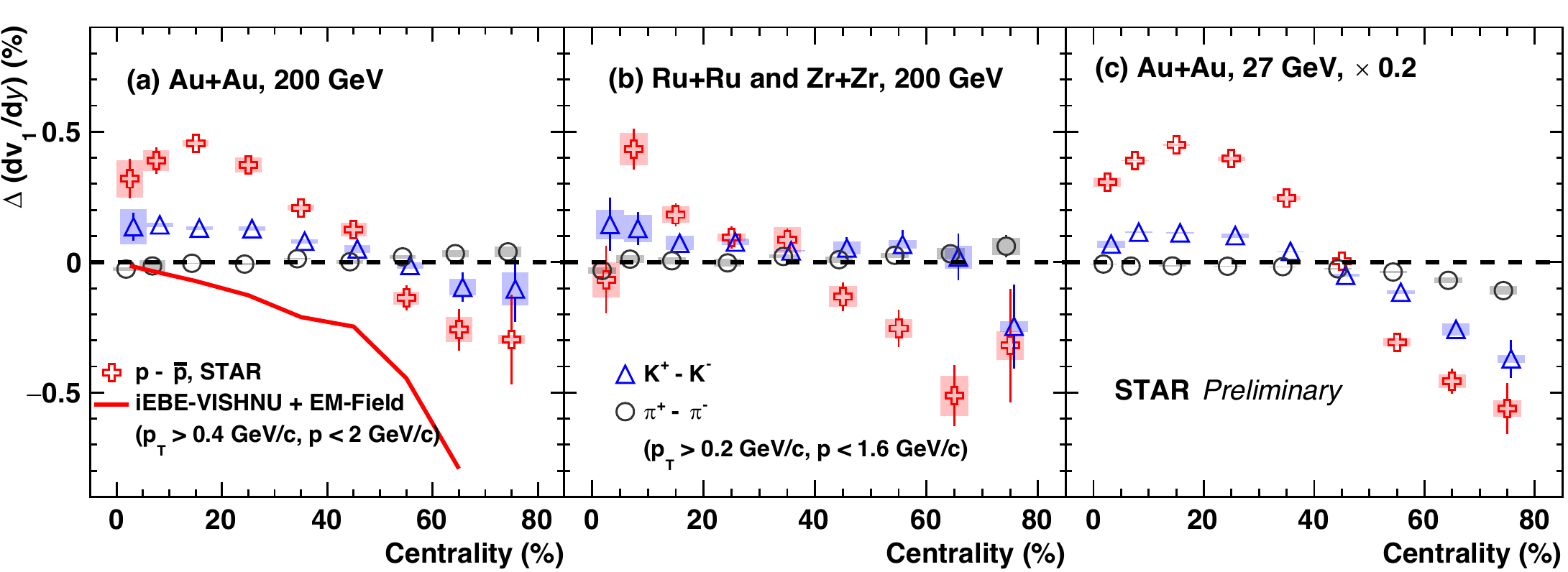}}
\caption{Slope difference ($\Delta dv_{1}/dy$) between positively and negatively charged pions, kaons and protons as a function of centrality for Au+Au collisions at $\sqrt{s_{NN}} =$ 27 and 200 GeV and isobar collisions at $\sqrt{s_{NN}} =$ 200~~\cite{Diyu:starv1} .  The systematic uncertainties are represented by shaded bands. The solid line is the electromagnetic field prediction for $v_{1}$ splitting between protons and anti-protons in Au+Au collisions at $\sqrt{s_{NN}} =$ 200 GeV~~\cite{Gursoy:2018yai} .}
\label{dv1_vs_cent}
\end{figure}

Figure~\ref{isobar_v1} presents $v_{1}(y)$ for protons and anti-protons in Au+Au collisions at $\sqrt{s_{NN}} =$ 27 and 200 GeV and isobar collisions at  $\sqrt{s_{NN}} =$ 200 GeV in the centrality range of 50–80$\%$.  Linear fits within -0.8 $< y <$ 0.8 (solid lines) that passing through (0, 0) is used to extract the slope ($\Delta dv_{1}/dy$). We observe a significant negative slope of proton and antiproton difference in the peripheral collisions which is inline with the prediction of dominance of the Faraday/Coulomb effect over the Hall and transported-quark effects, as transported quarks only provide positive contributions to the proton $\Delta dv_{1}/dy$~~\cite{Guo:2012qi} . The extracted $\Delta dv_{1}/dy$ values for each particle species using the same linear-function fit are plotted as a function of centrality and is presented in Fig.~\ref{dv1_vs_cent} for $\pi^{\pm}$, $K^{\pm}$, $p$ and $\bar{p}$ in Au+Au  at $\sqrt{s_{NN}} =$ 27 and 200 GeV and isobar collisions at  $\sqrt{s_{NN}} =$ 200 GeV.

 It is clear from Fig.~\ref{dv1_vs_cent}  that $\Delta dv_{1}/dy$ for protons shows decreasing trend, i.e positive to negative when going from central to peripheral collisions. The electromagnetic effect is weak in central collisions due to the lack of spectator protons. Therefore, the transported-quark effect can contribute to the  positive $v_{1}$ splitting. Towards the peripheral collisions the electromagnetic effect can be dominant and we see a sign change of $\Delta dv_{1}/dy$. The solid curve represented in the figure shows the quantitative calculation of electromagnetic-field contributions to the proton $\Delta dv_{1}/dy$ in Au+Au collisions at $\sqrt{s_{NN}} =$ 200 GeV~~\cite{Gursoy:2018yai} . Similar decreasing trend of $\Delta dv_{1}/dy$ is also observed for $K^{+}$ and $K^{-}$, but less significant compared to protons. This could be due to the fact that kaons have lower mean transverse momentum ($\langle p_{T} \rangle$) than protons and hence weaker electromagnetic field effects~\cite{Gursoy:2018yai} . 

The $v_{1}$ splitting between $\pi^{+}$ and $\pi^{-}$ is consistent with zero within uncertainty at $\sqrt{s_{NN}} =$ 200 GeV.  The transported quarks should give negative $\Delta dv_{1}/dy$ between $\pi^{+}$ and $\pi^{-}$. Also, the electromagnetic effect gives negative $\Delta dv_{1}/dy$. Since $\pi^{+}$ and $\pi^{-}$ numbers are almost symmetric at the top RHIC energy, the transported-quark effect is negligible. The electromagnetic effect can be diluted from neutral resonance decay. At $\sqrt{s_{NN}} =$ 27 GeV, we can see a small negative  $\Delta dv_{1}/dy$ at peripheral collisions which can have both transported-quark and electromagnetic effect.

\subsection{Summary and Outlook}
The study of charged-dependent $v_{1}$ can provide information about transported quark and electromagnetic (Hall, Faraday, and Coulomb etc.) effects in heavy-ion collisions. We have presented the $v_{1}$ measurements of $\pi^{\pm}$, $K^{\pm}$, $p$ and $\bar{p}$ for Au+Au collisions at $\sqrt{s_{NN}} =$ 27 and 200 GeV, and Ru+Ru and Zr+Zr collisions at $\sqrt{s_{NN}} =$ 200 GeV. The splitting between protons and anti-protons changes sign from positive value in central collisions to negative value in peripheral collisions. The positive value of $\Delta dv_{1}/dy$ in central collisions could be accommodated by transported quark contribution where as significant negative $\Delta dv_{1}/dy$ in peripheral collisions is consistent with expectation from dominance of  the Faraday/Coulomb effects over Hall effect.

\section{Suppressed Charmonium production in $pp$ Collisions at the LHC Energies}
\author{Captain R. Singh, Suman Deb, Raghunath Sahoo, and Jan-e Alam}	

\bigskip

\begin{abstract}
We investigate the possibility of formation of a deconfined QCD matter in $pp$ collisions at LHC. A 1$+$1D viscous hydrodynamical expansion is considered to study the evolution of the medium formed in $pp$ collisions. Here we present a theoretical study to investigate the presence of a QGP-like medium through charmonium suppression in such a small system. Our theoretical prediction for the normalized $J/\psi$ yield as a function of normalized multiplicity agrees well with ALICE data at mid-rapidity.
\end{abstract}

\subsection{Introduction}
High energy heavy-ion collisions offer an opportunity to explore the properties of hot and dense  thermalized and deconfined QCD matter known as Quark-Gluon Plasma (QGP), considering no QGP-like medium formation in proton-proton ($pp$) collisions. Therefore,  $pp$ collisions are usually considered as benchmark for investigating the existence of  such a medium in the heavy-ion ($AA$) collisions. Nevertheless, what if ultra-relativistic $pp$ collisions start showing similar behaviour like $AA$. Recent experimental findings for $pp$ collisions at $\sqrt{s}$ = 7 \& 13 TeV have provided some hints towards the existence of QGP-like medium. More precisely, collective phenomena and strangeness enhancement  observed in ultra-relativistic $pp$ collisions at the LHC energies press to retrospect small systems more comprehensively~\cite{multipart,2_Rituraj}. However, investigating such an exotic medium in these collisions is challenging, and the reasons are obvious from the physics perspective. Due to its unique features, quarkonium suppression is proposed as one of the important probes to look for the existence of a QGP-like medium in the heavy-ion collision. In this work Unified Model of Quarkonia Suppression (UMQS) is used to study the multiplicity ($dN_{ch} /d\eta$) dependent suppression of $J/\psi$ and $\psi$(2S) in $pp$ collisions at mid-rapidity~\cite{3_Rituraj,mts}.
\subsection{UMQS Model Formulation: in brief}
\begin{figure}[h!]
	\begin{center}
		\includegraphics[scale=0.27]{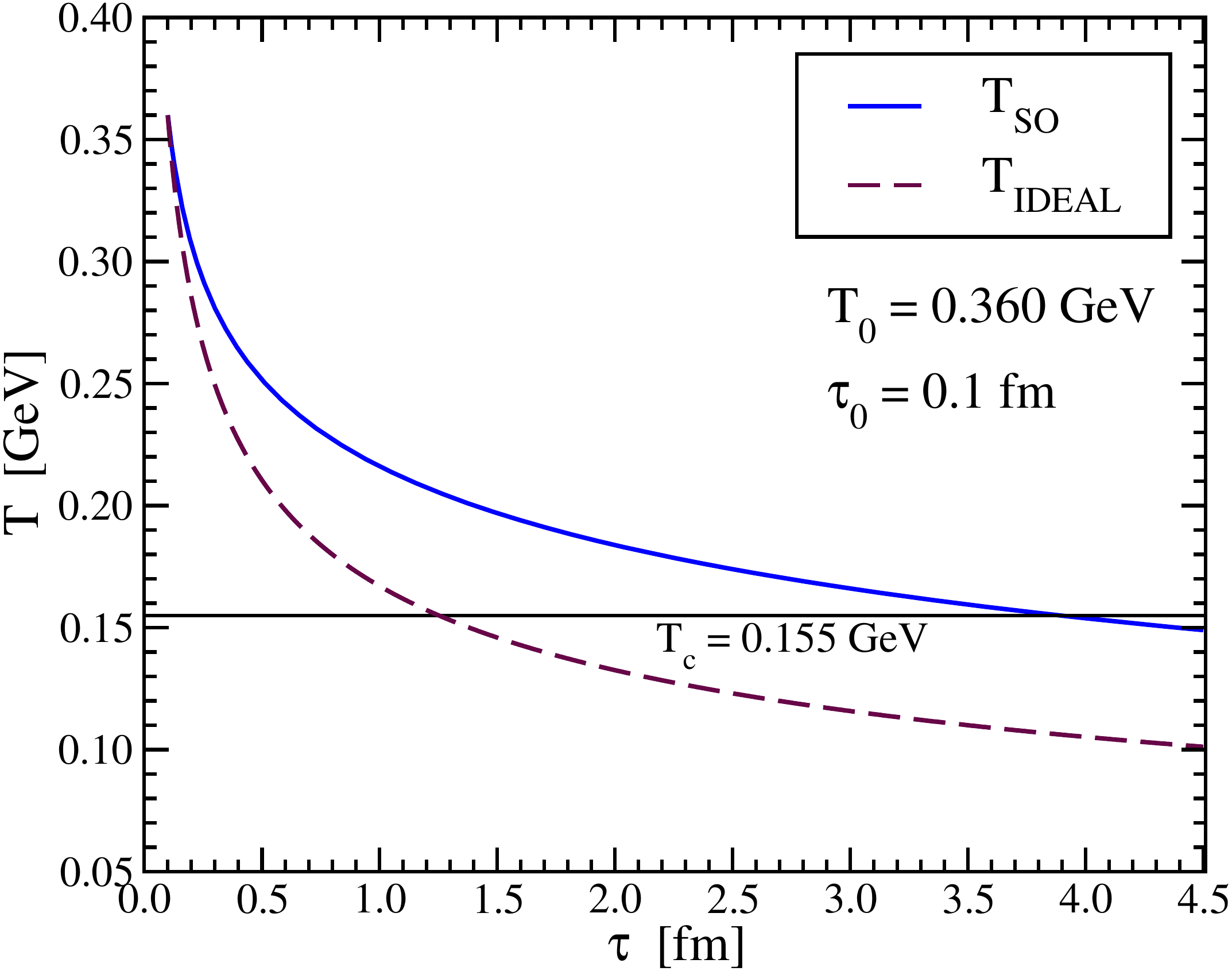}
        \includegraphics[scale=0.28]{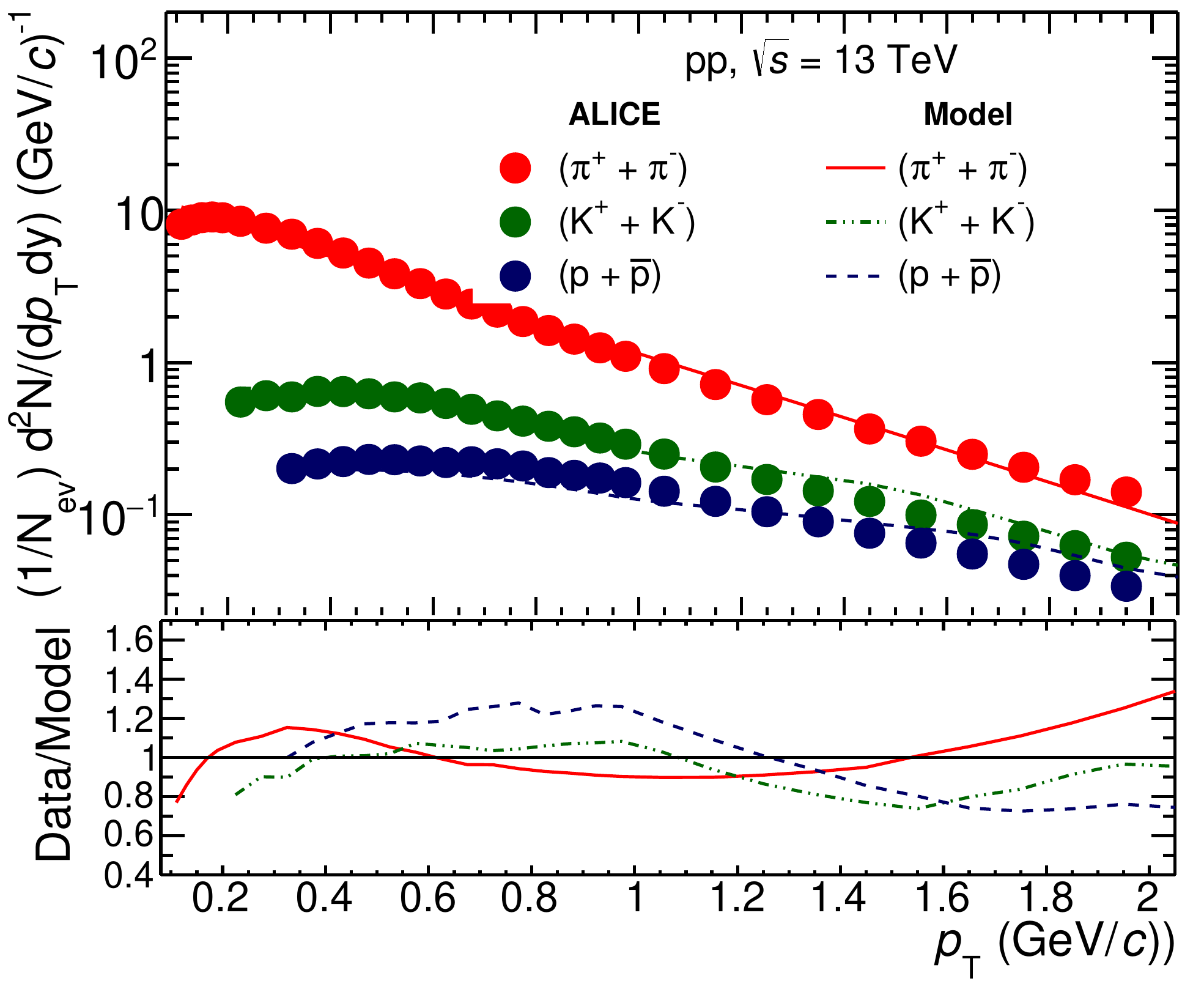}
		\caption{ {\bf Left:} The temperature cooling rate as function of proper time ($\tau$) is shown here for ideal ($T_{IDEAL}$) and second-order viscous hydrodynamics ($T_{SO}$). {\bf Right:} The $p_{T}$-spectra of identified particles obtained from hydrodynamical study are compared with experimental data.   
		}\label{rFig1}
		\end{center}
\end{figure}

\begin{figure}[h!]
	\begin{center}
		\includegraphics[scale=0.28]{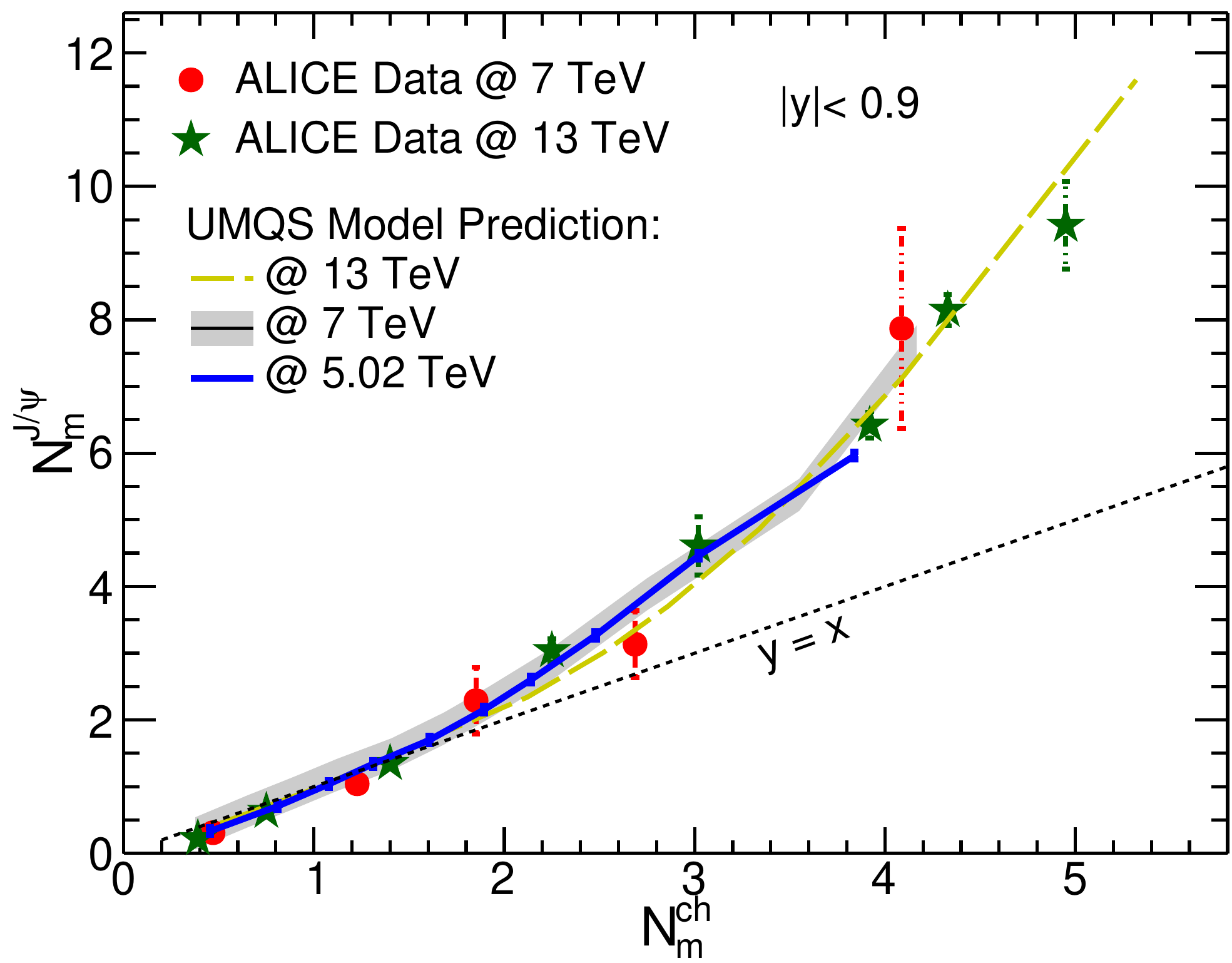}
        \includegraphics[scale=0.25]{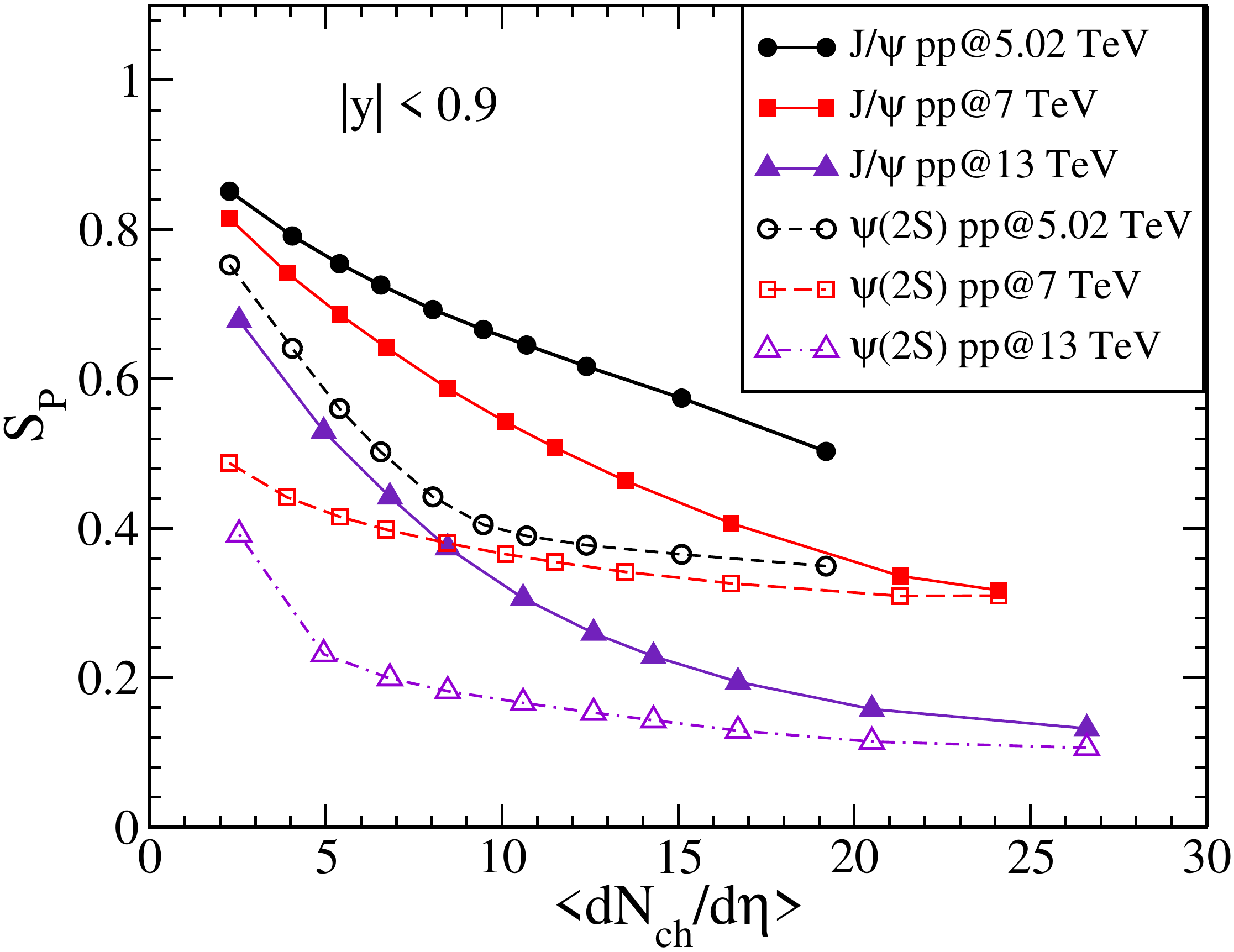}
		\caption{ {\bf Left:} Normalized yield as a function of normalized multiplicity at mid-rapidity is compared with ALICE data corresponding to V0M selection. A prediction for the same at $\sqrt{s}= 5.02$ TeV is also shown~\cite{3_Rituraj}. {\bf Right:} Model predictions for $J/\psi$ and $\psi$(2S) suppression as a function of multiplicity corresponding to various $pp$ collision energies~\cite{3_Rituraj}.   
		}\label{rFig2}
	\end{center}
\end{figure}

The UMQS formulation includes the in-medium dissociation and regeneration effects like color screening, gluonic dissociation, collisional damping, and regeneration due to the correlated $c\bar{c}$ pair~\cite{3_Rituraj}. It finally estimates the multiplicity ($dN_{ch}/d\eta$) dependent net charmonium yield in terms of survival probability, $S_{P}$. Charmonium kinematics in the QGP medium is governed by medium expansion rate, which can be obtained using hydrodynamics. To apply hydrodynamical evolution on a medium local thermalization is a necessary and sufficient condition. Therefore it is a must to check whether thermalization is achieved in $pp$ collisions or not. The initial temperature ($T_{0}$) obtained for $pp$ collisions at all the multiplicities is larger than the critical temperature ($T_{c}$) required for QCD phase transition~\cite{3_Rituraj}. Based on this motivation, we assume that the system formed in pp collisions achieve thermalization and find its consequences by contrasting the theoretical results with experimental data. The assumption of thermalization permits us to apply hydrodynamics to study its evolution. However, it is somewhat hard to estimate the initial thermalization time precisely. The attainment of thermalization in a classical system depends on two crucial factors: (i) the number of particles present in the system and (ii) the strength of interaction between the particles. A weakly interacting system with a high number density will take a long time to thermalize. On the other hand, a strongly interacting system with a small number of particles may quickly thermalize. However, from the first principle, it is impossible to state how many partons are required to produce a system which will thermalize within a time scale set by strong interaction $\approx 1/\Lambda_{QCD}$. 
In the absence of the first principle-based estimate of thermalization time of a strongly interacting partonic system and based on the above mentioned approach, we assume the initial thermalization time $\tau_{0}$ = 0.1 fm. Further, we obtained the temperature cooling rate using second-order viscous hydrodynamical equations under 1$+$1D evolution: 
\begin{equation}
\frac{dT}{d\tau} = -\frac{T}{3\tau} + \frac{T^{-3}\phi}{12a\tau}
\label{so1}
\end{equation}
\begin{equation}
\frac{d\phi}{d\tau} = -\frac{2aT\phi}{3b} - \frac{1}{2}\phi\left[\frac{1}{\tau} -\frac{5}{\tau}\frac{dT}{d\tau}\right] + \frac{8aT^{4}}{9\tau}
\label{so2}
\end{equation}
The $\phi$ appearing in Eqs. \ref{so1} and \ref{so2} is given by $\phi=4\eta/(3\tau)$. We have taken the initial value of $\phi$ that is $\phi(\tau_{0})=\phi_{0} =s_{0}/(3\pi\tau_{0})$ by using the relation $\eta/s = 1/4\pi$. The value of  $\phi_{0} = 0.1709$ $GeV^{4}$  obtained, is comparable with $\phi_{0} \sim 2p_{0}$ and $\phi_{0} < 4p_{0}$ as discussed in Ref.~\cite{6_Rituraj}.  It may be noted that we have taken $\eta/s = 1/4\pi$ to set only the initial condition for $\phi$ because its initial value is not known from QCD. Moreover, apart from $\eta$, the value of $\tau_{0}$ (which is not known precisely)  is  also required to get $\phi_{0}$, therefore, we have taken $\eta/s = 1/4\pi$ to estimate the initial value of $\phi$. The cooling rate obtained using second-order viscous hydrodynamical equations is shown in Fig.\ref{rFig1}. It shows including the viscosity of the medium cooling becomes slower and that can lead to the particles suppression if $T>T_{c} = 155$ MeV. 
In order to validate our theoretical prediction, we produced the identified soft-hadron spectra using 1$+$1D second-order viscous hydrodynamics. Fig.~\ref{rFig1}  {Right} shows, that proposed hydro-model quantitatively explains the transverse momenta ($p_{T}$) spectra of identified particle obtained from experiments~\cite{7_Rituraj}. The temperature (T) versus proper time ($\tau$) shown in Fig.\ref{rFig1} depicts that medium created in $pp$ collisions cool down below $T<T_{c}$ at $\tau = 3\sim4$ fm. In this time scale, the charmonium suppression is expected, as the kinematics of quarkonia gets modified in the presence of the QGP medium. The transport equation that governs the charmonium kinematics in the QGP medium is given as; 
\begin{equation}
 \frac{d N_{J/\psi}}{d\tau} = \Gamma_{F,nl} N_{c}~N_{\bar{c}}~[V(\tau)]^{-1} - \Gamma_{D,nl} N_{J/\psi}
\label{tq}
\end{equation}
here, $\Gamma_{F,nl}$ and $\Gamma_{D,nl}$ represent the formation and dissociation of the $c\bar{c}$ bound states, respectively. The solution of this transport equation gives the net number of charmonia by considering gluonic dissociation, collision damping, and regeneration mechanisms. Color screening is another mechanism that does not allow charmonium to form if $T>T_{D}$, (here $T_{D}$ is the dissociation temperature of charmonia). Using this approach, we obtain the normalized $J/\psi$ yield compared with experimental data and show that it is qualitatively explaining the normalized $J/\psi$ yield plotted against the normalized multiplicity yield (see Fig.~\ref{rFig2}). Finally, charmonia suppression is measured in terms of the survival probability ($S_{P}$) of the resonance state propagating through deconfined QCD matter. Fig.~\ref{rFig2} also shows that suppression of charmonium states increases with the multiplicity, and it further increases with increasing centre-of-mass energy from $\sqrt{s} =$ 5.02 to 13 TeV.
\subsection{Summary}
Our current study shows a QGP-like behaviour at all the multiplicity intervals in ultra-relativistic $pp$ collisions at $\sqrt{s} = 5.02, 7\; \&\; 13$ TeV LHC energies. The formation of a QGP-like medium in such a small system requires a new baseline to analyze QGP effects in $AA$ collisions at LHC energies.
\begin{itemize}
\item We have found that second-order temperature cooling enables effective suppression of charmonium at all the multiplicities.
\item In our study, we obtained more suppression for $\psi(2S)$ than $J/\psi$  and the effect of regeneration is also marginal for 
$\psi(2S)$. Therefore, higher resonances like $\psi(2S)$ could be a cleaner probe to investigate QGP in small systems.
\item As the cold nuclear environment is absent in the $pp$ collisions, any modification in the charmonium yield can be considered as a pure 
effect of a hot QCD medium.
\item These observations challenge the previous findings on QGP existence, where most of the QGP signatures are measured considering $pp$ as a baseline. Therefore,  it requires comparative theoretical and experimental studies aiming to investigate a QGP-like medium in $pp$ collisions. 
\end{itemize}

 \section{Spatial Diffusion of Heavy Quarks in Magnetic Field}
\author{Sudipan De, Sarthak Satapathy, Jayanta Dey, Chitrasen Jena, Sabyasachi Ghosh}	

\bigskip

\begin{abstract}
Heavy quarks are one of the most important tools to probe Quark Gluon Plasma (QGP) due to their large masses, which are significantly higher than the quantum chromodynamics (QCD) energy scale and the temperature at which QGP is created. In the recent years, off-central heavy ion collisions have gained a lot of attention owing to the possibility of the creation of strong magnetic fields of the order of 10$^{18}$ Gauss at the Relativistic Heavy-Ion collider (RHIC) and $\sim$10$^{19}$ Gauss at the Large Hadron Collider (LHC). In presence of this strong magnetic field, spatial diffusion coefficient of heavy quark splits into two components viz. transverse and longitudinal components. In the present work, we have estimated spatial diffusion of heavy quarks as a function of magnetic field and temperature.

\end{abstract}

\subsection{Introduction}

During the collisions of two nuclei at relativistic energies it is expected to form a hot and dense state of matter known as Quark Gluon Plasma (QGP) which is governed by light quarks and gluons~\cite{Shuryak:2004cy}. Heavy quarks, namely charm ($c$) and bottom $(b)$ quarks are considered one of the fine probes of quark gluon plasma (QGP). Due to the fact that their masses $(M)$ are significantly larger than the QCD energy scale ($\Lambda_\text{QCD}$) and the temperature ($T$) at which QGP is created i.e $ M >> \Lambda_\text{QCD}, T$. Unlike the light quarks they do not thermalize quickly and witness the entire evolution of the fireball. Currently, it is admitted that a very strong magnetic field is created at very early stage of heavy-ion collisions~\cite{Skokov:2009qp,Bzdak:2011yy}. The estimated values of the intensity of the strong magnetic field created at RHIC and LHC is of the order of $10^{18}$ to $10^{19}$ Gauss~\cite{Tuchin:2013ie}. The effects of such a strong magnetic field is discussed in different cases such as jet quenching coefficient $\hat{q}$~\cite{Banerjee:2021sjm}, elliptic flow~\cite{Das:2016cwd}, chiral magnetic effect~\cite{Kharzeev:2007jp} etc. The effect has also been studied to diffusion coefficients of charm quarks~\cite{Fukushima:2015wck,Finazzo:2016mhm}. As heavy quarks produced early in the heavy-ion collisions their dynamics can be affected by such strong magnetic field. 
In this article we aim to study the longitudinal and transverse components of diffusion coefficients of heavy quarks in a QGP medium in the presence of a background magnetic field $\vec{B}$ in $z$-direction.  

\subsection{Formulation}
Let us consider a background magnetic field $\vec{B} = B~\hat{k}$ pointing in the $z$-direction which magnetizes a relativistic fluid. In presence of magnetic field anisotropic properties of the fluid is considered which leads to multi-component structure of the electrical conductivity tensor such as, $\sigma_{xx}$, $\sigma_{xy} = -\sigma_{yx}$ and $\sigma_{zz}$. In this article we obtained diffusion coefficient in two different approaches, one is classical based Relaxation Time Approximation (RTA) method and another is Quantum Mechanical (QM) approximation. According to Einstein's relation the spatial diffusion coefficient $(D)$ can be expressed as a ratio of electrical conductivity $(\sigma)$ and susceptibility $(\chi)$ of the medium. These become anisotropic in the presence of magnetic field thus taking a $3\times 3$ matrix structure given by  
\bea
D_{ij} = \frac{\sigma_{ij}}{\chi},~~~i,j = x,y,z
\eea 
where $D_{ij}$ is the spatial diffusion matrix, $\sigma_{ij}$ is the electrical conductivity matrix, and  $\chi$ is the susceptibility. 
According to RTA approach we obtain
\bea
&&\sigma_{zz} = \sigma_\parallel = \frac{gq^2\beta}{3}\int\frac{d^3k}{(2\pi)^3}\frac{(k_z)^2}{\omega_k^2}\tau_cf\big[1-f\big]  \\
&&\sigma_{xx} = \sigma_{yy} = \sigma_\perp = \frac{gq^2\beta}{3}\int\frac{d^3k}{(2\pi)^3}\frac{(k_{x,y})^2}{\omega_k^2}\frac{\tau_c}{ 1 + \frac{\tau_c^2}{\tau_B^2}}f\big[1-f\big],
\label{eq:RTAsigmazz}
\eea
where, $f$ represents the distribution function such as,
\bea 
f = \begin{cases}
	\big[e^{\beta(\omega_k-\mu)} +1\big]^{-1},~{\rm~for~Fermions} \\
	\big[e^{\beta \omega_k} -1\big]^{-1},~{\rm~for~Bosons~~,}
\end{cases} 
\eea 
$\omega_k = \sqrt{\vec{k}^2 + m^2}$ is the energy with momentum $\vec{k}$ and mass $m$,  $\beta = T^{-1}$, where $T$ is the temperature, $\mu$ is the chemical potential, $q$ is the charge and $\tau_{c}$ is the relaxation time.
Here, $g$ is the degeneracy factor and $\tau_{B} = \frac{m}{q B}$ is the inverse of classical cyclotron frequency. $\sigma_{zz}$ (or $\sigma_\parallel$) is the parallel component of the electrical conductivity i.e. parallel to the magnetic field and $\sigma_{xx}$ (or $\sigma_\perp$) is the perpendicular component of the electrical conductivity. The absence of Landau quantization over energies in RTA expressions prompts us to call them as classical results. On the other hand, on applying Landau quantization of energies and quantizing the phase space part of the momentum integral we obtain

\bea
\chi_\text{QM} = \beta\sum_{n=0}^{\infty}(2-\delta_{n,0})\frac{qB}{2\pi}\int_{-\infty}^{+\infty}\frac{dk_z}{2\pi}~f(1-f).
\eea

\bea
\sigma_\perp^\text{QM} &=& \frac{2q^2}{T}\sum_{n=0}^\infty (2-\delta_{n,0})\frac{qB}{2\pi}\int_{-\infty}^{+\infty}\frac{dk_z}{2\pi}\frac{lqB}{\omega_l^2}\tau^\perp f(1-f)
\label{ecqm:perp}\\
\sigma_\parallel^\text{QM} &=& \frac{2q^2}{T}\sum_{n=0}^\infty (2-\delta_{n,0})\frac{qB}{2\pi}\int_{-\infty}^{+\infty}\frac{dk_z}{2\pi}\frac{k_z^2}{\omega_l^2} \tau^\parallel f(1-f),
\label{ecqm:pll}
\eea 
where the superscript QM denotes quantum theoretical results, $\tau^\parallel = \tau_c$, $\tau^\perp = \frac{\tau_c}{1 + \frac{\tau_c^2}{\tau_B^2}}$, $\tau_c$ is the relaxation time.

\subsection{Results}

\begin{figure}[h]
	\begin{center}
		\includegraphics[width = 5cm]{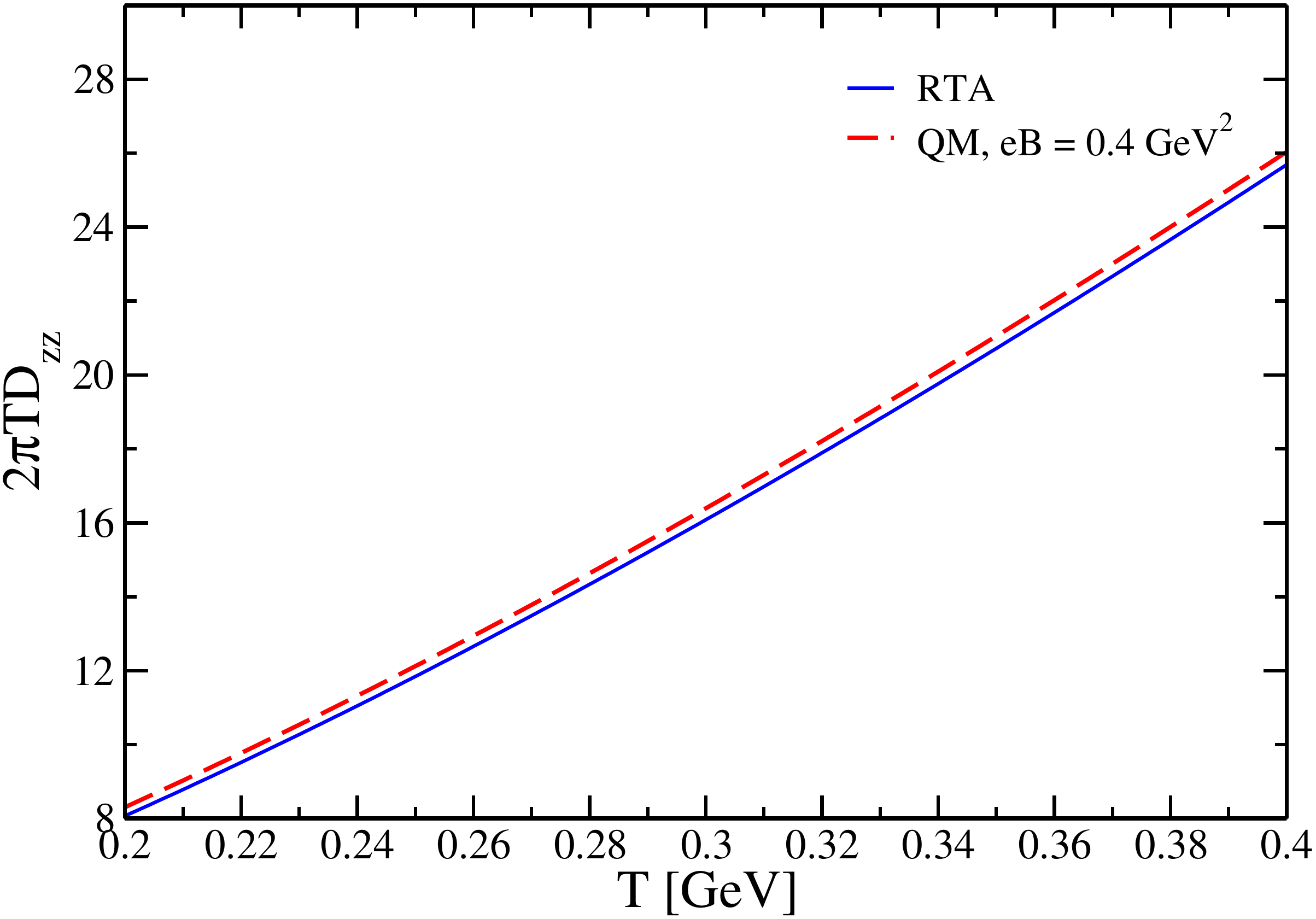}
		\includegraphics[width = 5cm]{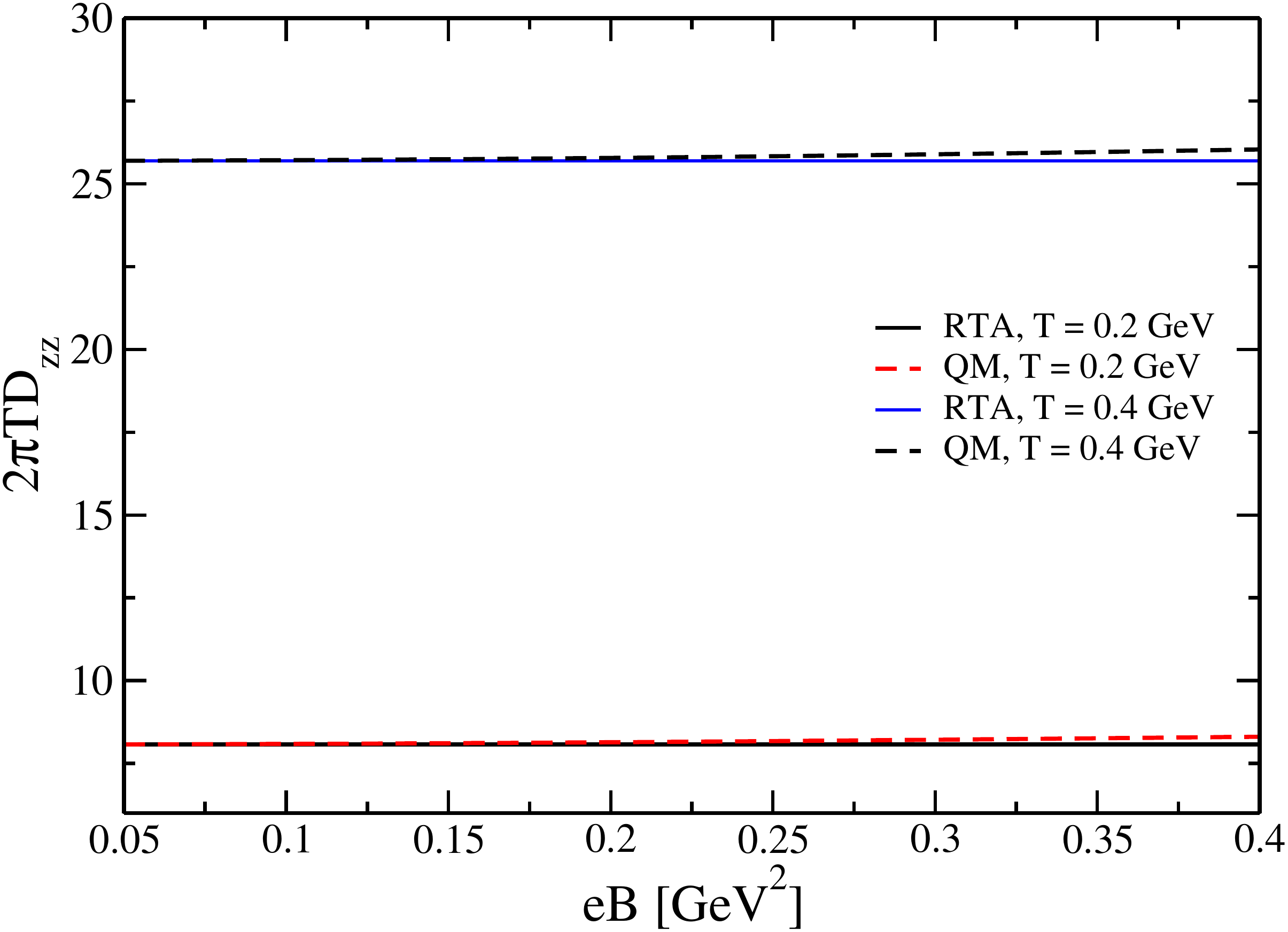}
		\includegraphics[width = 5cm]{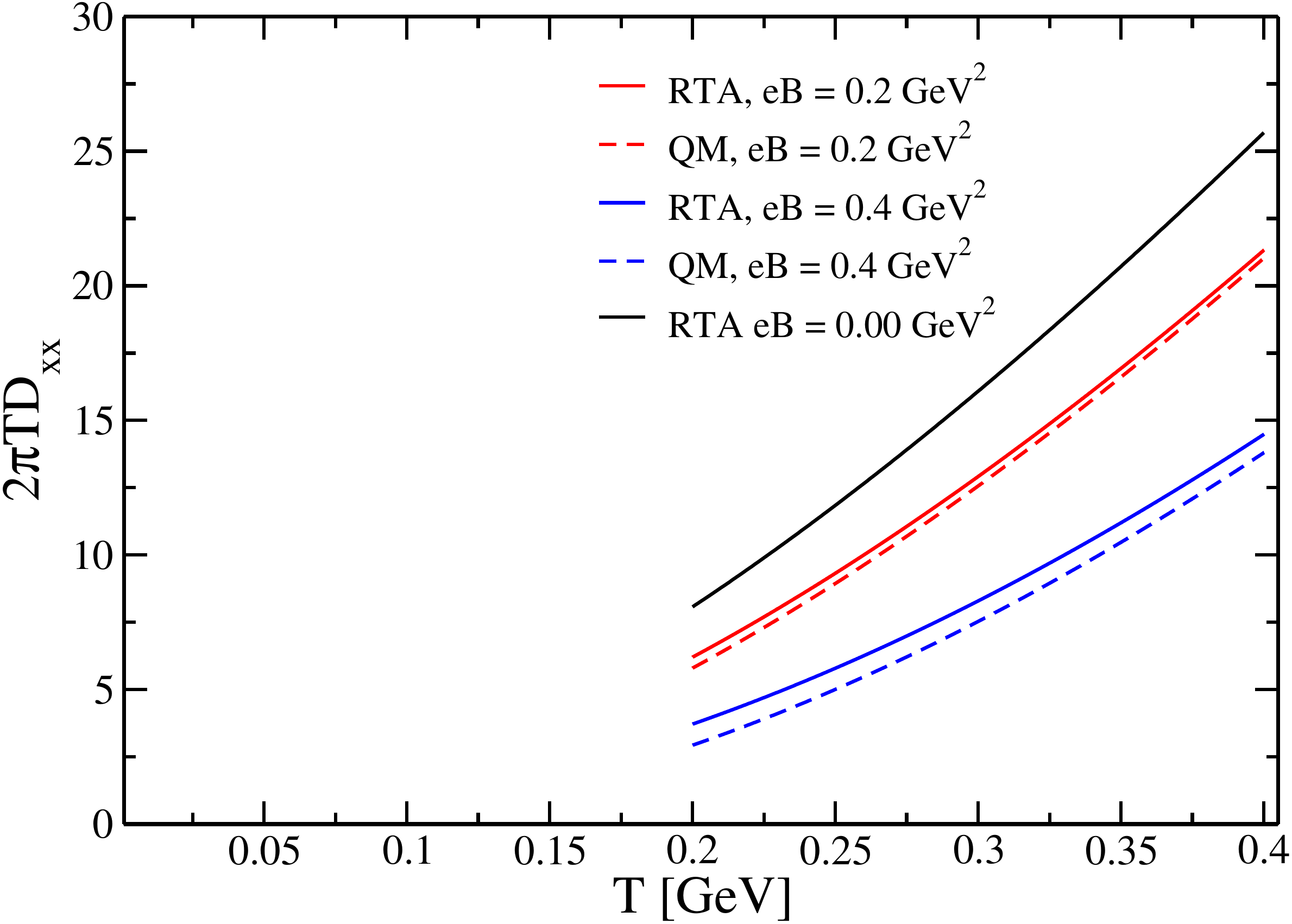}
		\includegraphics[width = 5cm]{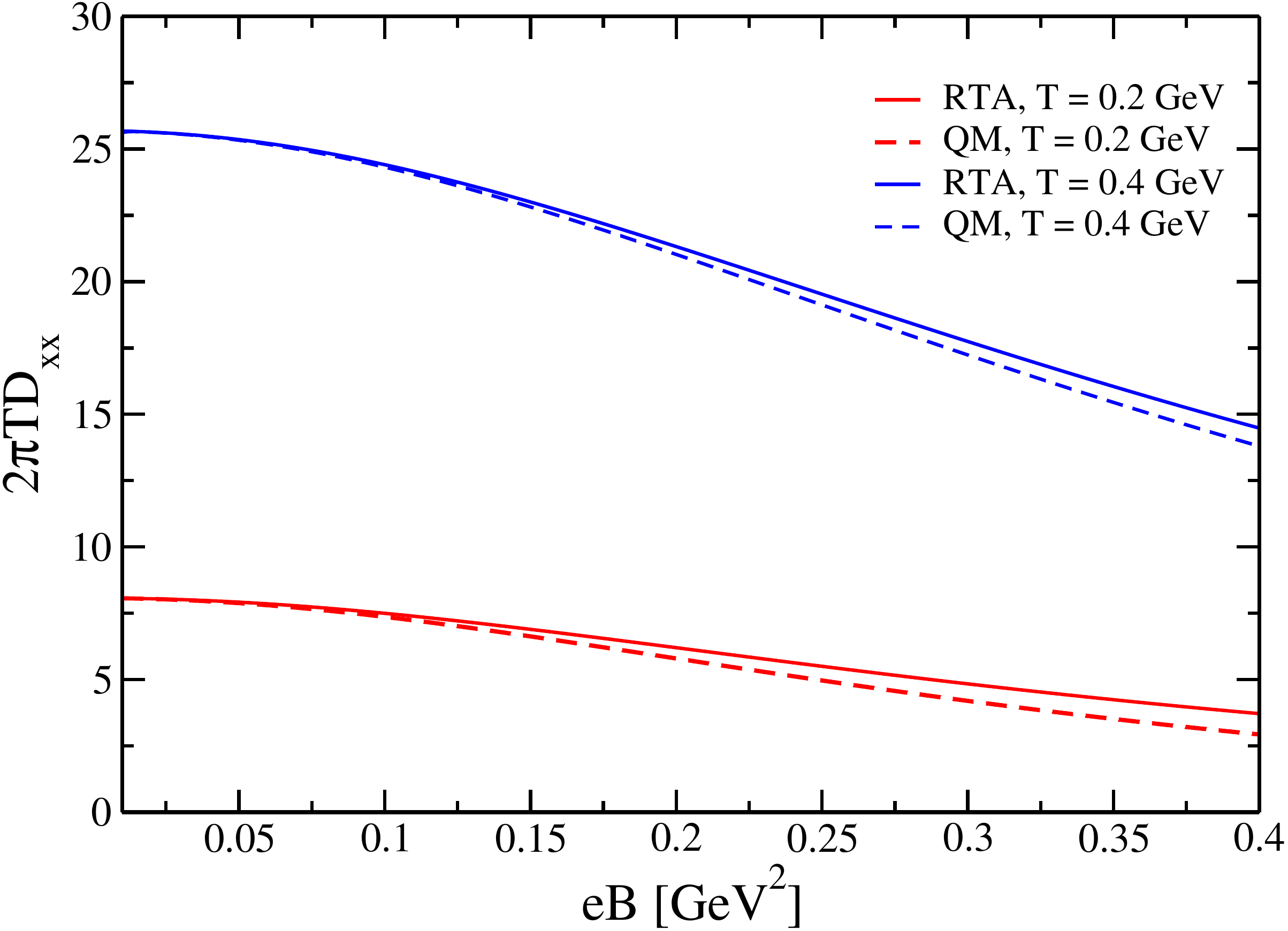}
		\caption{Upper panel: solid line shows the result from RTA calculation and dotted line shows results from QM approximation. Lower Panel: Black solid line represents the RTA result without magnetic field. Solid and dotted lines represent the results from RTA and QM respectively. Different colors represent the results at different magnetic fields as well as at different temperatures.
		}\label{Fig:diffusion}
	\end{center}
\end{figure}

The left figure of the upper panel of Fig.~\ref{Fig:diffusion} shows the results of longitudinal diffusion as a function of temperature by plotting $2\pi T D_{zz}$ with $T$ which basically shows how the heavy quarks can diffusively travel in a direction parallel to the background magnetic field. In this figure the solid lines are for the isotropic case calculated via RTA formalism for $\tau_c = 6~fm$ and $eB$ = 0.2 and 0.4 GeV$^2$. The dotted lines are the results of $2\pi T D_{zz}$ for QM case where Landau quantization were employed, thus providing an expression of spatial diffusion as a sum over Landau levels. The increasing trend of diffusion coefficient with temperature can be attributed to the fact that an increase in kinetic energy contributes to an increase in diffusion along the direction parallel to magnetic field.
The right figure of the upper panel of Fig.~\ref{Fig:diffusion} shows the results of longitudinal diffusion as a function of magnetic field by plotting $2\pi T D_{zz}$ with $eB$ at two different temperatures 0.2 GeV and 0.4 GeV for both RTA and QM formalisms. At low magnetic field the RTA and QM results coincide with each other whereas at high magnetic fields we see a clear deviation of QM results from RTA results which implies the universal behaviour of spatial diffusion at low magnetic field.  The longitudinal component of diffusion calculated via RTA formalism, does not depend on the magnetic field since along the direction of magnetic field the heavy quarks do not experience Lorentz force. Contrary to the RTA case, the QM results depend on the magnetic field. Here we see that for the QM case the diffusion increases with increasing magnetic field. To understand this we note that spatial diffusion is the ratio of electrical conductivity and static susceptibility. In the QM  approach the Landau level summation appears in the expression of $\sigma$ and $\chi$ due to which quantum effects come into play and thus we observe a deviation relative to RTA results. The left figure of the lower panel of Fig.~\ref{Fig:diffusion} shows the results of perpendicular component of spatial diffusion coefficient as a function of temperature for two different magnetic field strength, $eB$ = 0.2 GeV$^2$ and 0.4 GeV$^2$ along with the RTA results without presence of magnetic field. It is observed that in presence of strong magnetic field diffusion decreases. The right figure of the lower panel of Fig.~\ref{Fig:diffusion} shows the behaviour of $D_{xx}$ with magnetic field for different values of temperatures. At low magnetic field the RTA and QM results coincide implying the universal nature of diffusion at low magnetic field whereas at high magnetic field they are different due to Landau level summation present in QM expressions.   

The aim of our present work is not the comparison between classical and quantum estimation of heavy quark diffusion at finite magnetic field, rather addressed them systemically step by step, which was missing in literature. However, for application point of view, they should be used in the weak and strong magnetic field domains respectively.

\subsection{Summary}
The spatial component of diffusion coefficient for heavy quarks was calculated in presence of strong background magnetic field using the well known Einstein relation. As diffusion coefficient becomes anisotropic in presence of magnetic field, both parallel and perpendicular components of diffusion coefficient were obtained. It is found that diffusion coefficient increases with increasing temperature both for parallel and perpendicular components. This behaviour can be attributed to the increase in kinetic energy of heavy quarks with increasing temperature. The difference between RTA and QM results is higher at higher magnetic field due to Landau level summation present in QM expression. Parallel component of diffusion coefficient is independent of magnetic field in case of RTA formulation contrary to the results obtained in QM formalism. Perpendicular component of diffusion coefficient decreases with increasing magnetic field both for RTA and QM. In summary, different behaviour of parallel and perpendicular component of spatial diffusion coefficient of heavy quarks in presence of strong initial magnetic field may lead us to interesting directions in recent future.




\section{Weak Interaction driven bulk viscosity of hot and dense plasma}
\author{Sreemoyee Sarkar}	

\bigskip

\begin{abstract}
We present the formalism of bulk viscosity coefficient of baryonic matter in presence of trapped neutrinos.   In neutron star at temperature $T\sim 5$ MeV, neutrino mean free path remains  small in comparison to the dimension of the star which results in non-vanishing neutrino chemical potential.  This directs  modified URCA process for the beta equilibration rate to produce maximum bulk viscosity at temperature  larger than the neutrino-trapped temperature. The calculation has been performed considering beta non-equilibration  of modified URCA process. 
The resonant behaviour of the bulk viscosity is dependent on the particle interaction rate and thermodynamic susceptibilities of the medium. The susceptibilities have been calculated   considering free Fermi gas equation of state of hadrons. The bulk viscosity coefficient attains its maximum at  $T\sim 9$ MeV.  The bulk-viscous dissipation time scale for compression-rarefaction oscillation is found out to be $50$ millisecond.  These results imply the relevance of the formalism in high temperature, highly dense medium produced in binary neutron star merger. 
\end{abstract}

\subsection{Introduction}
Time scale for damping of stellar vibrations of neutron stars and  maximum rotation rate of millisecond pulsars are greatly influenced by the bulk viscosity of the stars. For a vibrating star density changes due to vibrational and rotational instabilities and these change the concentrations of different species mainly through URCA and modified URCA (MURCA) processes. This results in dissipation and the dissipation becomes  maximum once the rate of the above microscopic reactions become comparable to the rate of the oscillation of the chemical potential.


Over the last decade the theory of  bulk viscosity coefficient ($\xi$) has been studied  in the field of  heavy ion collision experiment \cite{DenicolPRL,Denicol} as well as in the isolated neutron stars and in the quark stars \cite{1968ApJ,Pethick:1979,Madsen1992,Alford:2010gw,Schmitt:2017efp}.  In Ref.\cite{Most:2021zvc} it has been shown that the value of average contribution of $\xi$ in  heavy ion collision experiment is comparable  to that of  in binary neutron star (BNS) merger.
Recently, authors in   Ref. \cite{Alford:2017rxf} have revealed the fact that  $\xi$, driven by MURCA process in the neutrino transparent regime gives rise to   dissipation time scale of the order of survival time period (millisecond) of  BNS merger.  In connection to this  the bulk viscosity coefficient due to URCA process in the neutrino trapped domain has been evaluated in Ref.\cite{Alford:2019kdw}. 
All these current findings motivate us to estimate  the bulk viscous dissipation  due to  MURCA process in the high temperature and high density plasma in presence of trapped neutrinos relevant for BNS merger. 

%

             \subsection{Bulk Viscosity of dense matter with trapped neutrinos}
 In this paper we consider a system of baryonic matter consists of neutrons, protons, electrons, neutrinos. In neutron star due to rotational and vibrational motion fluid elements undergo continuous compression and rarefaction and this leads to  bulk viscous dissipation.
We consider the  MURCA as beta equilibration process which contributes maximum in bulk viscosity at high temperature in comparison to the URCA process \cite{Alford:2010gw}. The MURCA processes are $n+N\leftrightarrow N+p+e+\bar\nu$, $N+p+e\leftrightarrow N+n+\nu_e$. 
 In chemical equilibrium the forward and backward reactions occur at the same rate and the sum of the incoming chemical potentials remain the same as that of the backward reaction. Volumetric fluid element oscillations give rise to non-equilibrium scenario when equality of forward and backward chemical equilibration does not happen. 
 
 In a neutron star with large temperature, typically  $T\sim 5$ MeV the mean free path of neutrinos remain small and they remain trapped in side the star and hence non-zero chemical potential. 
 Subtracting the final state chemical
    potential from the initial state the fluctuation in chemical potential is $\mu_{\Delta}=\sum_i\mu_i-\sum_f \mu_f\neq 0$. The time derivative of $\mu_{\Delta}$ gives a linear equation in $\mu_{\Delta}$, $\frac{d\mu_{\Delta}}{dt}=C\omega \frac{\delta n_{\star}}{\bar n_{\star}}cos (\omega t)+B n_{\star}\frac{dx}{dt}$. $\delta n_{\star}$ is the deviation of baryon density and $\delta x$ is the deviation of the baryon density fraction from the equilibrium value.  $C$ is defined as  the beta non-equilibration baryon density susceptibility and $B$ is the beta non-equilibration proton fraction susceptibility, $\omega$ is the oscillation frequency, $n_{\star}$ is the equilibrium baryon density. Susceptibilities are defined as, $C\equiv{n}_{*}\left.\frac{\partial\mu_\Delta}{\partial n_{*}}\right|_x$, 
$B \equiv\frac{1}{{n}_{*}}\left.\frac{\partial\mu_\Delta}{\partial x}\right|_{n_*}$.

The differential equation for the chemical fluctuation can be written as
\bea
\frac{dA}{d\phi}=d\cos(\phi)+ \frac{B \Gamma^{\leftrightarrow}}{\omega T}.
\label{mu_delta}
\eea
where, $\phi=\omega t$, $A=\mu_{\Delta}/T$, the prefactors $d\equiv \frac{C}{T}\frac{\delta n_{\star}}{ n_{\star}}$, $f\equiv \frac{B\tilde \Gamma T^{6}}{\omega}$, $\Gamma^{\leftrightarrow}=n^{\star}\frac{\partial x}{\partial t}$. The details of $\Gamma^{\leftrightarrow}$ we present later in this paper. The bulk viscosity of a given system is defined to be the response of the system to an oscillating compression and rarefaction of the medium. The energy dissipation rate per volume in the fluid due to oscillation is $\frac{d\epsilon}{dt}=\xi (\vec \nabla. \vec v)^2$, $v$ is the local fluid velocity of the conserved baryon density. In the hydrodynamic limit after averaging the energy dissipation rate per volume  over one time period the bulk viscosity can be expressed as, 
 \bea
\xi=\frac{TC}{\pi\omega B}\frac{\bar n_{\star}}{\delta n_{\star}}\int_0^{2\pi} A(\phi)cos(\phi)d\phi
\label{bulk_vis}
\eea
The detailed calculation of obtaining $\zeta$ can be found in Ref.\cite{Alford:2010gw}
$A(\phi)$ can be evaluated by solving Eq.(\ref{mu_delta}).
%

Now, we present the microphysics of re-equilibration rate  $\Gamma^{\leftrightarrow}\equiv \Gamma^{\rightarrow}-\Gamma^{\leftarrow}$, where, $ \Gamma^{\rightarrow}$ is the rate of the forward and $\Gamma^{\leftarrow}$ is the backward reaction. For MURCA process \cite{1983bhwdS},  
 \bea
&&\Gamma^{\rightarrow}-\Gamma^{\leftarrow}=
\int \frac{d^3p_n}{(2\pi)^3}\frac{d^3p_N}{(2\pi)^3}\frac{d^3p_N'}{(2\pi)^3}\frac{d^3p_p}{(2\pi)^3}
\frac{d^3p_e}{(2\pi)^3}
\frac{d^3p_{\nu_e}}{(2\pi)^3}\nn\\
&&(2\pi)^4|{\cal M}_{fi}^2|
[\delta^4(p_n+p_N-p_N'-p_e-p_{\nu_e})]{\cal P}
\label{int_rate}
\eea
The scattering matrix element ${\cal M}_{fi}$ is given in Ref.\cite{Yakovlev:2000jp}. ${\cal P}$ is the phase space factor and $I$ is the energy integral.
 \bea
 {\cal P}=[f_Nf_pf_ef_{\bar \nu_e}(1-f_N)(1-f_n) -f_Nf_n(1-f_N)(1-f_p)(1-f_e)(1-f_{\bar \nu_e})]
\eea
  After performing angular integration we obtain, 
 \bea
 \Gamma^{\rightarrow}-\Gamma^{\leftarrow}={\cal C}\left(I_4-I_5(\nu)-I_6(\nu)\right)
 \eea 
  where, $I_4$, $I_5$ and $I_6$ are given below \cite{Madsen1992},
  \bea
 I_4-I_5(\nu)
=\frac{1}{4!(1+e^{-x_{\bar\nu_e}+\frac{\mu_{\nu}}{T}})}\frac{1}{1+e^{(x_{\bar\nu_e}+\frac{\delta\mu}{T})}} \left[(x_{\bar\nu_e}+\frac{\delta\mu}{T})^4
+10\pi^2 (x_{\bar\nu_e}+\frac{\delta\mu}{T})^2+9\pi^4\right]
\label{1steq}
\eea
\bea
I_6(\nu)= \frac{1}{4!(1+e^{x_{\bar\nu_e}-\frac{\mu_{\nu}}{T}})}\frac{1}{1+e^{(x_{\bar\nu_e}+\delta\mu/T)}}\ \left[(x_{\bar\nu_e}+\frac{\delta\mu}{T})^4
+10\pi^2 (x_{\bar\nu_e}+\frac{\delta\mu}{T})^2+9\pi^4)\right].
\label{2ndeq}
  \eea
  We write $x_i=\beta(E_i-\mu_i)$,$i=n,N,p,e$ and $x_{\nu_e}=\beta(E_{\nu_e}-\mu_{\nu_e})$ for neutrinos  and  $x_{\nu_e}=\beta(E_{\nu}+\mu_{\nu_e})$ for anti-neutrinos. Now, after evaluating the beta equilibration rate, we solve the integro-differential Eq.(\ref{mu_delta}) to obtain ${\cal A}$. Inserting ${\cal A}$ in Eq.(\ref{bulk_vis}) we obtain  the final expression for $\xi$.
 

\subsection{Result}
\begin{figure}[tbp]
 \includegraphics[width=0.45\textwidth,clip=]{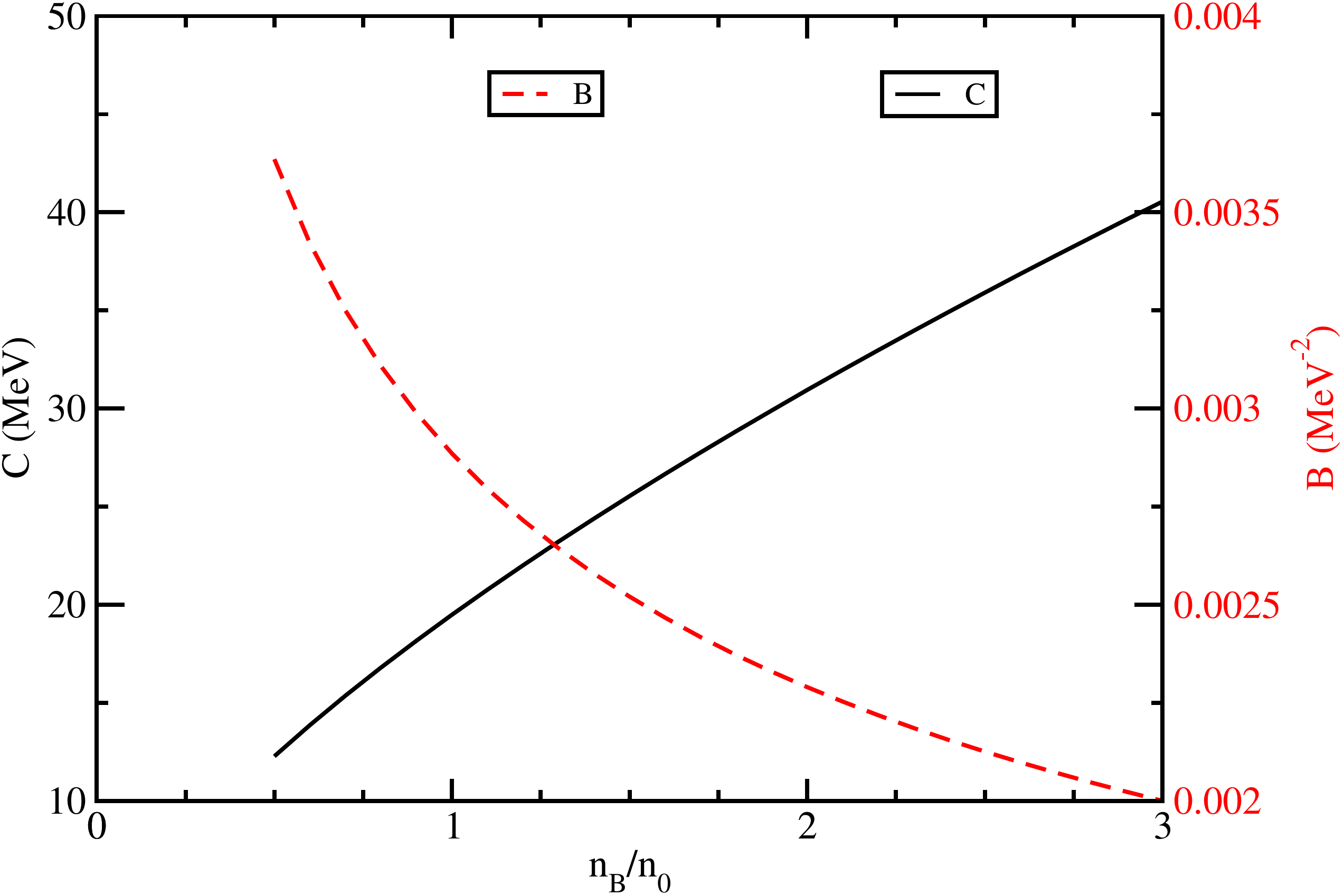}
 ~~~~~~~~~~~~~~\includegraphics[width=0.45\textwidth,clip=]{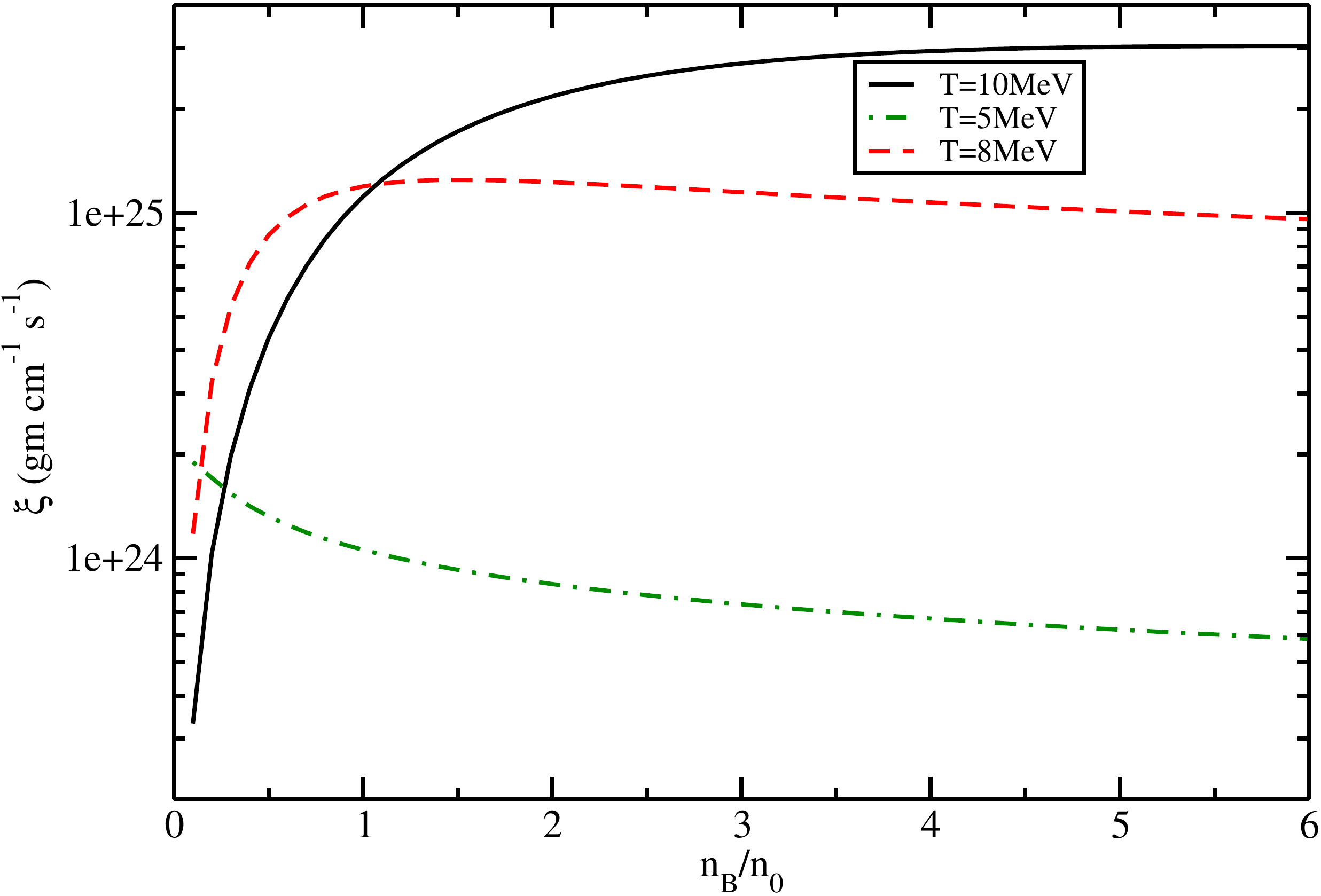}
 \caption{Left: Density variation of susceptibilities. Right: Variation of $\xi$ with density for different temperatures for neutrino-trapped matter}
 \label{density_dep}
 \end{figure}

\begin{figure}[tbp]
    \begin{center}
 \includegraphics[width=0.50\textwidth,clip=]{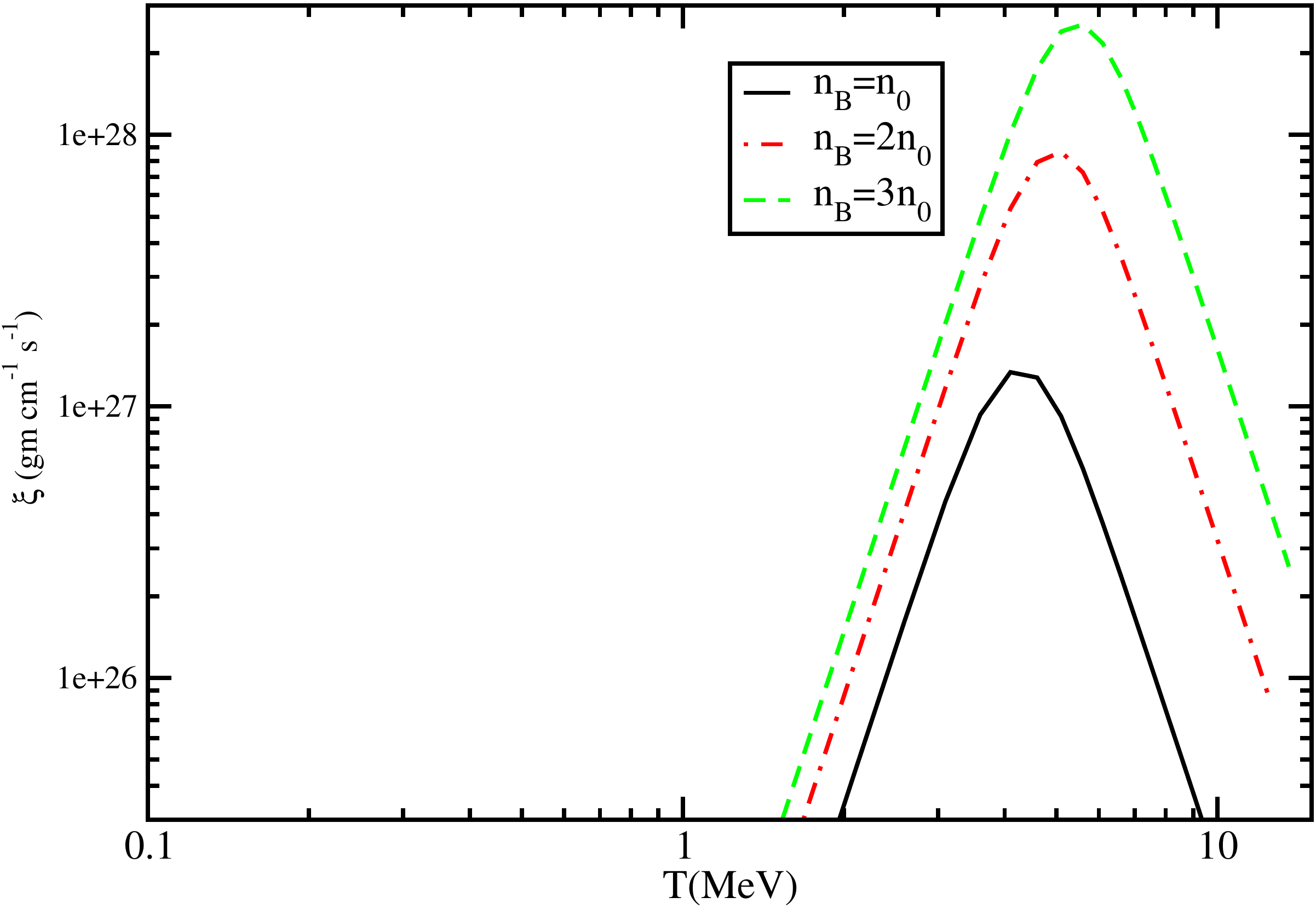}~~~~~~\includegraphics[width=0.50\textwidth,clip=]{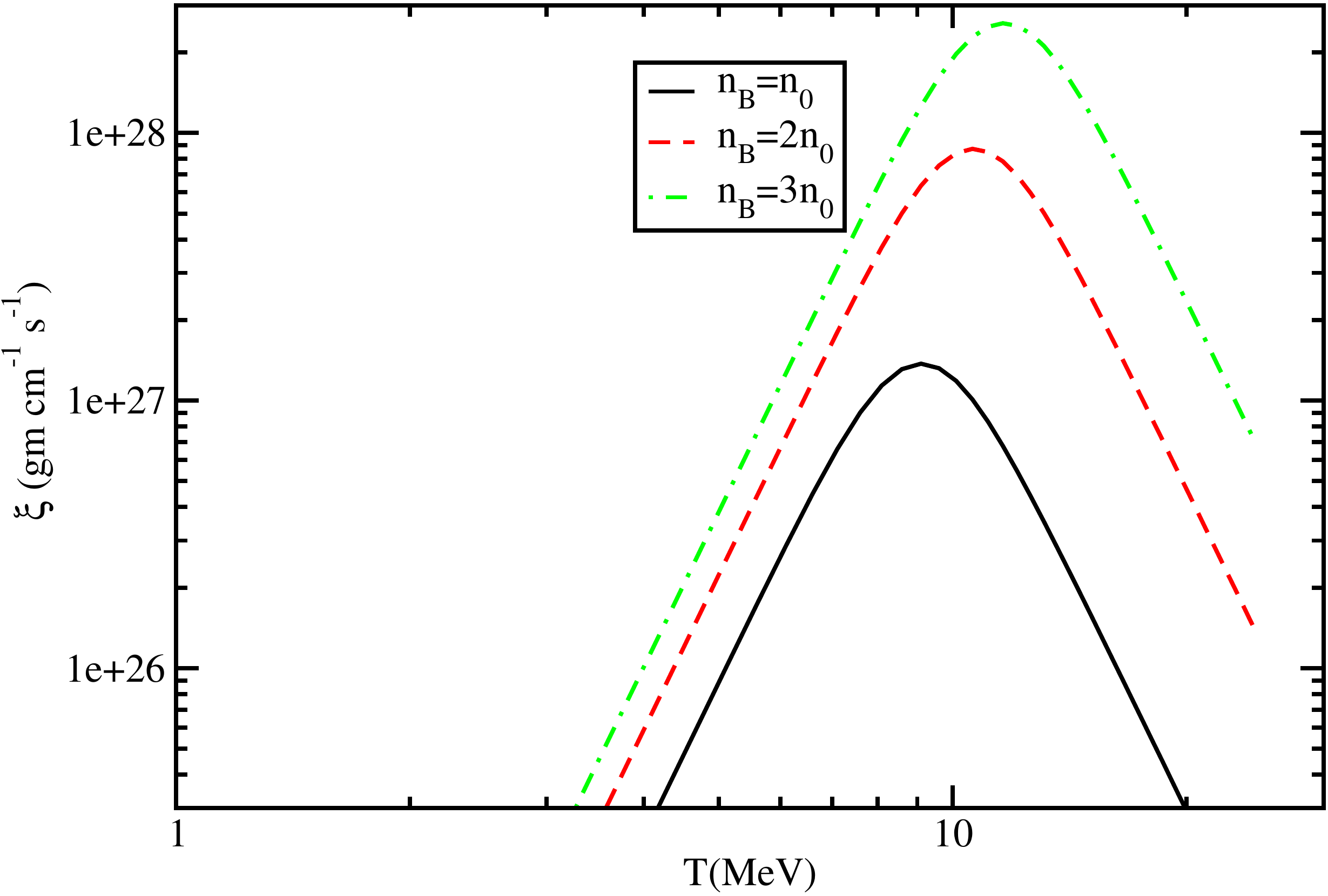}
 \caption{Left: Temperature variation of bulk viscosity for various density for neutrino-transparent matter. Right: Temperature variation of bulk viscosity for various density for neutrino-trapped matter}
 \label{temp_dep}
 \end{center}
 \end{figure}

To quantify the amount of dissipation due to bulk viscosity in the medium we first estimate the susceptibilities of the medium.  We compute the susceptibilities        in  free Fermi gas  model of neutron star. In this model the system is made up of noninteracting neutrons, protons and electrons \cite{2004AmJPh..72..892}. We calculate  the  susceptibilities considering beta equilibrium and  charge neutrality condition to obtain , $B=\frac{4m_N^2}{3(3\pi^2)^{\frac{1}{3}}n^{\frac{4}{3}}}$ and
$C=\frac{(3\pi^2n)^{\frac{2}{3}}}{6m_N}$. This has been obtained after expanding in the powers of $n_B/m_N^3$ ($n_B$ the total baryon density, $m_N$ mass of neutron). In the left panel of Fig. (\ref{density_dep})  we present the density variation of $B$ and $C$ in the medium at zero temperature. $C$ increases with density whereas $B$ decreases with density. From the plot it is evident that the variation of $C$ with density is much prominent in contrast to the variation of $B$ with density. From the right panel of Fig.(\ref{density_dep}) the density dependence of $\xi$ can be observed. Calculation of $\xi$ reveals that interaction rate has very little density dependence whereas the susceptibilities are density dependent only. Hence, the nature of the $\xi$ plots with density is equivalent to the combined nature of both $B$ and $C$. In Fig.(\ref{temp_dep}) we present the temperature  dependence of the bulk viscosity coefficient. In the left panel of Fig.(\ref{temp_dep}) temperature dependence of $\xi$ in neutrino transparent matter has been presented. The plots are for different baryonic densities $n_B=n_0$, $2n_0$, $3n_0$. In the right panel we  plot the same for neutrino-trapped matter. From both the plots it can be seen that bulk viscosity shows resonant behaviour and  the maximum of the curve appears  at different temperatures. For the neutrino transparent medium the maximum  appears at  $T\sim 4$MeV and for the neutrino trapped matter this happens at $T\sim 9$MeV. From both the plots one can infer that incorporating neutrino in the interaction rate changes the position of the maximum of the curves but does not change the maxima of the curve. In the same plot with change in density  the maxima of the curve increases since the height of the maxima is dependent on equation of state.
 For both temperature and density dependent plots we consider the frequency of density oscillation as $8.4$ kHz. The dissipation time scale is calculated considering $\tau=K n_B t^2_{dns}/(36\pi^2\xi_{max})$, $\tau\sim 50$ ms if we consider $K$ is the nuclear compressibility $\sim 250$ MeV, $t_{dns}\sim 1/\omega\sim 1ms$ and $\xi_{max}$ is the maximum of the bulk viscosity.  
\subsection{Discussion}
In this paper we have computed bulk viscosity coefficient of neutrino trapped baryonic matter in high temperature high density plasma.  For the neutron stars when the temperature is high enough $T\sim 5$ MeV mean free path of neutrinos remain small in comparison to the radius of the neutron star. Hence, we incorporate neutrinos in the medium along with neutrons, protons and electrons.  Bulk viscosity coefficient is significantly dependent upon the underlying equation of state of the medium.  We have not considered neutrinos in the equation of state. Apart from equation of state the viscosity coefficient also requires the rate of interaction of constituent particles in the medium. We have calculated the phase space of interaction rate for the MURCA process considering neutrinos in the medium.  The calculation considers explicit beta non-equilibration and does not consider the small amplitude oscillation approximation. The resonant behaviour of the bulk viscosity with temperature has its maximum at $T\sim 9$ MeV and the time scale during which  dissipation remains effective is found out to be $\tau\sim 50$ ms. These two indicate that neutrino trapped MURCA process could become a possible reaction for bulk viscous  dissipation in BNS merger since the maximum temperature attained in BNS merger is of the order of $T\sim 10$ MeV and survival time period of the merged object is millisecond. We will report more realistic calculation of bulk viscous dissipation in BNS merger soon with a comparative study of $\xi$ in heavy ion collision and BNS merger.   

 \section{ Study of  jet fragmentation functions at RHIC and LHC energies using the JETSCAPE framework}
\author{Vaishnavi Desai}	

\bigskip

\begin{abstract}
	Jet-medium interactions in the Quark-Gluon Plasma (QGP) created in high-energy heavy-ion collisions not only reduces the total energy of the reconstructed jets but also change the energy and momentum distributions among the jet constituents. 
	This work focuses on the modification of jet fragmentation function in relativistic heavy-ion collisions.
	Using the JETSCAPE framework, events produced in Au-Au collisions at  $\sqrt{ s_{NN}}$ = 200 GeV and Pb-Pb collisions at  $\sqrt{ s_{NN}}$ = 5.02 TeV are investigated to explore the dependence of modifications based on centrality and in combination with different energy loss modules such as 
	MATTER and LBT for partons with high and low virtuality respectively. The JETSCAPE framework is a modular and versatile Monte-Carlo event generation tool for the simulation of high energy nuclear collisions. Jet fragmentation function results based on MATTER and LBT indicate medium-induced
modifications in heavy-ion collisions.
\end{abstract}
%

%
%
\subsection{Introduction}
%
It is now widely accepted that jet quenching is a multi-stage phenomenon. \cite{1_Vaishnavi,2_Vaishnavi,3_Vaishnavi} Simulations of high energy heavy-ion collisions requires a unified framework that implements all stages of heavy ion collisions namely: 
initial state hard scattering, expansion of QGP, hadronization as well as the production of hard partons, their propagation, interaction with the dense medium and fragmentation into jets. 
To compare with high-statistics, event-by-event experimental data, requires a modular and extendable event generator, with state-of-the-art components modeling each aspect of the collision. The Jet Energy-loss Tomography with a Statistically and Computationally Advanced Program Envelope (JETSCAPE)\cite{4_Vaishnavi,5_Vaishnavi} is such a framework which is used in this work to study jet fragmentation function.

Pb-Pb and pp data at  $\sqrt{ s_{NN}}$ = 5.02 TeV recorded by the ATLAS collaboration\cite{6_Vaishnavi} and Au-Au data at $\sqrt{ s_{NN}}$ = 200 GeV recorded by the STAR collaboration\cite{6_Vaishnavi} are used to compare with the studies of jet fragmentation function. Centrality dependent comparisons between different energy loss modules, Modular All Twist Transverse-scattering Elastic-drag and Radiation (MATTER)\cite{6_Vaishnavi} and Linear Boltzmann Transport (LBT)\cite{7_Vaishnavi,11_Rohan} are carried out to explore the effect of each approach in the full history of parton evolution in heavy-ion collisions.
\subsection{Multi-stage Jet Evolution in JETSCAPE}
Simulation of jet events in Pb-Pb and Au-Au are performed within JETSCAPE (JS) framework. Initial hard partons generated by PYTHIA 8 \cite{11_Rohan} are fed into MATTER  for modeling high virtuality evolution, $Q^{2}>>\sqrt{\hat{q}E}$, 
and propogated with a virtuality-ordered splittings in the medium.
If the virtuality of a given parton drops below a specified separation scale,  $Q_{\mathrm{0}}$ , the low-virtuality energy loss modules, one of the LBT, MARTINI,\cite{12_Vaishnavi} or AdS/CFT\cite{13_Vaishnavi} take the parton over for time-ordered evolution. Switching between different energy loss modules is done independently for each parton. The separation scale  $Q_{\mathrm{0}}$ is set to 2 GeV. Colourless hadronization is performed by the Lund String Model as implemented in PYTHIA 8. The soft products in Au-Au collisions is generated using fluctuating TRENTO \cite{10_Rohan} initial conditions evolved hydrodynamically using the (3+1)D MUSIC\cite{11_Rohan,13_Rohan,14_Rohan} viscous hydrodynamic module.
\subsection{Jet Fragmentation Function}
The longitudinal momentum distribution of charged particles inside a reconstructed jets is calculated using the jet fragmentation function. It is measured as a function of both the charged-particle transverse momentum, $p_{\mathrm{T}}$ , and 
$z=\frac{p^{\mathrm{ch}}_{\mathrm{T}}} {p^{\mathrm{jet}}_{\mathrm{T}}}$  is the transverse momentum fraction between the charged particle and the jet.
The fragmentation function is calculated as
\begin{equation}
D(z) = \frac{1}{N_{\mathrm{jet}}} \frac{dN_{\mathrm{ch}}}{dz}
\end{equation}
and
\begin{equation}
D(p_{T}) = \frac{1}{N_{\mathrm{jet}}} \frac{dN_{\mathrm{ch}}}{dp_{T}}
\end{equation}
where  $N_{\mathrm{ch}}$ is the number of charged particles inside jets,  $N_{\mathrm{jet}}$ is the number of jets under consideration. 
D(z) and D$(p_{T})$ distributions are closely related to each other, differing, primarily, in the normalization by $p^{\mathrm{jet}}_{\mathrm{T}}$ in the definition of z. Therefore, a comparison of the modifications of the fragmentation functions as a function of $p^{\mathrm{jet}}_{\mathrm{T}}$
can show whether the size of modifications scales with charged-particle z or with $p_{T}$. The former would be expected for fragmentation effects, and the latter might indicate some scale in the QGP.
%
\subsection{Results and Discussion}
In this work, JETSCAPE results based on energy loss modules MATTER and LBT are discussed. In order to determine the centrality dependence of ratio of fragmentation function  on $p^{jet}_{T}$, the fragmentation functions from three $p^{jet}_{T}$ intervals are compared.
\begin{figure}[h]
\begin{center}
\includegraphics[width=.33\textwidth]{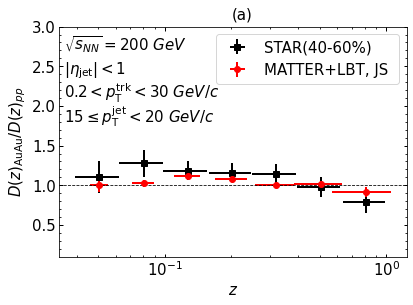} \hfill
 \includegraphics[width=.33\textwidth]{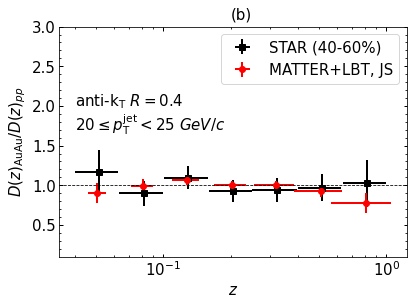}\hfill
\includegraphics[width=.33\textwidth]{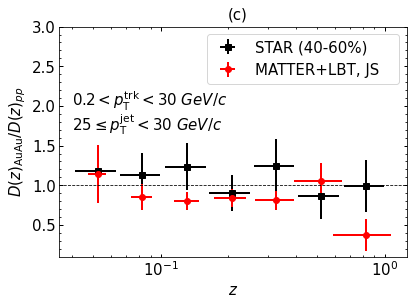}
\includegraphics[width=.33\textwidth]{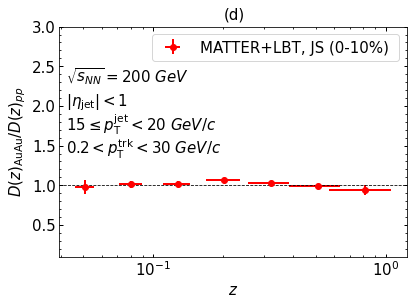}\hfill
\includegraphics[width=.33\textwidth]{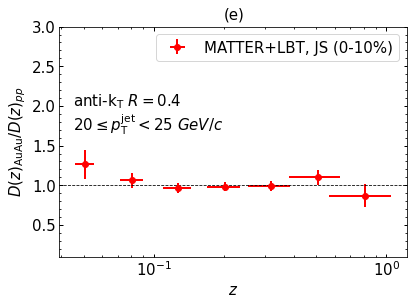}\hfill
\includegraphics[width=.33\textwidth]{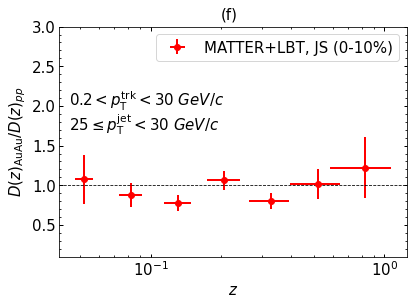}
\caption{(top panel) Shows the ratio of jet fragmentation function between 40-60\% central Au-Au collisions for three  $p^{jet}_{T}$ ranges for JETSCAPE calculations compared with STAR experimental data\cite{4_Vaishnavi}.
(bottom panel) Shows the ratio of jet fragmentation function between 0-10\% central Au-Au collisions for three  $p^{jet}_{T}$ ranges for JETSCAPE calculations.}
\end{center}
\label{figure1}
\end{figure}
Fig.43 (top panel) shows the ratio of jet fragmentation functions for 40-60\% Au-Au collisions to those in pp collisions. These calculations are extended to the most central 0-10\% Au-Au collisions as shown in Fig. 43 (bottom panel). The JETSCAPE results are consistent with STAR experimental data for mid-central collisions at 40-60\% as shown in Fig. 43 (a) and (b). In Fig. 43(c), MATTER + LBT under predicts the experimental data.  In Fig. 43(f), for 0-10\% centrality class, MATTER + LBT shows marginal suppression at low z, the ratio is close to unity at intermediate z and small enhancement at high z values. Fragmentation function shows no modifications in Au-Au collisions at 0-10\% and 40-60\% centrality classes for $15 \leq p^{\mathrm{jet}}_{\mathrm{T}} < 20 $ GeV and $20 \leq p^{\mathrm{jet}}_{\mathrm{T}} < 25 $ GeV as seen from Fig. 43 (a),(b),(d) and (e). 
%
%
Fig. 44 shows the ratio of the jet fragmentation function as a function of z and $p_{T}$ for Pb-Pb and pp collisions at 0-10\% (top panel) and 30-40\% (bottom panel) centrality classes respectively. 
In Fig. 44(a), no apparent modification of fragmentation function is seen. This could be attributed to lack of recoils in MATTER + LBT. In Fig. 44(b), JS results are in agreement with ATLAS data for $p_{T} > $ 10 GeV, this indicates that the size of modification scales with $p_{T}$, while it exhibits much smaller dependence on D(z) ratio. Enhancement at high $p_{T}$ region suggests re-distribution of energy inside the jet cone. In Fig. 44(c), at low-z region, JS under predicts the data whereas JS results are consistent in intermediate z region. In high z-region,  JS results are not consistent, partly due to statistical fluctuations and uncertainty. In Fig. 44(d), JS results shows consistency with the ATLAS data at intermediate to high $p_{T}$ region.
%
\begin{figure}[h]
   \begin{center}
    \includegraphics[width=.41\textwidth]{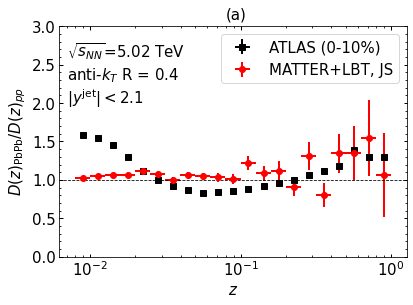}\hfill
    \includegraphics[width=.41\textwidth]{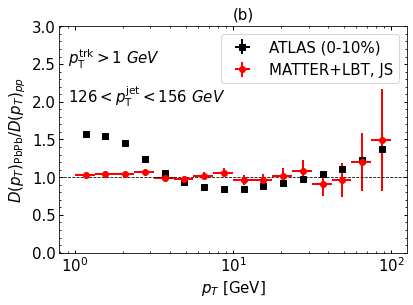}\hfill
    \includegraphics[width=.41\textwidth]{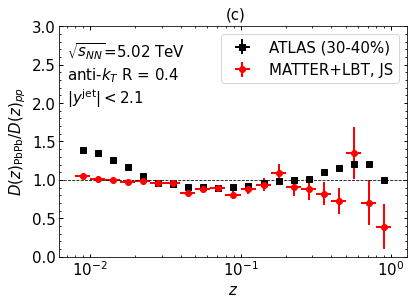}\hfill
   \includegraphics[width=.41\textwidth]{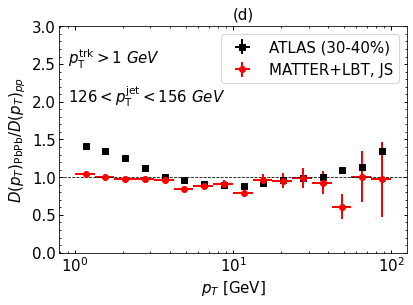}
\caption{The jet fragmentation distribution ratio of Pb-Pb to pp measured using the JS with MATTER + LBT at 0-10\% centrality (top panel) and 30-40\%  centrality (bottom panel) compared with the ATLAS experimental data\cite{3_Vaishnavi}.}
\end{center}
\end{figure}
%
%

\section{First-order stable and causal hydrodynamics from kinetic theory}
\author{Rajesh Biswas}	

\bigskip

\begin{abstract}
We derived a stable and causal relativistic first-order hydrodynamics from the relativistic Boltzmann equation. General hydrodynamic frame are introduced to incorporate the arbitrariness of hydrodynamics fields. The system interactions are included by momentum-dependent relaxation time in the relativistic Boltzmann equation. The hold the causality and stability of the first-order hydrodynamics, the system interactions play a crucial role along with the general frame.
\end{abstract}

\subsection{Introduction}

Relativistic Hydrodynamics is an effective theory; it
has been proved to be reasonably successful in describing
various systems' collective (long wavelength) behavior
ranging from relativistic heavy-ion collision to condensed matter system. Relativistic hydrodynamical theory can be expressed as the gradient expansion of the hydrodynamic variable around the local equilibrium state. The theory with zeroth-order gradients is called ideal hydrodynamics, and with first-order gradients is called first-order hydrodynamics, and so on. Relativistic Navier-Stokes equations show superluminal signal propagation as well as instability~\cite{Hiscock:1983zz,Hiscock:1987zz}. The second-order hydrodynamic theory is introduced to resolve the stability and causality problem of NS equations~\cite{Muller:1967zza,Israel:1979wp}. Recently, the solution to the stability and causality of first-order hydrodynamics have proposed by Bemfica, Disconzi, Noronha, and Kovtun (BDNK) theory~\cite{Bemfica:2017wps,Bemfica:2019knx,Kovtun:2019hdm} by introducing a general hydrodynamic frame. 

Here we derive a first-order stable and causal theory from the underlying microscopic kinetic theory for a general frame or general matching condition. We have calculated the transport coefficients that appeared in the constitutive relations. By studying the analysis of the modes in a linear regime, we show that microscopic interaction plays an important role along with general matching conditions to get a stable and causal, first-order theory. We have calculate the transport coefficients that appeared in the constitutive relations. By studying the modes analysis in linear regime, we show that to get a stable and causal, first-order theory microscopic interaction is play a important role  along with general matching conditions.

\subsection{General form of first-order field correction from Boltzmann equation}

To estimate the entire distribution function $f (x, p)~(=f_p)$, we starting from the relativistic Boltzmann equation 
\begin{equation}
p^{\mu}\partial_{\mu}f_{p}=C[f]=-\mathcal{L}[\phi]~,
\label{eq:RBE}
\end{equation}
here, $p^{\mu}$ is the particle four-momentum, $C[f]$ is the collision kernel. The distribution function can be written as $f_{p}= f_{p}^{0}+f_{p}^{0}\left(1\pm f_{p}^{0}\right)\phi_{p}$, here $f_{p}^{0}$ is the equilibrium distribution function and $\phi_{p}$ is the out-of-equilibrium deviation. The collision operator in linear order becomes $C[f]=-{\cal{L}}[\phi]=-\int d\Gamma_{p_1}d\Gamma_{p'}d\Gamma_{p'_1}f^{(0)}f_1^{(0)}(1\pm f'^{(0)})(1\pm f_{1}'^{(0)})\{\phi+\phi_1-\phi'-\phi'_1\}\mathcal{W}(p'p'_1|pp_1)$, with $d\Gamma_p=\frac{d^3p}{(2\pi)^3p^0}$ and $\mathcal{W}$ is the transition rate that depends on the cross-section of the interactions.

The out-of-equilibrium distribution function is expressed by the linear combination of the gradients of hydrodynamic variables with appropriate tensor combinations. To extract out the values of unknown coefficients, we expand it on a polynomial basis as Ref.~\cite{Biswas:2022cla}.
For the properties of linearized collision operator ($\mathcal{L}[1]$ and $\mathcal{L}[p^{\mu}]$), one can't determine the 
first two coefficients from scalar sector and one from vector sector. These coefficient are called the homogeneous solution and the rest of the coefficients are called inhomogeneous or interaction solutions. Here we introduced the general matching conditions
\begin{eqnarray}
\int dF_p\tilde{E}_{p}^{i}\phi=0,~\int dF_p\tilde{E}_{p}^{j}\phi=0,~\int dF_p\tilde{E}_{p}^{k}\tilde{p}^{\langle\mu\rangle}\phi=0~,
\end{eqnarray}
with $i\neq j$, $i,j,k$ are non-negative integers and $\tilde{E}_{p}=\frac{u\cdot p}{T}$. Employing the general matching conditions one can calculate  the homogeneous solutions in terms of inhomogeneous solutions.

We employ the relativistic Boltzmann equation to extract the inhomogeneous part of the out-of-equilibrium distribution function. Here we choose the Anderson-Witting type collision kernel~\cite{Relax} with momentum-dependent relaxation-time~(MDRTA). The relaxation time is expressed as $\tau_{R}=\tau_{R}^{0}\tilde{E}_{p}^{\Lambda}$, here $\tau_{R}^{0}$ is momentum independent~\cite{Dusling:2009df,Rocha:2021zcw,Mitra:2020gdk,Mitra:2021owk,Dash:2021ibx,Rocha:2022ind}. To ensure the microscopic conservation of energy-momentum and particle four-current, we propose a linearized collision operator~\cite{Rocha:2021zcw}
\begin{align}
{\cal{L}}_{\text{MDRTA}}[\phi]
=&\frac{\left(p\cdot u\right)}{\tau_R}f^{(0)}(1\pm f^{(0)})\bigg[\phi-\frac{\langle\frac{\tilde{E}_p}{\tau_R}\tilde{E}_p^2\rangle\langle\frac{\tilde{E}_p}{\tau_R}\phi\rangle-\langle\frac{\tilde{E}_p}{\tau_R}\tilde{E}_p\rangle\langle\frac{\tilde{E}_p}{\tau_R}\phi\tilde{E}_p\rangle}
{\langle\frac{\tilde{E}_p}{\tau_R}\rangle\langle\frac{\tilde{E}_p}{\tau_R}\tilde{E}_p^2\rangle-\langle\frac{\tilde{E}_p}{\tau_R}\tilde{E}_p\rangle^2}\nonumber\\
&-\tilde{E}_p\frac{\langle\frac{\tilde{E}_p}{\tau_R}\tilde{E}_p\rangle\langle\frac{\tilde{E}_p}{\tau_R}\phi\rangle-\langle\frac{\tilde{E}_p}{\tau_R}\rangle\langle\frac{\tilde{E}_p}{\tau_R}\phi\tilde{E}_p\rangle}
{\langle\frac{\tilde{E}_p}{\tau_R}\tilde{E}_p\rangle^2-\langle\frac{\tilde{E}_p}{\tau_R}\rangle\langle\frac{\tilde{E}_p}{\tau_R}\tilde{E}_p^2\rangle}-\tilde{p}_{\langle\nu\rangle}\frac{\langle\frac{\tilde{E}_p}{\tau_R}\phi \tilde{p}^{\langle\nu\rangle}\rangle}
{\frac{1}{3}\langle\frac{\tilde{E}_p}{\tau_R}\tilde{p}^{\langle\mu\rangle}\tilde{p}_{\langle\mu\rangle}\rangle}\bigg]~.
\label{MDRTAcoll}
\end{align} 

From the relativistic Boltzmann equation Eq.~\eqref{eq:RBE}, the first-order out-of-equilibrium distribution function written as~\cite{Rocha:2022ind}
\begin{align}
\phi_{\text{int}}^{(1)}=&
-\tau_R^0\tilde{E}_{p}^{\Lambda-1}\bigg[\tilde{E}_{p}^2\frac{\dot{T}}{T}+\tilde{E}_{p} \dot{\tilde{\mu}}+\left(\frac{\tilde{E}_{p}^2}{3}-\frac{z^2}{3}\right)(\partial\cdot u)
+\tilde{E}_{p}\tilde{p}^{\langle\mu\rangle}\left(\frac{\nabla_{\mu}T}{T}-\dot{u}_{\mu}\right)
\nonumber
\\&+\tilde{p}^{\langle\mu\rangle}\nabla_{\mu}\tilde{\mu}-\tilde{p}^{\langle\mu}\tilde{p}^{\nu\rangle}\sigma_{\mu\nu}\bigg],
\label{eq:phi_int1}
\end{align}
where $D=u^{\mu}\partial_{\mu}$, $\dot{A}=DA$, $\nabla^{\mu}=\Delta^{\mu\nu}\partial_{\nu}$, and $\sigma_{\mu\nu}=\nabla_{\langle{\mu}}u_{\nu\rangle}$.

Using the entire out-of-equilibrium distribution function upto first order, the dissipative currents can be written as~\cite{Bemfica:2017wps,Bemfica:2019knx,Kovtun:2019hdm}
\begin{align}
&\delta n^{(1)}, \delta \epsilon^{(1)}, \delta P^{(1)}=\nu_{1},\varepsilon_{1},\pi_{1} \frac{\dot{T}}{T} + \nu_{2} ,\varepsilon_{2},\pi_{2}\left(\partial \cdot u\right)+\nu_{3},\varepsilon_{3},\pi_{3} \dot{\tilde{\mu}}~,
\label{sc}\\
&W^{(1)\mu},V^{(1)\mu} = \theta_{1},\gamma_{1}\left[\frac{\nabla^{\mu}T}{T} - \dot{u}^{\mu}\right] +\theta_{3},\gamma_{3} \nabla^{\mu}\tilde{\mu}~.
\label{vec}
\end{align}
The details expression of the transport coefficients~($\epsilon_{i}, \nu_{i}, \theta_{i}$, etc.) given in Ref.~\cite{Biswas:2022cla}
\subsection{Resuts}
After derive the transport coefficients we studied the stability ad causality of theory in linear regime. We linearizing the conservation equations for small perturbations of fluid variables around their equilibrium $\epsilon(t,x)=\epsilon_0+\delta\epsilon(t,x),~n=n_0+\delta n(t,x),,P(t,x)=P_0+\delta P(t,x),~u^{\mu}(t,x)=(1,\vec{0})+\delta u^{\mu}(t,x)$ and find out the different modes.

For the shear channel the  group velocity of the perturbation turns out to be $v_g=\sqrt{\eta/\theta}$. Here $\eta=\tau_R^0T^2K_{\Lambda-1}/2$  and we define $\theta=-\theta_1=\tau_R^0 T^2 \left( J_{\Lambda+1}+\frac{\epsilon_0+P_0}{T^2}\frac{J_{k+\Lambda}}{J_k}\right)$ with  $ J_n=\int dF_p \tilde{p}^{\langle\mu\rangle}\tilde{p}^{\langle\nu\rangle}\tilde{E}_{p}^n$ and
$\Delta^{\alpha\beta\mu\nu}K_n=\int dF_p\tilde{p}^{\langle\mu}\tilde{p}^{\nu\rangle}\tilde{p}^{\langle\alpha}\tilde{p}^{\beta\rangle}\tilde{E}_{p}^n~$. So the causality criteria of the shear channel is $v_{g}<1$ and the stability criteria turns out to be $\theta >0$. 

Using Routh-Hurwitz criteria, we find the following conditions for stability of the non-hydro modes,
\begin{align}
&A_6>0~,~A_5>0~,~A_3^0>0~,
B_2=(A_4^0A_5-A_3^0A_6)/A_5>0~,
\label{stab3}
\end{align}
here $A_{i}$'s are the are function of the transport coefficients~\cite{Biswas:2022cla} appeared in Eqs.~\eqref{sc} and \eqref{vec}.  

\begin{figure}[h]
	\begin{center}
		\includegraphics[width=0.55\textwidth]{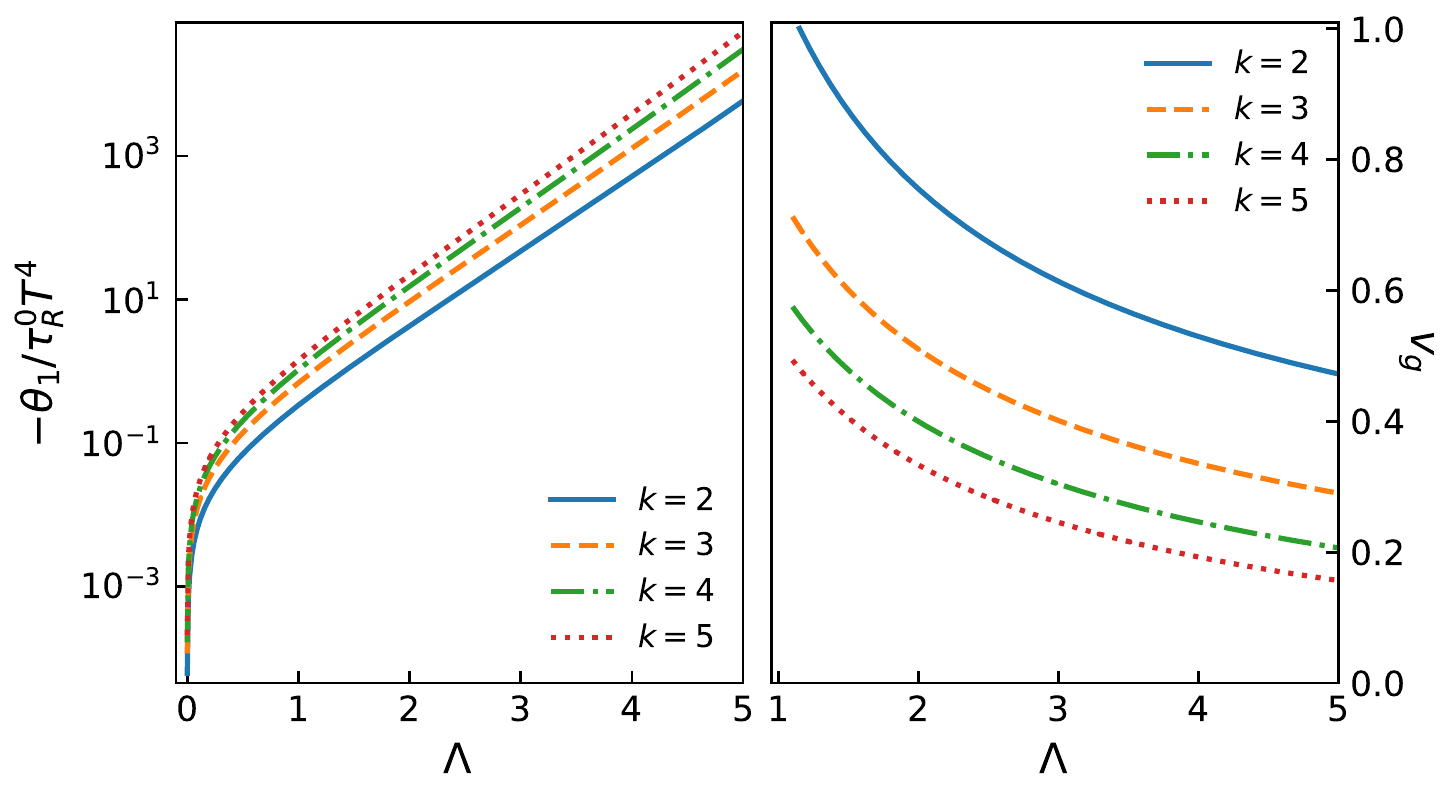}
		\includegraphics[width=0.4\textwidth]{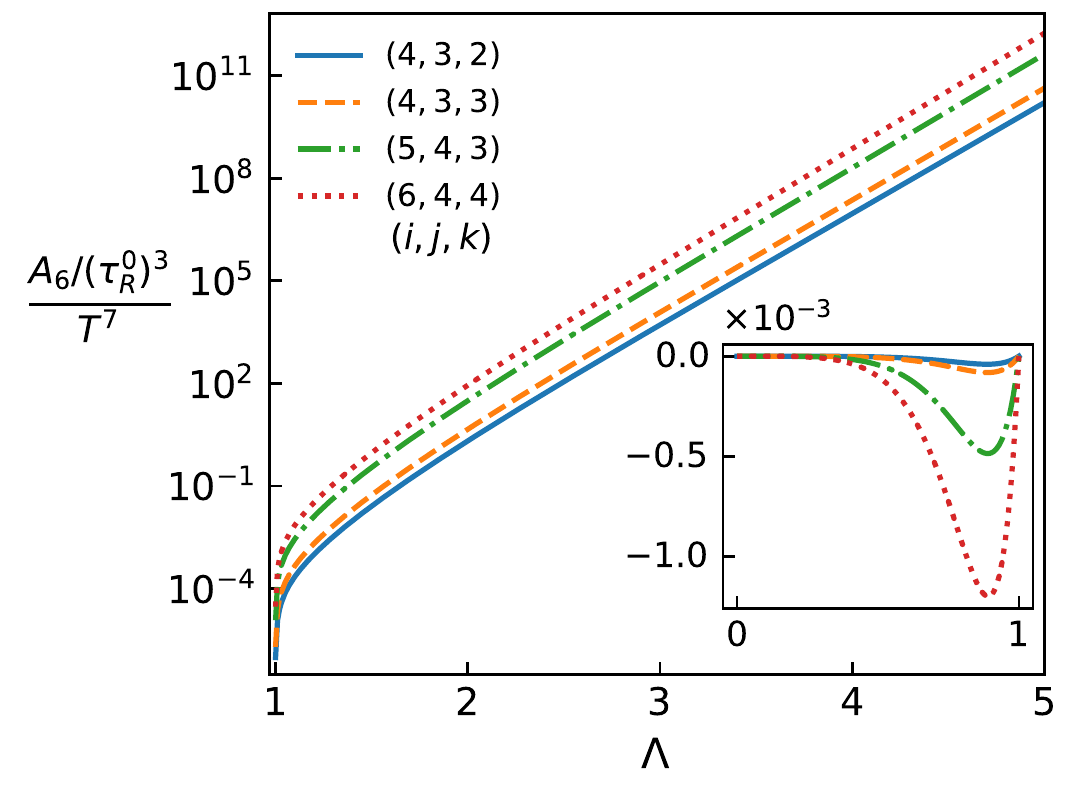}
		\caption{(Color online) The stability and causality criteria. Left panel two plots is for shear channel and right panel plot is for sound channel.}\label{Fig:eta}
	\end{center}
\end{figure}
From the first two plots of Fig.~\eqref{Fig:eta} it is clear that the shear channel is stable with different vector matching conditions~($k$). However, the group velocity of the shear channel becomes superluminal for $\Lambda<1$. Also, from the right plot, the stability criteria for sound channel~($A_6$) are not held. 

\subsection{Summary and conclusion}
In this work, we derive a first-order, relativistic stable, and causal hydrodynamic theory from the relativistic Boltzmann equation in general matching conditions. We propose a collision operator for MDRTA to obey the microscopic conservation equations. We have shown that to get a first-order stable and causal theory,
not only the general frame but also the system interactions play a crucial role. The conventional momentum-independent RTA leads to superluminal signal propagation in the general frame.

\section{Hydrodynamical Attractor and Signals from Quark-Gluon Plasma}
\author{Lakshmi J. Naik, Sunil Jaiswal, K. Sreelakshmi, Amaresh Jaiswal, and V. Sreekanth }	

\bigskip

\begin{abstract}
We study 
the analytical attractor
solutions of  third-order viscous hydrodynamics by considering thermal particle production from heavy-ion collisions within the longitudinal boost-invariant expansion. Using these analytical solutions, the allowed initial
states are constrained by demanding positivity and reality of energy density throughout the evolution. 
Further, we calculate the thermal dilepton spectra within the framework of hydrodynamic attractors. 
It has been observed that the evolution corresponding to attractor solution leads to
maximum production of thermal particles.
\end{abstract}


\subsection{Introduction}

Quark-Gluon Plasma (QGP), state of hot and dense deconfined nuclear matter has been realized in the 
relativistic heavy ion collision experiments at RHIC and LHC. Relativistic hydrodynamics has been extremely 
successful in describing the QGP created in these experiments and this has led to the development of
several causal dissipative theories of hydrodynamics. The formulation of relativistic viscous 
hydrodynamics usually proceeds with the assumption that system is in local thermodynamic equilibrium. 
However, hydrodynamical simulations have showed unexpected success in explaining the flow data from small collisional 
systems (far from equilibrium) and this has generated much interest in the foundational aspects of causal relativistic hydrodynamics.
We investigate an important aspect 
which manifests in 
causal boost invariant relativistic viscous hydrodynamics, the {\it hydrodynamical attractor}\cite{Heller:2015dha,Jaiswal:2019cju} 
and 
study its phenomenological consequences through thermal particle production\cite{Naik:2021yph}. We consider thermal dileptons and
photons, since their interaction with quark-gluon matter is less and can be detected easily. We calculate the
thermal particle production rate in the presence of first order Chapman-Enskog type viscous correction. The spectra
of particles are obtained by employing the recently developed analytical solutions of higher order
dissipative hydrodynamics under Bj\"orken expansion.

\subsection{Hydrodynamical attractor}
Considering longitudinal boost-invariant Bj\"orken flow\cite{Bjorken:1982qr} in the Milne coordinates $x^\mu = (\tau, r, \varphi, \eta_s)$, 
the evolution equations
for energy density and shear stress tensor can be written in the generic form\cite{Jaiswal:2019cju}:
\begin{eqnarray}
\frac{d\epsilon}{d\tau} &=& -\frac{1}{\tau}\left(\frac{4}{3}\epsilon -\pi\right), 
\label{bde1} \nonumber \\
  \frac{d\pi}{d\tau} &=& - \frac{\pi}{\tau_\pi} + \frac{1}{\tau}\left[\frac{4}{3}\beta_\pi 
  - \left( \lambda + \frac{4}{3} \right) \pi - \chi\frac{\pi^2}{\beta_\pi}\right],
  \label{bde2}
\end{eqnarray}
where $\pi \equiv -\tau^2 \pi^{\eta_s\eta_s}$, with $\tau$ and $\eta_s$ being the proper time and space-time rapidity
of the system. The 
coefficients $\gamma$ and $\chi$ appearing in the above equations are tabulated in Ref.~\refcite{Jaiswal:2019cju} for
the three different causal theories considered here.  In the present work, we consider the coefficients corresponding to third-order theory as it shows better agreement with the exact solution of kinetic theory~\cite{Jaiswal:2013vta}.
Also, since we consider a conformal system, 
we have $\beta_\pi = 4P/5$. 
The above equations can be solved analytically by considering
certain approximations for $\tau_\pi$. Analytical solutions of these equations were obtained in Ref.~\refcite{Jaiswal:2019cju} for a 
conformal system and by considering 
the relation $T\tau_\pi = 5(\eta/s) = \textrm{const.,}$  
for three
cases : $T$ is either a constant or has proper time evolution following ideal or Navier-Stokes solutions. 
These analytical solutions can be written in a generic form as :
\begin{eqnarray}
\bar{\pi}(\bar{\tau})&= & \frac{(k{+}m{+}\frac{1}{2}) M_{k+1,m}(w) 
- \alpha \, W_{k+1,m}(w)}{\gamma |\Lambda| \left[ M_{k,m}(w)
+\alpha \, W_{k,m}(w) \right]},\label{generic_pibar}\nonumber \\ 
\epsilon(\bar{\tau})&= & \epsilon_0 \left(\frac{w_{0}}{w}\right)^{\!\frac{4}{3} \left(|\Lambda|-\frac{k}{\gamma}\right)}e^{-\frac{2 }{3 \gamma} \left( w-{w_0} \right)} \left(\frac{M_{k,m}(w) 
+ \alpha \, W_{k,m}(w) }{M_{k,m}(w_0) + \alpha \, W_{k,m}(w_0)}\right)^{\frac{4}{3 \gamma}}, \label{generic_energy}
\end{eqnarray}
where $\bar{\pi} \equiv \pi/(\epsilon + P)$ is the normalized shear stress tensor. 
Here $M_{k,m}(w)$ and $W_{k,m}(w)$ are the Whittaker functions and $\alpha$ is the integration constant which encodes
the initial energy density $\epsilon_0$ and normalized shear stress tensor $\bar{\pi}_0$. 
The parameters of 
Whittaker functions appearing in the above equations are tabulated in 
Ref.~\refcite{Jaiswal:2019cju} for the three cases of $\tau_\pi$. 
From the above analytical solutions, the hydrodynamic attractor solution can be obtained for
$\alpha=0$ and repulsor curve correspond to $\alpha = \infty$. 

\subsection{Thermal particles from expanding QGP}

Thermal particles, such as dileptons and photons are emitted throughout the expansion of the QGP. 
The major source of dileptons in a QGP medium is from the $q\bar{q}-$annihilation process. 
The rate of dilepton production for this process is given by
\begin{equation}
    \frac{dN_{l^{+}l^{-}}}{d^4x d^4p}=g^2 \int  \frac{d^3\textbf{p}_1}{(2\pi)^3} 
 \frac{d^3\textbf{p}_2}{(2\pi)^3} f(E_1,T) f(E_2,T)
 v_{rel} \sigma(M^2)\delta^4(p-p_1-p_2).
\label{dil_rate1}
\end{equation}
$f(E,T)$ represents the viscous modified quark (anti-quark) distribution function and is given by
\begin{equation}
f(E, T)=f_{0}(E, T)\left(1+\frac{\beta}{\beta_{\pi}} 
\frac{p^{\alpha} p^{\beta} \pi_{\alpha \beta}}{2(u \cdot p)}\right),\label{mod_distfn}
\end{equation}
where $f_{0}(E, T) \approx e^{-E/T}$ denotes the ideal part in the Maxwell-Boltzmann limit, $\beta=1/T$ and $\beta_\pi=(\epsilon+P)/5$. 
We have considered the form of viscous correction due to Chapman-Enskog method. 
Substituting Eq.~\eqref{mod_distfn} in Eq.~\eqref{dil_rate1} and keeping the terms 
upto second order in momenta, we can write the total dilepton rate as 
$ \frac{d N_{l^{+}l^{-}}}{d^{4} x d^{4} p}=\frac{d N_{l^{+}l^{-}}^{0}}{d^{4} x d^{4} p}
 +\frac{d N_{l^{+}l^{-}}^{\pi}}{d^{4} x d^{4} p}$, 
where the ideal and viscous contributions respectively are obtained as\cite{Naik:2021yph}
\begin{eqnarray}
\frac{d N_{l^{+}l^{-}}^{0}}{d^{4} x d^{4} p}&=&\frac{1}{2} \frac{M^{2} g^{2} \sigma
\left(M^{2}\right)}{(2 \pi)^{5}} e^{-E / T}, \nonumber\\
\frac{d N_{l^{+}l^{-}}^{\pi}}{d^{4} x d^{4} p}&=&\frac{d N_{l^{+}l^{-}}^{0}}{d^{4} x d^{4} p}
\Bigg\{ \frac{\beta}{2\beta_{\pi}|\mathbf{p}|^{5}}
\Bigg[\frac{E|\mathbf{p}|}{2}\left(2|\mathbf{p}|^{2}-3 M^{2}\right)
+\frac{3M^{4}}{4} \ln \left(\frac{E+|\mathbf{p}|}{E-|\mathbf{p}|}\right)\Bigg]
p^{\alpha} p^{\beta} \pi_{\alpha \beta}\Bigg\}. \nonumber \\
\end{eqnarray}

Similarly, we calculate the photon production rate due to Compton scattering: $q(\bar{q}) g \rightarrow q(\bar{q}) \gamma$ 
and $q \bar{q}$-annihilation $q \bar{q} \rightarrow g \gamma$. The ideal and viscous contributions to photon rate are given as\cite{Naik:2021yph}
\begin{eqnarray}
E \frac{d N_{\gamma}^0}{d^{4} x d^{3} p} &=&  \frac{5}{9} \frac{\alpha_{e} \alpha_{s}}{2 \pi^{2}} f_0(E, T) T^{2} 
 \left[ \ln \left(\frac{12 E}{g^{2} T}\right)
 \!+\! \frac{C_{\textrm{ann}} \!+\! C_{\textrm{Comp}}}{2} \right], \nonumber\\
 E \frac{d N_{\gamma}^{\pi}}{d^{4} x d^{3} p}&=&E \frac{d N_{\gamma}^{0}}{d^{4} x d^{3} p}
\left\{\frac{\beta}{\beta_{\pi}} \frac{p^{\alpha} p^{\beta} \pi_{\alpha \beta}}{2 E}\right\},
\end{eqnarray}
respectively. The constants have the values $C_{\textrm{ann}} = -1.91613$, $C_{\textrm{Comp}} = -0.41613$
and $g=\sqrt{4\pi \alpha_s}$; with $\alpha_s$ denoting the strong coupling constant. 
Thermal particle yields are calculated by numerically integrating the rate equations obtained above over the 
space-time history of the collisions. Considering the Bj\"orken expansion, four-dimensional volume element
becomes $d^4 x = \pi R_A^2 d\eta_s \tau d\tau$, with $R_A$ being the radius of the colliding nuclei. 

\subsection{Results and discussions}
\begin{figure}
\centering
\begin{minipage}{.5\textwidth}
  \centering
  \includegraphics[width=.95\linewidth]{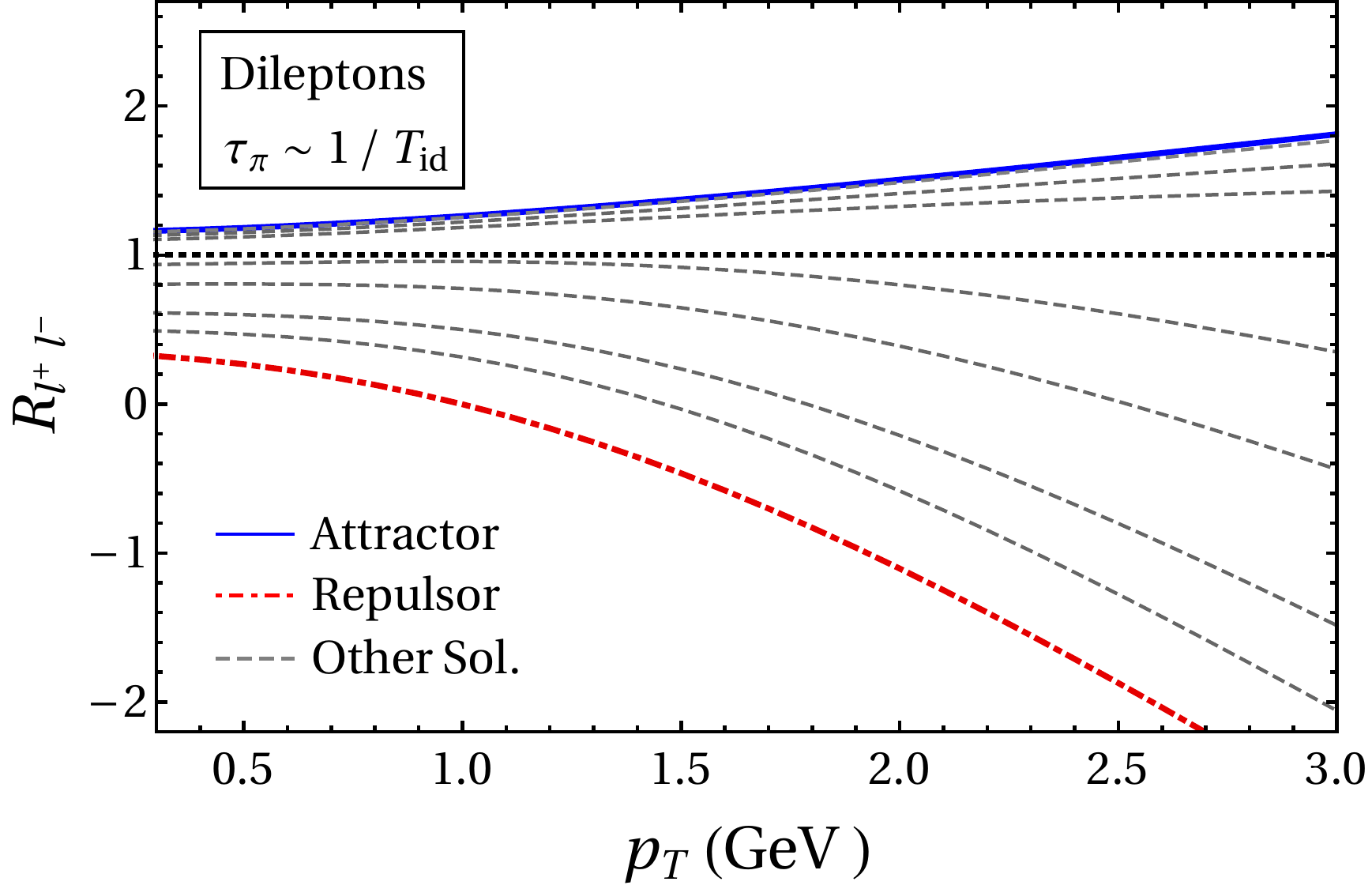}
  \caption[short]{Ratio of viscous to ideal thermal dilepton yields for $\tau_\pi \sim 1/T_{id}$ 
  approximation.}
  \label{ldilepton}
\end{minipage}%
\begin{minipage}{.5\textwidth}
  \centering
  \includegraphics[width=.95\linewidth]{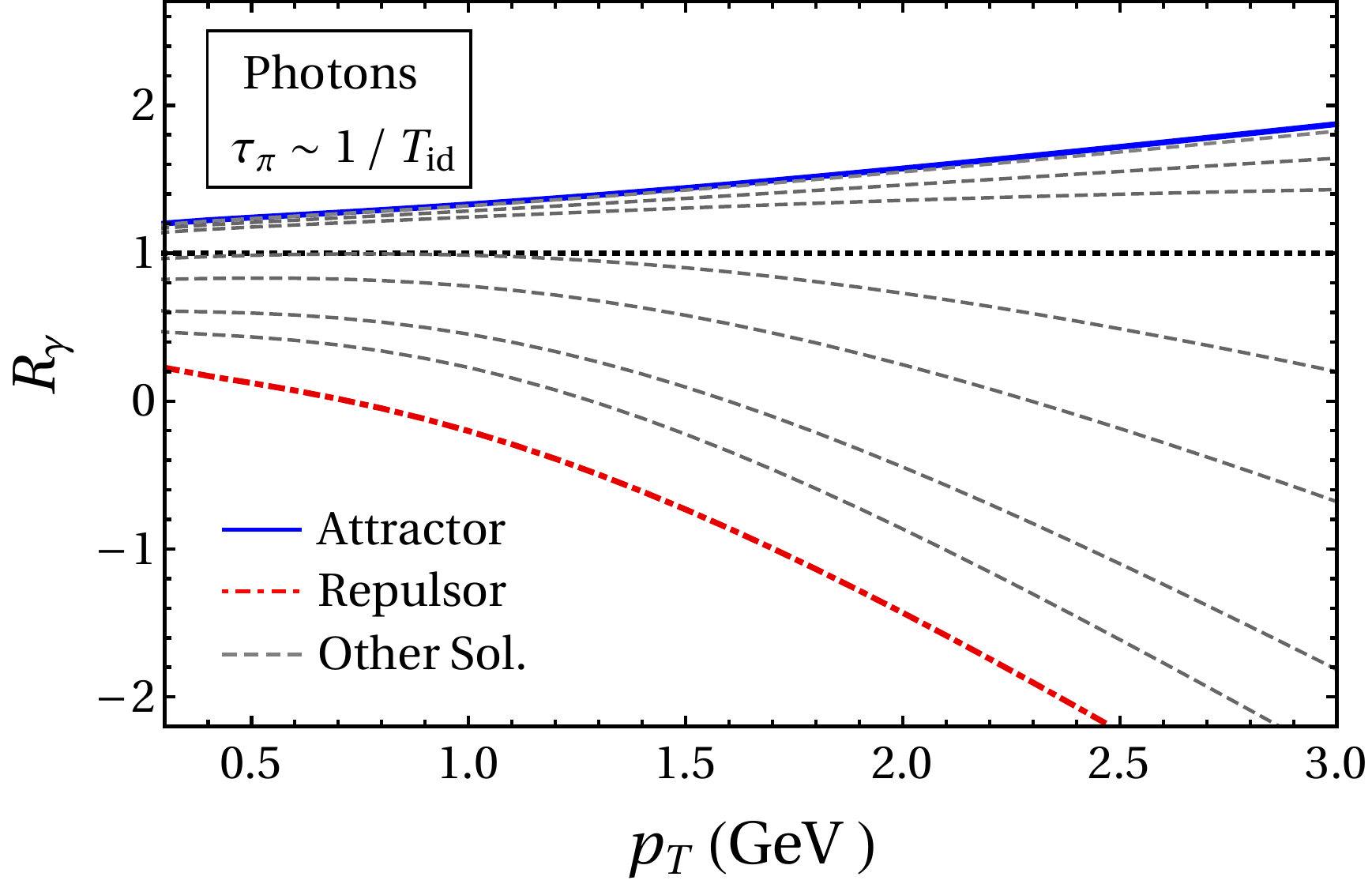}
  \caption[short]{Ratio of viscous to ideal thermal photon spectra for $\tau_\pi \sim 1/T_{id}$ 
  approximation.}
  \label{lphoton}
\end{minipage}
\end{figure}
We study the impact of viscous effects on thermal particle yields by constructing the ratios : 
$R_{l^{+}l^{-}}=\left(\frac{d N_{l^{+}l^{-}}}{d M^2 d^2 p_{T} d y}\right) \!\Big/\!
 \left(\frac{d N_{l^{+}l^{-}}^i}{d M^2 d^2 p_{T} d y}\right)$ and 
 $R_{\gamma} = \left(\frac{d N_{\gamma}}{d^{2} p_{T} d y}\right) \!\Big/\!
 \left(\frac{d N_{\gamma}^{i}}{d^{2} p_{T} d y}\right)$, where the superscript $i$ denotes the ideal yield 
 and is obtained by integrating the ideal rate expressions ($\delta f=0$) over the ideal Bj\"orken evolution. 
In Figs.~\ref{ldilepton} and \ref{lphoton}, we plot the ratio of viscous to ideal dilepton and photon yields 
 respectively for the $\tau_\pi \sim 1/T_{id}$ case. The dashed grey curves in the figures represent the ratios 
 corresponding to the viscous evolution for the $\alpha$ values ranging from $10$ to $5000$.
 We observe that the thermal particle yields are maximum 
for the attractor solution and minimum for the repulsor. Viscous contributions for non-zero values of $\alpha$
tend to approach the repulsor one with increase in $\alpha$ value. Also, it can be noted 
that the large $\alpha$ values suppress the spectra considerably over the entire $p_T$ regime.

\section{Dependence of anisotropic flow and particle production on particlization models and nuclear equation of state}
\author{Sumit Kumar Kundu, Yoshini Bailung, Sudhir Pandurang Rode, Partha Pratim Bhaduri, and Ankhi Roy}	

\bigskip

\begin{abstract}
We investigate the effect of particlization models on particle production for the various equation of states in heavy-ion collisions using the UrQMD event generator. We study anisotropic flow coefficients and particle ratios for mid-central (b=5-9 fm corresponds to approximately 10-40\% central) Au-Au collisions for beam energies 1A-158A GeV. UrQMD provides different equations of state in a hybrid mode: chiral EoS, hadron gas EoS, and bag model EoS. Three different particlization models to convert fluid dynamic description to the transport description using various hypersurface criteria are provided by the UrQMD event generator. The results are also compared with available experimental results.

\end{abstract}


\subsection{Introduction}

Different heavy-ion collision experiments around the globe are being operated to explain the QCD phase diagram in different temperature and baryon density regions. Large Hadron Collider (LHC)~\cite{lhc1} and Relativistic Heavy Ion Collider (RHIC)~\cite{STAR:2005gfr} experiments are exploring the high temperature and zero net baryon density region up to a greater extent. Several new experimental facilities like Facility for Anti-proton and Ion Research (FAIR)~\cite{Ablyazimov:2017guv} and Nuclotron-based Ion Collider fAcility (NICA)~\cite{Kekelidze:2016wkp} are trying to explore the intermediate temperature and non-zero net baryon density region of the QCD phase diagram. Until these experimental facilities become operational, several simulation packages and phenomenological models can be employed to provide predictions at the corresponding regime of temperature and baryon density.   

\subsection{UrQMD Model Description}

The ultrarelativistic quantum molecular dynamic (UrQMD)~\cite{Bass:1998ca,Bleicher:1999xi,Petersen:2008dd} model provides different hybrid modes and particlization scenarios to replicate the hydrodynamic effect. UrQMD involves a hybrid mode with the pure transport approach for a better understanding of hot and dense stages of collision. It provides three hybrid modes; Chiral EoS, Hadron Gas EoS, and Bag Model EoS. Chiral and Bag Model EoS have a partonic degrees of freedom, while Hadron Gas EoS involves a hadron degrees of freedom. Chiral, Hadron Gas, and Bag Model EoS include cross-over, no phase transition, and first-order phase transition respectively. Besides that, UrQMD provides three different freeze-out or particlization models for the fluid to the particle transition process. First is gradual freeze-out (GF) hypersurface, where particlization takes place slice by slice of 0.2 fm thickness each when the energy density of all cells of that slice falls below the critical energy density, which is five times the nuclear ground state energy density, i.e., 5$\epsilon_{0}$ ($\sim 730 MeV/fm^{3}$)~\cite{Huovinen:2012is}. The second is isochronous freeze-out (ICF), where particlization takes place at the time when the energy density of all cells drops below the critical energy density value. The third is the iso-energy density particlization scenario (IEF), where the hypersurface for particlization is developed numerically using the Cornelius routine~\cite{Huovinen:2012is} which is a freeze-out hypersurface finder code used to generate 3D hypersurface in four dimensions. 

\begin{figure}[th]
\centerline{\includegraphics[width=6cm]{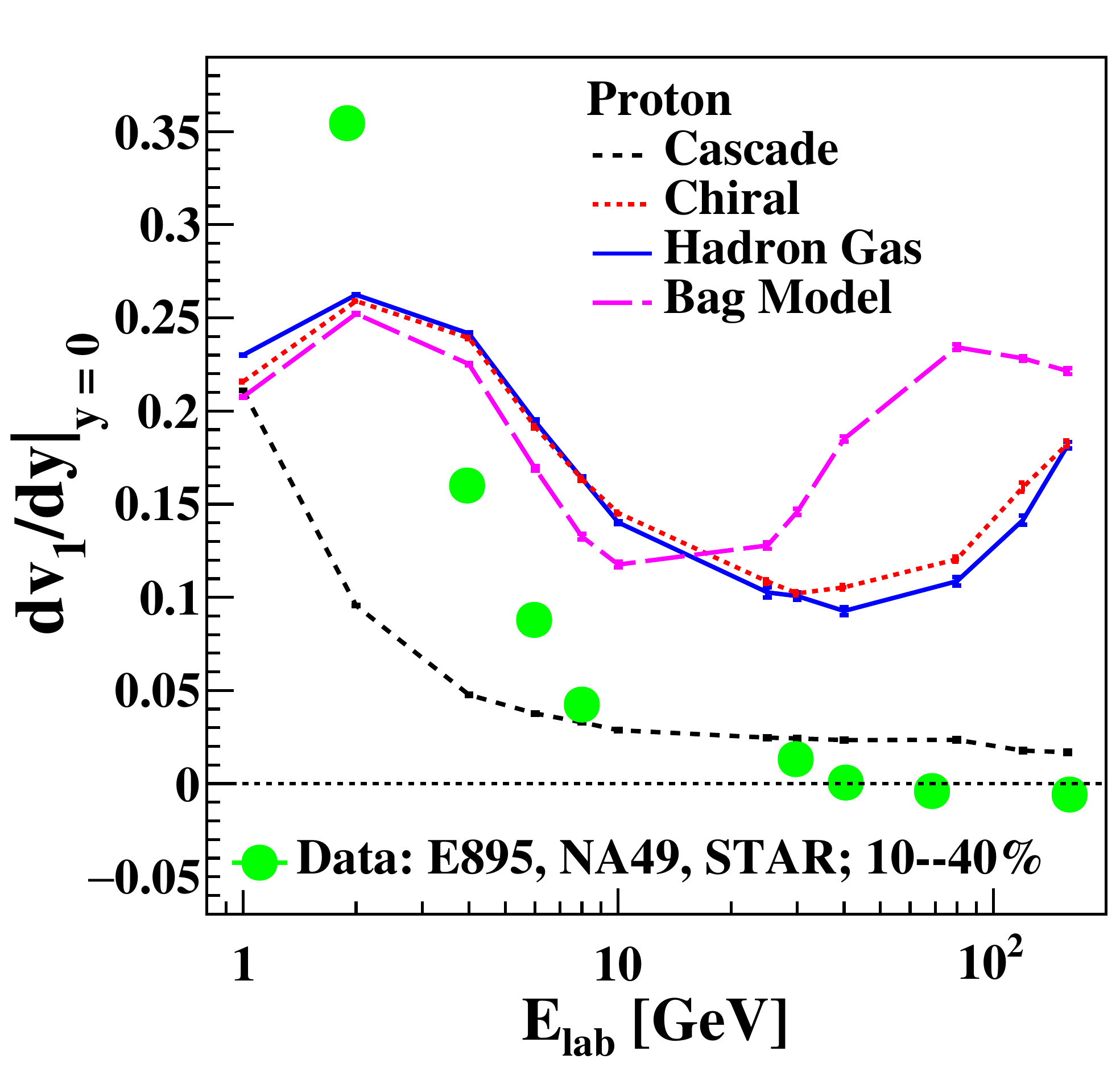}}
\caption{The slope of directed flow of protons w.r.t. the beam energy in a lab frame at midrapidity region for various configurations of UrQMD with default gradual freeze-out hypersurface for noncentral Au-Au collisions with E895~\cite{Liu:2000am} and STAR~\cite{Adamczyk:2014ipa} experimental measurements in Au-Au collisions and with NA49~\cite{Alt:2003ab} experimental measurements in Pb-Pb collisions. Reprinted from Ref.~\cite{Kundu:2021afz}}
\label{figure1}
\end{figure}

\subsection{Results and Discussion}
We show the effect of different EoS on the anisotropic flow and particle production~\cite{Kundu:2021afz} in noncentral Au-Au collisions for beam kinetic energies in the range 1A-158A GeV. Figure~\ref{figure1} shows the effect of the equation of states on the slope of the directed flow of protons. Up to 10A GeV, the employed equations of state show a similar outcome. However, Bag Model EoS shows a split, which may be due to the first-order phase transition. Overall, all EoS overestimate the experimental results. Other observables such as elliptic flow ($v_{2}$), particle ratios, and net-proton rapidity spectra are also studied, and for detailed information on the findings, the reader is referred to Ref.~\cite{Kundu:2021afz}. In every case, we see the sensitivity for different EoS. However, we cannot predict which EoS is most suitable for reproducing experimental measurements.

\begin{figure}[th]
\includegraphics[width=6cm]{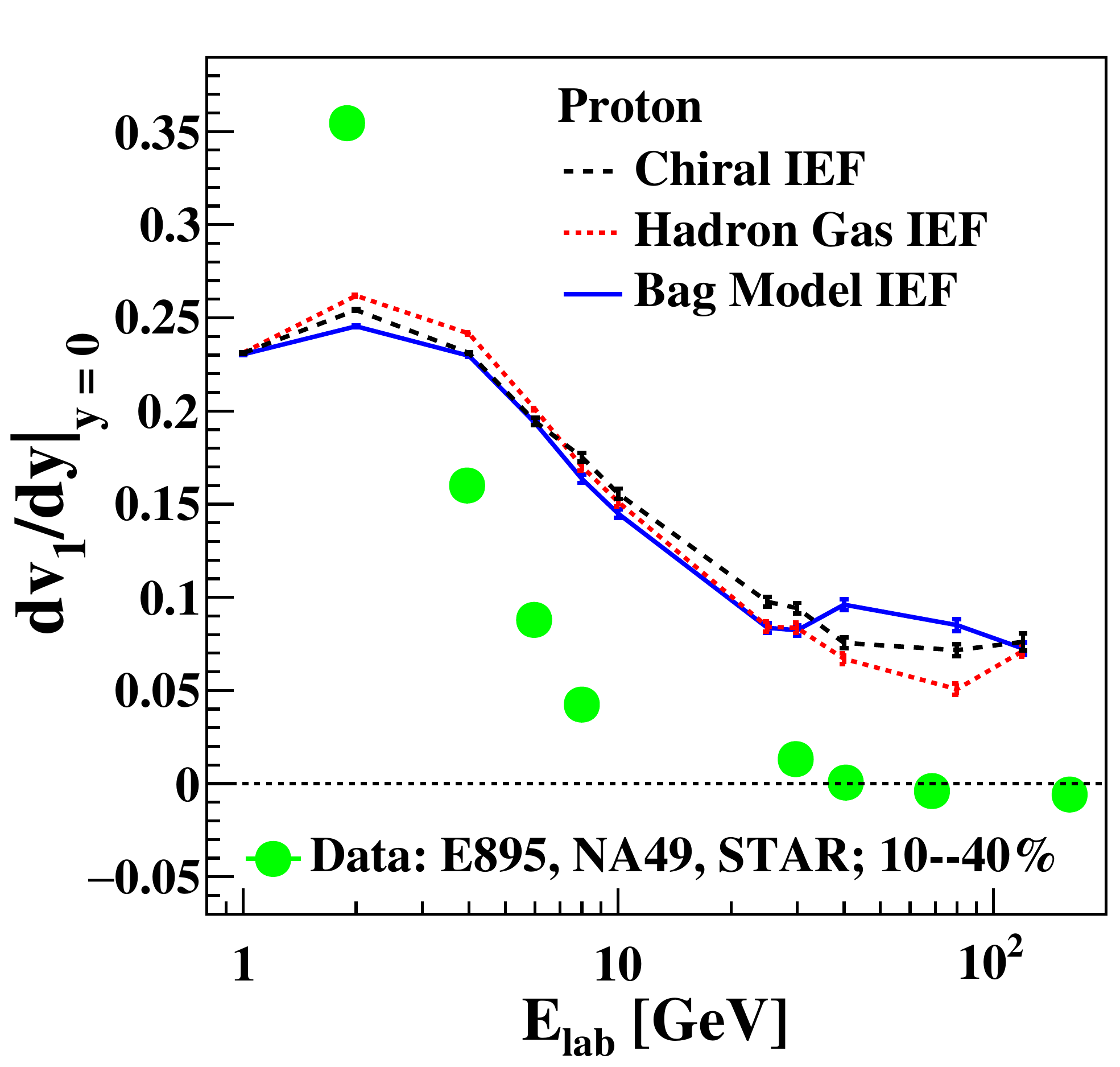}
\includegraphics[width=6cm]{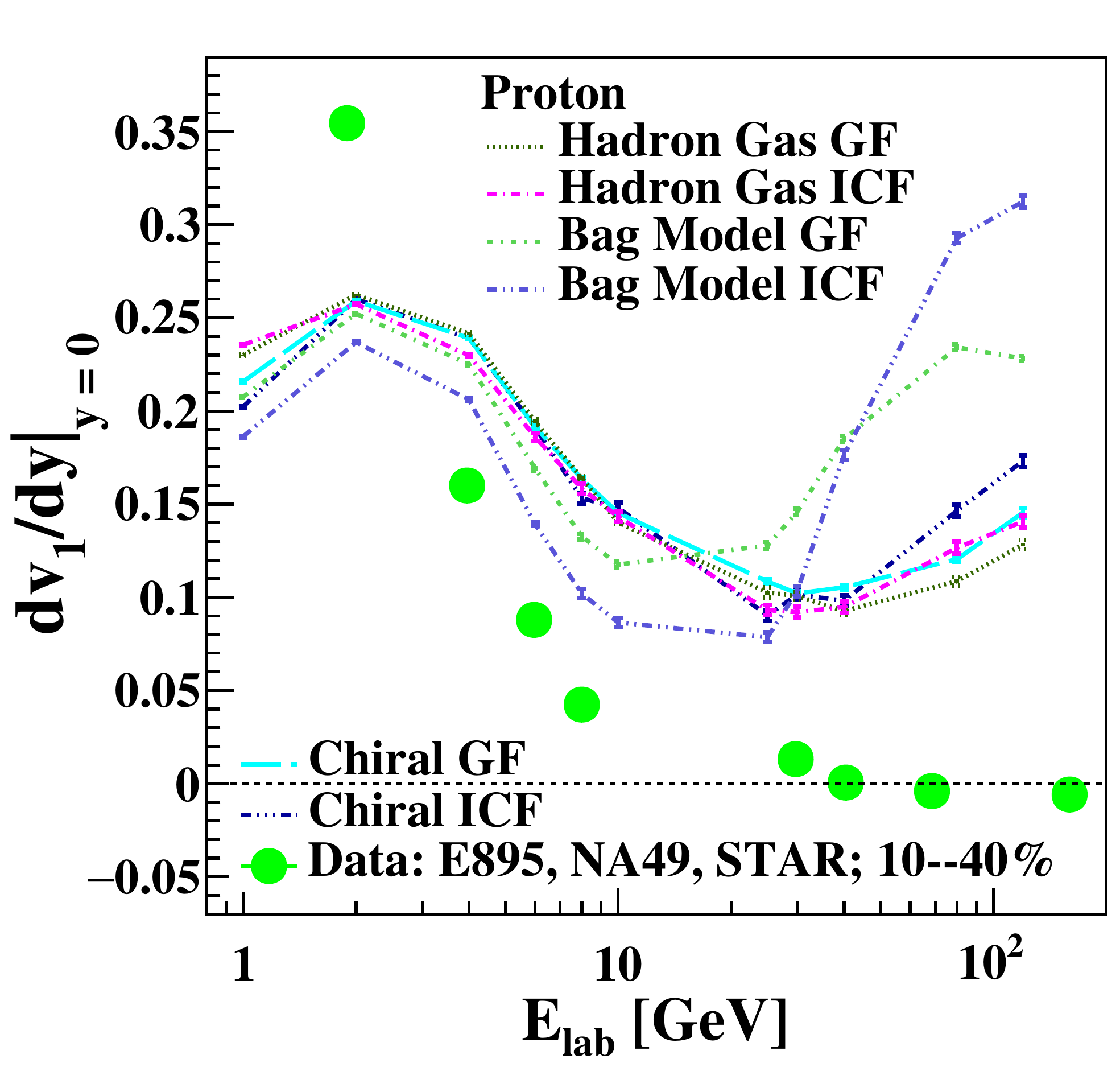}
\caption{The slope of directed flow of protons w.r.t. the beam energy in a lab frame at midrapidity region for various configurations of UrQMD for noncentral Au-Au collisions with E895~\cite{Liu:2000am} and STAR~\cite{Adamczyk:2014ipa} experimental measurements in Au-Au collisons and with NA49~\cite{Alt:2003ab} experimental measurements in Pb-Pb collisions. Reprinted from Ref.~\cite{Kundu:2022tec }}
\label{figure2}
\end{figure}

\begin{figure}[th]
\begin{center}
\includegraphics[width=6cm]{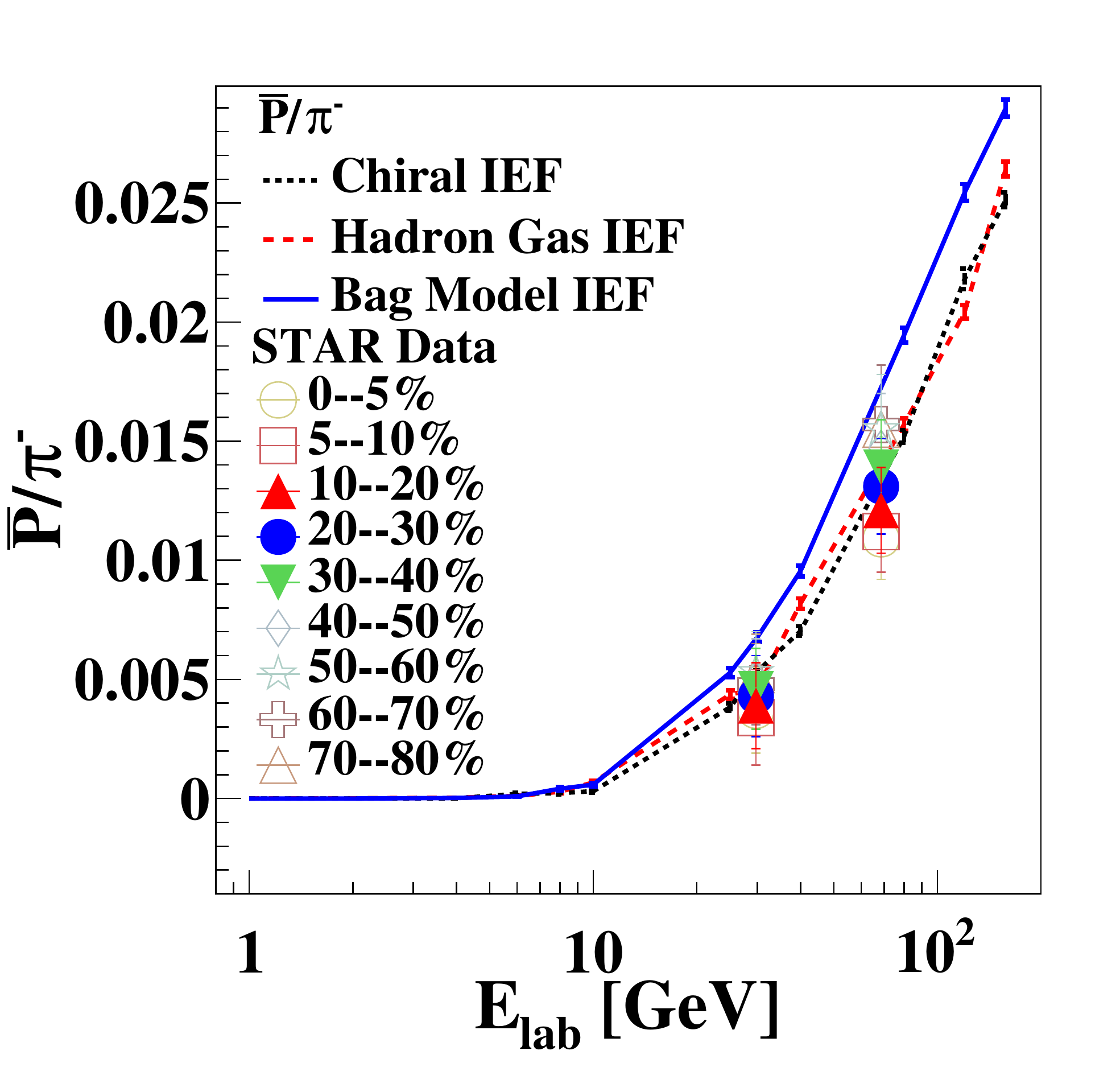}
\includegraphics[width=6cm]{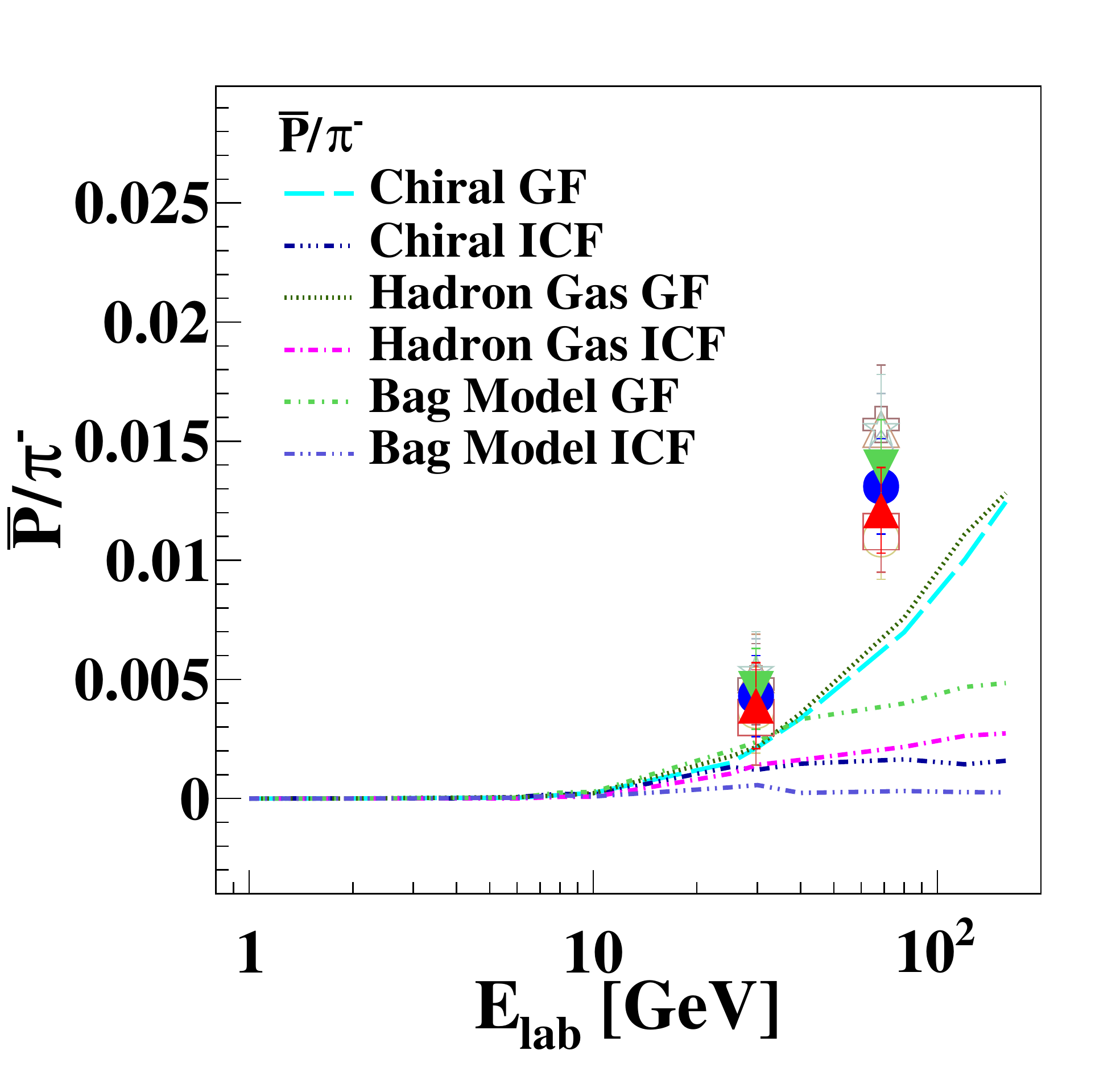}\\
\end{center}
\caption{$\overline{P}/\pi^{-}$ ratio w.r.t. the beam energy in a lab frame for various configurations of UrQMD for noncentral Au-Au collisions and comparison with STAR~\cite{Adamczyk:2017iwn} experimental measurements in Au-Au collisions for all available centralities. Reprinted from Ref.~\cite{Kundu:2022tec}.}
\label{figure4}
\end{figure}

Furthermore, in another study~\cite{Kundu:2022tec}, we attempt to investigate the effect of different particlization models provided by UrQMD. Figure~\ref{figure2} compares the slope of the directed flow of protons to see the effect of these particlization models. It is observed that Iso-Energy hypersurface (IEF) with any EoS predicts the experimental measurements much better compared to other freeze-out scenarios. The similar observations are seen in other measurements~\cite{Kundu:2022tec} and are shown in Figure~\ref{figure4} in case of ratio, $\overline{P}/\pi^{-}$. Observations are in line with ones made in the case of $dv_{1}/dy$ that iso-energy density particlization scenario explains the experimental measurements reasonably well.  

\subsection{Summary}

The IEF scenario reproduces the experimental results much better than the other two freeze-out scenarios. However, no such noticeable difference is observed for the different EoS. It will be interesting to employ these different variants of EoS and particlization models in the higher beam energy range.

\section{Conductivity of massless quark matter for its lowest possible relaxation time}
\author{Cho Win Aung, Thandar Zaw Win, Sabyasachi Ghosh}	

\bigskip

\begin{abstract}
	We have calculated microscopically electrical conductivity of massless quark matter by using relaxation time approximation of kinetic theory framework. The lowest possible quark relaxation time, tuned from the quantum lower bound of shear viscosity to entropy density ratio for massless matter, is used to obtain its corresponding (normalized by temperature T) electrical conductivity $\sigma/T=0.0135$. By comparing with earlier existing numerical values of electrical conductivity, we marked roughly $(0.25-15)\times 0.0135$ as strongly and beyond $20\times 0.0135$ as weakly quark gluon plasma domain.   
\end{abstract}

\subsection{Introduction}
According to the Biot and Savart law, when two heavy nuclei are doing peripheral or non-central collisions in heavy ion collisions (HIC) experiments, a huge magnetic field can be produced. Its values are expected approximately $m_\pi^2$ in RHIC energy and $10 m_\pi^2$ in LHC energy~\cite{Tuchin:2013ie}. The order of magnitude of this magnetic field ($1-10 m_\pi^2\approx 10^{18}-10^{19} G$) is perhaps the strongest magnetic field that has ever existed in nature. However, it will decay with time, whose detailed time profile (including space profile) are investigated in Refs.~\cite{Tuchin:2013ie,Skokov:2009qp,Voronyuk:2011jd}. The decay time with respect to QGP life time will basically fix whether QGP face strong~\cite{Tuchin:2013ie} or weak~\cite{B_t1,Voronyuk:2011jd} magnetic field which is not approached towards converging conclusion probably.

The electrical conductivity of QGP becomes an important quantity, which controls the decay profile~\cite{Tuchin:2013ie}. So present work has explored its existing numerical values, calculated from different microscopic models and their corresponding effective relaxation time scales. We have guidance on the shear relaxation time scale ($\tau_c$) from the experimental direction, which is expected to be close to its lower bound $\tau_c=5/(4\pi T)$ (for massless QGP) as shear viscosity ($\eta$) to entropy density ($s$) ratio of QGP is experimentally expected to be close to its quantum lower bound $\eta/s=1/(4\pi)$. Present work has tried to explore the numerical bands of the electric charge relaxation time scale in terms of the lower bound of shear relaxation time.

Article is organized as follows. After brief addressing of kinetic theory framework of electrical conductivity of relativistic matter in Sec.~(\ref{sec:Form}), we have estimated its values for massless QGP with lowest possible relaxation time in the Sec.~(\ref{sec:Res}), where estimations of earlier works are included. We have analysed possible numerical bands of conductivity in terms of corresponding band relaxation time in Sec.~(\ref{sec:sum}).  

\subsection{Framework}
\label{sec:Form}
The dissipative current density $J_D$ due to external electric field ${\tilde{E}}$ can be expressed in macroscopic relation $J^i_D=\sigma^{ij} ~\tilde{E}_j$ (Ohm's law) and microscopic relation
\bea
J_D^i&=& g \sum_{u,d} e_Q \int \frac{d^3 \bf{p}}{(2\pi)^3}\frac{p^i}{E} \delta f_\sigma~,
\label{0}
\eea
where $g=12$ is total degeneracy factor of quark, coming from its spin, particle-anti-particle and color degeneracy factors. The quark flavor are summed as electric charge of u and d quarks are different. The $\delta f$ is the deviation from equilibrium distribution function $f_0=1/[exp(\beta E) + 1]$ of quark. To know this $\delta f$, we will use the relaxation time approximation(RTA) of Boltzmann transport equation with force term
$F^i=e_Q \tilde{E}^i$ and we will get
\begin{align}
e_Q \tilde{E}^i\frac{\partial f}{\partial p^i}=-\frac{\delta f}{\tau_c}, & \Longrightarrow
\delta f =\Big[ e_Q \big(\frac{p^i}{E} \big)  \tau_{c} \beta f_0(1-f_0)\Big]\tilde{E}_i~.
\end{align} 
The Eq.(\ref{0}) becomes
\bea
J^i&=&  g \sum_{u,d} \int e_Q^2 \frac{d^3 \bf{p}}{(2\pi)^3} \frac{p^i p^j}{E^2} \tau_{c} \beta  f_0(1-f_0) \tilde{E}_j~.
\eea
Comparison with the macroscopic description, $J^i=\sigma^{ij}\tilde{E}_j$, we get
\bea
\sigma= \frac{g}{T} \sum_{u,d} e_Q^2 \int \frac{d^3 \bf{p}}{(2\pi)^3} \frac{p^2}{3E^2}  \tau_{c} f_0(1-f_0)~.
\label{el_RTA}
\eea
\subsection{Results and discussion}
\label{sec:Res}
For generating results, let us use massless limit $E=p$ in Eq.~(\ref{el_RTA})~\cite{Pramana} and then we get
\begin{eqnarray}
\sigma&=&\frac{10}{27}~ e^2 \tau_{c}T^2,
\label{1}
\end{eqnarray}
where 
$\sum e_Q^2=\big(+\frac{2}{3}e\big)^2+\big(-\frac{1}{3}e\big)^2=\big(\frac{5}{9} \big) e^2$~.
Now, lower bound of $\eta/s$ for massless medium can provide lower bound of (shear) relaxation time\cite{Pramana},
\begin{align}
\frac{\eta}{s}=\frac{\tau_c T}{5}=\frac{1}{4 \pi}, & \Longrightarrow \tau_{c}=\frac{5}{4\pi T}.
\label{2}
\end{align}
Using this lowest possible (shear) relaxation time, we will get conductivity
\begin{eqnarray}
\sigma(T)&=&\frac{25}{54}~ \frac{e^2}{\pi} T\approx 0.0135 T~.
\end{eqnarray}

\begin{figure}
	\centering
	\includegraphics[scale=0.4]{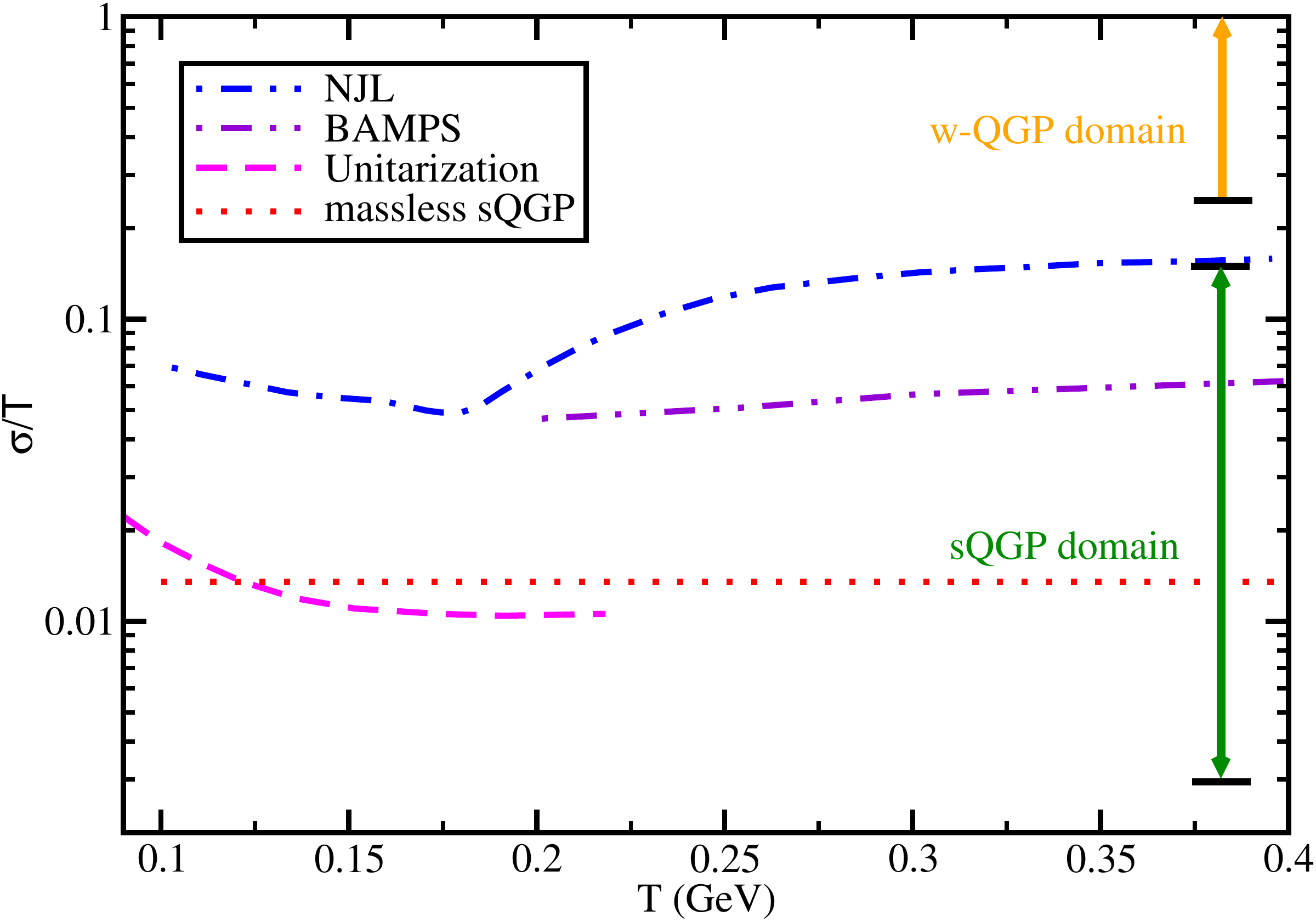}
	\caption{Normalised conductivity of different estimations with temperature, based on NJL (blue dash-dotted line), BAMPS (violet dash-double dotted line) and uniterization (pink dash line) methodologies with respect to the massless sQGP estimation (red dotted line)}
	\label{fig:my_label}
\end{figure}
Assuming $\sigma/T=0.0135$ as lowest reference point, we have plotted it by red dotted line in Fig.(\ref{fig:my_label}) and marked as massless strongly quark gluon plasma (sQGP). The results of electrical conductivity from Refs.\cite{BAMPS}~, \cite{NJL}~, \cite{Unit} are also included in the figure, which covers a large numerical band of electrical conductivity. Ref.~\cite{BAMPS} is based on the transport quark model and its estimated values are approximately 5 times larger than the massless sQGP values. Ref.~\cite{NJL} is based on an effective QCD model, called Nambu-Jona-Lasinio (NJL) model and its estimated values located in the range 5-15 times larger than the massless sQGP values. Interestingly hadronic conductivity estimation, based on unitarization methodology, is close to the massless sQGP values. On the other hand, lattice quantum chromodynamics (LQCD) calculations by Amato et al.~\cite{9_Gowthama} provide values from 0.003 to 0.015. It means that 4 times smaller than the massless sQGP values (0.0135) should have to be considered also within the numerical band of $\sigma/T$. Now this entire numerical band of $\sigma/T$ from 0.003 to 0.2 might be considered as sQGP domain for electrical conductivity because weakly QGP (w-QGP) estimation from perturbative QCD (pQCD) calculation~\cite{pQCD1,pQCD2} provides more than 20 times larger values of $\eta/s$ with respect to its quantum lower bound $1/(4\pi)\approx 0.08$. So corresponding (shear) relaxation time may be larger than $25/(\pi T)$ and using that relaxation time range for electrical conductivity, one can expect $\sigma/T\geq 20\times 0.0135$ as wQGP domain. Our exploration of sQGP and wQGP domain of $\sigma/T$ might be very important information for knowing the corresponding domain of decay profile of magnetic field, produced in heavy ion collision experiments.
%
%
\subsection{Summary}
\label{sec:sum}
In summary, with the help of the relaxation time approximation of the kinetic theory framework, we have calculated microscopically electrical conductivity of massless quark matter. Based on the knowledge of quantum lower bound for shear viscosity to entropy density ratio, one can get the lowest possible (shear) relaxation time for massless matter. Present work has tried to understand electric charge transportation in terms of that quantum bound. So, we have used that relaxation time to obtain electrical conductivity by temperature ratio $\sigma/T=0.0135$, which might be considered as a reference point for charge transportation of QGP. Comparing with earlier existing numerical values of electrical conductivity, we noticed that LQCD estimations of $\sigma/T$ can be 4 times smaller to 3 times larger than the reference value $0.0135$. When we considered some earlier model calculations like NJL, BAMPS, and unitarization, we find the ratio may go up to 15 times larger than $0.0135$. In this angle, we may assign roughly $\sigma/T=(0.25-15)\times 0.0135$ as our known range of microscopic calculations, which might be connected to the strongly quark-gluon plasma domain. On the other hand, beyond $20\times 0.0135$ as weakly quark gluon plasma domain because earlier perturbative QCD calculation by Arnold et al. indicate that range for shear viscosity to entropy density ratio. We believe that these strongly and weakly interacting domains of electrical conductivity will be very important inputs to investigate the corresponding detailed decay profile of magnetic field, produced in heavy ion collision experiments.   


\section{Exploring the numerical bands of electrical conductivity of quark gluon plasma and decay profile of magnetic field}
\author{Thandar Zaw Win, Cho Win Aung, Sabyasachi Ghosh}	

\bigskip

\begin{abstract}
	In heavy ion collision experiments, a huge magnetic field can be produced in peripheral collisions, owing to Ampere's law. The quark-gluon plasma, produced in the heavy ion collision experiments, will face this field, which will decay with time. This exponential decay profile will be controlled by the electrical conductivity of expanding quark-gluon plasma. Present work has explored that connection by using different earlier works, predicting the values of electrical conductivity for quark-gluon plasma.
\end{abstract}

\subsection{Introduction} 
When two heavy nuclei are colliding in peripheral collision in heavy ion collisions experiments (HIC), a huge magnetic field can be produced according to Biot Savart law. The field strengths can approximately be $m_\pi^2$ in RHIC energy and $10 m_\pi^2$ in LHC energy~\cite{Tuchin:2013ie}~. 
It can decay with time, whose detailed space-time profile are investigated recently in many references. The reader may see Refs.~\cite{Tuchin:2013ie,B_t1,Voronyuk:2011jd} and references therein. This decay time may~\cite{B_t1,Voronyuk:2011jd} or may not~\cite{Tuchin:2013ie} be smaller than lifetime of QGP. Fact is still a matter of debate.
The decay profile of the magnetic field~\cite{Tuchin:2013ie} can be controlled by the electrical conductivity of the QGP medium. In this context, the present work has explored its existing numerical values, calculated from different microscopic models, and then use them for generating corresponding decay profiles of magnetic field. 
  The article is organized as follows. Next, in the Formalism section (\ref{sec:Form}), we have briefly addressed the formalism of magnetic field decay, as prescribed by Tuchin. Then, we have generated their curves in the result section (\ref{sec:Res}) along with a detailed discussion. At the end, Sec~(\ref{sec:Sum}) provides summary of our findings.
  
  \subsection{Framework}
\label{sec:Form}
 Here, we are adopting the formalism, prescribed by Tuchin~\cite{Tuchin}~, which is briefly described below.
 
 The parameter that controls the strength of the matter effect on the field evolution is $\sigma\gamma b$ , where $\sigma$ is the electrical conductivity, $\gamma$ is the Lorentz boost factor, and b is the characteristic transverse size of matter. 
Let us start with Maxwell's equations for EM field created by a point charge e moving along the positive z-axis with velocity v~\cite{Tuchin}~: 
\begin{align}
    {\bf{\nabla \cdot B}} & = 0 ,\notag &
    {\bf{\nabla \times E}} & = - \frac{\partial{\bf{B}}}{\partial t} \\ 
    {\bf{\nabla \cdot D}} & = e\delta(z-vt)\delta(b), \notag &
    {\bf{\nabla \times H}} & = \frac{\partial {\bf{D}}}{\partial t} + \sigma E + ev\hat{z}\delta(z-vt)\delta(b)~. 
\end{align} 
Performing Fourier transform, we get the magnetic field part
\begin{equation} \label{Htrg}
    {H}(t,r) = \int_{-\infty}^{\infty} \frac{d\omega}{2\pi} \int_{-\infty}^{\infty}\frac{dk_z}{2\pi} \int \frac{d^2k_{\perp}}{(2\pi)^2} e^{-i\omega t + ik_zz + ik_{\perp}b} H_{\omega k}~,
\end{equation} 
whose solution becomes
\begin{equation}
 H_{\omega k}  = - 2\pi iev \frac{|k \times  \hat{z}|}{\omega^2\Tilde{\epsilon}\mu - k^2}\delta(\omega - k_zv)~. \label{Htr}
\end{equation} 

Substituting the Eq.(\ref{Htr}) into the Eq.(\ref{Htrg}) and taking integral over k, we get~\cite{Tuchin}
\begin{equation}
    H(t,r)  = \frac{e}{2\pi\sigma} \int_0^\infty \frac{J_1(k_{\perp}b)k_{\perp}^2}{\sqrt{1 + \frac{4k_{\perp}^2}{\gamma^2\sigma^2}}} e^{\left\{ \sigma\gamma^2x_-/2  (1-\sqrt{1 + \frac{4k_\perp^2}{\gamma^2\sigma^2}}  \right\}}dk_\perp~.\label{Htrn} 
\end{equation}
%
%
If $\gamma \sigma b \gg 1$, the Eq.(\ref{Htrn}) becomes~\cite{Tuchin},  
\begin{equation}\label{Etdr}
     H  = \frac{e}{2\pi} \frac{b\sigma (t)}{4x_-^2} e^{-\frac{b^2\sigma (t)}{4x_-}},
    \end{equation} 
where, $x_-~=~ t-z/v $ , hence the magnetic field is given by the solution (\ref{Etdr}). This Eq.~(\ref{Etdr}) will be our working 
formula for generating results.

 \subsection{Results}
 \label{sec:Res}
 %


    \begin{figure}[!h] 
    	\centering 
    		\includegraphics[scale=0.25]{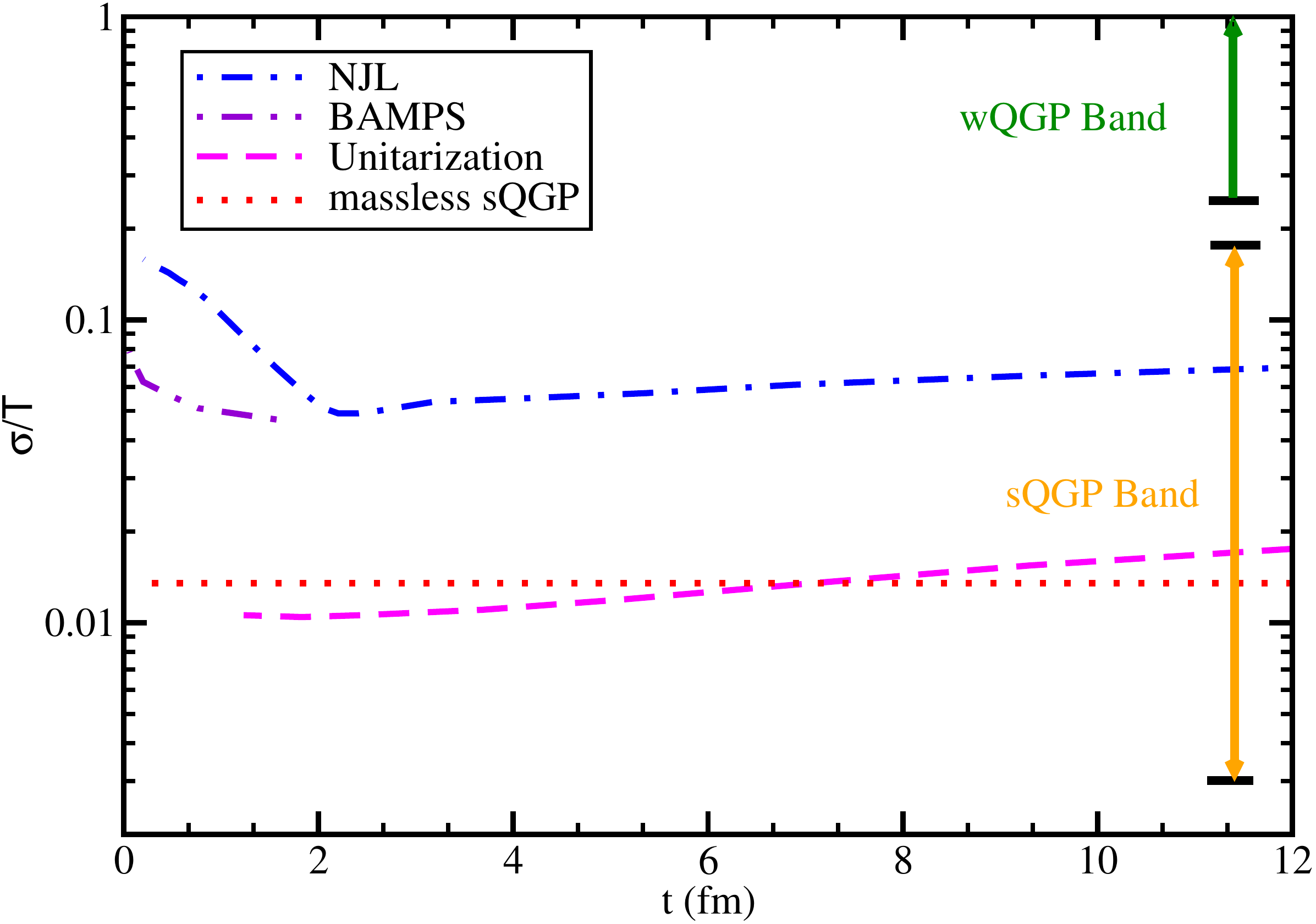}
    		\includegraphics[scale=0.25]{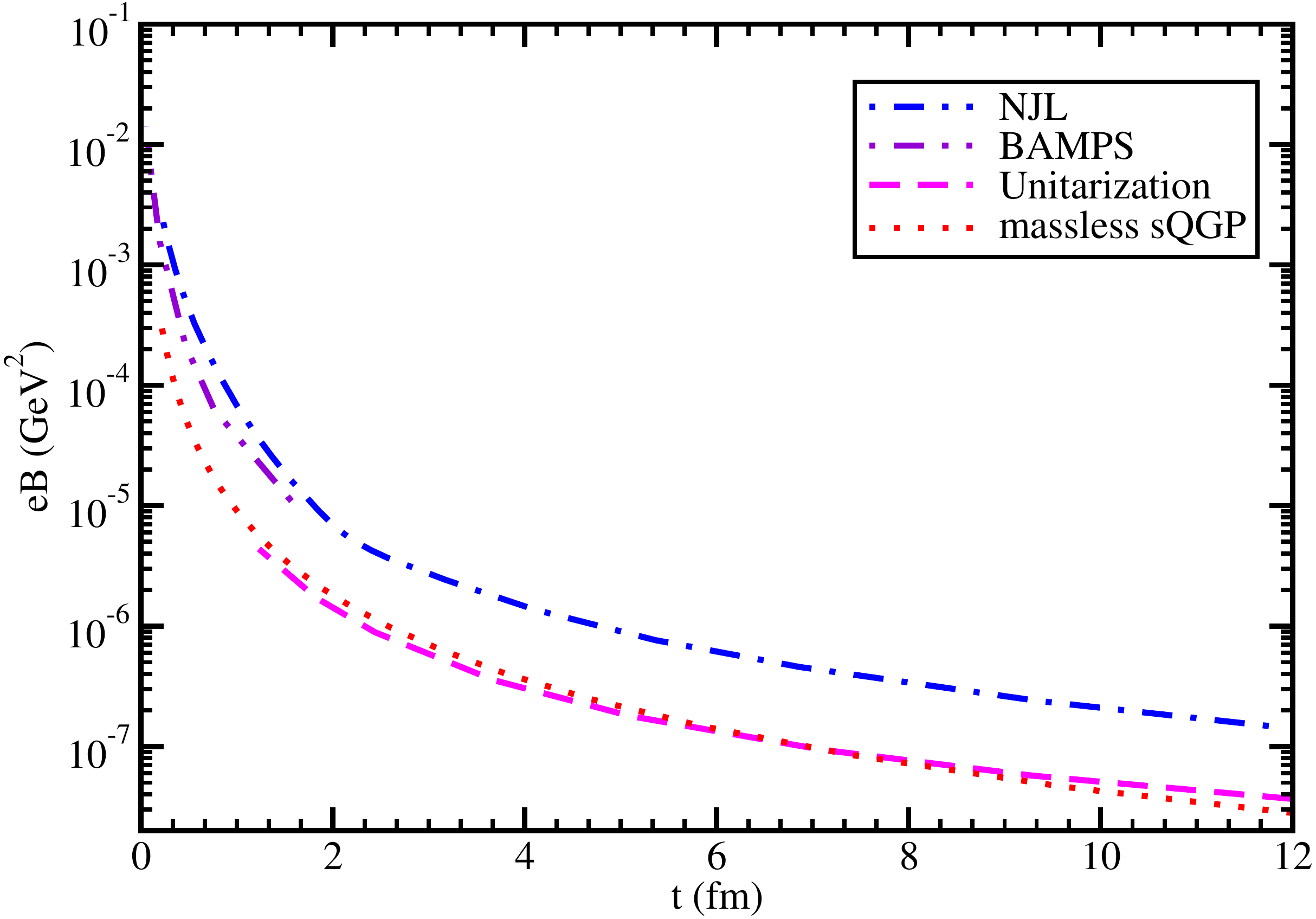}
    		\caption{The electrical conductivity (Left) and corresponding magnetic field decay (Right) of different calculations, based on NJL~\cite{Marty} (blue dash-dotted iine), BAMPS~\cite{BAMPS} (violet dash-double dotted line) and uniterization~\cite{Nicola} (pink dash line) methodologies, and for the massless sQGP case (red dotted line).} 
    		\label{fig:sig_T}
   \end{figure} 		
%
If we explore the long list of earlier references, which have provided estimations of electrical conductivity for QGP by using different microscopic model calculations, then we will get a rough numerical range of its values. Lattice quantum chromodynamics (LQCD) calculations~\cite{ding,Aarts:2016hap,Brandt,Amato} mostly provide temperature-independent values and electrical conductivity by temperature ratio remains within $0.003-0.03$ approximately. However, different microscopic model calculations can go towards larger values. To cover that broad numerical band, we have selected a few Refs.\cite{Marty}~, \cite{BAMPS}~,\cite{Nicola}~, which have provided temperature-dependent conductivity values. To convert from temperature to time, we use the standard cooling law of QGP $T^3t = T_i^3t_i\approx 0.0128$, where the initial temperature and time of expanding QGP are considered as ${T_i}=400$ MeV and ${t_i} = 0.2$ fm respectively.
We get the electrical conductivity in terms of time, as shown in the left panel of Fig.~(\ref{fig:sig_T}). If we notice the NJL model calculations, given in Ref.~\cite{Marty}~, then we can recognize the minimum values of $\sigma/T$ near transition temperature. So, the blue dash-dotted line in the left panel of Fig.~(\ref{fig:sig_T}), presenting $\sigma/T$ along the time axis, also exhibit a minimum value. According to the present version of simple T-t mapping via cooling law, we can identify $t\leq 2$ fm as expanding QGP and $t\geq 2$ fm as expanding hadronic matter up to freeze out time. This time ranges of QGP and hadronic matter is bit of sensitive on initial values of time, temperature. Hence, we should consider them as gross ranges instead of exact quantitative values. These ranges can also be seen from BAMPS~\cite{BAMPS} and unitarization~\cite{Nicola} results, whose predicted values are confined within QGP and hadronic matter temperature respectively. Let us try to understand these results, predicted from earlier microscopic model calculations, in terms of simple massless QGP conductivity $\sigma=\frac{10}{27}~ e^2 \tau_{c}T^2$ for lowest possible (shear) relaxation time $\tau_{c}=\frac{5}{4\pi T}$, which can be tuned by imposing quantum lower bound of fluid property~\cite{Pramana}~. So we may call this $\sigma=\frac{25}{54}~ \frac{e^2}{\pi} T\approx 0.0135 T$ as massless sQGP, whose time dependent expression will be $\sigma=0.0032/t^{1/3}$, as plotted by red dotted line in Fig.~(\ref{fig:sig_T})(Left). This result may be called as conductivity of massless strongly QGP (sQGP), where strongly interacting behaviour is linked with the lowest possible shear relaxation time.
If one compare existing literature for $\sigma/T$, then with respect to massless sQGP, then they will be within the ranges from 1/4 to 15 times larger than the massless sQGP. This band may be considered as sQGP bands of conductivity (marked in Fig.~(\ref{fig:sig_T})(Left)) because perturbative QCD (pQCD) calculations~\cite{pQCD1,pQCD2} give more than 20 times larger values of shear relaxation, marked as weakly QGP (wQGP) domain in Fig.~(\ref{fig:sig_T})(Left).

Now, considering this sQGP bands of $\sigma(t)$ in Eq.~(\ref{Etdr}), we will get magnetic field decay profile as shown in Fig.~(\ref{fig:sig_T})(Right), where $b=1$ fm, $z=0$ and $v=0.99$ are considered. Changing those $b$, $z$ and $v$, decay profile will be modified but our aim is to show corresponding bands of the decay profile by using existing sQGP conductivity band. 
 

\subsection{Summary}
\label{sec:Sum}
In summary, we have extracted the temperature-dependent conductivity data of earlier references and converted them to time-dependent data by using the standard cooling law of expanding QGP. Then, using those data in Tuchin's decay expression of the magnetic field, we have plotted different decay curves for constant values of other parameters like the Lorenz factor, transverse characteristic size, etc. In terms of massless sQGP values of conductivity, we have realized that the existing estimations of earlier work locate within 0.25 to 15 times larger values, which may be considered as its sQGP bands as wQGP cover 20 times larger and beyond values. Present work provided the numerical band of magnetic field decay profiles due to the sQGP conductivity band.   


\section{Causal hydrodynamics based on effective kinetic theory and particle yield from QGP}
\author{Lakshmi J. Naik and V. Sreekanth}	

\bigskip

\begin{abstract}
We investigate thermal particle production and evolution of QGP created 
in heavy ion collisions in presence of viscosities,
by employing the recently formulated second order dissipative hydrodynamics estimated
within the effective fugacity quasi-particle model of hot QCD medium. We employ the
viscous corrections to the parton distribution functions obtained from the Chapman-Enskog 
method in the relaxation time approximation.
We analyze the 
sensitivity of shear and bulk viscous pressures to the temperature dependence of relaxation time
within one dimensional boost invariant flow. 
Particle emission yields are quantified for the longitudinal expansion of QGP
with different temperature dependent relaxation times. 
Our results indicate that the particle spectra computed using this formalism is
well behaved and is sensitive to the relaxation times.
\end{abstract}



\subsection{Introduction}

Relativistic dissipative hydrodynamics has been successfully employed to study the evolution of 
Quark-Gluon Plasma (QGP), created in the relativistic heavy ion collisions at RHIC, LHC. The evolution of QGP has to be
studied with higher order viscous hydrodynamics as the first order Navier-Stokes theory shows acausal behaviour. Several causal
hydrodynamic theories have been developed and have been applied in the context of heavy ion collisions. Recently, a causal 
second order viscous
hydrodynamic formulation was developed\cite{Bhadury:2020ivo} by employing the effective fugacity
quasi-particle model (EQPM)\cite{Chandra:2009jjo} for the thermal QCD medium. This model incorporates the QCD medium interactions
into the analysis
through thermal effective fugacity parameters in the distribution functions. The hydrodynamic equations in this 
formulation have been 
estimated by employing the covariant kinetic theory for EQPM\cite{Mitra:2018akk}.
The viscous corrections to the distribution
functions have been determined by the Chapman-Enskog (CE) expansion in RTA. We intend to employ this
hydrodynamic framework to study the thermal dilepton and photon production from heavy ion collisions under boost invariant
Bj\"orken expansion of QGP.

\subsection{Second Order hydrodynamics based on EQPM}
We present the formalism to derive the second order causal hydrodynamics within the EQPM description of thermal QCD medium. 
EQPM incorporates the hot QCD medium effects into the analysis through temperature dependent effective fugacity parameters $z_k$ ($k\equiv (q,g)$ stands for quarks and gluons respectively). The equilibrium momentum distribution function within this model is given by\cite{Chandra:2009jjo}
\begin{equation} \label{QPM_f0}
 f_k^0 = \frac{z_k \exp[-\beta(u_\mu p_k^\mu)]}{1 \pm z_k \exp[-\beta(u_\mu p_k^\mu)]},
\end{equation}
where $\beta=1/T$ is the inverse of temperature. It must be noted that the quantity $z_k$ in this model represents the interaction between the partons and is not associated with any conservation laws.
Here, $p_k^\mu = (E_k,\vec{p}_k)$ is the bare four-momenta of the particles and this is related to the quasi-particle 
four-momenta through the dispersion relation $\tilde{p}_{g,q}^\mu = p_{g,q}^\mu + \delta \omega_{g,q} u^\mu;$ where
$\delta \omega_{g,q}= T^2 \partial_T \ln (z_{g,q})$ 
denotes the modified part of energy dispersion relation due to the interaction. The dissipative 
 hydrodynamic equations is derived by quantifying the non-equilibrium corrections in the system. The 
 viscous corrections to the momentum distribution functions are estimated by considering the effective Boltzmann equation
 within EQPM description in RTA\cite{Mitra:2018akk}
\begin{equation} \label{BE_RTA}
 \tilde{p}_k^\mu \partial_\mu f_k^0(x,\tilde{p}_k) + F_k^\mu \partial_\mu^{(p)} f_k^0
 = -\frac{\delta f_k}{\tau_R} \omega_k.
\end{equation}
 In the above equation, $\tau_R$ denotes the relaxation time and $\delta f_k$ is the non-equilibrium part of the distribution
function.  Presence of non-trivial dispersion relation in the theory gives the mean field force term $F_k^\mu = - \partial_\nu (\delta \omega_k u^\nu u^\mu)$ in the effective transport equation 
and the same is obtained 
from energy-momentum and particle flow conservation~\cite{Mitra:2018akk}. We use the form of distribution function 
derived from the Chapman-Enskog like expansion of effective Boltzman equation in RTA,
\begin{equation}
\delta f_q = \tau_R \bigg[ \tilde{p}_q^\mu\partial_\mu \beta 
 + \frac{\beta\, \tilde{p}_q^\mu\, \tilde{p}_q^\nu}{u\!\cdot\!\tilde{p}_q}\partial_\mu u_\nu 
 - \beta \Theta (\delta\omega_q) 
 - \beta \dot{\beta} \left(\frac{\partial(\delta\omega_q)}{\partial\beta}\right)\bigg] 
 f_{q}^0\bar{f}_{q}^0, 
\end{equation}
where $\bar{f}_{q}^0 = 1- f_{q}^0$. The expressions for $\pi^{\mu\nu}$ and $\Pi$ within the effective kinetic theory
are obtained as\cite{Mitra:2018akk}, 
\begin{eqnarray}
 \pi^{\mu\nu} &=& \sum_k g_k \Delta_{\alpha\beta}^{\mu\nu} \int d\tilde{P}_k 
 \tilde{p}_k^\alpha \tilde{p}_k^\beta \delta f_k  
 + \sum_k g_k \delta \omega_k \Delta_{\alpha\beta}^{\mu\nu} \int d\tilde{P}_k 
 \tilde{p}_k^\alpha \tilde{p}_k^\beta \frac{\delta f_k}{E_k}, \label{pi_eqn} \nonumber \\
 \Pi &=& -\frac{1}{3}\sum_k g_k \Delta_{\alpha\beta}\int d\tilde{P}_k 
 \tilde{p}_k^\alpha \tilde{p}_k^\beta \delta f_k  
 -\frac{1}{3} \sum_k g_k \delta \omega_k \Delta_{\alpha\beta} \int d\tilde{P}_k 
 \tilde{p}_k^\alpha \tilde{p}_k^\beta \frac{\delta f_k}{E_k}.   \label{Pi_eqn}
\end{eqnarray}
Following Ref.~\refcite{Bhadury:2020ivo}, the evolution equations for $\pi^{\mu\nu}$ and $\Pi$ are obtained as given below
\begin{eqnarray} \label{shear evolution def hydro}
\dot{\pi}^{\langle\mu\nu\rangle} + \frac{\pi^{\mu\nu}}{\tau_R} &=& 2 \beta_{\pi} \sigma^{\mu\nu} 
+ 2 \pi_{\phi}^{\langle\mu} \omega^{\nu\rangle\phi} 
- \delta_{\pi\pi} \pi^{\mu\nu} \theta 
- \tau_{\pi\pi} \pi_{\phi}^{\langle\mu} \sigma^{\nu\rangle \phi} 
+ \lambda_{\pi\Pi} \Pi \sigma^{\mu\nu}, \nonumber \\
\dot{\Pi} + \frac{\Pi}{\tau_R} &=& -\beta_{\Pi} \theta - \delta_{\Pi\Pi} \Pi \theta 
+ \lambda_{\Pi\pi} \pi^{\mu\nu} \sigma_{\mu\nu}.\label{bulk evolution def hydro}
\end{eqnarray}
Now, we solve the above equations for various temperature dependent relaxation times by considering the longitudinal
boost-invariant expansion of Bj\"orken\cite{Bjorken:1982qr}. The space-time coordinates are parameterized 
as $t=\tau\cosh\eta_s$ and $z=\tau\sinh\eta_s$, with
$\tau=\sqrt{t^2-z^2}$ and $\eta_s=\frac{1}{2}\ln \frac{t+z}{t-z}$ respectively being the 
proper time and space-time rapidity of the system. Four$-$velocity of the fluid is given by 
the ansatz $u^\mu = (\cosh\eta_s,0,0,\sinh\eta_s)$. Considering the boost-invariance, the 
above equations reduce to the following first order coupled equations in $\tau$ :\cite{Bhadury:2020ivo} 
\begin{eqnarray}    \label{energy_evolution}
 \frac{d \epsilon}{d\tau} &=& -\frac{1}{\tau}\left(\epsilon + P + \Pi - \pi \right),  \nonumber  \\
 \frac{d \pi}{d\tau} + \frac{\pi }{\tau_\pi} &=& \frac{4}{3}\frac{\beta_\pi}{\tau} 
 -\left( \frac{1}{3}\tau_{\pi\pi} + \delta_{\pi\pi}\right) \frac{\pi}{\tau} 
 + \frac{2}{3} \lambda_{\pi \Pi} \frac{\Pi}{\tau},  \label{pi_evolution}\nonumber \\
 \frac{d \Pi}{d\tau} + \frac{\Pi }{\tau_\Pi} &=& -\frac{\beta_\Pi}{\tau} 
 -\delta_{\Pi\Pi} \frac{\Pi}{\tau} + \lambda_{\Pi \pi} \frac{\pi}{\tau};   \label{Pi_evolution}
\end{eqnarray}
where $\pi=\pi^{00}-\pi^{zz}$. The expressions for the transport coefficients can be found in Ref.~\refcite{Bhadury:2020ivo}.
Now, we analyze the evolution of $\pi$ and $\Pi$ by choosing
three different forms of relaxation times : $\tau_R = 3(\eta/s)/T, 2(\eta/s)/T$ and $(\eta/s)/T$.
\begin{figure}
\centering
\begin{minipage}{.5\textwidth}
  \centering
  \includegraphics[width=.9\linewidth]{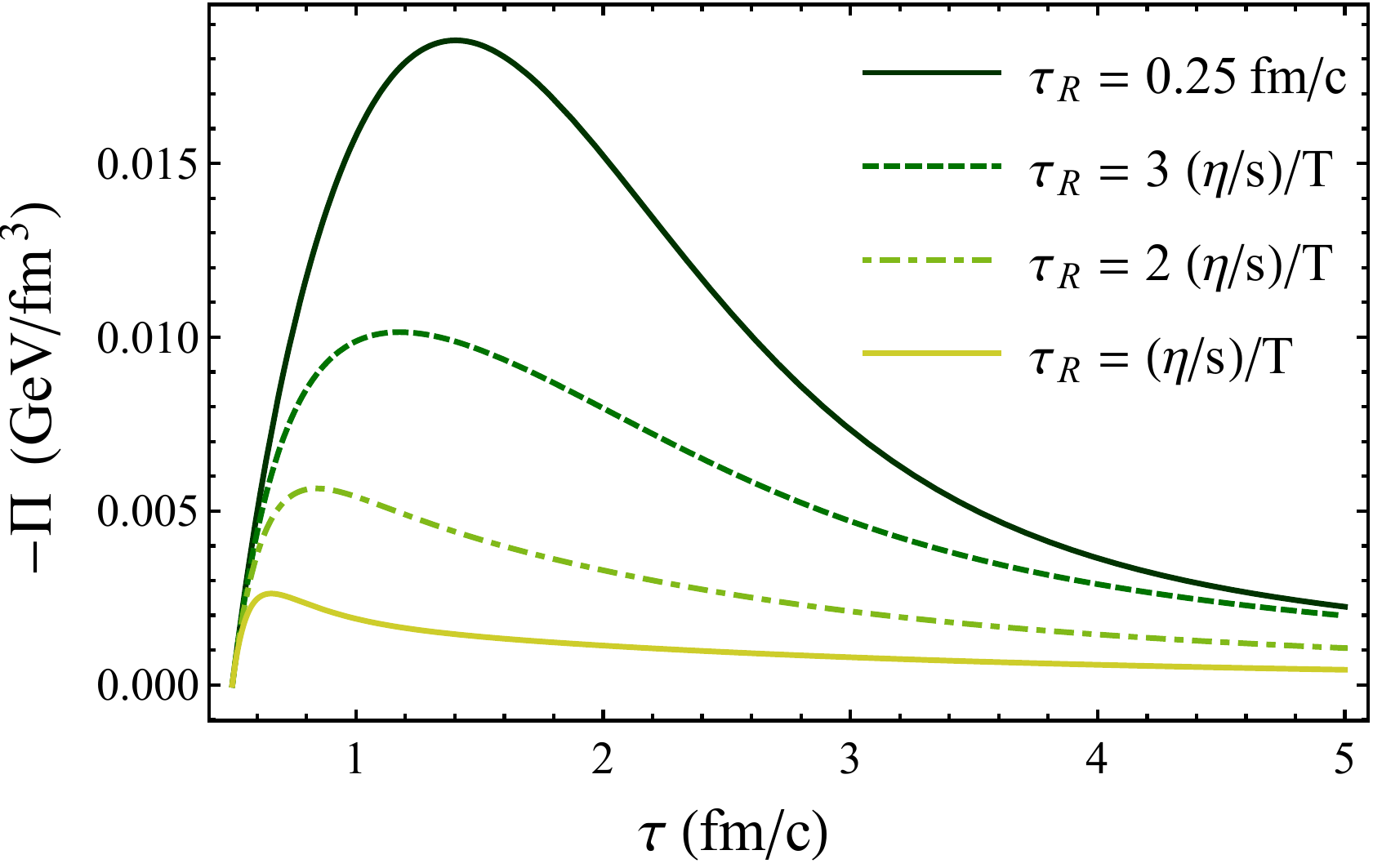}
  \caption{Proper time evolution of bulk \\ viscous pressure.}
  \label{bulk}
\end{minipage}%
\begin{minipage}{.5\textwidth}
  \centering
  \includegraphics[width=.9\linewidth]{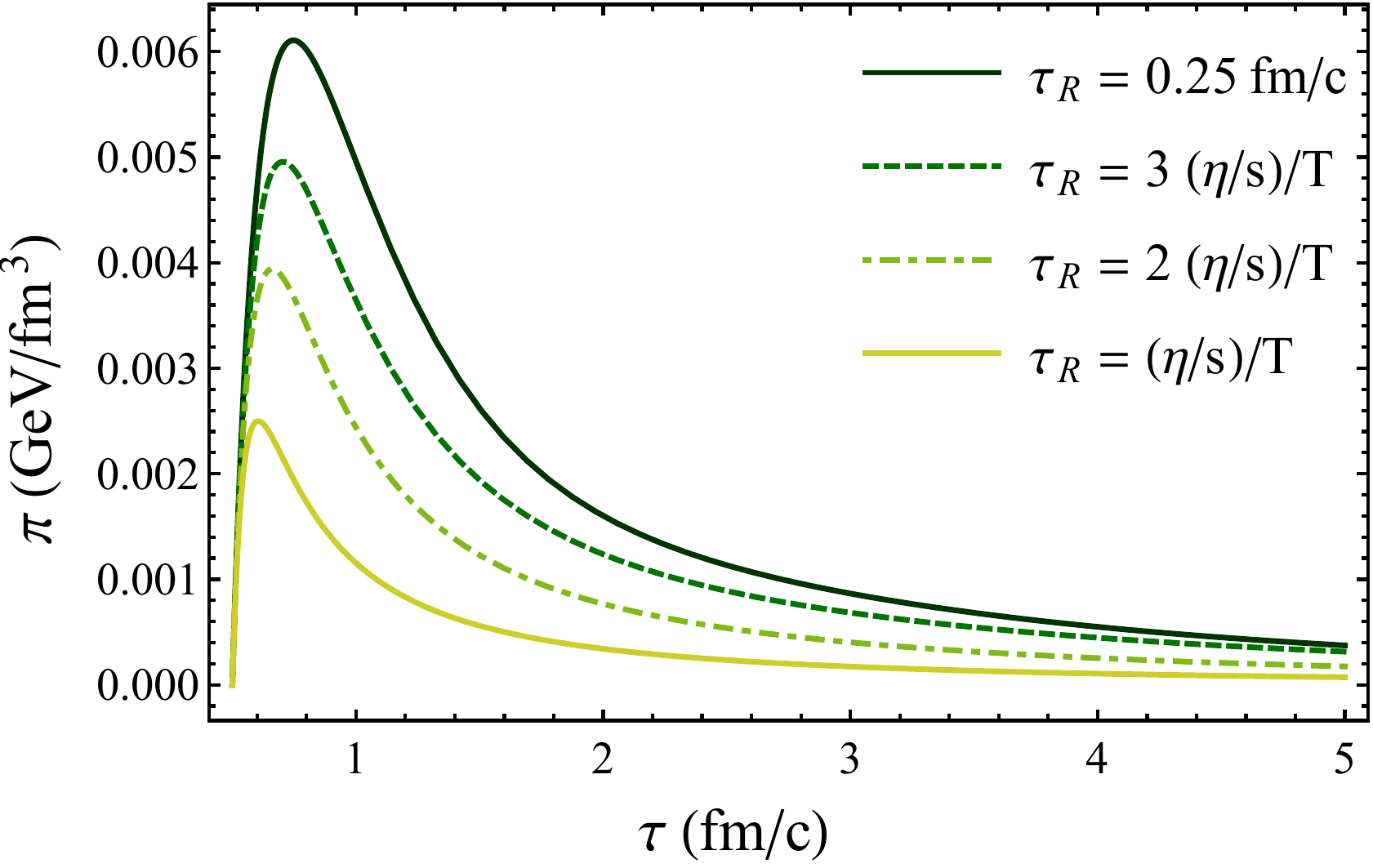}
  \caption{Proper time evolution of shear viscous pressure.}
  \label{shear}
\end{minipage}
\end{figure}
In Figs.~(\ref{bulk}) and (\ref{shear}), we plot the bulk and shear viscous pressures respectively
as a function of proper time for different temperature dependent relaxation times. 
We choose the initial temperature and proper time to be $T_0=0.31$ GeV and 
$\tau_0=0.5$ fm/c respectively. It is observed that both $\pi$ and $\Pi$ have strong dependency on the value of
$\tau_R$. The effect due to viscous contributions increases with increase in $\tau_R$. We also plot the evolution
corresponding to a constant relaxation time, $\tau_R = 0.25$ fm/c (as taken in Ref.~\refcite{Bhadury:2020ivo}) 
for comparison. It can be noted that the viscous effects are maximum for this constant value compared to the
temperature dependent cases considered here.

\subsection{Thermal particle spectra from heavy ion collisions}
In this section, we calculate the thermal particle spectra from heavy ion collisions by employing the second
order hydrodynamic formulation discussed above. We consider dileptons from $q\bar{q}-$annihilation process, 
$q\bar{q}\longrightarrow \gamma^{*} \longrightarrow l^{+}l^{-}$. The rate of dilepton production within EQPM in the presence of viscous 
modified distribution function $f(\vec{p}) = f_0(\vec{p}) + \delta f$ is given by
\begin{equation}
    \frac{dN}{d^4x d^4p} = \iint \frac{d^3\vec{p}_1}{(2\pi)^3} \frac{d^3\vec{p}_2}{(2\pi)^3}\,
 \frac{M_{\textrm{eff}}^2\,g^2 \,\sigma(M_{\textrm{eff}}^2)}{2 \omega_1 \omega_2}
f(\vec{p}_1) f(\vec{p}_2) \delta^4(\tilde{p}-\tilde{p}_1-\tilde{p}_2).
\end{equation}
We consider the viscous correction upto first order in momenta, obtained from Chapman-Enskog method :
\begin{eqnarray}\label{delta_f_2}
 \delta f &=& \delta f_\pi + \delta f_\Pi =\frac{\beta}{2 \beta_\pi (u \cdot \tilde{p})}
 \tilde{p}^\mu \tilde{p}^\nu \pi_{\mu\nu} + 
 \frac{\beta \Pi}{\beta_\Pi} \Big[\xi_1 - \xi_2 (u \cdot \tilde{p}) \Big],
\end{eqnarray}
where
\begin{eqnarray}  \label{coeff_xi}
 \xi_1 = \beta c_s^2 \frac{\partial \delta \omega_q}{\partial \beta} + \delta \omega_q; \,\,\,\,\,\,\,\,\,
 \xi_2 = \left( c_s^2 - \frac{1}{3} \right)  
 + \frac{\delta \omega_q}{3(u \cdot \tilde{p})^2 } \left[2(u \cdot \tilde{p})-\delta \omega_q\right].
\end{eqnarray}
Keeping the terms upto second order in momenta, we can write the total dilepton rate as 
sum of ideal and viscous contributions. The ideal part of dilepton rate, within EQPM is given by
 \begin{eqnarray}
 \frac{dN^{(0)}}{d^4x d^4p} 
 &=& \frac{z_q^2}{2} \frac{M_{\textrm{eff}}^2\,g^2\,\sigma(M_{\textrm{eff}}^2)}{(2\pi)^5}e^{-(u\cdot \tilde{p})/T}.
\end{eqnarray}
The contribution to the dilepton rate due to shear and bulk viscosities are obtained respectively as\cite{Naik:2022}
\begin{eqnarray}
 \frac{dN^{(\pi)}}{d^4x d^4p} &=& \frac{d N^{(0)}}{d^4 x d^4 p}
\Bigg\{\frac{\beta}{\beta_{\pi}} \frac{1 }{ 2|\vec{p}|^5 }
\Bigg[\frac{(u\cdot \tilde{p}) |\vec{p}|}{2}
\left(2|\vec{p}|^2-3 M_{\textrm{eff}}^2\right)\nonumber\\
&&+\frac{3}{4} M_{\textrm{eff}}^4  
\ln \left( \frac{(u\cdot \tilde{p}) + |\vec{p}|}{(u\cdot \tilde{p}) - |\vec{p}|} \right)\Bigg]
\tilde{p}^\mu \tilde{p}^\nu \pi_{\mu\nu}   \Bigg\},\nonumber \\
\frac{dN^{(\Pi)}}{d^4x d^4p} &=& \frac{dN^{(0)}}{d^4x d^4p} \frac{2\beta \Pi}{\beta_\Pi}
 \Bigg\{\beta c_s^2 \frac{\partial \delta\omega_q}{\partial\beta} 
 - \frac{2}{3}\delta\omega_q 
 - \left(c_s^2 -\frac{1}{3} \right)\frac{(u \cdot \tilde{p})}{2} \nonumber \\
 &&+\frac{\delta\omega_q^2}{6 |\vec{p}|^5} \Bigg[\frac{(u\cdot \tilde{p}) |\vec{p}|}{2}
\left(2|\vec{p}|^2-3 M_{\textrm{eff}}^2\right)+\frac{3}{4} M_{\textrm{eff}}^4  
\ln \left( \frac{(u\cdot \tilde{p}) + |\vec{p}|}{(u\cdot \tilde{p}) 
- |\vec{p}|} \right)\Bigg]\Bigg\},  
\end{eqnarray}
where $|\vec{p}| = \sqrt{(u\cdot \tilde{p})^2 - M_{\textrm{eff}}^2}$. Similarly, we determine the photon production rate 
expressions in the presence of modified distribution functions. The ideal, shear and bulk viscous contributions to the
photon production rate are obtained as\cite{Naik:2022}
\begin{eqnarray}
\omega_0 \frac{d N_\gamma^{(0)}}{d^3 p d^4 x} &=& \frac{5}{9} \frac{\alpha_{e} \alpha_{s}}{2 \pi^{2}}
T^2 f(\vec{p}) \left[ \ln \left(\frac{12\,(u\cdot \tilde{p})}{g^{2} T}\right)
 \!+\! \frac{C_{\textrm{ann}} \!+\! C_{\textrm{Comp}}}{2} \right] \nonumber\\
 \omega_0 \frac{d N_\gamma^{(\pi)}}{d^3 p d^4 x} &=& \omega_0 \frac{d N_\gamma^{(0)}}{d^3 p d^4 x}
 \left\{ \frac{\beta}{2 \beta_\pi (u \cdot \tilde{p})} \right\}, \nonumber\\
 \omega_0 \frac{d N_\gamma^{(\Pi)}}{d^3 p d^4 x} &=& \omega_0 \frac{d N_\gamma^{(0)}}{d^3 p d^4 x}
 \left\{ \frac{\beta \Pi}{\beta_\Pi} \Big[\xi_1 - \xi_2 (u \cdot \tilde{p}) \Big]\right\}.
\end{eqnarray}

\subsection{Results and discussions}
\begin{figure}
\centering
\begin{minipage}{.5\textwidth}
  \centering
  \includegraphics[width=.95\linewidth]{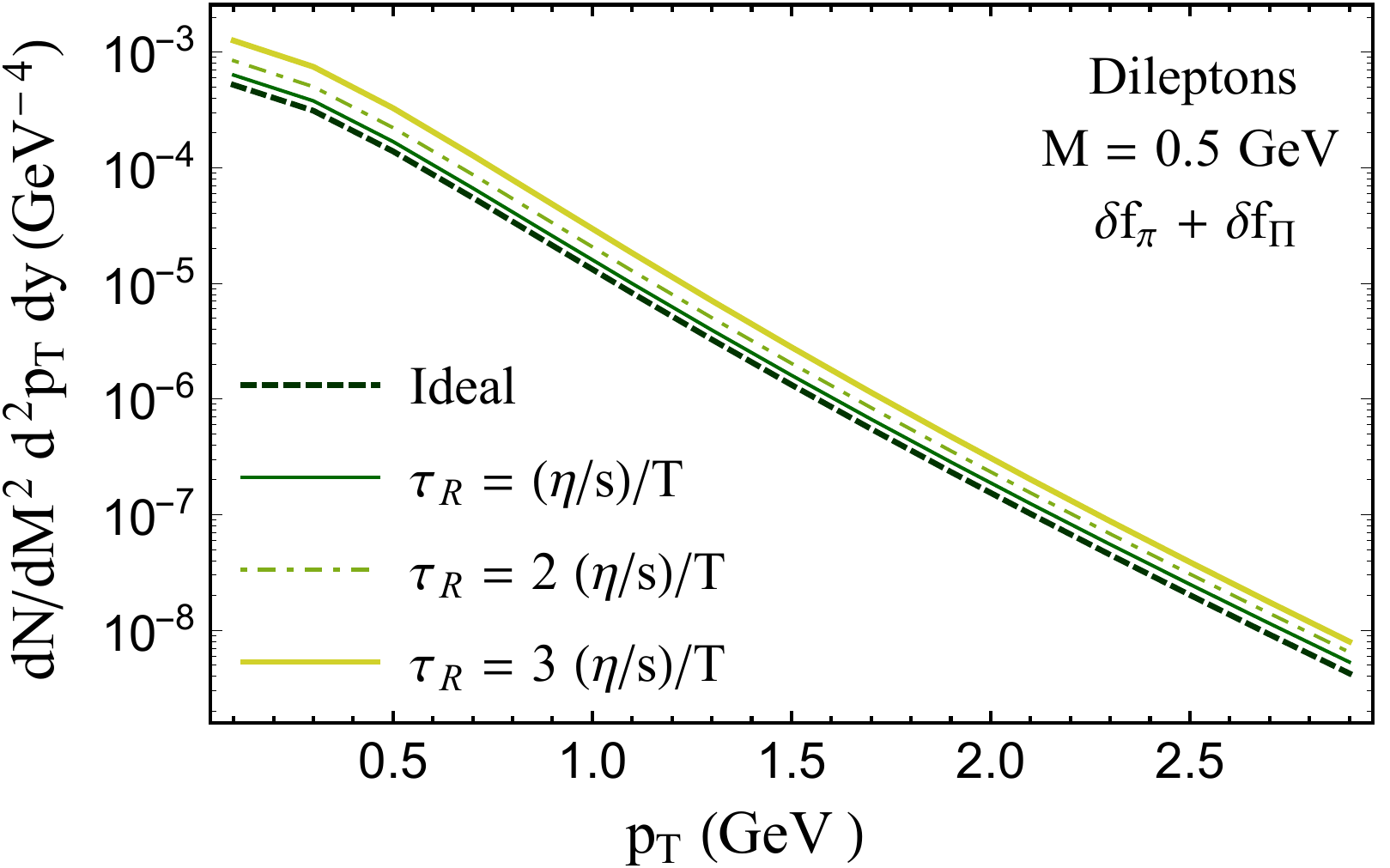}
  \caption[short]{Thermal dilepton spectra in the presence of viscous corrections for 
  different $\tau_R$ and with $M=0.5$ GeV.}
  \label{dilepton}
\end{minipage}%
\begin{minipage}{.5\textwidth}
  \centering
  \includegraphics[width=.95\linewidth]{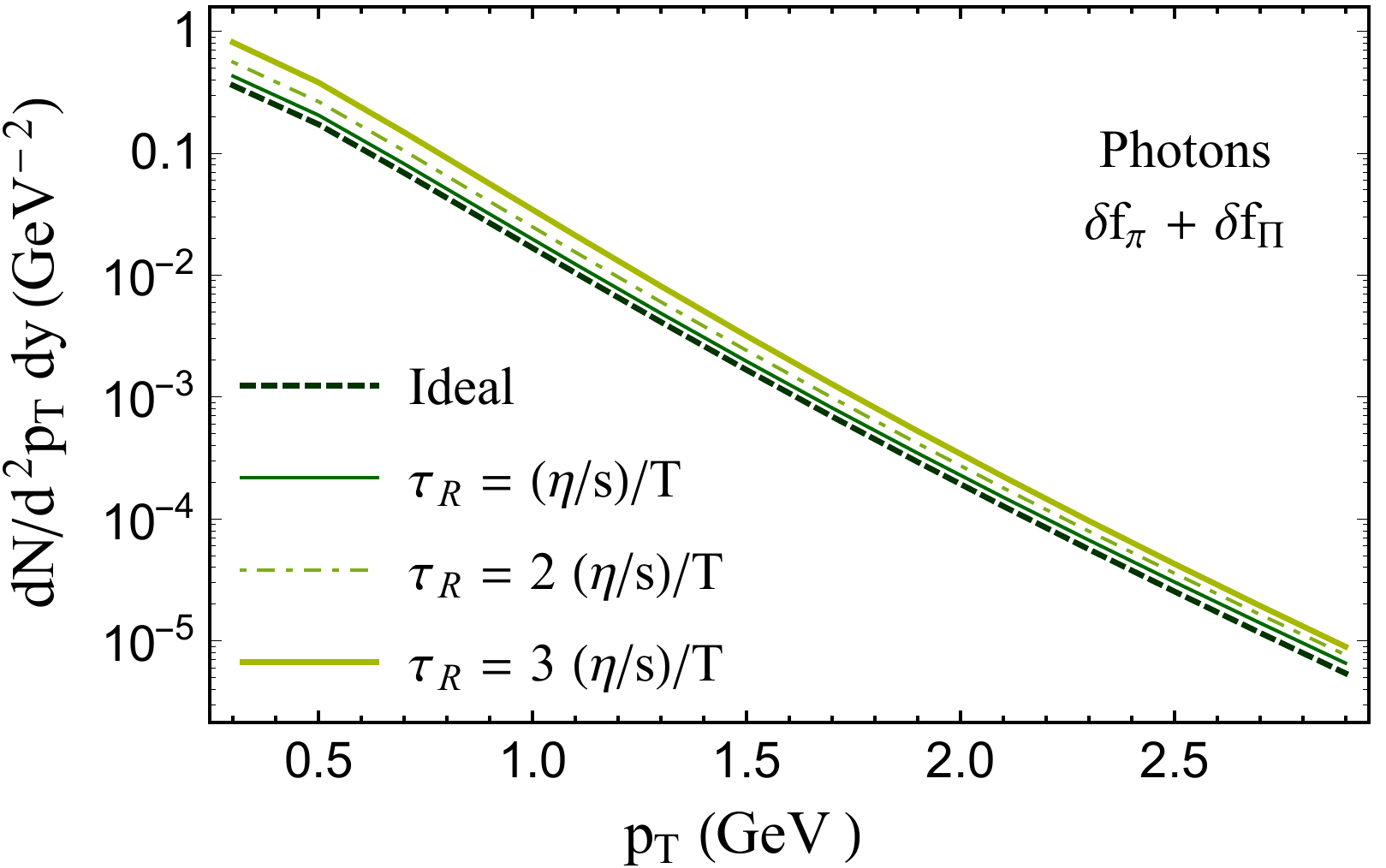}
  \caption[short]{Thermal photon spectra in the presence of viscous corrections by varying $\tau_R$.}
  \label{photon}
\end{minipage}
\end{figure}

We calculate thermal particle spectra
by numerically integrating the above particle rate expressions over the space-time history of the collisions under boost 
invariant Bj\"orken flow. We use the
viscous and temperature evolutions of the QGP obtained in the previous section. 
We choose the initial conditions :
$T_0=0.31$ GeV and $\tau_0=0.5$ fm/c and temperature is evolved till $T_c=0.17$ GeV. 
In Figs.~\ref{dilepton} and \ref{photon}, we plot the thermal dilepton and photon yields respectively
for different temperature dependent relaxation times. 
It can be seen that the particle yields increase with increase in 
magnitude of $\tau_R$. As expected 
(from figs.~\ref{bulk} 
and \ref{shear}), 
we observe maximum enhancement in the spectra with evolution corresponding to $\tau_R = 3(\eta/s)/T$ and minimum 
with $\tau_R = 3(\eta/s)/T$. Our results indicate that the 
particle spectra obtained by employing this second order hydrodynamics is well behaved and is sensitive to $\tau_R$. We note that the Bj\"orken expansion employed here overestimates the yields and a quantitative study requires employing $2+1$ or $3+1$ dimensional hydrodynamic simulations for the evolution.

\section{Relativistic Dissipative Magnetohydrodynamic from Kinetic Theory}
\author{Ankit kumar panda and Victor Roy}	

\bigskip

\begin{abstract}
We derive the second-order magnetohydrodynamics evolution equations of the dissipative stresses for both non-resistive and resistive cases from kinetic theory using the relaxation time approximation for the collision kernel. We found new transport coefficients that were not present in an earlier study by a different group using the 14-moment method. Also, we calculate the anisotropic transport coefficients pertaining to this. We further show the temperature and hadronic mass dependence of the two newly derived leading order transport coefficients $\delta_{VB}$ and $\delta_{\pi B}$.
\end{abstract}



\subsection{Introduction}
The magnetic field seems to play an important role in the working of
our present-day universe. They find a lot of applications, from the laboratory systems to the very large systems like the astrophysical objects. The strength of the magnetic field in these cases can vary by several orders of magnitudes; for example, Earth's magnetic fields on the surface have typical values $10^{-4}$T, whereas one of the strongest fields $\sim 10^{14}\mbox{-}10^{15}$T ever known can be found in the high energy heavy-ion collision experiments at RHIC and LHC. Along with the magnetic field produced in the initial stages of the heavy-ion collisions (primarily due to the spectators), a new form of very hot and dense matter known as quark-gluon plasma~(QGP) is also formed. QGP can also be found in astrophysical objects such as in the core of superdense neutron stars. Considering that the QGP evolves in a strong background magnetic field, we can study its dynamical evolution within the framework of the Relativistic magnetohydrodynamics~(RMHD).

It is well known that a straightforward extension of the Navier-Stokes equations for viscous fluids leads to unacceptable acausal behavior. Israel and Stewart~(IS) were the first who develop a causal and stable relativistic version of the dissipative hydrodynamics known as the second-order hydrodynamics, as it contains dissipative terms proportional to the second-order gradients of the fluid variables. IS equations are applicable for ordinary (uncharged) fluids. Only recently, second-order evolution equations for charged fluids, i.e., the second-order causal magnetohydrodynamics equations, were derived for non-resistive in  Ref.[\citen{Denicol:2018rbw}] and resistive case in Ref.[\citen{Denicol:2019iyh}] for a single component system of spinless particles (no antiparticle) using a 14-moment approximation. In our recent work, Refs.[\citen{Panda:2020zhr,Panda:2021pvq}] we consider the contribution from both particles and antiparticles. We derive the RMHD equations for the non-resistive and resistive cases using the Chapman-Enskog-like gradient expansion of the single-particle distribution function within relaxation time approximation~(RTA). Also, the anisotropic transport coefficients pertaining to this case have been evaluated. In this proceeding, we further study the temperature and hadronic mass dependence of the two newly derived leading order transport coefficients $\delta_{VB}$ and $\delta_{\pi B}$.

\subsection{Formalism }
In this section, we discuss the essential part of the formalism, for details see Refs.[\citen{Panda:2020zhr,Panda:2021pvq}]. The relativistic Boltzmann equation (RBE)  in the presence of a
non-zero force $\mathcal{F}^{\nu}$ is given by:
\begin{equation}\label{RBE}
  p^{\mu}\partial_{\mu}f +\mathcal{F}^{\nu}\frac{\partial}{\partial p^{\nu}}f= C[f],
\end{equation}
where  $f({\bf x,p},t)$ is the one particle distribution function characterising
the phase space density of the particles,  $C[f]$
is the collision kernel. Here we simply the collision kernel using the RTA, that is we choose $C[f]= -\frac{u.p}{\tau_c} \delta f $.
In this approach we calculate off-equilibrium distribution $f$ order-by-order as : 
	\begin{equation}\label{eq:fexpansion}
		f=\sum_{n=0}^{\infty}\left(-1\right)^n \left(\frac{\tau_c}{u.p}\right)^n\left(p^{\mu}\partial_{\mu}+qF^{\mu\nu}p_{\nu}\frac{\partial}{\partial p^{\mu}}\right)^n f_0.
	\end{equation}
We use $\mathrm{Kn} = \tau_c \partial_{\mu}$ ,$\chi = qB\tau_c /T$ and  $\xi=qE\tau_c/T$ as small parameters for the order-by-order expansion. Truncating this series upto the second-order we get:
	\begin{eqnarray}\label{eq:uptosecondorderf}
		f=f_0+\delta{f}^{(1)}+\delta{f}^{(2)}.
	\end{eqnarray}
We evaluate the dissipative part of the energy-momentum tensor (which includes the shear,
bulk viscosity, and diffusion) using $\delta \tilde{f}^{(1,2)}$ and $\delta \tilde{\bar{f}}^{(1,2)}$ in the following expressions,
\begin{eqnarray}
\label{firstshear}
  \pi^{\mu\nu}&=&\Delta^{\mu\nu}_{\alpha\beta}\int dp p^{\alpha}p^{\beta}\left(\delta \tilde{f}+\delta \tilde{\bar{f}}\right),\\
\label{firstbulk}
  \Pi&=&-\frac{\Delta_{\mu\nu}}{3}\int dp p^{\mu}p^{\nu}\left(\delta \tilde{f}+\delta \tilde{\bar{f}}\right),\\
  V^{\mu}&=&\Delta^{\mu}_{\alpha}\int dp p^{\alpha}\left(\delta \tilde{f}-\delta \tilde{\bar{f}}\right).
\end{eqnarray}
Using the above definitions along with the conservation equations for energy-momentum we get the magnetohydrodynamics evolution equations
for the dissipative quantities.
\subsection{Second order Evolution equations for dissipative stresses}
From the above formalism we derive all the set of equations for non-resistive and resistive mhd for the dissipative stresses. First for the non-resistive case we have :
\begin{eqnarray}
\nonumber
  \frac{\pi^{\mu\nu}}{\tau_c}&=&-\dot{\pi}^{\mu\nu}+2\beta_{\pi}\sigma^{\mu\nu}
  +2\pi^{\langle\mu}_{\gamma}\omega^{\nu\rangle\gamma}-\tau_{\pi\pi}
  \pi^{\langle\mu}_{\gamma}\sigma^{\nu\rangle\gamma} -\delta_{\pi\pi}\pi^{\mu\nu}\theta
  +\lambda_{\pi\Pi}\Pi\sigma^{\mu\nu}\\
  \nonumber
   &&-\tau_{\pi V}V^{\langle\mu}\dot{u}^{\nu\rangle}+\lambda_{\pi V}V^{\langle\mu}\nabla^{\nu \rangle}\alpha +l_{\pi V}\nabla^{\langle\mu}
   V^{\nu\rangle} +\delta_{\pi B}\Delta^{\mu\nu}_{\eta \beta}q B b^{\gamma \eta}g^{\beta \rho}
   \pi_{\gamma\rho}\\
   \label{eq:shear_evolution2} \nonumber
   &&-\tau_c qB\lambda_{\pi VB} V_{\gamma}b^{\gamma\langle\mu}\nabla^{\nu\rangle}\alpha
  -q\tau_c \delta_{\pi VB}   \nabla^{\langle\mu}\left(B^{\nu\rangle\gamma}V_{\gamma} \right)-\tau_c qB\tau_{\pi VB} \dot{u}^{\langle\mu}b^{\nu\rangle\sigma} V_{\sigma},\\ \nonumber
   \frac{\Pi}{\tau_c}&=&-\dot{\Pi}-\delta_{\Pi\Pi}\Pi \theta +\lambda_{\Pi\pi}\pi^{\mu\nu}
   \sigma_{\mu\nu}-\tau_{\Pi V}V\cdot \dot{u}-\lambda_{\Pi V}V\cdot \nabla \alpha
    -l_{\Pi V}\partial \cdot V-\beta_{\Pi}\theta \\ \nonumber
    &&+\tau_c \tau_{\Pi V B}\dot{u}_{\alpha} qBb^{\alpha\beta}V_{\beta}-\tau_c q
    \delta_{\Pi V B}\nabla_{\mu}\left(B b^{\mu\beta}V_{\beta} \right)-
    \tau_c qB\lambda_{\Pi VB} b^{\mu\beta}V_{\beta}\nabla_{\mu}\alpha,
    \label{eq: bulk_evolution2} 
\end{eqnarray}
\begin{eqnarray}
\label{eq:diffusionEvolution2}
\nonumber
  \frac{V^{\mu}}{\tau_c}&=&-\dot{V}^{\langle\mu\rangle}-V_{\nu}\omega^{\nu \mu}
  -\lambda_{VV}V^{\nu}\sigma^{\mu}_{\nu}-\delta_{VV}V^{\mu} \theta
  +\lambda_{V\Pi}\Pi\nabla^{\mu} \alpha-\lambda_{V\pi}\pi^{\mu \nu}\nabla_{\nu}\alpha
 \\ \nonumber
  && -\tau_{V\pi}\pi^{\mu}_{\nu}\dot{u^{\nu}}+\tau_{V\Pi}\Pi \dot{u^{\mu}} +l_{V\pi}\Delta^{\mu \nu}\partial_{\gamma}\pi^{\gamma}_{\nu}
  -l_{V\Pi}\nabla^{\mu}\Pi+\beta_V \nabla^{\mu} \alpha-q B \delta_{V B} b^{\mu\gamma}V_{\gamma}
 \\ \nonumber
  && +\tau_c q B l_{V\pi B} b^{\sigma \mu}\partial^{\kappa}\pi_{\kappa\sigma}  +\tau_c q B \tau_{V \Pi B}b^{\gamma\mu}\Pi \dot{u}_{\gamma}-\tau_c qB l_{V \Pi B} b^{\gamma\mu}
  \nabla_{\gamma}\Pi \\ \nonumber
  && -q\tau_c \delta_{VVB} B b^{\mu\nu}V_{\nu}\theta  -q \tau_c \lambda_{VVB}B
  b^{\gamma \nu}V_{\nu}\sigma^{\mu}_{\gamma}-q \tau_c \mathbf{\rho}_{VVB} B b^{\gamma \nu}V_{\nu}
  \omega^{\mu}_{\gamma} \\ \nonumber
  &&-\tau_c q \tau_{VVB}\Delta^{\mu}_{\gamma}D\left(Bb^{\gamma \nu}V_{\nu} \right),
\end{eqnarray}

Here we can see that along with all the usual transport coefficients in ordinary hydro we have  co-efficients like $\delta_{\pi B}$, $\lambda_{\pi VB}$, $ \delta_{\pi VB}$, $\tau_{\pi VB}$, $\tau_{\Pi V B}$,  $ \delta_{\Pi V B}$, $\lambda_{\Pi VB}$, $\delta_{VB}$ , $l_{V\pi B}$, $\tau_{V\pi B}$, $l_{V\Pi B}$, $\delta_{VVB}$, $\lambda_{VVB}$, $\rho_{VVB}$, $\tau_{VVB}$ are arising from the external magnetic field.\\
Similarly for resistive case we get :
\begin{eqnarray}
\frac{\Pi}{\tau_c}&=&-\dot{\Pi}-\delta_{\Pi\Pi}\Pi \theta +\lambda_{\Pi\pi}\pi^{\mu\nu}
		\sigma_{\mu\nu}-\tau_{\Pi V}V\cdot \dot{u}-\lambda_{\Pi V}V\cdot \nabla \alpha -l_{\Pi V}\partial \cdot V-\beta_{\Pi}\theta
		\nonumber
		\\ \nonumber
		&&-
		 qB\lambda_{\Pi VB} b^{\mu\beta}V_{\beta}V_{\mu} +\tau_c \tau_{\Pi V B}\dot{u}_{\alpha} qBb^{\alpha\beta}V_{\beta}- q
		\delta_{\Pi V B}\nabla_{\mu}\left(\tau_c B b^{\mu\beta}V_{\beta} \right) \\ \nonumber &&-q^2\tau_{c}\chi_{\Pi EE}E^{\mu}E_{\mu}.
		\label{eq:2bulkevolutionexp} \\ \nonumber
  \frac{V^{\mu}}{\tau_c}&=&-\dot{V}^{\langle\mu\rangle}-V_{\nu}\omega^{\nu \mu}
		+\lambda_{VV}V^{\nu}\sigma^{\mu}_{\nu}-\delta_{VV}V^{\mu} \theta
		+\lambda_{V\Pi}\Pi\nabla^{\mu} \alpha -\lambda_{V\pi}\pi^{\mu \nu}\nabla_{\nu}\alpha
	\\ \nonumber
	&&-\tau_{V\pi}\pi^{\mu}_{\nu}\dot{u^{\nu}}-q B \delta_{V B} b^{\mu\gamma}V_{\gamma} +\tau_{V\Pi}\Pi \dot{u^{\mu}}+l_{V\pi}\Delta^{\mu \nu}\partial_{\gamma}\pi^{\gamma}_{\nu}
		-l_{V\Pi}\nabla^{\mu}\Pi+\beta_V \nabla^{\mu} \alpha \nonumber
		\\ \nonumber
		&&  +\tau_c q B l_{V\pi B} b^{\sigma \mu}\partial^{\kappa}\pi_{\kappa\sigma}-q \tau_c \lambda_{VVB}B
		b^{\gamma \nu}V_{\nu}\sigma^{\mu}_{\gamma}+\tau_c q B \tau_{V \Pi B}b^{\gamma\mu}\Pi \dot{u}_{\gamma} \nonumber \\
		&&  -\tau_c qB l_{V \Pi B} b^{\gamma\mu}
		\nabla_{\gamma}\Pi-q\tau_c \delta_{VVB} B b^{\mu\nu}V_{\nu}\theta -q \tau_c \mathbf{\rho}_{VVB} B b^{\gamma \nu}V_{\nu}
		\omega^{\mu}_{\gamma} 
	  \nonumber \\ \nonumber
	  	&&+ \chi_{VE}  q E^{\mu}+q\Delta^{\mu}_{\alpha}\chi_{VE}D\left(\tau_c E^{\alpha}\right)-q \tau_c \rho_{VE} E^{\mu}\theta	- q \tau_{VVB}\Delta^{\mu}_{\gamma}D\left(\tau_cBb^{\gamma \nu}V_{\nu} \right).
		\label{eq:fulldiffusion}  \\ \nonumber
  \frac{\pi^{\mu\nu}}{\tau_c}&=&-\dot{\pi}^{\left<\mu\nu\right>}+2\beta_{\pi}\sigma^{\mu\nu}
		+2\pi^{\langle\mu}_{\gamma}\omega^{\nu\rangle\gamma}-\tau_{\pi\pi}
		\pi^{\langle\mu}_{\gamma}\sigma^{\nu\rangle\gamma} -\delta_{\pi\pi}\pi^{\mu\nu}\theta  +\lambda_{\pi\Pi}\Pi\sigma^{\mu\nu}\\ \nonumber
		&&-\tau_{\pi V}V^{\langle\mu}\dot{u}^{\nu\rangle}-\tau_c qB\tau_{\pi VB} \dot{u}^{\langle\mu}b^{\nu\rangle\sigma} V_{\sigma}+\lambda_{\pi V}V^{\langle\mu}\nabla^{\nu \rangle}\alpha -l_{\pi V}\nabla^{\langle\mu}
		V^{\nu\rangle}
		\nonumber \\
		&& +\delta_{\pi B}\Delta^{\mu\nu}_{\eta \beta}q B b^{\gamma \eta}g^{\beta \rho}
		\pi_{\gamma\rho} - qB\lambda_{\pi VB} V_{\gamma}b^{\gamma\langle\mu}V^{\nu\rangle}
		-q \delta_{\pi VB}   \nabla^{\langle\mu}\left(\tau_c B^{\nu\rangle\gamma}V_{\gamma} \right) \nonumber
		\\ \nonumber
		&&+ q^2\tau_{c} \chi_{\pi EE} \Delta^{\mu\nu}_{\sigma \rho } E^{\sigma}E^{\rho}. 
		\label{eq:2evolutionexp}
\end{eqnarray}

In resistive case we have $\chi_{\Pi EE}$, $\chi_{ V E}$, $\rho_{VE}$, $\chi_{\pi EE}$ as the new transport co-efficients.
\subsection{Anisotropic transport co-efficients}
As we have external electromagnetic field so the isotropic transport coefficients will now split in different directions of the applied force and hence become anisotropic.  
For anisotropic diffusion co-efficients are :
\begin{eqnarray} \nonumber
 \kappa_{\parallel}&=&\beta_{V}\tau_c,\\ \nonumber
\kappa_{\bot}&=&\frac{\beta_{V}\tau_c}{1+{\left(qB\tau_c\delta_{VB}\right)}^2},\\ \nonumber
\kappa_{\times}&=&\frac{\beta_{V} qB\tau_c^2\delta_{VB}}{1+{\left(qB\tau_c\delta_{VB}\right)}^2}=\kappa_{\bot}qB\tau_c\delta_{VB}.
\end{eqnarray}
We can see that there is no magnetic field dependence along the direction of B (magnetic field) which is expected and others are  $k_{\bot}$ and $k_{\times}$ which are the perpendicular component and the hall coefficient respectively.
 For the shear case we have:
\begin{eqnarray} \nonumber
 \eta_0&=&2\beta_{\pi}\tau_c,\\ \nonumber 
  \eta_1&=&\frac{2\beta_{\pi}\tau_c}{1+{\left(2qB\tau_c\delta_{\pi B}\right)}^2},\\ \nonumber
  \eta_2&=&\frac{4\beta_{\pi}qB\tau_c^2\delta_{\pi B}}{1+{\left(2qB\tau_c\delta_{\pi B}\right)}^2}=2\eta_1qB\tau_c\delta_{\pi B},\\ \nonumber
  \eta_3&=&\frac{2\beta_{\pi}\tau_c}{1+{\left(qB\tau_c\delta_{\pi B}\right)}^2},\\ \nonumber
   \eta_4&=&\frac{2\beta_{\pi}qB\tau_c^2\delta_{\pi B}}{1+{\left(qB\tau_c\delta_{\pi B}\right)}^2}={\eta_3 qB\tau_c\delta_{\pi B}}.
\end{eqnarray}
In this case $\eta_0$ is the coefficient along the direction of applied field, whereas $\eta_1$, $\eta_3$ are perpendicular components, with $\eta_2$ and $\eta_4$ being the hall coefficients. 

Lastly for anisotropic coefficients for conductivity we have:  
\begin{eqnarray}
	\nonumber{\label{eq:conductivity}}
		\sigma_E^{\parallel}&=&q^2 \tau_c \beta \beta_{V} ,\\
		\nonumber
		\sigma_{E}^{\bot}&=&\frac{q^2 \tau_c \beta \beta_{V}}{1+\left(qB \tau_c \delta_{VB}\right)^2},\\ \nonumber
		\sigma_{E}^{\times}&=&\frac{q^3 B \tau_c^2 \beta \beta_{V}  \delta_{VB}}{1+\left(qB \tau_c \delta_{VB}\right)^2}.
\end{eqnarray}	
	
Here also we have the same interpretation for the said notations as in case of the diffusion.
The leading contributors to the  anisotropy are $\delta_{\pi B}$ and $\delta_{VB}$ and its variation with mass and temperature has been studied below in Figs.[\ref{afig1},\ref{afig2}] .

\begin{figure}[h] 
 \centering
  \begin{minipage}{.5\textwidth}
   \begin{center}
  \includegraphics[width=1.\linewidth]{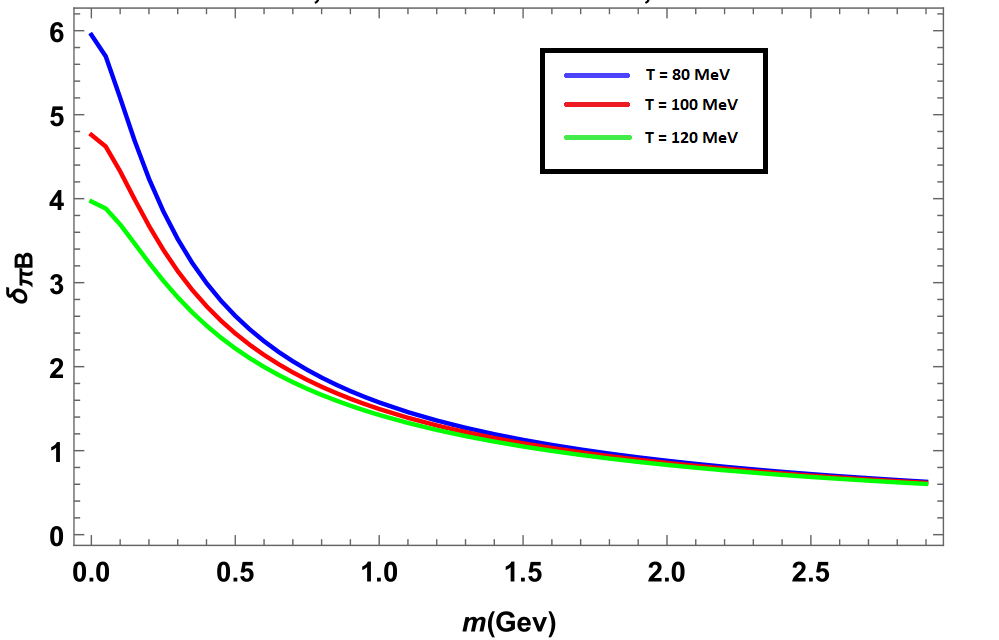}
 \end{center}
 \end{minipage}%
 \begin{minipage}{.5\textwidth}
  \begin{center}
   \includegraphics[width=1\linewidth]{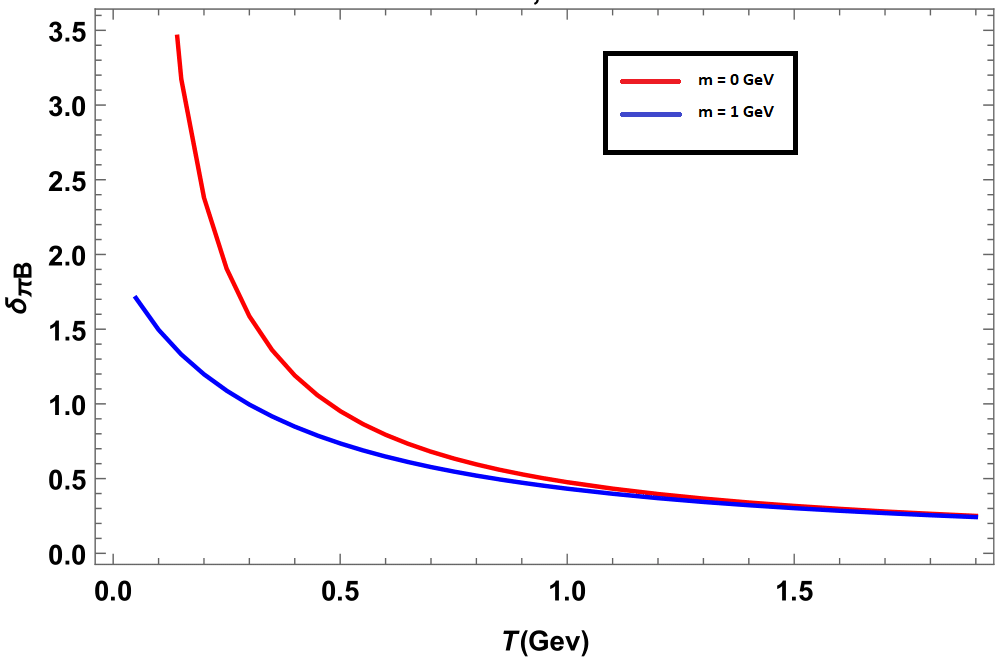}
   \end{center}
  \end{minipage}%
  \caption{Here  left plot shows variation of $\delta_{\pi B}$ with mass for different values of temperature. Right plot shows variation of $\delta_{\pi B}$ with temperature for different values of masses.}
  \label{afig1}
  \end{figure} 

\begin{figure}[h] 
 \centering
  \begin{minipage}{.5\textwidth}
   \begin{center}
  \includegraphics[width=1.0\linewidth]{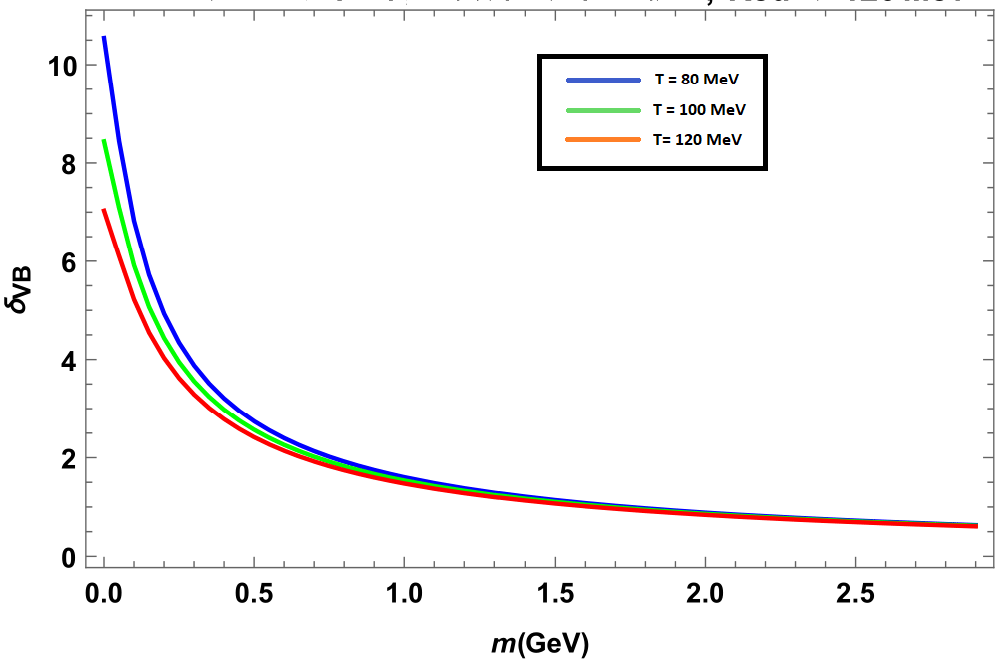}
 \end{center}
 \end{minipage}%
 \begin{minipage}{.5\textwidth}
  \begin{center}
   \includegraphics[width= 1.0\linewidth]{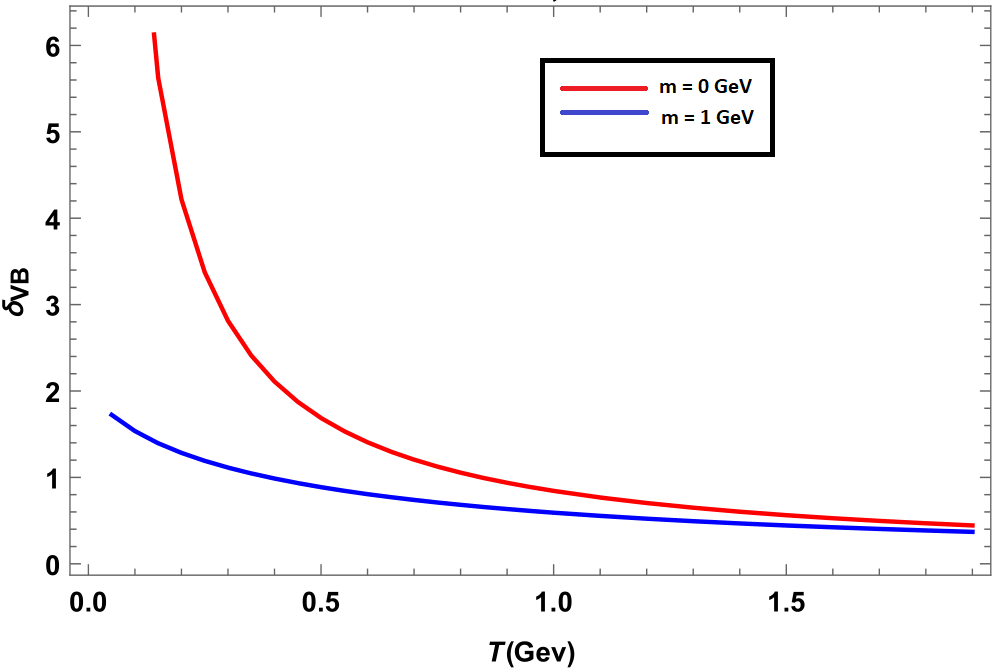}
   \end{center}
  \end{minipage}%
  \caption{Here left plot shows variation of $\delta_{V B}$ with mass for different values of temperature. Right plot shows variation of $\delta_{V B}$ with temperature for different values of masses.}
  \label{afig2}
  \end{figure} 

 Here we can see that $\delta_{\pi B}$ and $\delta_{VB}$ are sensitive to different temperature at low mass region Fig.[\ref{afig1} (left),\ref{afig2} (left)] and sensitive to different mass at low temperature region  in Fig.[\ref{afig1} (right),\ref{afig2} (right)].

\section{Parameters estimation of the Viscous Blast-Wave model using Machine learning techniques}
\author{Nachiketa Sarkar and Amaresh Jaiswal}	

\bigskip

\begin{abstract}
	Recently different statistical-based Machine learning techniques are being used vastly in the field of computational heavy-ion physics to overcome the need for immense computational power. We have developed a general machine learning code using the bayesian statistics that enables us to quantify the multi-parameters model by comparing multiple experimental observables simultaneously. Though this framework is universal and can be applied to any model or data set, in this study, we have implemented this framework in the Viscous Blast-Wave model, which has six parameters, including the $\eta/s$. We have calibrated the model to reproduce experimental data and extracted all the model parameters and their correlation simultaneously.
\end{abstract}

\subsection{Introduction}

The primary goal of the ultra-relativistic heavy-ion experiments is to inspect the properties of quark-gluon plasma (QGP) produced in such collisions. In the absence of a smoking-gun signal, a general approach to understand the QGP properties is to compare experimental results with model predictions. Numerical descriptions of heavy-ion physics are very complex and need huge CPU time, sometimes beyond our reach. Different phases of the heavy-ion collision, i.e., from collision geometry to final state hadron production, are described by different physics models. The model often takes multiple inputs and produces multiple observables that usually have complex correlations. Recently, efficient implementation of the bayesian techniques has been done in \cite{1_Nachiketa,2_Nachiketa} to overcome the massive computational challenges of quantifying the properties of QGP by extracting model parameters via comparing the model to data.  Following the prescription in \cite{1_Nachiketa,2_Nachiketa}, we have developed our parameter-estimation code using different machine learning techniques. This framework is wholly based on the data-driven approach and can be applied to any data or model. The present work is dedicated to estimating the Viscous Blast-Wave (VBW) model parameters. For the constraints of space, we will not discuss the VBW model here, details can be found in the ref. \cite{3_Nachiketa}. The VBW model has six input parameters, including freeze-out temperature $T_f$, the radial velocity $\beta_0$ at the freeze-out surface, and $\eta/s$. The other three parameters, $m$, $\kappa$, and $\alpha_0$ related to the freeze-out radius and flow harmonics. For detailed information about parameters see ref. \cite{3_Nachiketa}. Our goal is the quantitative-estimation of these parameters by simultaneous fitting the experimental $p_T$ spectrum of different hadrons for different centrality classes.

\subsection{Model Description and Result}

We have performed our analyses in two phases. In the first step, using the Gaussian Process(GP) emulator, we have built a surrogate model of the given physics model of interest, i.e., fast model prediction for any arbitrary set of parameters without explicitly executing the VBW-code. In the next phase, using bayesian statistics, we extracted the posterior probability distribution of the model parameters by comparing the experimental result.
 
\subsection{Gaussian Process Emulator}

\begin{figure}[t]
	\centering
	\includegraphics[width=15.0cm,height=15cm,keepaspectratio]{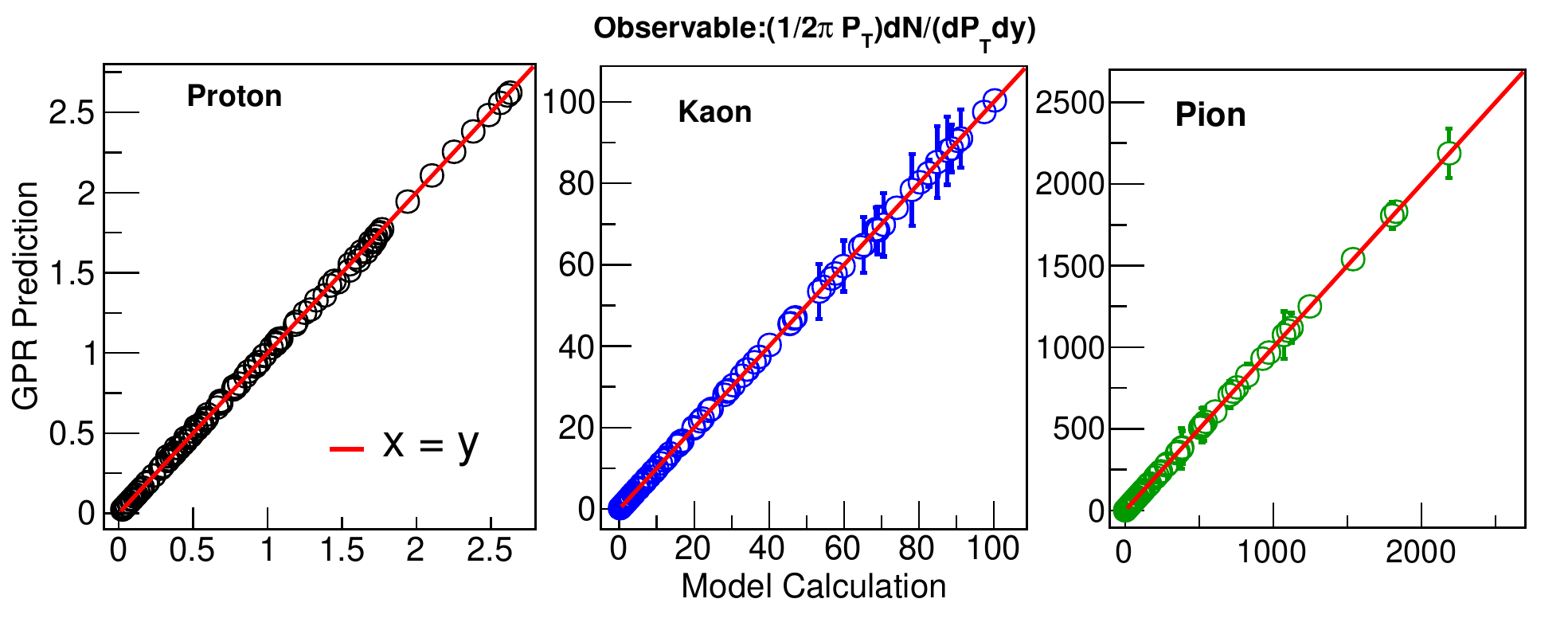}
	\caption{The emulator validation plots, represent the comparison between the emulator predictions and VBW model outputs for fifty randomly generated points in the parameter space.
	}
	\label{fig:ModelCompare}
\end{figure}
Gaussian Process (GP) \cite{4_Nachiketa}, is a supervised learning algorithm based on bayesian statistics. The posterior distribution is obtained from the prior distribution through the training data set. To create the training data set we generate 300 $(m=300)$ points $X[x_1,x_2....x_m]$ in the six-dimensional $(n=6)$ parameter space using Latin Hypercube sampling algorithm \cite{5_Nachiketa} and then produce $m$ model outputs $Y[y_1,y_2....y_m]$  at the each point by executing VBW-code. Finally, we trained the emulator using this data set. For a given test point $x_*$, the posterior distribution of the emulator outputs $y_*$ is the  multivariate gaussian distribution, completely defined by mean, $\mu$, and covariance matrix, $\Sigma$, i.e.,

\begin{equation}
\begin{aligned}
y_*  & \sim \mathcal{N} (\mu,\Sigma) \\
\mu =  & \sigma(X_*,X) \sigma(X_*,X)^{-1}y,\\
\Sigma = &\sigma(X_*,X_*)-\sigma(X_*,X)\sigma(X_*,X)^{-1}\sigma(X,X_*)
\end{aligned}
\end{equation}
Where, $\sigma$'s are the square matrices whose each element represents the covariance function between pairs of points. 
\begin{equation}
\sigma(X,X)=\begin{aligned}
\left (
\begin{array}{ccc}
\begin{array}{l}
\sigma(x_1,x_1) 
\end{array}
& \cdots & 
\begin{array}{l}
\sigma(x_1,x_m)
\end{array} \\
\vdots & \ddots & \vdots\\
\begin{array}{l}
\sigma(x_m,x_1) 
\end{array} &
\cdots & 
\begin{array}{l}
\sigma(x_m,x_m) 
\end{array} 
\end{array}
\right )
\end{aligned}
\end{equation}

In the present work, we have chosen the following covariance function,

\begin{equation}
\sigma(x,x \prime)=\sigma_{GP}^2 \exp \Bigg[-\sum_{k=1}^{n} \frac{(x_k-x_k\prime)^2}{2l_k^2}\Bigg]+\sigma_n^2\delta_{xx\prime}.
\end{equation}

$\sigma^2_{GP}$,  $l_k$, and $\sigma_n$ are known as $hyperparameters$,   related to the total variance of the GP, correlation length between pairs of points, and statistical noise, respectively. The $hyperparameters$ are tuned during the tanning process.
 
Finally, to validate emulator prediction, we generate fifty random points in the parameter space and compare emulator predictions with VBW model outputs at each point. We have presented our validation check in Figure \eqref{fig:ModelCompare}. It is clear from Figure \eqref{fig:ModelCompare}, that the emulator faithfully reproduces the VBW model outputs. One should note that, the errors in Figure \eqref{fig:ModelCompare}, represent emulator prediction error which depends on the relative position of the prediction and training point in the parameter space. The emulator error increase if the distance between the nearest training point and the prediction point increases. Thus some points which are far away from training points have larger errors than others.  

Here we like to mention that we have performed Principal Component Analyses (PCA) \cite{6_Nachiketa} to reduce the dimensionality. All the analyses are done on the reduced PCA space and finally performed the reverse transformation to get the results in the physical space.

 \subsection{Model Calibration}
After building a faithful emulator, the final task left is the model calibration, thereby extracting the posterior distributions of model parameters by comparing experimental data\cite{7_Nachiketa}. According to bayesian statistics, if $x_*$ represents the desired set of input that reproduces the experimental results, then the posterior distribution is,

 \begin{figure}[t]
	\centering
	\includegraphics[width=12.0cm,height=15cm,keepaspectratio]{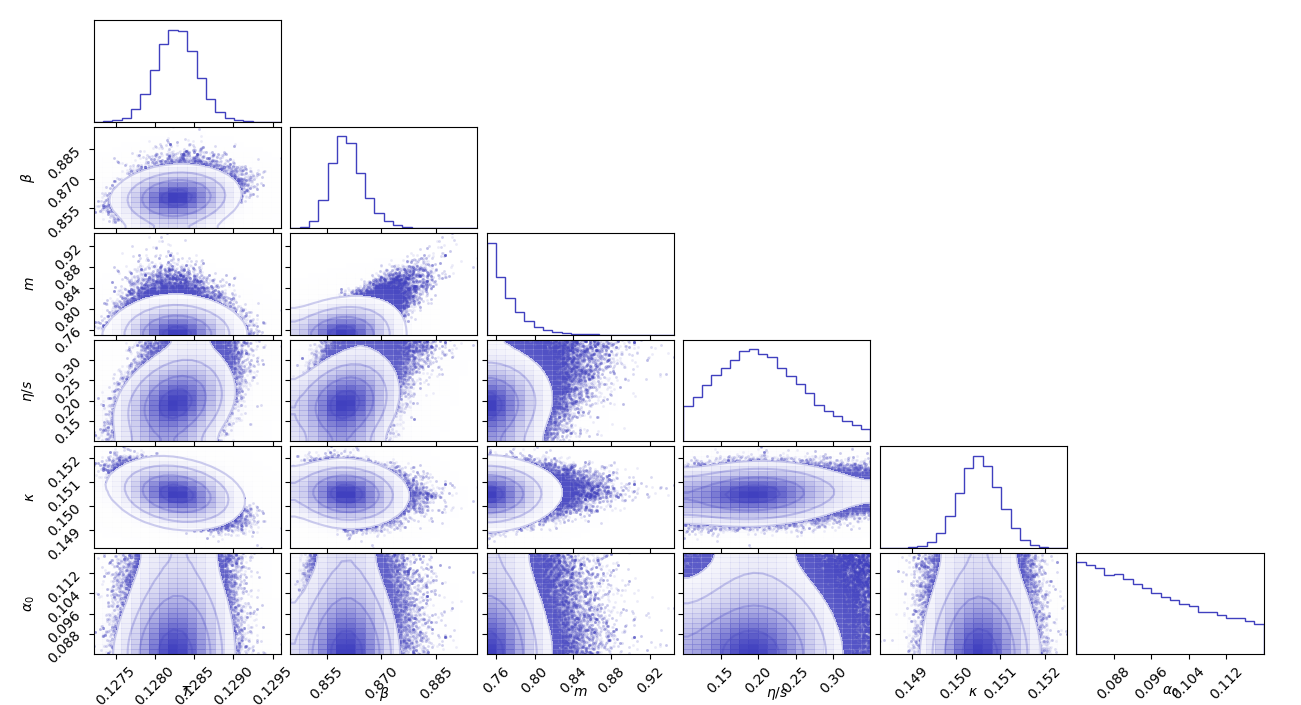}
	\caption{The corner plot generated from the MCMC samples represents the posterior marginal and joint probability distribution of the VBW model parameters.  }
	\label{CornerPlot}
\end{figure}

 \begin{figure}[t]
 	\centering
 	\includegraphics[width=8.0cm,height=15cm,keepaspectratio]{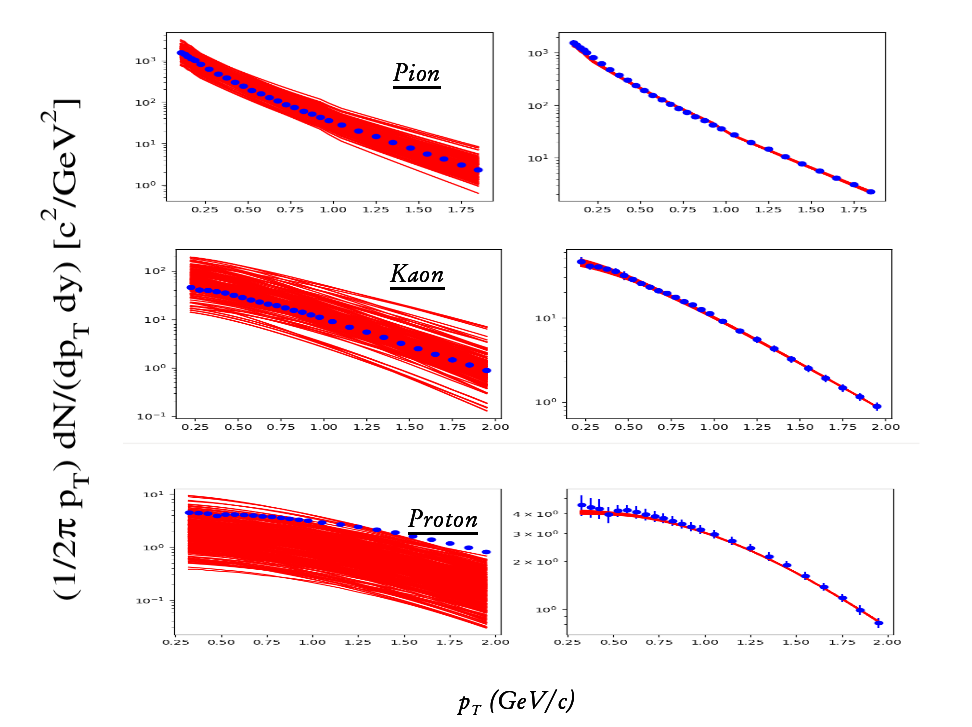}
 	\caption{Comparison of the transverse momentum distribution of the hadrons in the Pb+Pb collision for the most central events $(0-5\%)$ between the experimental results and emulator predictions with prior and posterior parameter sets. The symbols represent experimental data \cite{7_Nachiketa}. }
 	\label{fig:Fig3}
 \end{figure}
 
 \begin{equation}
 P(x_*|X,Y,y_{exp}) \propto P(X,Y,y_{exp}|x_*)P(x_*)
 \end{equation}  
 
 For the Bayesian analysis, we used Markov chain Monte Carlo (MCMC) based on affine-invariant ensemble sampling algorithm\cite{8_Nachiketa} employing the following likelihood and prior distribution function.
 
  \begin{equation}
 P(x_*|X,Y,y_{exp}) \propto \exp \Big[-1/2(z_*-z_{exp})^T\Sigma^{-1}_z(z_*-z_{exp})\Big] 
 \end{equation}  
 
 \begin{equation}
 	P(x_*) = \begin{cases}
 		1 &    min (x_i) \le x_i \le max(x_i), \forall i \\
 		0 & otherwise.
 	\end{cases}
 \end{equation}
  
  \begin{table}[pt]
  	   
  	\caption{Quantitative estimated parameters values of the VBW model from the posterior distribution. C.I. represents the confidence interval.}
  	{\begin{tabular}{@{}cccccccc@{}} \toprule
  			 C.I. & $T_F$ & $\beta_0$& m & $\eta/s$&$\kappa$& $\alpha_0$ &\\
  			  \colrule
  			99\%\hphantom{00} & \hphantom{0}126.6-129.0 & \hphantom{0}0.85-0.88 & 0.75- 0.84&0.16- 0.32&0.149-0.152 &0.08-0.12 \\
  		 
  			95\%\hphantom{00} & \hphantom{0}126.8-128.8 & \hphantom{0}0.84-0.87 & 0.75- 0.81&0.15- 0.30&0.149-0.151&0.08-0.12  \\ \botrule

  	\end{tabular}}
  		\label{table:Tab1}	
 
  \end{table}
  $z_*$ and $z_{exp}$ are the principal components of emulator outputs, $y_*$ and experimental data $y_{exp}$ respectively. $\Sigma_z$ is the covariance or the uncertainty matrix, for this work we take 
  $\Sigma_z= diag(\sigma^2_z z_{exp})$ with $\sigma^2_z$ = 0.10 \cite{2_Nachiketa}.
  
  The final results of our analyses are presented in Figure \eqref{CornerPlot} and \eqref{fig:Fig3}. Figure \eqref{CornerPlot} is the well-known corner plot, where the diagonal plots are the marginal distributions of the parameters and off-diagonal plots represent contour plots of the correlations between the pair of parameters. The estimated values of different parameters are given in Table \eqref{table:Tab1}. Figure \eqref{fig:Fig3}  represents the comparison between the experimental data with emulator predication for the prior and posterior parameter sets. Figures in the right column represent the visualization confirmation that the posterior parameters sets truly represent the experimental data.

\subsection{Summary and Future plan}
In summary, we have presented the Viscous Blast-Wave model parameters estimation analyses using bayesian technique and different machine learning tools. This is our preliminary analyses using the machine learning techniques. We are working to incorporate different flow coefficients as well as different centrality classes in our future analyses.

\section{Charge and heat transport properties of a weakly magnetized hot QCD matter at finite density}
\author{Shubhalaxmi Rath}	

\bigskip

\begin{abstract}
The effect of weak magnetic field on the transport of charge and heat in hot and dense QCD matter has been explored by calculating electrical ($\sigma_{\rm el}$) and thermal ($\kappa$) conductivities in kinetic theory approach. The interactions among partons have been encoded in their thermal masses. We have noticed that both $\sigma_{\rm el}$ and $\kappa$ decrease with magnetic field, whereas, these transport coefficients increase with chemical potential. Further, we have observed the effects of weak magnetic field and chemical potential on the Knudsen number. We have observed a reduction of the Knudsen number in a weak magnetic field, contrary to its enhancement at finite chemical potential. However, its value remains below unity, so, the hot and dense QCD matter remains in the equilibrium state in the presence of a weak magnetic field. 

\end{abstract}

\subsection{Introduction}
Strong evidences for the production of quark-gluon plasma (QGP) have been found in heavy ion collisions at Relativistic Heavy Ion Collider (RHIC) and Large Hadron Collider (LHC). High temperature and/or high density can facilitate the production of QGP. In addition, for noncentral collisions, strong magnetic fields are produced with strength varying between $eB=m_{\pi}^2$ ($\simeq 10^{18}$ Gauss) at RHIC and $eB=15$ $m_{\pi}^2$ at LHC. These magnetic fields are transient, so, there are two scenarios of magnetic field: strong magnetic field ($eB\gg T^2$) and weak magnetic field ($T^2\gg eB$). However, the electrical conductivity can noticeably elongate the lifetimes of such magnetic fields \cite{Tuchin:2013ie,Rath:PRD100'2019}. In the presence of a magnetic field, charge and heat transport coefficients acquire multicomponent structures \cite{Lifshitz:BOOK'1981}. Out of different components, our present analysis focuses on the study of electrical ($\sigma_{\rm el}$) and thermal ($\kappa$) conductivities using kinetic theory approach in the relaxation time approximation, where thermal masses of particles encode the interactions among them. Previously, thermal mass of quark in the strong magnetic field regime had been calculated and used in the study of transport coefficients \cite{Rath:PRD100'2019,Rath:PRD102'2020}. In weak magnetic field limit, we assume that the phase space and the single particle energies are not affected by the magnetic field through Landau quantization \cite{Feng:PRD96'2017}, rather, the main contribution of the magnetic field enters through the cyclotron frequency. As an application, we also study the Knudsen number ($\Omega$) to understand the local equilibrium property in the presence of weak magnetic field and finite chemical potential. 

\subsection{Charge transport coefficient}
The external electric field disturbs the system infinitesimally, thus resulting an induced electric current density, which is written as
\begin{eqnarray}\label{current}
J^\mu=\sum_f g_f \int\frac{d^3\rm{p}}{(2\pi)^3}
\frac{p^\mu}{\omega_f} [q\delta f_f(x,p)+{\bar q}\delta \bar{f_f}(x,p)]
~,\end{eqnarray}
where $g_f$ is the degeneracy factor of quark with flavor $f$, $g_f=6$. According to the Ohm's law, the longitudinal component of the spatial part of four-current density is directly proportional to the electric field with the proportionality factor being the electrical conductivity ($\sigma_{\rm el}$), 
\begin{equation}\label{Multicomponent structure}
J^i=\sigma_{\rm el}\delta^{ij}E_j
~.\end{equation}
The $\delta f_f$ is calculated from the relativistic Boltzmann transport 
equation in the relaxation time approximation, 
\begin{equation}\label{R.B.T.E.}
p^\mu\frac{\partial f_f(x,p)}{\partial x^\mu}+\mathcal{F}^\mu\frac{\partial f_f(x,p)}{\partial p^\mu}=-\frac{p_\nu u^\nu}{\tau_f}\delta f_f(x,p)
~,\end{equation}
where $f_f=\delta f_f+f_f^0$, $\mathcal{F}^\mu=qF^{\mu\nu}p_\nu=(p^0\mathbf{v}\cdot\mathbf{F}, p^0\mathbf{F})$ with $F^{\mu\nu}$ being the electromagnetic field strength tensor and the Lorentz force, $\mathbf{F}=q(\mathbf{E}+\mathbf{v}\times\mathbf{B})$. The relaxation time for quark (antiquark), $\tau_f$ ($\tau_{\bar{f}}$) is written \cite{Hosoya:NPB250'1985} as
\begin{equation}
\tau_{f(\bar{f})}=\frac{1}{5.1T\alpha_s^2\log\left(1/\alpha_s\right)\left[1+0.12(2N_f+1)\right]}
~.\end{equation}
For a spatially homogeneous distribution function with steady-state condition ($\frac{\partial f_f}{\partial \mathbf{r}}=0$ and 
$\frac{\partial f_f}{\partial t}=0$), and for $\mathbf{E}=E\hat{x}$ and $\mathbf{B}=B\hat{z}$, Eq. \eqref{R.B.T.E.} becomes
\begin{equation}\label{R.B.T.E.(3)}
\tau_fqEv_x\frac{\partial f_f}{\partial p_0}+\tau_fqBv_y\frac{\partial f_f}{\partial p_x}-\tau_fqBv_x\frac{\partial f_f}{\partial p_y}=f_f^0-f_f-\tau_fqE\frac{\partial f_f^0}{\partial p_x}
~.\end{equation}
To solve the above equation, the following ansatz has been used, 
\begin{equation}\label{ansatz_l}
f_f=f_f^0-\tau_fq\mathbf{E}\cdot\frac{\partial f_f^0}{\partial \mathbf{p}}-\mathbf{\Gamma}\cdot\frac{\partial f_f^0}{\partial \mathbf{p}}
~,\end{equation}
where $\mathbf{\Gamma}$ depends on magnetic field. Using ansatz \eqref{ansatz_l}, Eq. \eqref{R.B.T.E.(3)} becomes 
\begin{equation}\label{R.B.T.E.(4)}
\tau_fqEv_x\frac{\partial f_f}{\partial p_0}+\beta f_f^0\mathbf{\Gamma}\cdot\mathbf{v}-qB\tau_f\left(v_x\frac{\partial f_f}{\partial p_y}-v_y\frac{\partial f_f}{\partial p_x}\right)=0
~.\end{equation}
From Eq. \eqref{R.B.T.E.(4)} and ansatz \eqref{ansatz_l}, we calculate $\delta f_f$ 
and $\delta \bar{f_f}$ and then by substituting them in Eq. \eqref{current} 
and comparing with Eq. \eqref{Multicomponent structure}, we get $\sigma_{\rm el}$ \cite{Rath:2112.11802} as
\begin{equation}\label{E.C.}
\sigma_{\rm el}=\frac{\beta}{3\pi^2}\sum_f g_f q_f^2\int d{\rm p}\frac{{\rm p}^4}{\omega_f^2}\left[\frac{\tau_f f_f^0\left(1-f_f^0\right)}{1+\omega_c^2\tau_f^2}+\frac{\tau_{\bar{f}} \bar{f_f^0}\left(1-\bar{f_f^0}\right)}{1+\omega_c^2\tau_{\bar{f}}^2}\right]
.\end{equation}

\subsection{Heat transport coefficient}
The heat flow four-vector is the difference between the 
energy diffusion and the enthalpy diffusion, 
\begin{eqnarray}\label{heat flow}
Q_\mu=\Delta_{\mu\alpha}T^{\alpha\beta}u_\beta-h\Delta_{\mu\alpha}N^\alpha
,\end{eqnarray}
where $\Delta_{\mu\alpha}=g_{\mu\alpha}-u_\mu u_\alpha$ 
and the enthalpy per particle $h=(\varepsilon+P)/n$ 
with the particle number density $n=N^\alpha u_\alpha$, the 
energy density $\varepsilon=u_\alpha T^{\alpha\beta} u_\beta$ and the 
pressure $P=-\Delta_{\alpha\beta}T^{\alpha\beta}/3$. The spatial 
component of the heat flow is given by
\begin{eqnarray}\label{heat1}
Q^i=\sum_f g_f\int \frac{d^3{\rm p}}{(2\pi)^3} ~ \frac{p^i}{\omega_f}\left[(\omega_f-h_f)\delta f_f(x,p)+(\omega_f-\bar{h}_f)\delta \bar{f}_f(x,p)\right]
.\end{eqnarray}
According to the Navier-Stokes equation, we have
\begin{eqnarray}\label{heat2}
Q^i=-\kappa\delta^{ij}\left[\partial_j T - \frac{T}{\varepsilon+P}\partial_j P\right] 
,\end{eqnarray}
where $\kappa$ is thermal conductivity. With the help of ansatz \eqref{ansatz_l}, Eq. \eqref{R.B.T.E.} becomes
\begin{equation}\label{eq.1}
\tau_f\frac{\partial f_f^0}{\partial t}+\frac{\tau_f\mathbf{p}}{p_0}\cdot\frac{\partial f_f^0}{\partial \mathbf{r}}+\beta f_f^0\mathbf{\Gamma}\cdot\mathbf{v}+\tau_fqEv_x\frac{\partial f_f}{\partial p_0}-qB\tau_f\left(v_x\frac{\partial f_f}{\partial p_y}-v_y\frac{\partial f_f}{\partial p_x}\right)=0
.\end{equation}
From Eq. \eqref{eq.1} and ansatz \eqref{ansatz_l}, we calculate $\delta f_f$ and 
$\delta \bar{f_f}$. Substituting them in Eq. \eqref{heat1} and then comparing 
with Eq. \eqref{heat2}, we get $\kappa$ \cite{Rath:2112.11802} as
\begin{equation}\label{H.C.}
\kappa=\frac{\beta^2}{6\pi^2}\sum_f g_f\int d{\rm p}\frac{{\rm p}^4}{\omega_f^2} \left[\frac{\tau_f \left(\omega_f-h_f\right)^2 f_f^0\left(1-f_f^0\right)}{1+\omega_c^2\tau_f^2}+\frac{\tau_{\bar{f}} \left(\omega_f-\bar{h}_f\right)^2 \bar{f_f^0}\left(1-\bar{f_f^0}\right)}{1+\omega_c^2\tau_{\bar{f}}^2}\right]
.\end{equation}

\subsection{Knudsen number}
The Knudsen number ($\Omega$) gives the information about the local 
equilibrium property of the medium and is 
defined by the ratio of the mean free path ($\lambda$) to the 
characteristic length scale of the medium ($l$). Since $\lambda$ is 
related to the thermal conductivity ($\lambda=\frac{3\kappa}{vC_V}$), one 
can write $\Omega$ in terms of $\kappa$ as
\begin{eqnarray}
\Omega=\frac{\lambda}{l}=\frac{3\kappa}{lvC_V}
~,\end{eqnarray}
where $v$ is relative speed and $C_V=\partial (u_\mu T^{\mu\nu}u_\nu)/\partial T$ is specific heat at constant volume. In our calculation, we have set $v\simeq 1$ 
and $l=4$ fm. 

The aforementioned transport coefficients and application are studied by 
using the thermal masses of charged particles. The thermal mass 
(squared) of quark is given \cite{Peshier:PRD66'2002} by $m_{fT}^2=\frac{g^2T^2}{6}\left(1+\frac{\mu_f^2}{\pi^2T^2}\right)$. We have taken the chemical 
potentials for all flavors to be the same, {\em i.e.} $\mu_f=\mu$. 

\subsection{Results and discussions}
\begin{figure}[h]
\begin{center}
\begin{tabular}{c c}
\includegraphics[width=4.79cm]{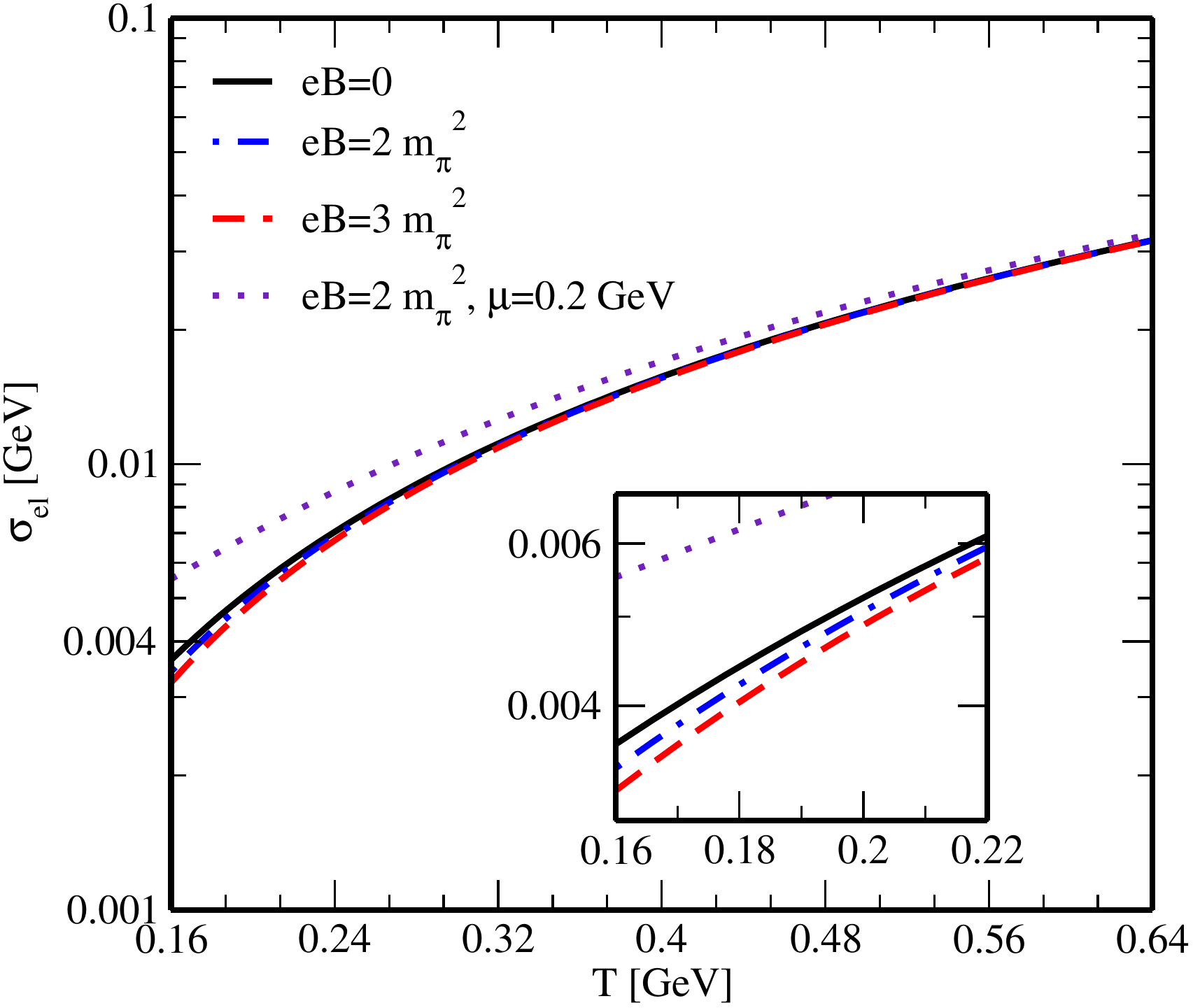}&
\hspace{0.249 cm}
\includegraphics[width=4.79cm]{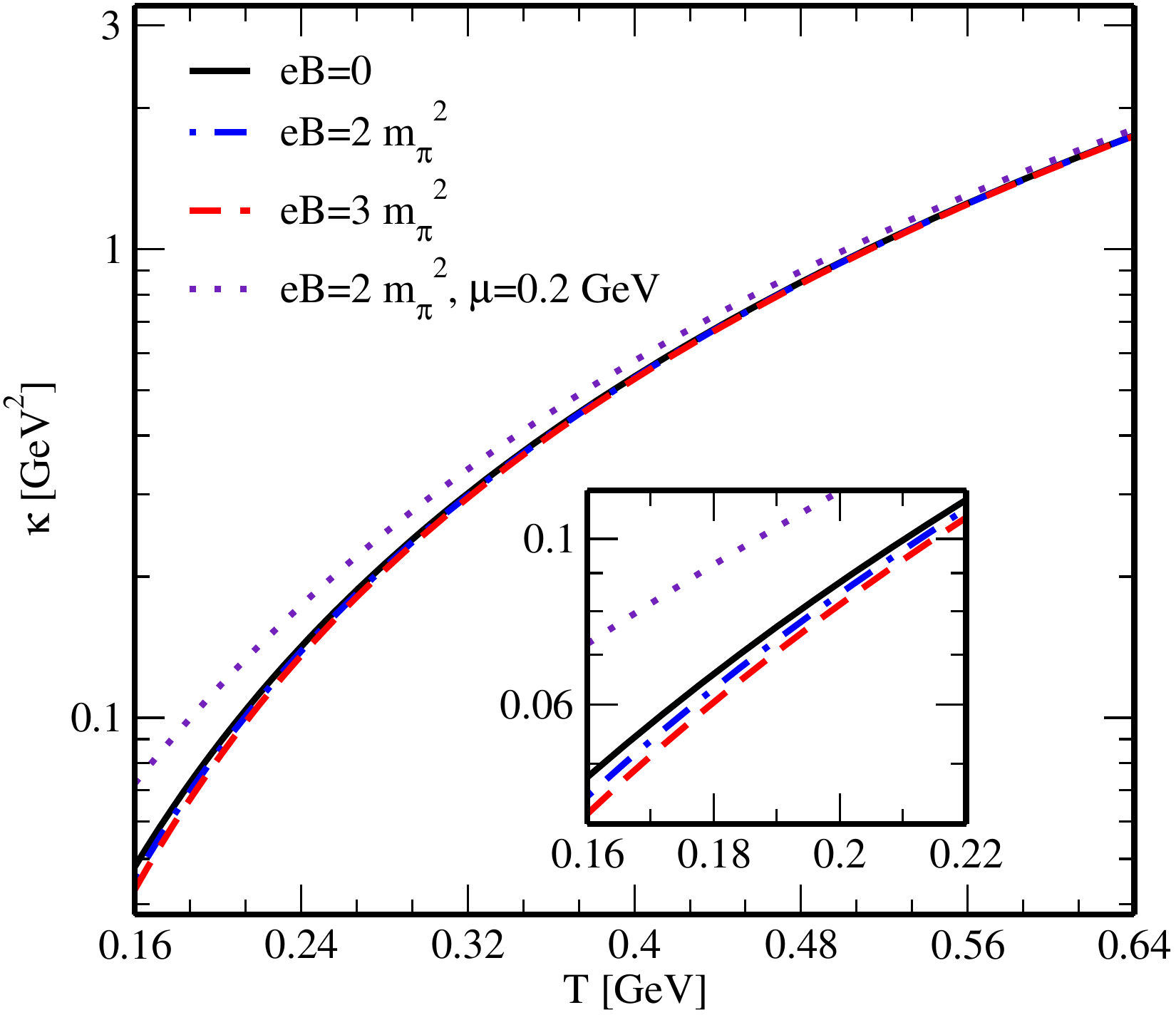} \\
a & b
\end{tabular}
\caption{Variations of (a) $\sigma_{\rm el}$ and (b) $\kappa$ with temperature for different weak magnetic fields and chemical potentials.}\label{Fig.1}
\end{center}
\end{figure}

From Fig. \ref{Fig.1}a, it is observed that, $\sigma_{\rm el}$ gets decreased 
in the presence of a weak magnetic field, which can be understood as 
follows: $\sigma_{\rm el}$ is directly proportional to the current along the 
direction of electric field. But, due to 
the emergence of magnetic field, the direction of moving quarks gets deflected 
and it causes a reduction of current along the direction of electric field, thus 
decreasing the electrical conductivity. On the other hand, an increase of 
$\sigma_{\rm el}$ is observed at finite chemical potential, which is mainly 
due to the increase of distribution function at finite chemical potential. Thus, 
weak magnetic field reduces the charge conduction in QCD medium, whereas finite 
chemical potential facilitates this. From Fig. \ref{Fig.1}b, it is found that the 
weak magnetic field reduces $\kappa$ too, whereas its increase is observed 
at finite chemical potential. The increase of $\kappa$ with temperature is mainly due to 
the increase of both enthalpy per particle and distribution function 
with temperature. Thus, like charge conduction, heat conduction also gets waned 
in a weak magnetic field, whereas finite chemical potential enhances this. 

\begin{figure}[h]
\begin{center}
\includegraphics[width=4.79cm]{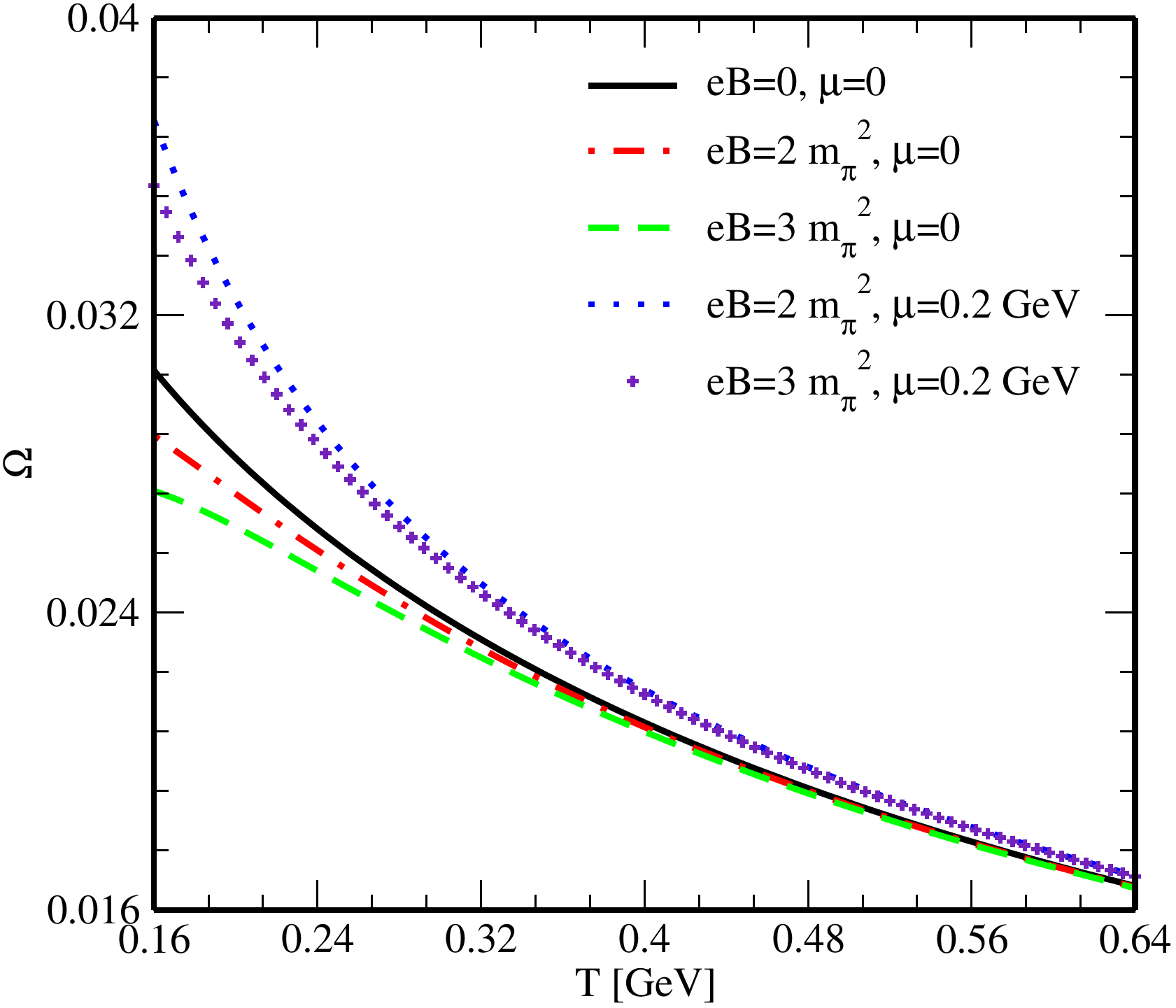}
\caption{Variation of $\Omega$ with temperature for different weak magnetic fields and chemical potentials.}\label{Fig.2}
\end{center}
\end{figure}

Figure \ref{Fig.2} depicts the variation of the Knudsen number 
with the temperature. The $\Omega$ retains its magnitude much 
below unity, but in the presence of weak magnetic field it 
gets decreased, whereas the emergence of chemical potential 
increases its magnitude. The behavior of $\Omega$ at finite 
$eB$ and finite $\mu$ corroborates the behavior of $\kappa$ in the 
similar environment. Throughout the 
variation, the Knudsen number remains much below unity, 
indicating that the macroscopic length scale prevails over the 
microscopic length scale. Thus, the hot QCD matter remains in 
local equilibrium state even in the presence of both weak 
magnetic field and finite chemical potential. 

\subsection{Conclusions}
In this work, we studied the effects of weak magnetic field and finite chemical potential on the electrical and thermal conductivities of hot and dense QCD matter, which were determined by following the kinetic theory. It is observed that both $\sigma_{\rm el}$ and $\kappa$ get decreased in a weak magnetic field, whereas the finite chemical potential increases their magnitudes. Further, we studied the Knudsen number and found that it retains its value below unity, so, the characteristic length scale remains larger than the mean free path and the hot QCD matter stays in the equilibrium state even in the presence of both weak magnetic field and finite chemical potential.

\section{Multiplicity dependent study of $\Lambda(1520)$ production in pp collisions at $\sqrt{s} $ = 5 and 13 TeV}
\author{Sonali Padhan (For the ALICE collaboration)}	

\bigskip

\begin{abstract}
We present the measurement of the baryonic resonance particle $\Lambda(1520)$ (mass = 1520 MeV/$c^{2}$) at mid-rapidity $(|y| < 0.5 )$ in pp collisions at 5.02 and 13 TeV as a function of charged-particle multiplicity. $\Lambda(1520)$ is reconstructed using its hadronic decay channel $\Lambda(1520)$ ($\bar\Lambda$( 1520)) $\rightarrow pK^-(\bar{p}K^{+} $) with branching ratio (BR = 22.5 $\pm$ 1\%). $\Lambda(1520)$ has a lifetime of around 13 fm/$\it{c}$, which lies between the lifetimes of $K^*$ and $\phi$ resonances. The multiplicity dependence of the $\Lambda(1520)/\Lambda$ ratio for pp collisions can serve as a baseline for heavy-ion collisions. 

\end{abstract}


\subsection{Introduction}
Hadronic resonances are short-lived particles having lifetime comparable to QGP fireball.
They are valuable tools to study the hadronic phase in ultrarelativistic heavy-ion collisions. Most of the hadrons' measured yields are constant between the chemical and kinetic freeze-out. In the case of resonance particles, their yields and transverse momenta may be affected by pseudo-elastic rescattering or regeneration effects due to interactions with hadrons in the hadron gas phase ~\cite{ab}. The rescattering suppresses the measured yield, while the regeneration enhances the resonance yield.

 The ratio of hadronic resonances to the stable particles' yields could provide information on the in-medium properties of the hadronic phase. The $\Lambda(1520)/\Lambda$ ratio in pp collisions can serve as a baseline for heavy-ion collisions. We present the transverse-momentum spectra, the integrated yields $(\rm d \it N/ \rm d \it y )$, the mean transverse-momentum $(\langle p_{\rm{T}}\rangle)$ and the $ \Lambda(1520)/\Lambda$ yield ratio and the multiplicity dependence of $\Lambda(1520)$ production as a function of the charged-particle multiplicity in pp collisions at $\sqrt{s}$ = 5.02 and 13 TeV with ALICE.

\subsection{Analysis Procedure}
This analysis is carried out by the data collected using the ALICE detector~\cite{cd}. The details on the performance of the ALICE detector can be found in ~\cite{ALICE_det}. The Inner Tracking System (ITS) is responsible for the tracking and vertex finding. The tracking and particle identification can be done by using the Time Projection Chamber (TPC). The Time of Flight (TOF) detector is used for the PID of relativcely high momentum particles. The identification of protons and kaons is carried out using the TPC dE/dx energy loss measurement. For event selection, it is required to have a reconstructed primary vertex within 10 cm of the interaction point.
\begin{figure}[h!]
 \centering

 {\includegraphics[width=0.4\textwidth]{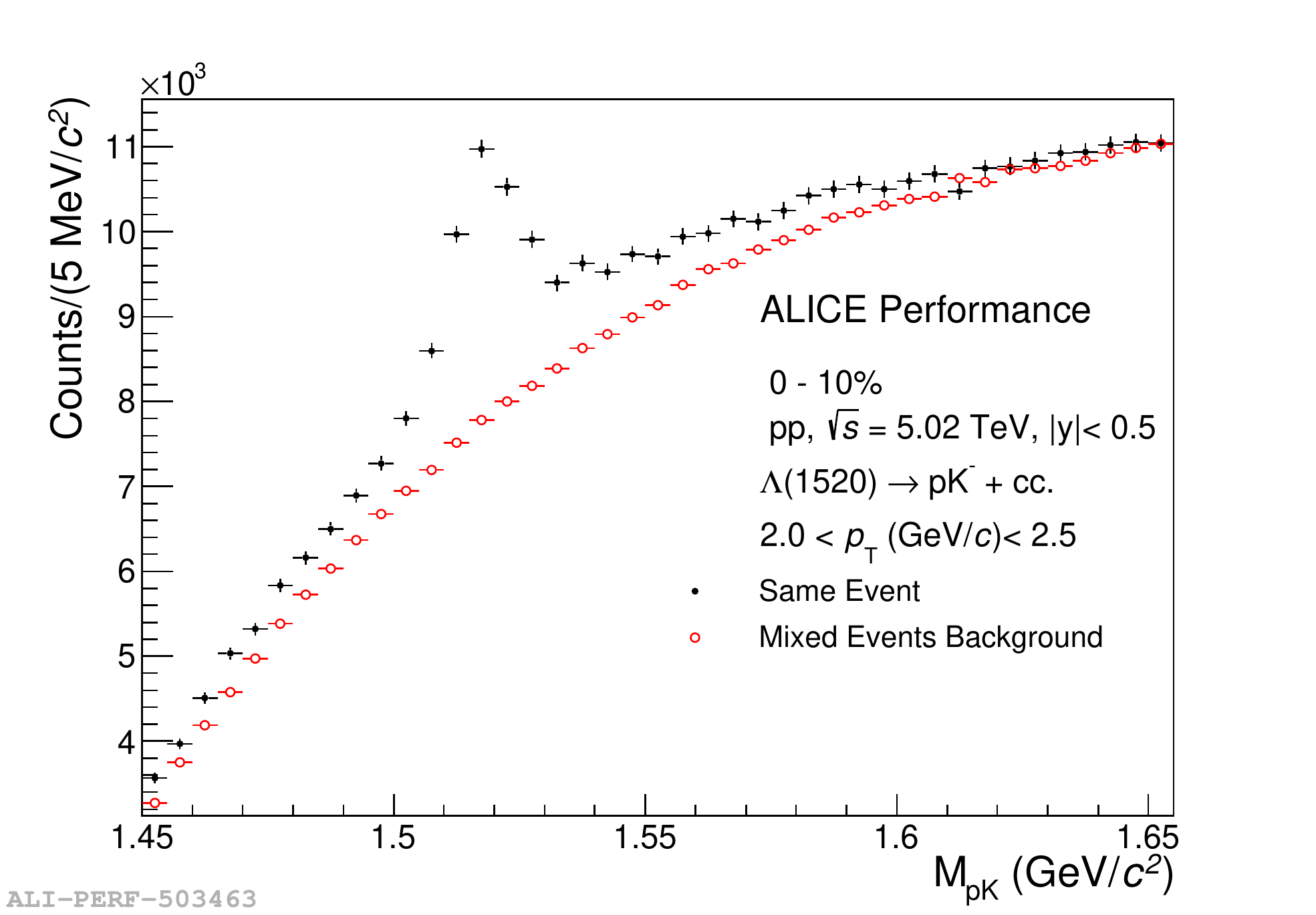}}
 {\includegraphics[width=0.4\textwidth]{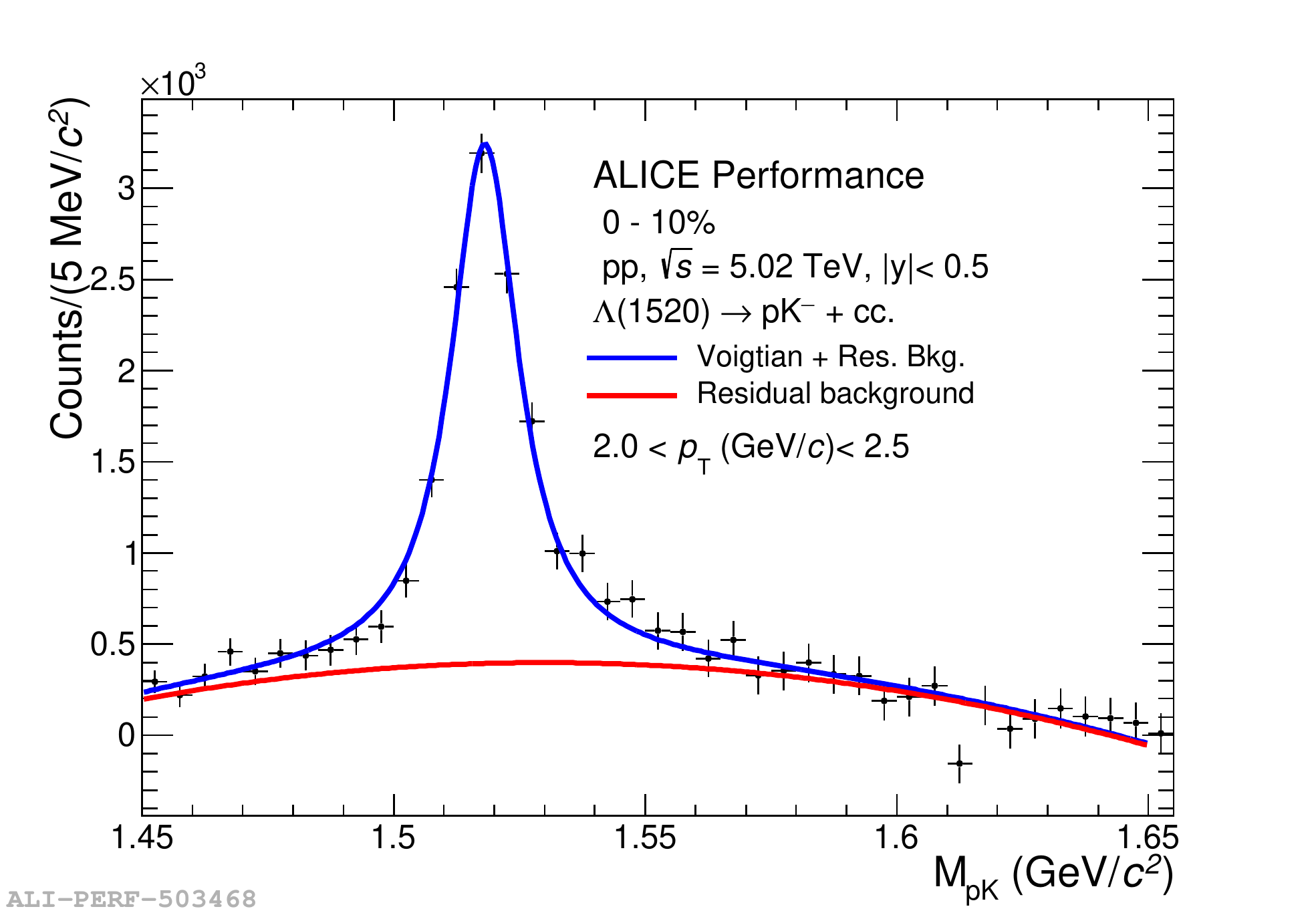}}
  
\caption{\label{invmass} The invariant mass distribution before (left panel) and after (right panel) subtracting the normalized mixed-event background distribution.The black histograms are the data. The signal is described by the blue line which is the Voigtian function plus a second order polynomial and the red line is the second order polynomial describing the residual background.}
\end{figure}

The $\Lambda(1520)$ production has been measured by invariant mass reconstruction of its decay daughters through the hadronic decay channel: $\Lambda(1520)$ ($\bar\Lambda$( 1520)) $\rightarrow pK^-(\bar{p}K^{+} $). $\Lambda(1520)$ resonance is reconstructed by computing the invariant mass spectrum of all the pK primary track pairs. Then the combinatorial background is subtracted, which is estimated by mixed-event techniques. The signal of the invariant mass distribution is fitted with a voigtian function ( convolution of non-relativistic Briet-Wigner and a gaussian detector resolution) plus a 2nd order polynomial for the residual background. The invariant mass plot of $\Lambda(1520)$ is given in Fig.~\ref{invmass}. The fully corrected $p_{\rm T}$ spectrum is obtained by correcting the raw yield with branching ratio, re-weighted efficiency, vertex correction factor, signal loss correction factor, and trigger efficiency factor.

\subsection{Results}

\begin{figure}[h!]
 \centering
 {\includegraphics[width=0.4\textwidth]{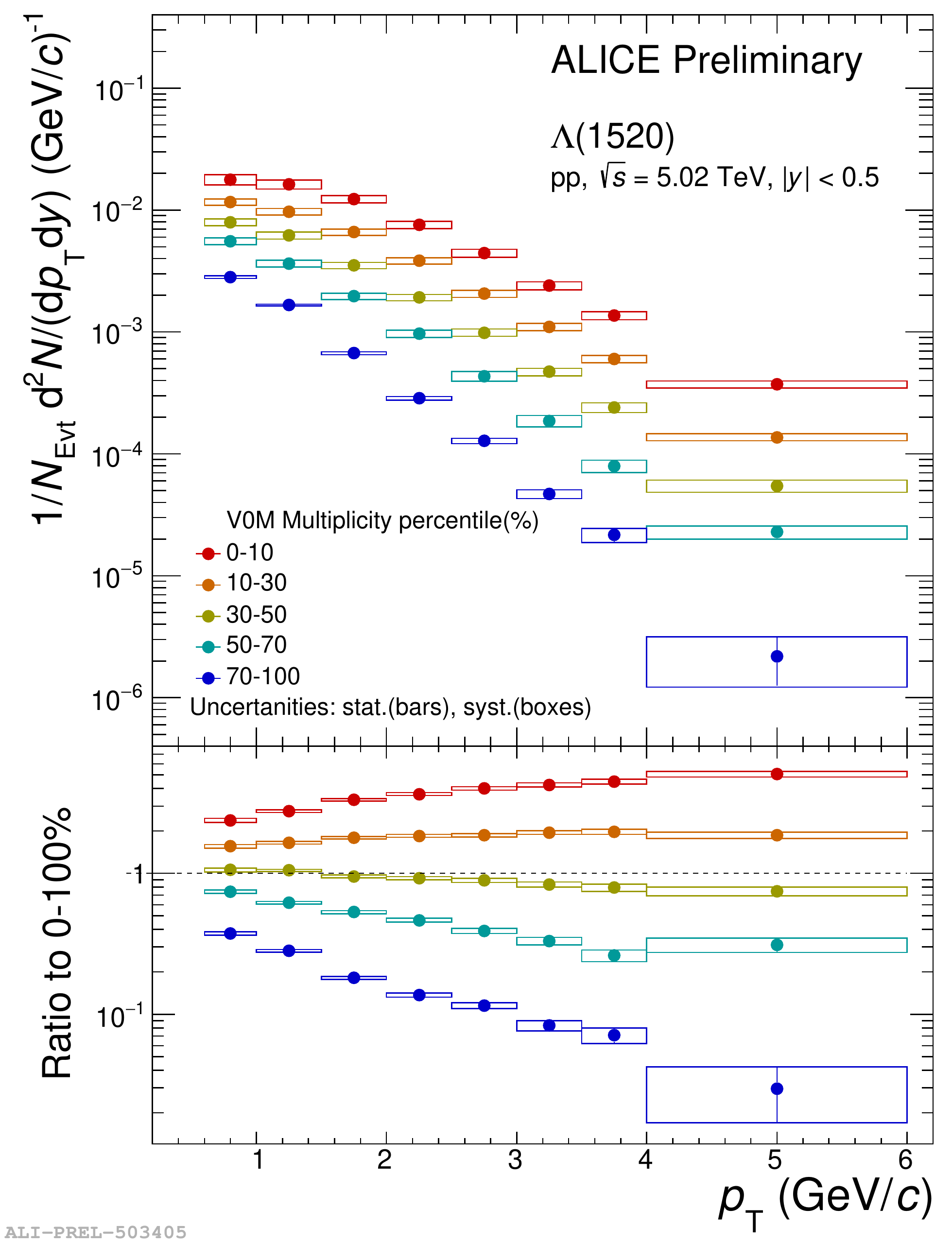}}
 {\includegraphics[width=0.4\textwidth]{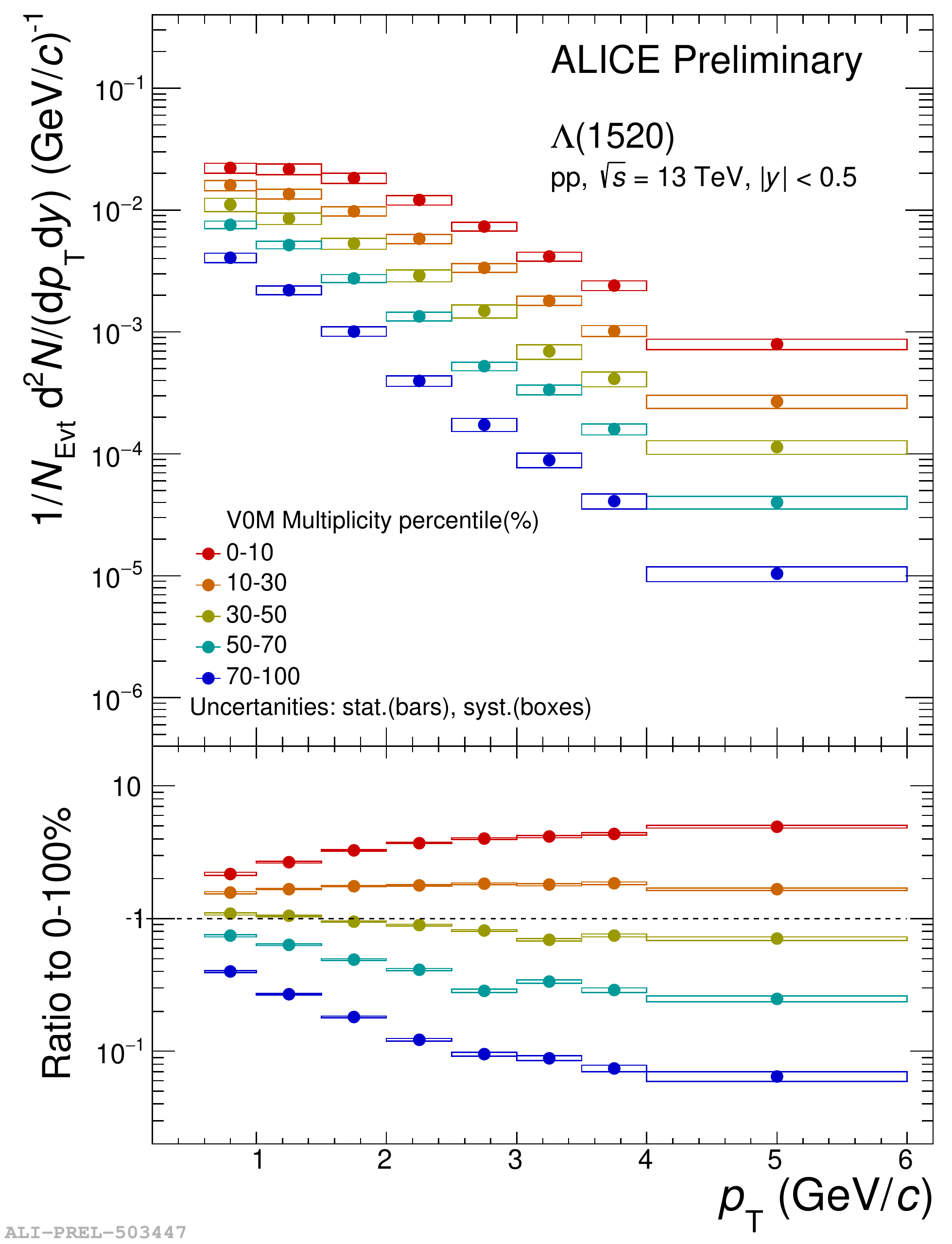}}
\caption{\label{correctspectra} $\Lambda(1520)$ $p_{\rm T}$ spectra for different multiplicity classes in mid rapidity pp collisions at $\sqrt{s} = $ 5.02 TeV (left panel)and 13 TeV (right panel).Bars show statistical errors and boxes show the systematic errors. }
\end{figure}

We present the $p_{\rm T}$ spectra of $\Lambda(1520)$ for five V0 multiplicity classes (0--10\%, 10--30\%, 30--50\%, 50--70\%, 70--100\%) as shown in Fig.~\ref{correctspectra}. The ratio to 0-100\% multiplicity class is shown in the bottom panel. We observed the hardening of $p_{\rm T}$ spectra with the increasing multiplicity classes, and similar behaviour is observed for other resonances in pp, p--Pb, and Xe--Xe collisions.  

\begin{figure}[h!]
 \centering

 {\includegraphics[width=0.4\textwidth]{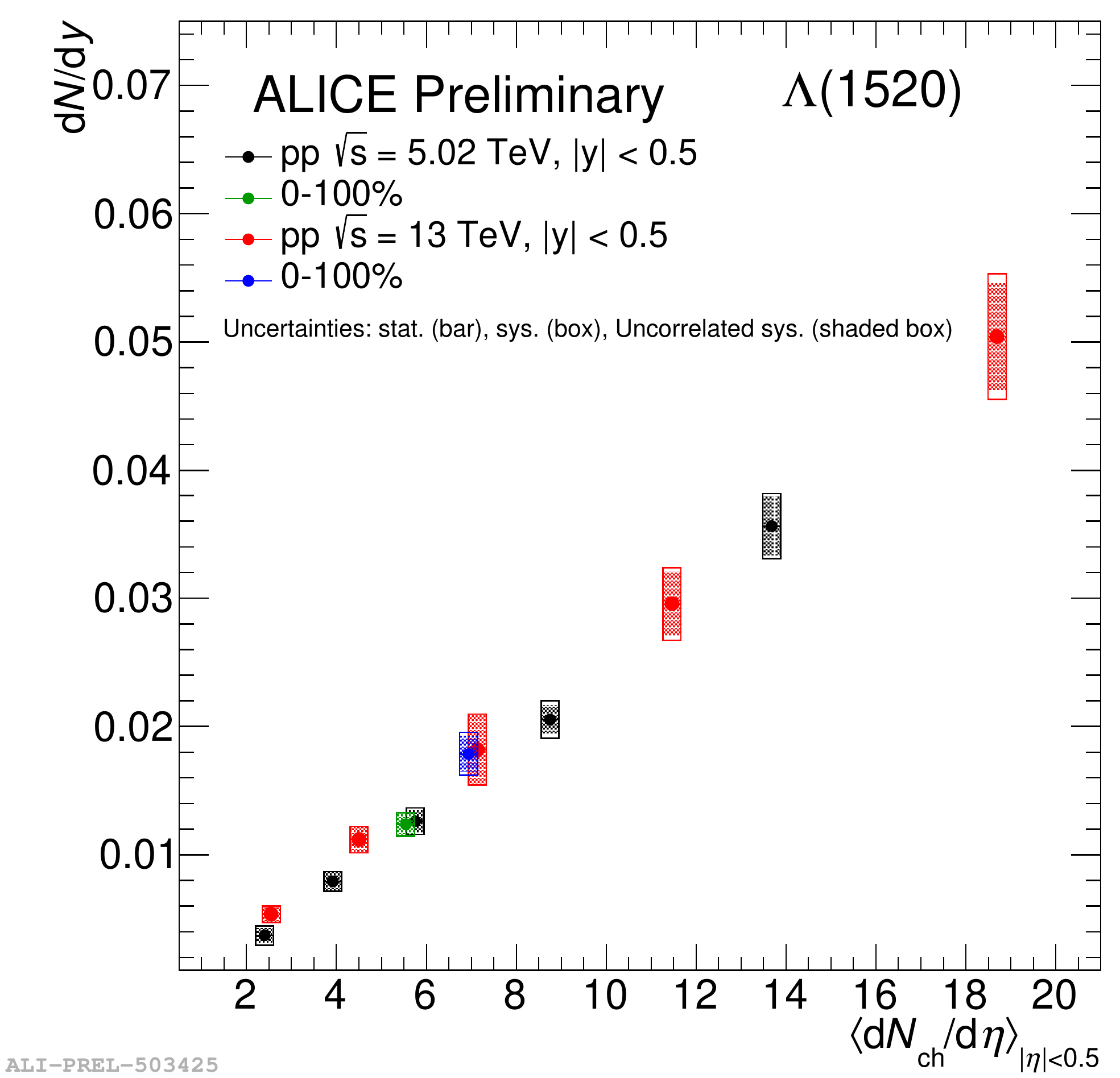}}
 {\includegraphics[width=0.4\textwidth]{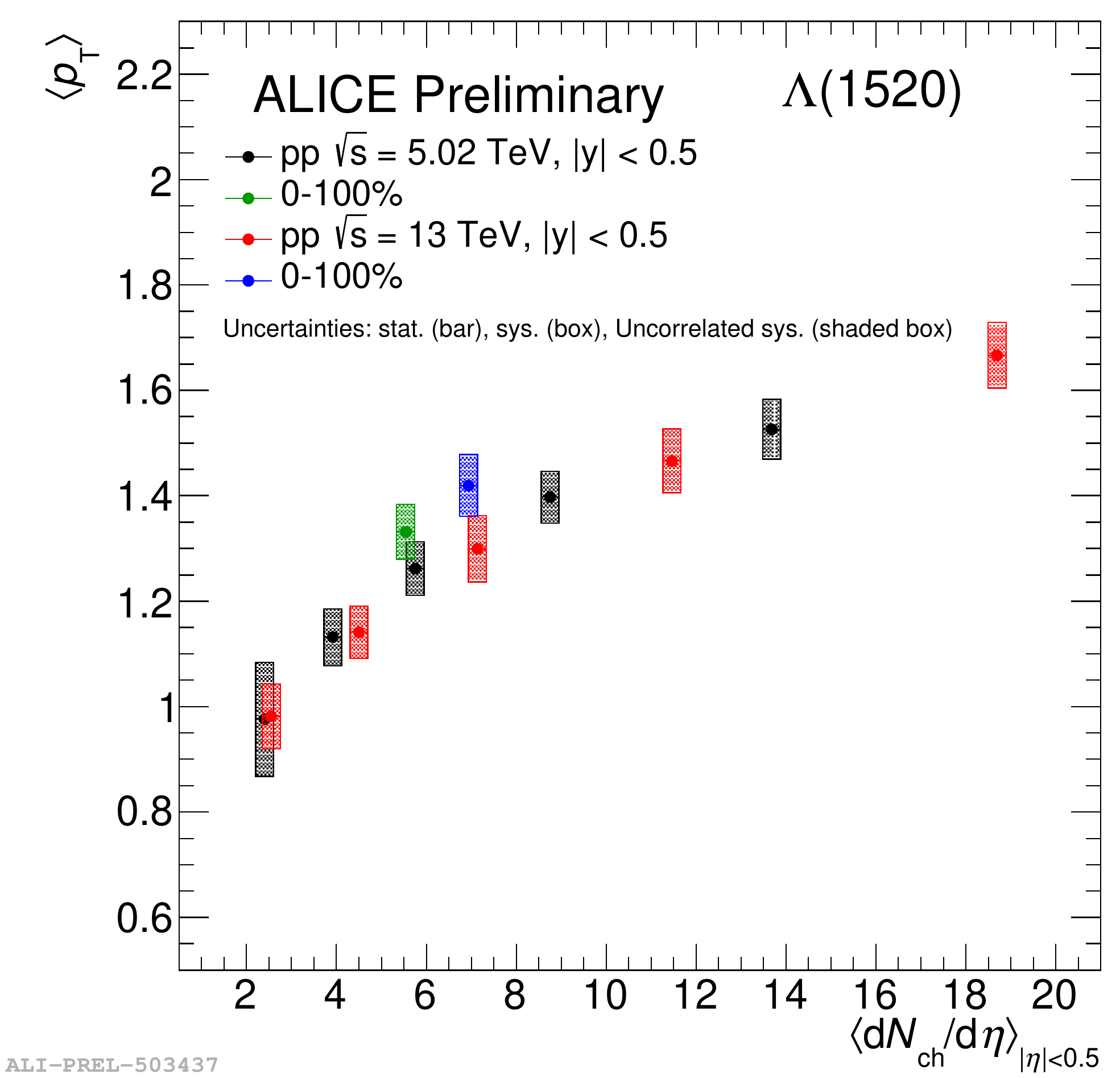}}
\caption{\label{meanpt}Integrated yield(left panel) and mean transverse momentum(right panel) of $\Lambda(1520)$ as a function of charged particle multiplicity in pp collisions at $\sqrt{s} = $ 5.02 TeV and 13 TeV in various multiplicity classes. Bars, boxes, and shaded boxes are the statistical errors, systematics errors and uncorrelated errors respectively . }
\end{figure}

A Levy-Tsallis function is fitted to the $p_{\rm T}$-spectra to obtain integrated yields dN/dy and average transverse momentum $\langle p_{\rm{T}}\rangle$.
$\Lambda(1520)$ $p_{\rm T}$ Integrated yield and mean transverse momentum have been calculated as a function of charged particle multiplicity in various V0 multiplicity classes as shown in Fig. ~\ref{meanpt}.  The $p_{\rm T}$ Integrated yield and mean transverse momentum increase with multiplicity and are independent of the collisions systems and energies.

\begin{figure}[h!]
 \centering

 {\includegraphics[width=0.4\textwidth]{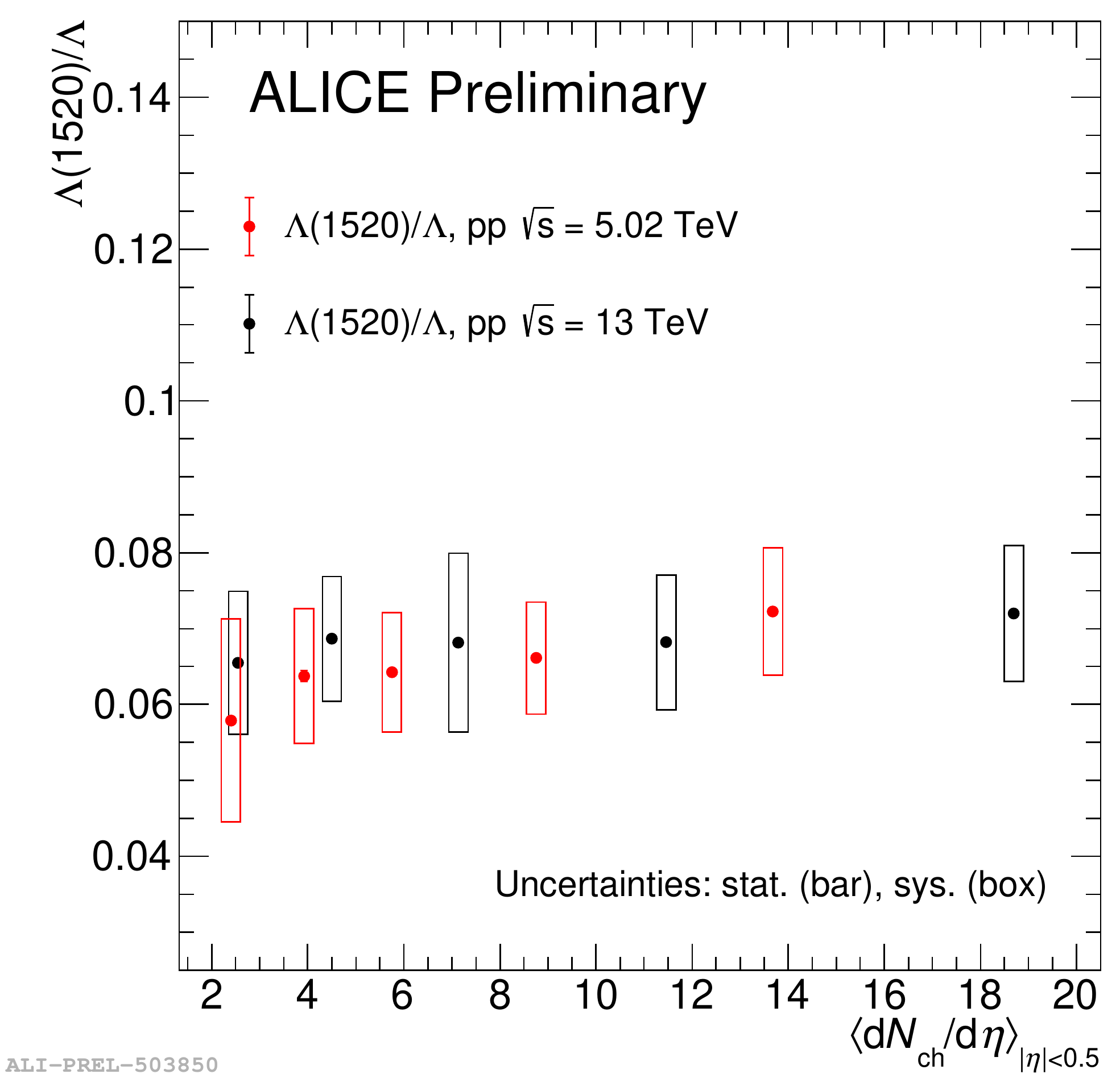}}
 {\includegraphics[width=0.41\textwidth]{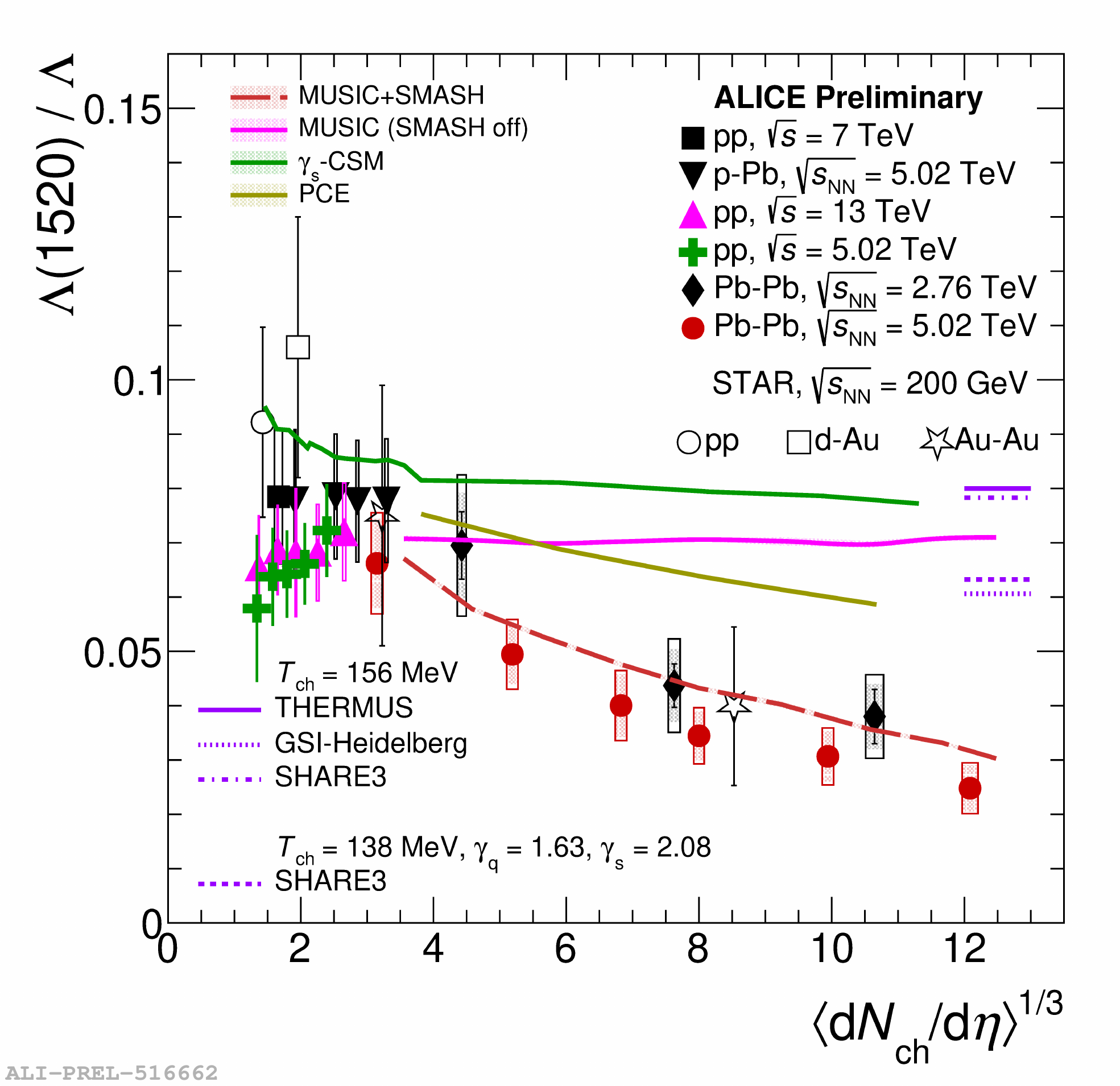}}
\caption{\label{ratioplot}$\Lambda(1520)/\Lambda$ ratio as a function of charged particle multiplicity in pp collisions at $\sqrt{s} = $ 5.02 and 13 TeV (left panel) and compared with previous measurement (right panel). }
\end{figure}

The $\Lambda(1520)/\Lambda$ ratio as a function of charged particle multiplicity in pp collisions at $\sqrt{s} = $ 5.02 and 13 TeV is shown in the Fig. ~\ref{ratioplot}(left panel). Right panel shows the comparison of this ratio with other systems and energies [~\cite{ij} ,  ~\cite{kl}] .
There is a suppression of the $\Lambda(1520)/\Lambda$ ratio in Pb--Pb collisions at $\sqrt{s_{\rm NN}}$ = 2.76 TeV as a function of centrality with respect to peripheral Pb--Pb collisions~\cite{ij}. But no such significance suppression is observed in case of p--Pb collisions at $\sqrt{s_{\rm NN}}$ = 5.02 TeV ~\cite{kl}. We observed that $\Lambda(1520)/\Lambda $ ratio is almost flat in pp collisions at $\sqrt{s} = $ 5.02 and 13 TeV and is independent of multiplicity.

\subsection{Conclusion}
The resent results on the measurement of baryonic resonance $\Lambda(1520)$ in pp collisions at $\sqrt{s} = $ 5.02 TeV and 13 TeV obtained from ALICE detector have been presented. Both $p_{\rm T}$ Integrated yield and mean transverse momentum increase with multiplicity and are independent of collisions systems and energies. The 
$\Lambda(1520)/ \Lambda$ ratio is flat as a function of charged particle multiplicity in pp collisions.

\section{Non-identical particle femtoscopy in Pb--Pb collisions at $\sqrt{s_{\rm NN}}=$ 5.02 TeV with ALICE}
\author{Pritam Chakraborty (for the ALICE Collaboration)}	

\bigskip

\begin{abstract}
	The pion-kaon femtoscopic correlation functions are obtained in Pb--Pb collisions at $\sqrt{s_{\rm NN}}=$ 5.02 TeV with ALICE at the LHC and femtoscopic parameters are extracted. The spherical harmonics representations of the correlation function are investigated. The results are compared with the predictions from (3+1)D Lhyquid + THERMINATOR 2 model.
\end{abstract}


\subsection{Introduction}

Femtoscopy has been used to measure the space--time dimensions of the particle-emitting source created in heavy-ion collisions using two-particle correlations. With non-identical particle femtoscopy, one can measure the size of source as well as the average pair-emission asymmetry that are directly related to the collectivity of the system and transverse mass of the particles. The measurement of femtoscopic correlations between charged pion and kaon pairs for different charge combinations obtained in Pb--Pb collisions at $\sqrt{s_{\rm NN}}=$ 5.02 TeV with ALICE at the LHC is presented.

\subsection{Formalism of Femtoscopy} \label{theory}
The femtoscopic correlation function (CF) can be constructed experimentally using Eq.~\ref{eq:C}
\begin{equation}
	C(k^*) = A(k^*)/B(k^*) \label{eq:C}
\end{equation}
where, $k^*$ is the pair-relative momentum, $A(k^*)$ and $B(k^*)$ are the distribution of particle-pairs selected from same events (signal) and different events (background), respectively. Theoretically, the CF can be interpreted as the convolution of emission probability of particle-pair, namely source function, $S(k^*,r^*)$ and the final state interaction between particles, $\Psi(k^*,r^*)$, as given in Eq:~\ref{eq:psi} \refcite{adam_primary}.
\begin{equation}
	C(k^*) = \int S(k^*,r^*)|\Psi(k^*,r^*)|^2d^3r^* \label{eq:psi}
\end{equation}

For pion-kaon pairs, $\Psi(k^*,r^*)$ includes the Coulomb and Strong interaction as given in Eq:~\ref{eq:wave} \refcite{adam_primary}.
\begin{equation}
	 \Psi(k^*,r^*)=\sqrt{A_C(\eta)}\left[e^{-ik^*r^*}F(-i\eta,1,i\zeta)+f_C(k^*)\frac{G(\rho,\eta)}{r^*}\right] \label{eq:wave}	 
\end{equation}
where, $\mathbf{r^*}$ and $\mathbf{k^*}$ are the pair-relative separation and half of pair-relative momentum at Pair Rest Frame (PRF, where total momentum of the pair is zero), respectively, $A_C$ is the Gamow factor, $\eta=1/k^*a_C$, F is the confluent hypergeometric function, G is the combination of the regular and singular s-wave Coulomb functions,
$\zeta=k^*r^*(1+cos\theta^*)$, $f_C$ is the strong scattering amplitude, $\theta^*$ is the angle between the pair relative momentum $k^*$ and relative position $r^*$ in PRF, $a_C$ is the Bohr radius of the pair which is equal to 248.52 fm for like-sign and -248.52 fm \refcite{adam_primary} for unlike-sign pion-kaon pair.

In this analysis, the spherical harmonics representations (SH) of the correlation function have been analysed since most of the femtoscopic information can be extracted using the lower harmonics of CF and lesser statistics of pair-particles compared to the conventional method. The coordinate system used in this analysis consists of 3 axes, $out$, $side$ and $long$, where, $out$ is along the pair-transverse momentum, $long$ is along the beam direction and $side$ is perpendicular to other two axes. Using $C_{0}^{0}$ and $ReC_{1}^{1}$, the system-size ($R_{\rm out}$) and the pair-emission asymmetry along the $out$ direction ($\mu_{\rm out}$), respectively, can be estimated. 

To extract the $R_{\rm out}$ and $\mu_{\rm out}$, the experimental correlation function is parameterised using Eq.~\ref{eq:psi}. The source function is assumed to be a 3 dimensional Gaussian function with sizes $R_{\rm out}$, $R_{\rm side}$ and $R_{\rm long}$ in $out$, $side$ and $long$ directions and the pair emission asymmetry $\mu_{\rm out}$ along $out$ direction as given in Eq. \ref{eq:sourcefcn} \refcite{adam_primary}. 
\begin{equation}
	S(\mathbf{r})=exp\left(-\frac{(r_{\rm out}-\mu_{\rm out})^2}{R^2_{\rm out}}-\frac{r^2_{\rm side}}{R^2_{\rm side}}-\frac{r^2_{\rm long}}{R^2_{\rm long}}\right)  \label{eq:sourcefcn}
\end{equation}
While parameterising S, it is assumed that $R_{\rm side} = R_{\rm out}$ and $R_{\rm long} = 1.3R_{\rm out}$ based on the results from identical particle 3D femtoscopy for pions from RHIC \refcite{adam_primary}.

\subsection{Analysis details} \label{ana}
\begin{figure}[tbph]
				\centering
				{\includegraphics[width=0.49\textwidth]{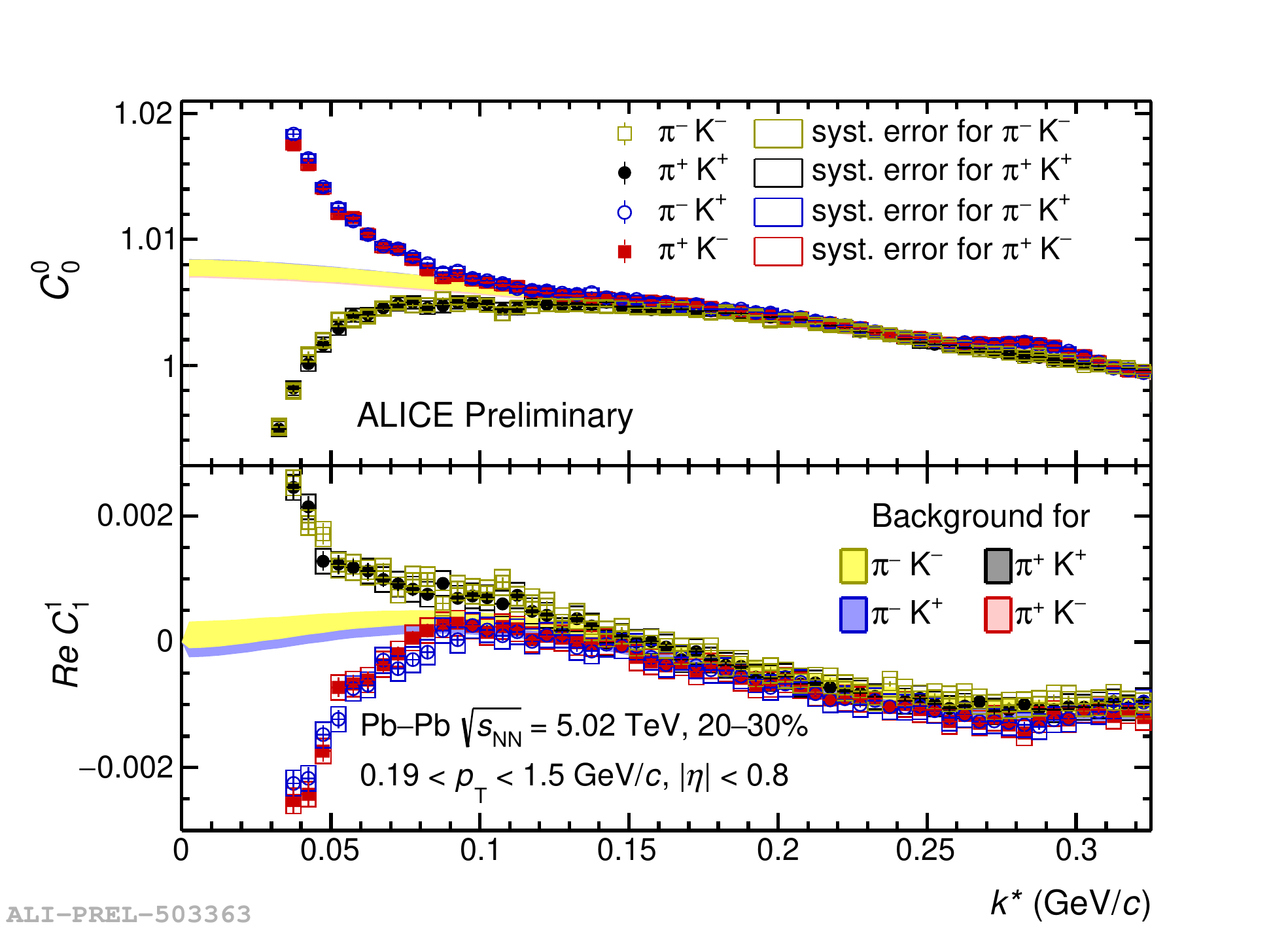}}
				{\includegraphics[width=0.49\textwidth]{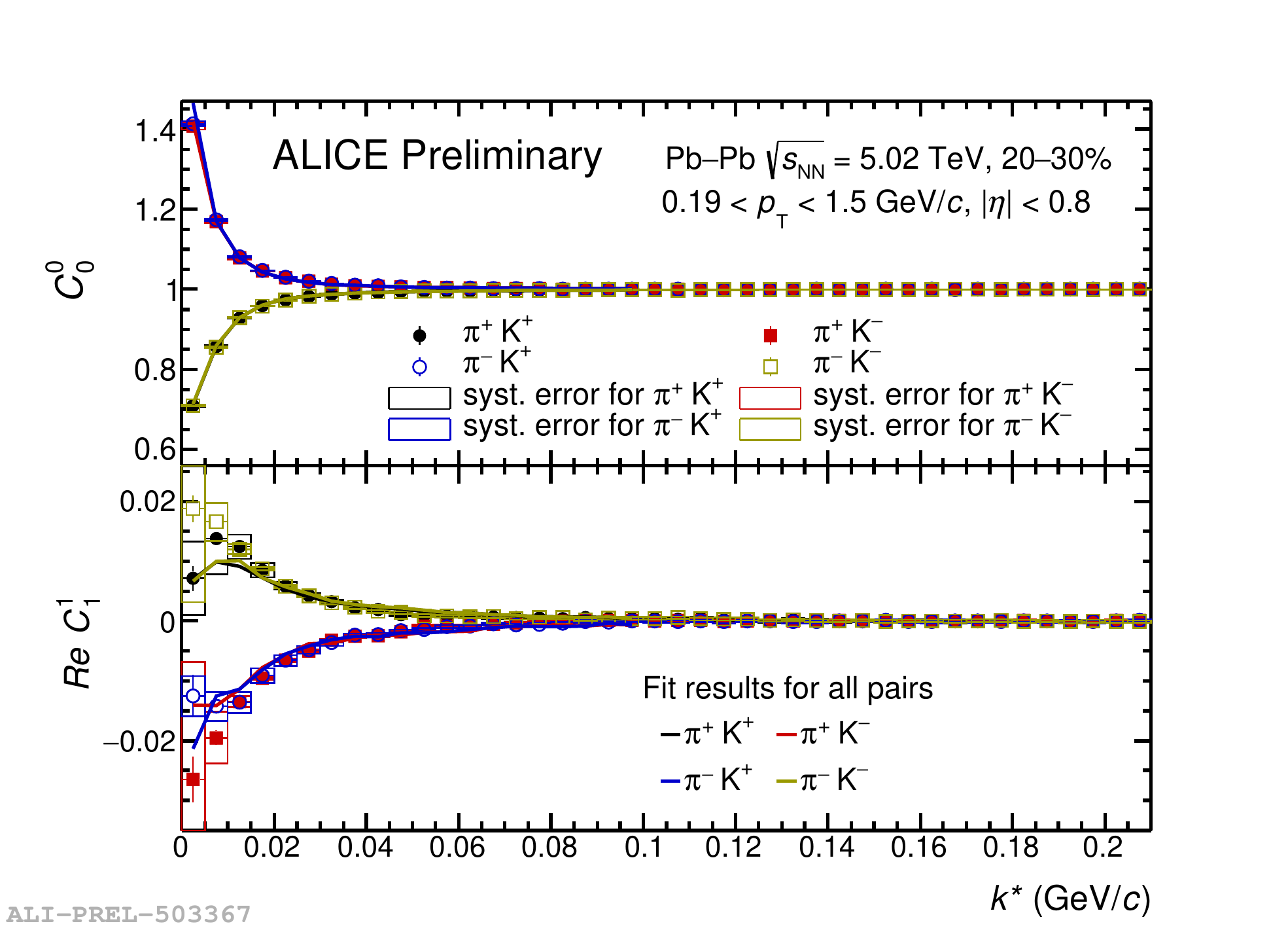}}
				\caption{Correlation functions for pion-kaon pairs with $6^{th}$ order polynomial function as the non-femtoscopic	background (left) and background minimised femtoscopic correlation functions for 20-30\% for all charge-pair combinations of pion and kaon, together with their fits calculated using CorrFit software (right) for Pb$-$Pb collisions at $\sqrt{s}_{\rm NN} = 5.02$ for 20-30\% centrality}
				\label{CF_fit}
				\end{figure}

The femtoscopic correlation functions of pion-kaon pairs, produced in Pb$-$Pb collisions at $\sqrt{s_{\rm NN}}=5.02$ TeV with ALICE at the LHC have been analysed for 0-5\% to 40-50\% central events. The events with vertex-z positions $\rm |{z_{vtx}}|<7.0$ cm are selected. The pions and kaons are selected within $|\eta|<0.8$ having $0.19<p_{\rm T} (\rm{GeV}/c)<1.5$ with Time Projection Chamber and Time Of Flight detectors. The CFs for all charged pion-kaon pairs for 20-30\% centrality are shown in the left plot of Fig.~\ref{CF_fit}. The $C_{\rm 0}^{\rm 0}$ for unlike and like-signed pion-kaon pairs go above and below 1, respectively at lower k* due to Coulomb interaction. Similarly, $Re C_{\rm 1}^{\rm 1}$'s deviation from the flat region at lower k*, signals the existence pair-emission asymmetry along the $out$ direction. However, at higher k* region, the $C_{\rm 0}^{\rm 0}$ and $Re C_{\rm 1}^{\rm 1}$ is not flat which implies the presence of non-femtoscopic background due to the elliptic flow, residual correlation functions, resonance decays etc. The background function is similar for all combinations of pion-kaon pairs and assumed to be $6^{th}$ order polynomial as shown in the left plot of Fig. ~\ref{CF_fit}.

The right plot of Fig.~\ref{CF_fit} shows the background minimised correlation functions with fit results, obtained using CorrFit package based on the process described in section \ref{theory}.

\subsection{Results and discussions} \label{result}

\begin{figure}[tbph]
				\centering
				{\includegraphics[width=0.49\textwidth]{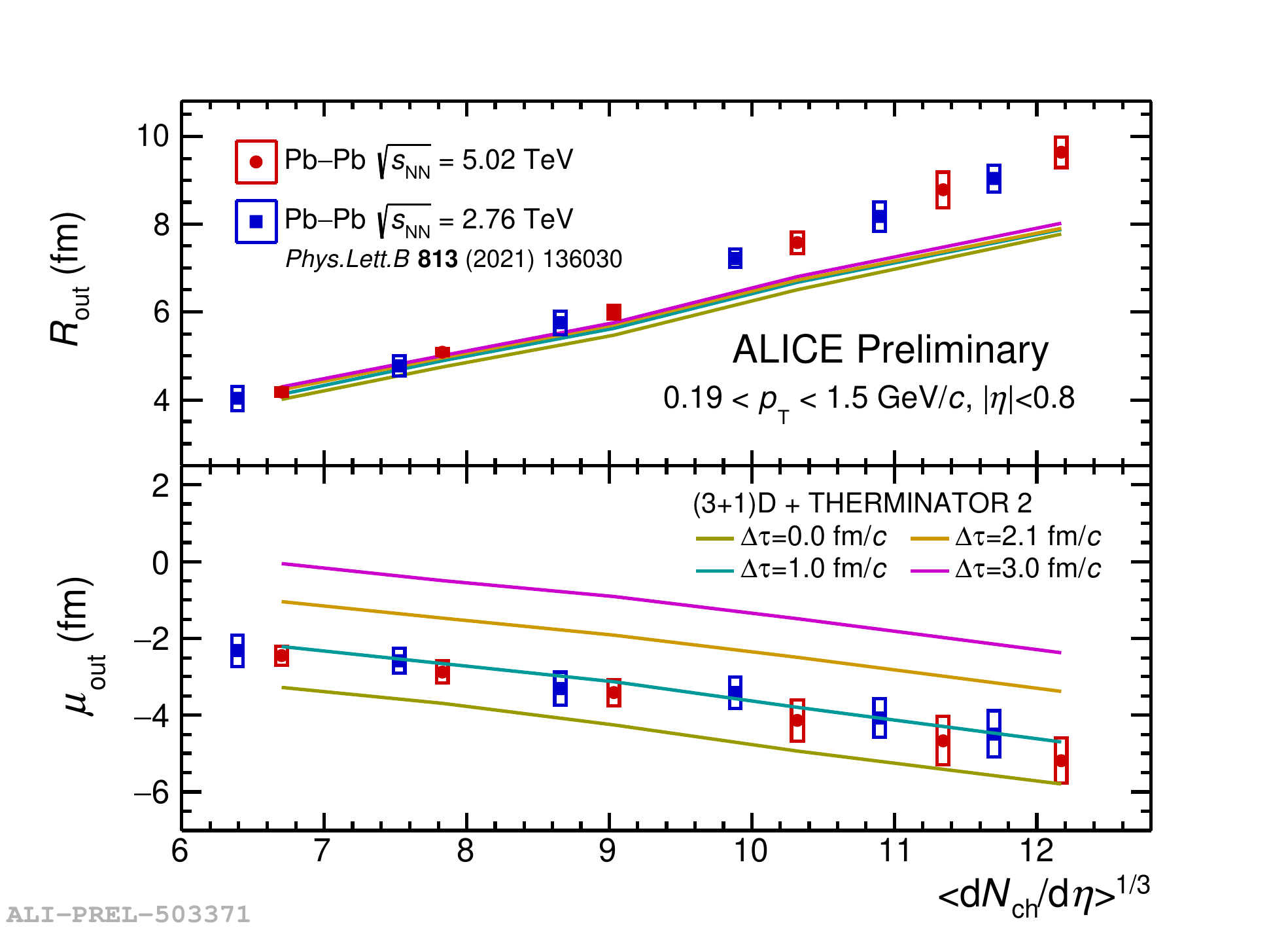}}
				{\includegraphics[width=0.49\textwidth]{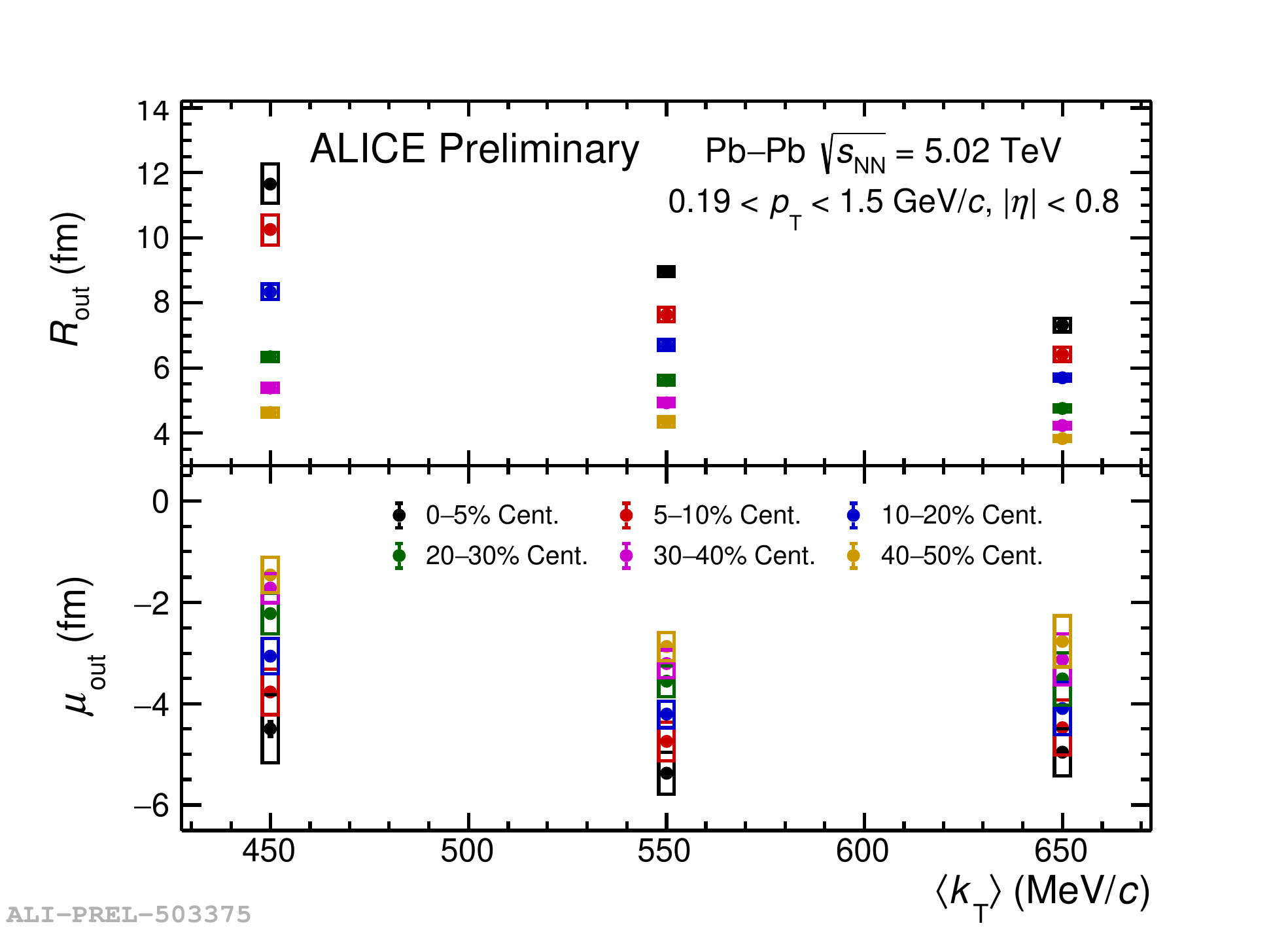}} 
				\caption{$R_{\rm {out}}$  and $\mu_{\rm {out}}$ vs. $\langle {\rm d}N_{ch}/{\rm d}{\eta}\rangle^{1/3}$ with predictions from (3+1)D + THERMINATOR 2 model (left)) and $R_{\rm {out}}$  and $\mu_{\rm {out}}$ vs. $\langle k_{\rm T}\rangle$ as the function of centrality (right)}
				\label{R_mu}
			\end{figure}

The $R_{\rm {out}}$ and $\mu_{\rm {out}}$, extracted from the background minimised femtoscopic correlation functions of pion-kaon pairs from 0-5\% to 40-50\% centrality are as the function of charge particle multiplicity density ($\langle {\rm d}N_{ch}/{\rm d}{\eta}\rangle^{1/3}$) in Fig.~\ref{R_mu}. The $R_{\rm {out}}$ is observed to be increasing with multiplicity due to the increase of system size with increasing number of participants. The $\mu_{\rm {out}}$ has been observed to be negative indicating that pions are emitted closer to the center of system than kaons and the finite value of $\mu_{\rm {out}}$ indicates the presence of radial flow in the system. The results are compatible with the pion-kaon femtoscopic analysis in Pb$-$Pb collisions at $\sqrt{s_{\mathrm NN}}=$ 2.76 TeV with ALICE and no energy dependence of $R_{\rm {out}}$  and $\mu_{\rm {out}}$ has been found so far. By comparing the results with predictions from (3 + 1)D viscous hydrodynamics coupled to decay code THERMINATOR 2 with different hypotheses of the extra delay ($\Delta\tau$) in emission for kaons, it is observed that the measured and predicted radii are in well agreement for peripheral events. The trend of measured $\mu_{\rm {out}}$ almost matches with the predicted ones with $\Delta\tau=1.0$ fm/{\textit{c}}, which indicates the presence of rescattering phase in the system along with the radial flow.

In the right plot of Fig.~\ref{R_mu}, the $R_{\rm {out}}$  and $\mu_{\rm {out}}$ are plotted as the function of average pair-transverse momentum ($\langle k_{\rm T}\rangle$) and centrality. The radii are found to be decreasing with $\langle k_{\rm T}\rangle$ in all centralities, indicating the presence of collectivity in the system. The $\mu_{\rm {out}}$  is lowest in the $k_{\rm T}$ range: 400-500 MeV/\textit{c} in every centrality. To understand the effect of $k_{\rm T}$ on $\mu_{\rm {out}}$, further investigation is needed.

\subsection{Summary}
The analysis of pion-kaon femtoscopic correlation function in Pb$-$Pb collision at $\sqrt{s_{\rm NN}}=$5.02 TeV shows that the source-size increases with multiplicity as the number of participants increase from low to high multiplicity events. Also, source-size decreases with increasing $k_T$ due to the collective expansion of the system. The observed negative pair emission asymmetry signals that pions are emitted closer to the center of system than kaons. The predictions of pair emission asymmetry from theoretical model indicates the presence hadronic rescattering phase in the system. Also, no beam-energy dependence of the femtoscopic parameters is found.

\section{Estimation of hadronic phase lifetime and  locating the QGP phase boundary}
\author{ Dushmanta Sahu, Sushanta Tripathy, Girija Sankar Pradhan, Raghunath Sahoo}	

\bigskip

\begin{abstract}
	Resonances are very useful probes to understand the various phases of the system evolution in ultra-relativistic collisions. A simple toy model is adopted to estimate the hadronic phase lifetime of the systems produced in ultra-relativistic collisions at RHIC and LHC by taking advantage of the short lifetime of $K^{*0}$, which is a resonance particle. With this model, we estimate the lower limit of the hadronic phase lifetime as a function of charged particle multiplicity for various collision systems and collision energies. On the other hand, $\phi$, a long-lived resonance, can be used to locate the Quark-Gluon Plasma (QGP) phase boundary. We fit the Boltzmann-Gibbs blast-wave function and estimate the effective temperature of $\phi$ mesons, which gives information about the location of the QGP phase boundary.
	
	 \end{abstract}

\subsection{Introduction}

Estimating the hadronic phase lifetime of any collision system is not trivial. Although we know that the hadronic phase lifetime would be in the order of a few fermis in high-energy heavy-ion collisions, a proper phenomenological study as a function of charged particle multiplicity is of utmost importance. Furthermore, this can be taken as inputs in event generators such as a multi phase transport model (AMPT), where hadronic phase lifetime is usually set randomly. We have used $K^{*0}$ produced in high energy collisions to estimate the hadronic phase lifetime \cite{Sahu:2019tch,Singha:2015fia}, which we have studied as a function of event multiplicity across various collision energies and system sizes. In addition, we have also used long-lived hadronic resonances like $\phi$ mesons to locate the QGP phase boundary with the information of effective temperature, which can be extracted from fitting the transverse momentum spectra with the Boltzmann-Gibbs blast wave function\cite{Sahu:2019tch}.

\subsection{Results and discussion}

Hadronic resonances, which are produced in high-energy collisions, can decay while traveling through the medium. The decay daughters can then interact with other particles in the medium and lose momentum, suppressing resonances during their invariant mass reconstructions. This process is called rescattering. In addition, resonances can be regenerated within the hadronic phase due to pseudoelastic collisions, enhancing the resonance yields. The suppression of $K^{*0}/K$ ratio thus hints at the domination of rescattering over the regeneration effect. This can help us to estimate the hadronic phase lifetime. A simple toy model like the nuclear decay formula can be taken and modified for this purpose. $K^{*0}/K$ ratio for low multiplicity pp collisions can be taken as the ratio at chemical freeze-out temperature, and the $K^{*0}/K$ ratio at different event multiplicities for different collision systems can be taken as the ratio at the kinetic freeze-out temperature. Thus, the hadronic phase lifetime can be estimated by the following relation \cite{Sahu:2019tch,Singha:2015fia},

\begin{equation}
[\rm{K^{*0}}/\rm{K}]_{kinetic} = [\rm{K^{*0}}/\rm{K}]_{chemical} \times e^{-\Delta t/\tau}.
\label{eq1}
\end{equation}

Here $\tau$ is $K^{\* 0}$ lifetime and $\Delta t$ is defined as the hadronic phase lifetime multiplied by the Lorentz factor.

\begin{figure}[h]
	\begin{center}
		\includegraphics[width = 6cm]{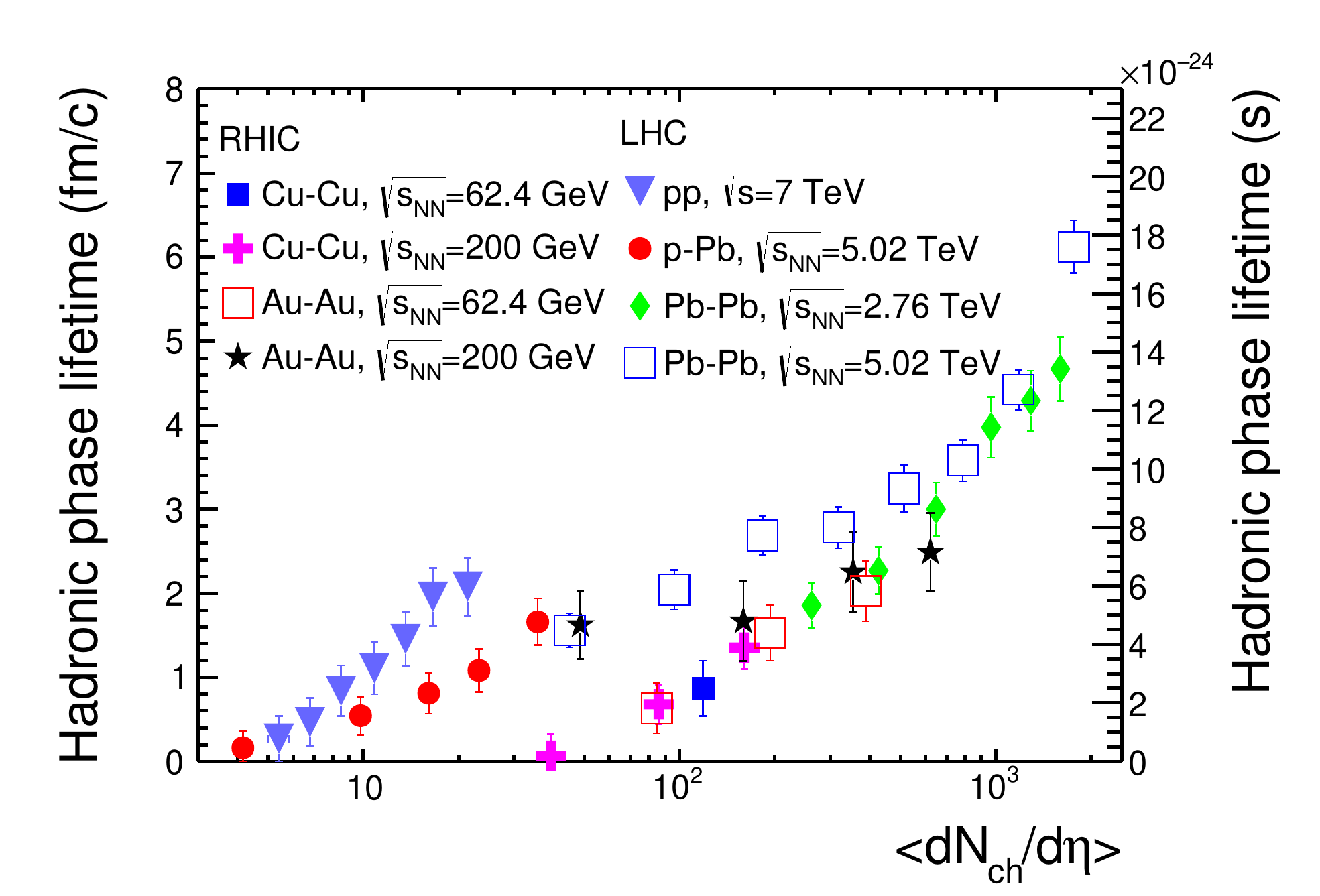}
		\caption{Hadronic phase lifetime as a function of charged particle multiplicity for RHIC and LHC energies.~\cite{Sahu:2019tch}. 
		}\label{Fig1}
	\end{center}
\end{figure}

In fig. \ref{Fig1}, we have plotted the hadronic phase lifetime as a function of charged particle multiplicity for various collision systems and collision energies. The detailed framework and results can be found in ref.\cite{Sahu:2019tch}. We observe strong dependencies of the hadronic phase lifetime on the multiplicity and the collision energies. We also observe that in high-multiplicity pp collisions, the hadronic phase lifetime is around two fm/c. However, the most central Pb-Pb collisions it is around six fm/c \cite{Sahu:2019tch}.

Resonances with a relatively higher lifetime might not go through the rescattering and regeneration processes. Thus, in contrast to $K^{*0}$, $\phi$ meson can act as a tool to locate the QGP phase boundary. The transverse momentum ($p_{\rm T}$) spectra of $\phi$ meson will not be distorted during the hadronic phase. Hence, by using the $p_{\rm T}$ spectra of $\phi$ meson, we can extract information about the location of the QGP phase boundary. To do this, we fit the $\phi$ meson $p_{\rm T}$ spectra with the Boltzmann-Gibbs blast-wave (BGBW) function up to $p_{\rm T} \sim$ 3 GeV/c \cite{Blast,PHENIX:2003wtu}. From this, we can get the freeze-out temperature ($T_{\rm th}$) and the average velocity of the medium ($\langle \beta \rangle$), which can then help us to estimate the effective temperature ($T_{\rm eff}$) by the formula \cite{Sahu:2019tch};

\begin{equation}
T_{\rm{eff}} = T_{\rm{th}} + \frac{1}{2}m\langle \beta \rangle^{2}.
\label{eq2}
\end{equation}

\begin{figure*}[h]
	\begin{center}
		\includegraphics[width = 4cm]{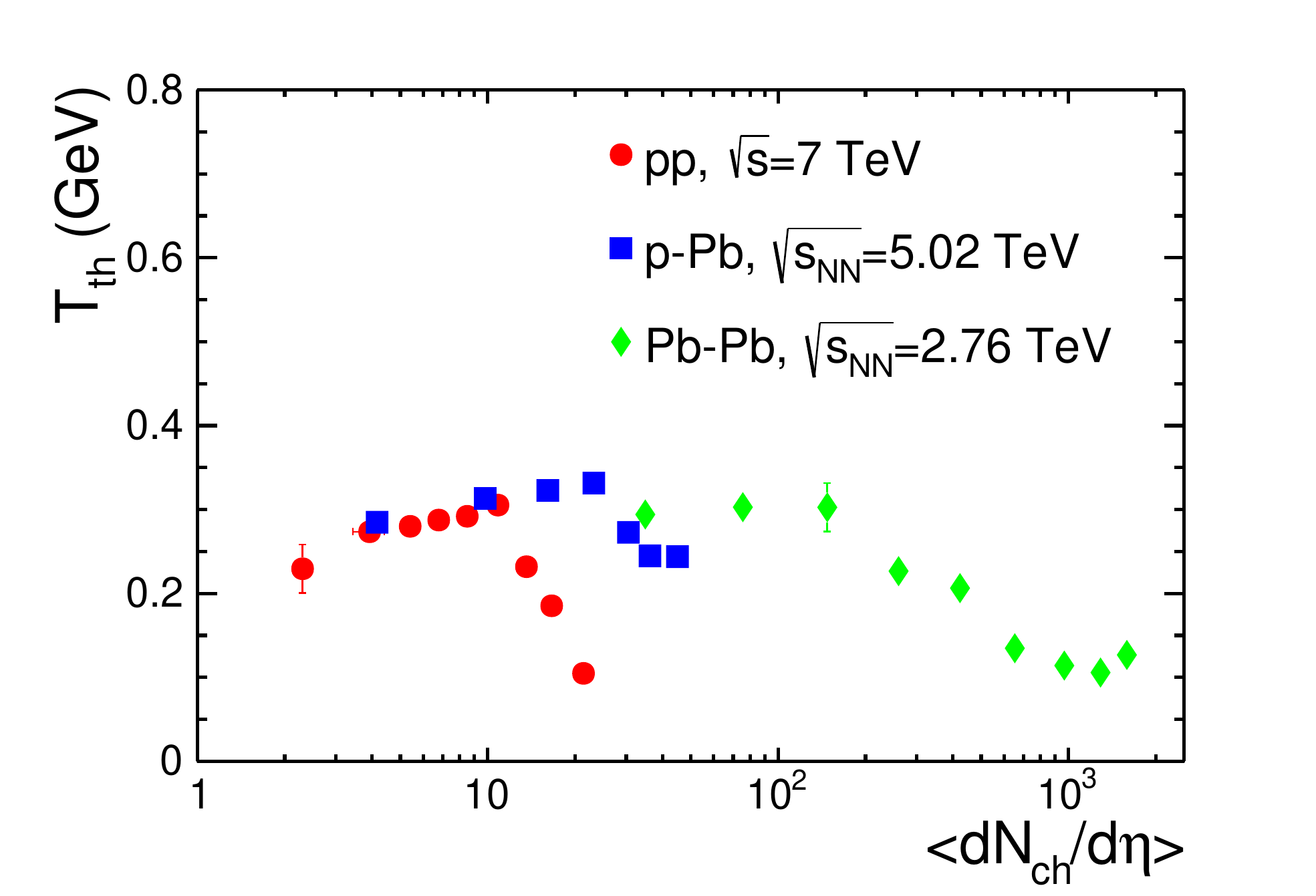}
		\includegraphics[width = 4cm]{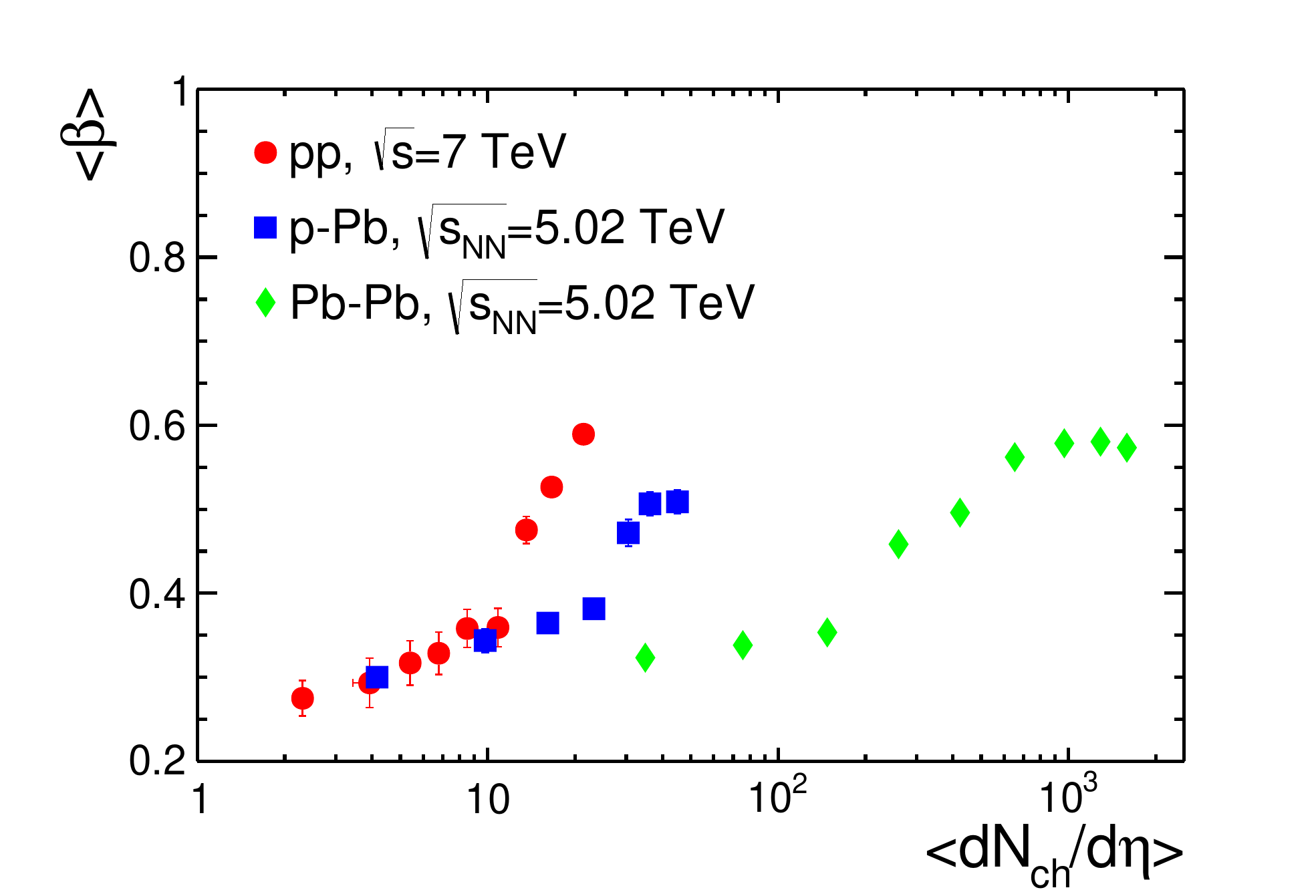}
		\includegraphics[width = 4cm]{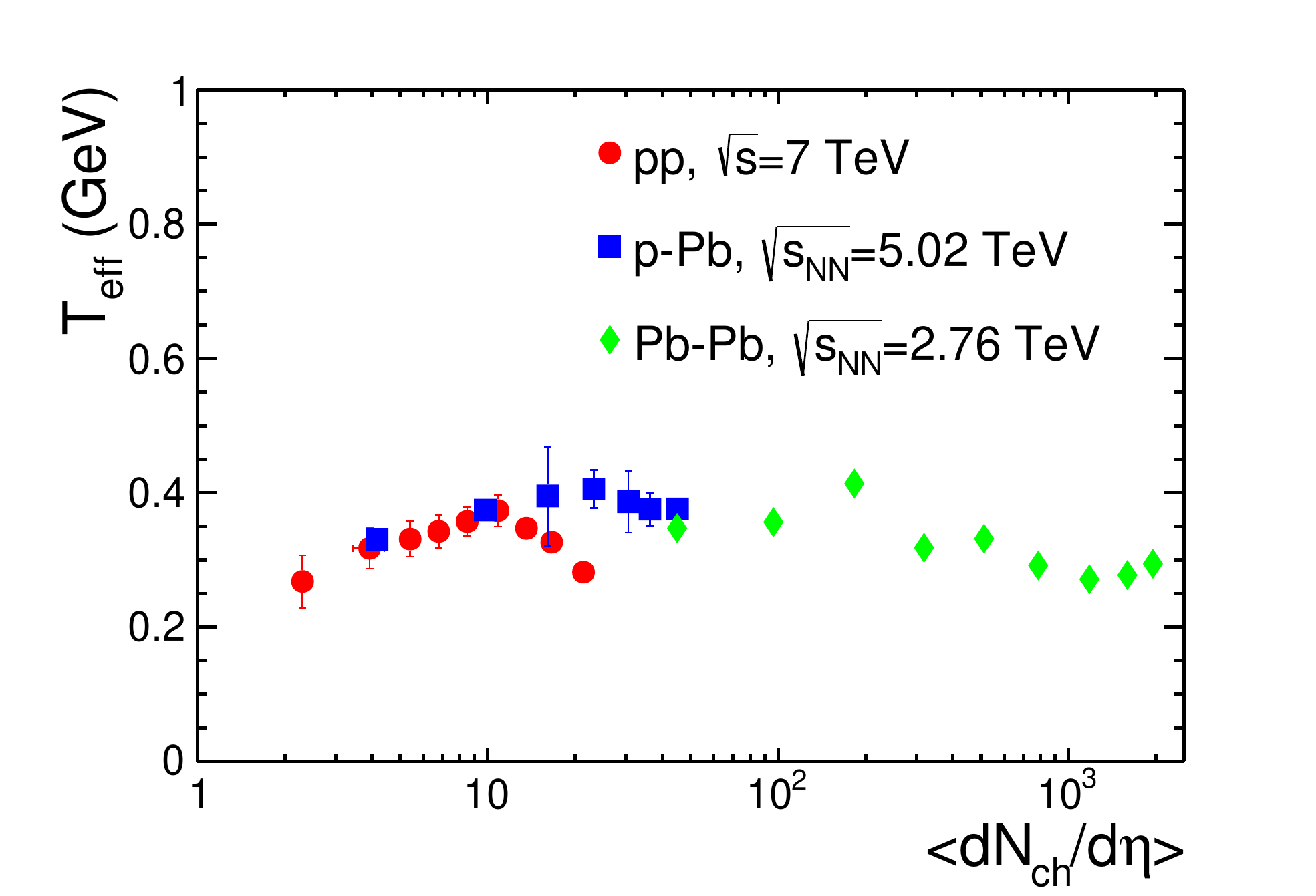}		
		\caption{Kinetic freezeout temperature, radial flow velocity and effective temperature of $\phi$ meson as functions of charged particle multiplicity for pp, p-Pb and Pb-Pb collisions at LHC energies.~\cite{Sahu:2019tch}. 
		}\label{Fig2}
	\end{center}
\end{figure*}

From the BGBW fits, we extract the values of $T_{\rm th}$ and $\langle \beta \rangle$ and have plotted them as functions of charged particle multiplicity in fig. \ref{Fig2}. We observe a sudden decrease and increase in $T_{\rm th}$ and $\langle \beta \rangle$ respectively at $\langle dN_{\rm ch}/d\eta \rangle$ $\sim$ 10 - 20. This can be because, for low charged-particle multiplicity, the system freezes out early. It means the system freezes out at high $T_{\rm th}$. However, as the charged-particle multiplicity increases, the system is supposed to have gone through a QGP phase, which results in the system taking a long time to attain the kinetic freeze-out, thus a lower $T_{\rm th}$. This is also the reason for a higher average velocity of the system after certain charged-particle multiplicity \cite{Sahu:2019tch}.
Moreover, as $\phi$ meson keeps the information of the QGP phase boundary intact, from the right-hand side panel of fig. \ref{Fig2}, we can observe that $T_{\rm eff}$, and the location of the QGP phase boundary, in turn, is independent or weakly dependent on charged-particle multiplicity. This observation is supported by earlier reports of the chemical freeze-out temperature independent of final-state charged-particle multiplicity.

\subsection{Summary}
In summary, we have presented a possible way to estimate the hadronic phase lifetime by considering a nuclear decay formula-like toy model and studied the hadronic phase lifetime as a function of final state event multiplicity. The hadronic phase lifetime is found to be strongly dependent on the charged particle multiplicity and the collision energy. In addition, we have also made an attempt to locate the QGP phase boundary by taking $\phi$ meson as our probe.

\section{First deep learning based estimator for elliptic flow in heavy-ion collisions}
\author{N. Mallick, S. Prasad, A. N. Mishra, R. Sahoo, and G. G. Barnaf\"oldi}	

\bigskip

\begin{abstract}
We propose deep learning techniques such as the feed-forward deep neural network (DNN) based estimator to predict elliptic flow ($v_2$) in heavy-ion collisions at RHIC and LHC energies. A novel method is designed to process the final state charged particle kinematics information as input to the DNN model. The model is trained with Pb-Pb collisions at $\sqrt{s_{\rm NN}} = 5.02$ TeV minimum bias events simulated with a multiphase transport model (AMPT). The trained model is successfully applied to estimate centrality and transverse momentum ($p_{\rm T}$) dependence of $v_2$ for both RHIC and LHC energies. A noise sensitivity test is also performed to estimate the systematic uncertainty of this method by adding the model response to uncorrelated noise. Results of the proposed estimator are compared to both simulation and experiment, which confirms the model's accuracy.
\end{abstract}


\subsection{Introduction}
Relativistic heavy-ion collisions have been studied extensively in experiments at the Large Hadron Collider (LHC), CERN, Switzerland, and Relativistic Heavy Ion Collider (RHIC), BNL, USA for quite some time now and have established the fact that an extremely hot and dense state of strongly interacting matter is produced in such collisions. This thermalized deconfined state of matter is known as the quark-gluon plasma (QGP). One of the crucial signatures for QGP is the presence of transverse collective exapansion~\cite{Heinz:2013th}. This transverse flow is anisotropic following the almond-shaped nuclear overlap region in noncentral heavy-ion collisions. The anisotropic flow of different orders can be expressed as the Fourier expansion coefficients of the produced final state particles' azimuthal momentum distribution as follows~\cite{Voloshin:1994mz}.

\begin{eqnarray}
\frac{{\rm d}N}{{\rm d}\phi} = \frac{1}{2\pi}\Bigg[1+\sum_{n=1}^{\infty}2v_n \cos{\left[n(\phi-\psi_n)\right]}\Bigg].
\label{eq1}
\end{eqnarray}

Here, $v_n = \langle \cos{[n(\phi-\psi_n)]}\rangle$ is the nth order harmonic flow coefficient, $\phi$ is the azimuthal angle and $\psi_n$ is the nth harmonic symmetry plane angle~\cite{Voloshin:1994mz}. Existence of finite anisotropic flow, mainly the elliptic flow ($v_2$) in heavy-ion collisions is observed in experiments~\cite{STAR:2003wqp,ALICE:2010suc}. For the first time, we propose a machine learning based deep learning estimator for elliptic flow in heavy-ion collisions~\cite{Mallick:2022alr}. Machine learning algorithms are well known for their capabilities to exploit features and correlation from data for mapping complex nonlinear functions. In recent works~\cite{Biro:2021zgm,Pang:2016vdc,Xiang:2021ssj}, deep neural network based studies are quite successful in heavy-ion physics. The motivation of this work is to prepare a deep learning framework to estimate $v_2$ from final state charged particle kinematics information and also learn the centrality and transverse momentum dependence of $v_2$ for RHIC and LHC energies. For this study, we have used AMPT with string melting mode (AMPT version 2.26t9b)~\cite{Lin:2004en} to obtain the required data sets.

\subsection{Deep learning estimator}

The primary input to the model comes from the binned ($\eta-\phi$) distribution of all the charged particles in an event. Additionally, these ($\eta-\phi$) layers are weighted with transverse momentum ($p_{\rm T}$), mass of the charged particles, and a term related to collision energy, \textit{i.e.}, $\log(\sqrt{s_{\rm NN}/s_0})$ separately.  $\sqrt{s_0} = 1~ \rm GeV$ makes $\sqrt{s_{\rm NN}/s_0}$ unit-less. The grid size is $32\times32$ for each layer, which makes the total number of input features 3072. All charged particles with transverse momentum cut, $0.2 < p_{\rm T}<5.0$ GeV/c in pseudorapidity, $|\eta|<0.8$ are considered for the training of the minimum bias regression model.  The input is normalized using the {\it L2-Norm} to make a more meaningful representation for the training algorithm. This step is crucial as it helps faster convergence of the regression estimator and keeps the model's coefficients small. 



\begin{figure}[ht!]
\centering
\includegraphics[scale=0.32]{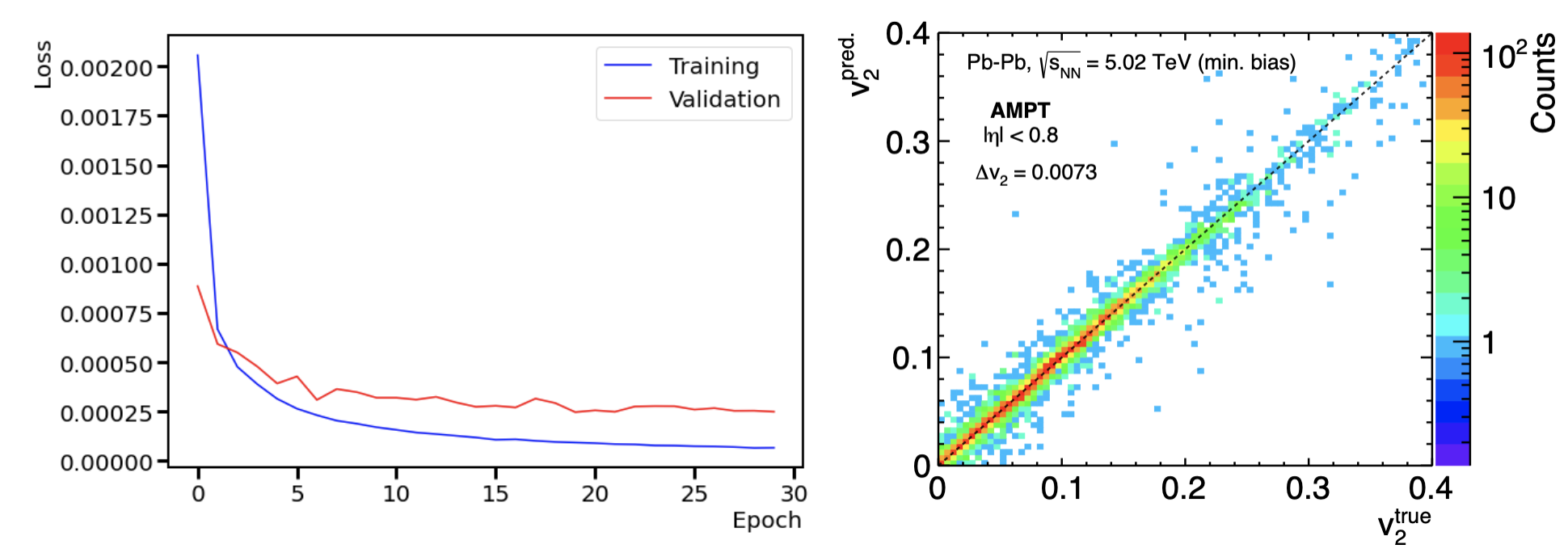}
\caption{(Color online) Left: The evaluation of loss function against the epochs during the training and validation runs. Right: Prediction of $v_2$ from the DNN estimator versus the true values from simulation for 10K minimum bias events of Pb-Pb collisions at $\sqrt{s_{\rm NN}} = 5.02$ TeV. (Fig. 72 and 73~\cite{Mallick:2022alr})}
\label{fig1}
\end{figure}

The feed-forward deep neural network used in this work has four dense layers having 128-256-256-256 nodes each. All the layers have ReLU activation. The output layer has a single node as $v_2$ with linear activation. The role of activation function is to bring nonlinearity to the network. The network is trained with Pb-Pb collisions at $\sqrt{s_{\rm NN}} = 5.02$ TeV minimum bias events using the \textit{adam} optimizer with mean-squared-error (MSE) loss function~\cite{Mallick:2022alr}. The model is trained with a maximum of 60 epochs with a fixed batch size of 32. An early stopping callback is used with a patience level of 10 epochs to reduce overfitting. In Fig.~\ref{fig1}, the left plot shows the performance of the DNN model during the training and validation runs as a function of epochs. The right plot shows the comparison of the model prediction with the true value of $v_2$. The model seems to perform quite nicely when subjected to a noise sensitivity test by adding uncorrelated noise to the feature space~\cite{Mallick:2022alr}. The systematic uncertainty is estimated by taking the MAE from this exercise for a given centrality bin.

\subsection{Results and discussion}

The trained model is successfully applied to predict the centrality dependence of $v_2$ for Pb-Pb collisions at $\sqrt{s_{\rm NN}} = 5.02, 2.76$ TeV and Au-Au collisions at $\sqrt{s_{\rm NN}} = 200$~GeV as shown in Fig.~\ref{fig2}. The experimental results from ALICE~\cite{ALICE:2016ccg} and PHENIX~\cite{PHENIX:2018lfu} are also shown. There is a good agreement between AMPT values and DNN predictions, as seen in the bottom ratio plots. In Fig.~\ref{fig2}, the quadratic sum of statistical and systematic uncertainty is shown in a solid red band in the upper panel. In contrast, on the bottom panel, a solid band and a dashed band are used separately for the statistical and systematic uncertainty, respectively. By training the model with minimum bias Pb-Pb collisions at $\sqrt{s_{\rm NN}} = 5.02$ TeV, we allow the machine to learn physics for a larger and more complex system. The DNN model is thus shown to preserve the centrality and energy dependence of $v_2$~\cite{Mallick:2022alr}. 

\begin{figure}[ht!]
\centering
\includegraphics[scale=0.37]{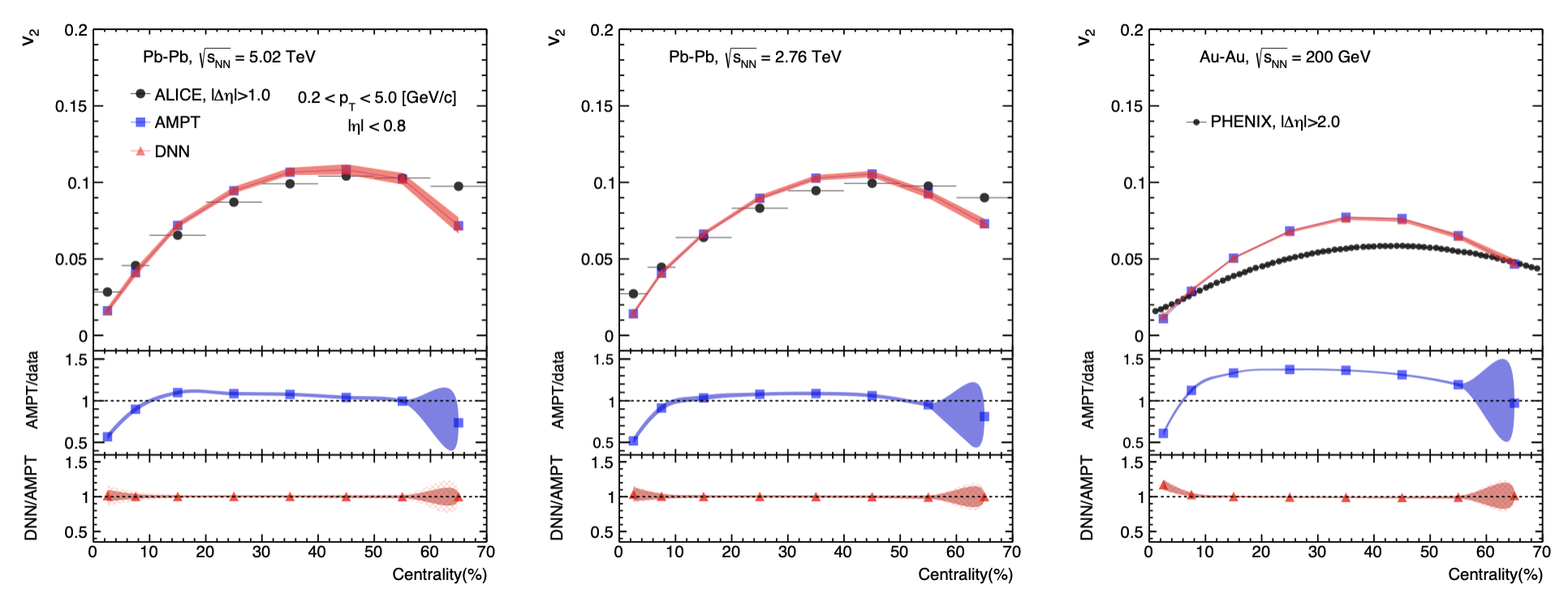}
\caption{(Color online) Prediction of $v_2$ versus centrality in Pb-Pb collisions at $\sqrt{s_{\rm NN}} = 5.02$~TeV, $2.76$~TeV and Au-Au collisions at $\sqrt{s_{\rm NN}} = 200$~GeV from AMPT using the DNN estimator. Experimental results are added for comparison. (Fig. 6~\cite{Mallick:2022alr})}
\label{fig2}
\end{figure}

\subsection{Summary}
We propose a DNN-based estimator for elliptic flow in heavy-ion collisions at RHIC and LHC energies. The model is trained with Pb-Pb collisions at $\sqrt{s_{\rm NN}} = 5.02$ TeV minimum bias events simulated with AMPT by taking input from particle kinematics information. The model is shown to preserve the centrality, energy, and transverse momentum dependence of $v_2$~\cite{Mallick:2022alr}. When subjected to uncorrelated noise added to the simulation, the model seems to preserve the prediction accuracy up to a reasonable extent. Current work is being taken up for testing and implementing this model with experimental data.


\section{Thermoelectric response in a thermal QCD medium with chiral quasiparticle masses}
\author{Debarshi Dey and Binoy Krishna Patra}	

\bigskip

\begin{abstract}
The lifting of the degeneracy between $L$- and $R$-modes
of massless flavors in a weakly magnetized 
thermal QCD medium leads to a novel phenomenon of chirality dependence of 
the thermoelectric tensor, whose diagonal and non-diagonal elements are 
the Seebeck and Hall-type Nernst coefficient, respectively. 
Both coefficients in $L$-mode have 
been found to be greater than their counterparts in $R$-mode, however the disparity
is more pronounced in the Nernst coefficient.
\end{abstract}

\subsection{Introduction}

The deconfined hot QCD medium created in relativistic 
heavy ion collision experiments may be exposed to magnetic fields ($B$) arising from non-central nucleus-nucleus collisions\cite{Skokov:2009qp,tuchin:PRC82'2010}.  
Depending on the time-scale of evolution, magnetic field could be strong 
or weak. Whereas a strong $B$ provides the ground 
for probing the topological properties of QCD 
vacuum\cite{Kharzeev:2007jp,fukushima:PRL120'2018}, a weak 
$B$ leads to some novel phenomenological consequences, such as 
the lifting of degeneracy between left and right handed quarks. Our aim 
is to explore the consequence of this mass splitting of chiral quasiparticle modes on the thermoelectric 
response of the medium.  

In the weak $B$ regime ($m_{0}^2\ll eB\ll T^2$), the thermoelectric 
response of the thermal QCD medium is quantified by the Seebeck ($S$)
and Nernst ($N|\bm{B}|$) coefficients, which, respectively are
the measures of induced electric fields in the longitudinal and 
transverse directions with respect to the direction 
of temperature gradient.  Large fluctuations in the initial energy density in the 
heavy-ion collisions\cite{Schenke:PRL108'2012} translate to significant temperature gradients between 
the central and peripheral regions of the produced fireball, providing the ideal ground to study thermoelectric phenomena.

\subsection{Dispersion Relations for quarks in weak $B$: Chiral modes}
The dispersion relation of quarks is obtained from the zeros of the inverse
resummed quark propagator, which is related to the bare propagator and the self energy via the Dyson-Schwinger equation:
\begin{equation}
S^{-1} (P)= S^{-1}_0 (P) - \Sigma (P),
\end{equation}
 The one loop quark self energy is then given by
\begin{align*}\label{self_energy}
\Sigma(P) =  g^2 C_F T\sum_n\int\frac{d^3k}{(2\pi)^3}\gamma_{\mu}\Bigg(\frac{\slashed{K}}{(K^2-m_f^2)} - \frac{\gamma_5[(K.b)\slashed{u}-(K.u)\slashed{b}]}{(K^2-m_f^2)^2}(|q_fB|)\Bigg)\gamma^{\mu}\frac{1}{(P-K)^2},
\end{align*}
with $u^{\mu}= (1,0,0,0)$ and $b^{\mu}=(0,0,0,1)$ being the fluid four-velocity in the rest frame of the heat bath and the direction of magnetic field, respectively.
 The full propagator (and the self energy) can eventually  be expressed in terms of 
projection operators $P_L =(\mathbb{I}-\gamma_5)/2$ and $P_R =(\mathbb{I}+\gamma_5)/2$ as

\vspace{-7mm}
\begin{equation}\label{prop_eff}
S(P) = \frac{1}{2}\Big[ P_L\frac{\slashed{L}}{L^2/2}P_R+\frac{1}{2}P_R\frac{\slashed{R}}{R^2/2}P_L\Big],
\end{equation}
where, $L$ and $R$ are combinations of structure constants that appear in the general expression of self energy in a tensor basis\cite{Karmakar:EPJC79'2019}. The $p_0=0$, ${\bf{p}}\rightarrow 0$ limit of the denominator of the effective propagator yields the quasiparticle masses as\cite{Das:PRD97'2018,Pushpa:PRD97'2022}
\vspace{-7mm}
\begin{align}\label{mass}
&m_L^2=\frac{L^2}{2}\arrowvert_{p_0= 0, {|\bf{p}|}\rightarrow 0} = m_{th}^2 + 4g^2 C_F M^2,\\
&m_R^2=\frac{R^2}{2}|_{p_0= 0, {|\bf{p}|}\rightarrow 0} = m_{th}^2 - 4g^2 C_F M^2,
\end{align}
thus lifting the degeneracy. Here,
\vspace{-2mm}
\begin{align}
M^2 &= \frac{|q_fB|}{16\pi^2}\left(\frac{\pi T}{2m_f} -\text{ln} 2+ \frac{7\mu^2\zeta(3)}{8\pi^2T^2}\right),\\
m_{th}^2&=\frac{1}{8}g^2C_F\left(T^2+\frac{\mu^2}{\pi^2}\right).
\end{align}

\subsection{The thermoelectric coefficients}
 We make use of the Boltzmann transport equation to 
calculate the infinitesimal deviation of the single particle distribution function from equilibrium caused by the temperature gradient. 
\begin{equation}
p^{\mu}\frac{\partial f_i(x,p)}{\partial 
	x^{\mu}}+ q_iF^{\mu\nu}p_{\nu}\frac{\partial f_i(x,p)}{\partial p^{\mu}}=\left(\frac{\partial f_i}{\partial t}\right)_{\text{coll}}
\label{BTE}
\end{equation}
The highly non linear collision integral on the R.H.S. can be linearized using the relaxation time 
approximation which reads (suppressing the flavor index $i$)
\begin{equation}
\left(\frac{\partial f}{\partial t}\right)_{\text{coll}}\simeq -\frac{p^{\mu}u_{\mu}}{\tau}\delta f=-\frac{f-f_0}{\tau}=-\frac{\delta f}{\tau},
\end{equation}
where, $\tau$ is the relaxation time\cite{Hosoya:NPB250'1985} and $f_0$ is the Fermi-Dirac distribution 
function. $\delta f$ is then used to calculate the induced current which is set to zero (enforcing 
the equilibrium condition) to evaluate the relevant response functions\cite{Callen} 
(Seebeck and Nernst coefficients). 

We use the following Ansatz\cite{Feng:PRD96'2017} to solve for $\delta f$ from Eq.\eqref{BTE}
\vspace{-1mm}
\begin{equation}
f^{L/R}=f_0^{L/R}-\tau q\bm{E}\cdot\frac{\partial f_0^{L/R}}{\partial \bm{p}}-\bm{\chi}.\frac{\partial f_0^{L/R}}{\partial \bm{p}}\label{ansatz},
\end{equation}

\vspace{-2mm}
where, the effect of $\mathbf{B}$ is encoded in $\bm{\chi}$. 
\vspace{-1mm}
The induced 4-current of the system consisting of $u$ and $d$ quarks is given by:
\vspace{-1mm}
\begin{equation}
J^{\mu}=\sum_{a=u,d}q_ag_a\int \frac{d^3\mbox{p}}{(2\pi)^3\epsilon_a}p^{\mu}\left[\delta f_a-\overline{\delta f_a}\right],\label{current_def}
\end{equation}

\vspace{-1mm}
where, $\overline{\delta f}$ denotes the contribution from antiparticles. For $\bm{\nabla T}=\frac{\partial T}{\partial x}\bm{\hat{x}}+\frac{\partial T}{\partial y}\bm{\hat{y}}$ the equilibrium condition $J_x=J_y=0$ finally yields:
\begin{equation}
\begin{pmatrix}
E_x\\
E_y
\end{pmatrix}=\begin{pmatrix}
S & N|\bm{B}|\\
-N|\bm{B}| & S
\end{pmatrix}\begin{pmatrix}
\frac{\partial T}{\partial x}\\[0.3em]
\frac{\partial T}{\partial y}
\end{pmatrix},
\end{equation}

with
\vspace{-1mm}
\begin{align*}
S=&-\frac{\sum\limits_{a=u,d}\!\!\!(C_1)_a\cdot \!\!\!\sum\limits_{a=u,d}\!\!\!(C_3)_a+\!\!\sum\limits_{a=u,d}\!\!\!(C_2)_a\cdot\! \!\!\sum\limits_{a=u,d}\!\!\!(C_4)_a}{\left(\sum\limits_{a=u,d}(C_1)_a\right)^2+\left(\sum\limits_{a=u,d}(C_2)_a\right)^2},\\[0.3em]
N|\bm{B}|&=\frac{\sum\limits_{a=u,d}\!\!\!(C_2)_a\cdot\! \!\!\sum\limits_{a=u,d}\!\!\!(C_3)_a-\!\!\sum\limits_{a=u,d}\!\!\!(C_1)_a\cdot\! \!\!\sum\limits_{a=u,d}\!\!\!(C_4)_a}{\left(\sum\limits_{a=u,d}(C_1)_a\right)^2+\left(\sum\limits_{a=u,d}(C_2)_a\right)^2}.
\end{align*}

\vspace{-1.9mm}
where,
\vspace{-2mm}
\begin{align*}
C_{1/2}=q&\int \mbox{dp}\,p^4 \frac{\tau(\omega_c\tau)^{0/1}}{\epsilon^2(1+\omega_c^2\tau^2)}\big\{f_0(1-f_0)
\pm\bar{f_0}(1-\bar{f_0})\big\},\\
C_{3/4}=\beta& \int \mbox{dp}\,p^4\frac{\tau (\omega_c\tau)^{0/1}}{\epsilon^2(1+\omega_c^2\tau^2)}\big\{\pm(\epsilon+\mu)\bar{f_0}(1-\bar{f_0})-(\epsilon-\mu)f_0(1-f_0)\big\}.
\end{align*}
\subsection{Results}


\begin{figure}[htb]
\centering
\begin{minipage}{.5\textwidth}
  \centering
  \includegraphics[scale=0.25]{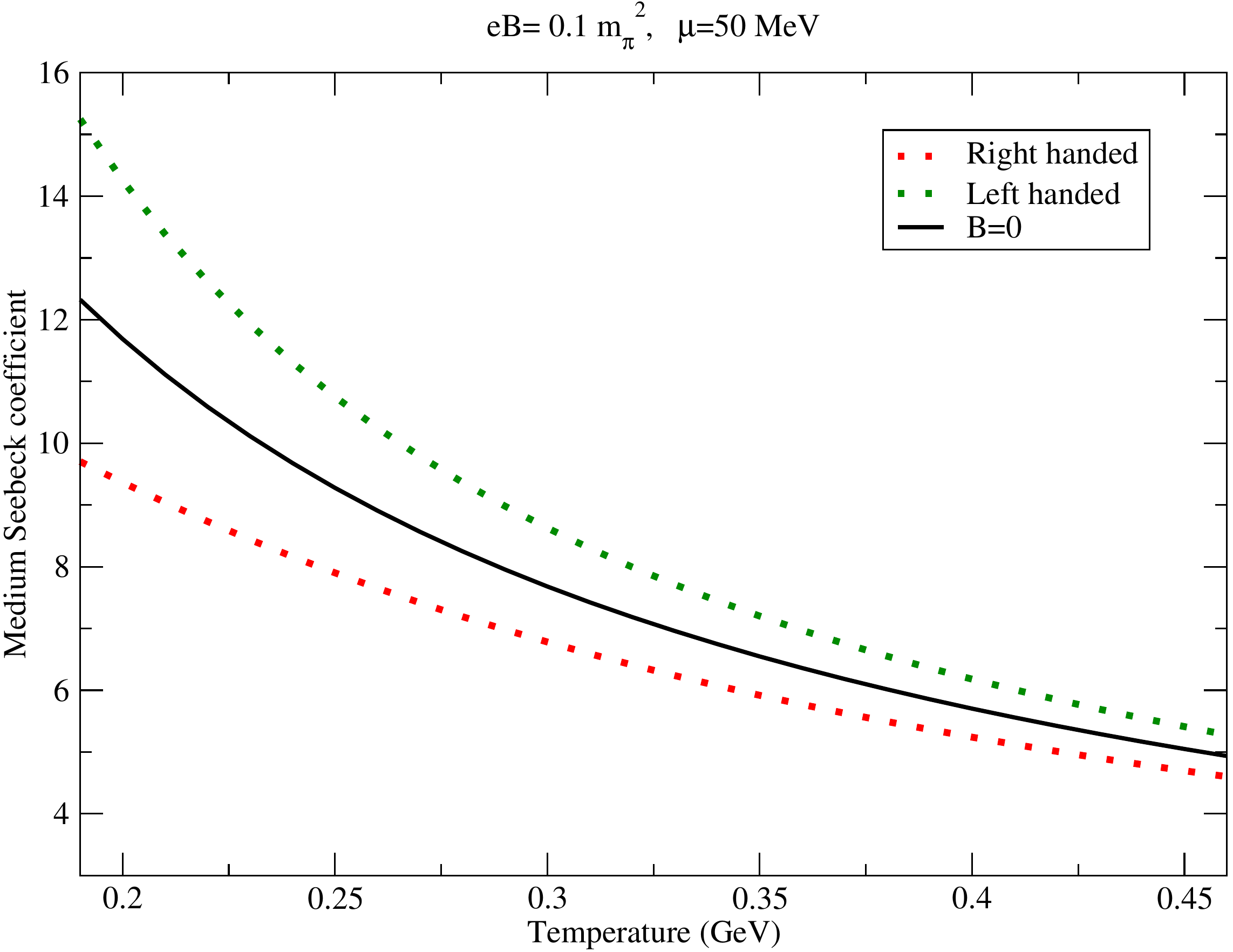}
 \label{seebeck_med_p}
\end{minipage}%
\begin{minipage}{.5\textwidth}
  \centering
  \includegraphics[scale=0.25]{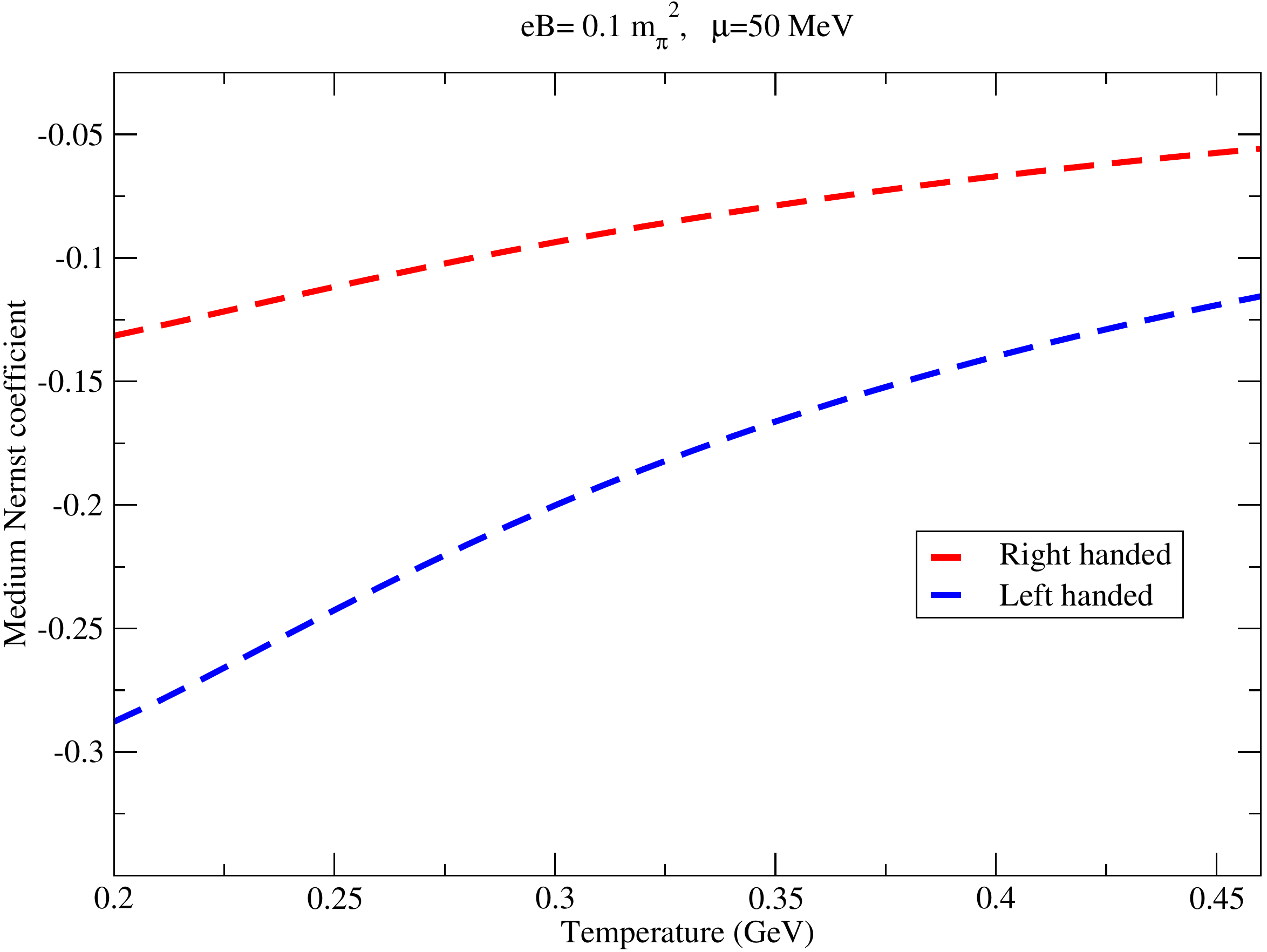}
\label{nernst_med}
\end{minipage}
\vspace{-1mm}
\caption{Variation of Seebeck (left) and Nernst
	coefficients (right) with
	temperature.}\label{seebeck_med_p}
\end{figure}
\vspace{-4mm}
Figures \eqref{seebeck_med_p} left and \eqref{seebeck_med_p} right show the variation of Seebeck and Nernst coefficients 
of the medium with temperature. The coefficient magnitudes show 
similar trends as far as variation with temperature is considered; for both the modes, the magnitudes 
decrease with temperature. Also, for both the coefficients, the $L$ mode elicits a larger comparative 
response.
\vspace{-4mm}
\begin{table}[H]
\begin{center}
	\begin{tabular}{c c c}
		Coefficient & \multicolumn{2}{c}{\% change between $L$ and $R$ modes}\\
		\hline \hline
		& Max & Min\\
		\hline
\textbf{Seebeck} & 57.1 & 13.9\\
\textbf{Nernst} & 118.6 & 105.7\\
\hline
	\end{tabular}
\vspace{-2mm}
\caption{Comparative response strength of $L$ and $R$ modes.}
\label{table}
\end{center}
\end{table}
The difference between the $L$ and $R$ mode responses is amplified significantly for the Nernst coefficient, as can be seen from Table \ref{table}. Also, the temperature sensitivity of the Seebeck coefficient is much greater than that of the Nernst coefficient. For comparison, the $B=0$ case is also shown 
for the Seebeck coefficient. The Nernst coefficient is, however zero for $B=0$, as should be the case.
\section{Charge and heat transport in hot quark matter with chiral dependent quark masses}
\author{Pushpa Panday and Binoy Krishna Patra}	

\bigskip

\begin{abstract}
The generation of nondegeneracy in the mass of left-handed (L) and right-handed (R) chiral modes of quarks is noticed in a weak magnetic field (B), which is in contrast to the case of the strong magnetic field. Therefore, we have studied the impact of nondegenerate mass on longitudinal and Hall components of charge and heat transport coefficients.
\end{abstract}



\vspace{-4mm}
\subsection{Introduction}
The generation of fast decaying and strong magnetic field during the initial stage of heavy-ion collisions \cite{k_tuchin} brings the new aspects for the study of quark-gluon plasma. The lifetime of magnetic field gets enhanced due to the charge properties of the medium \cite{Tuchin, Mclerran, tuchin}. Therefore, the study of transport coefficients in weak magnetic field is of great interest, where these coefficients serve as input parameters for the hydrodynamical study of QGP. Furthermore, the lifting up of degeneracy in mass of chiral modes of quarks in the ambience of weak magnetic field can give more insight to the study of medium properties of QGP.
\vspace{-4mm}
\subsection{Effective Mass of quark in weakly magnetized thermal medium}
The effective quark mass for $f$th 
flavor in presence of magnetic field can be written in terms of current quark mass and medium generated mass as \cite{bannur}
\begin{equation}
	m^2_f = m_{f0}^2 + \sqrt{2}m_{f0}m_{fth,B}+m_{fth,B}^2.
\end{equation}
 $m_{fth,B}$ can be obtained by taking the static limit of denominator of the dressed quark propagator in magnetic 
field. The inverse of dressed quark propagator using Schwinger-Dyson can be written in terms of bare inverse propagator $(S^{-1}(P))$ and quark self energy $(\Sigma(P))$ as
\begin{equation}\label{inverse_prop}
	S^{*-1}(P) = S^{-1}(P) - \Sigma(P).
\end{equation}
 The one-loop quark self energy upto $\mathcal{O}(q_fB)$ 
in hot and weakly magnetized medium can be written as
\vspace{-3mm}
\begin{align}\label{self_energy}
	\Sigma(P) =  g^2 C_F T\sum_n\int\frac{d^3k}{(2\pi)^3}\gamma_{\mu}\Bigg(\frac{\slashed{K}}{(K^2-m_{f0}^2)}
	-& \frac{\gamma_5[(K.b)\slashed{u}-(K.u)\slashed{b}]}{(K^2-m_{f0}^2)^2}(q_fB)\Bigg)\times&\nonumber\\
	\gamma^{\mu}\frac{1}{(P-K)^2}.
\end{align}
We will employ the general covariant structure of quark self energy \cite{Das:PRD97'2018} in terms of chiral projection operator to obtain the effective quark propagator as
\begin{equation}
	S^*(P) = \frac{1}{2}\left[ P_L\frac{\slashed{L}}{L^2/2}P_R+P_R\frac{\slashed{R}}{R^2/2}P_L\right].
\end{equation}
The static limit ($p_0 =0,|{\bf{p}}|\rightarrow 0$) of $L^2/2$ and $R^2/2$ will give the different medium generated mass for L and R mode thus lifting up the degeneracy in mass.
\begin{align}\label{qaurk_mass}
	&m_{L}^2 = m_{th}^2 + 4g^2 C_F M^2,\\
	&m_{R}^2 = m_{th}^2 - 4g^2 C_F M^2.
\end{align}

\subsection{Charge and heat transport coefficients}
\subsection{Ohmic and Hall conductivity}
The Boltzmann transport equation governs the evolution
of phase space density $f(x,p)$ associated with partons in our system. The relativistic Boltzmann transport equation (RBTE) in presence of external electromagnetic field ($F^{\mu\nu}$) is \cite{landau}
\begin{equation}\label{1}
	p^{\mu}\partial_{\mu}f(x,p) + q F^{\mu\nu}p_{\nu}\frac{\partial f(x,p)}{\partial p^{\mu}} = C[f],
\end{equation}
where, $f$ is 
the distribution function deviated slightly 
from equilibrium distribution function ($f_0$) with $f = f_0 + \delta f$ ($\delta f<<f_0$). $C[f]$ is the collision integral whose general form consists of absorption and
emission terms in phase space volume element. This leads to complicated nonlinear integro-differential equation which can be solved easily under relaxation-time approximation (RTA).  To solve RBTE for static and spatially uniformed medium, we take the following \textit{ansatz} of $f(p)$ as \cite{Feng:PRD96'2017}
\begin{equation}\label{6}
	f(p) = f_0 - \tau q{\bf{E}}.\frac{\partial f_0}{\partial{\bf{p}}} - {\bm{\xi}}.\frac{\partial f_0}{\partial{\bf{p}}},
\end{equation}
and hence $f(p)$ for quarks simplifies to
\begin{equation}\label{13}
	f(p) = f_0 - \frac{qEv_x\tau}{(1+ \omega_c^2\tau^2)}\left(\frac{\partial f_0}{\partial\varepsilon}\right) + \frac{qEv_y\omega_c\tau^2}{(1 +\omega_c^2\tau^2)}\left(\frac{\partial f_0}{\partial\varepsilon}\right).
\end{equation}
Similarly, we can solve for antiquarks. The induced current due to external electromagnetic field is $j^i  = \sigma_{\text{Ohmic}}\delta^{ij}E_j + \sigma_{\text{Hall}}\epsilon^{ij}E_{j}$, where $\sigma_{\text{Ohmic}}$ is the response along the direction of electric field and $\sigma_{\text{Hall}}$ is the transverse response to the electric field. Further, the induced current can be written in terms of deviation $\delta f$ and $\delta\bar{f}$ as
\begin{equation}\label{64}
	{\bf{j}} = g_f\int\frac{d^3{p}}{(2\pi)^3}{\bf{v}} \left(q\delta f(p) + \bar{q}\delta\bar{f}(p)\right).
\end{equation}
Hence, $\sigma_{\text{Ohmic}}$ and $\sigma_{\text{Hall}}$ comes out to be
\begin{equation}\label{15}
	\sigma_{{\text{Ohmic}}} = \frac{1}{6\pi^2T}\sum_fg_fq^2_{f}\tau_f\int dp\frac{p^4}{\varepsilon_f^{2}}\frac{1}{(1 +\omega_c^2\tau_f^2)}[f^0_f(1-f^0_f)+\bar{f^0_f}(1-\bar{f^0_f})],
\end{equation}
\vspace{-3mm}
\begin{equation}\label{16}
	\sigma_{{\text{Hall}}} = \frac{1}{6\pi^2T}\sum_fg_fq^2_{f}\tau_f^2\int dp\frac{p^4}{\varepsilon_f^{2}}\frac{\omega_c}{(1 +\omega_c^2\tau_f^2)}[f^0_f(1-f^0_f)-\bar{f^0_f}(1-\bar{f^0_f})],
\end{equation}
\subsection{Thermal and Hall-type thermal conductivity}
We will express the RBTE in terms of gradients of flow velocity and temperature under RTA as
\begin{align}\label{23}
	p^{\mu}\partial_{\mu}T\left(\frac{\partial f}{\partial T}\right) + 
	& p^{\mu}\partial_{\mu} (p^{\nu}u_{\nu})\left(\frac{\partial f}{\partial p^0}\right)
	+q\Bigg(F^{0j}p_{j}\frac{\partial f}{\partial p^0}+ F^{j0}p_{0}\frac{\partial f}{\partial p^{j}} + \nonumber \\
	& F^{ij}p_{j}\frac{\partial f}{\partial p^{i}}+F^{ji}p_{i}\frac{\partial f}{\partial p^{j}}\Bigg)= -\frac{p^{\mu}u_{\mu}}{\tau}\delta f.
\end{align}
We will choose the \textit{ansatz} for $\delta f$ as $({\bf{p}}.\bm{\chi})\frac{\partial f_0}{\partial\varepsilon}$, where $\bm{\chi}$ is related to thermal driving forces and magnetic field in the medium. 
The spatial heat flow in terms of $\delta f$ and $\delta\bar{f}$ can be written as
\begin{align}\label{20}
	{\bf{Q}} = \sum_f g_f \int\frac{d^3p}{(2\pi)^3}\frac{{\bf{p}}}{\varepsilon_f}\left[\left(\varepsilon_f - h_f\right)\delta f_f + \left(\varepsilon_f + \bar{h}_f\right)\delta\bar{f}_f\right]
\end{align}
Comparing Eq.\eqref{20} with ${\bf{Q}} = -\kappa_0 T{\bf{Y}} - \kappa_1 T({\bf{Y\times c}})$, where ${\bf{Y}} = \frac{\bm{\nabla}T}{T}-\frac{\bm{\nabla}P}{nh}$ and ${\bf{c=\frac{B}{|B|}}}$, thermal ($\kappa_0$) and Hall-type thermal conductivity ($\kappa_1$) obtained to be as
\begin{equation}\label{41}
	\kappa_0 = \sum_f\frac{g_f\tau_f}{6\pi^2 T^2}\int dp\frac{p^4}{\varepsilon_f^2}\Bigg[\frac{(\varepsilon_f - h_f)^2}{(1+\omega_c^2\tau_f^2)}f_f^0(1-f_f^0) + \frac{(\varepsilon_f + \bar{h}_f)^2}{(1+\omega_c^2\tau_f^2)}\bar{f}_f^0(1-\bar{f}_f^0)\Bigg] ,
\end{equation}
\begin{equation}\label{42}
	\kappa_1 = \sum_f\frac{g_f\tau_f^2}{6\pi^2 T^2}\int dp\frac{p^4}{\varepsilon_f^2}\Bigg[\frac{(\varepsilon_f - h_f)^2\omega_c}{(1+\omega_c^2\tau_f^2)}f_f^0(1-f_f^0) - \frac{(\varepsilon_f + \bar{h}_f)^2\omega_c}{(1+\omega_c^2\tau_f^2)}\bar{f}_f^0(1-\bar{f}_f^0)\Bigg].
\end{equation}
\subsection{Results and Discussion}
\begin{figure}[htb]
\centering
\begin{minipage}{.5\textwidth}
  \centering
  \includegraphics[scale=0.25]{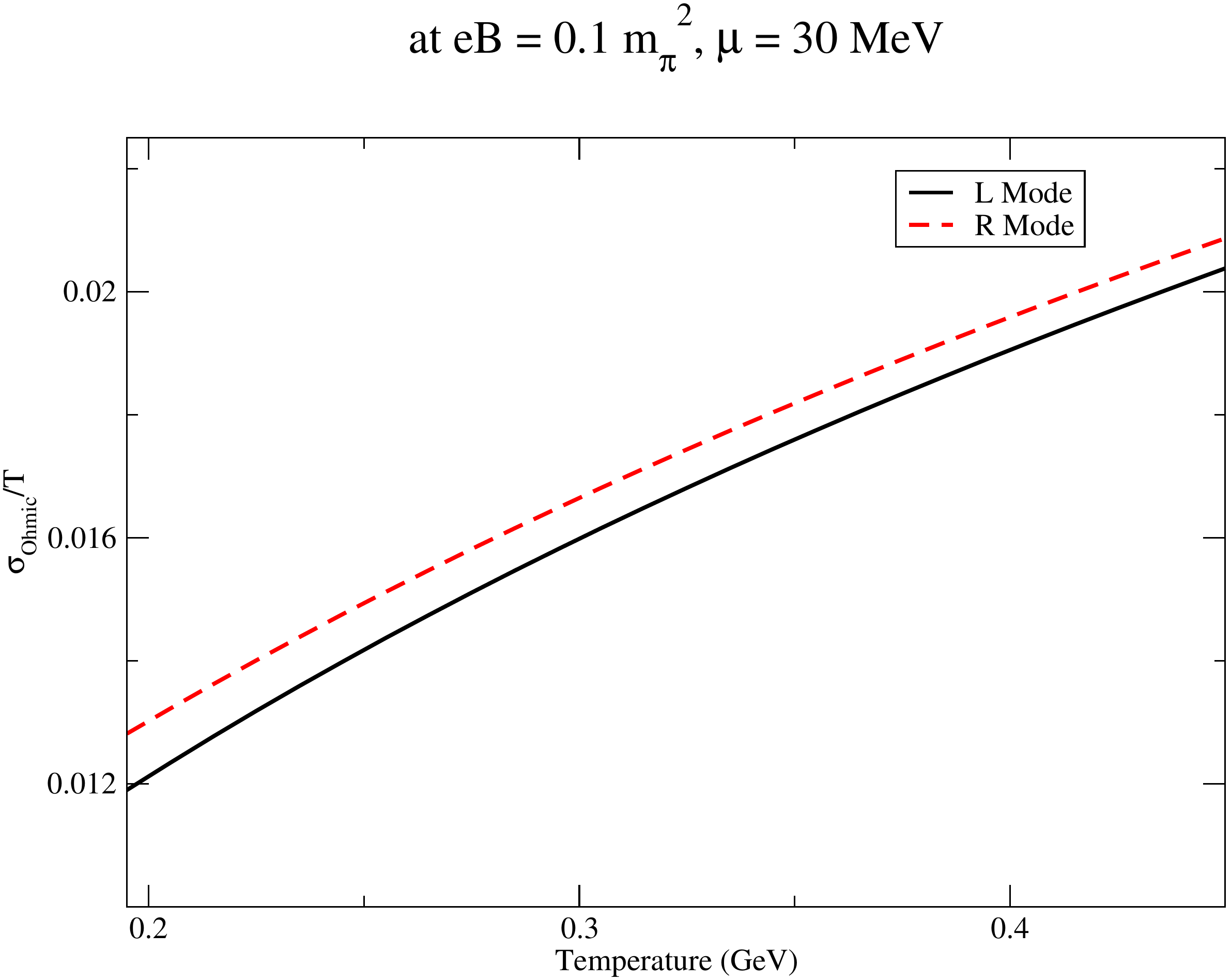}
  \label{ohmic}
\end{minipage}%
\begin{minipage}{.5\textwidth}
  \centering
  \includegraphics[scale=0.25]{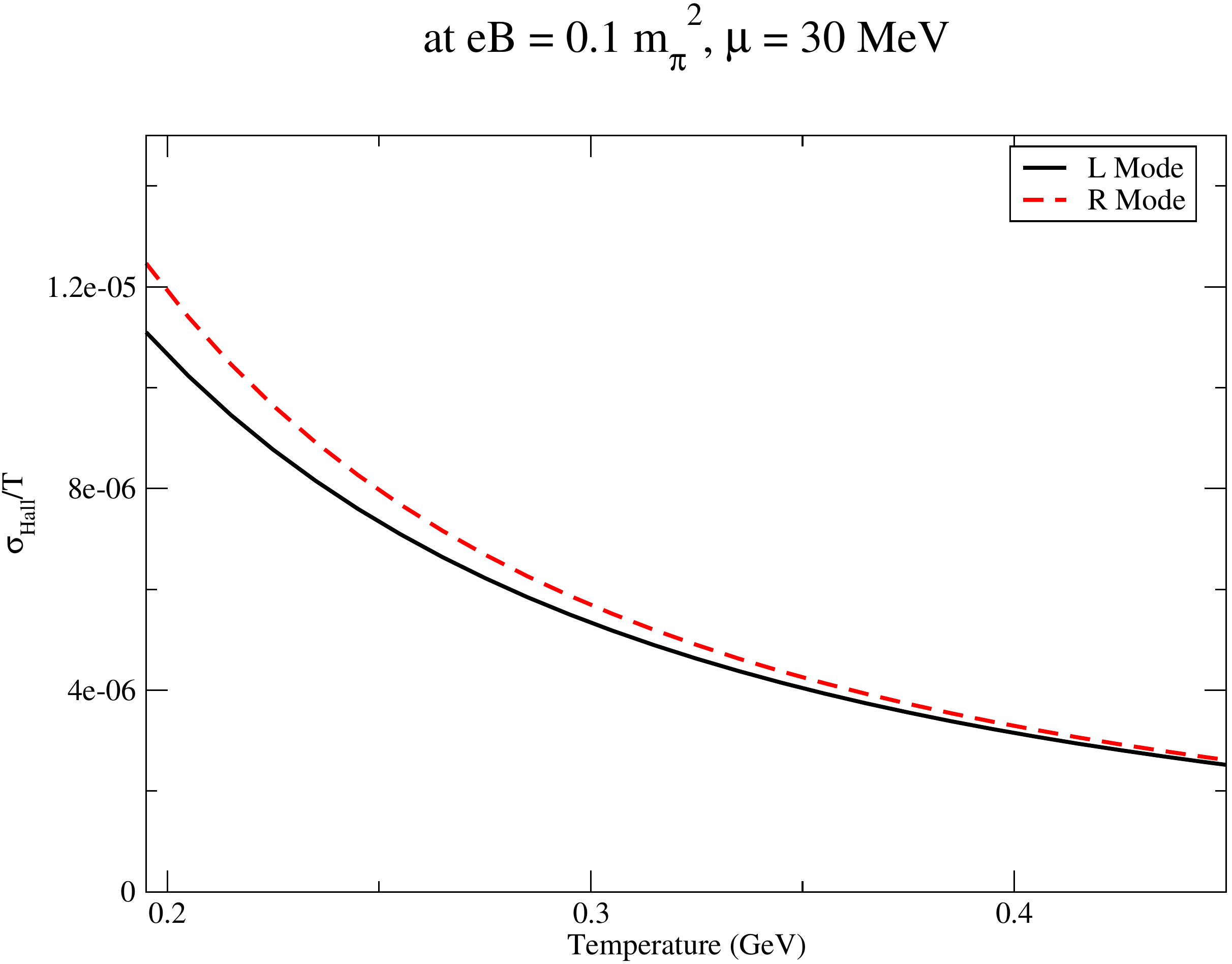}
  \label{hall}
\end{minipage}
\caption{Variation of $\sigma_{{\text{Ohmic}}}/T$ (a) and $\sigma_{\text{Hall}}/T$ (b) with temperature.
		}\label{charge}
\end{figure}
\vspace{-2mm}
\begin{figure}[htb]
\centering
\begin{minipage}{.5\textwidth}
  \centering
  \includegraphics[scale=0.25]{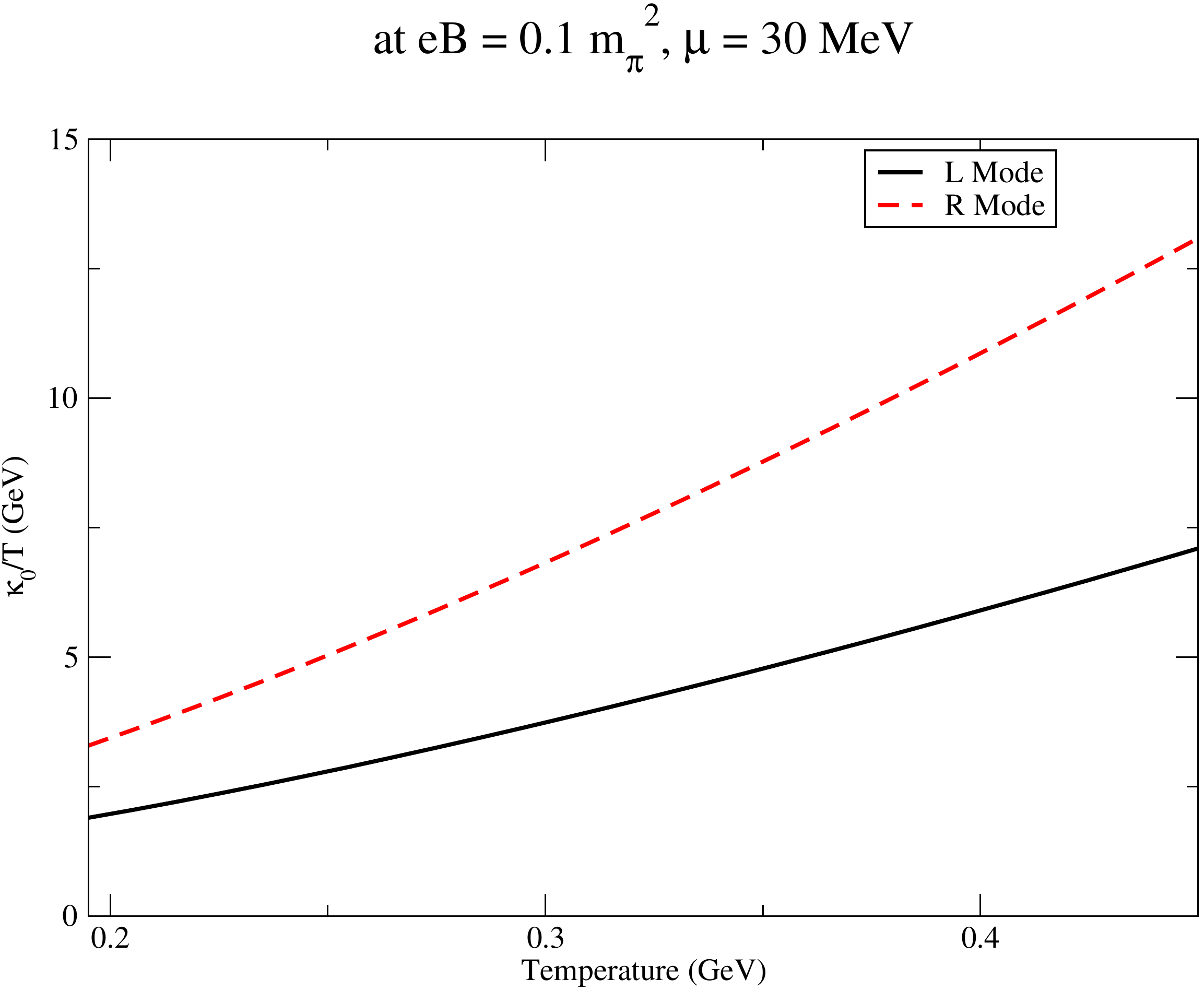}
\label{ohmic}
\end{minipage}%
\begin{minipage}{.5\textwidth}
  \centering
  \includegraphics[scale=0.25]{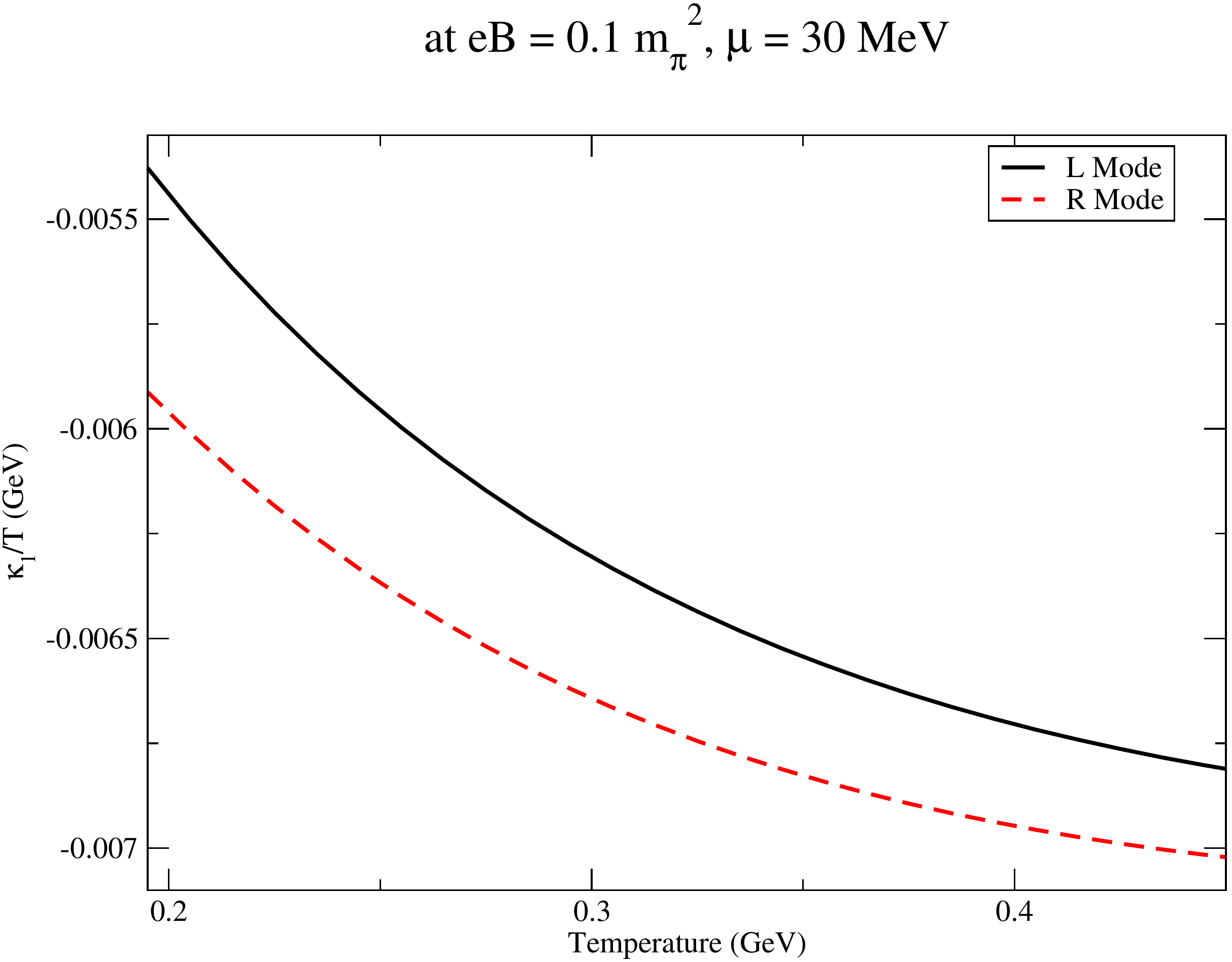}
  \label{hall}
\end{minipage}
\caption{Variation of $\kappa_0/T$ (a) and $\kappa_1/T$ (b) with temperature.
	}\label{heat}
\end{figure}
\vspace{2mm}
The magnitude of R mode $\sigma_{{\text{Ohmic}}}/T$ and $\sigma_{{\text{Hall}}}/T$ is higher than L mode as shown in Fig. \eqref{charge}. This can be attributed to the different mass of L and R mode. The normalized Ohmic conductivity increases with temperature due to the Boltzmann factor in the distribution function whereas normalized Hall conductivity decreases with temperature due to the factor $\omega_c\tau$ in the numerator of Eq.\eqref{16}. Similarly, the variation of $\kappa_0/T$ and $\kappa_1/T$ with temperature is shown in Fig. \eqref{heat} where magnitude for R mode is higher than L mode. $\kappa_0/T$ increases with temperature due to the factor $(\varepsilon-h)^2$, $(\varepsilon+h)^2$ and distribution function in Eq.\eqref{41}. Considering the absolute value of the ratio $\kappa_1/T$, we infer that $\kappa_1/T$ also increases with temperature due to $(\varepsilon+h)^2$ factor in the numerator of Eq.\eqref{42}. Moreover, Hall component for both charge and heat transport will vanish for zero magnetic field.




\section{NLO quark self-energy and dispersion relation using the hard thermal loop resummation}
\author{Sumit, Najmul Haque, and Binoy Krishna Patra}	

\bigskip

\begin{abstract}
Using the hard-thermal-loop (HTL) resummation in real-time formalism, we study the next-to-leading order (NLO) quark self-energy and corresponding NLO dispersion laws. We calculate the momentum integrals in the transverse part of the NLO quark self-energy numerically and plot them as a function of the ratio of momentum and energy. Using that, we plot the transverse contribution of NLO dispersion laws.
\end{abstract}

\subsection{Introduction}

The standard perturbative loop expansion in quantum chromodynamics (QCD) encounters several issues at finite temperatures. One of those issues is that the physical quantities, for example, dispersion laws, become gauge-dependent. Another critical point is that Debye screening in the chromoelectric mode does occur in the leading order in the one-loop calculation, but chromomagnetic screening does not.\cite{Gross:1980br}.

The issue related to the gauge dependence of the gluon damping rates, which have been calculated up to one-loop order, particularly at zero momentum, in different gauges and schemes, has been studied, and different outcomes have been obtained~\cite{Kobes:1987bi}. Later, it was concluded that the lowest order result is not complete, and higher-order diagrams can contribute to lower orders in powers of the QCD coupling~\cite{Pisarski:1988vd}. 
Braaten and Pisarski developed a systematic theory for an effective perturbative expansion that sums the higher-order terms into effective propagators and effective vertices~\cite{Braaten:1989kk} and is known as hard-thermal loop (HTL) resummation. Using the effective HTL propagators and vertices, the transverse part of the gluon damping rate $\gamma_{t}(0)$ at vanishing momentum was calculated~\cite{Braaten:1990it}, and it was found to be finite, positive, and gauge independent.
		
Using the HTL resummed propagators and vertices, the pressure and quark number susceptibilities up to three-loop order have been studied using the thermodynamic potential\cite{Haque:2014rua}. 
Using the HTL summation, it has been found that massless quarks and gluons acquire the thermal masses of order $gT$, $m_q$, and $m_g$ respectively~\cite{Weldon:1982bn}, which shows that for the lowest order $gT$ in effective perturbation, the infrared region is `okay.' However, the static chromomagnetic field does not screen at the lowest order; instead, it gets screened at the next order, so-called magnetic screening~\cite{Gross:1980br}. Thus, if we want to demonstrate the infrared sector of the HTL perturbative expansion,  we need to go beyond the leading-order calculations.

\subsection{NLO Formalism}\label{Sec:nlo_formalism_}
The lowest order dispersion relation can be summarized in the equation:
\begin{equation}\label{disp_two}
p_0 \mp p - \Sigma_{\pm}(P) = 0 .
\end{equation}
For on shell quarks, we write the (complex) quark energy $p_0 \equiv \Omega(p)$  as
\begin{equation}\label{disp_decom}
\begin{aligned}
\Omega(p)= \Omega^{(0)}(p) + \Omega^{(1)}(p) + \cdots \hspace{2mm} .
\end{aligned}
\end{equation}
A similar kind of approach is also relevant for self-energy $\Sigma$, as well
\begin{equation}\label{sigma_nlo}
\begin{aligned}
\Sigma(P)= \Sigma_\text{HTL}(P)+ \Sigma^{(1)}(P)+\cdots ,
\end{aligned}
\end{equation}
where $\Sigma_\text{HTL}$ is the lowest-order quark self-energy having $gT$ order, whereas $\Sigma^{(1)}$ the NLO contribution of quark self-energy, with order $g^{2}T$. 

\begin{figure}[tbh]
	\centering
	\includegraphics[scale=1,keepaspectratio]{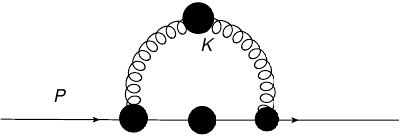}
	\qquad
	\includegraphics[scale=1,keepaspectratio]{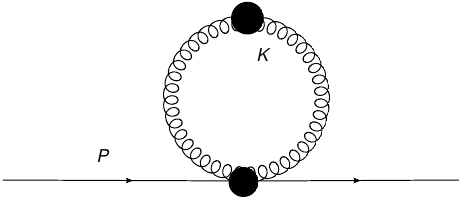}	
\caption{Feynman graphs for the NLO HTL resummed quark self-energy $\Sigma_{\pm}^{(1)} $. The black blobs indicate HTL effective quantities. All momenta are soft.}
	\label{Figure1}
\end{figure}

Thus, eq.~\eqref{disp_two} will take the form as
\begin{equation}
\Omega^{(0)}_{\pm}(p)+\Omega_{\pm}^{(1)}(p) +\cdots= \pm p + \left.\Sigma_{\mathrm{HTL} \pm}\left(\omega, p\right)\right|_{\omega\rightarrow\Omega_{\pm}(p)}+\left.\Sigma_{\pm}^{(1)}\left(\omega, p\right)\right|_{\omega\rightarrow\Omega_{\pm}(p)}+\ldots \hspace{2mm} . 
\end{equation}
After simplification, for slow-moving quarks ($p \sim gT$), we get

\begin{equation}\label{disp_nlo}
\Omega_{\pm}^{(1)}(p)=\frac{{\Omega^{(0)}_{\pm}}^{2}(p)-p^{2}}{2 m_{q}^{2}} \Sigma_{\pm}^{(1)}\left(\Omega^{(0)}_{\pm}(p), p\right) . 
\end{equation}
To calculate the NLO quark self-energy, we have to consider two one-loop graphs with effective vertices, shown in figure~\ref{Figure1}. 

The final compact expressions for the NLO one-loop HTL-summed quark self-energy as\cite{Abada:2014bma}
\ba
\Sigma_{\pm}^{(1)}(P)&=& -\frac{i g^{2} C_{F}}{2} \int \frac{d^{4} K}{(2 \pi)^{4}}\left[F_{\pm ; 0}^{\mathrm{SR}}(P, K)+F_{\pm ; 0}^{\mathrm{AS}}(P, K)+2 F_{\pm ;--}^{\mathrm{SR}}(P, K)+F_{\pm ;--}^{\mathrm{AS}}(P, K)\right.\nn
&+&\left.F_{\pm ;-+}^{\mathrm{AS}}(P, K)+F_{\pm ;--;--}^{\mathrm{SR}}(P, K)+F_{\pm ;--;+-}^{\mathrm{AS}}(P, K)+G_{\pm ;--}^{\mathrm{S}}(P, K)\right].\label{sigma_final}
\ea


\subsection{Evaluation of NLO quark self-energy}

In order to evaluate eq.~\eqref{sigma_final}, we need retarded transverse $D_{T}^{R}(k,k_0,\varepsilon)$, retarded longitudinal $D_{L}^{R}(k,k_0,\varepsilon)$ gluon propagators which we have derived\cite{Sumit:2022snb}. The retarded transverse gluon propagator comes out to  be
\bas
D_{T}^{R{(-1)}}(k,k_0,\varepsilon) &=& -\left[\frac{4k_0^{2}}{k^{2}} + \left(k^2 - k_0^{2}\right) \left\{1 - \frac{k_0}{k^3} \ln \frac{(k_0 - k)^2 + \varepsilon^2}{(k_0+k)^2 + \varepsilon^2}\right\}\right]\nn
&-&  i \left[\frac{2k_{0}}{k^3} \left(k_0^{2} - k^2 \right) \left\{\tan^{-1} \left(\frac{\varepsilon}{k_0 - k}\right) - \tan^{-1} \left(\frac{\varepsilon}{k_0 + k}\right) \right\} - \varepsilon \Theta(k_0) \right],
\eas

Here, we show how the terms in eq.~\eqref{sigma_final} have been evaluated. For example, in the third term of eq.~\eqref{sigma_final}, we will get the following lines of discontinuity.
\ba
k_{0}&=&0 ; \quad k_{0}=\pm k ; \nn
k_0 &=& p_0 \pm \sqrt{p^2+k^2-2pkx} ;   \nn
k&=&k_{t} \equiv \frac{1}{2} \frac{p_{0}^{2}-p^{2}}{p_{0}-x p}=\frac{1}{2 t} \frac{1-t^{2}}{1-x t} \sqrt{\frac{t}{1-t}-\frac{1}{2} \ln \left(\frac{1+t}{1-t}\right)}\label{diverg_pts_1}
\ea


\begin{figure}[tbh]
	\centering
	{\includegraphics[scale=.53,keepaspectratio]{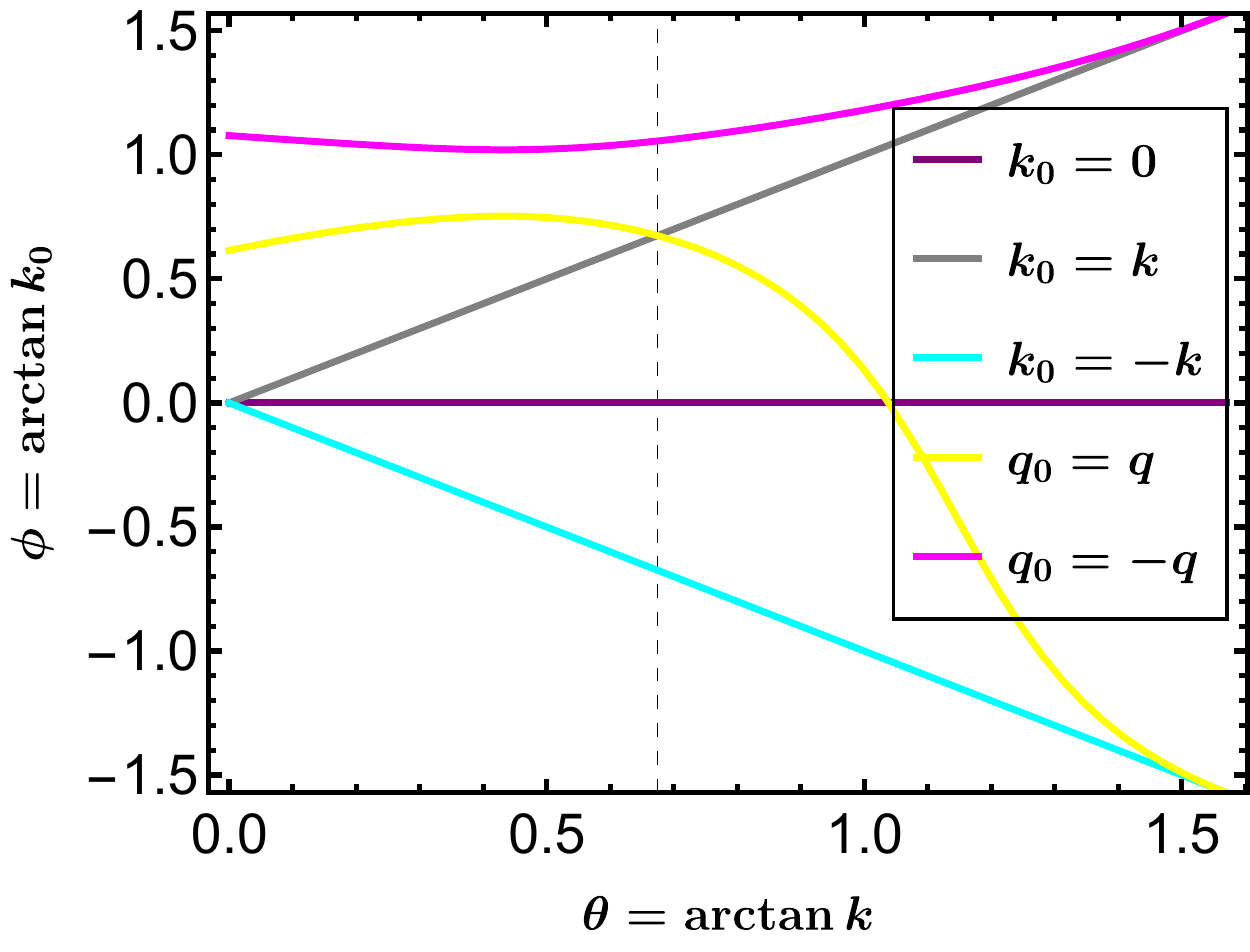}}
\caption[.]{Domains in ($k,k_0$) plane at which the third term of eq.~(\ref{sigma_final}) has sharp jumps. Here we used $ t = \frac{p}{p_{0}} = 0.45 $ and $x = 0.8 $}
\label{fig7}
\end{figure}  

These domains are shown in figure~\ref{fig7}. We numerically evaluated this term in each of the domains of the figure~\ref{fig7} and summed up the results\cite{Sumit:2022snb}. Similar technique has been used to evaluate the other terms\cite{Sumit:2022snb} of eq.~\eqref{sigma_final}.

\subsection{Results and Discussion}
\begin{figure} [tbh]
	\centering
	\includegraphics[scale=0.46,keepaspectratio]{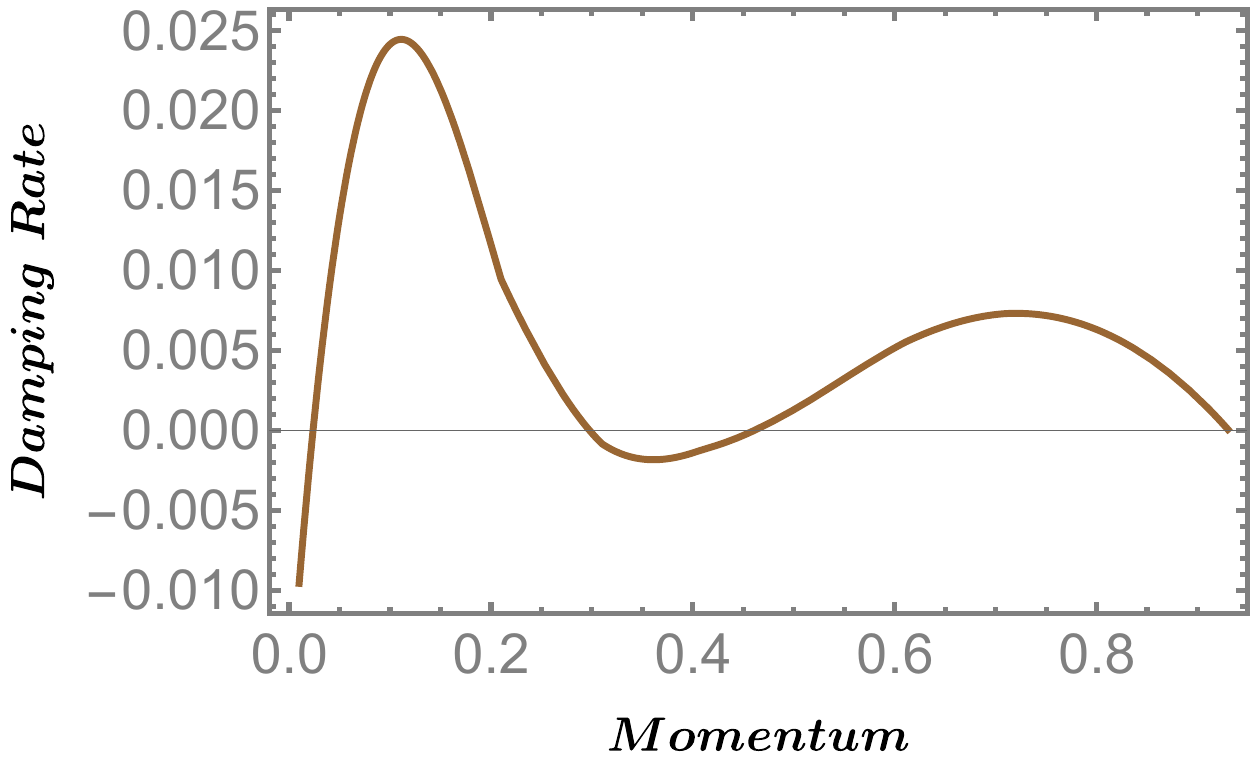} 
	\qquad
	\includegraphics[scale=0.46,keepaspectratio]{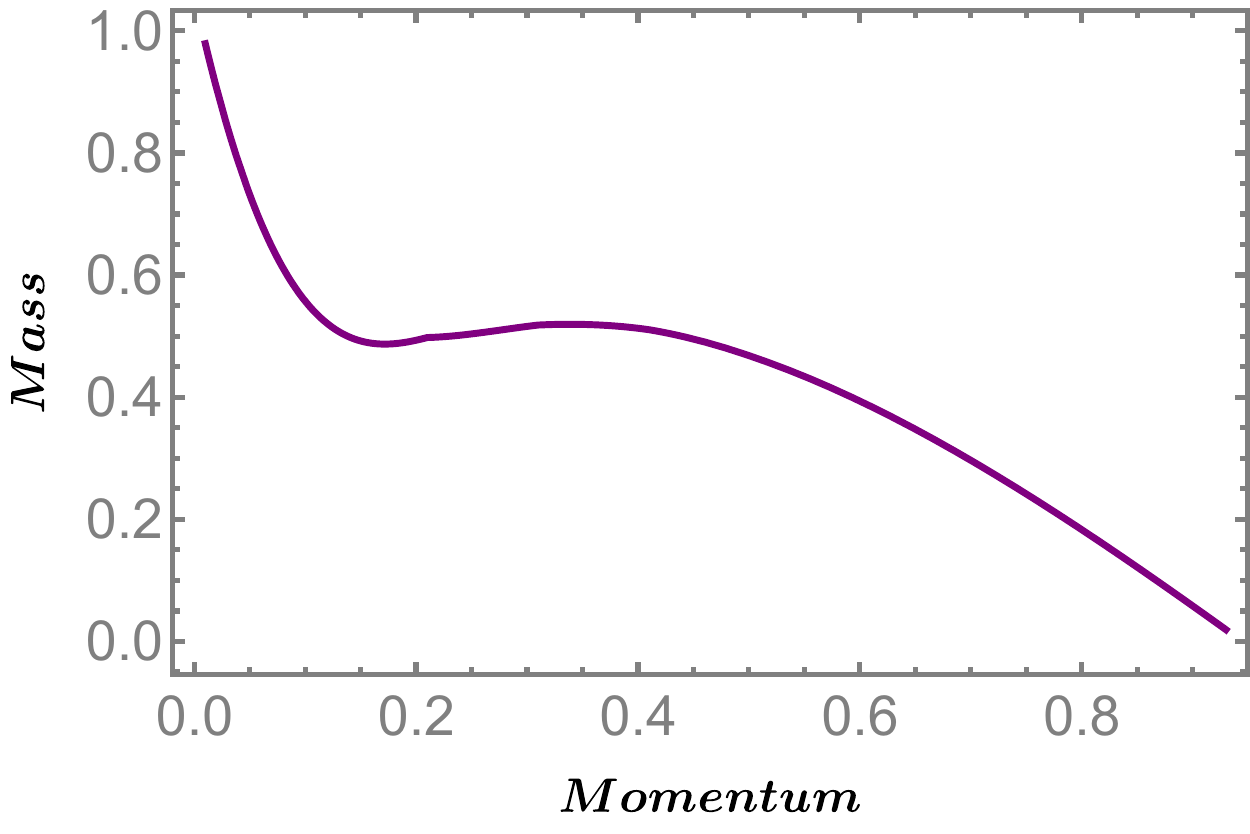}
\caption{Damping rate and quark energy variation with soft momentum $p$ for `+' mode with a coefficient $4\pi g$.}
	\label{fig10}
\end{figure}

\begin{figure} [tbh]
	\centering
	\includegraphics[scale=0.46,keepaspectratio]{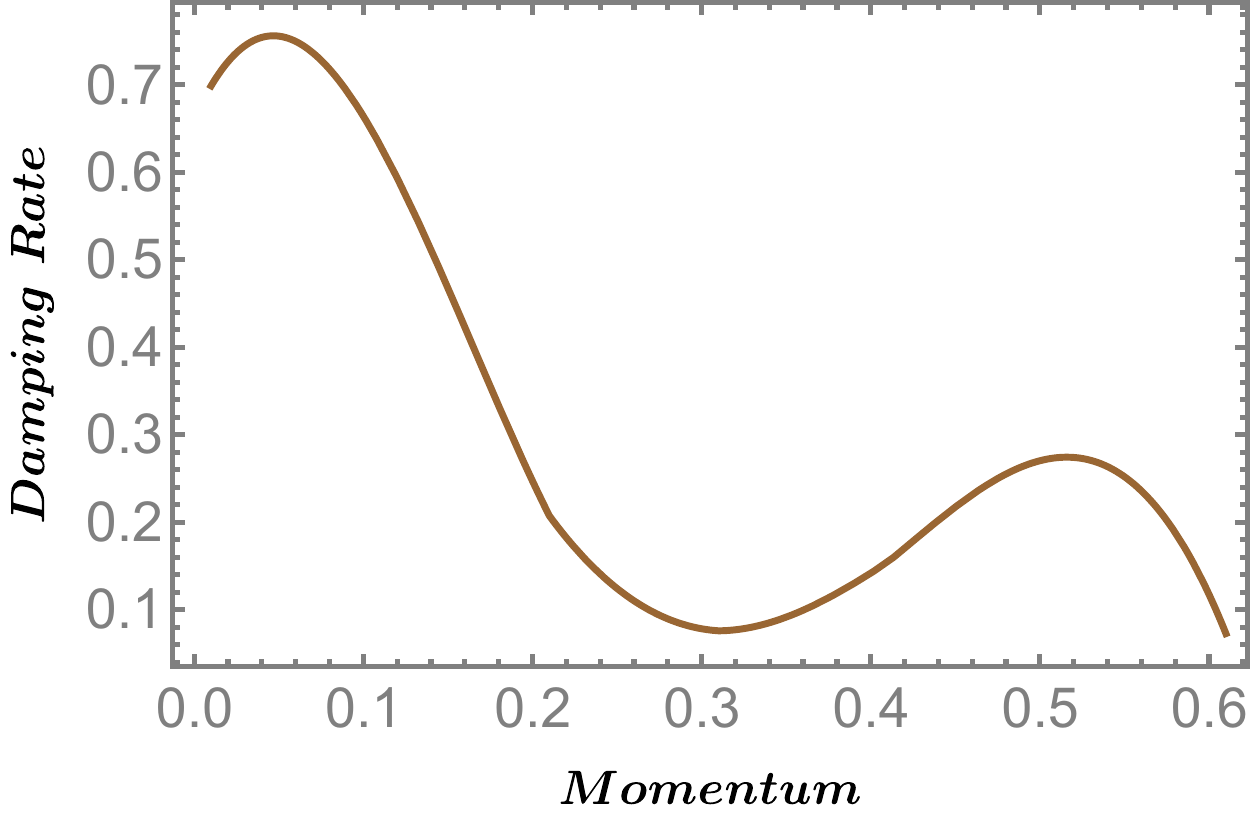} 
	\qquad
	\includegraphics[scale=0.46,keepaspectratio]{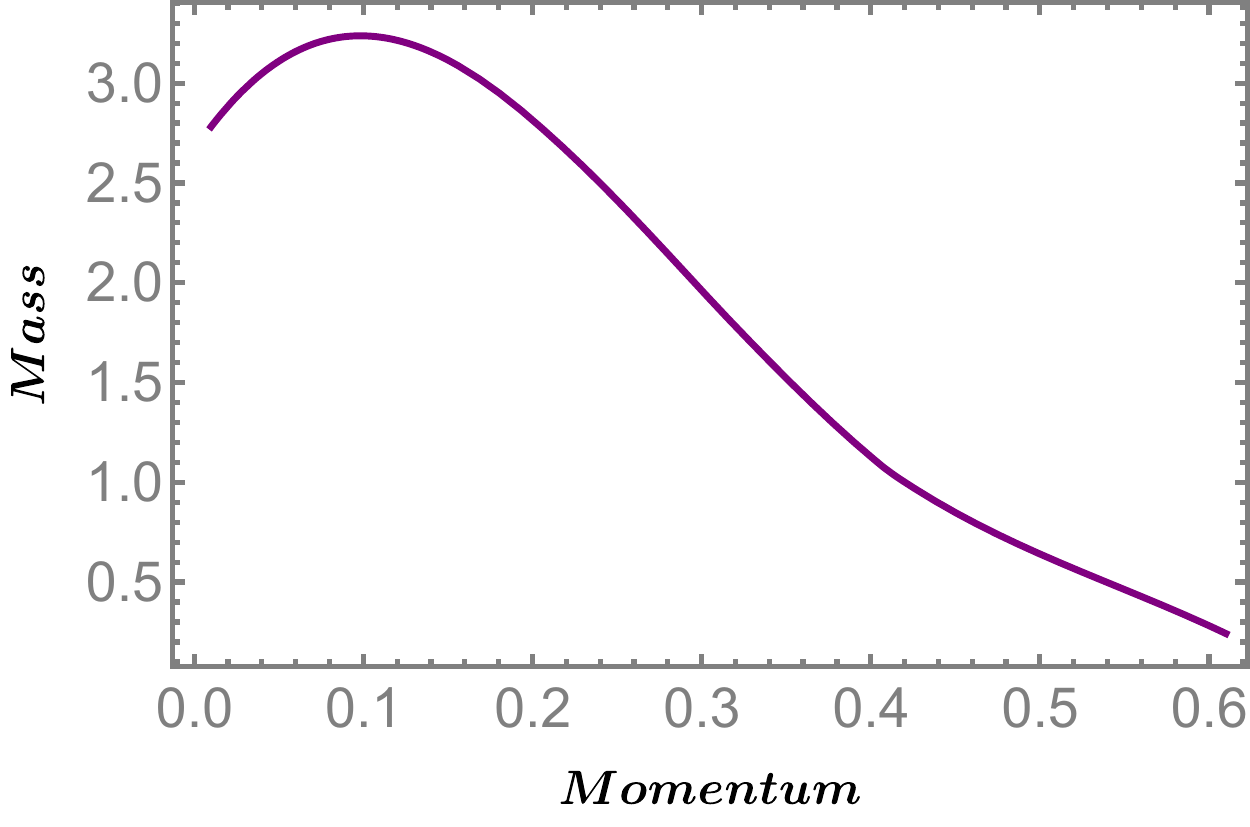}
\caption{Damping rate and quark energy variation with soft momentum $p$ for `-' mode with a coefficient $4\pi g$.}
	\label{fig12}
\end{figure}

All the terms of NLO quark self-energy in eq.~\eqref{sigma_final} have a non-trivial dependence on $\varepsilon $. So, we have checked the stability for each term by plotting them as a function of $m = -\log_{10}\varepsilon.$ This is an essential task because different terms have different stability regions, and if one does the integration beyond those regions, then numerical values lose reliability. After calculating each term's transverse part separately, we added all terms to plot the transverse contribution of NLO quark self-energy w.r.t ratio of momentum and energy. Using eq.~\eqref{disp_nlo}, we plotted the transverse contribution of NLO damping rate and NLO mass for each quark mode shown in Figure~\ref {fig10} and figure~\eqref{fig12}. The important outcome of the results shown is that one can handle the instabilities that arise, at least in part arising from the gluon propagator's transverse component.\cite{Sumit:2022snb} The infrared divergence in the longitudinal part\cite{Sumit:2022snb} can be handled by introducing an infrared cut-off or resumming a specific class of diagrams.

 \section{Overview of experimental results}
 \author{Nihar Ranjan Sahoo}	
 
 \bigskip

 \begin{abstract}
 	In this conference, variety of topics related to the QCD and its hot-dense medium---known as Quark-Gluon Plasma---were discussed. Different experimental results pertaining to the particle production and bulk properties of the medium to the hard probes of the QGP medium were presented. In this proceedings, I summarize succinctly all these results from the LHC and RHIC experiments shown at this conference. 
\end{abstract}

\subsection{Introduction}
In heavy-ion collisions, neutrons and protons inside heavy ions melt, due to extreme temperature and pressure, into a state of matter where the color degrees of freedom plays an important role---this state of matter is known as Quark-Gluon Plasma(QGP). Dedicated heavy-ion experiments at the Large Hadron Collider (LHC) at CERN and Relativistic Heavy-Ion Collider (RHIC) at BNL observed different signatures of QGP created in heavy-ion collisions. In this proceedings, I discuss some recent results presented at this conference.

{\it Hard probes}: The hard probes such as jets and heavy-flavors are produced in the early stages of the heavy-ion collisions and hence carry full space-time evolution of the medium. The J/$\psi$, $\Upsilon$, and D-mesons are measured in the heavy-ion experiments to study the energy loss of heavy-flavors by comparing with its vacuum reference in proton+proton collisions and their collectivity in the QGP medium. The ratio of production yield in nucleus-nucleus collisions and the production cross-section in proton-proton collisions normalized by the average number of nuclear thickness function is known as the nuclear modification factor, $R_{\rm AA}$. The $R_{\rm AA}$ as a function of transverse momentum ($p_{\rm T}$) enables a way to measure the energy loss in heavy-ion collisions. In the ALICE experiment~\cite{ALICE:2021rxa}, a clear difference of $R_{\rm AA}$ between D-mesons, J/$\psi$, and charged pions at low $p_{\rm T}$ suggesting the colour charge and the quark mass dependence of in-medium parton energy loss. The measurement of second order coefficient, $v_{2}$, called elliptic flow, provides information about the heavy quark interaction with the medium and their thermalization. D-mesons show a finite elliptic flow as a function of $p_{T}$ whereas CMS results~\cite{CMS:2019uhg} show zero $\Upsilon$(1S) $v_{2}$ and non-zero J/$\psi$ $v_{2}$ suggesting different medium effects for charmonia and bottomonia. However, different model comparisons suggest that most of the models fail the simultaneous description of D-meson $R_{\rm AA}$ and $v_{2}$ in central and peripheral collisions~\cite{ALICE:2021rxa}. We need a further study to constrain different model parameters, interaction processes, and hadronization mechanisms in the hot-dense QCD medium. \\

Jets are collimated spray of hadrons fragmented from an energetic parton (quark or gluon) originating from a hard scattering. Jet measurement is an important tool to study the energetic parton in heavy-ion and proton-proton collisions. In this conference, jet-fragmentation function in p+p and p+Au collisions were presented at $\sqrt s$=13 TeV and $\sqrt S_{\rm NN}$ = 5.02 TeV, respectively. In $p+p$ collisions, jet fragmentation function is softer in high-multiplicity events than the minimum bias events in the ALICE experiment~\cite{ProttoyDejaniTalk}. A further detailed study is ongoing to investigate the multiplicity dependence of jet production and fragmentation at the LHC. \\

{\it Collectivity in heavy-ion collisions:} Different flow harmonic measurements in heavy-ion collisions measure the collectivity in the QGP. The elliptic flow measurement in the STAR Beam Energy Scan pase-I (BES-I) experiment observes the scaling of $v_{2}$ as a function of $p_{T}$ with the number of constituent quarks (NCQ) for different hadrons~\cite{STAR:2013cow}. This observation concludes that elliptic flow develops in the early stage of heavy-ion collision where partonic degrees of freedom play an important role. The recent BES phase-II high statistics data at $\sqrt S_{\rm NN}$ = 19.6 GeV show the NCQ scaling holds better for anti-particles than for particles~\cite{ChitrasenJena}. The light nuclei and hyper-nuclei $v_{2}$ measurements show that a systematic deviation from mass number scaling of $v_{2}$ ($v_{2}/A$) as a function $p_{\rm T}/A$ where $A$ is the mass number of nuclei. The first order coefficient of the Fourier-expansion of momentum azimuthal distribution, known as the directed flow $v_{1}$, provides sensitive information on early nuclear collisions. In the STAR experiment, observation of hyper-nuclei $^{3}_{\Lambda} H$ and $^{4}_{\Lambda} H$ $v_{1}$ at $\sqrt{ s_{\rm NN}}$ = 3 GeV is seen~\cite{ChitrasenJena}. More detailed studies and analyses for other energies from BES-II are underway in the STAR experiment. \\

{\it Particle production in heavy-ion collisions:} In heavy-ion collisions, different species of particles are produced at both mid- and forward-rapidity. The preliminary result from inclusive photon yield measurement was presented at forward rapidity in $p$+Pb collisions at $\sqrt {S_{\rm NN}}$=5.02 TeV using Photon Multiplicity Detector (PMD) in the ALICE experiment~\cite{AbhiModak}. The pseudorapidity dependence of photon yield are comparable with that of the charged particles at forward rapidity at this collision energy. On the other hand, in the ALICE experiment, the yield ratio of $\Lambda$(1520) over $\Lambda$ as function of average multiplicity for $p+p$ collisions at $\sqrt s$=5.02 and 13 TeV are comparable, and the ratio lies between 0.05-0.07~\cite{SonaliPadhan}. In the STAR experiment, the ratio of $\phi/K^{-}$ and $\phi/\Xi^{-}$ in Au+Au collisions at $\sqrt {S_{\rm NN}}$= 3 GeV from fixed target configuration and compared with different statistical model predictions. The anti-hyper-hydrogen-4 ($^{4}_{\bar \Lambda} H$) state is observed in the STAR experiment by combining all heavy-ion data from Au+Au, (U+U) Ru+Ru, and Zr+Zr collisions at $\sqrt {S_{\rm NN}}$=200 (193) GeV in STAR~\cite{ChitrasenJena}.

{\it Correlations and fluctuations:} In heavy-ion collisions, the two-particle correlation function provides the information about the distribution of separation of emission sources and the final-state interactions. At the LHC and RHIC,  the $\Lambda\Lambda$ correlation measurements where performed. The STAR experiment results and model comparison suggest that the $\Lambda\Lambda$ interaction is weak in nature~\cite{STAR:2014dcy}. However, this contradicts the observation by the recent CMS measurement~\cite{RaghunathPradhan}. 

Search for the QCD critical point is one of the main goals of the BES program at RHIC. The recent net-proton number $\kappa\sigma^{2}$ values as a function of beam energy shows 3.1$\sigma$ non-monotonic variation in central Au+Au collisions~\cite{STAR:2020tga,LokeshKumar}. In addition, the recent STAR measurement at $\sqrt{ s_{\rm NN}}$ = 3 GeV with fixed-target configuration, the net-proton number $\kappa\sigma^{2}$ value is negative (-0.85$\pm$0.09(stat)$\pm$0.82(sys))~\cite{STAR:2021fge}. At this collision energy, the results are consistent with the baryon number conservation. High precision measurements with large acceptance are ongoing using BES-II data for incisive understanding in the direction of QCD critical point search by the STAR collaboration.


{\it Summary and outlook:}
Many results were presented at this conference primarily on the particle production, freeze-out parameters obtained in heavy-ion collsions, correlation measurements, collectivity in heavy-ion collisions and the small system, hard probes, etc. I try to shed light on those results briefly in this proceedings. At its nascent stage, this conference foresees a reassuring future sequel providing a suitable environment and forum for the discussions between young and senior researchers in Goa, India.


%
\newpage
\section*{Acknowledgments}
This writeup is a compilation of the contributions presented at the "Hot QCD Matter 2022 conference" held from May 12–14, 2022, in Goa, India. This first national conference on "HOT QCD Matter 2022" was jointly organized by the School of Physical Sciences, Indian Institute of Technology Goa, and the School of Physical and Applied Sciences, Goa University. This conference's academic activities took place inside the Goa University, situated at the heart of Goa, from 12th to 14th May 2022. This conference attracted 62 participants from different institutes and universities all over India. Prof. B. K. Mishra, Director, IIT Goa, and Prof. H. B. Menon, Vice-Chancellor, Goa University, inaugurated the conference and delighted the event with their inspirational speeches. An energetic and vibrant group of Ph.D. students made this event lively by active contribution to the lively discussion. The scientific program consisted of 24 invited plenary talks by renowned scientists from all over India and 24 contributed talks by Ph.D. students. After each presentation, there was a discussion session to discourse on that topic. At the end of the conference, two overview talks were scheduled to summarize all the presentations and discussions. The six contributed talks (three from experiment and three from theory) presented by the Ph.D. students were facilitated as the best presentations. The conference organizers, Santosh K. Das, Prabhakar Palani, Jhuma Sannigrahi, and Kaustubh R.S. Priolkar, are greatly indebted to the sponsors' generous support: IOP Publishing, Balani Infotech (Library and information services) and San Instruments. The organizers would like to acknowledge all the administrative staff from IIT Goa and Goa University associated with this conference. They worked behind the screen with lots of dedication and enthusiasm to ensure that "HOT QCD Matter 2022" ran smoothly, technically as well as socially. Saumen Datta's research is supported by the Department of Atomic Energy, Government of India, under Project Identification No. RTI 4002. Santosh K. Das and Marco Ruggieri acknowledge the support by the National Science Foundation of China (Grant Nos. 11805087 and 11875153). Santosh K. Das acknowledges the support from DAE-BRNS, India, Project No. 57/14/02/2021-BRNS. Pooja acknowledges IIT Goa and MHRD for funding her research. J. Prakash would like to acknowledge support from IIT Goa and MHRD for funding this project. Debjani Banerjee would like to acknowledge the DST INSPIRE research grant [DST/INSPIRE Fellowship/2018/IF180285] and the ALICE project grant [SR/MF/PS-02/2021-BI (E-37125)], and the computing facility at Bose Institute. Prottoy Das would like to acknowledge the Institutional Fellowship of  Bose Institute, the ALICE project grant [SR/MF/PS-02/2021-BI (E-37125)], and the computing server facility at Bose Institute. Rohan V S would like to acknowledge JETSCAPE collaboration and simulation framework. The support from the DST project No. SR/MF/PS-02/2021-PU (E-37120) is acknowledged by Lokesh Kumar. Sudipan De and Sarthak Satapathy acknowledge financial support from the DST INSPIRE Faculty research grant (IFA18-PH220), India. Sreemoyee Sarkar would like to thank Dr. Rana Nandi for fruitful discussions of the topic. Vaishnavi Desai would like to acknowledge JETSCAPE collaboration, its simulation framework, Goa University for seed money grant support and Param computing facility. Lakshmi J. Naik thanks the Department of Science and Technology, Govt. of India for the INSPIRE Fellowship. Sumit Kumar Kundu would like to acknowledge the financial support provided by the Council of Scientific and Industrial Research (CSIR) (File No. 09/1022(0051)/2018-EMR-I). Cho Win Aung and Thandar Zaw Win deeply thank to DIA-Programme  funded by Ministry of Education, India. Ankit Kumar Panda acknowledges the CSIR-HRDG financial support. Victor Roy acknowledges support from the DAE, Govt. of India and the DST Inspire faculty research grant (IFA-16-PH-167), India. Shubhalaxmi Rath is thankful to the Indian Institute of Technology Bombay for the Institute postdoctoral fellowship. Sonali Padhan and Pritam Chakraborty would like to thank the Department of Science and Technology (DST), Government of India, for supporting the present work. S.P. acknowledges the doctoral fellowship from UGC, Government of India. 
Raghunath Sahoo acknowledges the financial support under the CERN Scientific Associateship and the financial grants under DAE-BRNS Project No. 58/14/29/2019-BRNS. Aditya Nath Mishra and  Gergely G\'abor Barnaf\"oldi gratefully acknowledge the Hungarian National Research, Development and Innovation Office (NKFIH) under the contract numbers OTKA K135515, K123815, and NKFIH 2019-2.1.11-TET-2019-00078, 2019-2.1.11-TET-2019-00050 and Wigner Scientific Computing Laboratory (WSCLAB, the former Wigner GPU Laboratory). The authors gratefully acknowledge the MoU between IIT Indore and WRCP, Hungary, under which this work has been carried out as a part of the techno-scientific international cooperation. Nihar Ranjan Sahoo is supported by the Fundamental Research Funds of Shandong University and National Natural Science Foundation of China, Grant number:12050410235.


\end{document}